\pdfoutput=1

\documentclass[
	11pt, 
	fleqn, 
	a4paper, 
]{LegrandOrangeBook}

\hypersetup{
	pdftitle={Title}, 
	pdfauthor={Author}, 
	pdfsubject={Subject}, 
	pdfkeywords={Keyword1, Keyword2, ...}, 
	pdfcreator={LaTeX}, 
}

\definecolor{ocre}{RGB}{243, 102, 25} 
\definecolor{unipvred}{RGB}{160, 57, 83}

\chapterimage{orange1.jpg} 
\chapterspaceabove{6.5cm} 
\chapterspacebelow{6.75cm} 

\usepackage{physics}
\usepackage{quantikz}
\usepackage{tikz-cd}
\tikzset{operator/.append style={rounded corners = 1mm}}

\usepackage{epigraph}
\usepackage{lipsum}
\usepackage{tikz}
\usepackage{multirow}
\usepackage{mathtools}
\usepackage{bm}
\usepackage{printlen}
\usepackage{xfrac}
\usepackage[normalem]{ulem}

\usepackage[ruled]{algorithm2e}
\SetKwComment{Comment}{>}{}

\usepackage{subcaption}

\usepackage{pgfplots}
\usepackage{xcolor}
\pgfplotsset{compat=newest}
\usepgfplotslibrary{groupplots}
\usepgfplotslibrary{dateplot}

\definecolor{g85}{gray}{0.333}
\definecolor{g170}{gray}{0.666}

\definecolor{g102}{gray}{0.4}
\definecolor{g16}{gray}{0.0636}

\definecolor{g204}{gray}{0.8}
\definecolor{g28}{gray}{0.11}
\definecolor{g40}{gray}{0.16}

\definecolor{g32}{gray}{0.127}
\definecolor{g162}{gray}{0.6366}
\definecolor{g65}{gray}{0.25}
\definecolor{g211}{gray}{0.827}

\definecolor{g242}{gray}{0.949}
\definecolor{g60}{gray}{0.235}
\definecolor{g105}{gray}{0.411}

\definecolor{g17}{gray}{0.066}
\definecolor{g230}{gray}{0.901}
\definecolor{g196}{gray}{0.768}
\definecolor{g110}{gray}{0.431}

\usepackage{amsthm}

\usepackage{amsmath}
\DeclareMathOperator*{\argmax}{arg\,max}
\DeclareMathOperator*{\argmin}{arg\,min}

\newcommand{\vecparam}{\bm{\theta}}
\newcommand{\bmt}{\bm{\theta}}
\newcommand{\bmx}{\bm{x}}
\newcommand{\bmw}{\bm{w}}
\newcommand{\bmo}{\bm{\omega}}
\newcommand{\uparam}{U(\vecparam)}
\newcommand{\ee}[1]{\mathbb{E}\qty[#1]}
\newcommand{\vv}[1]{\text{Var}\qty[#1]}
\newcommand{\hs}[2]{\langle #1, #2 \rangle_{HS}}

\newcommand{\fmap}{\mathcal{F}}
\newcommand{\varans}{V}

\newcommand{\border}[2]{\gategroup[#1,steps=#2,style={dashed,rounded corners,fill=blue!0, inner xsep=0pt, inner ysep = 0pt}, background]{{$\mathbb{I}$}}}

\newcommand{\circuit}[1]{\textsc{C}_{\textsc{#1}}}

\newcommand{\xox}[1]{\sigma_x {#1} \sigma_x}
\newcommand{\yoy}[1]{\sigma_y {#1} \sigma_y}
\newcommand{\zoz}[1]{\sigma_z {#1} \sigma_z}

\newcommand{\sx}{\sigma_x}
\newcommand{\sy}{\sigma_y}
\newcommand{\sz}{\sigma_z}

\makeatletter
\newsavebox{\@brx}
\newcommand{\llangle}[1][]{\savebox{\@brx}{\(\m@th{#1\langle}\)}%
  \mathopen{\copy\@brx\kern-0.5\wd\@brx\usebox{\@brx}}}
\newcommand{\rrangle}[1][]{\savebox{\@brx}{\(\m@th{#1\rangle}\)}%
  \mathclose{\copy\@brx\kern-0.5\wd\@brx\usebox{\@brx}}}
\makeatother

\newcommand{\brows}[1]{%
  \begin{bmatrix}
  \begin{array}{@{\protect\rotvert\;}c@{\;\protect\rotvert}}
  #1
  \end{array}
  \end{bmatrix}
}
\newcommand{\rotvert}{\rotatebox[origin=c]{90}{$\vert$}}
\newcommand{\rowsvdots}{\multicolumn{1}{@{}c@{}}{\vdots}}

\usepackage{nomencl}
\makenomenclature

\usepackage{etoolbox}
\renewcommand\nomgroup[1]{%
  \item[\bfseries
  \ifstrequal{#1}{A}{Abbreviations}{%
  \ifstrequal{#1}{M}{Mathematical symbols}{%
  \ifstrequal{#1}{O}{Other symbols}{}}}%
]}

\renewcommand{\nompreamble}{This list list summarise the abbreviations and symbols that are used throughout the work.}

\DeclareMathAlphabet{\mathcal}{OMS}{cmsy}{m}{n}

\begin{document}

\thispagestyle{empty} 

\begin{tikzpicture}[remember picture, overlay]
    \node (bg) [opacity = 0.6, inner sep=0pt, yshift = -2cm] at (current page.center) {\includegraphics[width=\paperwidth]{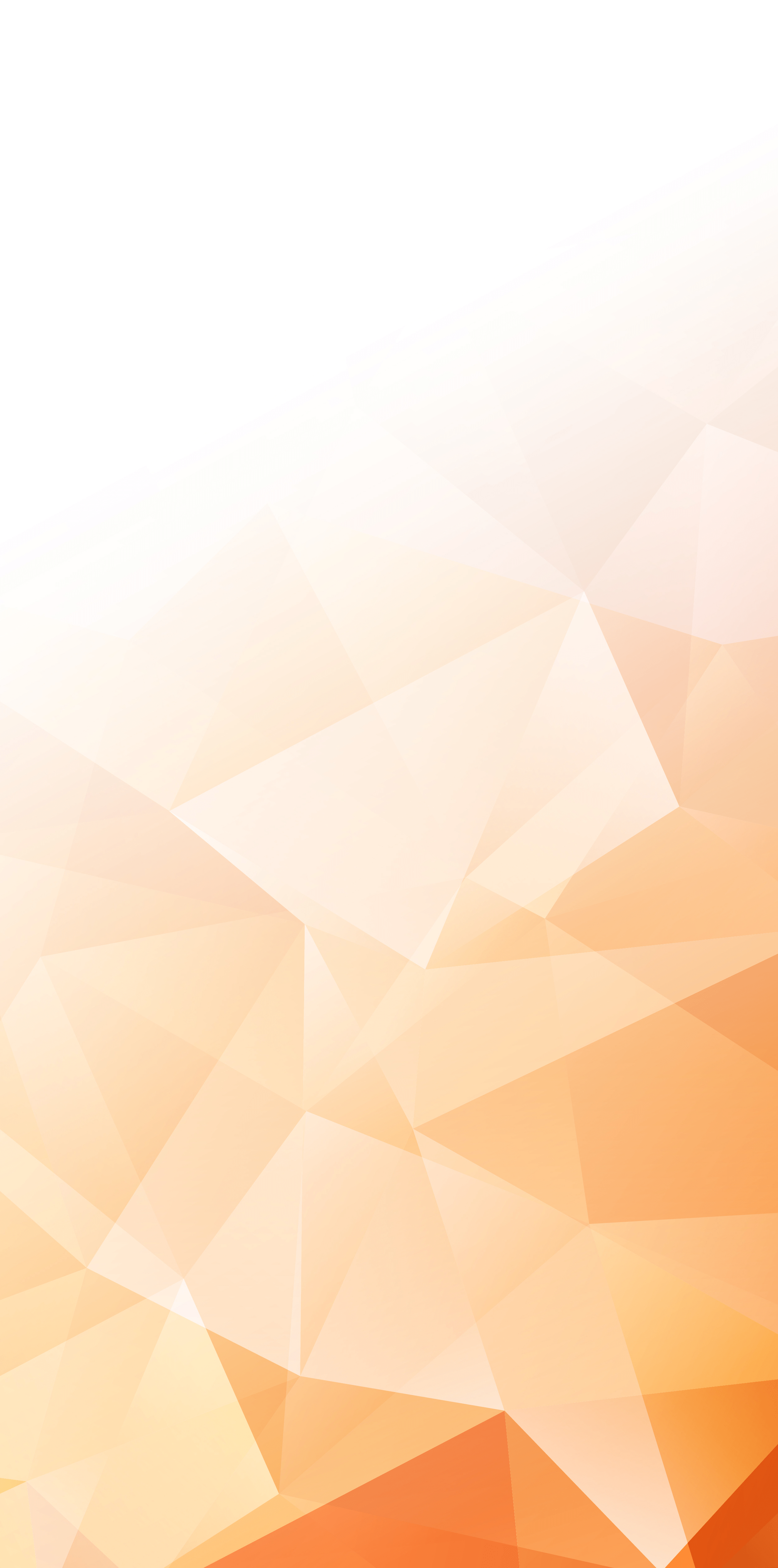}}; 
    \node (Logo) [inner sep=0pt, yshift = 10cm] at (current page.center) {\includegraphics[width=0.7\paperwidth]{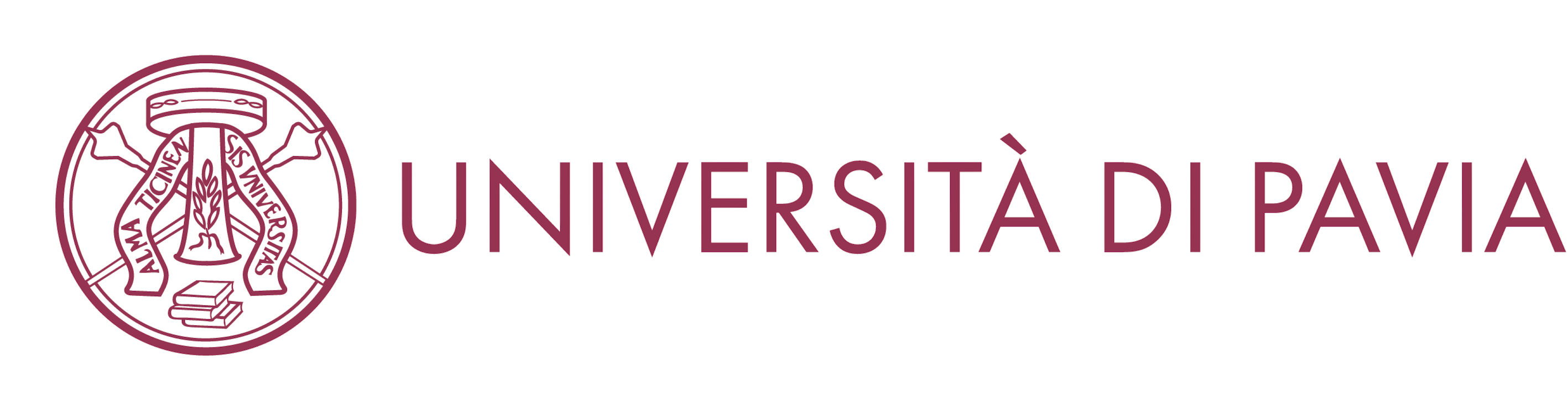}}; 
    \node (Dep) [above of= Title, yshift = 5.5cm, inner sep=1.25cm, rectangle, minimum height=0.2\paperheight, minimum width=\paperwidth, text width=0.8\paperwidth] at (current page.center) {
        \centering\sffamily 
  	{\Large Dottorato di Ricerca in Fisica --- XXXV ciclo\par} 
   
    };
    \node (Title) [anchor=center, inner sep=1.25cm, rectangle, fill=unipvred!10!white, fill opacity=0.0, text opacity=1, minimum height=0.2\paperheight, minimum width=\paperwidth, text width=0.8\paperwidth] at (current page.center) {
        \centering\sffamily 
  	{\bfseries Ph.D. Thesis\par} 
	\vspace{16pt} 
	{\Huge\bfseries Variational quantum algorithms for machine learning \par} 
	\vspace{16pt} 
	{\LARGE Theory and Applications \par} 
	\vspace{24pt} 
	{\huge\bfseries Stefano Mangini\par} 
    };
    \node (Info) [below of= Title, yshift = -8cm, inner sep=1.25cm, minimum height=0.2\paperheight, minimum width=\paperwidth, text width=0.8\paperwidth] {
        \centering\sffamily 
  	{\Large Supervisor: Prof. Chiara Macchiavello\par} 
	\vspace{36pt} 
	{\small Submitted to the Graduate School of Physics in partial fulfillment of the requirements for the degree of Dottore di Ricerca in Fisica (Doctor of Philosophy in Physics) at the\par} 
        \vspace{16pt} 
 	{\Large University of Pavia\par} 
	\vspace{72pt} 
	{\today \par} 
    };
\end{tikzpicture}

\newpage


\thispagestyle{empty} 

~\vfill 

\noindent Copyright \copyright\ 2023 Stefano Mangini\\ 

\noindent \LaTeX~Template \href{https://www.latextemplates.com/template/legrand-orange-book}{``The Legrand Oragne Book''} v3.1 by Vel and Mathias Legrand.\\

\noindent Quantum circuits drawn with the \href{https://ctan.org/pkg/quantikz}{quantikz} \LaTeX~package.\\

\noindent Thesis background image by kjpargeter, initial covers by Harryarts, on \href{https://www.freepik.com/}{Freepik}.\\
\noindent Chapter cover images generated with \href{https://midjourney.com/}{Midjourney}, a Generative AI for images.

\newpage
\thispagestyle{empty}
\begin{flushright}
        \vspace*{5cm}
        \textit{A tutta la mia famiglia.}
\end{flushright}
\let\cleardoublepage\clearpage

\chapterimage{bg0bis.jpg}
\addcontentsline{toc}{chapter}{Acknowledgements}
\chapter*{Acknowledgements}
I remember when, on January 2019, already a few months after I moved to Pavia to start my Ph.D., I was still in the guest room with Leonardo, the colleague with whom I would eventually have shared the office, and the highs and lows of this journey, hopelessly waiting for the assignment of a proper office that was still lacking. While trying to figure out where we ended up and what it means to ``research'', we also got to know each other cheerfully and innocently read the news about a weird health situation in China, without any concern whatsoever that such a situation would become our situation. Or better, everyone's situation. 
\begin{figure*}[ht]
    \centering
    \includegraphics[width=0.45\textwidth]{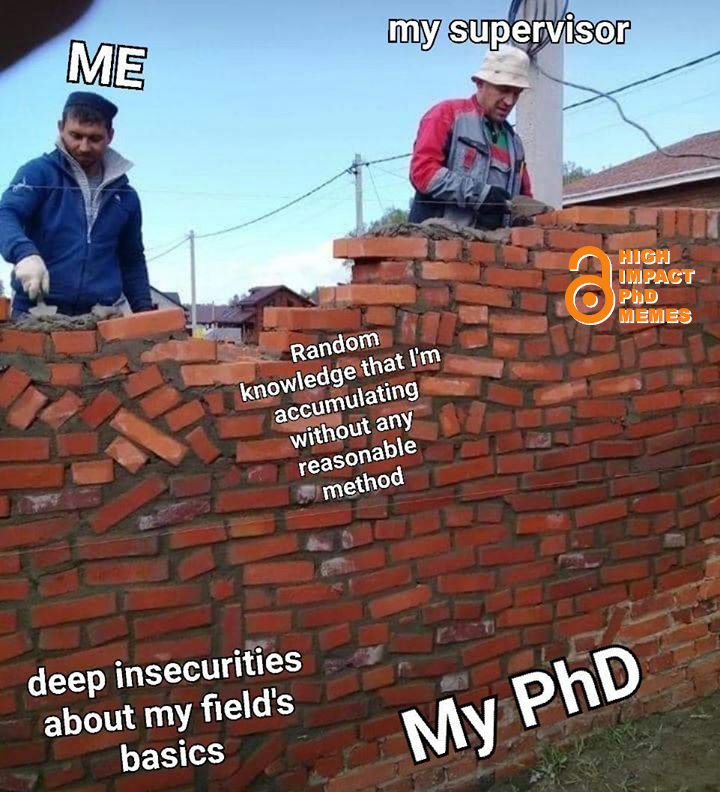}\hfill
    \includegraphics[width=0.45\textwidth]{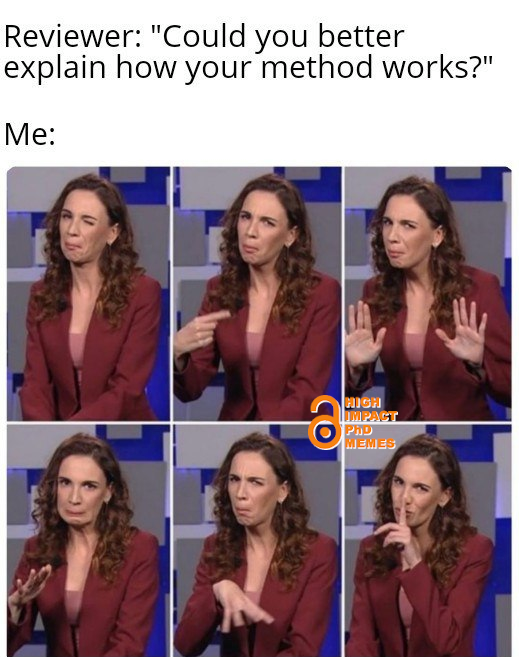}\\
    \includegraphics[width=0.45\textwidth]{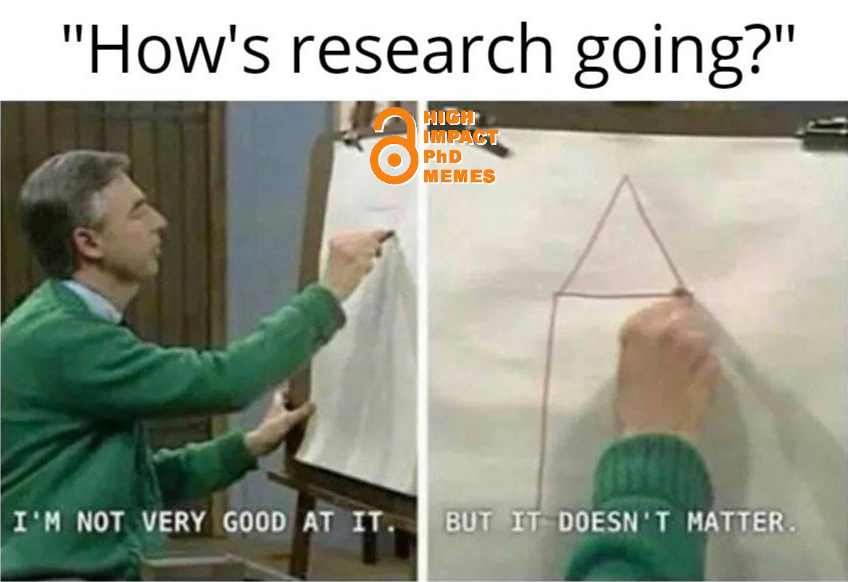}\hfill
    \includegraphics[width=0.45\textwidth]{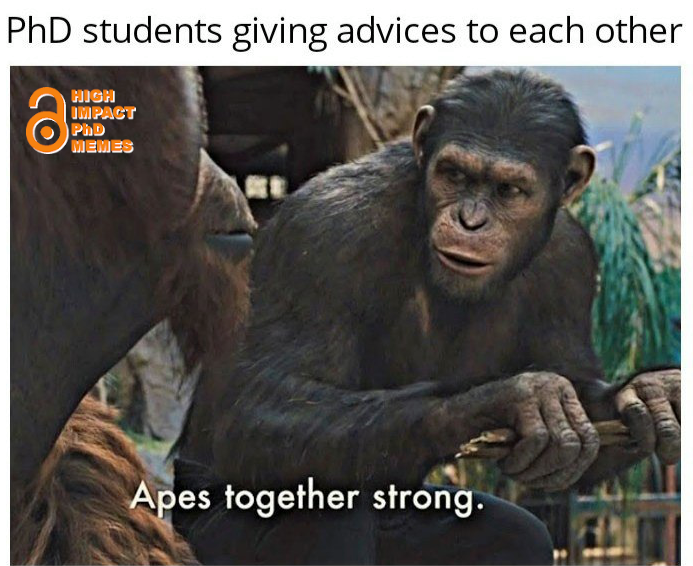}
    \caption[A selection of very high-impact memes]{A careful selection of high quality memes from ``High Impact PhD Memes'' that were hung on the walls in the office by Matteo Lugli, our \textit{de facto} meme interior designer.}
    \label{fig:memes}
\end{figure*}

Needless to say, Covid-19 essentially cut down in half the enjoyable part ---assuming there is one--- of the Ph.D., as the first half of it was essentially a sequence of lockdown, boring remote conferencing, and lone investigations. The original idea of spending the Ph.D. travelling the world for conferences and research visits was brutally set aside. But finally, as the health situation recovered and the possibility to travel was restored, I could participate in an amazing (un)conference in Lapland and a Quantum Hiking in Val d'Aosta (I cannot be grateful enough to Sabrina Maniscalco and Lorenzo Maccone for organising such cool and inspiring events), and then also to London for a research period abroad, and eventually to sunny Malta for a summer school. 

At the end of the day, I am more than happy with how things ended up. During these years many things happened to the World (even a bit too much) and to myself personally, and, as I repeatedly experience over and over when something ends, what remains are the people we have met and the experiences we have shared. All the rest is just sand castles.

I cannot but start thanking all the people in the Quantum Information Group (QUIT) at the University of Pavia, my academic family for the last few years, for the coffee breaks, lunch (with or without ``schiscetta''), and dinners, and for the teachings and anecdotes of what it means to be a true ``quittino'', a title I am very proud to carry. I thank Alessanrdo Bisio and Alessandro Tosini (aka ''gli Alessanndri''), Max Sacchi, and Paolo Perinotti. I thank Lorenzo Maccone for the fruitful discussions about science, mountains, and all sorts of things, for always being kind, and for organising the Quantum Hiking Conference, one of the most unique experiences I have had. I deeply thank my supervisor Chiara Macchiavello for giving me space and freedom to follow my interests and experiment with research topics and methods, and for always being kind and open to discussions. 

I wish to thank Dario Gerace and Daniele Bajoni for the great collaboration and for being always available for interesting discussions, and to Francesco Scala for being such a cool roommate for the QTML2022 conference in Naples, and for the stimulating chats we had on quantum machine learning. A big thank goes to Francesco Tacchino for helping me out at the start of the Ph.D., for the very fruitful and happy collaboration that succeeded, and for the chats and thoughts we shared down the road. Although we overlapped in person for just a few months in Pavia, my Ph.D. may have started in a completely different route if it was not for you, thanks!

I am happy to thank all the people with whom I spent many great days and nights: Giovanni ``il Giò'' Chesi for teaching all of us how to ``calm down now'', to Matteo Lugli for providing our office with top-notch memes on Ph.D. life, Davide Rolino for his unbelievable politeness, Simone Roncallo for the few but inspiring discussions, Simanraj for his very entertaining and weird stories, Lorenzo Trezzini for sharing great memories of Trieste, to Francesca Brero and Margherita Porru for offering what is now known as the best coffee in the Department, and eventually to Marco Erba for introducing me to Pavia's and QUIT's world and for letting me experience ``atypical adventures''.

Also, I am very happy to thank all the fantastic people I have had the honour of meeting in London, where I spend a wonderful four months in the Spring of 2022 for an internship in Quantinuum. I must thank Mattia Fiorentini for giving me the opportunity of joining the team and for being such a friendly person, and Marcello Benedetti for what I consider the most stimulating scientific discussions I have had in these years, and especially for being such a cheerful and humane person. Also, a big thanks to all the Quantum Lads: to Chris, Conor, Mateusz, Enrique, and Miguel for all the amazing chats we have had in front of a pint, at Enrique's farewell party, or while eating the street food from near the office. To Sam Duffield for being my Englishman of trust and introducing me to the Brits' lifestyle, to David Amaro not only for being my best vegan buddy, but for the amazing discussions on society, and the wonderful day we spent together in Greenwich. To Kirill, my neighbour near Victoria, for all the days we spent together in the office, for introducing me to ``Secret Hitler'', and overall for being such a nice friend there in London. And finally to Luuk, for all the stories about drunk people in Dublin, for the exciting discussions we had in the office, but ultimately for being such a wonderful person and a close friend. Although it was only for a relatively brief period, the time spent in London was one of the few highs of the Ph.D., and this is especially because of all of you. Moreover, it only rained once in four months and I even saw The Queen's Platinum Jubilee, probably the last one for a long long time. What else could I have wanted?

Finally, I am obliged to thank my amazing friends Claudio and Leonardo, with whom I had the pleasure of sharing the office these years! It was an honour to have you by my side, in what is easily the best office in the Department, thanks to the people inside (``i Dottorandini''), and of course also the classy balcony. The discussions we had in the office on the most disparate topics (quantum, society, mathematics, teaching, music, ...), the pizzas we ate at ``La Botticella'', and the coffees we drank at the bar: these memories are the most important outcome of the Ph.D., memories that I am truly grateful to bring with me. A very special thank goes to Leonardo, my most faithful companion of these years, with whom I shared everything (even a sofa), and without whom there would be no thesis or Ph.D. to talk about. Despite Covid, breakups, and stress, we were together fighting against the odds. Thank you for being there and for making these three years worth remembering! 

There are many other people that I came across and which I would like to thank, but I will do that when I see you in person! For the moment, take a look at the figure above and enjoy a selected choice of memes about Ph.D. life that accompanied me and my colleagues in these years.

\vspace*{0.5cm}
At last, the biggest and deepest thank you goes to my Family. In a world where everything changes, you remain the same, and we remain together.
\vspace*{0.5cm}

\hfill\today

\chapterimage{bg0bis.jpg}
\addcontentsline{toc}{chapter}{List of Publications}
\chapter*{List of Publications}

\noindent
These manuscripts are part of the thesis:
\vspace*{0.1cm}

\begin{enumerate}
    \item \textbf{Mangini, S.}, Tacchino, F., Gerace, D., Macchiavello, C., \& Bajoni, D. (2020). \href{https://iopscience.iop.org/article/10.1088/2632-2153/abaf98/meta}{Quantum computing model of an artificial neuron with continuously valued input data}. \textit{Machine Learning: Science and Technology}, \textbf{1}(4), 045008~\cite{ManginiCQN2020}.
    
    \item Tacchino, F., \textbf{Mangini, S.}, Barkoutsos, P. K., Macchiavello, C., Gerace, D., Tavernelli, I., \& Bajoni, D. (2021). \href{https://ieeexplore.ieee.org/abstract/document/9364892}{Variational Learning for Quantum Artificial Neural Networks}. \textit{IEEE Transactions on Quantum Engineering}, \textbf{2}, 1-10~\cite{TacchinoVariationalQNN2021}.
    
    \item \textbf{Mangini, S.}, Tacchino, F., Gerace, D., Bajoni, D., \& Macchiavello, C. (2021). \href{https://iopscience.iop.org/article/10.1209/0295-5075/134/10002/meta}{Quantum computing models for artificial neural networks}. \textit{Europhysics Letters}, \textbf{134}(1), 10002~\cite{ManginiQNN}.
    
    \item \textbf{Mangini, S.}, Marruzzo, A., Piantanida, M., Gerace, D., Bajoni, D., \& Macchiavello, C. (2022). \href{https://link.springer.com/article/10.1007/s42484-022-00070-4}{Quantum neural network autoencoder and classifier applied to an industrial case study}. \textit{Quantum Machine Intelligence}, \textbf{4}(2), 13~\cite{ManginiAutoencoder2022}.
    
    \item \textbf{Mangini, S.}, Maccone, L., \& Macchiavello, C. (2022). \href{https://epjquantumtechnology.springeropen.com/articles/10.1140/epjqt/s40507-022-00151-0}{Qubit noise deconvolution}. \textit{EPJ Quantum Technology}, \textbf{9}(1), 1-30~\cite{ManginiDeconvolution2022}.
    
    \item Ballarin, M., \textbf{Mangini, S.}, Montangero, S., Macchiavello, C., \& Mengoni, R. (2023). \href{https://quantum-journal.org/papers/q-2023-05-31-1023/}{Entanglement entropy production in Quantum Neural Networks}. \textit{Quantum}, \textbf{7}, 1023~\cite{BallarinEntQNN_2022}.
    
    \item \textbf{Mangini, S.}, Benedetti, M. \textit{Manuscript in preparation}~\cite{ManginiBenedettiINPREPARATION}.

    \end{enumerate}

\vspace*{0.3cm}
\noindent
These manuscripts are not part of the thesis:
\vspace*{0.1cm}

    \begin{enumerate}
    \setcounter{enumi}{7}
    
    \item Skolik, A., \textbf{Mangini, S.}, Bäck, T., Macchiavello, C., \& Dunjko, V. (2023). \href{https://epjquantumtechnology.springeropen.com/articles/10.1140/epjqt/s40507-023-00166-1}{Robustness of quantum reinforcement learning under hardware errors}. \textit{EPJ Quantum Technology}, \textbf{10}(1), 1-43~\cite{SkolikNoiseQRL}.

    \item Benatti, F., Mancini, S., \& \textbf{Mangini, S}. (2019). \href{https://www.worldscientific.com/doi/abs/10.1142/S0219749919410090}{Continuous variable quantum perceptron}. \textit{International Journal of Quantum Information}, \textbf{17}(08), 1941009~\cite{ManginiBenattiCQN2019}.
    
    \item Scala, F., \textbf{Mangini, S.}, Macchiavello, C., Bajoni, D., \& Gerace, D. (2022). \href{https://ieeexplore.ieee.org/abstract/document/9892080}{Quantum variational learning for entanglement witnessing}. In \textit{2022 International Joint Conference on Neural Networks (IJCNN) (pp. 1-8)}. IEEE~\cite{ScalaEntanglementWitness2022}.
    
    \item Di Sipio, R., Huang, J. H., Chen, S. Y. C., \textbf{Mangini, S.}, \& Worring, M. (2022). \href{https://ieeexplore.ieee.org/abstract/document/9747675}{The dawn of quantum natural language processing}. In \textit{ICASSP 2022-2022 IEEE International Conference on Acoustics, Speech and Signal Processing (ICASSP) (pp. 8612-8616)}. IEEE~\cite{DiSipioDawnQNLP2022}.
\end{enumerate}

\vspace*{0.5cm}
\noindent

Chapter~\ref{ch:QC_and_VQAs} and Chapter~\ref{ch:QML} are loosely based on manuscripts 3 and 7 with substantial additions; Chapter~\ref{ch:CQN} is based on manuscript 1~\cite{ManginiCQN2020}; Chapter~\ref{ch:VariationalQN} is based on manuscript 2~\cite{TacchinoVariationalQNN2021}; Chapter~\ref{ch:Autoencoder} is based on manuscript 4~\cite{ManginiAutoencoder2022}; Chapter~\ref{ch:entanglement} is based on manuscript 6~\cite{BallarinEntQNN_2022}; Chapter~\ref{ch:NoiseDeconvolutionChapter} is based on manuscript 5~\cite{ManginiDeconvolution2022}.

\chapterimage{bg0bis.jpg}
\addcontentsline{toc}{chapter}{Summary}
\chapter*{Summary}

The topic of the present Ph.D. thesis is Quantum Computation and Information processing, with a specific focus on the relatively new fields of Variational Quantum Algorithms and Quantum Machine Learning. These disciplines not only offer a unique perspective on studying quantum information processing tasks, but also have the potential to provide a useful computational quantum advantage even with the currently available first-generation small and noisy quantum computing devices.

Variational Quantum Algorithms involve a hybrid quantum-classical computational loop, where a quantum computer is used only for some specific ideally quantum-native subroutines, and a classical computer runs an optimisation procedure on variational parameters to minimise a cost function whose minimum corresponds to the solution to the problem to be solved. This framework shares the same foundational idea as state-of-the-art Deep Learning models, where complex parametric models are tuned via optimisation methods to solve various tasks. The intersection of quantum computing and machine learning has led to the development of quantum machine learning, an interdisciplinary area that explores the benefits of combining quantum computation and artificial intelligence.

This thesis provides a comprehensive analysis of the state of the art of the field, including numerous original contributions, from the study of quantum models for artificial neurons to the characterisation of entanglement created in quantum architectures for neural networks, up to discussing the effect of measurement noise on a more quantum information perspective. 

The first chapters are devoted to a careful review of the basics of quantum computing and a thorough discussion of variational quantum algorithms. Then the discussion is moved to quantum machine learning, where an introduction to the elements of machine learning and statistical learning theory is followed by a review of the most common quantum counterparts of machine learning models. 

Afterward, multiple novel contributions to the field are presented. A newly introduced model for a quantum perceptron is discussed, along with applications to pattern recognition and classification tasks. Such a model is then generalised to include strategies based on variational protocols to reduce the circuital footprint of the proposed architecture, and also analyse its performances where multiple optimisation strategies are considered. Subsequently, a quantum algorithm comprising a quantum autoencoder followed by a quantum classifier is presented to first compress and then label classical data coming from an industrial power plant, thus providing one of the first attempts to integrate quantum computing procedures in a real-case scenario of an industrial pipeline.

The analysis is then broadened to a more quantum information perspective, by first studying the entanglement features of quantum neural networks. Specifically, tensor networks are employed to study the entanglement entropy in parameterized quantum circuits of up to fifty qubits and show that the entanglement generated in such architectures reaches that of typical random quantum states under various measures. Finally, the focus is shifted from quantum machine learning to that of quantum noise, and a noise deconvolution technique is presented to remove a wide class of noises when performing arbitrary measurements on qubit systems. 

The thesis then ends with conclusions where loose ends are discussed, and final remarks are exposed. Overall, the thesis provides a well-balanced investigation of multiple scientific domains, including quantum physics and computer science, through theoretical exploration, computational simulations, and experimental verification on already available quantum computing devices.


\pagestyle{empty} 
\tableofcontents 
\listoffigures 
\listoftables 


\nomenclature[A]{PQC}{Parameterised Quantum Circuit}
\nomenclature[A]{VQA}{Variational Quantum Algorithm}
\nomenclature[A]{QML}{Quantum Machine Learning}
\nomenclature[A]{BP}{Barren Plateau}
\nomenclature[A]{ML}{Machine Learning}
\nomenclature[A]{PTM}{Pauli Transfer Matrix}
\nomenclature[A]{CPTP}{Completely Positive Trace Preserving}

\nomenclature[M]{\(\mathcal{H}\)}{Hilbert space}
\nomenclature[M]{\(\bmx\)}{Input sample belonging to input data space, $\bmx \in \mathcal{X}$}
\nomenclature[M]{\(y\)}{Output sample belonging to output data space $y \in \mathcal{Y}$}
\nomenclature[M]{\(\mathcal{X}\)}{Input data space, usually $\mathcal{X} \subset \mathbb{R}^d$}
\nomenclature[M]{\(\mathcal{Y}\)}{Output data space, usually $\mathcal{Y} \subset \mathbb{R}$}
\nomenclature[M]{\(\mathcal{Z}\)}{Data space given by pairs inputs and outputs, $\mathcal{Z} = \mathcal{X} \times \mathcal{Y}$}
\nomenclature[M]{\(S\)}{Training set consisting of pairs of inputs and outputs}
\nomenclature[M]{\(d\)}{Dimension of the input vectors, $\bmx \in \mathcal{X} \subset \mathbb{R}^d$}
\nomenclature[M]{\(p\)}{Dimension of the trainable parameter vectors, $\bmw,\, \bmt \in \subset \mathbb{R}^d$}
\nomenclature[M]{\(m\)}{Number of data samples in the training set $\mathcal{S} \subset \mathcal{Z}^m$}
\nomenclature[M]{\(\bmt\)}{Trainable parameters of a parameterised quantum circuit in a variational quantum algorithm}
\nomenclature[M]{\(\bmw\)}{Trainable parameters of a parametric classical machine learning model}

\printnomenclature


\pagestyle{fancy} 

\cleardoublepage 


\chapterimage{bg1.png} 
\chapterspaceabove{6.75cm} 
\chapterspacebelow{7.25cm} 

\chapter{Introduction}\index{Introduction}
\epigraph{\textit{Nature isn’t classical, dammit, and if you want to make a simulation of Nature,
you’d better make it quantum mechanical, and by golly it’s a wonderful problem
because it doesn’t look so easy.}}{Richard Feynman, 1982~\cite{Feynman1982}}

As far as I can tell from my still brief experience of research into Quantum Computation during these years of Ph.D., it is not possible to start a discussion on this topic without mentioning the (moral) father of such a discipline, the mighty Richard Feynman who, in 1982, roughly at the same time of other scientists of those years~\cite {PreskillQC40yearslater}, firstly proposed the idea of building a quantum-mechanical computing device for simulating Nature. Thus, since I don't feel in the position to interrupt this grand tradition, please take a few moments to enjoy for the umpteenth time Feynman's quote on quantum computing, which you can find at the start of the page. 

With a leap forward 40 years into the future, quantum computing is starting to become a solid reality, with the first small-scale prototypes of quantum computers being actively developed and tested, and with a universal fault-tolerant quantum computing device hopefully to appear within a few decades in the future, although the road to this goal is strewn with major technical, experimental, and theoretical challenges. The current generation of quantum computers has been dubbed NISQ, for Noisy Intermediate-Scale Quantum~\cite{Preskill2018NISQ}, which indicates devices consisting of small quantum processing units consisting of just tens to few hundreds of carriers of quantum information, the \textit{qubits}, and that these qubits are imperfect, subject to noise that diminishes their quantum properties, hence their computational relevance.

With the impossibility of timely orchestrating those elements that make quantum computers unique and powerful objects, namely \textit{superposition}, \textit{entanglement}, and \textit{interference}, many of the celebrated quantum algorithms that were proposed in the previous decades with provable computational speedups, Shor's being the most convincing example~\cite{NielsenChuang}, currently remain well out of reach. However, the imperfect nature of current quantum devices prompted the creation of seemingly imperfect quantum algorithms that trade off provable guarantees of quantum speedups with reasonable intuitions of classical hardness and the careful human craftsmanship of algorithms with the heavy lifting provided by powerful classical optimisation techniques. However, as we shall see in a few paragraphs, this situation does not only concern quantum computing but rather the field of applied computer science as a whole. 

These relatively new types of quantum procedures are called Variational Quantum Algorithms and are specifically thought to be executable already on available computing devices, and thus take full advantage of the current generation of quantum computing. The variational paradigm prescribes the use of a hybrid quantum-classical computational loop in which a quantum computer is used in tandem with a classical computer, where the former is used only for some specific ideally quantum-native subroutines, and the latter instead to run an optimisation procedure on some variational parameters to minimise a cost function whose minimum (maximum) corresponds to the solution to the problem to be solved. This framework for NISQ-friendly quantum algorithms gained much momentum over the last few years, and it is considered one of the best candidates to achieve some form of useful computational quantum advantage, even though the debate on this expectation is far from being settled. 

However, it is very important to recall that quantum computation and quantum information science first remain scientific disciplines that deserve thorough and enthusiastic scientific research on their own, and regardless of possible technological applications, with quantum computers to be considered, at least at this moment, mainly scientific tools. In this respect, borrowing the words from Simone Severini: ``\textit{The paradigm with which quantum computers work is totally different from the one with which classical computers work. Therefore, it is too simplistic to make comparisons in terms of speed and efficiency. The most sensible analogy is not with a classical computer, but rather with a telescope: the quantum computer should be regarded as an instrument that allows one to look further}''~\cite{WiredSimoneQuantum}.

Even more than Quantum Computing and Quantum Technologies in general, over the past decade, Artificial Intelligence and Deep Learning have garnered increasing scientific and technological attention. These fields offer a variety of tools that can handle a diverse range of tasks, from achieving superhuman performance on board games~\cite{MasteringGo} to controlling nuclear reactors~\cite{DegraveMagneticControlTokamak2022}, and even having astonishingly human-like conversational abilities~\cite{ChatGPT}. State-of-the-art Deep Learning models operate by optimising large, complex, and often opaque parametric models to minimise a problem-dependent loss function. Although a complete theoretical understanding of these models' inner workings is still lacking and the subject of active research, their widespread success is undoubtedly motivated by their empirical success in practical applications.

In recent years, the fields of quantum computing and machine learning joined forces and stimulated the development of Variational Quantum Algorithms, which, as we mentioned above, rely on optimising complex parametric models (parametric quantum circuits instead of artificial neural networks) using optimisation methods like gradient descent to minimise a problem-specific cost function. This union was so fruitful that it was given the name Quantum Machine Learning, an interdisciplinary area that explores the interplay and benefits of combining quantum computation and artificial intelligence~\cite{DunjkoWittekReview2020}. In fact, the overlap between these two fields predates Variational Quantum Algorithms, as linear algebra-based quantum subroutines have been proposed as accelerators in classical machine learning algorithms for some time already~\cite{Biamonte2017QML}, and it is possible to find resources about quantum computation and neural networks even from decades ago~\cite{GuptaQNN_2001, Lewenstein_Perceptron_1994}.

The topic of this thesis lies in the most recent incarnation of quantum machine learning, namely the use of variational quantum circuits as machine learning models. It aims to provide a comprehensive analysis of the state of the art of the field, as well as the discussion of numerous original contributions, from the study of quantum models for artificial neurons to the characterisation of entanglement created in common quantum models for neural networks, up to discussing the effect of measurement noise on a more quantum information perspective. As we shall see in the following chapters, this crossover proves very valuable and most importantly interesting both on a theoretical and practical level. Indeed, studies on these topics are stimulating as they require specialised knowledge from multiple disciplines from quantum physics and computer science, and permit a very well-balanced investigation between theoretical exploration, computational simulations, and also experimental verification on already available quantum computing devices. 

The rest of the thesis is organised as follows. In Chapter~\ref{ch:QC_and_VQAs} we start reviewing the basics and introducing the notation of quantum computing, to then move towards a thorough discussion on variational quantum algorithms. Chapter~\ref{ch:QML} is instead dedicated exclusively to quantum machine learning, where an introduction to the elements of machine learning and statistical learning theory is followed by a review of the most common quantum counterparts of machine learning models. 

In Chapter~\ref{ch:CQN} we discuss a newly introduced model for a quantum perceptron, which is a quantum algorithm mimicking the behaviour of a classical artificial neuron, and show how it can be used to implement pattern recognition and classification tasks. This model is then generalised in Chapter~\ref{ch:VariationalQN}, where various strategies based on variational protocols are described to reduce the circuital footprint of the quantum neuron model. Afterwards, in Chapter~\ref{ch:Autoencoder} we propose a quantum algorithm comprising a quantum autoencoder followed by a quantum classifier to first compress and then label classical data coming from an industrial power plant, thus providing one of the first attempts to integrate quantum computing procedures in a real-case scenario of an industrial pipeline.

In Chapter~\ref{ch:entanglement} we expand the perspective and analyse the entanglement features of quantum neural networks. Specifically, we use tensor network tools to study the entanglement entropy in parameterized quantum circuits of up to fifty qubits and show that the entanglement generated in such architectures reaches that of typical random quantum states under various measures. Finally, in Chapter~\ref{ch:NoiseDeconvolutionChapter} we step out from quantum computation and machine learning and rather focus on the topic of noise from a quantum information viewpoint and present a noise deconvolution technique to remove a wide class of noises when performing arbitrary measurements on qubit systems. 

Finally, in Chapter~\ref{ch:Conclusion} we try to draw some inspiring conclusions about this journey on variational algorithms, noisy quantum computers, and machine learning. In the appendices, some calculations or additional details regarding the related topics discussed in the main text are reported. 

Without further ado, let us start!

\chapterimage{bg2.png} 
\chapterspaceabove{6.75cm} 
\chapterspacebelow{7.25cm} 

\chapter{Quantum Computing and Variational Quantum Algorithms}\index{Quantum Computing and VQAs}
\label{ch:QC_and_VQAs}
\epigraph{\textit{When life gives you lemons, make lemonade!}}{Proverbial phrase, and often Scott Aaronson.}
\vspace*{0.cm}
\startcontents[chapters]
\printcontents[chapters]{}{1}[3]{}
\vspace*{1cm}

In this chapter, we discuss Quantum Computing and Variational Quantum Algorithms. We start by introducing the necessary tools and definitions of quantum computation and then move on to discuss recent results on variational quantum algorithms, which are a class of algorithms specifically suited for currently available near-term quantum devices. Extended discussions of such topics can be found in~\cite{ManginiQNN, CerezoQMLPerspective_2022, CerezoPQC2021Review, Bharti2021NIQSReview, TillyVQE_2022}.

\section{Basics of Quantum Computation}
Quantum Computation and Quantum Information are fields of research that study how physical quantum systems can be used to process information and perform computations. In analogy with the classical setting, the basic element of quantum information is called quantum bit or \textit{qubit}\footnote{A term which is surprisingly recent, just as young as me at the time of writing, introduced for the first time in 1995 by Benjamin Schumacher~\cite{SchumacherQubit1995}.}, an object describing the behaviour of a two-level quantum system, for example a particle having access to the ground state and first excited state of an energy potential. Independently of its actual physical realisation, the quantum \textit{state} of a qubit is mathematically described as a vector living in a two-dimensional vector space over the complex numbers, namely $\mathbb{C}^2$. 

\subsection{Single qubit systems and operations}
Let $\mathcal{H} = \mathbb{C}^2$ denote the \textit{Hilbert} space of the qubit, that is the vector space $\mathbb{C}^2$ equipped with the standard inner product given by the Euclidean dot product for vectors with complex entries. Dirac's bra-ket notation of quantum mechanics prescribes the use of the so-called \textit{kets} $\ket{\cdot}$ to indicate vectors (i.e. quantum states) in this space, namely $\ket{\psi} \in \mathcal{H}$, where $\psi$ indicates a state. Similarly, one uses \textit{bras} $\bra{\cdot}$ to indicate the conjugate transpose of a vector
$$
\bra{\psi} = \big(\ket{\psi}\big)^\dagger,\quad \ket{\psi} \in \mathcal{H}\,,
$$
where the dagger operation $A^\dagger = (A^*)^T$ is the composition of complex conjugation $A^*$ and transposition $A^T$. With this notation, the inner product between two states is denoted with the juxtaposition of a bra with a ket, as follows
$$
\braket{\cdot}{\cdot}: \mathcal{H}\times \mathcal{H} \rightarrow \mathbb{C},\quad \ket{\psi},\ket{\phi} \in \mathcal{H},\quad \braket{\phi}{\psi} \in \mathbb{C}\,.
$$

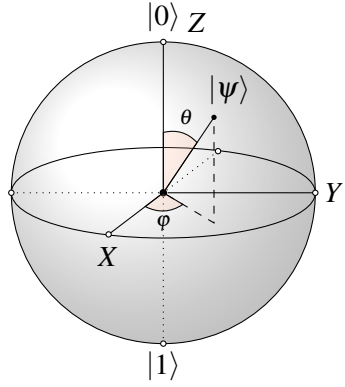
\begin{figure}[ht]
    \centering
    \begin{tikzpicture}

    \def\r{2}
    \coordinate (orig) at (0,0);
    
    \shade[ball color = gray!0, fill opacity = 0.4] (0,0) circle (\r cm);
    \draw[] (orig) circle (\r cm);
    \draw[] (-\r,0) arc (180:360:2 and 0.6);
    \draw[] (\r,0) arc (0:180:2 and 0.6);

    \draw (0,0) node[circle, fill, inner sep=1] (orig) {} -- (\r/3,\r/2) node[circle, fill, inner sep=0.7, label={[xshift=0.2cm]above:$\ket{\psi}$}] (a) {};
    \draw[dashed] (orig) -- (\r/3, -\r/5) node (phi) {} -- (a);

    \coordinate (x1) at (-0.7\r, -0.55\r);
    \draw (orig) -- ++(-0.7\r, -0.55\r);
    \node[below of=x1, yshift=0.7cm] {$X$};
    \draw (orig) -- ++(\r, 0) node[right] (x2) {$Y$};
    \draw (orig) -- ++(0, \r) node[above] (x3) {$\ket{0}$} node[above right, xshift=0.5em] {$Z$};
    \coordinate (z1) at (0, -\r);
    \node[below of=z1, yshift=0.7cm] {$\ket{1}$};
    \draw[dotted] (orig) -- ++ (-\r, 0);
    \draw[dotted] (orig) -- ++ (0, -\r);
    \draw[dotted] (orig) -- ++ (0.7\r, 0.55\r);

    \fill[fill=white,draw=black] (x1) circle (1pt);
    \fill[fill=white,draw=black] (\r,0) circle (1pt);
    \fill[fill=white,draw=black] (0,\r) circle (1pt);
    \fill[fill=white,draw=black] (-\r,0) circle (1pt);
    \fill[fill=white,draw=black] (0,-\r) circle (1pt);
    \fill[fill=white,draw=black] (+0.7\r, +0.55\r) circle (1pt);

    \draw[fill=ocre!50, fill opacity=0.2] (0,0) -- ($(orig)!8mm!(x3)$) to[bend left] ($(orig)!8mm!(a)$)  -- cycle;
    \draw[fill=ocre!50, fill opacity=0.2] (0,0) -- ($(orig)!3mm!(x1)$) to[bend right] ($(orig)!3mm!(phi)$)  -- cycle;

    \node at (0.3, \r/2) {$\scriptstyle \theta$};
    \node at (0, -0.4) {$\scriptstyle \varphi$};

    \end{tikzpicture}
    \caption[Bloch sphere representation of a qubit]{Bloch sphere representation of a qubit. A pure state of a qubit can be represented as a point on a unit sphere whose poles are the orthogonal basis states $\ket{0}$ and $\ket{1}$. The angles $(\theta, \varphi)$ are defined in Eq.~\eqref{eq:qubit_psi_bloch}. The six points arising from the intersection of the unit sphere with the three orthogonal axes are the eigenstates of the Pauli matrices $X$, $Y$, and $Z$, defined in Eq.~\eqref{eq:Paulimats}}
    \label{fig:bloch_sphere}
\end{figure}

Let $\ket{0}$ and $\ket{1}$ denote an orthonormal basis of $\mathcal{H}$, and call it the \textit{computational basis} of the space. Then, any (pure) state can be expressed as a linear combination of the computational basis states with complex coefficients
\begin{equation}
\label{eq:qubit_state}
    \ket{\psi} = \alpha\ket{0} + \beta \ket{1}, \quad \alpha,\,\beta \in \mathbb{C}\,.
\end{equation}
with a normalisation condition $\braket{\psi}{\psi}=1$ which constraints the complex coefficients to satisfy $\alpha\alpha^* + \beta\beta^* = \abs{\alpha}^2 + \abs{\beta}^2 = 1$\footnote{Normalisation of states is a useful requirement that makes Born's rule for probabilities of measurement outcomes easier to compute.}. Also, states that differ only for a global phase factor, like $\ket{\psi}$ and $e^{i\delta}\ket{\psi}$, represent the same \textit{physical} state because overall phases are not observable, in that measurements involving phase-shifted states will yield the same results. Thus, out of the initial four real parameters, a (pure) state of a qubit can be written in terms of just two parameters $(\theta, \varphi)$ as
\begin{equation}
\label{eq:qubit_psi_bloch}
    \ket{\psi}= \cos{\frac{\theta}{2}}\ket{0}+e^{i\varphi}\sin{\frac{\theta}{2}}\ket{1}\,.
\end{equation}
This equation is called \textit{Bloch sphere} representation of the qubit, because it makes it evident that the state of a qubit can be visualised as a point on the unit sphere with coordinates $(\theta, \varphi)$, where the poles are the orthogonal basis states $\ket{0}$ and $\ket{1}$, see Fig.~\ref{fig:bloch_sphere}.

It is often useful to reason in terms of vector components instead of states. Representing the basis states $\ket{0}$ and $\ket{1}$ as column vectors, Eq.~\eqref{eq:qubit_state} can be rewritten as
\begin{equation}
    \ket{\psi} = \alpha\ket{0}+\beta\ket{1} = \alpha\begin{bmatrix}1 \\ 0 \end{bmatrix} + \beta\begin{bmatrix}0 \\ 1 \end{bmatrix} = \begin{bmatrix}\alpha \\ \beta \end{bmatrix}\, \quad \text{with}\quad \ket{0} = \begin{bmatrix}1 \\ 0 \end{bmatrix},\,\,
    \ket{1} = \begin{bmatrix}0 \\ 1 \end{bmatrix}\,.
\end{equation}

\paragraph{Single qubit operations}
Operations on qubits are linear transformations $U:\mathcal{H}\rightarrow\mathcal{H}$ mapping quantum states to quantum states, and these can be represented by unitary matrices that map unit complex vectors to other unit complex vectors preserving the norm. The unitary evolution of (closed) quantum systems implies that not only the norm of quantum states is preserved, but also the \textit{reversibility} of quantum computation, since for any operation $U$ also the inverse one $U^\dagger$ with $U U^\dagger=\mathbb{I}$ is a valid quantum operation. This is different from classical computation, where operations on bits can be, in general, irreversible.

A set of operators of particular relevance for describing qubit systems are the Pauli matrices $\{X, Y, Z\}$, which in the computational basis $\{\ket{0}, \ket{1}\}$ have matrix representation\footnote{Operators and their matrix representation are not the same things, as one can define and use operators without invoking their matrix representation, which is specific to the chosen basis. In quantum computing however, the basis is always considered to be the \textit{computational basis}, and, with a slight abuse of notation, in the following we use the same symbol to denote both the operator and its matrix representation.}
\begin{equation}
\label{eq:Paulimats}
    X = \begin{bmatrix} 0 & 1 \\ 1 & 0\end{bmatrix},
    \quad
    Y = \begin{bmatrix} 0 & -i \\ i & 0\end{bmatrix},
    \quad
    Z = \begin{bmatrix} 1 & 0 \\ 0 & -1\end{bmatrix}\,.
\end{equation}
In hindsight, the computational basis is actually \textit{defined} to consist of the eigenstates of the $Z$ operator, which acts on such basis as $Z\ket{0}=\ket{0}$ and $Z\ket{1}=-\ket{1}$. The $X$ operator is also called the ``NOT" operation, again in accordance with the classical computation terminology, as its action is to flip the state of the qubit $X\ket{0}=\ket{1}$ and $X\ket{1}=\ket{0}$. Together with the identity $\mathbb{I}$, these matrices form a basis of the space of $2\times 2$ complex matrices. Other useful properties are that the Pauli matrices are unitary ($AA^\dagger = \mathbb{I}$), Hermitian ($A=A^\dagger$), involutory ($A^2=\mathbb{I}$), traceless ($\Tr[A]=0$), and the following commutation relations hold $[\sigma_i, \sigma_j] = 2i\varepsilon_{ijk}\sigma_k$, where $\sigma_i, \sigma_j \in \{X, Y, Z\}$, $\varepsilon_{ijk}$ is the Levi-Civita tensor, and summation is implied over repeated indices. The Pauli matrices have eigenvalues $\lambda \in \{\pm 1\}$, with corresponding eigenstates
\begin{equation}
\label{eq:Paulieigvals}
\begin{array}{lll}
Z\ket{0} = +\ket{0}  &  X\ket{+} = +\ket{+}  &  Y\ket{+i} = +\ket{+i}\\[0.2em]
Z\ket{1} = -\ket{1}  &  X\ket{-} = -\ket{-}  &  Y\ket{-i} = -\ket{-i}\\[0.4em]
\text{with} & \displaystyle{\ket{\pm } \coloneqq \frac{\ket{0}\pm \ket{1}}{\sqrt{2}}} & \displaystyle{\ket{\pm i} \coloneqq \frac{\ket{0}\pm i\ket{1}}{\sqrt{2}}}
\end{array}
\end{equation}
These states are highlighted in the Bloch sphere of Fig.~\eqref{fig:bloch_sphere}, and are located at the intersection of the unit sphere with the three orthogonal axes, which then represent the ``directions" of the Pauli matrices. 

In the quantum computing jargon, operations on qubits are called \textit{gates}, in analogy with the terminology used in classical computation to indicate elemental logical operations on bits. A major role in variational quantum algorithms is played by Pauli rotations gates, which are parametrized operations defined via exponentiation of the Pauli as follows
\begin{align}
    \label{eq:pauli_rotations}
    &R_{X}(\theta) \coloneqq e^{-iX\theta/2} \equiv \cos \frac{\theta}{2}\mathbb{I} - i\sin\frac{\theta}{2}X = 
    \begin{bmatrix}
    \cos\frac{\theta}{2}   & -i\sin\frac{\theta}{2}\\
    -i\sin\frac{\theta}{2} & \cos\frac{\theta}{2}
    \end{bmatrix}\\
    &R_{Y}(\theta) \coloneqq e^{-iY\theta/2} \equiv \cos \frac{\theta}{2}\mathbb{I} - i\sin\frac{\theta}{2}Y = 
    \begin{bmatrix}
    \cos\frac{\theta}{2}   & -\sin\frac{\theta}{2}\\
    \sin\frac{\theta}{2} & \cos\frac{\theta}{2}
    \end{bmatrix}\\
    &R_{Z}(\theta) \coloneqq e^{-iX\theta/2} \equiv \cos \frac{\theta}{2}\mathbb{I} - i\sin\frac{\theta}{2}Z = 
    \begin{bmatrix}
    e^{-i\theta/2}  & 0\\
    0 & e^{i\theta/2}
    \end{bmatrix}
\end{align}
where the passage from the exponential to the trigonometric formula is easily obtained via the definition of the exponential function and the involutory property ($A^2=\mathbb{I}$) of Pauli matrices
\begin{equation}
\label{eq:trig_form_exp}
    e^{-i\omega A} \coloneqq \sum_{k=0}^{\infty}\frac{(-i \omega A)^k}{k!} = \sum_{k \in \text{even}} \frac{\omega^k}{k!}\,\mathbb{I} - \sum_{k \in \text{odd}}\frac{\omega^k}{k!}A = \cos{\omega}\mathbb{I} - \sin{\omega}\,A\,.
\end{equation}
These gates are called rotation operations because they act on a qubit rotating it around the respective Pauli axis in the Bloch sphere representation by an amount indicated by the angle parameter. Also, these operations are ubiquitous in Variational Quantum Algorithms because they are the most straightforward way to introduce free parameters in a quantum computation, and are also usually easy to implement on real quantum hardware. 

In addition to the aforementioned Pauli and Pauli rotation gates, another fundamental operation is the Hadamard gate, denoted by $H$, which is used to create \textit{superposition} states as it maps the ground state $\ket{0}$ to the \textit{superposition} state $\ket{+}=H\ket{0}$. In Table~\ref{tab:1q_gates} we report the most common single qubit gates, along with their circuital representation when depicted as gates in a quantum circuit and their matrix representation in the computational basis. All other single-qubit gates can be expressed in terms of these operations. Indeed, the most general single-qubit operation, that is a general $2\times2$ unitary matrix, can be written (up to global phase) as 
\begin{equation}
\label{eq:qubit_rotation}
    U(\theta,\varphi,\lambda) = 
    \begin{bmatrix}
    \cos\frac{\theta}{2} & -e^{i\lambda}\sin\frac{\theta}{2} \\
    e^{i\varphi}\sin\frac{\theta}{2} & e^{i(\varphi+\lambda)}\cos\frac{\theta}{2}
    \end{bmatrix}\,,
\end{equation}
and there exists some angles such that it can be decomposed with a sequence of rotations $U(\theta,\varphi,\lambda) = e^{i\alpha}R_Z(\beta)R_Y(\gamma)R_Z(\delta)$~\cite{NielsenChuang}. Note that other equivalent decompositions into elementary rotations are possible. 

\begin{table}[ht]
    \centering
    \setlength{\defaultaddspace}{10pt}
    \addtolength{\tabcolsep}{5pt} 
    \begin{tabular}{ccc}
    \toprule
    \textbf{Name} & \textbf{Symbol/Circuital rep.} & \textbf{Matrix representation} \\
    \midrule
    Pauli-$X$ & \begin{tikzcd} & \gate{X} &\qw \end{tikzcd} & $\begin{bmatrix}0&1\\1&0\end{bmatrix}$\\\addlinespace
    Pauli-$Y$ & \begin{tikzcd} & \gate{Y} &\qw \end{tikzcd} & $\begin{bmatrix}0&-i\\i&0\end{bmatrix}$\\\addlinespace
    Pauli-$Z$ & \begin{tikzcd} & \gate{Z} &\qw \end{tikzcd} & $\begin{bmatrix}1&0\\0&-1\end{bmatrix}$\\\addlinespace
    Pauli rotation-$X$ & \begin{tikzcd} & \gate{R_X(\theta)} &\qw \end{tikzcd} & 
    $\begin{bmatrix}
    \cos\frac{\theta}{2} & -i\sin\frac{\theta}{2}\\
    -i\sin\frac{\theta}{2} & \cos\frac{\theta}{2}
    \end{bmatrix}$\\\addlinespace
    Pauli rotation-$Y$ & \begin{tikzcd} & \gate{R_Y(\theta)} &\qw \end{tikzcd} &
    $\begin{bmatrix}
    \cos\frac{\theta}{2}   & -\sin\frac{\theta}{2}\\
    \sin\frac{\theta}{2} & \cos\frac{\theta}{2}
    \end{bmatrix}$\\\addlinespace
    Pauli rotation-$Z$ & \begin{tikzcd} & \gate{R_Z(\theta)} &\qw \end{tikzcd} &
    $\begin{bmatrix}
    e^{-i\theta/2}  & 0\\
    0 & e^{i\theta/2}
    \end{bmatrix}$\\\addlinespace
    Phase gate & \begin{tikzcd} & \gate{P(\theta)} &\qw \end{tikzcd} &
    $\begin{bmatrix}
    1  & 0\\
    0 & e^{i\theta}
    \end{bmatrix}$\\\addlinespace
    T gate & \begin{tikzcd} & \gate{T} &\qw \end{tikzcd} &
    $\begin{bmatrix}
    1  & 0\\
    0 & e^{i\pi/4}
    \end{bmatrix}$\\\addlinespace
    Hadamard & \begin{tikzcd} & \gate{H} &\qw \end{tikzcd} & $\frac{1}{\sqrt{2}}\begin{bmatrix}1&1\\1&-1\end{bmatrix}$\\
    \bottomrule
    \end{tabular}
    \caption[Summary table of two-qubits gates]{Summary of the most important single-qubit operations. Here are shown the name of the operations, the abbreviations, the circuital representations when implemented as a \textit{gates} in a quantum circuit, and the matrix representations in the computational basis. Note that the phase gate $P$ and $R_Z$ are related by just a global phase, but it is useful to keep them separately.}
    \label{tab:1q_gates}
\end{table}

\subsection{Multi qubits systems and two-qubits operations}
Usually, more qubits together are used to implement a quantum information processing task or a computation. The Hilbert space for a system composed of multiple qubits is built considering the \textit{tensor product} of the single-qubit Hilbert spaces. For example, a two-qubit system lives in the Hilbert space $\mathcal{H}=\mathcal{H}_1 \otimes \mathcal{H}_2=(\mathbb{C}^2)^{\otimes 2} = \mathbb{C}^4$, and its state can be expressed as a linear combination of the four computational basis states
\begin{equation}
    \ket{\psi} = \alpha\ket{00} + \beta\ket{01} + \gamma\ket{10} + \delta\ket{11}\,, \quad \alpha,\beta,\gamma,\delta \in \mathbb{C}\,, \abs{\alpha}^2+\abs{\beta}^2+\abs{\gamma}^2+\abs{\delta}^2 = 1
\end{equation}
where the basis states $\qty{\ket{00}, \ket{01}, \ket{10}, \ket{11}}$ arise from considering tensor products of the single qubits basis states, namely
\begin{equation}
    \ket{00}\coloneqq \ket{0} \otimes \ket{0} = \begin{bmatrix}1 \\ 0 \end{bmatrix} \otimes \begin{bmatrix}1\\0\end{bmatrix} = \begin{bmatrix}
        1 \cdot \begin{bmatrix} 1 \\ 0 \end{bmatrix} \\[0.5cm]
        0 \cdot \begin{bmatrix} 1 \\ 0 \end{bmatrix}
    \end{bmatrix} = 
    \begin{bmatrix} 1 \\ 0 \\ 0 \\ 0 \end{bmatrix},\,
    \ket{01} = \begin{bmatrix} 0 \\ 1 \\ 0 \\ 0 \end{bmatrix},\,
    \ket{10} = \begin{bmatrix} 0 \\ 0 \\ 1 \\ 0 \end{bmatrix},\,
    \ket{11} = \begin{bmatrix} 0 \\ 0 \\ 0 \\ 1 \end{bmatrix}.
\end{equation}

In the same way, an $n$-qubit (pure) state is a vector in $\mathcal{H}=(\mathbb{C}^2)^{\otimes n}=\mathbb{C}^{2n}$ and it can be expressed in full generality as a normalized superposition of the $2^n$ basis states
\begin{equation}
    \ket{\psi} = \sum_{i=0}^{2n-1} c_i \ket{i}\,, \quad c_i \in \mathbb{C}\,, \quad \sum_{i=0}^{2^n-1}\abs{c_i}^2 = 1
\end{equation}
where the state $\ket{i}$ is a shorthand to denote multi-qubit computational basis states, where $i$ is the decimal representation of the binary string (or \textit{bit-string}) of zeros and ones composing the basis\footnote{For example, the two-qubit state $\ket{11}$ is denoted as $\ket{11} \rightarrow \ket{3}$, as $11_{2}$ in basis 2 is $3_{10}$ in basis $10$ (the subscript denotes the basis).}.\\

\paragraph{Two-qubit operations} As for single qubits, an operation on a $n$-qubit state can be represented by a unitary matrix in $\qty(\mathbb{C}^2)^{n} \times \qty(\mathbb{C}^2)^{n}$, and any such matrix is a valid multi-qubit quantum gate. However, instead of considering general unitaries, one usually constructs multi-qubit gates starting from single- and two-qubit ones, since these form a so-called \textit{universal} set of gates. We elaborate more on this concept at the end of the section and now proceed to describe common two-level gates. 

First, two independent single-qubit operations acting on two different qubits are cast in the form of a single two-qubit gate via the tensor product operation. For example, the action of two Pauli-$Z$ gates in parallel on a system of two qubits is described by the operator
\begin{equation}
    Z \otimes Z = \begin{bmatrix}1&0\\0&-1\end{bmatrix} \otimes \begin{bmatrix}1&0\\0&-1\end{bmatrix} = \begin{bmatrix} 1 \cdot \begin{bmatrix}1&0\\0&-1\end{bmatrix} & \phantom{-}0\cdot\begin{bmatrix}1&0\\0&-1\end{bmatrix} \\[0.5cm] 0\cdot\begin{bmatrix}1&0\\0&-1\end{bmatrix} & -1\cdot\begin{bmatrix}1&0\\0&-1\end{bmatrix} \end{bmatrix}
    = \begin{bmatrix}1&0&0&0\\0&-1&0&0\\0&0&-1&0\\0&0&0&1\end{bmatrix}\,.
\end{equation}

At the basis of every quantum computation are the two-qubit controlled operations that are used to create \textit{entangled} states. These operations act on the state of a \textit{target} qubit conditionally on the state of another qubit, called \textit{control}. The prototypical gate in this class is the controlled-NOT operation (CNOT or C$X$), which does nothing ---i.e., acts with the identity--- if the control qubit is in the state $\ket{0}$, and acts with the $X$ gate on the target if the control is in the $\ket{1}$ state instead. Its definition and matrix representation is
\begin{align}
    \text{CNOT} \coloneqq \dyad{0} \otimes \mathbb{I} + \dyad{1} \otimes X &= 
    \begin{bmatrix}1\\0\end{bmatrix}\overbrace{\begin{bmatrix}1&0\end{bmatrix}}^{\bra{0}} \otimes \begin{bmatrix}1&0\\0&1\end{bmatrix}
    + \begin{bmatrix}0\\1\end{bmatrix} \overbrace{\begin{bmatrix}0&1\end{bmatrix}}^{\bra{1}} \otimes \begin{bmatrix}0&1\\1&0\end{bmatrix} \\
&=\begin{bmatrix}1&0\\0&0\end{bmatrix}\otimes\begin{bmatrix}1&0\\0&1\end{bmatrix}+\begin{bmatrix}0&0\\0&1\end{bmatrix}\otimes\begin{bmatrix}0&1\\1&0\end{bmatrix}\\ 
& = \begin{bmatrix}1&0&0&0\\0&1&0&0\\0&0&0&1\\0&0&1&0\end{bmatrix}\,.
\end{align}

Similarly, the controlled-Z (C$Z$) operation does nothing if the control qubit is $\ket{0}$ and applies $Z$ to the target qubit otherwise, it is defined as $\text{C}Z \coloneqq \dyad{0}\otimes\mathbb{I}+\dyad{1}\otimes Z$. It can be checked that the CZ has the same action if the control and target are exchanged, since its overall effect is to add a minus sign to $\text{C}Z\ket{11}=-\ket{11}$, while leaving the remaining three states in the computational basis unchanged. Indeed, its graphical representation, shown in Fig.~\ref{tab:2q_gates}, is symmetrical and does not distinguish between a target and a control. In general, one can define a controlled version of any single qubit operation $U$, via $\text{C}U \coloneqq \dyad{0}\otimes\mathbb{I}+\dyad{1}\otimes U$. 

One last very important operation is the SWAP gate, whose action is to exchange the state of two qubits $\text{SWAP}\ket{\psi}\otimes\ket{\phi} = \ket{\phi}\otimes \ket{\psi}\,, \forall \ket{\psi},\ket{\phi} \in \mathbb{C}^2$. It can be checked that this operation can be implemented using a sequence of three CNOTs with alternating target and control, as shown in Fig.~\ref{tab:2q_gates}. In Fig.~\ref{tab:2q_gates} we summarise some common two qubits gates, showing their circuital form and matrix representation in the computational basis.\\
\begin{table}[ht]
    \centering
    \setlength{\defaultaddspace}{10pt}
    \addtolength{\tabcolsep}{5pt} 
    \begin{tabular}{ccc}
    \toprule
    \textbf{Name (abbr.)} & \textbf{Symbol/Circuital rep.} & \textbf{Matrix representation} \\
    \midrule
    Controlled-NOT (CNOT) & \begin{tikzcd} & \ctrl{1} &\qw \\  & \targ{} &\qw \end{tikzcd} & $\begin{bmatrix}1&0&0&0\\0&1&0&0\\0&0&0&1\\0&0&1&0\end{bmatrix}$\\\addlinespace
    Controlled-Z (C$Z$)     & \begin{tikzcd} & \ctrl{1} & \qw \\  & \control{} & \qw \end{tikzcd} & $\begin{bmatrix}1&0&0&0\\0&1&0&0\\0&0&1&0\\0&0&0&-1\end{bmatrix}$\\\addlinespace
    Controlled-phase (C$P$)   & \begin{tikzcd} & \ctrl{1} &\qw \\  & \gate{P(\varphi)} &\qw \end{tikzcd} & $\begin{bmatrix}1&0&0&0\\0&1&0&0\\0&0&1&0\\0&0&0&e^{i\varphi}\end{bmatrix}$\\\addlinespace
    Controlled-$U$ (C$U$)   & \begin{tikzcd} & \ctrl{1} &\qw \\  & \targ{} &\qw \end{tikzcd} & $\begin{bmatrix}1&0&0&0\\0&1&0&0\\0&0&U_{00}&U_{01}\\0&0&U_{10}&U_{11}\end{bmatrix}$\\\addlinespace
    Swap (SWAP)           & \begin{tikzcd} & \swap{1} &\qw \\  & \targX{} &\qw \end{tikzcd} = \begin{tikzcd} & \ctrl{1} & \targ{} & \ctrl{1} & \qw \\ & \targ{} & \ctrl{-1} & \targ{} & \qw \end{tikzcd} & $\begin{bmatrix}1&0&0&0\\0&0&1&0\\0&1&0&0\\0&0&0&1\end{bmatrix}$\\\addlinespace
    \bottomrule
    \end{tabular}
    \caption[Summary table of two-qubits gates]{Summary of the most important two qubits operations. The name of the operation is shown, the symbol used to refer to the operation when implemented as a \textit{gate} in a quantum circuit, and the matrix representation of the gate in the computational basis.}
    \label{tab:2q_gates}
\end{table}

\paragraph{One and two-qubit gates are universal}
\label{sec:1q2q_universality}
It can be proven that any $n$-qubit unitary can be written as a product of just two-qubit operations, and further that any such two-level unitary can be approximated \textit{efficiently} with a composition of single-qubit gates and two-qubit controlled operations~\cite{NielsenChuang, BarencoGates1995}. This means that a restricted set of operations, namely single qubit gates and a controlled operation like the CNOT, is sufficient to implement an arbitrary $n$-qubit computation, and for this reason, such a pool of operators is called a \textit{universal} gate set. 

Examples of universal gate sets are $\{H, T, \text{CNOT}\}$ or $\{R_x, R_y, R_z, \text{CNOT}\}$, but there exist several different ways of combining single-qubit operations and two-qubit interactions to achieve universality. A closely related concept is that of \textit{native} or \textit{basis} gates, which are the set of physical transformations that can actually be performed on quantum computing hardware, and depends on the specific technology used to build the hardware (superconducting, ion traps, neutral atoms, ...). We briefly touch upon this theme in Sec.~\ref{sec:QuantumCircuits} when discussing the quantum circuit model. 

\subsection{Density matrix formalism}
So far we have described quantum states as vectors in a Hilbert space, yet there exists an equivalent, and often more appropriate, description of quantum states via operators, which go by the name of \textit{density matrices}. This alternative formalism arises quite naturally in quantum mechanics when one wants to account for uncertainties in the knowledge of a quantum state. 

Consider a set of quantum states $\{\ket{\psi_i}\}$ and a process that selects a state $\ket{\psi_i}$ from such a set with probability $p_i \in [0,1]$, so that all probabilities sum up to one $\sum_i p_i = 1$. Given that each state is drawn probabilistically from the set, one can describe this statistical uncertainty by considering a weighted \textit{mixture} of the states in the set. Formally, given an ensemble of states with corresponding probabilities $\{p_i, \ket{\psi_i}\}_i$, the \textit{density matrix} which describes the quantum state of the system is given by the convex combination
\begin{equation}
    \rho = \sum_i p_i \dyad{\psi_i}\,\quad \sum_{i}p_i = 1\,.
\end{equation}
Such a state is called \textit{mixed} state, and is used to describe the statistical uncertainty one could have about the state of a quantum system. These are opposed to the \textit{pure} states described so far, which are instead used to describe systems whose state is known exactly, with no statistical uncertainty associated with it. 

By definition, let $\mathcal{H}$ be the Hilbert space associated with a quantum system, density matrices are linear squared operators on $\mathcal{H}$ that are positive semidefinite $\rho \geq 0 $ with unit trace $\Tr[\rho]=1$, and are used to represent the quantum state of the system. The density matrix notation can also easily take into account pure states, represented by projectors $\rho = \dyad{\psi}$. Indeed, one defines the \textit{purity} of a quantum state $\rho$ as $\Tr[\rho^2]$, which is highest when the state is pure
\[\Tr[\rho^2] = \Tr[\dyad{\psi}^2] = \Tr[\dyad{\psi}] = 1\,,\]
and reaches the minimum value of $1/d$ when the quantum system is in the so-called \textit{completely mixed} state $\rho = \mathbb{I}/d$, where $d$ is the dimension of the Hilbert space of the system\footnote{This fact can be easily proven by considering an eigendecomposition of the density matrix, and using Lagrange multipliers to minimize the purity $\Tr[\rho^2] = \sum_{i=1}^d p_i^2$ under the constraint $\sum_{i=1}^d p_i = 1$.} ($d=2^n$ for a system of $n$ qubits).

As discussed in Chapter~\ref{ch:NoiseDeconvolutionChapter}, density matrices are of fundamental importance when dealing with \textit{quantum channels}, a generalisation of the unitary evolution for open quantum systems, and the backbone of the mathematical description of noise processes acting on quantum systems.\\

\paragraph{Bloch representation of a single qubit mixed state}
\label{par:ch_QC_BlochSphere}
We argued earlier that the Pauli matrices together with the identity form a basis of the space of $2\times 2$ complex matrices, and in fact the state of a qubit can be expanded in this basis as the linear combination
\begin{equation}
\label{eq:bloch_state}
    \rho = \frac{\mathbb{I} + r_x X + r_y Y + r_z Z}{2}\,.
\end{equation}
where the condition $\Tr[\rho]=1$ along with the fact that the Pauli matrices are traceless imposes that the identity appears with coefficient $1/2$, and $r_x, r_y, r_z \in \mathbb{R}$ because of the positivity of the density matrix and the hermiticity of the Pauli matrices. If the state is pure, then
\begin{equation}
\label{eq:bloch_vector_pure}
    1 = \Tr[\rho^2] = \Tr[\qty(\frac{\mathbb{I} + r_x X + r_y Y + r_z Z}{2})^2] = \frac{1 + r_x^2 + r_y^2 + r_z ^ 2}{2} \implies r_x^2 + r_y^2 + r_z ^ 2 = 1.
\end{equation}
This is the equation of a sphere of radius one, and this is, indeed, another way to derive the Bloch sphere representation of a qubit of Fig.~\eqref{fig:bloch_sphere}. On the contrary, if the state is not pure then the purity is less than one, which is equivalent to the condition
\begin{equation}
\label{eq:bloch_vector_mixed}
\Tr[\rho^2] \leq 1 \implies r_x^2 + r_y^2 + r_z^2 = \vec{r}^2 \leq 1\,,
\end{equation}
where we have introduced the \textit{Bloch vector} $\bm{r} \coloneqq \qty(r_x,\, r_y,\, r_z)$, consisting of the coordinates of the quantum state along the Pauli axes. For example, the ground state $\rho = \dyad{0}$ has Bloch vector $\bm{r}=(0,\,0,\,1)$, since it can be written in terms of the Pauli matrices as $\rho = (\mathbb{I} + Z) / 2 = \dyad{0}$. One can check that a general quantum state of the form in Eq.~\eqref{eq:qubit_psi_bloch} has Bloch vector $\bm{r} = (\cos\varphi \sin\theta,\, \sin\varphi\sin\theta,\, \cos\theta)$\,.

Equations~\eqref{eq:bloch_vector_pure} and~\eqref{eq:bloch_vector_mixed} then tell that pure states live on the surface of the Bloch sphere of Fig.~\ref{fig:bloch_sphere}, while mixed states correspond to points inside the sphere. The centre of the sphere is the completely mixed state $\rho = \mathbb{I}/2$, with the corresponding Bloch vector $\bm{r}=(0,\,0,\,0)$.\\

\paragraph{Evolution}
While pure states evolve after the application of a unitary operation $U$ as $\ket{\psi} \xrightarrow{U} U\ket{\psi}$, the state obtained by acting with a unitary operation on a quantum system described by density matrix $\rho$ is instead $\rho \xrightarrow{U} U\rho U^\dagger$.\\

\paragraph{Reduced density matrices}
One is often interested in studying the state of just a subsystem (for example, a single qubit) of a larger quantum system consisting of multiple parts. Let $\rho_{AB}$ be the density matrix of a bipartite quantum system with Hilbert space $\mathcal{H}_A \otimes \mathcal{H}_B$, where $A$ and $B$ denote the subsystem we are interested in. One defines the \textit{reduced density matrix} of the system $A$ the density operator $\rho_A$ given by the partial trace over the uninteresting degrees of freedom of $B$, namely
\begin{equation}
    \rho_A \coloneqq \Tr_B[\rho_{AB}] = \sum_{i=1}^{d_B} (\mathbb{I}_A \otimes \bra{\psi_i}_B)\, \rho_{AB}\,  (\mathbb{I}_A \otimes \ket{\psi_i}_B) \equiv \sum_{i=1}^{d_B} \mel{\psi_i}{\rho_{AB}}{\psi_i}_B
\end{equation}
where $\Tr_B[\cdot]$ indicates the partial trace operation as defined above, $d_B = |\mathcal{H}_B|$ is the dimension of the Hilbert space of $B$, $\{\ket{\psi_i}_B\}_i$ is an orthonormal basis in $\mathcal{H}_B$. 

One can check that the reduced density operator defined above is a valid quantum state as it is positive semi-definite $\rho_A \geq 0$ and with trace equal to one $\Tr[\rho_A] = 1$\,.

\subsection{Measurements and expectation values}
\label{sec:measure_and_expectation}
Measurements are processes that extract classical information from quantum states. These usually occur at the end of a quantum computation (but also during it, depending on the task) to read out the final outcome of the implemented data processing task.
The measurement process in quantum mechanics is a very delicate topic at a fundamental level since it involves the problem of the quantum to classical transition, but here we take an operational formulation following the standard probabilistic interpretation of quantum mechanics stemming from \textit{Born's rule}, used to determine probabilities of measurement outcomes. 

The probability that a quantum system described by quantum state $\ket{\psi} = \sum_i c_i \ket{i}$ when measured in the computational basis reduces to state $\ket{k}$ is given by
\begin{equation}
    \label{eq:BornRule}
    p_k = \abs{\braket{k}{\psi}}^2 = \abs{c_k}^2\,.
\end{equation}

In addition to measurements in the computational basis, a measure can be used to infer some physical properties of the quantum system under investigation. Since these properties can only take real values (as opposed to complex), the set of physical quantities to which an observer has access, which are referred to as \textit{observables}, are represented mathematically by Hermitian operators $O = O^\dagger$, which have real eigenvalues. Examples of common observables are the Pauli matrices $X, Y$, and $Z$. The \textit{expectation value} of an observable $O$ on a quantum system described by pure state $\ket{\psi}$ is
\begin{equation}
    \label{eq:observable_expectation_values_pure}
    \expval{O} \coloneqq \mel{\psi}{O}{\psi}\, .
\end{equation}
More generally, in the density matrix formalism, expectation values instead read
\begin{equation}
    \label{eq:observable_expectation_values_mixed}
    \expval{O} \coloneqq \Tr[O \rho]\, ,
\end{equation}
where this formula clearly reduces to the first one for pure states $\Tr[O\dyad{\psi}] = \mel{\psi}{O}{\psi}$. Since observables are Hermitian operators, $O$ admits a spectral decomposition $O = \sum_{i}o_i\dyad{o_i}$, and so Eq.~\eqref{eq:observable_expectation_values_mixed} can be also expressed explicitly as
\begin{equation}
    \label{eq:observable_expectation_values_mixed2}
    \Tr[O\rho] = \Tr[\sum_{i}o_i\dyad{o_i}\,\rho] = \sum_{i}o_i \mel{o_i}{\rho}{o_i}\, = \sum_{i} o_i~p_i\,.
\end{equation}
where $p_i = \mel{o_i}{\rho}{o_i}$ is the probability of measuring the state $\rho$ in $\ket{o_i}$. 

In real experiments, the estimation of the expectation value of an observable is a resourceful multi-step process that requires the ability to repeatedly prepare and measure the quantum state, and then combine the measurement outcomes via classical post-processing. In the quantum computing jargon, a single act of measure is called \textit{shot}, and the overall number of measurement shots used in an experiment, often indicated with $M$, affects the statistical accuracy of the estimation, which scales as $\order{1/\sqrt{M}}$. In Algorithm~\ref{alg:expec_values} we summarise the steps needed to estimate the expectation value of an observable $O$ on a quantum state $\rho$, and we now proceed to explain them in detail.
\begin{algorithm*}
\label{alg:measurement_scheme}
\caption{Estimate expectation value of an observable}\label{alg:expec_values}
\KwData{Quantum state $\rho \in \mathbb{C}^d\times\mathbb{C}^d$; observable $O$, number of shots $M$.}
\KwResult{$\bar{O} \approx \expval{O} = \Tr[O\rho]$ with associated statistical error $\order{1/\sqrt{M}}$.}
\For{$m=1, \hdots, M$}{
Prepare $\rho$;\\
Apply a change of basis operation $U$ on $\rho$, such that $U^\dagger O U = \sum_{i=0}^{d-1}o_i\dyad{i}$ is diagonal in the computational basis;\\
Do projective measurement on computational basis $\{\ket{i}\}_{i=0}^{d-1}$, find state $\ket{k}$;\\
Store result $r_m = o_k$;\\
}
Classically compute the sample mean $\bar{O} \coloneqq \frac{1}{M}\sum_{m=1}^M r_m$.
\end{algorithm*}

First, it is required that the experimenter has access to many copies of the quantum state $\rho$, or can prepare it efficiently multiple times. The preparation of the quantum state is then followed by a \textit{change of basis} that makes the observable $O$ diagonal in the computational basis, or equivalently, expresses $\rho$ in the eigenbasis of the observable. This is usually a required step, as common quantum computing hardware can only perform projective measurements on the computational basis, and so one has to reframe every estimation procedure on this basis. Since $O$ is Hermitian, there is a unitary $U$ that diagonalizes the observable as $O \rightarrow O' =  U^\dagger O U = \sum_i o_i\dyad{i}$. Then, by rotating the quantum state $\rho \rightarrow \rho' = U^\dagger \rho U$ and measuring the diagonal ---in the computational basis--- observable $O'$, one correctly obtains the desired expectation value
\begin{equation}
    \sum_{i=0}^{d-1} o_i \mel{i}{U^\dagger \rho U}{i} = \Tr[\sum_{i=0}^{d-1} o_i \dyad{i} U^\dagger \rho U] = \Tr[U^\dagger O U\, U^\dagger \rho U] = \Tr[O\rho] = \expval{O}\,,
\end{equation}
where the first term on the left is the definition of $\Tr[O'\rho']$ from Eq.~\eqref{eq:observable_expectation_values_mixed2}, which, as required, only involves measurement along the computational basis $\{\ket{i}\}$. 

As an example, consider the simple case of measuring the expectation value of the Pauli-$X$ operator on a single-qubit quantum state $\rho$. Using $Z = H X H$, it is easy to check that a rotation of the qubit with a Hadamard gate $\rho \rightarrow \rho'=H\rho H$, followed by a measurement of the experimentally accessible Pauli-$Z$ operator, correctly yields the desired expectation value $\expval{X}$, in fact
\[
 \expval{Z}_{\rho'} = \Tr[Z\, \rho'] = \Tr[Z\, H\rho H] = \Tr[H Z H\,\rho] = \Tr[X \rho] = \expval{X}_{\rho}\,.
\]
We remark that the change of basis step is only a practical requirement to circumvent the hardware constraints that only allow for measurements on the computational basis, but adds nothing fundamental to the computation. In addition, note that the operator $U$ that diagonalizes the observable $O$ has to be calculated classically, before the measurement procedure begins. 

After preparation of the state and subsequent change of basis, the system is finally measured in the computational basis. Suppose that upon measurement the state is found in state $\ket{k}$ corresponding to the eigenvalue $o_k$: this number constitutes the result of the measurement. By repeating the entire preparation and measurement procedure $M$ times, one gathers a sample of measurement results $\{r_1,\, \hdots,\, r_M\}$, where each result $r_m \in \{o_0,\, o_1,\, \hdots,\, o_{d-1}\}$ is the eigenvalue obtained on the $m$-th measurement shot. The values $r_m$ are \textit{independent} ---because they come from independent measurement events--- random variables whose statistical properties are defined by 
\begin{equation}
\label{eq:outcomes_moments}
    \mathbb{E}[r_m] = \mathbb{E}[O] \coloneqq \expval{O}\,,\quad 
    \text{Var}[r_m] = \text{Var}[O] \coloneqq \expval{O^2} - \expval{O}^2\,, \quad \forall~m=1,\hdots, M
\end{equation}
where the expectation values and variances of the random variables $r_m$ are evaluated over the probability distribution of measurement outcomes given by Born's rule~\eqref{eq:BornRule}. The sample mean $\bar{O} \coloneqq \sum_{m=1}^M r_m / M$ is an  unbiased estimator of the true expectation value $\expval{O}$, with variance
\begin{align}
    \mathbb{E}[\bar{O}] &= \frac{1}{M}\sum_{m=1}^M \mathbb{E}[r_m] = \expval{O}\,,\\
    \text{Var}[\bar{O}] &= \frac{1}{M^2}\sum_{m=1}^M \text{Var}[r_m] = \frac{1}{M}\text{Var}[O]\,,\label{eq:outcomes_var}
\end{align}
where we have used Eqs.~\eqref{eq:outcomes_moments}, and the fact that the $r_m$ are independent random variables to move the variance inside the sum in the second equation. 

Assuming that the variance of the observable is bounded and independent of the system size, as it happens for observables made of tensor product of Pauli matrices\footnote{This can be easily proved as follows. Let $ O = \sigma_1 \otimes \sigma_2 \otimes \cdots \otimes \sigma_n$ be an observable made of tensor products of single-qubit Pauli matrices $\sigma_i \in \{\mathbb{I}, X, Y, Z\}$. Since Pauli matrices are involutory, $\sigma_i^2 = \mathbb{I}$, then $O^2 = \mathbb{I}_{n}$. In addition, since the eigenvalues of $O=\sum_i o_i \dyad{o_i}$ are either $o_i \in \{\pm 1\}$, expectation values $\expval{O}$ are bounded
\begin{equation}
\label{eq:bound_pauli}
    \abs{\expval{O}} = \abs{\Tr[O\rho]} = \abs{\sum_i o_i~\mel{o_i}{\rho}{o_i}} \leq \sum_i \abs{o_i} \mel{o_i}{\rho}{o_i} = \sum_i \mel{o_i}{\rho}{o_i} = 1\,,
\end{equation}
where we have used that $\mel{o_i}{\rho}{o_i} \geq 0$ are positive numbers because $\rho \geq 0$ is a positive operator, the eigenstates $\{\ket{o_i}\}_i$ are a basis of the space, and the state is normalized $\Tr[\rho]=1$. Thus, finally one has $0 \leq \text{Var}[O] = \Tr[O^2\rho] - \Tr[O\rho]^2 = 1 - \Tr[O\rho]^2 \leq 1\,,$ where again we have used $\Tr[\rho]=1$ and $0 \leq \Tr[O\rho]^2 \leq 1$ from~\eqref{eq:bound_pauli}.}, then
\begin{equation}
    \text{Var}[O] = \expval{O^2} - \expval{O}^2 = \Tr[O^2\rho] - \Tr[O\rho]^2 \in \order{1}\,,
\end{equation}
and so the statistical error associated to the empirical mean $\bar{O}$ scales as $\text{Std}[\Bar{O}] \in \order{1/\sqrt{M}}$. Alternatively, suppose one wants to calculate the expectation value with precision $\varepsilon$ with a failure probability of at most $\delta$, then by Chebyshev's inequality
\begin{equation}
    P\qty(|\bar{O}-\expval{O}| > \varepsilon ) \leq \frac{\text{Var}[\bar{O}]}{\epsilon^2} = \delta \implies M \in \order{\frac{1}{\delta \varepsilon^2}}\,,
\end{equation}
which states that the number of measurements $M$ to reach a target accuracy $\varepsilon$ scales quadratically with the desired precision. A similar scaling can also be obtained with other statistical inequalities, for example via Hoeffding's inequality~\cite{HuangNISQLinear_2019}.

We conclude by noticing that in practical scenarios one is often interested in measuring either simple observables made of tensor products of Pauli matrices, often referred to as \textit{Pauli strings} 
\begin{equation}
     O = \sigma_1 \otimes \sigma_2 \otimes \cdots \otimes \sigma_n\,,
\end{equation}
where $n$ is the number of qubits in the system, or in weighted sums of these Pauli strings 
\begin{equation}
    \label{eq:obs_pauli_strings}
    O = \sum_{k=1}^P \gamma_k~O_k = \sum_{k=1}^P \gamma_k~\sigma_1^{(k)} \otimes \sigma_2^{(k)} \otimes \cdots \otimes \sigma_n^{(k)}\,,
\end{equation}
where the coefficients are real to ensure Hermiticity $\gamma_k \in \mathbb{R}$, and $\sigma_i^{(k)} \in \{\mathbb{I}, X, Y, Z\}$ are single-qubit Pauli matrices. In this last case, very prominent in quantum chemistry applications~\cite{GuillePOVM_2021}, the usual approach is to estimate each Pauli string separately $\expval{O_k}$, and then combine the results classically with the coefficients $\gamma_k$, via $\expval{O} = \sum_k \gamma_k \expval{O_k}$. 

Clearly, measuring a single Pauli string is much simpler, and the estimation is associated with a lower variance compared with the case of more complex observables when a fixed budget of measurement shots is allowed. Various strategies have been proposed in the literature to optimize the measurement resources needed to estimate expectation values of the form in~\eqref{eq:obs_pauli_strings}, most of them based on 
a clever grouping of commuting Pauli strings that can be measured at the same time~\cite{Bharti2021NIQSReview}, or for example using adaptive methods to reduce the statistical fluctuations associated with the estimation~\cite{GuillePOVM_2021}.

\subsection{The quantum circuit model\label{sec:QuantumCircuits}}
As discussed earlier, a generic quantum computation can be expressed in terms of single- and two-qubit operations, which makes it easy to represent it graphically via a sequence of a few circuital symbols, already introduced in Tables~\ref{tab:1q_gates} and ~\ref{tab:2q_gates}.

In quantum circuit diagrams, time evolves from left to right, single wires are used to indicate separate quantum systems (qubits in our case), and the single- and two-qubit gates are represented with the symbols shown in the corresponding summary tables. Unless otherwise stated, qubits are assumed to start in the ground state, that is all qubits are in the $\ket{0}$ state when the computation is started. At last, at the end of the circuit some or all the qubits may be measured, in which case an appropriate symbol is shown on the corresponding wire. Unless explicitly said otherwise, every measurement is assumed to be in the computational basis.

A simple example of a quantum circuit is the following, which creates and measures the three-qubit \textit{maximally entangled} GHZ state $\ket{\psi_{\text{GHZ}}} = (\ket{000} + \ket{111}) / \sqrt{2}\,$.
\[
\begin{quantikz}[column sep = 0.2cm, row sep = 0.1cm]
    \lstick{$\ket{0}$} & \gate{H} & \ctrl{1} & \qw      & \meter{}\\
    \lstick{$\ket{0}$} & \qw      & \targ{}  & \ctrl{1} & \meter{}\\
    \lstick{$\ket{0}$} & \qw      & \qw      & \targ{}  & \meter{}
\end{quantikz}
\]        

\subsubsection{Executing a quantum circuit on a real quantum device} 
\label{ssec:real_execution}
The quantum circuit model is a useful tool to describe the logical action of quantum computation through a pictorial representation. Furthermore, although quantum circuits are the standard format for submitting instructions to a real quantum computer for execution, they are by no means a faithful description of what actually happens on the device. 

In fact, for a logical quantum circuit to be executed on a real device, some classical preprocessing steps are needed. First, it is necessary to identify a subset of the available physical qubits on the machine that matches the connectivity required by the two-qubits gates in the circuit to be run. Secondly, one has to express every gate in the original circuit in terms of the so-called \textit{native basis gates}, which are the set of physical operations that can actually be implemented on the device. These strongly depend on the actual technology used to build the quantum computer, and in Table~\ref{tab:HW_basis_gates} we report the native gates available on some common quantum computing hardware. Following the discussion in Sec.~\ref{sec:1q2q_universality}, these single and two-qubit operations provide different universal sets of gates that can implement any quantum computation.

\begin{table}[h]
    \centering
    \begin{tabular}{cccc}
    \toprule
    Manufacturer & Technology & Native gates & Refs. \\
    \midrule
    IBM    & Superconducting & $\qty{\sqrt{X}, X, R_Z, \text{CNOT}}$ & \cite{IBMQuantum} \\
    Google & Superconducting & $\qty{U(\theta, \varphi, \lambda),\, \text{Sycamore gate}}$ & \cite{GoogleBasisGates, GoogleSupremacy2019}\\
    IonQ   & Trapped-ions & $\qty{R_X,\, R_Y,\, \text{M{{\o}}lmer-S{\o}rensen} }$ & \cite{IONQBasisGates, IONQBenchmarking11qubit2019}\\
    \bottomrule
    \end{tabular}
    \caption[Examples of native gates on real quantum hardware]{Examples of single- and two-qubits gates available on some current quantum computing hardware. The $U(\theta, \phi, \lambda)$ operation is the general single-qubit rotation of~\eqref{eq:qubit_rotation}, and the ``Sycamore gate" is a two-qubit gate similar to a combination of a SWAP gate and a controlled phase rotation. The M{{\o}}lmer-S{\o}rensen gate~\cite{MolmerSorensenGate} is a two-qubit gate of the form $XX(\phi) = e^{-i \phi X\otimes X / 2}$ common in ion-trap based architectures. The single-qubit rotation gates on available on IONQ devices are variations of $R_X$ and $R_Y$ rotations, for a precise definition see~\cite{IONQBasisGates}.}
    \label{tab:HW_basis_gates}
\end{table}

This process of rewriting the circuit into an appropriate form goes by the name of circuit \textit{compilation}, or \textit{transpilation}, and its result is to output a new quantum circuit which is (approximately) equivalent to the original one, but that can be readily executed on the machine. As discussed in the next section, current quantum hardware is limited in both size and performance, and a quantum circuit may lend itself to better execution on specific backends, due to a more favourable mapping in terms of required qubit-qubit connection topology, and/or available gates.

In the next section, we provide a brief overview of the state of the art of current quantum computing technology, discussing both the technological aspects as well as their scientific relevance. 

\subsection{The NISQ era of quantum computation}
\label{sec:nisq}
\begin{figure}[ht]
    \centering
    \includegraphics[width=\textwidth]{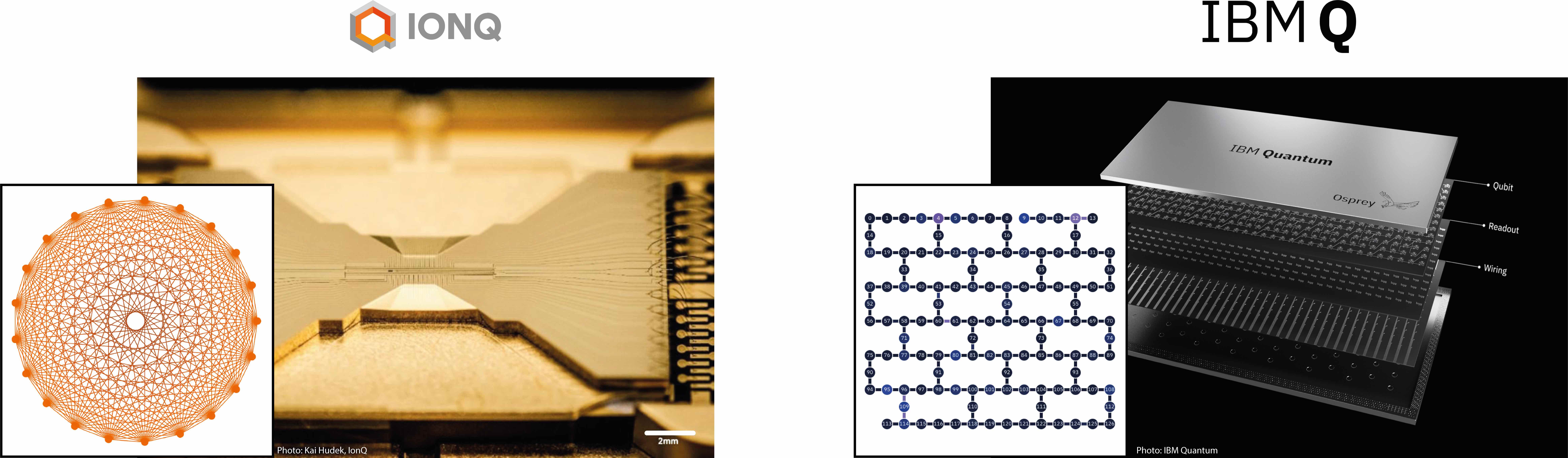}
    \caption[Examples of currently available quantum computers]{Examples of currently available quantum computers based on two different technologies: trapped-ions (IonQ)~\cite{IONQ} and superconducting circuits (IBM Quantum)~\cite{IBMQuantum}. In the panels next to the devices, examples of the usual connectivity between qubits found in the respective technologies are shown, specifically the \texttt{ Aria} device with $n=21$ qubits with an all-to-all connectivity~\cite{IONQ_Aria}, and IBM's \texttt{ibm\_washington} device with $n=127$ qubits leveraging a grid-like connectivity~\cite{IBMQuantum_Eagle}.}
    \label{fig:nisq_chips}
\end{figure}

Although enormous technological progress made it possible to build the first generations of quantum computers, these are still far from the realisation of universal fault-tolerant quantum computers, which are ideal quantum computers in which errors in the computation can be detected and corrected, through a procedure called Quantum Error Correction (QEC)~\cite{NielsenChuang}. 

Instead, the current era of quantum computation has been dubbed \textit{Noisy Intermediate-Scale Quantum}~\cite{Preskill2018NISQ}, \textit{NISQ} for short, which is used to denote near-term quantum devices which are imperfect and not error corrected, hence are strongly affected by noise, and can only leverage a limited number of qubits, and thus are also limited in scale. Several experimental platforms are being used to develop quantum computing solutions~\cite{TacchinoSimulationReview2020, Braket}, based, for example, on superconducting circuits~\cite{IBMQuantum, GoogleSupremacy2019}, trapped-ions~\cite{IONQ, Quantinuum}, neutral atoms~\cite{Henriet2020_Pasqal, QuEra}, photonic chips~\cite{Xanadu, Qandela} and others\footnote{In this list we omit computing platform based on quantum annealing, like D-Wave's machines~\cite{Dwave}, as they cannot implement universal quantum computation.}. In Fig.~\ref{fig:nisq_chips} we report two examples of currently available quantum computers based on trapped ions and superconducting circuits, which are arguably the leading technologies for constructing quantum computers in the NISQ era. In the figure, both the physical Quantum Processing Units (QPUs), and a graphical representation of the layout of the qubits on such chips.

Although the pace at which these machines are developed suggests a quantum variant of the classical Moore's law to hold, NISQ devices still face serious constraints, which we briefly discuss in what follows. First of all, state of the art quantum computers are limited in size, consisting of about dozens to a few hundred qubits depending on the technology, with first generations of QPUs with more than a thousand qubits only planned to appear in the following years~\cite{IBMQuantum_Roadmap}. In order to reach a clear sign of quantum advantage and surpass the regime of classical simulability, it is necessary to be able to efficiently scale up these machines. For example, it is estimated that roughly 20 million \textit{physical} qubits ---as opposed to \textit{logical}\footnote{By \textit{physical} qubit one refers to a single noisy hardware implementation of a two-level quantum system corresponding to a qubit. These are opposed to \textit{logical} qubits, which are instead ideal versions of the qubits that are immune to errors. A single logical qubit can be represented using multiple physical qubits along with quantum error correction codes~\cite{NielsenChuang}.}--- are required to implement a non-trivial application of Shor's algorithm, arguably the most convincing example of exponential speedup of quantum computation over classical one~\cite{Gidney2021_Shor}. Quantum computers of this size and capabilities will probably require new breakthroughs both on the experimental and algorithmic side, and will only appear in the next decades.

In addition, as we already discussed previously in Sec.~\ref{ssec:real_execution}, the types of operation that can be implemented on a device are often quite limited. This is especially true for two-qubit gates, which can only be applied directly to qubits that are physically close on the machine, so that they can be made to interact. The \textit{connectivity} between qubits on a device, as shown in Fig.~\ref{fig:nisq_chips}, imposes a serious constraint on the class of algorithms that can be run on a specific machine. While the limited connectivity is a clear issue for superconducting-based quantum computers, other platforms, like trapped-ions, implement an all-to-all interaction that makes it possible to operate two-qubit gates on any pair of qubits on the device. However, operations (both single- and two-qubit gates) on these machines can be orders of magnitudes slower than on superconducting chips, with the time to implement a quantum gate ranging in the order of microseconds $t \sim 1-100\mu$s for the former~\cite{Bruzewicz2019_IonTrapReview} and nanoseconds $t \sim 1-100$ns for the latter~\cite{IBMQuantum}.

The most compelling problem with near-term devices is noise that affects the qubits inside the computer, which strongly limits the computing capabilities of the hardware. Errors in quantum computers occur due to undesired interaction of the quantum systems with the external environment, as well as due to faulty execution of the operations inside the computer. The former type of noise is commonly measured by \textit{coherence times}, which characterise the \textit{quality} of the qubit by assessing the time it takes for it to lose its quantum properties due to unwanted interaction with the external environment. For example, common values of coherence times for current superconducting architectures are of about $T \sim 10-100\,\mu$s~\cite{IBMQuantum}, while trapped-ions usually have longer times of about $T \sim 1-100\,$s~\cite{IONQ}, corresponding to higher quality qubits. We will discuss more in detail how to describe noise in quantum systems in Chapter~\ref{ch:NoiseDeconvolutionChapter}. 

As for errors introduced by an imperfect realisation of the quantum gates ---especially two-qubit gates---, these are measured in terms of gate error rates via a procedure called Randomised Benchmarking~\cite{MagnesanRB_2012, HelsenRB_Review_2022}. Current error rates are of the order of $0.01\%$ and $1\%$ for single and two-qubit gates respectively for superconducting circuits, and roughly one order of magnitude less for ion traps.

It is then clear that multiple sources concur to the realisation of an effective device, namely the number of qubits, the connectivity map, and the error rates. For this reason, new figures of merit such as Quantum Volume (QV) have been introduced to provide a unifying measure to assess the capability of near-term quantum computing devices~\cite{IBMQuantumVolume_2019, Bharti2021NIQSReview}. Roughly speaking, the QV is a single real number that measures the largest square-shaped (i.e. number of qubits equal to the depth of the circuit) random quantum circuit that the quantum computer can execute successfully. It can be understood as the number of ``effective" qubits available on the machine, and it is formally defined as~\cite{IBMQuantumVolume_2019}
\begin{equation}
\label{eq:quantum_volume}
    \log _2 QV = \argmax_n \text{min}(n, D(n))
\end{equation}
where $n$ is the number of qubits and $D(n)$ is the depth of the quantum circuit, that is the minimum number of steps needed for quantum gates to be applied in parallel to implement a quantum circuit. At the time of writing, the highest Quantum Volumes reported for superconducting and trapped-ions architectures are, respectively, $QV = 512$, corresponding to running square circuits of size $\log_2 QV = 9$, for IBM's superconducting chip \texttt{Prague} with a total of $n=27$ qubits~\cite{IBM_QV_Highest}, and $QV = 8192$ ($\log_2 QV = 13$) for Quantinuum's trapped-ions based \texttt{System Model H1}~\cite{Quantinuum_QV_Highest} which has a total of $n=20$ qubits.

Every technology has its own advantages and drawbacks, and it is not yet clear which one will provide the most effective solution towards scalable and fault-tolerant quantum computers. In the meantime, near-term devices not only constitute a necessary step towards the implementation of large-scale universal quantum computers but also provide new tools to investigate the limits of quantum mechanics and provide a new paradigm for performing computation. Indeed, the topic of the next section is to discuss what NISQ devices can be used for, thus introducing \textit{variational quantum algorithms}, which are a class of quantum algorithms specifically tailored for current quantum computing devices.

\section{Variational Quantum Algorithms}
\label{sec:VQAs}
Variational Quantum Algorithms (VQAs) are the leading proposal to exploit current quantum computing platforms, based on the hope that it is possible to achieve a meaningful \textit{quantum advantage} already in the non error-corrected regime, before standard quantum algorithms, like Shor's factoring, can be realised at scale. 
Moreover, and regardless of any quantum advantage, the research on variational quantum algorithms is of interest on its own, both on a fundamental level because it conceives a new paradigm of hybrid computation where classical and quantum resources are used in tandem to achieve a task, and on a practical level because it fosters the development of new software and hardware solutions for quantum computing by providing compelling use cases.

NISQ devices are limited in size and inherently noisy, and the idea of variational quantum algorithms is to circumvent the problem simply by using them as little as possible, only for the bare minimum, while outsourcing all remaining tasks to a classical computer. This idea then identifies a class of hybrid quantum-classical algorithms in which quantum and classical computational resources are used in combination to solve a task.

The second ingredient of variational algorithms, to which they owe their name\footnote{The term ``variational" as originally proposed by Peruzzo et al.~\cite{PeruzzoVQE2014} comes from the Rayleigh-Ritz \textit{variational} principle to compute lowest energy eigenvalues. The term variational in turn comes from the idea of \textit{varying} the parameters of a trial model in order to solve a task, a concept that is at the core of modern machine learning (ML) as well. The connection between VQAs and ML was developed later as the field gained momentum, and now there is a fruitful exchange of ideas and concepts between these two topics.}, make them also very similar to the highly successful field of machine learning. In fact, variational quantum algorithms are optimisation-based procedures that tackle a problem by first encoding its solution as the minimum of an appropriately defined \textit{cost function} $C(\bm{\theta})$ depending on some tunable parameters $\bm{\theta}$, and then iteratively \textit{vary} these parameters, usually via gradient-based methods, to find the minimum of the function, hence the solution. The incredible results achieved by deep learning in recent years have shown how versatile and powerful learning-based procedures can be~\cite{LeCun2015DeepL, mnih2015human, DegraveMagneticControlTokamak2022}. Variational quantum algorithms then introduce free adjustable parameters inside a quantum computation in order to cope with the serious constraints imposed by current NISQ hardware. In this way, however, the success and/or runtime guarantees of standard textbook quantum algorithms like Grover's and Shor's are lost, as the optimisation procedures in VQAs are usually highly non-convex, and so theoretical analysis of the performances can only go so far. 

In all, VQAs essentially trade off guarantees of success with the feasibility of execution, in the hope of retaining a quantum advantage of some sort. 

\subsection{The basis of variational quantum algorithms}
At the basis of any variational algorithm is the definition of a cost, or \textit{loss}, function that measures how well the algorithm is performing, and an \textit{ansatz} circuit, which is a guess quantum circuit with tunable parameters that should be able to represent a good solution to the problem. 

Let $U(\bm{\theta})$ be the unitary of the parameterised quantum circuit (PQC) ansatz, and $\vecparam \in \mathbb{R}^p$ the vector of the parameters. The goal of VQAs is to find an optimal set of parameters that minimises the cost function $C(\vecparam): \mathbb{R}^p \rightarrow \mathbb{R}$, namely
\begin{equation}
\label{eq:vqa_argmin}
    \bm{\theta}_{\text{opt}} = \argmin_{\vecparam} C(\vecparam)\, ,
\end{equation}
where the cost is some function of the expectation values of a set of observables $\{O_k\}$ measured on the parameterized state generated by the variational circuit, that is
\begin{equation}
\label{eq:vqe_general_loss}
    C(\vecparam) = \sum_{k} f_k(\Tr[O_k\, \rho_{\vecparam}]) = \sum_{k} f_k\qty(\Tr[O_k\,  \uparam \dyad{0} \uparam^\dagger])\, .
\end{equation}
This definition highlights that the cost function depends explicitly on the parameters $\vecparam$, and implicitly actually also on the set of observables $\{O_k\}_k$ and the specific shape of the circuit ansatz $U(\cdot)$, so that $C = C(\bm{\theta}; \{O_k\}_k; U)$~\cite{CerezoPQC2021Review}. While this definition is fully general, in practical scenarios one usually considers simpler costs given by the expectation value of a single observable\footnote{Note that this simplified loss is obtained whenever the general cost in Eq.~\eqref{eq:vqe_general_loss} depends linearly on the expectation values, that is $f_k(\expval{O_k}) = c_k \expval{O_k}$. Then, by linearity of the trace and considering the overall observable $O = \sum_{k} c_k O_k$, one obtains Eq.~\eqref{eq:vqa_usual_loss}. This is the case of quantum chemistry applications, as discussed earlier at the end of Sec.~\ref{sec:measure_and_expectation}.}
\begin{equation}
\label{eq:vqa_usual_loss}
    C(\bm{\theta}) = \expval{O}_{\bm{\theta}} = \Tr[O\, \uparam\dyad{0}\uparam^\dagger]\,.
\end{equation}
Ideally, the loss function should fulfil certain desirable properties to be considered a good candidate for a cost in a variational quantum algorithm. Firstly, it should be classically hard to compute but \textit{efficiently} calculable with a quantum device, for if it were not, this would negate any hope of quantum advantage. Secondly, the cost function should be \textit{trainable}, that is there exists a procedure capable of finding the minimum of the cost with some guarantees of success. This latter condition is particularly relevant for variational algorithms, as the loss landscape of these procedures is often found to be very flat and thus difficult to navigate via standard optimization methods, a phenomenon which goes by the name of \textit{barren plateaus}, see Sec.~\ref{sec:ch_VQAs_BP}.
Finally, two other useful requirements are that the global minimum of the loss function should only be attained when the true solution to the problem is found, and not otherwise; and that the cost function should always convey a measure of the suitability of the proposed solution, so that smaller values of the cost always correspond to better solutions. When the latter two conditions are met, the cost function is often referred to as \textit{faithful} and \textit{operationally meaningful}, respectively~\cite{CerezoPQC2021Review}.

Hence, given a properly defined cost function, variational algorithms then proceed by combining quantum and classical resources in an iterative loop as follows: 
\begin{itemize}
    \item[(\textit{i})] On the quantum computer: estimate the cost function $C(\vecparam)$ for the current values of the parameters $\vecparam$ via repeated measurements;
    \item[(\textit{ii})] On the classical computer: input the outcome in a classical optimisation algorithm that proposes a new value for the parameters $\bm{\theta}'$, so that the cost is lower $C(\vecparam')<C(\vecparam)$;
    \item[(\textit{iii})] repeat steps (\textit{i})-(\textit{ii}) until stop conditions are met (convergence, execution time, ...).
\end{itemize} 

\begin{figure}[t]
    \centering
    \includegraphics[width=0.9\textwidth]{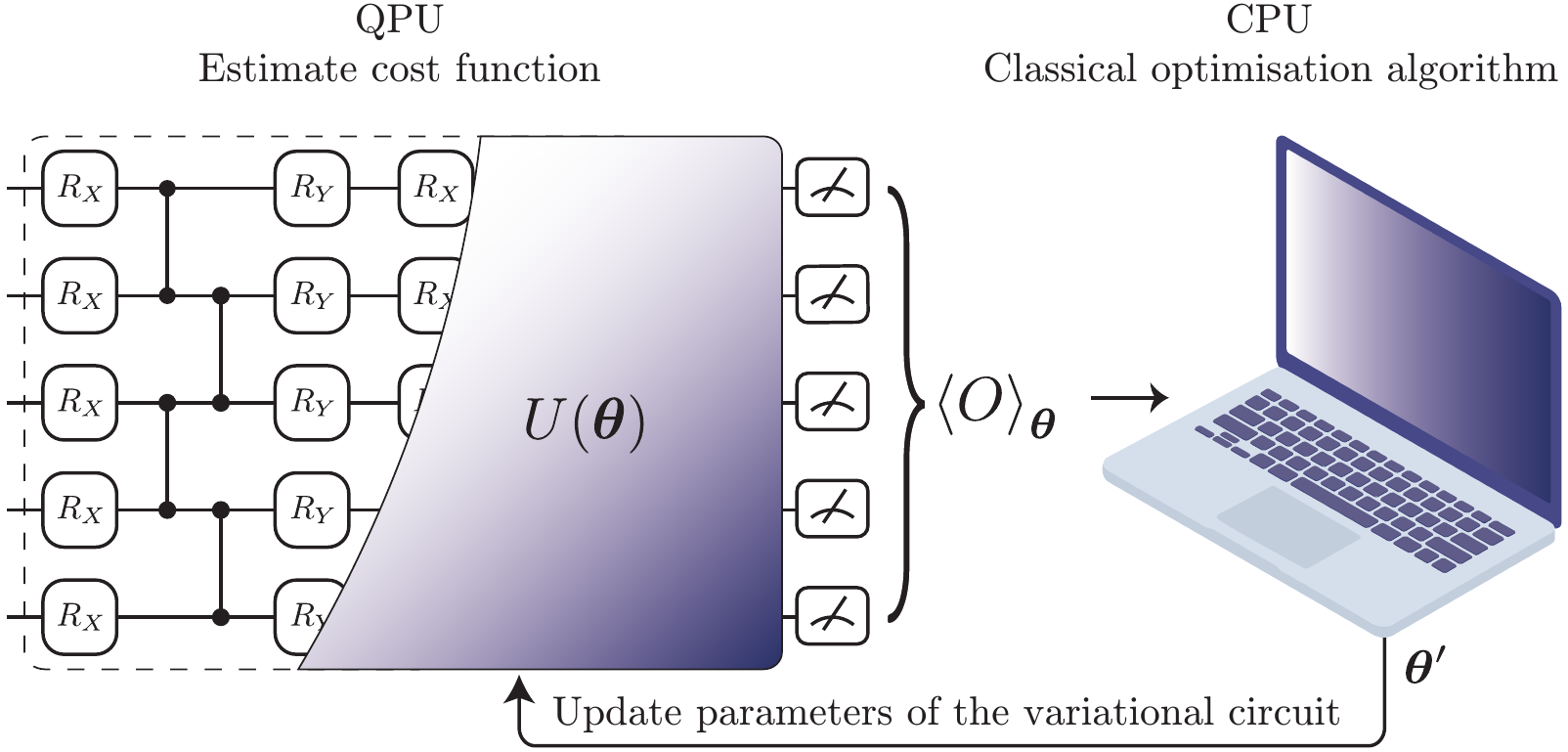}
    \caption[Schematic representation of variational quantum algorithms]{Schematic representation of variational quantum algorithms. First, on the left, a parameterised quantum circuit with parameters $\vecparam$ is used to prepare a variational trial state $\ket{\psi} = U(\vecparam)\ket{0}$ and a cost function $C(\vecparam) = \expval{O}_{\vecparam} = \mel{\psi_{\vecparam}}{O}{\psi_{\vecparam}}$ is measured on a Quantum Processing Unit (QPU). Then, on the right, the result is passed to a classical processing unit (CPU), that runs an optimisation algorithm to find new values for the parameters $\bm{\theta}'$ corresponding to a lower cost. The loop is repeated until convergence to a minimum of the cost is reached or stopping criteria are met.}
    \label{fig:vqa_main_fig}
\end{figure}

These steps are schematically represented in Fig.~\ref{fig:vqa_main_fig}, which shows the usual way of picturing the hybrid quantum-classical loop of variational quantum algorithms. 

Two questions remain to be answered. First, what is the explicit form of the parameterised quantum circuit $U(\vecparam)$ and how can one choose it? Secondly, what are the usual strategies for performing the optimisation loop? Let us now proceed by addressing these two topics. 

\subsection{Parameterised quantum ansätze}
\label{ssec:ch_VQA_ansatze}
Just as in standard parametric models, also in variational quantum algorithms the functional form of the model has to be fixed \textit{a priori}, and this is done by considering a specific parameterised quantum circuit $U(\vecparam)$. Such a choice is referred to as an \textit{ansatz} circuit, and while there is no unique prescription for constructing a good one, some guiding principles can be used to identify good candidates. 

It is clear that the choice of ansatz plays a key role in determining the effectiveness of the variational algorithm. For example, the model may be too simple to express the target solution, or it may be so complex that it makes optimisation difficult, thus favouring convergence to sub-optimal solutions. Any such ansatz prevents a good result from the variational algorithm. We hereby anticipate that these concepts of expressibility and trainability are indeed related, as will be discussed in detail in Sec.~\ref{sec:Udes_Bp_Exp}. In addition, the most important ---and almost trivial--- condition that an ansatz should satisfy is that it should be feasible in practice, in that there is a near-term quantum computing device capable of implementing it.

Generally, all ans\"{a}tze share the common feature of being defined in terms of a repeated structure of similar blocks arranged in sequential layers. For example, many variational ans\"{a}tze can be expressed as 
\begin{equation}
\label{eq:layered_ansatz}
    U(\vecparam) = \prod_{\ell=1}^L W_\ell\, U_\ell(\vecparam_\ell) = W_L U_L(\vecparam_L)\, \cdots\,  W_1 U_1(\vecparam_1)\, ,
\end{equation}
where $U_\ell(\vecparam_\ell)$ are parameterised gates, usually single-qubit Pauli rotations, and $W_\ell$ on the other hand are fixed operations usually consisting of two-qubits entangling gates, such as CNOTs or C$Z$. The number of layers $L$ in the circuit is an important hyperparameter in the definition of the ansatz and decides the overall depth of the quantum circuit. The circuit shown in Fig.~\ref{fig:vqa_main_fig} is an example of such a layered structure. 

In the following, we summarise the most common strategies used in the literature to propose ans\"{a}tze.\\

\paragraph{Hardware Efficient Ansätze (HWE)} 
This class of ansätze contains all those parameterised circuits that are specifically thought to be easily run on current quantum hardware~\cite{KandalaHWE_2017}, and are often used as a starting point for exploring a variational approach to solving a problem, especially when it is not straightforward to incorporate knowledge about the task to be solved in a circuital form. All the gates used in the ansatz are either taken from the native basis set of the device, or can be implemented with only a few of them. Most importantly, two-qubit operations act only on those pairs of qubits that are physically connected on the device, thus respecting the connectivity map available on the hardware. 

Various properties of these circuits have been studied in the literature, regarding their expressibility~\cite{SimPQCs2019}, entanglement~\cite{BallarinEntQNN_2022}, performances on benchmark problems~\cite{Hubregtsen2021EvaluationParameterizedQuantum}, as well as trainability properties~\cite{McCleanBarren2018, CerezoBarrenLocalCost2021}.\\

\paragraph{Problem-inspired Ansätze} 
\label{ph:problem_inspired_ansatze}
Whenever the ansatz is built by leveraging the knowledge of the underlying physics of the problem to be solved, one says that the ansatz is problem-inspired. This is often the case for quantum chemistry applications, and the most prominent example in this class is the Unitary Coupled Cluster (UCC) ansatz, specifically thought for dealing with electronic structure calculations, also used in the seminal work by Peruzzo et al.~\cite{PeruzzoVQE2014} on variational quantum algorithms. Various generalisations of the UCC ansatz for several quantum chemistry problems have been introduced, and one can find detailed information in the reviews~\cite{TillyVQE_2022, UCCReview_2021, CerezoPQC2021Review, Bharti2021NIQSReview}. Another example of a problem-inspired ansatz for quantum chemistry is the Hamiltonian Variational Ansatz inspired by the adiabatic state preparation theorem~\cite{WeckerHVA_2015}. 

Regarding machine learning applications of variational quantum algorithms, and inspired by the classical machine learning literature on geometric deep learning~\cite{BronsteinGDL_2021}, many recent proposals focus on constructing so-called \textit{equivariant} quantum circuits, that are parameterised quantum circuits that encode the symmetries of the problem to be solved~\cite{skolik2022equivariant, MeyerSimmetriesQML_2022, LaroccaGQML2022, NguyenQML_2022}.

By focusing specifically on a restricted class of relevant models, problem-inspired ansätze usually perform better and are easier to optimise with respect to problem-agnostic ones, even though this advantage may come at a cost of a more difficult implementation on real hardware.\\

\paragraph{QAOA-like Ansätze} 
The Quantum Approximate Optimisation Algorithm (QAOA)~\cite{FarhiQAOA_2014} is a widely studied parameterised ansatz to deal with combinatorial optimisation problems on near-term devices. The variational ansatz is given by a layered alternating structure
\begin{equation}
\label{eq:qaoa}
    U(\bm{\beta}, \bm{\gamma}) = \prod_{\ell=1}^L e^{-i \beta_\ell\, H_M}\, e^{-i \gamma_\ell\, H_P}\,,
\end{equation}
where $H_M$ is called mixer Hamiltonian, $H_P$ is the problem Hamiltonian whose ground state encodes the solution to the combinatorial problem at hand, $L$ is a hyperparameter that defines the accuracy of the procedure, and $(\bm{\gamma}, \bm{\beta})$ are two sets of parameters to be optimised. 

This ansatz essentially implements a Trotterised adiabatic evolution of degree $L$, that slowly evolves an easy to prepare ground state of the mixer Hamiltonian $H_M$, to the target ground state of the problem Hamiltonian $H_P$. 
It has been shown that the alternating structure of QAOA implements a universal dynamics for quantum computation~\cite{LloydQAOA_2018}, and an extension of QAOA to deal with more general classes of Hamiltonians has been proposed with the name of Quantum Alternating Operator Ansatz~\cite{HadfieldQAOA_2019}.\\

\paragraph{Adaptive Ansätze}
The optimisation of the variational angles alone is often not enough to obtain a well performing algorithm. In this case, one can relax the condition of fixing the circuit ansatz upfront, and instead \textit{adaptively} optimise its structure along with that of the variational parameters. Examples are the RotoSolve algorithm~\cite{Ostaszewski2019CircuitLearning} that optimizes both the angle and the axis of single qubit rotation gates, ADAPT-VQE algorithms~\cite{GrimsleyAdaptVQE_2019, TangeQubitAdaptVQE_2021} that procedurally grow an ansatz by adding gates to the circuit form a pool of operators, and other approaches that try to add but also remove gates to keep the circuit shallow~\cite{BilikisVANS_2021}. The adaptive generation of superior ans\"{a}tze comes at the cost of optimisation problems that are hard to solve in general, but that can be addressed using approximate heuristics like Evolutionary Algorithms~\cite{RattewEvolVQE_2019}.

\subsection{Optimisation of variational quantum algorithms}
Once a cost function and an ansatz have been defined, one can then proceed with the optimisation loop to find the optimal solution of Eq.~\eqref{eq:vqa_argmin}. A simple yet effective strategy to solve optimisation problems of differentiable functions is gradient-descent methods, widely used and studied in the classical machine learning literature. These procedures are first-order methods that need access to the first derivatives of the function, as opposed to zeroth- and second-order ones, that instead require only function evaluations or access to the Hessian, respectively. 

Be $\vecparam^{(t)}$ the value of the variational parameter at a time step $t$, and $C\qty(\vecparam) \in \mathbb{R}$ the scalar cost function to be minimised. Gradient descent methods iteratively change the value of the parameters by moving against the direction of the gradient of the function evaluated at that point, namely
\begin{equation}
\label{eq:gradient_descent}
    \vecparam^{(t+1)} = \vecparam^{(t)} - \eta \nabla_{\vecparam} C(\vecparam)\big|_{\vecparam^{(t)}}\, .
\end{equation}
where $0< \eta \ll 1$ is a hyperparameter called \textit{learning rate} that is used to tune the step size of the algorithm. As long as the learning rate is small enough, this update rule will propose a new value for the parameters corresponding to a lower cost. This can be seen by Taylor expanding the cost at step $t+1$ around the parameters of the previous step $t$, as
\begin{eqnarray}
C\qty(\vecparam^{(t+1)}) &=& C\qty(\vecparam^{(t)} - \eta \nabla_{\vecparam} C\qty(\vecparam^{(t)})) \\
&\approx& C\qty(\vecparam^{(t)}) + \nabla_{\vecparam}C\qty(\vecparam^{(t)})\qty(-\eta \nabla_{\vecparam}C\qty(\vecparam^{(t)})) + \order{\eta^2} \\
&=& C\qty(\vecparam^{(t)}) - \eta \norm{\nabla_{\vecparam} C\qty(\vecparam^{(t)})}^2_2 \leq C\qty(\vecparam^{(t)})\, .
\end{eqnarray}
where $\norm{\cdot}_2$ denotes the 2-norm (or Euclidean norm) of a vector, and the last inequality comes from the fact that both the learning rate and the gradient norm are positive quantities. 

The basic gradient descent update rule in Eq.~\eqref{eq:gradient_descent} can be easily improved and generalised, for example taking into account information about the second derivatives of the function (e.g. BFGS~\cite{BGFS_Nocedal2006}), employing a stochastic approximation of the exact gradient (e.g. SPSA~\cite{Spall1998overview}), or adding \textit{momentum} terms and adaptive learning rates (e.g. ADAM~\cite{KingmaAdam_2014}). The research on numerical optimisation methods is very vast and active, providing useful tools for training variational quantum algorithms~\cite{Stokes2020Quantumnatural, WierichsOptimisation_2020}.

\subsubsection{Parameter-shift rule}
Gradient-based methods require access to the first (partial) derivatives of the target function, which can be approximated numerically with the (central) finite-difference formula
\begin{equation}
\label{eq:finite_difference}
    \partial_i C(\vecparam) \coloneqq \frac{\partial C(\vecparam)}{\partial\theta_i} \approx \frac{C(\vecparam + \varepsilon \bm{e}_i) - C(\vecparam - \varepsilon \bm{e}_i)}{2\varepsilon}\,,\quad 0< \varepsilon \ll 1\,
\end{equation}
where $\bm{e}_i = (0,\,\hdots, 1,\, \hdots,\, 0)$ is a unit vector with entries zero everywhere except for a one in the $i$-th position, and the equality sign is recovered in the limit of vanishing displacement $\varepsilon \rightarrow 0$. It turns out that a similar but \textit{exact} formula holds for derivatives of parameterised quantum circuits, a fact which is now commonly referred to as \textit{parameter-shift rule}~\cite{Schuld2019Gradients, Mitarai2018Learning, Wierichs2022generalparameter, MariHighGradients_2021, CerezoBarrenLocalCost2021}. This rule is a direct consequence of the rotation-like nature of the parameterised gates used in variational circuits, and we hereby prove this formula following the derivation presented in ref.~\cite{MariHighGradients_2021}. 

Common parameterised operations used in variational quantum circuits are rotation-like gates of the form
\begin{equation}
\label{eq:exp_param_gate}
    V_j(\theta_j) = e^{-i \theta_j P_j / 2}\,,
\end{equation}
where $\theta_j$ is a variational angle, and $P_j$ is the Hermitian generator of the gate. When $P_j$ is involutory ($P_j^2=\mathbb{I}$), then the exponential can be recast in trigonometric form as (see Eq.~\eqref{eq:trig_form_exp})
\begin{equation}
\label{eq:trig_param_gate}
    V_j(\theta_j) = \cos\frac{\theta_j}{2}\,\mathbb{I} - i \sin\frac{\theta_j}{2}\,P_j\,.
\end{equation}
Note that many operations can be expressed in this way, including single-qubit rotations and more generally any rotation generated by tensor products of Pauli matrices.

Let $C(\vecparam) = \Tr[O U(\vecparam)\rho U(\vecparam)^\dagger]$ be the cost function to be differentiated, and $V_j(\theta_j)$ be the parameterised gate depending on the variable $\theta_j$ with respect to which we want to calculate the derivative. Consider a bipartition of the circuit $U(\vecparam)$ containing all the gates that act before ($U_B$) and after ($U_A$) the operation of interest  $V_j(\theta_j)$ takes place\footnote{If the same parameter appears more than once in the circuit, one can check that the full derivative is given by shifting independently each parameterised gate, and then summing all contributions~\cite{Schuld2019Gradients}.}
\begin{equation}
\label{eq:qc_bipartition}
    U(\vecparam) = U_A V_j(\theta_j) U_B\, ,
\end{equation}
where the dependence on the remaining variational parameters in $U_B$ and $U_A$ is suppressed for ease of notation.
The cost function can then be written as
\begin{equation}
\label{eq:single_param_cost}
    C(\vecparam) = \Tr[O U_A V_j(\theta_j) U_B \rho U_B^\dagger V^\dagger_j(\theta_j) U^\dagger_A] = \Tr[O_A V_j(\theta_j) \rho_B V_j(\theta_j)^\dagger]
\end{equation}
where $O_A = U_A^\dagger O U_A$ and $\rho_B = U_B \rho U_B^\dagger$, and we isolated the dependence of the cost on the variable of interest $\theta_j$. Substituting the trigonometric formula~\eqref{eq:trig_param_gate} in the expression above one obtains
\begin{align}
\label{eq:trigonometric_cost}
    C(\vecparam) &= \Tr\qty[O_A \qty(\cos\frac{\theta_j}{2}\,\mathbb{I} - i \sin\frac{\theta_j}{2}\,P_j) \rho_B \qty(\cos\frac{\theta_j}{2}\,\mathbb{I} + i \sin\frac{\theta_j}{2}\,P^\dagger_j)]\nonumber\\
    &= \Tr \bigg[\cos^2\frac{\theta_j}{2} O_A\rho_B + i\sin\frac{\theta_j}{2}\cos\frac{\theta_j}{2}\qty(O_A\rho_B P^\dagger_j -O_AP_j\rho_B)+\sin^2\frac{\theta_j}{2}O_A P_j\rho_B P_j^\dagger \bigg] \nonumber\\
    &= \frac{1+\cos\theta_j}{2}\Tr[O_A\rho_B] + \frac{1-\cos\theta_j}{2}\Tr[O_AP_j\rho_BP_j^\dagger] + \frac{i}{2}\sin\theta_j\Tr[O_A\qty(\rho_BP_j^\dagger - P_j\rho_B)]\nonumber\\
    &= C_0 + C_1\cos(\theta_j) + C_2\sin(\theta_j)
\end{align}
where $C_0$, $C_1$, and $C_2$ are real numbers that do not depend on the parameter $\theta_j$, defined as
\begin{equation}
\begin{aligned}
    & C_0 = \frac{\Tr[O_A\qty(\rho_B + P_j\rho_B P_j^\dagger)]}{2},\quad C_1 = \frac{\Tr[O_A\qty(\rho_B - P_j\rho_B P_j^\dagger)]}{2},\\
    & \quad \quad \quad \quad \quad \quad C_2 = \frac{\Tr[O_A\qty(\rho_B - P_j\rho_B P_j^\dagger)]}{2}\, ,
\end{aligned}
\end{equation}
Equation~\eqref{eq:trigonometric_cost} is particularly interesting because it shows that the cost function is essentially a trigonometric polynomial with respect to each individual variational parameter. Then, using the following identities for the derivatives of sine and cosine functions
\begin{equation}
\begin{aligned}
    \frac{d\cos x}{dx} &=& \frac{\cos(x+s)-\cos(x-s)}{2\sin s}\\
    \frac{d\sin x}{dx} &=& \frac{\sin(x+s)-\sin(x-s)}{2\sin s}\\
\end{aligned}\quad\quad\quad \forall s\neq m\pi,\, m \in \mathbb{Z}\,,
\end{equation}
the derivative of the cost in Eq.~\eqref{eq:trigonometric_cost} with respect to $\theta_j$ can be written as
\begin{align}
    \frac{\partial C(\vecparam)}{\partial \theta_j} &= C_1 \frac{d \cos\theta_j}{d \theta_j} + C_2 \frac{d\sin\theta_j}{d\theta_j}\nonumber \\
    &= \frac{C_0 + C_1\cos(\theta_j+s) + C_2\sin(\theta_j+s)}{2\sin s} - \frac{C_0 + C_1\cos(\theta_j-s) + C_2\sin(\theta_j-s)}{2\sin s}\nonumber \\
    & = \frac{1}{2\sin s}\qty[C(\vecparam + s\bm{e}_j)- C(\vecparam-s\bm{e}_j)]\, ,
\end{align}
where in the last line we recognised that both numerators are instances of the cost~\eqref{eq:trigonometric_cost} with an appropriate redefinition of the variational parameter. By setting $s=\pi/2$ in the expression above, one eventually arrives at the usual formulation of the parameter-shift rule~\cite{Schuld2019Gradients, Mitarai2018Learning, CerezoHigherOrder2021}
\begin{equation}
    \label{eq:parameter_shift_rule}
    \frac{\partial C(\vecparam)}{\partial \theta_j} = \frac{1}{2}\qty[C\qty(\vecparam + \frac{\pi}{2}\bm{e}_j) - C\qty(\vecparam - \frac{\pi}{2}\bm{e}_j)]\, ,
\end{equation}

As argued earlier, the parameter-shift rule~\eqref{eq:parameter_shift_rule} is indeed very similar to the finite-difference formula~\eqref{eq:finite_difference}, with the big difference that while the former is an \textit{exact} relation between the derivative of a function and its value at specific points, the latter is only an \textit{approximation}. 

In general, a parameter-shift rule~\eqref{eq:parameter_shift_rule} can be derived when the parameterised operation is of the form~\eqref{eq:exp_param_gate} and the generator of the rotation has only two unique eigenvalues, which is related to the requirement that the generator is involutory~\cite{CrooksParamShift2019, Schuld2019Gradients}. Whenever these conditions are not met, one can use generalisations of the parameter-shift rules that overcome the issue for example by decomposing the parameterised gate as a product of rotation-like operations~\cite{CrooksParamShift2019}, or expanding the generator of the unitary evolution in the Pauli basis and evaluating the gradients of each component with a stochastic approach~\cite{BanchiGeneralization2021}.

The parameter-shift rule is a useful tool because it gives a precise recipe on how to calculate gradients of variational circuits directly on real quantum hardware, simply by composing the result of two appropriately defined quantum circuits. For example, a single-step of the gradient descent algorithms in Eq.~\eqref{eq:gradient_descent} requires measuring the outcome of 2$p$ different circuits, where $p$ is the number of parameters $\vecparam \in \mathbb{R}^p$. While such linear scaling of resources is not bad per se, this procedure can still be daunting on near-term quantum devices with limited access, especially when the number of parameters is large and many iterations are required to reach a good solution. For example, suppose that the optimisation of a variational circuit with $p$ parameters requires $T$ applications of the gradient-descent update rule, and that each expectation value is estimated on quantum hardware with $M$ measurements shots, then the total number of circuit executions on the quantum hardware scales like $\order{pTM}$, which can be quite big already for modest instances ($p\sim T\sim 10^2$, $M\sim 10^3$). 
As a comparison, gradients in classical machine learning algorithms can be calculated much more efficiently, within a single computational step, via automatic differentiation techniques and backpropagation. This is done by storing intermediate values of the computation while it takes place, and then combining these values at the end via the chain rule to calculate derivatives of composed functions. Unfortunately, since it is not possible to measure intermediate states in a quantum computation without disturbing the system, there is no straightforward way of applying backpropagation to variational quantum algorithms.

Finally, note that gradients of parameterised quantum circuits can also be calculated with an ancilla-based measurement scheme similar to a Hadamard test~\cite{Schuld2019Gradients, TillyVQE_2022}.

\subsubsection{Higher order derivatives}
The parameter shift rule can be applied iteratively to calculate also higher-order derivatives of the cost function~\cite{CerezoHigherOrder2021, MariHighGradients_2021}. For example, second mixed derivatives read
\begin{eqnarray}
    \frac{\partial^2 C(\vecparam)}{\partial\theta_j\partial\theta_i} &=& \frac{1}{2}\qty[\frac{\partial}{\partial\theta_j}C\qty(\vecparam + \frac{\pi}{2}\bm{e}_i) - \frac{\partial}{\partial\theta_j}C\qty(\vecparam - \frac{\pi}{2}\bm{e}_i)]\nonumber\\
    &=&  \frac{1}{4}\bigg[C\qty(\vecparam + \frac{\pi}{2}\bm{e}_i +\frac{\pi}{2}\bm{e}_j) - C\qty(\vecparam + \frac{\pi}{2}\bm{e}_i -\frac{\pi}{2}\bm{e}_j)\nonumber \\
    && \quad\quad\quad - C\qty(\vecparam -\frac{\pi}{2}\bm{e}_i +\frac{\pi}{2}\bm{e}_j) + C\qty(\vecparam -  \frac{\pi}{2}\bm{e}_i -\frac{\pi}{2}\bm{e}_j)\bigg],
\end{eqnarray}
which assume the rather simple form for diagonal elements $i=j$ 
\begin{equation}
    \label{eq:hess_param_shift}
     \frac{\partial^2 C(\vecparam)}{\partial\theta_i^2} = \frac{1}{2}\qty[C(\vecparam + \pi\bm{e}_i) - C(\vecparam)]\, ,
\end{equation}
since one can easily check that $C(\vecparam+\pi\bm{e}_i) = C(\vecparam - \pi\bm{e}_i)$, which is due to to the $2\pi$-periodicity of parameterised gates~\eqref{eq:exp_param_gate}.

In general, any derivative of a parameterised quantum circuit for which the parameter shift holds, can be expressed as a linear combination of circuit executions
\begin{equation}
    \frac{\partial^{\alpha_1 + \hdots +\alpha_M} C(\vecparam)}{\partial\theta_1^{\alpha_1}\hdots\partial\theta_M^{\alpha_M}} = \frac{1}{2^{\alpha_1 + \hdots +\alpha_M}}\sum_{m = 1}^{2^{\alpha_1 + \hdots +\alpha_M}} s_m\,C(\tilde{\vecparam}_m)\,,
\end{equation}
where $s_m \in \{\pm 1\}$ are signs, and $\tilde{\vecparam}_m$ are parameters obtained by accumulating shifts along different directions.  

\subsubsection{Discussion}
The easy access to the derivatives of the cost function provided by the parameter-shift rule opens up the possibility of using a plethora of gradient-based optimisation methods, like vanilla gradient descent, BFGS or ADAM, but these are not the only viable route. In fact, also zeroth-order methods like Nelder-Mead~\cite{NelderMead} and COBYLA~\cite{Cobyla} have been widely used in the literature to optimise parameterised quantum circuits~\cite{TillyVQE_2022, Bharti2021NIQSReview}, as well as more quantum-native approaches developed specifically for variational algorithms. Examples are the Rotosolve algorithm~\cite{Ostaszewski2019CircuitLearning} that offers an analytic solution to the optimisation problem, or the Quantum Natural Gradient (QNG) method, that, inspired by its classical counterpart, proposes an update of the parameters that takes into account the metric of the space of the quantum states created by the parameterised quantum model, rather than the euclidean metric of parameter space~\cite{Stokes2020Quantumnatural}.

With every approach having its own perks and disadvantages, no consensus has been reached at the moment regarding the best practices for optimising variational quantum algorithms, and the performances are usually studied on a case-by-case basis.

\subsection{Barren plateaus and unitary designs}
\label{sec:Udes_Bp_Exp}

\begin{figure}[ht]
    \centering
    \includegraphics[width=\textwidth]{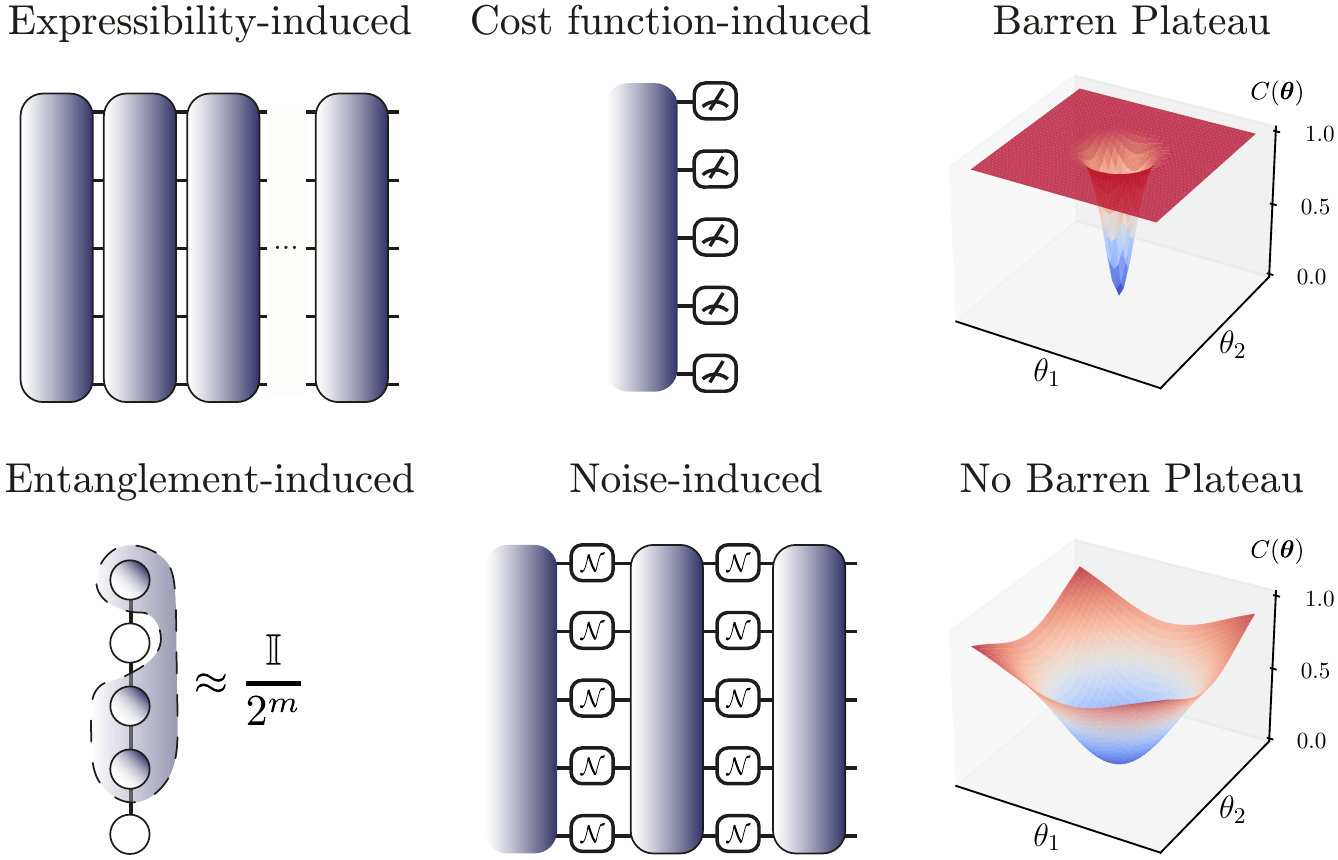}
    \caption[Sources of Barren Plateaus]{Summary of the sources of Barren Plateaus (BP) in the optimisation landscape of variational quantum algorithms. Trainability issues arise in variational algorithms whenever the circuit is too deep and expressive, global observables are used in the cost function, too much entanglement is created inside the circuit, or too much noise is present in the circuit. In these cases, as the number of qubits grows large, the optimisation landscape becomes exponentially flat almost everywhere, hence the name barren plateau.}
    \label{fig:barren_plateau_pictorial}
\end{figure}

In the previous section, we discussed how to optimise variational circuits, but now we discuss why this procedure may turn out to be a particularly arduous task in practice. Indeed, various theoretical challenges hinder the optimisation of variational algorithms, a phenomenon dubbed in the literature as the \textit{barren plateau} (BP) problem. At the current state of the art, by barren plateaus one usually refers to a number of different scenarios where it can be shown that variational quantum algorithms suffer from trainability issues related to vanishing gradients and very flat optimisation landscapes, hence their name. 

The emergence of barren plateaus was first observed in~\cite{McCleanBarren2018}, where the occurrence of vanishing gradients was related to the way parameterised quantum ans\"{a}tze behave in the large qubit regime upon assignment of random parameters, as they resemble high-dimensional random unitary matrices. A large set of results in measure theory in high-dimensional spaces underline the phenomenon of concentration of measure, an example being Levy's lemma, according to which smooth functions on these spaces strongly concentrate around their mean values, and are essentially constant on the space~\cite{Ledoux2001concentration}. Since the average value of gradients of parameterised circuits is often found to be zero, then by concentration of measure they are zero almost always, for most random choices of the parameters, hence any optimisation method will likely fail to find any interesting minimising direction. Formally, given a PQCs acting on a system of $n$ qubits, it can be theoretically argued and numerically confirmed that~\cite{McClean2016VQAs, CerezoBarrenLocalCost2021, Holmes2021connecting}
\begin{align}
    &\mathbb{E}_{\vecparam}\qty[\frac{\partial C(\vecparam)}{\partial\theta_k}] = 0, \\
    &\text{Var}_{\vecparam}\qty[\frac{\partial C(\vecparam)}{\partial\theta_k}] \in \order{b^{-n}}, \quad b >0\,\label{eq:bp_summary}.
\end{align}

That is, upon random assignment of the variational parameters, the expectation value of the gradients of the cost function is zero, and their variance is exponentially vanishing with the number of qubits. Then, as the number of qubits grows large, the gradients become exponentially small on average, and thus one would need an exponential number of measurement shots to reliably estimate their value on real hardware, for example via parameter-shift rules. Such impossibility of properly training parameterised circuits clearly hampers the applicability of any quantum variational procedure suffering from a barren plateau (concentration of measure) phenomenon. 

Trainability issues in variational quantum algorithms have been studied extensively in the literature, and the emergence of barren plateaus in the optimisation landscape parameterised circuits has been found in various settings, that can be grouped into four major classes~\cite{SupanutSubteltiesQML2021}\footnote{As briefly mentioned here when discussing Entanglement-induced BP and then carefully explained in Ch.~\ref{ch:entanglement}, the sources of barren plateaus can be reduced to three, not four, since Expressibility-induced BP and Entanglement-induced BP are intimately connected to each other through randomness.}:
\begin{enumerate}
    \item \textbf{Expressibility-induced BP}~\cite{McCleanBarren2018, Arrasmith2021equivalence}: this is the case explained above, where the untrainability stems from the randomness induced by deep random ans\"{a}tze initialised with random parameters. This causes an exponential flattening of the loss landscape and its gradients due to concentration of measure.
    \item \textbf{Cost function-induced BP}~\cite{CerezoBarrenLocalCost2021}: the use of a global observable, that is, an observable acting nontrivially on all qubits in the circuit, in the cost function~\eqref{eq:vqa_usual_loss}, is also linked to vanishing gradients. Intuitively, this can be understood as a consequence of measuring quantities in an exponentially big Hilbert space, which effectively results in a wash-out of information. 
    
    Most importantly, while expressibility-induced barren plateaus are obtained only for deep-enough (or better, random-enough) ans\"{a}tze, the use of a global observable will likely result in exponentially vanishing gradients irrespectively of the depth of the circuit, and a barren plateau will occur even at shallow depths $L \in \order{1}$, where $L$ is the number of layers in the ansatz~\eqref{eq:layered_ansatz}. 
    
    On the contrary, it was shown that a layered ansatz leveraging local operations on neighbouring qubits, when used with a local cost function, that is a loss defined in terms of expectation values of observables acting non-trivially only on a small subset –one or two– of qubits, can be trainable at least for shallow depths $L \in \order{\log{n}}$, where $n$ is the number of qubits.

    For completeness, in Appendix~\ref{sec:app_global_cost} we report the toy model proposed in ref.~\cite{CerezoBarrenLocalCost2021} to show the occurrence of barren plateaus even for depth-one circuits when a global cost is used, and how the issue can be mitigated by leveraging an equivalent local version of the cost.  
    
    \item \textbf{Entanglement-induced BP}~\cite{MarreroBarrenEntanglement2021}: it was also shown the creation of large entanglement inside a circuit can be detrimental to variational algorithms. As for the expressibility-induced case, also entanglement-induced BP arise as a consequence of the behaviour of random quantum states and matrices. Essentially, whenever the ansatz is expressible enough to create highly-entangled states, then any reduced density operator on $m$ qubits will be very close to the maximally mixed state
    \begin{equation}
        \rho = \Tr_{n-m}[\dyad{\psi_{\vecparam}}] \approx \frac{\mathbb{I}}{2^m}, \quad \ket{\psi_{\vecparam}} = U(\vecparam)\dyad{\bm{0}}U(\vecparam)^\dagger\,\,\,\textrm{highly entangled},
    \end{equation}
    and thus little information can be retrieved from measurements of observables on reduced systems. This source of barren plateaus motivates the study performed in Chapter~\ref{ch:entanglement}, where the production of entanglement in common variational quantum circuits is studied. In particular, we anticipate that entanglement-induced BP and expressibility-induced BP both stem from the resemblance of the parameterised quantum circuit to random unitary matrices, and these two sources are indeed one and the same~\cite{SackBPShadows2022}, as extensively discussed in Chapter~\ref{ch:entanglement}.
    
    \item \textbf{Noise-induced BP}~\cite{WangNoiseinducedBarrenPlateaus2021}: noise that occurs in the quantum circuit can be on its own the cause of the emergence of flat loss landscapes and barren plateaus. Indeed, it was shown that local Pauli errors happening throughout the circuit perturb the quantum state by moving it close to the fixed point of the noise map\footnote{The fixed point of a noise channel $\mathcal{E}[\rho]$ is the state that is unchanged after application of the map, namely $\mathcal{E}[\rho]=\rho$. It is interesting to notice that every quantum noise channel has a fixed point~\cite{NielsenChuang}. The definition of noise channels and their properties can be found in Chapter~\ref{ch:NoiseDeconvolutionChapter}.}, that is the maximally mixed state. By this mechanism, parameterised circuits whose depth scales linearly with the number of qubits $L \in \order{n}$ will again suffer from barren plateaus. 
    
    It is interesting to notice that this class of barren plateaus is conceptually different from all the cases treated previously, since the former are generally probabilistic statements regarding the expectation value of the gradients and their variance, which is exponentially suppressed, when random initialisation of the parameters are considered. The latter instead is not a probabilistic statement on the variance of the gradient, but rather is the gradient itself that is vanishing, due to the loss landscape becoming essentially flat \textit{everywhere} in parameter space. On the contrary, in the previous cases the loss landscape is flat \textit{almost} everywhere, in fact there are exceptional optimal points (called in the literature \textit{narrow gorges}) where the loss is very steep and optimisation is possible~\cite{Arrasmith2020effect, CerezoBarrenLocalCost2021}.
\end{enumerate}

In the following, we show how to derive Equations~\eqref{eq:bp_summary} for the case of expressibility-induced barren plateaus, as they are the most studied and common source of trainability issues in variational algorithms, and are also the most amenable to a concise exposition. The analysis of the problem requires knowledge of random unitary matrices, which we briefly introduce below. Then, after showing how to use these results to prove the emergence of barren plateaus, we discuss how this issue is related to the concept of \textit{expressibility} of quantum circuits, which is a measure of how well a parameterised ansatz addresses the full unitary space. 

\subsubsection{Haar measure and random unitary matrices}
\label{ssec:ch_VQAs_Haar}
Expressibility-induced barren plateaus arise whenever a parameterised quantum circuit resembles a random unitary matrix when random parameters are assigned to the circuit. Equations~\eqref{eq:bp_summary} are then derived by characterising the statistics of random unitaries, and studying what happens when they are used inside a variational optimisation routine. In order to evaluate average values, namely integrals, of functions of random unitary matrices, one requires a probability measure that weights the elements of the unitary group and that is used inside the integrals. The most natural distribution one can think of is the \textit{uniform} distribution, and the \textit{Haar measure} is the mathematical tool that formally defines a uniform probability measure on compact groups~\cite{Meckes2019Haar}. 

Let $\mathcal{U}(d)$ be the group of $d \times d$ unitary matrices, the Haar measure $d\mu(U)$ on the group $\mathcal{U}(d)$ is the \textit{unique} translationally invariant measure, i.e. for any integrable function $f(\cdot)$ and unitary matrix $V \in \mathcal{U}(d)$ it holds
\begin{equation}
    \int d\mu(U) f(V U) = \int d\mu(U) f(U V) = \int d\mu(U) f(U),\quad \forall V \in \mathcal{U}(n)\,.
\end{equation}
The property of translational invariance, which is also referred to as right- and left-invariance, encodes the fact that the Haar measure represents the uniform measure of the group. Thus, we say that a unitary matrix is Haar distributed to indicate a unitary matrix that is sampled uniformly from the space of unitary matrices. In addition, the volume element is normalised, namely
\begin{equation}
    \int d\mu(U) = 1\,.
\end{equation}

We will use the following simplified notation to indicate expectation values and integration over Haar-distributed random unitary matrices
\begin{equation}
    \mathbb{E}_{U}[f(U)] \coloneqq \int d\mu(U) f(U)\, .
\end{equation}

Explicit formulas exist for evaluating integrals over random unitary matrices~\cite{HaarPuchala2017, FukudaRTNI2019}, and here we recall two useful identities regarding the expectation values for low degree polynomials of random unitary matrices~\cite{CerezoBarrenLocalCost2021, Holmes2021connecting, McCleanBarren2018, Huang2020Power}
\begin{align}
\mathbb{E}_{U}[U A U^\dagger] &= \int d\mu(U)\,U A U^\dagger = \frac{\Tr[A]\, \mathbb{I}}{d}\, \label{eq:1haar}\\[1em]
\mathbb{E}_{U}[A U B U^\dagger C U D U^\dagger] &= \int d\mu(U)\,A U B U^\dagger C U D U^\dagger \nonumber\\
&= \frac{\Tr[BD]\Tr[C]A + \Tr[B]\Tr[D]AC}{d^2-1} \label{eq:2haar} \\ & \quad -\frac{\Tr[BD]AC + \Tr[B]\Tr[C]\Tr[D]A}{d(d^2-1)}\nonumber \\[1em]
\mathbb{E}_U\qty[\Tr[UAU^\dagger B]\Tr[UCU^\dagger D]] &= \int d\mu(U)\,\Tr[UAU^\dagger B]\Tr[UCU^\dagger D] \nonumber \\
&= \frac{\Tr[A]\Tr[B]\Tr[C]\Tr[D] + \Tr[AC]\Tr[BD]}{d^2-1} \label{eq:2haar_alternative} \\ & \quad -\frac{\Tr[AC]\Tr[B]\Tr[D] + \Tr[A]\Tr[C]\Tr[BD]}{d(d^2-1)}\nonumber
\end{align}
Here $A$, $B$, $C$ and $D$ are unitary matrices in $\mathcal{U}(d)$, where $d=2^n$ for matrices acting on a system of $n$ qubits.

Strictly related to uniform random matrices, is the concept of unitary designs, which are ensembles of unitaries that replicate properties of the Haar distribution up to a given moment. Formally, let 
$\mathbb{U} = \{U_1, \hdots, U_K\}$ be an ensemble of unitary matrices $U_i \in \mathcal{U}(d)$, and $P_{t,t}(U_i)$ a polynomial of degree at most $t$ in the entries of $U_i$, and degree at most $t$ in the entries of $U_i^\dagger$. Then, the ensemble $\mathbb{U}$ is said to be a unitary $t$-design if~\cite{DankertUnitaryDesings2009}
\begin{equation}
\label{eq:t_designs}
    \mathbb{E}_{\mathbb{U}}[P_{t,\,t}(U)] \coloneqq \frac{1}{K} \sum_{i=1}^K P_{t,\,t}(U_i) = \int d\mu(U)P_{t,\,t}(U)\,,
\end{equation}
that is, averaging over the discrete set $\mathbb{U}=\{U_1, \hdots, U_K\}$ is \textit{equivalent} to averaging over the full unitary group with respect to the Haar measure. Intuitively, unitary designs are sets of operators that are random enough to match moments of the Haar distribution up to a given degree. 

In particular, looking back at Eqs.~\eqref{eq:1haar} and~\eqref{eq:2haar}, one can obtain the same expectation values by averaging over a unitary $2$-design instead of considering the full unitary group. This is by definition of unitary $2$-design, and noticing that both $UAU$ and $A U B U^\dagger C U D U^\dagger$ are polynomials $P_{2,2}(U)$ of degree of most $t=2$ in the entries of $U$ and $U^\dagger$.

Examples of unitary designs are the Pauli group, defined as the set of all possible tensor products of Pauli matrices, which is a unitary $1$-design, and the Clifford group, defined as the set of operators that normalises the Pauli group, that can be shown to be a unitary 3-design (hence also a 1- and 2-design)~\cite{ZakClifford3des}. The Pauli group $P_n$ on $n$ qubits is defined as all possible combinations of Pauli matrices with coefficients $\pm 1$ and $\pm i$ 
\begin{equation}
\label{eq:PauliGroup}
    P_n \coloneqq \qty{\pm 1, \pm i} \times \qty{ \sigma_1 \otimes \sigma_2 \otimes \hdots \otimes \sigma_n \,|\, \sigma_i \in \{\mathbb{I}, X, Y, Z\} },
\end{equation}
while the Clifford group $\mathcal{C}_n$ consists of set of unitary operators that maps the Pauli group to the Pauli group under conjugation (and up to a phase), namely
\begin{equation}
    \label{eq:CliffordGroup}
    \mathcal{C}_n \coloneqq \{U \in \mathcal{U}(2^n)\,|\, \forall\sigma \in P_n \implies U\sigma U^\dagger \in P_n \}  / \mathcal{U}(1).
\end{equation}

\subsubsection{Barren plateaus in the optimisation of PQCs}
\label{sec:ch_VQAs_BP}
The theory of random unitary matrices, and specifically unitary 2-designs, is at the core of the barren plateau phenomenon, as it will be now clear. 

Let $U(\vecparam) \in \mathcal{U}(2^n)$ a parameterised quantum circuit with variational parameters $\vecparam \in \mathbb{R}^P$ acting on a system of $n$ qubits, and $C(\vecparam) = \Tr[O U(\vecparam)\rho U(\vecparam)^\dagger]$ a loss function to be minimised. As done previously for deriving the parameter-shit rule~\eqref{eq:qc_bipartition}, given a parameter $\theta_k \in \vecparam = (\theta_1, \hdots, \theta_P)$, consider a bipartition of the circuit happening at the position of the parameterised gate depending on the parameter $\theta_i$
\begin{equation}
    U(\bm{\theta}) = U_A V_k(\theta_k) U_B\,,
\end{equation}
where $V_k(\theta_k) = e^{-i\theta_kP_k/2}$ is again a parameterised Pauli rotation, and for simplicity of notation we suppressed the dependence of $U_R$ and $U_L$, on the remaining variational parameters $U_{R,L} = U_{R,L}(\theta_1, \hdots, \theta_{k-1}, \theta_{k+1}, \hdots, \theta_P)$. The derivative of the unitary $U(\vecparam)$ with respect to $\theta_k$ then amounts to
\begin{equation}
    \partial_k U(\vecparam) = U_A\, \partial_k \qty(e^{-i\theta_kP_k/2}) \, U_B = -\frac{i}{2} U_A\, P_k\, U_B\, ,
\end{equation}
and similarly for $U(\vecparam)^\dagger$, $\partial_k U(\vecparam) = iU_L^\dagger\,P_k\,U_R^\dagger / 2$. Then, the derivative of the cost with respect to parameter $\theta_k$ can be written as
\begin{align}
    \partial_k C(\vecparam) &= \Tr[O\, \qty(\partial_k U(\vecparam))\,\rho\, U(\vecparam)^\dagger] + \Tr[O\, U(\vecparam)\,\rho\, \qty(\partial_k U(\vecparam)^\dagger)]\\
    &= -\frac{i}{2}\Tr[O\, U_A P_k U_B\, \rho\, U_B^\dagger U_A^\dagger] + \frac{i}{2}\Tr[O\, U_A U_B\,\rho\, U_B^\dagger P_k U_A^\dagger]\\
    &= -\frac{i}{2}\Tr[U_A^\dagger O U_A\,\qty[P_k, U_B\rho U_B^\dagger]] = -\frac{i}{2}\Tr[O_A\,\qty[P_k, \rho_B]]\,,
\end{align}
where again we have defined the evolved observable and state $O_A  = U_A^\dagger O U_A$ and $\rho_B = U_B\rho U_B^\dagger$, respectively. Note that this is an alternative expression for the derivative of the cost function, as opposed to the one derived above when discussing the parameterised-shift role. One can check that all the calculations that follow can be applied identically if using the expression~\eqref{eq:parameter_shift_rule}.

Suppose now that one of the parameterised circuits, either $U_A$ or $U_B$, is complex enough so that the unitaries generated by assigning random parameters to the circuit closely resemble random unitary matrices. More formally, given a set of parameter vectors $\{\vecparam_1, \vecparam_2, \hdots, \vecparam_K\}$, suppose that corresponding set of unitaries $\mathbb{U}_{A,B} = \{U_{A,B}(\vecparam_1), \hdots, U_{A,B}(\vecparam_K)\} = \{U_{A,B}^{(1)}, \hdots, U^{(K)}_{A,B}\}$ form a 1-design, as defined in~\eqref{eq:t_designs}. Then, the expected value of the derivative $\partial_k C(\vecparam)$ over a random assignment of the parameter vector can be evaluated as follows
\begin{align}
\label{eq:A1des}
    \mathbb{E}_{\mathbb{U}_A}\qty[\partial_k C(\vecparam)] &= -\frac{i}{2} \mathbb{E}_{\mathbb{U}_A}\qty[\Tr[O_A\,\qty[P_k, \rho_B]]] = -\frac{i}{2}\Tr[\mathbb{E}_{\mathbb{U}_A}[U_A^\dagger \,O\,U_A]\,\qty[P_k, \rho_B]]\\
    &= -\frac{i}{2}\Tr[\frac{\mathbb{I}\Tr[O]}{2^n}\,\qty[P_k, \rho_B]] = -\frac{i\Tr[O]}{2^{n+1}}\Tr[P_k\rho_B - \rho_B P_k] = 0\,,
\end{align}
where in the first line we exchanged the trace and the expectation value since they are both linear operations, and in the second line we first made use of the formula for first moments randomly distributed unitary matrices~\eqref{eq:1haar}, and then used the fact that the trace of a commutator is zero, due to cyclicity of the trace. The same result is obtained if $U_B$ is a $1$-design instead, in fact
\begin{align}
\label{B1des}
    \mathbb{E}_{\mathbb{U}_B}\qty[\partial_k C(\vecparam)] &= -\frac{i}{2} \Tr[O_A\,\qty[P_k, \mathbb{E}_{\mathbb{U}_B}[U_B\rho U_B^\dagger]]] = -\frac{i}{2}\Tr[O_A\,\qty[P_k, \frac{\mathbb{I\Tr[\rho]}}{2^n}]] = 0\,
\end{align}
where the last equality follows from the vanishing commutator $[P_k, \mathbb{I}]=0$. Thus, one can summarise Eqs.~\eqref{eq:A1des} and~\eqref{B1des} as follows.

\begin{theorem}[Zero average gradient for random PQCs]
\label{th:1des_grad}
    Let $U(\vecparam)$ represent a parameterised quantum circuit with variational parameters $\vecparam \in \mathbb{R}^M$ acting on a system of $n$ qubits, and let $C(\vecparam) = \Tr[O\, U(\vecparam)\rho U(\vecparam)^\dagger]$ be a cost function depending on the variational parameters via gates of the form $V_k(\theta_k) = e^{-i\theta_k P_k / 2}$. For any parameter $\theta_k,\, k=1,\,\hdots,\, M$, consider the bipartition of the circuit $U(\vecparam) = U_A(\vecparam)V_k(\theta_k)U_B(\vecparam)$, and let $\mathbb{U}_{A,B} = \{U_{A,B}(\vecparam_1), \hdots, U_{A,B}(\vecparam_K)\}$ denote the set of unitaries generated when some randomly generated parameters $\{\vecparam_1, \hdots, \vecparam_K\}$ are assigned to the circuit sections $U_{A,B}$. Then, if at least one of $\mathbb{U}_A$ or $\mathbb{U}_B$ forms a 1-design, the expected value of any partial derivative of the cost function when random parameters are assigned to the quantum circuit is zero
\begin{equation}
    \mathbb{E}_{\mathbb{U}_{A,B}}\qty[\frac{\partial C(\vecparam)}{\partial\theta_k}] = 0 \quad \forall~k, \quad \textrm{if either $\mathbb{U}_A$, or $\mathbb{U}_B$, or both, are at least a 1-design.} 
\end{equation}
\end{theorem}

Note that no particular assumptions are made on the actual probability distribution used to sample the parameters $\theta_k$, the only requirement is that the corresponding set of unitaries $\mathbb{U}_{A,B}$ form 1-designs. In practical scenarios, it is common practice to initialise parameterised quantum circuits with random parameters uniformly distributed in $\theta_k \sim \text{Unif}[0, 2\pi]$. In this case, one can also use the trigonometric nature of the cost function shown before~\eqref{eq:parameter_shift_rule} to show that derivatives are indeed biased towards zero
\begin{align}
    \mathbb{E}_{\theta_k \sim \text{Unif}[0, 2\pi]}\qty[\partial_k C(\vecparam)] = &= \int_0^{2\pi} d\theta_k \, \partial_k\qty(C_0 + C_1\cos\theta_k + C_2\sin\theta_k) \\
    & = \int_0^{2\pi} \qty(-C_1\sin\theta_k + C_2\cos\theta_k) = 0\, . 
\end{align}

So far, we have reproduced the first result of the barren plateau phenomenon described in~\eqref{eq:bp_summary}, and now we move on to estimating the variance of the cost function derivatives when random parameters are assigned to the circuit. Since the variance is a function of second degree with respect to the circuit, the calculations are more involved, as they entail computing expectation values over 2-designs. We are now interested in studying the following quantity
\begin{align}
    \text{Var}[\partial_k C(\vecparam)] &\coloneqq \mathbb{E}\qty[(\partial_k C(\vecparam))^2] - \mathbb{E}[\partial_k C(\vecparam)]^2 = \mathbb{E}\qty[(\partial_k C(\vecparam))^2]\\
    &= -\frac{1}{4}\mathbb{E}\qty[\Tr[U_A^\dagger O U_A\,\qty[P_k, U_B\rho U_B^\dagger]]^2]\, ,
\end{align}
which can be computed using Eq.~\eqref{eq:2haar} and~\eqref{eq:2haar_alternative} for second degree polynomials of random unitary matrices, assuming that the ensembles $\mathbb{U}_{A,B}$ are now random enough to be also 2-designs. We report the full calculation in Appendix~\ref{sec:app_variance_gradients}, and hereby state only the final result.

\begin{theorem}[Vanishing gradients for random PQCs]
\label{th:2des_grad}
In the same conditions of Th.~\ref{th:1des_grad}, but assuming that the ensembles of unitaries $\mathbb{U}_{A,B}$ are now random enough to be also 2-designs, then the following holds:
\begin{gather}
    \text{If $\mathbb{U}_A$ is at least a 2-design, then $\forall~k$:}\nonumber\\
    \quad\text{Var}_{\mathbb{U}_A}[\partial_k C(\vecparam)] = -\frac{1}{4}\frac{1}{2^{2n}-1}\qty(\Tr[O^2]-\frac{\Tr[O]^2}{2^n})\Tr[\qty[P_k, U_B\rho U_B^\dagger]^2]\label{eq:A2des_total}
\\[0.5em]
    \text{If $\mathbb{U}_B$ is at least a 2-design, then $\forall~k$:}\nonumber\\
    \quad\text{Var}_{\mathbb{U}_B}[\partial_k C(\vecparam)] = -\frac{1}{4}\frac{1}{2^{2n}-1}\qty(\Tr[\rho^2]-\frac{1}{2^n})\Tr[\qty[U_A^\dagger O U_A, P_k]^2]\label{eq:B2des_total}\\[0.5em]
    \text{If $\mathbb{U}_{A,B}$ are both at least 2-designs, then $\forall~k$:}\nonumber\\
    \quad\begin{aligned}
    \text{Var}_{\mathbb{U}_{A, B}}[\partial_k C(\vecparam)] &= -\frac{1}{4}\frac{2}{2^{2n}-1}\qty(\Tr[\rho^2]-\frac{1}{2^n})\bigg(\frac{\Tr[O^2]\Tr[P_k]^2+\Tr[O]^2\Tr[P_k^2]}{2^{2n}-1}\\
    &\quad\quad\quad - \frac{\Tr[O^2]\Tr[P_k^2]+\Tr[O]^2\Tr[P_k]^2}{2^n(2^{2n}-1)} - \frac{\Tr[O^2]\Tr[P_k^2]}{2^n} \bigg)
    \end{aligned}
\end{gather}
In practical instances when for example the parameterised gates are generated by Pauli rotations $P_k$, the measured observable $O$ is a Pauli string, and the input state is a pure state $\rho=\dyad{0}$, then the variance vanish exponentially with the number of qubits
\begin{equation}
    \text{Var}[\partial_k C(\vecparam)] \in \order{2^{-n}}\,.
\end{equation}
\end{theorem}
As shown in the Appendix, despite the minus sign in the expressions above, one can check that the variances are correctly positive values. 

By combining the results of Theorem~\ref{th:1des_grad} and Theorem~\ref{th:2des_grad} together with Chebyshev's inequality, one eventually arrives at
\begin{equation}
    \text{Pr}\qty(\abs{\partial_k C(\vecparam)} > \delta) \leq \frac{\text{Var}[\partial_k C(\vecparam)]}{\delta} \in \order{2^{-n}}\,
\end{equation}
which states that the probability of having a non-negligible gradient in a random parameterised quantum circuit vanishes exponentially with the number of qubits. Thus, as it is not possible to determine any interesting minimising direction, it is not possible to train variational circuits efficiently in the large qubit regime $n\gg 1$, where one expects to find a quantum advantage of sort. 

\subsubsection{Discussion and mitigation of barren plateaus}
The emergence of randomness-induced barren plateaus is tightly connected to the concentration on measure phenomenon in high-dimensional spaces, and indeed it was clearly shown that vanishing gradients (i.e. barren plateaus) are the flip side of the exponential concentration of the cost function around the mean~\cite{Arrasmith2021equivalence}, which makes the loss landscape flat almost everywhere, except for so-called \textit{narrow gorges} corresponding to the minimum of the function. If, by chance, one initialises the parameters of the quantum circuit inside or close to a narrow gorge, then optimisation is possible. Moreover, note that while barren plateaus are usually defined in terms of vanishing derivatives, gradient-free optimisation methods are not a solution to the problem~\cite{Arrasmith2021effectofbarren}.

Several mitigation strategies have been proposed to alleviate trainability issues related to barren plateaus, for example initialising the parameters in the quantum circuit such that the circuits initially acts as an identity~\cite{Grant2019Initialization}, correlating parameters inside the quantum circuit~\cite{Volkoff2020Correlating}, using heuristics to initialise parameters already close to an optimal solution~\cite{ZhouQAOABP}, leveraging classical algorithms based on tensor-network pre-training~\cite{RudolphSynergyTN_PQC} or recurrent neural networks~\cite{Verdon2019Learning} to propose good values for the parameters, optimise the parameters procedurally in a layer-wise fashion~\cite{SkolikLayerwise2021}, opting for local---instead of global--- cost functions~\cite{CerezoBarrenLocalCost2021}, and eventually restricting the expressibility of the circuit ansatz by carefully choosing problem-inspired ans\"{a}tze~\cite{skolik2022equivariant, LaroccaGQML2022, MeyerSimmetriesQML_2022}. 

While all these strategies may seem unrelated to each other, most of them work by imposing some type of constraint on the parameterised quantum circuit, so that it is far from being a general random circuit and resembling a unitary design. Indeed, while trainability issues may arise as a consequence of many different factors (see Fig.~\ref{fig:barren_plateau_pictorial}), all of these sources can be tamed by appropriately controlling the expressibility of the circuit, which comes at the cost of either using very specific problem-inspired ans\"{a}tze, or strongly limiting the depth of the quantum circuit to include only logarithmically many operations $L \in \order{\log n}$~\cite{CerezoBarrenLocalCost2021, Pesah2020ConvBarren, CongQCNN2019}. The use of shallow circuits does not only avoid randomness-induced barren plateaus, but also noise-induced ones, as errors cannot accumulate and grow to the extent of causing a flattening of the loss landscape.

\subsection{Expressibility of PQCs}
Up until now we frequently used the term \textit{expressibility} of parameterised quantum circuits to intuitively convey a measure of their ability to represent a general (random) unitary matrix. This statement can be made formal by defining the expressibility with the superoperator~\cite{SimPQCs2019, Holmes2021connecting}
\begin{equation}
\label{eq:expressibility_haar}
    A^{(t)}_{\mathbb{U}}(\cdot) \coloneqq \int_{\text{Haar}} d\mu(V)~V^{\otimes t} (\cdot) (V^{\dagger})^{\otimes t} - \int_{\mathbb{U}} dU~U^{\otimes t} (\cdot) (U^{\dagger})^{\otimes t}
\end{equation}
where the first integral is evaluated over the Haar distribution on the unitary group $\mathcal{U}(d)$, and the second one is over the uniform distribution on the ensemble of unitaries $\mathbb{U}$ to be characterised. Small values of $A^{(t)}_{\mathbb{U}}(\cdot)$ means high expressivity. Indeed, if the ensemble $\mathbb{U}$ is a $t$-design, then $A^{(t)}_{\mathbb{U}}(X)=0$ will be zero for all operators $X$. The expressibility then measures the ``power" of an ensemble of unitaries $\mathbb{U}$ in terms of its generality, that is measuring how faithful it is at reproducing the same statistical moments of the Haar distribution.  

With this definition, it is possible to derive a formal connection between the expressibility and the barren plateau phenomenon described previously. In fact, authors in~\cite{Holmes2021connecting} provided a generalisation of the randomness-induced barren plateau phenomenon, by extending its validity to circuits forming \textit{approximate} rather than \textit{exact} 2-designs. In particular, it was shown that the variance of gradients of an arbitrary ansatz can be upper bounded by
\begin{equation}
    \label{eq:expr_bp_holmes}
\begin{aligned}
    & \text{Var}[\partial_k C(\vecparam)] \leq \text{Var}_{\text{2-des}}[\partial_k C(\vecparam)] + f\qty(\varepsilon^O_{\mathbb{U}}, \varepsilon^\rho_{\mathbb{U}})\\
    & \varepsilon^O_{\mathbb{U}} :=  \norm{A^{(2)}_{\mathbb{U}}\qty(O^{\otimes 2})}_2\,,\quad \varepsilon^\rho_{\mathbb{U}} :=  \norm{A^{(2)}_{\mathbb{U}}\qty(\rho^{\otimes 2})}_2
\end{aligned}
\end{equation}
where the first term on the right is the variance obtained if the circuit --- or better, parts of it as shown in Th.~\ref{th:2des_grad} --- was an exact 2-design, and the second term is a function that depends on the expressibility of the circuit, specifically through some observable ($O$) and state ($\rho$) dependent quantities. The notation $\norm{\cdot}_2$ denotes the \textit{Frobenius norm} for operators, defined as $\norm{A}_2 \coloneqq \sqrt{\Tr[A^\dagger A]}$. We refer to~\cite{Holmes2021connecting} for the complete statement of the result, including the explicit form of the function $f(\cdot, \cdot)$. 

For our discussion, it is sufficient to note that if the circuit is an exact 2-design, then the expressibility term $f\qty(\varepsilon^O_{\mathbb{U}}, \varepsilon^\rho_{\mathbb{U}})$ vanishes and the inequality becomes an equality, thus obtaining again the exponentially vanishing gradients of Theorem~\ref{th:2des_grad}. On the contrary, if the circuit is not very expressible and $f\qty(\varepsilon^O_{\mathbb{U}}, \varepsilon^\rho_{\mathbb{U}}) \notin \order{2^{-n}}$, then the upper bound allows for non-vanishing gradients. Thus, in conclusion, this result implies that while highly expressive ansätze have vanishing gradients and hence are harder to train, imperfectly expressive guarantee a viable solution to restore trainability. This is in line with recent proposals in the literature, already briefly mentioned in Sec.~\ref{ph:problem_inspired_ansatze}, that advocate for the use of problem-inspired constrained ansätze for variational quantum algorithms.

Finally, we note that while the definitions in Eq.~\ref{eq:expr_bp_holmes} turn out useful for theoretical analysis, an alternative formulation in terms of so-called \textit{frame potentials} can be used for numerical evaluations of the expressibility, as discussed in Chapter~\ref{ch:entanglement}. 

\section{Conclusions}
In this chapter, we gave an overview of the state of the art of quantum computing, specifically focused on the class of quantum algorithms that can be run on near-term devices.

We started by recalling the basic building blocks of quantum computation --- qubits, gates, and measurement --- and then proceeded by describing the current situation on the practical realisation of quantum computing machines. As opposed to ideal universal fault-tolerant quantum computers, current quantum devices are referred to as \textit{Noisy Intermediate-Scale Quantum} (NISQ) computers, to remark that they are imperfect devices which are not error-corrected and are of limited size. 

Meanwhile future experimental and theoretical advancements pave the way toward the construction of large-scale noise-resilient quantum computers, near-term devices give the opportunity to experiment with quantum information processing, and also invite researchers to explore a new paradigm for computation based on hybrid quantum-classical algorithms, namely Variational Quantum Algorithms (VQAs).

In the next chapter, we will introduce a sub-field of variational quantum algorithms called Quantum Machine Learning (QML), which has gained significant attention over the past years, and presents itself as one of the most interesting application for near-term devices.

\chapterimage{bg3.png} 
\chapterspaceabove{6.75cm} 
\chapterspacebelow{7.25cm} 

\chapter{Quantum Machine Learning}\index{Quantum Machine Learning}
\label{ch:QML}
\epigraph{\textit{‘Isn’t it a shame that with the tremendous amount of work you have done you haven’t been able to get any results?’ Edison turned on me like a flash, and with a smile replied: ‘Results! Why, man, I have gotten a lot of results! I know several thousand things that won’t work.’}}{Thomas Edison, to one of its associate~\cite{Dyer190Edison_Quote}}
\vspace*{0.cm}
\startcontents[chapters]
\printcontents[chapters]{}{1}[3]{}
\vspace*{1cm}

In this chapter, we explore the field of Quantum Machine Learning (QML), one of the leading proposals to achieve a meaningful quantum advantage already with current near-term devices based on variational quantum algorithms. We start by giving a broad overview of the field in the Introduction~\ref{sec:ch_QML_Intro}, and then proceed by introducing the main elements of classical machine learning in Sec.~\ref{sec:ch_QML_Classical}. These concepts will be used to present quantum versions of learning models in Sec~\ref{sec:ch_QML_QML}, which is dedicated to variational Quantum Machine Learning. In-depth analyses and reviews on various aspects of Quantum Computing and Machine Learning can be found in the following references~\cite{ManginiQNN, BookSchuldQML, CerezoQMLPerspective_2022, BenedettiPQCReview2019, Ciliberto2018, DunjkoWittekReview2020, DunjkoBriegelReview2018, Biamonte2017QML, Dawid2022_LectureNotesQML, Carleo2019_MLPhysics, DeWolf_QuantumLearning2017}.

\section{Introduction}
\label{sec:ch_QML_Intro}

So far, we have discussed how the current generation of noisy quantum computers impelled the use of a new paradigm of computation which uses classical and quantum resources in tandem to perform a computation. At the core of variational quantum algorithms is the optimisation process, which tunes the trainable parameters of a parameterised quantum circuit in order to minimise an appropriately chosen cost function.

Needless to say, this approach of optimising --- or \textit{training} --- a parametric model to solve a complicated problem also accurately describes the field of Machine Learning (ML), especially its latest version called Deep Learning (DL)~\cite{LeCun2015DeepL, Goodfellow2016DeepL}. Indeed, state of the art Deep Learning prescribes the use of massive parametric models with billions of tunable parameters, a recipe which has shown incredible success over the past decade in a variety of tasks, ranging from controlling nuclear fusion reactors~\cite{DegraveMagneticControlTokamak2022} to generating human-level original digital art~\cite{Ramesh_DALL.E2_2022}.

The strong connection between Variational Quantum Algorithms and Deep Learning made it possible to borrow many of the concepts from the well-established field of classical machine learning and apply them to variational quantum algorithms, so much so that this field is also often referred to with the more captivating name of ``Quantum Machine Learning" (QML)~\cite{BenedettiPQCReview2019, AbbasPowerQNN2021, CerezoQMLPerspective_2022}, even though a border between the two can be drawn, as discussed later. However, it is important to stress that the field of Quantum Machine Learning has been around since earlier than Variational Quantum Algorithms gained momentum over the last few years~\cite{DunjkoWittekReview2020, Lewenstein_Perceptron_1994, GuptaQNN_2001, Wittek2014quantum,  SchuldQuestQuantumNeural2014, Schuld_IntroQML2015}. 

Earlier works at the boundaries of machine learning and quantum physics include quantum algorithms with provable speedups devoted to solving linear algebra tasks relevant for some machine learning models~\cite{lloyd2013quantum, Rebentrost2014, Biamonte2017QML}. The interest in these approaches however progressively declined over recent years, giving way to the rise of variational quantum algorithms as best representatives of quantum machine learning. This happened for a variety of reasons, including the lack of good-enough experimental hardware to run these linear algebra-based quantum subroutines --- for example, based on the HHL procedure for matrix inversion~\cite{Harrow_HHL_2009}---, subtleties in the assumptions that hinders the applicability of these algorithms in real cases~\cite{Aaronson2015Fine}, and, arguably the most important, the discovery that many of these algorithms can be \textit{dequantised}, in that there exists a class of quantum-inspired classical algorithms that has the same runtime complexity~\cite{Tang2019Dequantization}. Nonetheless, while dequantisation proved the absence of \textit{exponential speedups} of some quantum algorithms over their classical counterparts, important \textit{polynomial} improvements may still be attained in practical scenarios~\cite{Arrazola2020Quantuminspired, DunjkoWittekReview2020, CotlerDequantisation2021}. 

\subsection{The four-fold way of Quantum Machine Learning}
\label{ssec:ch_QML_CQ}
\begin{figure}[ht]
    \centering
    \includegraphics{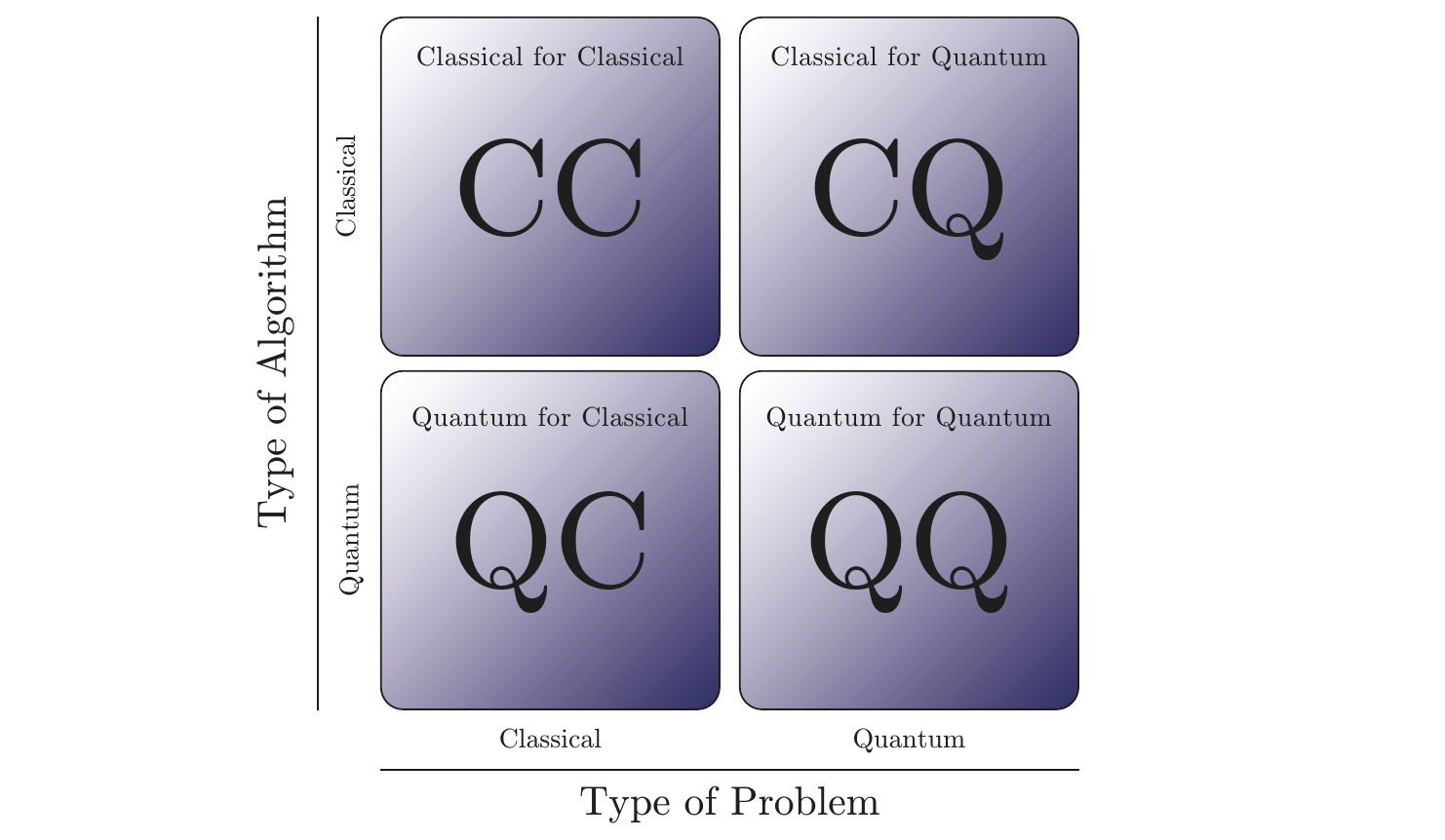}
    \caption[Variants of Quantum Machine Learning applications]{Quantum Machine Learning (QML) aims at studying the interplay between (classical) Machine Learning and Quantum Computation. Broadly speaking, the field can be schematically divided into four main areas, ``Classical for Classical" (CC), ``Classical for Quantum" (CQ), ``Quantum for Classical" (QC) and ``Quantum for Quantum" (QQ), depending on the type of algorithm used (hence the computing device, being classical or quantum), and the type of problem to be solved. See the main text for detailed explanation on such division. Such schematic representation of QML applications is customary in the literature, and can be found, for instance, in~\cite{BookSchuldQML, Mengoni2019Kernel}.}
    \label{fig:QML_main_intro}
\end{figure}

The most general definition of Quantum Machine Learning is that a research field whose aim is to investigate the interplay between (classical) Machine Learning (ML) --- and more generally, Artificial Intelligence (AI) --- and Quantum Computing, with the hope that a fruitful exchange between these two subjects can lead to mutual computational benefits and improved theoretical understandings. 

The field is quite broad and it requires specialised knowledge from various domains of science, including mathematics, computer science, statistics, and of course quantum physics. However, usual investigations regarding Machine Learning and Quantum Computation can be factorised in four main areas, almost unrelated to each other, which are usually represented as shown in Fig.~\ref{fig:QML_main_intro}. 

Indeed, depending on the type of problem to be solved, and the type of algorithm, hence the computing device, to solve it, the field of Quantum Machine Learning can be divided in:

\begin{enumerate}
    \item \textbf{Classical for Classical (CC)} This area actually has actually no quantum part to it, and indicates those purely classical cases when a classical machine learning algorithm is used to solve a classical problem, that is task defined on objects and/or datasets which does not come from a quantum process. Examples are using Deep Neural Networks to classify images~\cite{GoodfellowBengioDL} or mastering board games~\cite{MasteringGo};
    \item \textbf{Classical for Quantum (CQ)} This area indicates those studies which aim to use classical machine learning procedures to deal with problems in the quantum physics domain~\cite{Dawid2022_LectureNotesQML, Carrasquilla2020_ML4Quantum, Carleo2019_MLPhysics}. Examples are using neural networks to represent quantum states~\cite{Carleo602}, or using reinforcement learning techniques to compile quantum circuits for a quantum hardware~\cite{Moro2021};
    \item \textbf{Quantum for Classical (QC)} This area instead refers to using quantum resources or algorithms, for example variational quantum algorithms, to analyse or process classical information, that is data that come from a classical source~\cite{TacchinoQN2019, BenedettiPQCReview2019}. Examples are the use of quantum subroutines to speed up linear algebra tasks in classical machine learning algorithms~\cite{lloyd2013quantum, Kerendinis2020}, the use of parameterised quantum circuits to implement reinforcement learning agents for classical problems~\cite{skolik2022quantum, jerbi2021parametrized, SkolikNoiseQRL}, or lastly the use of quantum computers to solve optimisation problems in finance~\cite{Egger2020, HermanQuantumFinanceSurvey_2022};
    \item \textbf{Quantum for Quantum (QQ)} This last area is arguably the most compelling yet unexplored application of Quantum Machine Learning, regarding the use of quantum processors to learn or study properties of quantum systems. Indeed, while it is reasonable to assume that quantum computing devices should be particularly suited to learning properties of quantum systems, at the moment this area is the most challenging to investigate, not only for the unavailability of reliable quantum computers to run the algorithms, but especially for the lack of so-called \textit{quantum data}, on which the algorithms should be applied~\cite{CerezoQMLPerspective_2022}. Whilst a clear definition of quantum data is yet to be found, one generally refers to either: in a weaker sense, to classical information extracted from a quantum system, for example via a measurement process; or, in a stronger sense, to sets of quantum states or objects stored on a quantum memory or device, and that can be accessed or generated on demand. Examples of this area are the use of convolutional quantum neural networks to identify phase transitions or devise error correction schemes~\cite{CongQCNN2019}, and the use of variational algorithms to learn quantum circuits to prepare mixed states~\cite{EzzelMixedCompiling_2020}.
\end{enumerate}

Thus, as clear from the --- somewhat arbitrary --- division above, the subdomains of Quantum Machine Learning can be very broad and diverse, both in terms of topics and goals. In fact, as described in the review~\cite{Wittek2014quantum}, while some investigations are more applied and specifically devoted to \textit{quantum-enhanced} version of machine learning, others aim at a more theoretical exchange of ideas between machine learning and quantum physics, for example via \textit{quantum-inspired} algorithms for classical machine learning~\cite{Tang2019Dequantization, StoudenmireSupervisedTensor_2016}, or devising \textit{quantum-generalised} versions of machine learning models and tasks. Examples of this last case are generalisations of neural networks to the quantum domain~\cite{Killoran2019CVQN, Beer2020QNN, ManginiCQN2020, SchuldQuestQuantumNeural2014}, or the use of formalism of quantum mechanics to describe natural language processing (NLP) tasks~\cite{CoeckeQNLP_2020, WiebeQLP_2019}. Needless to say, the diversity of QML is also a consequence of the great diversity and very rapid development of classical Machine Learning and Artificial Intelligence.

In this work, we are mainly interested in the use of quantum resources, specifically variational quantum algorithms, to analyse classical or quantum data, thus corresponding to the ``QC" and ``QQ" cases described above and in Fig.~\ref{fig:QML_main_intro}. Despite the variety of approaches, the current leading proposal for quantum machine learning is based on variational quantum algorithms, which, thanks to the use of limited quantum circuits in tandem with classical computers, can be run on near-term quantum devices.

In the following section, we introduce some basic concepts and definitions of (classical) machine learning, which will then be used later to introduce their quantum variants or generalisations.

\section{Classical Machine Learning}
\label{sec:ch_QML_Classical}

Machine Learning (ML) is a subdomain of the broad field of Artificial Intelligence (AI), and it is a research field which investigates how to devise algorithms capable of discovering hidden patterns in data automatically, providing no ---or very little--- expert knowledge about the data to be analysed. The ``discovery" process is called \textit{learning} or \textit{training}, and usually consists of some form of optimisation procedure where a penalty (reward) function, called \textit{loss} function, depending on the data and on the task to be solved, is minimised (maximised).

Machine learning is usually divided into three main paradigms, \textit{supervised}, \textit{unsupervised} and \textit{reinforcement} learning, depending on how the algorithm interacts with the data to be analysed, and the corresponding task to be solved. These can be summarised as follows: 

\begin{enumerate}
    \item \textbf{Supervised learning:} the algorithm is asked to reproduce the mapping between a set of inputs and desired outputs, and then extrapolate the acquired knowledge also on other data that was not used during training. In this case, the dataset provided to the algorithms consists of a pair of numbers, the input to be processed, and the correct result that the machine should output. The name \textit{supervised} conveys the fact that, while training, the algorithm has a target output, chosen by a human, that it has to reproduce. Regression over a series of two-dimensional points is an example of this type of learning.
    
    This is arguably the most common and pedagogical use case of machine learning, even though in recent years it has been overshadowed by other approaches, due to the difficulty of meeting the criteria of having very large sets of \textit{annotated} datasets, that are data accompanied with the corresponding correct label. Indeed, with the advent of the big data era, the acquisition of input data is often automated and efficient, but the annotation of the dataset usually requires human intervention, which may cause a bottleneck.
    
    \item \textbf{Unsupervised learning:} in this case the algorithm has only access to a set of inputs, with no additional information about desired outputs. The goal, in this case, is to find patterns in the inputs, and extract relevant information and understanding about the whole dataset fed to the machine. The term unsupervised refers to the fact that the algorithm requires no human supervision to function.
    
    Traditional examples of unsupervised learning are clustering algorithms, which divide input data into groups sharing similar features, or Principal Component Analysis (PCA), used to compress a large-dimensional dataset into a lower-dimensional representation, depending on the direction of the highest variance of the inputs. These procedures are often used as a preprocessing step to simplify the dataset before this is used as input to other (machine learning) algorithms.

    State of the art research however points out how this paradigm of learning, when accompanied by very large parametric models and computational power, can reach amazing performances. An obvious example is Large Language Models (LLMs), which are advanced deep learning algorithms that are capable of understanding written language~\cite{Brown_GPT3_2020, ChatGPT}.
    
    \item \textbf{Reinforcement learning:} in this case the machine has no dataset at all to work with. Rather, this is \textit{generated} by the algorithm itself while running, specifically through so-called interaction with the environment. Indeed, in the reinforcement learning framework, the algorithm, also referred to as an \textit{agent}, interacts with an \textit{environment} by performing \textit{actions}. The environment then responds to the agent by giving it a \textit{reward} if the performed action was good, where ``good" is appropriately defined by the task to be solved. 

    Prototypical examples of reinforcement learning are algorithms capable of playing games, such as Chess or Go. In cases like these, there is no single obvious solution, and the algorithm is trained by trial end error, repeatedly playing the game, and rewarding it when it wins until it has learned a good-enough strategy. Again, state of the art research showed that reinforcement learning powered by large deep learning models can achieve super-human performances~\cite{MasteringGo}.
\end{enumerate}

While these three paradigms summarise most of the machine learning methods, state of the art research on the field also comprises new approaches, like semi-supervised learning~\cite{vanEngelen2019_SemiSupervisedLearning}, self-supervised learning~\cite{Doersch_SelfSupervised_2017}, continual learning~\cite{Parisi2019_ContinualLearning} and transfer learning~\cite{Weiss_TransferLearning_2016}. 

Before moving on, it is very important to stress that the most important feature of learning models is the so-called \textit{generalisation}. As the name suggests, generalisation indicates the ability of a model (or a human, for that matter) to use the knowledge acquired over a restricted set of observations on a larger set of data, that was never seen earlier and without losing performances. This is indeed the most striking feature of machine learning models, which are empirically seen to generalise very well, and justify their recent enormous success. Generalisation performances are what ultimately distinguish machine learning (especially supervised learning) from standard fitting techniques. 

In the following, we examine the basic definitions and tools of machine learning, particularly in the context of supervised learning, which is not only the most common and easiest way to introduce machine learning concepts, but is also the learning scenario used for the quantum machine learning models discussed in this work. 

\subsection{Basics of (supervised) Machine Learning}
\label{ssec:ch_QML_Basics}

The goal of machine learning is to discover patterns from a set of limited observations of a given problem, and extrapolate such knowledge to other previously unseen instances of the problem. We now introduce the main components of machine learning borrowing the terminology from statistical learning theory, which is the branch of machine learning devoted to its theoretical understanding and mathematical formulation. A thorough discussion on statistical learning theory is beyond the scope of the present work, and we refer to~\cite{MohriMLBook, ShalevBenDavid2014} for in-depth treatment of these topics.\\ 

\subsubsection{Training dataset}
The set of observations the learning algorithm has access to is called \textit{training set}, and in the supervised learning scenario it consists of a series of pairs of input data, accompanied by a corresponding desired output. In full generality, let $\mathcal{X}$ denote the input space from which inputs are drawn, and $\mathcal{Y}$ the space of the outputs. Let $\mathcal{Z} = \mathcal{X} \times \mathcal{Y}$ be the data space given by pairs of inputs and outputs, then the training set is usually defined as a set of \textit{identically independently distributed} \textit{iid} random variables
\begin{equation}
\label{eq:training_set_definition}
    S \coloneqq \qty{ z_i = (\bmx_i, y_i)~|~ z_i \sim \mathcal{D} }_{i=1}^m,\quad S \in \mathcal{Z}^m,
\end{equation}
where $\bmx_i \in \mathcal{X}$ are inputs, $y_i \in \mathcal{Y}$ are outputs, $\mathcal{D}$ is a probability distribution over the data domain $\mathcal{Z}$ from which samples $z_i \in \mathcal{Z}$ are sampled, and $m$ is the number of samples in the training dataset. 

In the vast majority of cases, inputs are vectors with real entries, $\bmx \in \mathcal{X} \subset \mathbb{R}^d$, and outputs are either real numbers $y \in \mathcal{Y} \subset \mathbb{R}$, or integers $y \in \mathcal{Y} \subset \mathbb{N}$. In the former case, the problem is referred to as \textit{regression} problem, while the latter is an example of a \textit{classification} problem, since the desired outputs are discrete, are called \textit{labels} in this context. However, there is no restriction on the nature of the inputs and the outputs, and the data space $\mathcal{Z} = \mathcal{X} \times \mathcal{Y}$ varies depending on the problem to be solved and the learning model used. For the sake of generality, in the following we keep using arbitrary domains for the data.\\

\subsubsection{Hypothesis class}
A learning model $\mathcal{M}$ is defined as a family of functions which maps the input data space $\mathcal{X}$ to the output data space $\mathcal{Y}$, namely
\begin{equation}
\label{eq:HypothesisClass}
    \mathcal{M} \subset \qty{h: \mathcal{X}\mapsto \mathcal{Y}}\,.
\end{equation}
This set of functions is called \textit{hypothesis class}, and represents the set of functions that the chosen model can implement. For example, this represents the set of all possible mappings that a given neural network (see Sec.~\ref{ssec:ch_QML_NeuralNetworks}) with a fixed architecture could implement. 

More often than not, the hypothesis class consists of a parametric model whose tunable parameters can be tuned to change the type of function implemented by the model. For simplicity, suppose the tunable parameters are real numbers $\bmw \in \mathbb{R}^p$, then the hypothesis class can be rewritten also as
\begin{equation}
\label{eq:ParameterisedHypothesisClass}
    \mathcal{M} \coloneqq \qty{ h_{\bmw}: \mathcal{X} \mapsto \mathcal{Y}~|~ h_{\bmw} = h(\cdot~;~\bmw),\, \bmw \in \mathbb{R}^p}\,,
\end{equation}
where $h_{\bmw} = h(\cdot~;~\bmw)$ is the specific parameterised function depending on trainable parameters $\bmw$.\\

\subsubsection{Empirical Risk Minimisation}
Given the training set and a hypothesis class, we now need to introduce a measure of the performance of the model. As with variational quantum algorithms, such measure is defined in terms of an \textit{objective function} (or \textit{loss function}, or \textit{cost function}) which measures how well a model $h \in \mathcal{M}$ is performing over the dataset $S$, and depends heavily on the task to be solved. The introduction of the loss function as a way to measure the fitness of the model effectively renders the training process an optimisation procedure. 

Let $\ell: \mathcal{Y} \times \mathcal{Y} \mapsto \mathbb{R}$ be a loss function, the performances of a model $h$ in an hypothesis class $\mathcal{M}$ are measured by the \textit{empirical risk} over the training dataset $S$, defined as
\begin{equation}
    \label{eq:empirical_risk}
    L_S(h) \coloneqq \frac{1}{m}\sum_{i=1}^m \ell(\hat{y}_i,\, y_i)\,, \quad \textrm{with}\quad \hat{y}_i = h(\bmx_i)\,,
\end{equation}
where $\hat{y}_i(\bmw)$ is the prediction of the model when evaluated on input $\bmx_i$, and $y_i$ is the desired output corresponding to such input, as prescribed by the training set $S$~\eqref{eq:training_set_definition}.
Alternatively, if the hypothesis class is determined by a parametric model as in Eq.~\eqref{eq:ParameterisedHypothesisClass}, the equivalent definitions hold
\begin{equation}
\label{eq:empirical_risk_2}
    L_S(\bmw) \coloneqq \frac{1}{m}\sum_{i=1}^m \ell(\hat{y}_i,\, y_i)\,, \quad \textrm{with}\quad \hat{y}_i(\bmw) = h_{\bmw}(\bmx_i) = h(\bmx_i;\, \bmw)\,.
\end{equation}
The optimal model is then defined as the one that minimises (or maximises) the empirical risk
\begin{equation}
    \label{eq:optimal_model_ERM}
    h_{\text{opt}} = \argmin_{h \in \mathcal{M}}~L_S(h) \quad \text{or equivalently}
    \quad  \bmw_{\text{opt}} = \argmin_{\bmw}~L_S(\bmw)\,,
\end{equation}
where the second equation above is equivalent to the definition of the optimal solution for variational quantum algorithms, as discussed in Eq~\eqref{eq:vqa_argmin}.  

The empirical risk~\eqref{eq:empirical_risk_2} is also called \textit{training error}, as it measures the fitness of the model on the available data contained in the training set. Such learning paradigm of defining the optimal model as the one having the lowest error on the training set is called \textit{Empirical Risk Minimisation}
(ERM)~\cite{ShalevBenDavid2014}. Empirical risk is opposed to the \textit{true risk}, which is defined as the average value of the loss evaluated over the actual probability distribution $\mathcal{D}$ from which observations $z_i = (\bmx_i, y_i) \sim \mathcal{D}$ are sampled, that is
\begin{equation}
    \label{eq:true_risk}
    L_{\mathcal{D}}(h) \coloneqq \mathbb{E}_{z \sim \mathcal{D}}\qty[\ell\qty(\hat{y};\, y)]\,, \quad \textrm{with}\quad \hat{y} = h(\bmx),\,\, z = (\bmx, y)\,.
\end{equation}

It is important to note that true risk~\eqref{eq:true_risk} represents the actual quantity of interest that characterises the real performance of a learning model when dealing with the task. Indeed, by definition, the true risk (or \textit{expected risk}) measures the performances of the learner on the whole probability distribution defining the task, not just on a restricted set of samples. However, this distribution is not known to the learner and the expected loss cannot be calculated directly, even though bounds on it can be derived, as discussed later in Sec.~\ref{ssec:ch_QML_Generalisation}. 

Thus, it is reasonable to use Empirical Risk Minimisation to select the optimal model, because it is the one that minimises the error over the available information of the problem to be solved, namely the one contained in the training dataset.\\

\subsubsection{Loss functions and Learning}
\label{ssec:loss_function}
Supervised machine learning tasks can be grouped into two main classes, namely \textit{Regression} and \textit{Classification} problems, which differ essentially for the type of output that the learners have to reproduce.

\textit{Regression} problems are those problems where the output data space is the real line, $\mathcal{Y} \subset \mathbb{R}$, and the goal of the learner is thus to learn a function mapping the inputs $\bmx_i \in \mathcal{X}$ to their corresponding outputs in the training set $y_i \in \mathcal{Y}$. The go-to loss function used in this scenario is the Mean Squared Error (MSE), defined as
\begin{equation}
    \label{eq:mse}
    L_S(h) = \frac{1}{m}\sum_{i=1}^{m} (\hat{y}_i - y_i)^2\,,
\end{equation}
where again $\hat{y}_i = h(\bmx_i)$ is the prediction of the model $h$ on input $\bmx_i$. A straightforward generalisation to the case of multidimensional outputs $\mathcal{Y} \subset \mathbb{R}^d$ is obtained by substituting the scalar squared difference with the Euclidean norm of the difference, namely $\ell(\hat{\bm{y}}_i,\, \bm{y}_i) \coloneqq \norm{\hat{\bm{y}}_i-\bm{y}_i}^2_2$.

\textit{Classification} tasks instead are those for which the outputs, called \textit{labels} in this context, are integers values $\mathcal{Y} \subset \mathbb{N}$. Thus, in this case, the goal of the learner is to split the inputs into separate classes, by assigning to each input in the dataset the correct label. For simplicity, let's consider the case of a binary classification task where the labels can only take two distinct values $\mathcal{Y} = \{0, 1\}$. The prediction of the learning model is then a real number $\hat{y}_i \in [0,1]$ that encodes the probability that the input $\bmx_i$ belongs to the class $0$, or $1$. The standard loss function used in this case is the so-called Crossentropy, defined as
\begin{equation}
    \label{eq:crossentropy}
    L_S(h) = \frac{1}{m}\sum_{i=1}^m - \qty( \hat{y}_i\log y_i + (1 - \hat{y}_i) \log(1-y_i))\,.
\end{equation}
which measures the difference between the predicted probability and the ground truth. Generalisations to \textit{multiclass} classification problems (that is those where the number of classes is greater than two) are straightforward.

The introduction of a loss function makes the training, that is the procedure by which an optimal model is selected out of the hypothesis class, an optimisation procedure, as clear from the ERM approach of Eq.~\eqref{eq:empirical_risk_2}. Specifically, whenever the hypothesis class is a parametric model, training consists in adjusting the trainable parameters to minimise the empirical loss $L_S(\bmw)$. As with variational quantum algorithms, such minimisation is implemented via variants of gradient descent~\eqref{eq:gradient_descent}, which we report also here for simplicity
\begin{equation}
\label{eq:gradient_descent_ML}
    \bmw^{(t+1)} = \bmw^{(t)} - \eta \nabla_{\bmw} L_S(\bmw)\big|_{\bmw^{(t)}}\, .
\end{equation}

Gradient descent-like update roles are particularly suited to neural network architectures, since there is an efficient strategy, called \textit{backpropagation}, to compute the partial derivatives of the empirical loss with respect to each parameter in the model. However, as shown later in the section regarding linear models~\ref{ssec:ch_QML_Kernel}, there are cases where performing gradient-descent in the parameters space is not needed, as the optimal parameters can be found using an analytical closed-form solution.

\subsubsection{Generalisation}
\label{sec:ch_QML_generalisation}
The success of machine learning models, especially state of the art Deep Learning ones, is rooted in their \textit{generalisation} performances, that is their ability to effectively use the knowledge extracted from the limited set of observations in the training dataset, also to new observations, which were not used during the training procedure. Generalisation conveys the desirable requirement that the learner has truly \textit{understood} something of the problem to be solved, whereas it didn't just learn by heart the patterns in the training set.\\

\paragraph{Test set}  As we argued earlier, the true measure of performance (and generalisation) of a learner is given by the expected loss $L_{\mathcal{D}}(h)$, which is however impossible to access because the probability distribution of the samples $\mathcal{D}$ is not known. 

A practical solution to the estimation of generalisation performances of a model is to take the available corpus of observations $\{z_i\}_i \in \mathcal{Z}^{m+m'}$, and split it into two separate datasets: the training set $S \in \mathcal{Z}^m$, and a \textit{test} set $T \in \mathcal{Z}^{m'}$. The former, defined before, is used during the training procedure to select the optimal model $h_\text{opt}$ via minimisation of the empirical risk $L_S(h_\text{opt})$; while the latter is used only at the end of training, to measure the performance of the model on previously unseen observations, called test error $L_T(h_\text{opt})$, which is exactly a measure of the generalisation capabilities of the model. Generally, it is common practice to use about $80\%$ of the available data to build the training set, and the remaining $20\%$ for the test set\footnote{Actually, the best practice for implementing a safe training procedure involves the use of three distinct datasets: training, test and \textit{validation} set, where the latter is used to monitor the generalisation error of the model \textit{while} the model is training.}.\\

\paragraph{Generalisation bounds}
\label{ssec:ch_QML_Generalisation}
It is desirable for a learner $h$ that its training error $L_T(h)$ and generalisation error $L_\mathcal{D}(h)$ are close, so that the performance shown on the training dataset are representative of those obtained also on new data drawn from the same distribution.

This is true when the number of samples $m$ in the training dataset is large, equivalent to saying that the model has access to a large amount of information about the problem to be solved. Given an hypothesis $h$, one can show that the probability that the true risk $L(h)$~\eqref{eq:true_risk} and the empirical risk $L_S(h)$~\eqref{eq:empirical_risk_2} are different goes to zero as the size $m$ of the training set $S \sim \mathcal{D}^m$ goes to infinity, namely
\begin{equation}
    \label{eq:asymptotic_generalisation_gap}
    \lim_{m \rightarrow \infty} P\qty(\abs{L_{\mathcal{D}}(h) - L_S(h)} \geq \epsilon) =
    \lim_{m \rightarrow \infty} P\qty(\abs{\mathbb{E}_{z \sim \mathcal{D}}\qty[\ell(z)] - \frac{1}{m}\sum_{i=1}^m \ell(z_i)} \geq \varepsilon) \rightarrow 0\,.
\end{equation}
This result is an application of the law of large numbers, which states that the average loss $L_S(h)$ tends to its expected value $L_{\mathcal{D}}(h)$~\eqref{eq:true_risk} as the sample size goes to infinity. Note that, for ease of notation, we used the simplified expressions for the risk $\ell(z) = \ell(h(x), y)$ and $\ell(z_i) = \ell(h(x_i), y)$.

While interesting, the result in Equation~\eqref{eq:asymptotic_generalisation_gap} holds only asymptotically and gives no information about real case scenarios when the training set has a finite size. Luckily, one of the greatest achievements of statistical learning theory was to show that guarantees on the generalisation performances of a learning model can be obtained also in the finite size case~\cite{VapnicChervonenkisPAC, ValiantPAC, BlumerPAC, ShalevBenDavid2014}. Indeed, making use of concentration inequalities~\cite{BoucheronBookConcentrationInequalities}, one can derive probabilistic statements about the generalisation performances of a hypothesis class $\mathcal{M}$, given a training dataset $S \in \mathcal{Z}^m$ consisting of \textit{iid} samples $z_i$ drawn from a probability distribution $\mathcal{D}$. These statements are referred to as \textit{generalisation bounds}, and roughly take the following form~\cite{Reid2010, ShalevBenDavid2014, Caro2021encodingdependent}. 

\begin{definition}[A general Generalisation Bound]
Let $\mathcal{M}$ be an hypothesis class and $\mathcal{D}$ a probability distribution, for all $\delta \in (0,1)$ with probability $1-\delta$ over randomly drawn samples $S \sim \mathcal{D}^m$, and for all $h \in \mathcal{M}$ it holds that
\begin{equation}
    \label{eq:generalisation_bound}
    L_\mathcal{D}(h) \leq  L_S(h) + f(\mathcal{M}, m, \delta)\,,
\end{equation}
where $f(\mathcal{M}, m, \delta)$ is a function that depends on the sample size $m$, the probability of error $\delta$, and the hypothesis class $\mathcal{M}$, specifically through measures of its \textit{complexity} (or \textit{capacity}), which is a measure of the expressible power or richness of the hypothesis class. Examples of such complexity measures are the VC-dimension~\cite{VapnicChervonenkisPAC} or the Rademacher Complexity~\cite{ShalevBenDavid2014}.
\end{definition}
Generalisation bounds are statements about the predictive power of a learning model, expressing its generalisation performances in terms of two contributions: the empirical error obtained on the available data, and the flexibility of the hypothesis class. Roughly, if the complexity of the model can be controlled, and the empirical error is low, then one can be confident that the true error will be also small, and so some form of generalisation will take place. These bounds are interesting because, based on the available information on the performance of the learner, $L_S(h)$, it is possible to bound the true quantity of interest, namely the generalisation error $L_\mathcal{D}(h)$, even though this is not directly accessible. In a certain sense, generalisation bounds can be seen as a mathematical formulation of Occam's razors: out of many possible explanations (the models), the simplest one (low complexity of the model) should be preferred.

There exists a plethora of generalisation bounds in the statistical learning literature, each one stemming from different assumptions and complexity measures. The interested reader can find detailed information in~\cite{ShalevBenDavid2014, MohriMLBook}, as well as in Appendix~\ref{app:proofs_QML}, where we show a concrete example of a generalisation bound for linear models based on Rademacher Complexity. For our discussion, it is sufficient to know that such bounds exist, since, as shown in Sec~\ref{sec:ch_QML_GeneralisationQNN}, we will discuss a specific scenario in which it is possible to derive a generalisation bound also for some type of quantum machine learning models, specifically data-reuploading quantum neural networks.\\

\paragraph{Overfiting}
\label{ssec:ch_QML_overfitting}
\begin{figure}[ht]
    \centering
    \includegraphics[width=\textwidth]{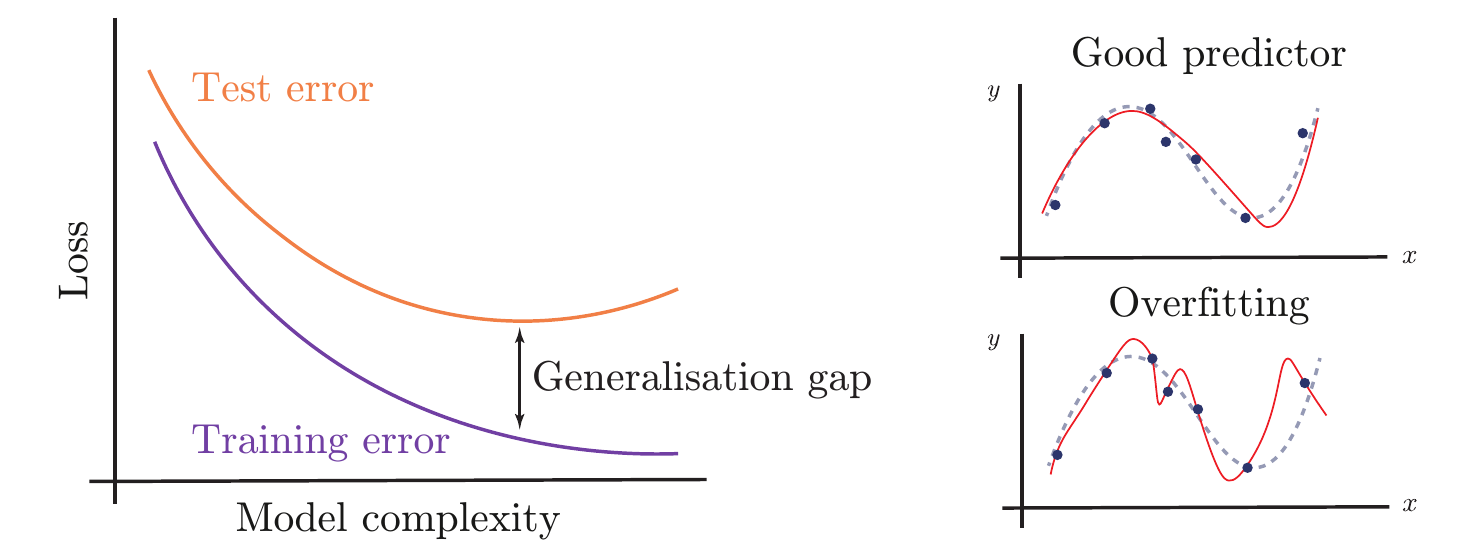}
    \caption[Generalisation and overfitting in machine learning]{Generalisation and overfitting in supervised machine learning. There is a trade-off between the minimisation of the training error and the complexity of the learning model: if the model is not complex enough, it won't be able to solve the problem (large training and test error), however, if the model is too expressible (complex) it will fit precisely the training data (low training error), at the cost of compromising its generalisation (high test error). The difference between the loss attained on the training and test set is called \textit{generalisation gap}. On the right, is an explicit example of overfitting in fitting a sinusoidal function (dashed line) given only a few data samples (the training dataset, blue dots): while a good model is capable of reproducing the desired sinusoidal behaviour of the data, and an overfitting model has zero error on the training data but miss the sinusoidal behaviour. Such a representation of the generalisation properties of a model is common in the machine learning literature, and can be found for example in~\cite{Hastie2009Statistical, GoodfellowBengioDL}.}
    \label{fig:generalisation_overfitting}
\end{figure}

Practically, generalisation is not guaranteed to happen in machine learning models unless specific actions, like regularisation techniques, are used to enforce it. Whenever the learner shows remarkable performances on the training set but fails to do the same on the test set, the model is said to be \textit{overfitting} the training data. In this scenario, the learning model has specialised to reproduce exactly the mapping in the training dataset but not on other samples, thus showing poor generalisation.

Such a scenario is graphically depicted in the left panel of Fig.~\ref{fig:generalisation_overfitting}. As the complexity of the hypothesis class increased the test error of the model usually follows an ``U"-shaped curve, which is a clear sign that the mode model has started overfitting the training data. Indeed, a first increase in complexity is needed to give the model enough flexibility to deal with the problem, but there is a moment after which the model is so expressible ---i.e. complex--- that it can explain every feature in the training data (even noisy data) so well that its knowledge cannot be generalised any more to other samples, and the error on the test set thus start increasing. An example of a good and bad learner is shown in the right panels of Fig.~\ref{fig:generalisation_overfitting}, for the task of reproducing a sinusoidal signal: whereas a good model (red line) is able to reproduce with good accuracy the required input-output mapping of the training data (blue dots) without compromising generalisation, an overfitting model reproduce exactly the training data at the cost of introducing data-dependent artefacts that miss the salient features of the underlying true problem. The trade-off between being to accurately model the pattern in the training data without compromising the generalisation performances is known as the \textit{bias-variance trade-off}.

In order to combat overfitting one resorts to regularisation techniques, which are set of procedures that impose constraints on the model to limit its complexity, with the hope that these help preserve generalisation. Examples are early-stopping, which stops the gradient-descent minimisation of the training loss before this reaches zero, at which point the model would have presumably overfitted the data, or the use of penalty terms in the loss function to favour simpler models over more complex ones~\cite{Hastie2009Statistical}.

\subsection{Machine learning models}

We now give two examples of machine learning models, specifically linear models (and kernel methods) and neural networks, of which generalised quantum versions have been proposed in the literature.

\subsubsection{Linear models and kernel methods}
\label{ssec:ch_QML_Kernel}

Linear models correspond to the parametric hypothesis class~\eqref{eq:ParameterisedHypothesisClass}
\begin{eqnarray}
    \label{eq:linear_model_class_main}
    \mathcal{M}_{\bmw} = \qty{ h_{\bmw}: \mathcal{X} \subset \mathbb{R}^d \mapsto \mathcal{Y} \subset \mathbb{R}~|~ h_{\bmw}(\bmx) = \bmw \cdot \bmx,~\bmw \in \mathbb{R}^d}\,,
\end{eqnarray}
where we added the subscript $\mathcal{M}_{\bmw}$ to make it explicit that the hypothesis class is implemented by a parametric model depending \textit{linearly} on parameters $\bmw \in \mathbb{R}^d$. Linear models like the one in~\eqref{eq:linear_model_class_main} can be used for regression tasks with the empirical risk given by the mean squared error
\begin{equation}
    L_S(\bmw) = \frac{1}{m}\sum_{i=1}^m (y_i - h_{\bmw}(\bmx_i))^2 = \frac{1}{m}\sum_{i=1}^m (y_i - \bmw \cdot \bmx_i)^2\,.
\end{equation}
By grouping inputs and outputs in the following matrix from
\begin{equation}
    \label{eq:linear_matrix_notation}
    \bm{X} \coloneqq
    \brows{\bmx_1 \\ \bmx_2 \\ \rowsvdots \\ \bmx_m}
    \in \mathbb{R}^{m \times d},\,
    \quad 
    \bm{y} \coloneqq 
    \brows{y_1 \\ y_2 \\ \rowsvdots \\ y_m}
    \in \mathbb{R}^m,
\end{equation}
the empirical mean squared error can be rewritten more compactly as
\begin{equation}
    \label{eq:linear_mse_vectorial}
    L_S(\bmw) = \frac{1}{m}(\bm{y} - \bm{X}\bmw)^2 = \frac{1}{m}\norm{\bm{y} - \bm{X}\bmw}_2^2\,,
\end{equation}
The optimal model is the one minimising the empirical loss
\begin{equation}
    \bmw_\text{opt} = \argmin_{\bmw} \norm{\bm{y} - \bm{X}\bmw}_2^2 
\end{equation}
and has a closed form expression, known as \textit{least square estimator}, which amounts to~\cite{Dawid2022_LectureNotesQML}
\begin{equation}
    \label{eq:opt_linear}
    \bmw_\text{opt} = (\bm{X}^\intercal \bm{X})^{-1} \bm{X}^\intercal \bm{y}\,.
\end{equation}
In this case, we assumed that the matrix $\bm{X}^\intercal\bm{X}$ is non-singular, hence admits an inverse. In general, the solution to the least square minimisation problem~\eqref{eq:linear_mse_vectorial} is given by the Moore-Penrose pseudoinverse, a generalisation of the inverse of a matrix, denoted as $\bmw_\text{opt} = \bm{X}^{+}\bm{y}$~\cite{BelkinFitFear, HastieSurprises2022, Goodfellow2016DeepL, BenIsraelGeneralisedInverse}\footnote{When there are more samples than parameters $n > p$, the learning model is said to be \textit{underparameterised}, and the related system of equations $\bm{X}\bmw = \bm{y}$ is \textit{underdetermined}, which means that a solution may not exist. In this case, if $\bm{X}$ has full column rank, then $(\bm{X}^\intercal \bm{X})^{-1}$ exists and the Moore-Penrose pseudoinverse reduces to $\bm{X}^{+} = (\bm{X}^\intercal\bm{X})^{-1}\bm{X}^\intercal$, as in Eq.~\eqref{eq:opt_linear} for ordinary least squares. On the contrary, if there are more parameters than training data $n < p$, the model is said to be \textit{overparameterised} and the system $\bm{X}\bmw = \bm{y}$ \textit{overdetermined}, that is multiple solutions exist. In this case, if $\bm{X}$ has full row rank, then $(\bm{X}\bm{X}^\intercal)^{-1}$ exists and the Moore-Penrose presudoinverse reduces to $\bm{X}^{+} = \bm{X}^\intercal (\bm{X}\bm{X}^\intercal)^{-1}$. At last, if $n=p$ and $\bm{X}^{-1}$ exists, then the solution is simply $\bmw_\text{opt} = \bm{X}^{-1}\bm{y}$. The introduction of a regularisation term as in Eq.~\eqref{eq:ridge_regression} is not only useful to enforce generalisation, but also to ameliorate issues related to the singularity of the data matrices.}.

A more comprehensive treatment of linear regression is given by introducing a regularisation term in the loss~\eqref{eq:linear_mse_vectorial} which penalises solutions with large norm, and also avoids subtleties related to the invertibility of the data matrices. This approach is known as \textit{ridge regression}, and the optimal solution is defined as
\begin{equation}
    \label{eq:ridge_regression}
        \bmw_\text{opt} = \argmin_{\bmw}\qty{ \frac{1}{m}\norm{\bm{y} - \bm{X}\bmw}_2^2 + \frac{\lambda}{m}\norm{\bmw}^2_2 }\,.
\end{equation}
where $\lambda > 0$ is a parameter that controls the trade-off between minimising the empirical risk and preferring lower-norm solutions. Similarly to the previous case, the optimal parameters can be written explicitly as
\begin{equation}
    \label{eq:wopt_mse}
    \bmw_\text{opt} = (\bm{X}^\intercal\bm{X} + \lambda\mathbb{I})^{-1}\bm{X}^\intercal\bm{y}\,,
\end{equation}
where the matrix $(\bm{X}^\intercal\bm{X} + \lambda\mathbb{I})$ is non singular. Moreover, by making use of the matrix equality $(\bm{X}^\intercal\bm{X}+\lambda \mathbb{I})\bm{X}^\intercal = \bm{X}^\intercal(\bm{X}\bm{X}^\intercal + \lambda\mathbb{I})$~\cite{Dawid2022_LectureNotesQML, MatrixCookbook, BenIsraelGeneralisedInverse}, the optimal model can be eventually written in a more convenient form
\begin{equation}
    \label{eq:linear_opt_dual}
    \begin{aligned}
    h_\text{opt}(\bmx) & = \bmw_\text{opt} \cdot \bm{x} = \bm{x}^\intercal \bmw_\text{opt} \\
    & = \bm{x}^\intercal \bm{X}^\intercal\, (\bm{X}\bm{X}^\intercal + \lambda\mathbb{I})^{-1}\bm{y}\,.
    \end{aligned}
\end{equation}
which is often referred to as the solution of the \textit{dual} problem of linear ridge regression. Let's now analyse the terms in this expression. First, one can easily check that the elements of the matrix $\bm{X}\bm{X}^\intercal \in \mathbb{R}^{m\times m}$ are the inner products of the training input vectors, namely 
\begin{equation}
    \label{eq:GramMatrix}
    K_{ij} \coloneqq [\bm{X}\bm{X}^\intercal]_{ij} = \bmx_i \cdot \bmx_j\,
\end{equation}
where $K = K_{ij}$ is usually called \textit{kernel}, or \textit{Gram}, matrix. Then, noticing that
\begin{gather}
    \label{eq:optimal_params_ridge}
    \bmx^\intercal \bm{X}^\intercal  = 
    [\bmx \cdot \bmx_1,~\bmx \cdot \bmx_2,~\hdots,\,~\bmx \cdot \bmx_m]    
    \in \mathbb{R}^{m}\,, \\
    \hat{\bm{\alpha}} \coloneqq (K + \lambda \mathbb{I})^{-1}\bm{y} \in \mathbb{R}^{m}\,.
\end{gather}
where $\hat{\bm{\alpha}}$ are often called \textit{dual} variables of the optimal weights $\bmw_{\text{opt}}$~\eqref{eq:wopt_mse}, one can finally rewrite the optimal predictor~\eqref{eq:linear_opt_dual} as
\begin{equation}
    \label{eq:linear_opt_kernel}
    h_\text{opt}(\bmx) = (\bmx^\intercal \bm{X}^\intercal) \cdot \hat{\bm{\alpha}} = \sum_{i=1}^m \hat{\alpha}_i\, (\bmx \cdot \bmx_i) = \bmx \cdot \qty(\sum_{i=1}^{m} \hat{\alpha}_i\,\bmx_i)\,.
\end{equation}
This expression makes it evident that the optimal model depends on the data only through the inner products among data samples, and that it consists of a linear combination of the inner product between the new sample $\bmx$, with all the input data in the training set $S \in \mathcal{Z}^m$. In addition, as clear from the last equality in~\eqref{eq:linear_opt_kernel}, the optimal weights can be expressed as a linear combination of such training samples.\\

\paragraph{Feature maps}
The derivation above follows identically even when we allow the input data $\bmx_i \in \mathcal{X} \subset \mathbb{R}^{d}$ to go through an arbitrary function $\bm{\phi}: \mathcal{X} \mapsto \mathcal{F} \subset \mathbb{R}^s$. Indeed, consider the arbitrary mapping
\begin{equation}
    \label{eq:linear_model_feature}
    \bmx \mapsto \bm{\phi}(\bmx) = \qty( \phi_1(\bmx),\, \phi_2(\bmx),\,\hdots,\,\phi_s(\bmx) )\,,
\end{equation}
this is called \textit{feature map}, and the inputs now belong to a new space called \textit{feature space}, in this case $\mathcal{F} \subset \mathbb{R}^s$. The linear model now acts on such feature vectors as $h(\bmx) = \bmw \cdot \bm{\phi}(\bmx)$. The construction of the optimal predictor derived before can be applied identically also in this case, simply by substituting the data matrix $\bm{X}$ with the feature matrix
\begin{equation}
    \label{eq:feature_matrix}
    \bm{F} \coloneqq
    \brows{\bm{\phi}(\bmx_1) \\ \bm{\phi}(\bmx_2) \\ \rowsvdots \\ \bm{\phi}(\bmx_m)}
    \in \mathbb{R}^{m \times s}\,,
\end{equation}
with the optimal predictor now being
\begin{align}
    \label{eq:linear_opt_kernel_feature}
    h_\text{opt}(\bmx) &= \sum_{i=1}^m \hat{\alpha}_i\, (\bm{\phi}(\bmx) \cdot \bm{\phi}(\bmx_i)),\quad \hat{\bm{\alpha}} = (\bm{F}\bm{F}^\intercal + \lambda \mathbb{I})^{-1}\bm{y}\\
    & = \sum_{i=1}^m \hat{\alpha}_i\, \kappa(\bmx, \bmx_i),\quad\quad\quad\,\,\,\, \hat{\bm{\alpha}} = \qty([\kappa(\bmx_i, \bmx_j)]_{ij} + \lambda \mathbb{I})^{-1}\bm{y}\,.\label{eq:kernel_model_0}\,.
\end{align}
In the last line we defined the \textit{kernel function} $\kappa: \mathcal{X} \times \mathcal{X} \mapsto \mathbb{R}$, which takes two inputs and outputs the scalar value $\kappa(\bmx, \bmx_i) \coloneqq \bm{\phi}(\bmx) \cdot \bm{\phi}(\bmx_i)$, from which one can define again the associated kernel matrix $K \in \mathbb{R}^{m \times m},\, K_{ij} \coloneqq \kappa(\bmx_i, \bmx_j)$, evaluated on the training points. 

The idea of using a feature map is to enrich the expressibility of the parameterised model by introducing a nonlinear dependence on the input data. Specifically, while the regression or classification task may not be solvable (or have high error) in the original data space $\mathcal{X}$, it may be much easier to solve in an appropriately chosen feature space~\cite{Dawid2022_LectureNotesQML}. Additionally, as clear from the expression~\eqref{eq:kernel_model_0}, it is important to remark that the optimal model \textit{only} depends on the data samples directly but \textit{only} on inner products. This turns out useful because there are cases where a closed form expression for $\kappa(\cdot, \cdot)$ exists, and it is thus not necessary to transform the data with feature map $\bm{\phi}(\cdot)$ and then computing the inner product in the feature space $\mathcal{F}$. This simplification is known as \textit{kernel trick}.\\

\paragraph{Kernel machines} 
Equation~\eqref{eq:kernel_model_0} may suggest that instead of considering the hypothesis class of linear models~\eqref{eq:linear_model_class}, one could start directly from considering the parameterised class of predictors of the form
\begin{equation}
    h(\bmx) = \sum_{i=1}^m \alpha_i\, \kappa(\bmx, \bmx_i)\label{eq:kernel_models}\,,
\end{equation}
where $\kappa : \mathcal{X} \times \mathcal{X} \rightarrow \mathbb{R}$ is a general but fixed \textit{kernel} function, the parameters $\bm{\alpha} \in \mathbb{R}^m$ are to be optimised, and the model thus consists of a linear combination of kernel evaluations of the new data $\bmx$ with those in the training set $\{\bmx_i\}_{i=1}^m$. Models like the one in Eq.~\eqref{eq:kernel_models} are called \textit{kernel methods}, as they are based on the choice of an appropriate kernel function, which is a measure of similarity between data samples. Moreover, if the kernel function is symmetric and positive definite (known as Mercer's conditions), then there exist a \textit{feature} map $\bm{\phi}: \mathcal{X} \mapsto \mathcal{F}$ and a \textit{feature} Hilbert space $\mathcal{F}$ such that the kernel function is equivalent to the inner product of vectors in such feature space~\cite{MohriMLBook, ShalevBenDavid2014}
\begin{equation}
    \kappa(\bmx, \bmx') = \braket{\phi(\bmx)}{\phi(\bmx')}_{\mathcal{F}}
\end{equation}
where $\braket{\cdot}{\cdot}_{\mathcal{F}}$ is the inner product on the Hilbert space $\mathcal{F}$, which is called the Reproducing Kernel Hilbert Space (RKHS) of the kernel function $\kappa: \mathcal{X} \times \mathcal{X} \rightarrow \mathbb{R}$. Interestingly, it turns out that kernel predictors of the form~\eqref{eq:kernel_models} are very general and powerful, since they arise as a result of common machine learning optimisation problems. This is a consequence of the renowned \textit{Representer theorem} of statistical learning, which shows that for supervised learning tasks that minimise an empirical risk in an RKHS, the optimal solutions can be expressed simply as a linear combination of kernel evaluations of with the training data~\cite{MohriMLBook, ScholkopfLearningKernelsSupport2002, Schuld2019FeatureSpace}. 

While this seems obscure, this result is important because it guarantees that one can formulate the search of an optimal model in a possibly infinite-dimensional RKHS, simply as a search of kernel expansion coefficients $\{\alpha_i\}_{i=1}^m$ as in~\eqref{eq:kernel_models}. For example, suppose one wants to minimise a linear model with a feature map 
\begin{equation}
    \label{eq:linear_to_kernel_example}
    h(\bmx) = \braket{\bmw}{\bm{\phi}(\bmx)}_{\mathcal{F}}
\end{equation}
where $\bm{\phi}: \mathcal{X} \rightarrow \mathcal{F}$, and the dimension of the feature space $\mathcal{F}$ is large, possibly even infinite. By virtue of the Representer theorem, one knows that the optimal predictor can be expressed as $h_\text{opt}(\bmx) = \sum_{i=1}^m \alpha_i \kappa(\bmx, \bmx_i)$ where $\kappa(\bmx, \bmx') = \braket{\bm{\phi}(\bmx)}{ \bm{\phi}(\bmx')}_{\mathcal{F}}$ is the kernel induced by the feature map. Thus, one only has to search for the $m$ real parameters $\alpha_i$ instead of looking directly for $\bmw_\text{opt}$ in the large dimensional space $\mathcal{F}$. An example of infinite dimensional feature space is that corresponding to the so-called Radial Basis Function (RBF) kernel $\kappa(\bmx, \bmx') = \exp(-\norm{\bmx-\bmx'}^2/2 \sigma^2)$, which is induced by an infinite feature map $\bm{\phi}(\bmx) = [\phi_1(\bmx),\, \phi_2(\bmx),\, \hdots]$~\cite{Dawid2022_LectureNotesQML}. Instead, an example of finite but large feature space is that of quantum states, which will be the topic of the next sections.

A common application of a kernel method is \textit{Kernel Ridge Regression} (KKR), where the kernel model~\eqref{eq:kernel_models} is trained with the regularised squared loss
\begin{equation}
    \label{eq:kernel_ridge_regression}
    L_S(h) = \sum_{i=1}^{m} (y_i - h(\bmx_i))^2 + \lambda \norm{h}_{\mathcal{F}}^2 = (\bm{y} - K \bm{\alpha})^2 + \lambda \bm{\alpha}^\intercal K\bm{\alpha}\,,
\end{equation}
where $K_{ij} = \kappa(\bmx_i, \bmx_j)$ is the kernel, or Gram, matrix evaluated on the training samples. Note that the first term is the usual squared loss~\eqref{eq:mse}, and the second term is a regularisation term similar to the one used in standard ridge regression~\eqref{eq:ridge_regression}, and depends on the norm of the kernel model in the corresponding RKHS\footnote{If the kernel $\kappa$ is a legitimate symmetric positive definite kernel, then it induces an inner product in a Reproducing Kernel Hilbert Space (RKHS) $\mathcal{F}$, so that $\kappa(\bmx, \bmx') = \braket{\phi(\bmx)}{\phi(\bmx')}_{\mathcal{F}}$. Thus the kernel model can be written as
\begin{equation}
    h(\bmx) = \sum_{i=1}^m \alpha_i\, \kappa(\bmx, \bmx_i) = \braket{\phi(\bmx)}{\sum_{i=1}^m \alpha_i \, \phi(\bmx')}_{\mathcal{F}} = \braket{\phi(\bmx)}{\Phi}_{\mathcal{F}}
\end{equation}
which is a ``linear model" in the RKHS of the kernel function, and the model ``parameters" are given by the state $\ket{\Phi} = \sum_{i=1}^m \alpha_i \ket{\phi(\bmx')}$. Then, the norm of the model in the RKHS is defined as $\norm{h}_{\mathcal{F}} = \sqrt{\braket{h}{h}_{\mathcal{F}}}$, and so $\norm{h}_{\mathcal{F}}^2 = \braket{\Phi}{\Phi} = \sum_{i,j=1}^{m}\alpha_i \alpha_k \braket{\phi(\bmx_i)}{\phi(\bmx_j)} = \sum_{i,j=1}^{m}\alpha_i \alpha_k \kappa(\bmx_i, \bmx_j) = \bm{\alpha}^\intercal K \bm{\alpha}$, where $K_{ij} = \kappa(\bmx_i, \bmx_j)$ is the kernel matrix.}. The empirical loss is convex with respect to the parameters, and the minimum can be found analytically, achieved with parameters $\hat{\bm{\alpha}} = \qty(K + \lambda \mathbb{I})^{-1}\bm{y}$. 

Another ubiquitous example of kernel methods for classification ---rather than regression--- task is Support Vector Machines (SVM), which minimize the so-called \textit{hinge loss} instead of the mean squared error~\cite{MohriMLBook, ShalevBenDavid2014, Dawid2022_LectureNotesQML}. To summarise, the general idea of kernel methods, is to define an appropriately chosen measure of similarity between samples (i.e. the kernel function $\kappa(\bmx, \bmx')$), and then search just for the expansion coefficients. The supervised learning task of finding the optimal parameters essentially translates to that of finding a proper kernel function for the task to be solved.


\subsubsection{Neural Networks}
\label{ssec:ch_QML_NeuralNetworks}
\begin{figure}[ht]
    \centering
deep    \includegraphics{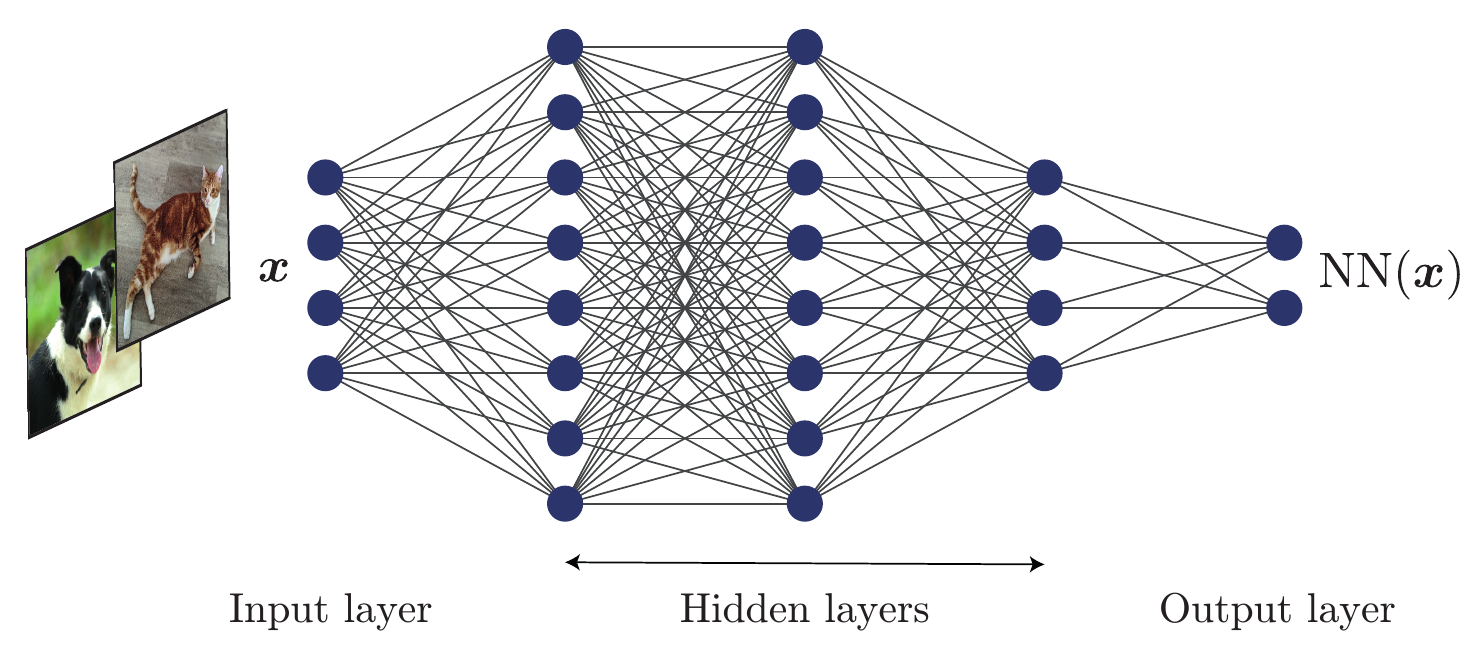}
    \caption[A feedforward neural network]{Graphical representation of a feedforward Neural Network. Inputs are fed to the network in the Input layer and then processed by a series of \textit{hidden} layers, up until the end of the network is reached, corresponding to the output layer. Each node in the network is called \textit{neuron} and implements the transformation~\eqref{eq:single_neuron}, while neurons aligned vertically constitute a layer, whose action is reported in Eq.~\eqref{eq:layer_neural}. The overall action of the network is given by concatenating each layer, as reported in Eq.~\eqref{eq:neural_network}. In this specific case, the neural network accepts inputs $\bmx \in \mathbb{R}^4$, has 3 hidden layers of variable sizes (8, 8 and 6), and outputs a 2-dimensional vector $\text{NN}(\bmx) \in \mathbb{R}^2$. The graph for the neural network was generated using~\cite{LeNail2019}, and similar representations are customary in standard literature on the subject, e.g.~\cite{Hastie2009Statistical, GoodfellowBengioDL, Goodfellow2016DeepL}.}
    \label{fig:ffNN}
\end{figure}

A second very important example of a machine learning model are Neural Networks (NN). While they inherit their name from early mathematical models for biological neurons in the brain~\cite{Rosenblatt1958}, modern neural networks have very little in common with how information is processed in the brain, but the name eventually stuck in the machine learning literature. Broadly speaking, by neural networks one refers to a broad class of parametric models where information is processed by single units, often referred to as neurons, and then transferred to other units in the network, which is composed of several of these neurons connected to each other according to a specific connectivity pattern. Depending on the specifics of the network, several different versions of neural networks have been proposed, and in the following we introduce the most common example, feed-forward Neural Networks (ffNN), graphically represented in Fig.~\ref{fig:ffNN}. 

Feed-forward Neural Networks are called this way because information flows only in one direction in the network, where inputs are progressively processed by a series of \textit{layers} of the network, until a final output is reached. A single unit in the neural network (drawn as a coloured node in Fig.~\ref{fig:ffNN}) implements the mapping
\begin{equation}
    \label{eq:single_neuron}
    \bmx \xrightarrow{\,neuron\,} \sigma(\bmw \cdot \bmx + b)
\end{equation}
where $\bmw \in \mathbb{R}^p$ and $b \in \mathbb{R}$ are parameters to be optimised, which in the neural network jargon are called \textit{weights} and \textit{biases} respectively, and $\sigma: \mathbb{R} \rightarrow \mathbb{R}$ is a generic \textit{nonlinear} function which is called \textit{activation function}. 

A set of neurons acting on the same input data form a \textit{layer} of the neural network, and are represented by the vertically aligned nodes of Fig.~\ref{fig:ffNN}. Let $\bmx \in \mathbb{R}^d$ and let the first layer be composed of $h$ neurons, then the overall action of the layer on the input can be written as
\begin{equation}
    \label{eq:layer_neural}
    \bmx \xrightarrow{\,layer\,} \sigma(W\bmx + \bm{b}),\quad W \in \mathbb{R}^{h \times d},\,\, \bm{b} \in \mathbb{R}^b
\end{equation}
where $W$ is a matrix whose rows are the weights of the neurons in the layer, and similarly for the bias vector $\bm{b}$, and with a slight abuse of notation we impose that the activation function $\sigma$ acts \textit{elementwise} on the entries of a vector, namely $\sigma(\bm{v}) = [\sigma(v_1),\,\sigma(v_2),\,\hdots]$.

In general, let $W^{(l)} \in \mathbb{R}^{h_l \times h_{l-1}}$ and $\bm{b}^{(l)}$ denote the weights and biases of the $l$-th layer of the neural network, an $L$-layers (excluding the input layer) feedforward neural network consists of the following parameterised hypothesis class
\begin{equation}
\begin{aligned}
    \mathcal{M}_\text{NN} &= \left\{ \text{NN}(\bmx) = \sigma \qty( W^{L}\,\sigma\qty(W^{(L-1)}\,\sigma\qty(\hdots \sigma\qty(W^{(1)}\bmx + \bm{b}^{(1)}) \hdots) + \bm{b}^{(L-1)}) + \bm{b}^{(L)})\right. \\
    & \left. \quad\quad\quad\quad\quad\quad\quad\quad\quad\quad\quad\quad\quad\quad\, \big|~ W^{(l)} \in \mathbb{R}^{h_l \times h_{l-1}},\,\, \bm{b}^{(l)} \in \mathbb{R}^{h_l},\,\, l=1,\,\hdots,\, L\right\}\,, 
\end{aligned}
\end{equation}
which consists of a nested application of the transformation~\eqref{eq:layer_neural} using different parameters, but using the same activation function $\sigma$, even though this last constraint is not necessary. By defining the activation vector at layer $l$ as $a^{(l)}(\bmx) = \sigma(W^{(l)}a^{(l-1)}(\bmx) + \bm{b}^{(l)})$, one can write the action of the neural network more compactly as
\begin{equation}
    \label{eq:neural_network}
    \text{NN}(\bmx) = a^{(L)}(\bmx) = \sigma\qty(W^{(L)}a^{(L-1)}(\bmx) + \bm{b}^{(L-1)})\,,
\end{equation}
with $a^{(0)}(\bmx) = \bmx$ being the trivial input layer containing simply the input vector.

Noteworthy, in addition to the trainable parameters given by the weights and biases, a neural network is specified by other additional parameters specifying its architecture, like the number of layers $L$ (or depth of the network) and the activation function, that have a direct impact on the type of functions that the network can implement. As for the activation function, the key requirement is that it has to be nonlinear, as if this is not the case one can easily prove that the whole neural network collapse to a single-layer architecture implementing a simple affine transformation of the input $\bmx \mapsto W\bmx + \bm{b}$. Standard choices for the activation function are the sigmoid or the Rectified Linear Unit (ReLu)
\begin{equation}
    \label{eq:activation_functions}
    \sigma(x) = \frac{e^x}{e^x+1},\,\quad \quad \sigma(x) = \text{ReLu}(x) = 
    \begin{cases}
    x \quad \text{if } x > 0 \\
    0 \quad \text{if } x \leq 0
    \end{cases}\,.
\end{equation}

A neural network can be trained in a supervised fashion by minimising the empirical mean squared loss over the training set $S = \{(\bmx_i, y_i)\}_{i=1}^m$~\eqref{eq:training_set_definition}. Let $\bm{W} = \{W^{(l)},\, \bm{b}^{(l)}~|~ l=1,\hdots,\,L\}$ denote the set of all trainable parameters in the neural network, the optimal model is implicitly defined by
\begin{equation}
    \label{eq:mse_nn}
    L_S(\text{NN}) = \frac{1}{m}\sum_{i=1}^m \qty( y_i - \text{NN}(\bmx_i))^2,\quad
    \bm{W}_\text{opt} = \argmin_{\bm{W}}~ L_S(\text{NN})\,.
\end{equation}
In this case however, there is no way analytical solution to this optimisation problem because the loss function is highly non-convex with respect to the parameters, as opposed to the previous case in Sec.~\ref{ssec:ch_QML_Kernel} where there was a linear dependence on the trainable parameters. 

Instead, neural network models are trained with gradient descent~\eqref{eq:gradient_descent_ML} via \textit{backpropagation}, which is a very efficient method for calculating gradients of composed functions as neural networks. Indeed, one can compute the derivative of any parameter inside the network by a repeated application of the chain rule: starting from the output of the network, one proceeds backwards by ``peeling-off" layers and accumulating gradients, up until the desired weight has been reached~\cite{Dawid2022_LectureNotesQML, Goodfellow2016DeepL}. The backpropagation algorithm permits a very efficient computations of gradients in neural networks and paved the way towards the adoption of large-scale Deep Learning models, with state of the art ones now leveraging up to hundreds of billions of parameters~\cite{ChatGPT}. At last, it is worth noticing that the backpropagation algorithm is a specific example of automatic differentiation (AD), which is a set of techniques aiming at calculating the gradients of a computation algorithmically.

\section{Quantum Machine Learning}
\label{sec:ch_QML_QML}

In the previous section, we introduced the main concepts and tools of machine learning, and also discussed two prototypical examples of learning models: kernel methods and neural networks. In this section, we first introduce the idea of Quantum Machine Learning models distinguishing them from Variational Quantum Algorithms, and then provide examples of quantum versions of the two classical models explained previously.

At the core of variational quantum algorithms and machine learning is the optimisation procedure, by which the algorithms are progressively adjusted to reach good performances. On the other hand, a striking difference is that machine learning models are algorithms that learn from observations (the training set), and in fact the presence of data is the hallmark of any machine learning model. Thus, a reasonable definition of a quantum machine learning model could be 
\begin{definition}[(Informal) Quantum Machine Learning model]
    \label{def:qml_model}
    A Quantum Machine Learning model is an algorithm that uses also, but not exclusively, quantum computational resources to solve a problem defined in terms of data. 
\end{definition}
Far from being a rigorous statement, such definition is admittedly general and omits a proper definition of ``data", still it suffices to describe many of the approaches described in Sec.~\ref{sec:ch_QML_Intro}. 

Specifically, in this work we are focused on near-term quantum algorithms, namely variational algorithms, and thus, by the definition above, a near-term quantum machine learning model is a variational quantum algorithm where the cost function to be optimised is defined in terms of a set of observations or data. For example, a variational quantum algorithm for estimating the ground state energy of a molecule is not a quantum machine learning model, but a parameterised quantum circuit to implement a classification task of a set of observations is. In this sense, quantum machine learning models are a subset of variational quantum algorithms. However, while such characterisation can be useful to draw a boundary between QML and VQAs, these are often used interchangeably to indicate a quantum computation with tunable parameters that needs optimisation.

In the following, we introduce two quantum machine learning models that have been proposed in the literature: quantum classifiers and quantum kernel machines, which can be seen as quantum counterparts of classical linear models, and quantum neural networks which, as the name suggests, are inspired by classical neural networks. 

\subsection{Linear quantum models: quantum classifiers and kernel methods}
\label{sec:ch_QML_ExplicitKernel}
\begin{figure}
    \centering
    \includegraphics{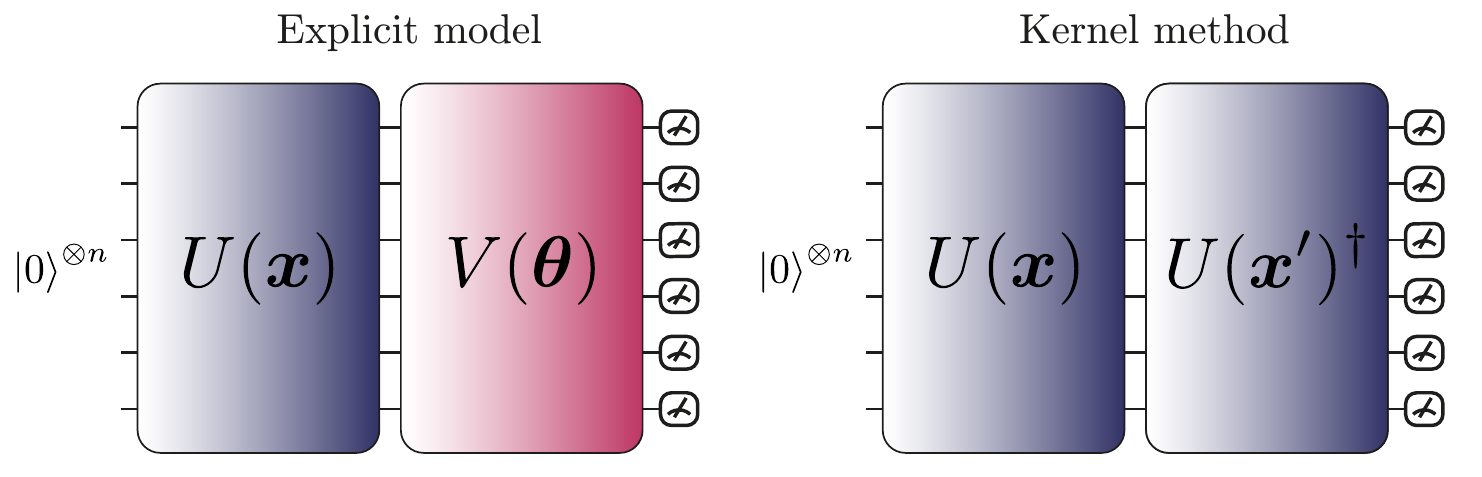}
    \caption[Linear quantum models]{Circuit representation of linear quantum models: an \textit{explicit model} on the left, and a kernel method (also called \textit{implicit} model) on the right. In the former, a data encoding unitary operation is followed by a trainable variational block before the measurement of an observable takes place. In quantum kernel methods, the quantum computer is used to calculate the kernel function (i.e. inner products), while the tunable parameters are instead purely classical.}
    \label{fig:explicit_kernel_quantum}
\end{figure}

The counterpart of linear models in the quantum domain are often called \textit{linear quantum models}, because they effectively implement a linear separation in the Hilbert space of the quantum computer. Also in this case, such models are also strongly connected to quantum kernel methods, a connection which we explore in the following sections.

\subsubsection{Explicit models}
\label{ssec:ch_QML_Explicit}

A quantum machine learning model deals with data, and thus it is necessary to find a way to load such information onto the quantum computer, so that it can be used as inputs for a quantum algorithm. The data to be analysed could be either \textit{classical}, for example a set of input vectors $\bmx_i \in \mathcal{X} \subset \mathbb{R}^d$, or \textit{quantum}, such as a set of quantum states $\ket{\psi_i} \in \mathcal{H} = \mathbb{C}^{2n}$, where $\mathcal{H}$ is the Hilbert space of a quantum system made of $n$ qubits. In the following, we restrict our attention to the case of classical data, even though the discussion could be adapted to quantum data as well with minor modifications\footnote{Specifically, instead of considering the parameterised encoding unitary $\ket{\psi_{\bmx}} = U(\bmx)\ket{0}^{\otimes n}$ depending on the input data $\bmx$, one can consider the unitary $U_i$ that prepares the desired input state when acting on the ground state $\ket{\psi_i} = U_i \ket{0}^{\otimes n}$. In this last case however, it is important to note the circuit implementation $U_i$ to implement a general (possibly Haar random) quantum state $\psi_i$ can be exponentially ---with respect to the number of qubits--- deep. Thus, ideally, either the input quantum states are efficiently preparable on the quantum computer, or there is a controllable physical evolution that outputs the desired quantum state to be then processed by an appropriate quantum computing device}.

Suppose we are given a set of classical data $\{\bmx_i\}_{i=1}^m$, the idea is to map these data to the quantum Hilbert space of the quantum computer, by a procedure which is often called \textit{feature embedding}. Specifically, this is done by using a parameterised unitary operation $U(\bmx)$ depending on the classical data to be loaded, so that the feature embedding consists of the map
\begin{equation}
\begin{aligned}
    \label{eq:feautre_embedding}
    & \phi: \mathcal{X} \rightarrow \mathcal{F},\quad \mathcal{X} \subset \mathbb{R}^d,\,\, \mathcal{F} = \mathcal{H} = \mathbb{C}^{2n}\,, \\
    & \bmx \mapsto \ket{\phi(\bmx)} = U(\bmx)\ket{0}^{\otimes n}\quad \text{or}\quad \bmx \mapsto \rho(\bmx) = \dyad{\phi(\bmx)}\,,
\end{aligned}
\end{equation}
where $\mathcal{F}$ is the feature space, as used in Sec.~\ref{ssec:ch_QML_Kernel} when discussing linear models is feature spaces, in this case given by the qubits Hilbert space $\mathcal{F} = \mathcal{H}$, and $\ket{\phi(\bmx)}$ and $\rho(\bmx)$ are the input quantum state and density matrix, respectively.

The parameterised block $U(\cdot)$ that actually maps the data to a quantum state is general, and various ansätze can be used to accomplish this task. A natural example, which has the benefit of being easily implementable on actual quantum computers, are parameterised Pauli rotations that use one qubit per dimension of the input data $d = n$, defined as
\begin{equation}
    \label{eq:angle_pauli_embedding}
    U(\bmx) = \bigotimes_{i=1}^n R_{P_i}(x_i),\quad \bmx = [x_1, x_2, \hdots, x_n],\,\, P_i \in \{X, Y, Z\}\,,
\end{equation}
where $R_P(x)$ are Pauli rotations~\eqref{eq:pauli_rotations} around one of the Pauli axis $P \in \{X, Y, Z\}$. This approach of encoding the data as a rotational angle in a parameterised Pauli gate goes by the name of \textit{angle embedding}. In addition, more complicated ansatz can be used involving multiple parameterised operations as well as two-qubits entangling gates, as those discussed in Chapter~\ref{ch:entanglement}. It is fundamental to remark that, whenever inputs are encoded in a circuit as angles of parameterised rotations, these have to be rescaled in an appropriate angular range, for example $\bmx \in \mathcal{X} \subset [0, 2\pi[^d$, before feeding them to the circuit. If this is not the case then inputs differing by even multiples of $2\pi$ would be mapped on the same quantum state, since $U(\bmx + 2\pi) = U(\bmx)$.

After the encoding phase, a variational unitary $V(\bmt)$ acts on the system, thus yielding the parameterised state $\ket{\phi_{\bmt}(\bmx)} = V(\bmt)\ket{\phi(\bmx)} = V(\bmt)U(\bmx)\ket{0}^{\otimes n}$. An observable $O$ is measured at the end of the computation, and the final result is then
\begin{align}
    f_{\bmt}(\bmx) = \expval{O}_{\bmt;\, \bmx} & = \mel{\phi_{\bmt}(\bmx))}{O}{\phi_{\bmt}(\bmx)} = \mel{\phi(\bmx)} {\underbrace{V(\bmt)^\dagger \,O\, V(\bmt)}_{O_{\bmt}}}{\phi(\bmx)} \\
    &= \Tr[O_{\bmt}\rho(\bmx)] = \hs{O_{\bmt}}{\rho(\bmx)}\label{eq:qml_linear_model}\,,
\end{align}
where in the last line we used the density matrix $\rho(\bmx) = \dyad{\phi(\bmx)}$, and then expressed the expectation value in terms of the Hilbert-Schmidt inner product of operators $\langle A, B \rangle_{HS} \coloneqq \Tr[A^\dagger B]$. A graphical representation of the quantum circuit implementing this evolution is shown in Fig.~\ref{fig:explicit_kernel_quantum}.

The last expression in Eq.~\eqref{eq:qml_linear_model} clearly exhibits the \textit{linear} nature of this model in the Hilbert space $\mathcal{H}$ of the quantum system\footnote{Note that for ease of exposition we are ignoring some subtleties related to the definition of the feature quantum space, as this can be the space of states $\ket{\phi(\bmx)}$ or that of Hermitian operators (density matrices) $\rho(\bmx)$. For a detailed discussion on this topic, we refer to~\cite{SchuldKernel2021}.}, in that the outcome of the circuit is a simple inner product between two Hermitian operators: the data-dependent state $\rho(\bmx)$ and the trainable observable $O_{\bmt}$, which is the quantum analogue of the trainable weights $\bmw$ of a classical linear model~\eqref{eq:linear_model_class_main}. However, note that while the results of the circuit depend \textit{linearly} on such parameterised observable (and also on the quantum state), the dependence on the parameters and the input data is clearly nonlinear, usually trigonometric, since Pauli rotations are used to parameterise the encoding block $U(\bmx)$ and the variational ansatz $V(\bmt)$. 

As with a regular machine learning model, the output of the quantum circuit $\hat{y}_i = f_{\bmt}(\bmx_i) = \hs{O_{\bmt}}{\rho(\bmx_i)}$ can then be used in a loss function to drive the training procedure, as discussed previously in Sec.~\ref{ssec:loss_function}. Moreover, whenever the required conditions hold one can use the parameter shift role~\eqref{eq:parameter_shift_rule} to calculate the gradients of the circuit and use gradient descent to find the optimal value of the parameters. 

The quantum machine learning model we have just described often goes by the name of \textit{explicit} model~\cite{Jerbi2022BeyondKernel, Schuld2019FeatureSpace, Mengoni2019Kernel, Havlicek2019QSVM}, because the optimal observable $O_{\bmt_{\text{opt}}}$ --- actually, the optimal parameters $\bmt_\text{opt}$ --- is searched directly in the Hilbert space via optimisation of the variational parameters, and the classification or regression problem is implemented by measuring it on a quantum computer. Explicit models are opposed to \textit{implicit} models (or quantum kernel methods), which is the topic of the next section.

\subsubsection{Quantum kernel (or implicit) models}
\label{ssec:ch_QML_kernel}
We have discussed in Section~\ref{ssec:ch_QML_Kernel} that linear models and kernel methods are strictly related, since kernel models can be seen as linear models in the so-called Reproducing Kernel Hilbert Space (RKHS) of the kernel function, and vice versa. These relations also hold for quantum kernel methods (often called \textit{implicit models}~\cite{Schuld2019FeatureSpace}), which are parameterised predictors of the form~\cite{Mengoni2019Kernel, SchuldKernel2021, Havlicek2019QSVM}
\begin{eqnarray}
    \label{eq:qml_quantum_kernel}
    f(\bmx) = \sum_{i=1}^m \alpha_i\, \kappa(\bmx, \bmx_i) = \sum_{i=1}^m \alpha_i\, \Tr[\rho(\bmx)\rho(\bmx_i)]\,.
\end{eqnarray}
As with their classical counterparts~\eqref{eq:kernel_models}, quantum kernel models make predictions using a linear combination of kernel evaluations between the new sample $\bmx$ and those in the training set $\{\bmx_i\}_{i=1}^m$, where now the kernel function $\kappa(\bmx, \bmx') = \Tr[\rho(\bmx)\rho(\bmx')] = \abs{\braket{\phi(\bmx)}{\phi(\bmx')}}^2$ is given by the inner product in the feature quantum space. Models like these use the quantum computer only to compute the kernel, while the parameters $\{\alpha_i\}_{i=1}^m$ remain purely classical. A quantum circuit for evaluating the kernel function (i.e. inner product) for pure states is shown in the right panel of Fig.~\ref{fig:explicit_kernel_quantum}, but other strategies exist, for example leveraging the so-called \textsc{SWAP} test.

As we briefly mentioned when discussing classical kernel models, models of the form~\eqref{eq:qml_quantum_kernel} are proven to be optimal by the Representer theorem, in the sense that any model in the RKHS that minimises an empirical risk can be expressed as a simple linear combination of kernel evaluations with the training set. Moreover, the complexity of the optimisation problem is sensibly reduced, since rather than optimising a parameterised observable $O_{\bmt} \in \mathbb{C}^{2n \times 2n}$ which can be a quite daunting task for $n \gg 1$, one only has to find the $m$ real parameters $\{\alpha_i\}_{i=1}^m$.

Finally, note that by linearity of the trace, the model~\eqref{eq:qml_quantum_kernel} implicitly defines a linear model of the form~\eqref{eq:qml_linear_model} where the observable is given by a linear combination of the feature quantum states from the training set, namely
\begin{equation}
    \label{eq:implicit_obs_kernel}
    f(\bmx) = \Tr[\rho(\bmx) \qty(\sum_{i=1}^m \alpha_i\,\rho(\bmx_i))] = \Tr[\rho(x)\, O_S].
\end{equation}
Using the inner products between quantum states as kernel, these quantum kernel machines can then be used within regular ridge regression~\eqref{eq:ridge_regression} or for classification tasks in the form of quantum support vector machines~\cite{Schuld2019FeatureSpace, Havlicek2019QSVM}.

\subsubsection{Explicit or Implicit?}
\label{ssec:ch_QML_explicit_vs_implicit}

We have seen two examples of quantum machine learning models, explicit and implicit models, both belonging to the class of linear quantum models since they can be expressed as the inner product of the input quantum states with an observable $f(\bmx) = \Tr[O\rho(\bmx)] = \hs{O}{\rho(\bmx)}$. But what is the difference between the two, and are there any reasons to prefer one over the other? Two main distinctions can be made, regarding the optimisation and the classification performances of these models.

Let's first discuss the optimisation properties of these two models. Implicit models like~\eqref{eq:kernel_models} require $\order{m^2}$ queries to the quantum computer to estimate the kernel matrix of inner products between the $m$ samples in the training dataset\footnote{We hereby only count the number of values to be estimated on the quantum computer, ignoring the actual number of measurements needed to estimate each expectation value.}, and additional $\order{m^3}$ classical post-processing steps to compute the optimal weights via inversion of the Gram matrix~\eqref{eq:optimal_params_ridge} for ridge regression~\cite{Jerbi2022BeyondKernel, MohriMLBook, ScholkopfLearningKernelsSupport2002}. On the other hand, variational training of explicit models is cheaper, because it is expected to terminate after a number of iterations proportional to the number of training samples, and thus it requires $\order{p\, m}$ calls to the quantum computer, where $p$ is the number of parameters in the trainable observable $O_{\bmt}, \bmt \in \mathbb{R}^p$. Thus, the most efficient strategy depends on the scenario: if a moderate amount of parameters are sufficient to fit a large amount of training data, then variational training may be the solution. On the contrary, if the required number of parameters scales with the number of data, then the two methods are equivalent in terms of computational resources. Clearly, the number of parameters in the variational block $V(\bmt)$ to parameterise a general observable $O_{\bmt}$, which is needed to explore the whole space of observables to look for the optimal one, scales exponentially with the number of qubits. In practical scenarios, however, one restricts the expressibility of the model by selecting a specific parameterised ansatz with a limited amount of parameters.

Regarding the classification performances, implicit models are guaranteed by the Representer theorem to achieve the lowest possible empirical risk on the training set, when compared to explicit models trained with the same feature encoding. On the other hand, as argued in ref.~\cite{Jerbi2022BeyondKernel}, the latter may be desirable in terms of generalisation error, because their limited expressivity could help avoid overfitting. Hence, the use of constrained observables in explicit models ---instead of the optimal observable of implicit models, see eq.~\eqref{eq:implicit_obs_kernel}--- may be advantageous to learning performances.

At last, we remark that if one is concerned with quantum advantages or speedups in these types of machine learning models, these can only be achieved if the kernel function $\kappa(\bmx, \bmx') = \Tr[\rho(\bmx)\rho(\bmx')]$ is hard to compute classically. Indeed, if this is not the case, then the quantum kernel model~\eqref{eq:qml_quantum_kernel} can be simulated classically without the need for a quantum computer, and thus no advantages can be attained.

\subsection{Data reuploading models and Quantum Neural Networks}
\label{sec:ch_QML_Reuploading}
A second class of quantum machine learning models are data reuploading quantum circuits, often referred to as Quantum Neural Networks (QNNs). Differently from linear quantum models, the output of a quantum neural network cannot be written as an inner product between a data-dependent quantum state and a trainable observable, because these circuits use a repeated structure where data encoding blocks are interleaved trainable unitaries, so that the action of these two elements cannot be separated anymore. 

Notably, it was recently shown that models in this class can be mapped to linear quantum models using approximate strategies or leveraging post-selection and gate teleportation, even though these constructions require additional conditions that cannot be easily met in practice~\cite{Jerbi2022BeyondKernel}. Thus, while such a result suggests a unifying theoretical framework for describing quantum learning models, in practical instances (that are those that one would execute on an actual near-term quantum computer), linear models and data reuploading ones behave rather differently, as we shall see.

Data reuploading quantum circuits have appeared independently multiple times in the literature~\cite{Theis2020Expressivity, Schuld2020Encoding, Perez2020Reuploading}, with all of them underlining the tight connection between input redundancy inside a quantum circuit, and the capability of the latter of expressing more complicated ----specifically, higher-frequency trigonometric--- functions on the input data. In a sense, this need of loading the inputs multiple times in the circuits, hence the name data reuploading, can be seen as an analogue of feeding the same input to multiple neurons in the first layer of a classical neural network, as depicted in Fig.~\ref{fig:ffNN}. 

Indeed, provided that input data enters the circuit via rotation-like gates (see below for a more rigorous statement), it is possible to show that the outcome of any parameterised quantum circuit depending on data can be written as a truncated Fourier series, or as a Generalised Trigonometric Polynomial (GTP)~\cite{Caro2021encodingdependent, Wierichs2022generalparameter, Theis2020Expressivity, Schuld2020Encoding}.
\begin{theorem}[Data-dependent parameterised quantum circuits are truncated Fourier series~\cite{Theis2020Expressivity, Schuld2020Encoding, Caro2021encodingdependent}]
\label{th:qnn_fourier}

Let $U_{\bmt}(\bmx)$ be the unitary matrix of a parameterised quantum circuit depending on some input data $\bmx \in \mathcal{X} \subset [0, 2\pi[^d $ and trainable parameters $\bmt$. If data coordinates $x_i \in \bmx$ enter the quantum circuit via data encoding operations of the form $U(x_i) = \exp(-i\,x_i\,H)$ where $H$ is an Hermitian operator, then the following holds
\begin{equation}
    \label{eq:qnn_gtp_summary}
    f_{\bmt}(\bmx) = \expval{O}_{\bmx;\,\bmt} = \mel{\bm{0}}{U_{\bmt}(\bmx)^\dagger\,O\,U_{\bmt}(\bmx)}{\bm{0}} = \sum_{\bmo \in \Omega} c_{\bmo}~e^{-i\, \bmo \cdot \bmx}\,,
\end{equation}
where $\Omega = \qty{\bmo \in \mathbb{R}^{d}}$ is the \textit{frequency spectrum} associated with the quantum circuit and depends solely on the number of data encoding operations present in the circuit, as well as their generators, a piece of information denoted as \textit{data encoding strategy}. The expansion coefficients $\{c_{\bmo} \in \mathbb{C}\}$ instead depend on the structure of the circuit (including the data encoding strategy), the trainable parameters $\bmt$, and finally also the observable $O$. For any frequency $\bmo \in \Omega$, also $-\bmo \in \Omega$. Also, the coefficients satisfy $c_{\bmo} = c^{*}_{-\bmo}$, which ensures that the Fourier expansion correctly evaluates to a real number.  

As an example, parameterised quantum circuits that encode the input data via Pauli rotations of the form~\eqref{eq:angle_pauli_embedding} belong to this class of model, hence admit a Fourier expansion. 
\end{theorem}

The Fourier representation in Eq.~\eqref{eq:qnn_gtp_summary} beautifully summarises the class of functions that parameterised quantum models depending on data via parameterised rotations can implement, namely trigonometric functions\footnote{It is important to note that we are neglecting classical preprocessing step of the input data. That is, the data $\bmx \in \mathcal{X}$ are fed directly as parameters to the data encoding operations $U(\bmx)$ without any classical preprocessing. If classical preprocessing is used, $\bm{y} = g(\bmx)$, then the PQC is clearly a Fourier series of $\bm{y}$, not of the original data $\bmx$. Classical preprocessing can be used to change the functional dependence of the expectation value on the original input~\cite{Mitarai2018Learning, ShinExponentialEncoding}.}\textsuperscript{,}\footnote{We already had an hint of this fact in Eq.~\eqref{eq:trigonometric_cost}, when we discussed how to derive the parameter-shift rule for parameterised quantum circuits. Also in that case, when fixing all the parameters but one, we saw that the circuit effectively implements a trigonometric function of that remaining parameter with frequency spectrum $\Omega = \{0, 1\}$.}. A clear example of such circuits are those that encode data via Pauli rotations of the form~\eqref{eq:angle_pauli_embedding}.

\begin{figure}[t]
    \centering
    \includegraphics{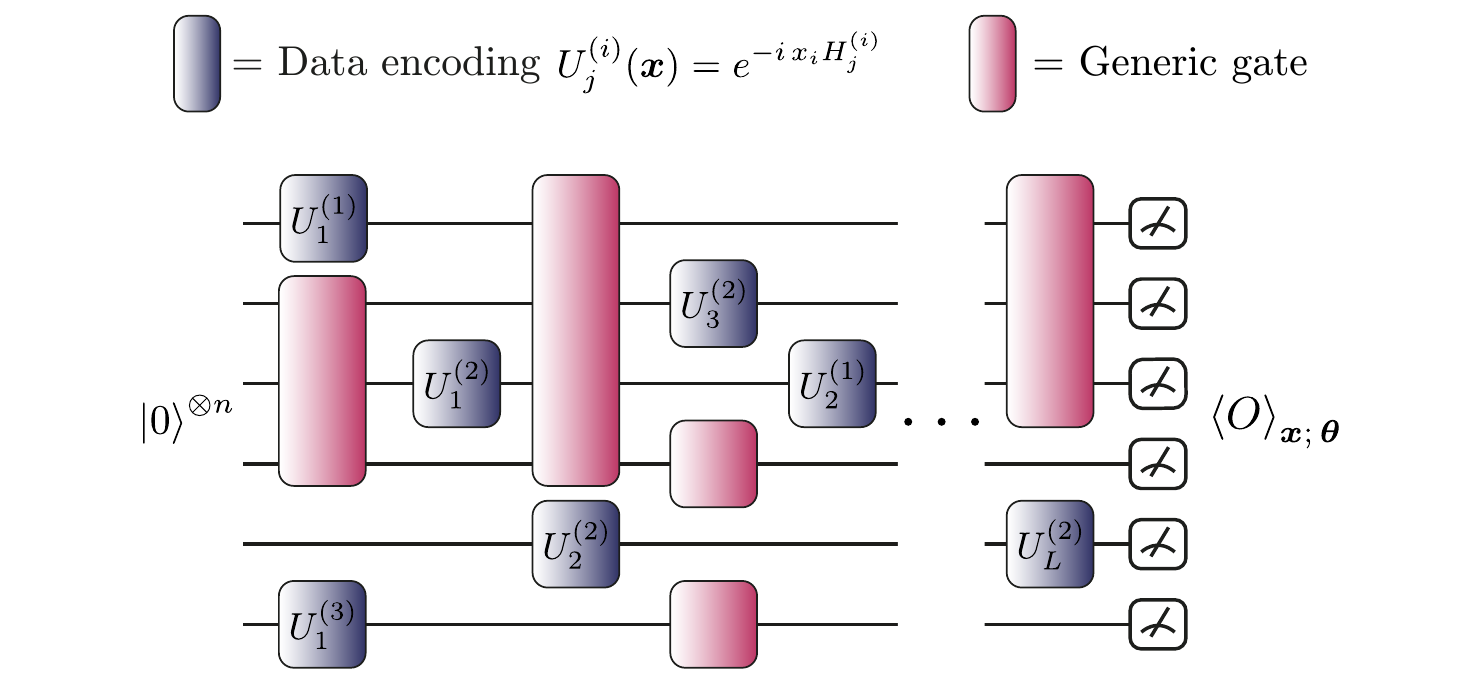}
    \caption[Parameterised quantum circuit as a Fourier series]{A generic quantum circuit can be expressed as a truncated Fourier series~\eqref{eq:qnn_gtp_summary} over the input data $\bmx$, provided that the coordinates of the input are loaded in the quantum circuit via gates of the form $U(x_i) = \exp(-i\,x_i\,H)$. This figure is a custom reproduction of Fig.~3 in~\cite{Caro2021encodingdependent}.}
    \label{fig:general_pqc_fourier}
\end{figure}

Interestingly, a Fourier expansion of the form~\eqref{eq:qnn_gtp_summary} holds for \textit{any} quantum circuit provided that inputs are loaded via an angle-embedding scheme, irrespectively of the choice of the gates in the circuit, their position, or their number, as shown in Fig.~\ref{fig:general_pqc_fourier}. Such formulation clearly demonstrates that the Fourier-like nature of the circuit is a consequence of the data encoding strategy, and also this implicitly determines the set of frequencies $\bmo \in \Omega$ that the quantum model has access to. Instead, the trainable parameters, along with the observable and the structure of the circuit, control which frequencies can actually be expressed by the circuit, by tuning the expansion coefficients $\{c_{\bmo}\}$. \\
 
\subsubsection{Deriving the Fourier expansion} 
In the following, we derive the Fourier expansion~\eqref{eq:qnn_gtp_summary} along the lines of the elegant treatment in ref.~\cite{Caro2021encodingdependent}. Different proofs and further discussions can be found also in refs.~\cite{Theis2020Expressivity, Schuld2020Encoding, Perez2020Reuploading}. 

As for the previous quantum machine learning models, also in this case we distinguish between encoding gates $U(\bmx)$, depending on the input data $\bmx \in \mathcal{X} \subset \mathbb{R}^d$, and trainable operations $\{V_l(\bmt_l)\}$, depending on trainable parameters $\{\bmt_l\}_l$. However, as we shall see, the explicit form of such trainable gates is not necessary to derive the desired Fourier-like expansion of the quantum circuit. In fact, any gate non depending on data, be it fixed or parameterised, is effectively absorbed into the definitions of the expansion coefficients, possibly in a very intricate way, without playing any major role in the mathematical derivation. 

Let $\bmx = [x_1,\, x_2\, \hdots,\, x_d] \in \mathbb{R}^d$ be the usual input vector, we consider encoding unitary operations that encode the coordinates of the input via evolutions of the form
\begin{equation}
    \label{eq:qnn_encoding_gates}
    U^{(i)}_j(\bmx) = \exp(-i~x_i~H_{j}^{(i)}) = \exp(-i~\bm{e}_i \cdot \bmx~H_{j}^{(i)})\,,
\end{equation}
where $H_j^{(i)}$ is the $j$-th Hermitian operator encoding the $i$-th coordinate $x_i$, and $\bm{e}_i$ is a unit vector having zeros everywhere except on the $i$-th component, so that $\bm{e}_i \cdot \bmx = x_i$. As the notation suggests, we allow for each data coordinate to be uploaded in the circuit multiple times throughout the circuit, even with different generators. 

Let's consider the action of the data encoding block~\eqref{eq:qnn_encoding_gates} on a generic quantum state with density matrix $\rho$. Let $U(\bmx) = \exp(-i~\bm{e}\cdot\bmx~H)$ with $H = \sum_{k}\lambda_k\dyad{\lambda_k}$ the spectral decomposition of the generator, by expanding state $\rho$ on the eigenbasis of $H$, one obtains
\begin{align}
    \rho \rightarrow \rho(\bmx) & = U(\bmx)\, \rho \, U(\bmx)^\dagger = U(\bmx) \qty(\sum_{kl}\rho_{kl}\, \dyad{\lambda_k}{\lambda_l}) U(\bmx)^\dagger \nonumber\\
    & = \sum_{kl} e^{-i\,(\lambda_k - \lambda_l)\,\bm{e}\cdot\bmx}\,  \rho_{kl}\, \dyad{\lambda_k}{\lambda_l}\,.
\end{align}
This expression can be simplified by grouping together those indexes $(k,l)$ that give rise to the same frequency difference $\lambda_k - \lambda_l$. In order to do so, it is convenient to introduce the frequency spectrum associated with the generator, that is the set of all possible differences of its eigenvalues multiplied by the unit vector
\begin{equation}
    \label{eq:spectrum_generator}
    \Omega(H) = \qty{(\lambda_k - \lambda_l)\bm{e}~|~\forall \lambda_k, \lambda_l \in \text{eigvals}(H)}\,.
\end{equation}
 Note that, by definition of the frequency spectrum, for any frequency $\bmo \in \Omega(H)$, also the negative frequency belongs to the spectrum $-\bmo \in \Omega(H)$. It is possible to group together those terms that correspond to the same frequency difference, that is defining
 \begin{equation}
    \label{eq:qnn_rho_one}
     \rho_{\bmo} = \sum_{(k,l) \in I(\bmo)} \rho_{kl} \dyad{\lambda_k}{\lambda_l}\quad \text{with} \quad I(\bmo) \coloneqq \qty{(i,j)~|~(\lambda_i - \lambda_j)\bm{e} = \bmo}\,.
 \end{equation}
With these definitions, the evolved quantum state can be finally rewritten as
\begin{equation}
    \label{eq:qnn_enc_one}
    \rho(\bmx) = \sum_{\bmo \in \Omega(H)} e^{-i \bmo \cdot \bmx}\, \rho_{\bmo}\,,
\end{equation}
with the expectation value of an observable $O$ on this state thus being
\begin{equation}
    \label{eq:qnn_fourier_one}
    \expval{O} = \Tr[O \rho(\bmx)] = \sum_{\bmo \in \Omega(H)} e^{-i\, \bmo \cdot \bmx}\, \Tr[O \rho_{\bmo}] = \sum_{\bmo \in \Omega(H)} c_{\bmo}\, e^{-i\, \bmo \cdot \bmx}\,,
\end{equation}
where we defined the coefficients $c_{\bmo} = \Tr[O \rho_{\bmo}]$. Since the original state $\rho$ is Hermitian, using the definitions~\eqref{eq:qnn_rho_one} one can check that $\rho_{\bmo} = \rho^\dagger_{-\bmo}$, hence the coefficients satisfy $c_{\bmo} = c^{*}_{-\bmo}$, as needed to ensure that $\expval{O}$ is real-valued as expected. We have thus recovered the Fourier representation of the circuit~\eqref{eq:qnn_gtp_summary} for the case of a single encoding step on a generic quantum state. The same derivation can be applied straightforwardly when more encoding gates are applied, and also when other operations, like the trainable unitaries, act on the system. 

Indeed, starting from the former case, one can see that the action of any \textit{non-data-encoding} unitary only amounts to a change of basis which can be absorbed inside the operators $\rho_{\bmo}$. In fact, suppose a unitary $W$ acts on the quantum state $\rho(\bmx)$~\eqref{eq:qnn_enc_one}, this is consequently changed to
\begin{align}
    \rho(\bmx) \rightarrow V \rho(\bmx) V^\dagger &= V \qty(\sum_{\bmo \in \Omega(H)} e^{-i\, \bmo \cdot \bmx}\, \rho_{\bmo}) V^\dagger = \sum_{\bmo \in \Omega(H)} e^{-i\, \bmo \cdot \bmx}\, V\rho_{\bmo}V^\dagger \\
    & = \sum_{\bmo \in \Omega(H)} e^{-i\, \bmo \cdot \bmx}\, \tilde{\rho}_{\bmo}\,,
\end{align}
where the action of the unitary can be just absorbed in the operators $\{\rho_{\bmo}\} \rightarrow \{V\rho_{\bmo}V^\dagger\}$, hence the in coefficients $\{c_{\bmo}\}$, without changing the frequency spectrum of the circuit $\Omega$ or its Fourier representation. Thus, we can concentrate on the action of the data encoding gates only.

Indeed, let $U_2(\bmx) = \exp(-i~\bm{e}_2\cdot\bmx~H_2)$ be another encoding operation (for clarity, in what follows we add the subscript ``1'' to indicate the previous encoding operation), the state $\rho(\bmx)$ is evolved according to
\begin{align}
    \rho(\bmx) \rightarrow \rho'(\bmx) & = U_2(\bmx) \rho(\bmx) U_2(\bmx)^\dagger = \sum_{\bmo_1 \in \Omega(H_1)} e^{-i\, \bmo_1 \cdot \bmx}\, U_2(\bmx) \rho_{\bmo_1} U_2(\bmx)^\dagger\\
    & = \sum_{\bmo_1 \in \Omega(H_1)} e^{-i\, \bmo_1 \cdot \bmx}  \sum_{\bmo_2 \in \Omega(H_2)} e^{-i\, \bmo_2 \cdot \bmx} \rho_{\bmo_1, \bmo_2}\label{eq:qnn_rho_two}
\end{align}
where in the second line we expressed the operators $\{\rho_{\bmo_1}\}$ in the eigenbasis of the generator $H_2$, and then re-indexed the sum in terms of the corresponding frequency spectrum $\Omega(H_2)$, namely
\begin{align}
    U_2(\bmx) \rho_{\bmo_1} U_2(\bmx)^\dagger & = U_2(\bmx) \qty(\sum_{kl} \qty[\rho_{\bmo_1}]_{kl} \dyad{\lambda^{(2)}_k}{\lambda^{(2)}_l}) U_2(\bmx)^\dagger \\
    & = \sum_{kl} e^{-i\, \qty(\lambda_k^{(2)} - \lambda_l^{(2)})\bm{e}_2\cdot \bmx}\qty[\rho_{\bmo_1}]_{kl} = \sum_{\bmo_2 \in \Omega(H_2)} e^{-i\,\bmo_2 \cdot \bmx}\rho_{\bmo_1, \bmo_2}
\end{align}
where the frequency spectrum $\Omega(H_2)$ is defined as before~\eqref{eq:spectrum_generator}, and the sum was re-indexed accordingly.

One more ingredient is needed to simplify the expression~\eqref{eq:qnn_rho_two}, namely how to compose the frequency spectrums $\Omega(H_1)$ and $\Omega(H_2)$ arising from the two different encoding operations. This can be done by the so-called Minkowski sum between the two sets of frequencies, defined as
\begin{equation}
    \label{eq:total_spectrum}
    \Omega = \Omega(H_1) + \Omega(H_2) \coloneqq \qty{\bmo_1 + \bmo_2~|~\forall \bmo_1 \in \Omega(H_1),\, \forall \bmo_2 \in \Omega(H_2)}\,.
\end{equation}
With this, the quantum state can thus be written as
\begin{equation}
    \label{eq:qnn_fourier_last_proof}
    \rho'(\bmx) = \sum_{\bmo_1 \in \Omega(H_1)}  \sum_{\bmo_2 \in \Omega(H_2)} e^{-i\, (\bmo_1 + \bmo_2)\, \cdot \bmx}\, \rho_{\bmo_1, \bmo_2} = \sum_{\bmo \in \Omega} e^{-i\, \bmo \cdot \bmx}\, \rho_{\bmo}\,,
\end{equation}
where, as before, the operators $\{\rho_{\bmo_1, \bmo_2}\}$ corresponding to the same frequency $\bmo = \bmo_1 + \bmo_2$ were summed together, and the sum was then re-indexed according to the frequencies in the joint spectrum $\Omega$. From this, one can calculate the expectation value of the observable $\expval{O} = \Tr[O \rho'(\bmx)]$, thus obtaining again Eq.~\eqref{eq:qnn_fourier_one}.

The same derivation can be applied multiple times for all data encoding gates in the circuit, which have the net effect of adding more frequencies in the accessible spectrum $\Omega$. All other operations instead just impact the coefficients in the series. Thus, we showed that \textit{any} quantum circuit that encodes input data via evolutions of the form~\eqref{eq:qnn_encoding_gates} can be expressed as a truncated Fourier series~\eqref{eq:qnn_gtp_summary} of the inputs, as desired. \qed \\

\subsubsection{A single-qubit data reuploading circuit}
\label{ssec:ch_QML_1q_qnn}
Although the Fourier expansion~\eqref{eq:qnn_fourier_last_proof} is a powerful and concise statement about the output of a quantum circuit, an inexperienced user might find it difficult to use it on real instances, mainly because of the rather opaque composition rule~\eqref{eq:total_spectrum} by which the total frequency spectrum $\Omega$ is defined. To make things clearer, let's then consider a simple example of a single-qubit data-reuploading circuit for a univariate input data $x \in \mathbb{R}$, namely
\begin{equation}
    \label{eq:single_qubit_gtp}
     f_{\bmt}(x) = \mel{0}{U_{\bmt}(x)^\dagger\,O\,U_{\bmt}(x)}{0},\quad 
    U_{\bmt}(x) = V_L(\bmt_L)U_L(x)\, \cdots\, V_1(\bmt_1)U_1(x)\,V_0(\bmt_0)\,,
\end{equation}
where the encoding gates are $U_l(x) = \exp(-i\,x\,P_l),\, P_l \in \{\sfrac{X}{2}, \sfrac{Y}{2}, \sfrac{Z}{2}\}$ are Pauli rotations. The eigenvalues of every Pauli matrix are $\text{eigvals}(P_l) = \{\pm \sfrac{1}{2}\}$, so by the definition~\eqref{eq:spectrum_generator}, the frequency spectrum associated to any such matrix is
\begin{equation}
    \Omega(P_l) = \qty{\lambda_k - \lambda_l~|~\forall \lambda_k, \lambda_l \in \qty{\pm \sfrac{1}{2}}} = \{-1, 0, 1\}, \quad P_l \in \qty{X, Y, Z} \times \frac{1}{2}\,.
\end{equation}
When $L$ encoding gates $U_l(x)$ are used load the data, the total spectrum~\eqref{eq:total_spectrum} will be 
\begin{equation}
\label{eq:single_qubit_total_spectrum}
\begin{aligned}
    \Omega & = \Omega(P_1) + \hdots + \Omega(P_L) = \qty{\omega_1 + \cdots + \omega_L ~|~ \omega_1 \in \Omega(P_1),\, \hdots,\, \omega_L \in \Omega(P_L)}\\
    & = \{\omega_1 + \cdots + \omega_L~|~\omega_i \in \qty{-1, 0, 1}\}\\
    & = \qty{-L,\, -(L-1),\, \hdots,\, -1,\, 0,\, 1,\, \hdots,\, L-1,\, L}\,,
\end{aligned}
\end{equation}
The size of the spectrum $\abs{\Omega} = 2L+1$ directly depends on the number of times the input $x$ appears in the circuit, as each Pauli encoding gate effectively increases the accessible frequency spectrum by adding higher order \textit{integer} frequencies. 

Noteworthy, an integer-valued spectrum is a general feature of Pauli encodings, which is due to their generators having eigenvalues $\pm\sfrac{1}{2}$. Indeed, this also holds for the more complicated case of multivariate inputs and multiple qubits: whenever Pauli rotations are used to encode the data $\bmx \in \mathbb{R}^d$ in the quantum circuit, the generated spectrum is integer-valued $\bmo \in \Omega \subset \mathbb{Z}^d$.

Finally, given the spectrum~\eqref{eq:single_qubit_total_spectrum}, the action of a data reuploading single-qubit quantum circuit using Pauli encodings can then be expressed as
\begin{equation}
    f_{\bmt}(x) = \sum_{n=-L}^{L} c_{n}(\bmt)\, e^{-i\, n \,x}\,,
\end{equation}
which is a truncated Fourier series of degree $L$ of the input data $x$.

With a little imagination, one can convince himself that an equivalent result can be obtained also for higher-dimensional inputs on bigger circuits with multiple qubits. Essentially, whenever a data coordinate $x_i$ is encoded in the circuit, more frequencies of that coordinate appear in the Fourier expansion of the circuit~\eqref{eq:qnn_gtp_summary}. 

In conclusion, one can \textit{design} the frequency spectrum accessible by the circuit by changing the data encoding strategy, namely the number of encoding gates per coordinate and their generators. The size of the accessible spectrum, hence the complexity of the Fourier expansion, depends uniquely on the data encoding strategy, and by considering different eigenvalues of the encoding generators one can create rather different frequency spectrum~\cite{ShinExponentialEncoding, PetersOverfitting}.\\

\subsubsection{Quantum Neural Networks}
\label{ssec:ch_QML_QNN}
\begin{figure}[t]
    \centering
    \includegraphics{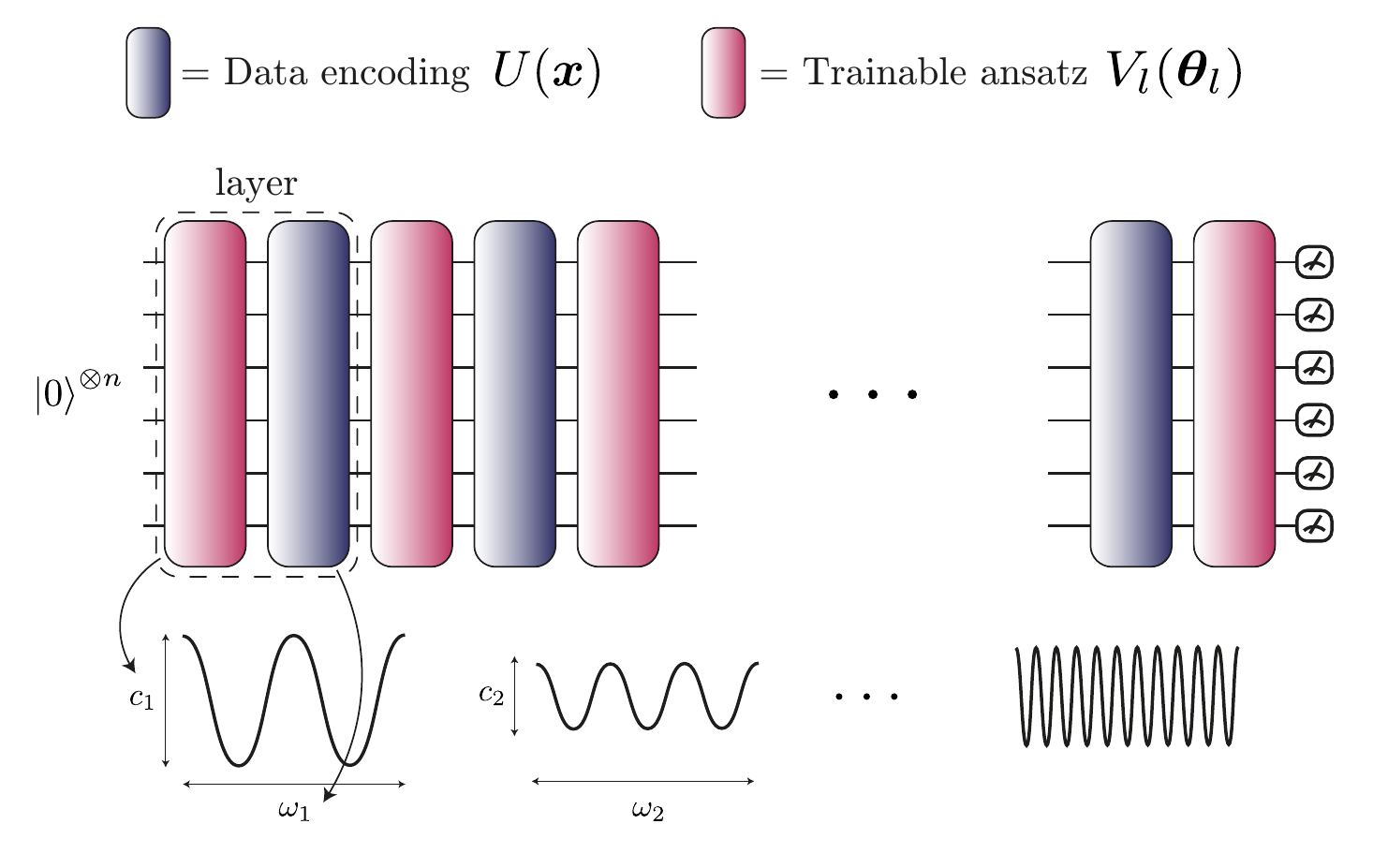}
    \caption[Quantum Neural Network]{Data reuploading quantum circuit in the form of Quantum Neural Network~\eqref{eq:qnn}, consisting of multiple layers of encoding gates $U(\bmx)$ and trainable unitaries $V(\bmt)$. Leveraging on the Fourier analysis developed so far~\eqref{eq:qnn_gtp_summary}, we know that the data encoding blocks define which frequencies $\{\omega_i\}$ are accessible by the quantum neural network, while the variational unitaries control the coefficients $\{c_i\}$ of the Fourier expansion. This figure is a custom reproduction of Fig.~1 in~\cite{Schuld2020Encoding}.}
    \label{fig:data_reuploading_qnn}
\end{figure}

We have just seen that uploading the data multiple times inside a circuit is of paramount importance to enrich the class of functions that a quantum machine learning model can express. 

Building on such intuition, a class of quantum models that appeared prominently in the literature are so-called \textit{Quantum Neural Networks} (QNNs). Although this term is also often used to indicate any variational quantum circuit using a machine learning jargon, it seems reasonable to indicate with quantum neural networks specifically those parameterised circuits which depend on data (see definition~\ref{def:qml_model}) and that use a repeated structure of encoding and variational layers to add input redundancy, thus increasing the expressivity of the model. More formally, QNNs usually take the following form
\begin{equation}
    \label{eq:qnn}
    U_{\bmt}(\bmx) = \prod_{l = 1}^{L} U(\bmx)\, V_l(\bmt_l) \,,
\end{equation}
which is a repeated structure of data encoding blocks $U(\bmx)$ and trainable operations $V_l(\bmt)$, and $L$ is the number of \textit{layers} in the quantum neural networks. A graphical representation of this circuit is shown in Fig.~\ref{fig:data_reuploading_qnn}.

Such circuital architectures made of repeated layers of similar operations make them similar to how feed-forward neural networks~\ref{fig:ffNN} are built, and thus ``justifies'' the name \textit{quantum neural networks}. Needless to say, classical and quantum neural networks are completely different objects, with different properties and corresponding hypothesis classes, and such juxtaposition only holds on at a conceptual level.

\subsection{Generalization of QML models}
\label{sec:ch_QML_GeneralisationQNN}

We have just shown that the action of a parameterised quantum model can be written concisely in closed form as a truncated Fourier series. Such a formulation is very helpful because it exposes the class of functions that the quantum model can implement, and we can thus use the formalism of statistical learning not only to define the hypothesis class implemented by data-reuploading circuits, but also derive statements about their generalisation performances. Indeed, in this section we give an example of how concepts from classical statistical learning can be applied to characterise quantum learning models, thus hinting at the fruitful exchange that there can be between these two fields and consequently highlighting the multidisciplinary nature of subjects like quantum machine learning.

Let $\rho_{\bmt}(\bmx) = U_{\bmt}(\bmx)\dyad{\bm{0}}U_{\bmt}(\bmx)^\dagger$ be the quantum state generated by a data-reuploading quantum circuit satisfying the assumptions of Th.~\ref{th:qnn_fourier}, and $O$ the observable estimated on such state. The hypothesis class implemented by such quantum neural network is
\begin{equation}
    \label{eq:HypothesisClass_qnn}
    \mathcal{M}_\text{QNN} = \qty{\bmx \mapsto f(\bmx) = \Tr[O\, \rho_{\bmt}(\bmx)] = \sum_{\bmo \in \Omega} c_{\bmo}\, e^{-i\, \bmo \cdot \bmx}~\big|~ \{c_{\bmo}\}_{\bmo}~\text{such that}~\abs{f(\bmx)} \leq \norm{O}_{\infty}}\,,
\end{equation}
where the constraint on the coefficients is a consequence of the output of the circuit being the expectation value of an observable, which is upper bounded by its largest eigenvalue
\begin{equation}
    \label{eq:bound_observable_norm}
    \abs{\expval{O}} = \abs{\Tr[O \rho]} \leq \norm{{\rho}}_{1} \norm{O}_\infty = \norm{O}_{\infty}
\end{equation}
where we used Hödler's inequality to upper bound the expectation value in term of the \textit{Shatten p-norms} $\norm{\cdot}_p$ of the operators\footnote{Given an operator $A$, its Shatten $p$-norm is equal to the vector $p$-norm of its singular values $\norm{A}_p = \norm{\bm{s}(A)}_p$ where $\bm{s}(A)$ is the vector of the singular values of $A$. A direct proof of the bound~\eqref{eq:bound_observable_norm} can also be obtained without resorting to operator norms, but by direct computation as in eq.~\eqref{eq:bound_pauli}, namely
\begin{equation}
\abs{\expval{O}} = \abs{\Tr[O\rho]} = \abs{\sum_i o_i \underbrace{\mel{o_i}{\rho}{o_i}}_{\geq 0,~\rho \geq 0}} \leq \sum_{i}\abs{o_i}\mel{o_i}{\rho}{o_i} \leq \max_i \abs{o_i}\, \underbrace{\sum_{i}\mel{o_i}{\rho}{o_i}}_{= \Tr[\rho] = 1} = \max_i \abs{o_i}\,.
\end{equation}}~\cite{Bhatia1996matrix, Watrous_TheoryQI2018}, 
and $\norm{\rho}_1 = 1$ because it is a density matrix, and $\norm{O}_{\infty} \coloneqq \max\qty{\abs{o_i}~|~O = \sum_i o_i \dyad{o_i}}$ is the maximum absolute eigenvalue of the observable. 

As shown and discussed in Appendix~\ref{app:qnn_generalization}, the hypothesis class $\mathcal{M}_\text{QNN}$ can be further simplified, so much so that it can be interpreted as a classical linear model~\eqref{eq:linear_model_feature} using a specific Fourier-like feature map. Indeed, a data-reuploading QNN can be equivalently rewritten as
\begin{equation}
    \label{eq:HypothesisClass_qnn_linear}
    \mathcal{M}_\text{QNN} = \qty{\bmx \mapsto f(\bmx) = \bm{w}_{\bmt} \cdot \bm{\phi}(\bmx) ~\big|~ \bm{w}_{\bmt},\bm{\phi}(\bmx) \in \mathbb{R}^{\abs{\Omega}},\,\norm{\bm{w}_{\bmt}}_2 \leq 2\norm{O}_{\infty}}\,,
\end{equation}
where $\bm{w}_{\bmt} \in \mathbb{R}^{\abs{\Omega}}$ are parameters that depend in a highly non-trivial way on the trainable parameters $\bmx$, and $\bm{\phi}: \mathcal{X} \rightarrow \mathbb{R}^{\Omega}$ is a feature map that takes an input $\bmx$ and maps it to a larger vector whose entries are Fourier features of the input, see Eq.~\eqref{eq:qnn_trig_features}. Essentially, this formulation tells us that data-reuploading quantum circuits can be seen as ``linear'' models in the Fourier features.

Interestingly, now that the hypothesis class is in the form of a classical ``linear'' model, one can use the results from classical statistical learning to characterise the generalisation performances of a data-reuploading quantum neural network. Indeed, as proved in Appendix.~\ref{app:qnn_generalization} by adapting results for classical linear models, one can derive generalisation bounds of the form~\eqref{eq:generalisation_bound} also for data-reuploading circuits. In fact, using the concepts and notation introduced earlier in Sec.~\ref{sec:ch_QML_generalisation}, one can prove the following theorem.

\begin{theorem}[Generalisation Bound for Quantum Neural Networks (see also Theorem 6 in Ref.~\cite{Caro2021encodingdependent})] 
\label{th:gen_qnn_main}
Be $\mathcal{Z} = \mathcal{X} \times \mathcal{Y} \subset [0,2\pi]^d \times \mathbb{R}$ a data space of inputs and outputs. Consider a data reuploading quantum circuit whose encoding scheme generates an integer-valued spectrum $\Omega$, whose model class is $\mathcal{M}_{\textrm{QNN}} := \{\bm{x} \mapsto \Tr[\rho(\bm{x};\bm{\theta})\, O] = \sum_{\bmo \in \Omega} c_{\bmo} e^{-i \bmo \cdot \bmx}\}$. Be $\ell: \mathbb{R} \times \mathbb{R} \rightarrow [0,c]$ an $L$-Lipschitz loss function and define $\mathcal{G}_{\textrm{QNN}} := \{z = (x, y) \mapsto \ell(hf(x), y)\, |\, f \in \mathcal{M}_{\textrm{QNN}}\}$. For any $\delta > 0$ and probability measure $\mathcal{D}$ over $\mathcal{Z}$, with probability at least $1-\delta$ over the drawn of a training set $S \in \mathcal{Z}^m$ of size $m$, for all $g \in \mathcal{G}_{\textrm{QNN}}$:
\begin{equation}
    \label{eq:qnn_gen_main}
    L_\mathcal{D}(g) - L_{S}(g) < 4\|O\|_{\infty}\,L\, \sqrt{\frac{|\Omega|}{m}}\, + 3c\,\sqrt{\frac{\log 2/\delta}{2m}}
\end{equation}
\end{theorem}
A complete derivation of Theorem~\ref{th:gen_qnn_main} can be found in Appendix~\ref{app:qnn_generalization}, where we first show how to use Rademacher complexity to prove a generalisation bound for classical linear models, and then show how this result directly translates to data-reuploading quantum neural networks~\cite{ManginiBenedettiINPREPARATION}. Various encoding-dependent generalisation bounds, like the one reported in Theorem~\ref{th:gen_qnn_main}, were first proved in ref.~\cite{Caro2021encodingdependent}, where the authors use different measures of complexity (Rademacher Complexity and Covering Numbers) to discuss generalisation performances of data-reuploading circuits. We remark that, even though our derivation is based on the Rademacher complexity to measure the complexity of the model, the derivation reported in Appendix~\ref{app:qnn_generalization} follows an arguably simpler and more straightforward strategy, based on the realisation that quantum neural networks can be seen as linear models in a Fourier space~\eqref{eq:HypothesisClass_qnn}. 

The gist of the generalisation bound~\eqref{eq:qnn_gen_main} is to show that the generalisation error, namely the difference between the expected (test) error and the empirical (training) error scales as
\begin{equation}
\label{eq:qnn_generalization}
    \textrm{Generalisation Error of QNN} \leq \order{\sqrt{\frac{|\Omega|}{m}}}\,,
\end{equation}
that is, if the Fourier spectrum $\Omega$ of the circuit is too large (which corresponds to a high complexity of the model) then there are poor guarantees that the model will generalise. However, as expected, an increased training set size $m$ helps in overcoming the issue.

The generalisation bound~\eqref{eq:qnn_gen_main} is specific to data-reuploading quantum circuits that admit a Fourier representation, where a strong accent is posed on the accessible spectrum defined by the data encoding strategy. However, many other results have appeared in the quantum machine literature aiming at characterising the complexity and generalisation performances of quantum models. Examples are works dedicated to the study of the complexity and performances of linear (implicit or explicit) variational models~\cite{Gyurik2023structuralrisk, KublerBiasQuantumKernel, Du2022ComplexityMeasureVQA, Huang2020Power}, based on a quantum information-theoretic approach linking the generalisation performances of a quantum model to the mutual information between the data-embedded quantum states and the classical data space~\cite{BanchiGeneralization2021}, or generalisation bounds based on the Fisher information~\cite{AbbasPowerQNN2021}. An extended discussion of such results is far beyond the scope of the present work, but we refer to~\cite{Caro2021encodingdependent} for a concise summary of recent results on the generalisation of quantum machine learning models.

\subsection{The power of quantum machine learning}
\label{sec:ch_QML_DicussionQML}
Part, if not all, of the reasons for the great success of classical machine learning, is that these methods perform incredibly well in practice and across a wide variety of domains, even reaching super-human performances in some cases. Although the theory behind learning models is rich, interesting and far from being completely understood, it is the practical benefits that drive the adoption of these learning models for data-intensive tasks.

One would then expect (near-term) quantum machine learning to hold these promises and even surpass its ``ordinary'' classical counterpart in terms of performance. However, despite much recent effort in exploring this topic, the question regarding the power of quantum machine learning model is still far from being settled, with no clear signs of quantum advantage. However, some investigations point out directions where a quantum advantage of some sort could be attained.

In terms of model complexity, quantum neural networks were shown to be richer than classical neural networks of comparable size~\cite{AbbasPowerQNN2021}, thus hinting at a possible way to ensure an advantage. However, not only theoretical analysis on the capacity of learning models does not directly translate to practical benefits, but generalisation bounds of the form~\eqref{eq:generalisation_bound} suggest care when using complex models, as these can incur in generalisation performances. Moreover, we have seen that data reuploading circuits essentially are to truncated Fourier series in disguise, and thus admit a simple and fully classical interpretation. Consequently, authors in ref.~\cite{SchreiberSurrogates2022} underline how quantum advantages with these models can only be achieved at training rather than prediction time, since a \textit{surrogate} classical model mimicking the prediction of the quantum circuit can be constructed efficiently via a discrete Fourier transform\footnote{This result can be seen as an application of \textit{Nyquist–Shannon sampling theorem} to reconstruct a continuous periodical signal given a discrete set observations. The output of a data-reuploading circuit is a truncated Fourier series~\eqref{eq:qnn_gtp_summary}, which is a periodic signal of the inputs composed of multiple waves having different frequencies. By sampling the signal ---that is, measuring the output of the circuit--- at inputs which are distant less than $1/2\omega_\text{max}$, where $\omega_\text{max}$ is the largest frequency in the signal, then one can reconstruct classically the complete action of the circuit. The procedure is efficient as long as the maximum frequency in the spectrum of the circuit scales polynomially with the number of qubits, even though this is not always the case~\cite{PetersOverfitting, ShinExponentialEncoding}.}.

Regarding quantum kernel methods, authors in ref.~\cite{Huang2020Power} show that there may be cases where these methods have lower prediction error than their classical counterparts, at least on some engineered datasets, as discussed also in~\cite{Liu2020rigorous}. Specifically, based on the available training data, the authors introduce a geometric test between the kernel functions implemented by the quantum and classical models to check whether there is room for a quantum advantage. If such geometric difference is small, then classical methods will perform similarly or even better. If, on the other hand, the geometric difference is large, then one can construct classification tasks (training data and labels) on which quantum models have better prediction errors.

As for quantum data, that is allowing for the possibility of storing and processing quantum information coming for example from experiments, quantum-native learning models can be shown to perform better than classical ones~\cite{HuangAdvantageExperiments_2022, Huang2020Power, HuangInfoBounds2021, CerezoQMLPerspective_2022}, even though classical algorithms learning on data can still perform well, better than classical non-learning procedures, when given access to quantum datasets~\cite{HuangProvablyEfficient_2022, Huang2020Power}. At last, regarding quantum machine learning applications for classical data on future fault-tolerant quantum computers, there are hopes that QML could provide polynomial speedups for algebra-based subroutines~\cite{CerezoQMLPerspective_2022}.

Researchers should not be discouraged by the lack of clear advantages, both because the study of quantum machine learning is inherently interesting from a theoretical point of view~\cite{SchuldPerspectiveQML_2022}, and because the field is still in its early stages and there is much work to be done to fully understand and harness the power of quantum algorithms for machine learning. As was the case for classical machine learning, it took many decades and several ``winter periods'' to transform the first proposals of learning algorithms in the 1950~\cite{Rosenblatt1958} into today's surprisingly powerful artificial intelligence models. With the strong hope for a shorter incubation period, in the coming years it will be interesting to see how the field develops and if quantum machine learning can live up to its potential.

\section{Conclusions}
\label{sec:ch_QML_Conclusion}
In this chapter, we have gone through a long journey about classical and quantum learning models. We have started with a general definition of quantum machine learning, the topic of the chapter and this entire thesis work, and we argued that different declinations of this field exist, thus explaining the four-fold way of QML~\ref{fig:QML_main_intro}. Moreover, we underlined our focus here is on near-term applications of quantum machine learning based on variational quantum algorithms, especially for analysing classical data. 

We then moved to laying the basics of classical machine learning, introducing the main concepts and tools and discussing two common models, namely linear models and kernel methods, and neural networks. With these tools, we proceeded to the main part of this chapter where we discussed examples of quantum machine learning models, showing how they relate to their classical counterparts. Moreover, we showed how results from classical statistical learning theory can be applied also to quantum models, which is a very helpful tool to characterise their performances. At last, we concluded with a birds-eye view of state-of-the-art QML, talking over how and where quantum computers could bring advantages for learning tasks.

In the following chapters, we will present explicit examples of variational algorithms implementing machine learning models, starting from models of quantum neurons and also showing how these can be applied to analyse data from real use cases. Then, we will discuss the entanglement properties of common parameterized quantum neural networks clinging on the relation between randomness and simulability of the circuit, and finally mention a technique that is not directly related to quantum machine learning, but can be used to mitigate the noise happening in quantum measurement procedures.


\chapterimage{bg4.png} 
\chapterspaceabove{6.75cm} 
\chapterspacebelow{7.25cm} 

\chapter{Quantum computing model of an artificial continuous neuron}\index{Quantum Perceptron}
\label{ch:CQN}
\startcontents[chapters]
\printcontents[chapters]{}{1}{}
\vspace*{1cm}

In this chapter\footnote{The content of this chapter is based on the author's work~\cite{ManginiCQN2020}, and all the figures in this chapter are taken from, or are adaptations of, the figures present in such work.} we discuss a newly introduced model for a quantum perceptron, that is a quantum algorithm mimicking the behaviour of a classical neuron, the building block of artificial neural networks, see Sec.~\ref{ssec:ch_QML_NeuralNetworks}. Specifically, we show how the design for the implementation of a previously introduced quantum artificial neuron~\cite{TacchinoQN2019}, which fully exploits the use of superposition states to encode binary valued input data, can be further generalised to accept continuous- instead of discrete-valued input vectors, without increasing the number of qubits. This further step is crucial to allow for a direct application of gradient descent-based learning procedures, which would not be compatible with binary-valued data encoding.

\section{Introduction}
\label{sec:Intro}
Quantum computers hold the promise to greatly enhance the computational power of not-so-distant in future computing machines~\cite{GoogleSupremacy2019, Preskill2018NISQ}. In particular, improving machine learning techniques by means of quantum computers is the essence of the thriving field of Quantum Machine Learning, as discussed in Sec~\ref{sec:ch_QML_Intro}. Several models for the quantum computing version of artificial neurons have been proposed~\cite{Schuld_IntroQML2015, WiebeSvorePerceptrons2016, CaoGuerreschiGuzik_Peceptron_2017, TacchinoQN2019, Torrontegui2019Perceptron,  KristensenNeuron2021, Killoran2019CVQN}, together with novel quantum machine learning techniques implementing classification tasks~\cite{Havlicek2019QSVM, Schuld2019FeatureSpace, Schuld_Implementing_2017}, quantum autoencoders~\cite{RomeroAutoencoder2017, LamataAutoencoderAdder2018}, quantum convolutional networks~\cite{HendersonQuanvolutional2020, CongQCNN2019} and quantum Boltzmann machines~\cite{Amin2018Boltzmann, XiaoBoltzmann2020} to give a non-exhaustive list. In this context, quantum signal processing leverages the capabilities of quantum computers to represent and elaborate exponentially large arrays of numbers, and it could be used for enhanced pattern recognition tasks, going beyond the capabilities of classical computing machines~\cite{KerendinisQCNN2019}. In these regards, the development of artificial Neural Networks dedicated for quantum computers~\cite{SchuldQuestQuantumNeural2014} is of fundamental importance, due to the preponderance of this type of classical algorithms in image processing~\cite{RojasBook_NN_1996}. 

In the commonly accepted terminology of graph theory, (feed-forward) neural networks are directed acyclic graphs (DAG), that is a collection of nodes where information flows only in one direction, without any loop, as shown previously in Fig.~\ref{fig:ffNN}. Each node is called artificial neuron, since it represent a very simplistic mathematical model for a natural neuron, and consists of an object that takes some input data, processes them using some internal parameters (\textit{weights}), and eventually gives an output value. In their simplest form, these are called McCulloch-Pitts neurons~\cite{McCullochPitts1943} and only deal with binary values, while in the most common and most useful form, named Perceptrons~\cite{Rosenblatt1958}, they accept real, continuously-valued inputs and weights. 

Continuous inputs are not possible in conventional digital computers, and these are usually represented using bitstrings: for instance, a grey scale image pixel is rendered with integer numbers ranging from 0 to 255 using 8-bit binary strings. Some approaches propose to use a similar representation in quantum computers by assigning several qubits per value~\cite{Li2013, LeFRQI2010, LatorreImage2005}. However, these approaches are particularly wasteful, especially in light of the fact that quantum mechanical wavefunctions can be inherently represented as continuously valued vectors.

Recent work has introduced a model for a quantum circuit that mimics a McCulloch-Pitts neuron~\cite{TacchinoQN2019}, and in this chapter we generalise this model to the case of a quantum circuit that also accepts continuous-value input vectors. We thus present a model for a continuous quantum neuron which, as we shall see, can be used for pattern recognition in grey-scale images without the need to increase the number of qubits to be employed. This  represents a further memory advantage with respect to classical computation, where an increase in the number of encoding bits is required to deal with continuous numbers. We employ a phase-based encoding, and show that it is particularly resilient to noise.
 
Differently from classical perceptron models, artificial quantum neurons as described, e.g., in Ref.~\cite{TacchinoQN2019} can be used to classify linearly non separable sets. In the continuously valued case, we thus harness the behaviour of our quantum perceptron model to show its ability to correctly classify several notable cases of linearly non separable sets. Furthermore, we test this quantum artificial neuron for digit recognition on the MNIST dataset~\cite{Farhi2018Classification}, with remarkably good results. We further stress that the present generalisation of the binary-valued artificial neuron model is a crucial step towards the use of gradient descent-based optimisation techniques (see Eq.~\eqref{eq:gradient_descent_ML}) that cannot be applied to the oversimplified integer-valued McCulloch-Pitts neuron model.

\section{Continuously valued quantum neuron model}
\label{sec:CQN}

\label{subsec:Algo}
\begin{figure}[ht]
\centering
\def\layersep{2.4cm}
\begin{tikzpicture}[shorten >=1pt,draw=black!50, node distance=\layersep]
    \tikzstyle{every pin edge}=[<-,shorten <=1pt]
    \tikzstyle{neuron}=[circle, fill=black!25, minimum size=30pt, inner sep=2pt]
    \tikzstyle{input neuron}=[neuron, fill=green!50];
    \tikzstyle{output neuron}=[neuron, fill=red!50];
    \tikzstyle{annot} = [text width=4em, text centered]
    \node[input neuron, pin=left:$i_0$] (i0) at (0, 0) {$w_0$};
    \node[input neuron, pin=left:$i_1$] (i1) at (0,-1.5) {$w_1$};
    \node[input neuron, pin=left:\raisebox{0.55 em}{$\vdots_{\ }$}](i2) at (0,-3) {\raisebox{0.55 em}{$\vdots$}};
    \node[input neuron, pin=left:\raisebox{0.55 em}{$\vdots_{\ }$}] (i3) at (0,-4.5) {\raisebox{0.55 em}{$\vdots$}};
    \node[input neuron, pin=left:$i_{d-1}$] (i4) at (0,-6) {$w_{d-1}$};
      
    \node[output neuron] (O) at (\layersep, -3) {$\sum i_k w_k $};
    
    \node[output neuron,pin={[pin edge={->}]right:Output}, minimum size=33pt] (O2) at (4, -3) {};
    \draw[thick] (3.6, -3.3) .. controls (4.3, -3.3) and (3.7, -2.7) .. (4.4, -2.7);
    \draw[->] (3.5, -3) --(4.5, -3);
    \draw[->] (4, -3.5) --(4, -2.5);
     
    \path[->] (O) edge (O2);     
    \path[->] (i0) edge (O);
    \path[->] (i1) edge (O);
    \path[->] (i2) edge (O);
    \path[->] (i3) edge (O);
    \path[->] (i4) edge (O);
     
    \node[annot,above of= O2, node distance=1.2cm] {Activation function};
    \node[annot] at (-1.3, 0.8) {\small Inputs};
    \node[annot] at (0.1, 0.8) {\small Weights};
\end{tikzpicture}
\caption[Scheme of a classical perceptron model]{Scheme of a classical perceptron model. The artificial neuron evaluates a weighted sum between the input vector, $\bm{i}$, and the weight vector, $\bm{w}$, followed by an activation function which determines the actual output of the neuron, see Eq.~\eqref{eq:single_neuron}.}
\label{fig:classical_perceptron}
\end{figure}
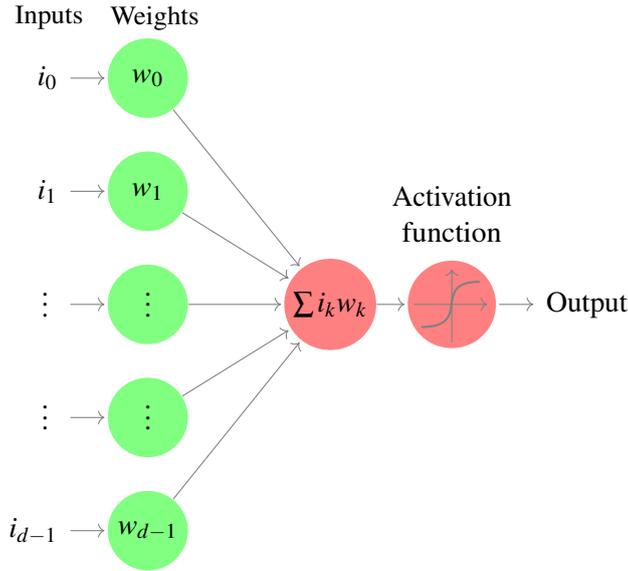

Let us consider a perceptron model with real-valued input and weight vectors, respectively indicated with $\bm{i} = (i_0, \dots, i_{d-1})$ and $\bm{w} = (w_0, \ldots, w_{d-1})$, with $i_k, w_k \in \mathbb{R}$. A schematic representation of a classical perceptron model is depicted in Fig.~\ref{fig:classical_perceptron}, whereas its mathematical formulation was already introduced previously in Eq.~\eqref{eq:single_neuron}, and also discussed in detail for the case of binary McCulloch-Pitts neurons in Sec.~\ref{sec:VAR_quantum_neuron} in the next chapter.

Similarly, we define a model of a quantum neuron capable of accepting continuously valued input and weight vectors, by extending a previous proposal for the quantum computing model of an artificial neuron only accepting binary valued input data~\cite{TacchinoQN2019}, of which we give an extended review in the next Chapter~\ref{ch:VariationalQN}. In order to encode data on a quantum state, we make use of a phase encoding. Given an input $\bm{\theta} = \qty(\theta_0, \ldots, \theta_{d-1})$ with $\theta_i \in [0, \pi]$, consisting of the classical data to be analyzed, we consider the vector:
\begin{equation}
\bm{i} = \qty(e^{i \theta_0}, e^{i \theta_1}, \ldots, e^{i\theta_{d-1}})\ , 
\label{eq:quantum_input}
\end{equation}
which we will be referring to as the input vector in the following. For data not lying in the interval $[0, \pi]$ but more generally in $[a,\, b]$, a normalisation scheme can be used to transform the data in the appropriate range, for example $\theta_i \rightarrow \pi(\theta_i-a)/(b-a)$. Explicit examples will be given later. With the input vector in Eq.~\eqref{eq:quantum_input}, we define the corresponding input quantum state of $n=\log_2 d$ qubits
\begin{equation}
\ket{\psi_i} = \frac{1}{2^{n/2}}\sum_{k=0}^{2^{n}-1} i_k \ket{k}\,,
\label{eq:input}
\end{equation}
where the states $\ket{k}$ denote the computational basis states of $n$ qubits ordered by increasing binary representation, namely $\ket{00\ldots 0},\, \ket{00\ldots 1},\, \ldots,\, \ket{11\ldots 1}$. Since we are dealing with an artificial neuron, we have to encode another vector, namely that of the weights $\bm{\varphi} = \qty(\varphi_0,\, \hdots,\, \varphi_{d-1})$ with $\varphi_i \in [0,\pi]$, and the corresponding vector
\begin{equation}
\bmw = \qty( e^{i \varphi_0},\, e^{i \varphi_1},\, \ldots, e^{i \varphi_{d-1}} )
\label{eq:quantum_weight}
\end{equation}
which in turn defines the weight quantum state
\begin{equation}
\ket{\psi_w} = \frac{1}{2^{n/2}}\sum_{k=0}^{2^{n}-1} w_k \ket{k}\ .
\label{eq:weight}
\end{equation}
Note that~\eqref{eq:input} and~\eqref{eq:weight} have the same structure, in that they consist of an equally weighted superposition of all the computational basis states, although with varying phases. By means of such encoding scheme, we can fully exploit the exponentially large dimension of the $n$ qubits Hilbert space, i.e., by only using $n$ qubits it is possible to encode and analyse data of dimension $d=2^n$. Due to global phase invariance, the number of actual independent phases is $2^n-1$, but this does not spoil the overall efficiency of the algorithm, as we will see. We also note that states of the form $\frac{1}{2^{n/2}}\sum_i e^{i\alpha_i}\ket{i}$, as those in~\eqref{eq:input} and~\eqref{eq:weight}, are known as locally maximally entanglable (LME) states, as introduced in~\cite{KrausLMEStates}.

Having defined the input and weight quantum states, their similarity is estimated by considering the inner product
\begin{equation}
\braket{\psi_w}{\psi_i} = \frac{1}{2^n}\ \sum_{k,j=0}^{2^n-1} i_k w^*_j \braket{j}{k} = \frac{1}{2^n}\ \bm{i} \cdot \bmw^* = \frac{1}{2^n}\qty( e^{i(\theta_0 - \varphi_0)} + \cdots + e^{i(\theta_{2^n-1} - \varphi_{2^n-1})} )\,,
\label{eq:innerprod}
\end{equation}
which corresponds to evaluating the scalar product between the input vector in Eq.~\eqref{eq:quantum_input} and the complex conjugate of the weight vector $\bmw^*$ in Eq.~\eqref{eq:quantum_weight}, thus implementing a similar processing of the classical perceptron algorithm. Since probabilities in quantum mechanics are represented by the squared modulus of wavefunction amplitudes, we consider $\abs{\braket{\psi_w}{\psi_i}}^2$, which can be calculated explicitly as (see App.~\ref{app:CQN_App_A}):
\begin{equation}
\label{eq:mod}
\abs{\braket{\psi_w}{\psi_i}}^2 = \frac{1}{2^n} + \frac{1}{2^{2n-1}}\sum_{i<j}^{2^n-1}\cos \qty(\qty(\theta_j - \varphi_j) - \qty(\theta_i - \varphi_i))\ .
\end{equation}
One can easily check that $\abs{\braket{\psi_w}{\psi_i}}^2=1$ for $\theta_i=\varphi_i~\forall i$, since the two states would coincide in such case. The trigonometric formula in Equation~\eqref{eq:mod} represents the \textit{activation function} implemented by the proposed quantum neuron. Even if it does not remind any of the activation functions conventionally used in classical machine learning techniques, such as the Sigmoid or ReLu functions shown in~\eqref{eq:activation_functions}, its nonlinearity suffices to accomplish classification tasks, as we will discuss in the following sections. 

\subsection{Some properties: colour invariance and noise resilience}
\label{ssec:color_invariance}
From Eq.~\eqref{eq:mod}, we define the activation function of the quantum artificial neuron as
\begin{equation}
\label{eq:mod2}
    f(\bm{\theta},\bm{\varphi}) = \abs{\braket{\psi_w}{\psi_i}}^2 \ .
\end{equation}
Keeping the weight vector $\bm{\varphi}$ fixed, suppose two different input vectors are passed to the quantum neuron, namely $\bm{\theta}$ and $\bmt' = \bmt +\bm{\Delta}$, with $\bm{\Delta} = (\Delta,\, \hdots,\, \Delta)$. One can easily check that whatever the value of $\Delta$, both input vectors will give rise to the same activation function, that is $f(\bmt, \bm{\varphi}) = f(\bmt', \bm{\varphi})$. Hence, two input vectors only differing by a constant, albeit real valued, quantity will be equally classified by such model of quantum perceptron. Hence, in the context of image classification, we can state that the present algorithm has a built-in colour translational invariance. This should not come as a surprise, since the activation function actually depends of the \textit{differences} between phases. In fact, the artificial neuron tends to recognise as similar any dataset that displays the same overall differences, instead of perfectly coincident states.

Next, we assume that the input and weight vectors do coincide, but only up to some noise corrupting the input vector, such that $\bm{\theta} = \bm{\varphi} + \bm{\Delta}$, where $\bm{\Delta}=(\Delta_0,\, \Delta_1,\, \hdots,\, \Delta_{2^n-1})$ represents the small variations, now assumed to be different on each coordinate. Substituting the above values in Eq.~\eqref{eq:mod2}, we obtain
\begin{equation}
f(\bm{\theta},\bm{\varphi}) = f(\bm{\Delta}) = \frac{1}{2^n} + \frac{1}{2^{2n-1}}\sum_{i<j}^{2^n-1}\cos(\Delta_j-\Delta_i)\,.
\end{equation}
For simplicity of calculation, assume the noise factors are distributed according to a uniform distribution in the interval $\Delta_i \sim \text{Unif}[-a/2, a/2],\, a \in \mathbb{R}$. Then, the activation function averaged over the probability distribution of $\Delta_i$ can be calculated as (see App.~\ref{app:CQN_App_B})
\begin{equation}
\label{eq:noise}
\mathbb{E}_{\bm{\Delta}}\qty[f(\bm{\Delta})] = \frac{1}{2^n}+\frac{2^n-1}{2^{n-1}} \qty(\frac{1-\cos(a)}{a^2})\,.
\end{equation}
Input data lie in the interval $[0,\pi]$, thus a reasonable noise is of the order of a some small fraction of $\pi$, which implies $a<1$. Specifically, in the case of small noise, Eq.~\eqref{eq:noise} reduces to
\begin{equation}
\label{eq:noiseapprox}
\mathbb{E}_{\bm{\Delta}} \qty[f(\bm{\Delta})] = 1 - \frac{2^n-1}{2^n}\frac{a^2}{12} + \order{a^4}\quad \text{for}~a \ll 1 \,.
\end{equation}
Thus, the output of the quantum neuron is only slightly perturbed by the presence of noise corrupting an input vector which would otherwise have a perfect activation. As shown in Appendix~\ref{app:CQN_App_B}, a similar result can be derived for any input vectors, not only those having perfect activation, and also for Gaussian ---instead of uniform--- noise. In this respect, one can find a more recent and comprehensive analysis on the effect of Gaussian noise on the parameters of variational quantum algorithms in Ref.~\cite{SkolikNoiseQRL}.

Having outlined the main steps defining the quantum perceptron model for continuously valued input vectors, we now proceed to build a quantum circuit that allows implementing it on a qubit-based quantum computing hardware.

\subsection{Quantum circuit model of a continuously valued perceptron}
\begin{figure}[!ht]
\centering
\begin{quantikz}[column sep = 0.4cm, row sep = 0.5cm]
\lstick[wires=4]{Encoding \\ qubits} &[1em] \lstick{$\ket{0}$} & \gate[style={text width={width("$AAA$")}}, 5, nwires={3}]{U_i} & \gate[style={text width={width("$AAA$")}}, 5, nwires={3}]{U_w} & \ctrl{1} & \qw \\
& \lstick{$\ket{0}$} &             &             & \control{0} & \qw \\
& \lstick{\raisebox{0.5em}{$\vdots$}}  &             &             & \raisebox{0.5em}{\vdots} & \\
& \lstick{$\ket{0}$} &             &             & \ctrl{1} & \qw \\
& \lstick{$\ket{0}$} &             &             & \ctrl{1} & \qw \\
\lstick{Ancilla} & \lstick{$\ket{0}$} &   \qw       &        \qw  & \targ{0}  & \meter{} & \cw & \rstick{$\abs{\langle\psi_i|\psi_w\rangle}^2$}
\end{quantikz}
\caption[Quantum circuit for the quantum perceptron model.]{Quantum circuit model of a perceptron with continuously valued input and weight vectors.}
\label{fig:CQN_CQN}
\end{figure}
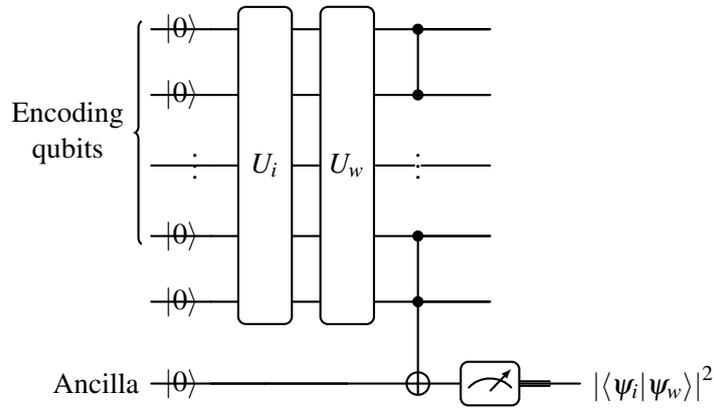

A quantum circuit implementing the quantum neuron model described above is schematically represented in Fig.~\ref{fig:CQN_CQN}. It consists of thee main parts: the first section of the circuit, denoted as $U_i$, transforms the initial quantum state $\ket{\bm{0}} = \ket{0}^{\otimes n}$ to the input quantum state $\ket{\psi_i}$ defined in Eq.~\eqref{eq:input}; then the operation $U_w$ performs the inner product of Eq.~\eqref{eq:innerprod} between the input and weight quantum state; and finally a multi-controlled CNOT gate targeting an ancillary qubit is used to extract the final result of the computation, namely Eq.~\eqref{eq:mod}. We now proceed by explaining in detail how each of these transformations can be implemented in practice.
 
\paragraph{Data encoding} The $U_i$ operation creates the quantum input state $U_i\ket{\bm{0}}=\ket{\psi_i}$ \eqref{eq:input}, and it can be implemented by means of a brute-force approach. First of all, we apply a layer of Hadamard gates, $H^{\otimes n}$, which creates the balanced superposition state $H^{\otimes n}\ket{\bm{0}} = \ket{+}^{\otimes n}$, with $\ket{+} = (\ket{0}+\ket{1})/\sqrt{2}$. The quantum state $\ket{+}^{\otimes n}$ consists of the equally weighted superposition of all the states in the $n$-qubits computational basis, hence we can target each of them and add the appropriate phase to it in order to obtain the desired result~\eqref{eq:input}. Such transformation is implemented by the diagonal (in the computational basis) unitary matrix
\begin{equation}
\label{eq:Ui_matrix}
U(\bmt) \coloneqq 
\begin{bmatrix}
e^{i\theta_0} & 0 & \cdots & 0 \\
0 & e^{i\theta_1}  & \cdots &  \\
\vdots  & \vdots  & \ddots & \vdots  \\
0 & 0 & \cdots & e^{i\theta_{2^n-1}}
\end{bmatrix}\,,
\end{equation}
whose action is to phase shift each state of the computational basis as $\ket{i} \rightarrow e^{i\theta_i}\ket{i}$, with phases $\theta_i \in \mathbb{R}$, that are (in general) independent from each other. One can decompose such overall unitary in smaller pieces 
\begin{equation}
   U(\bm{\theta})=\prod_{i=0}^{2^n-1} \tilde{U}_i(\theta_i)\,,
\end{equation}
where each $U_i(\theta_i)$ acts as $\tilde{U}_i(\theta_i)\ket{i} = e^{i\theta_i}\ket{i}$, while leaving all the other states in the computational basis unchanged. These unitaries can be constructed with an appropriate combination of Pauli-$X$ gates and a multi-controlled phase shift gates, $C^{n-1}P(\theta)$, where the phase shift $P(\theta) = \text{diag}\qty(1,\, e^{i\theta})$, was already introduced earlier in Tab.~\ref{tab:1q_gates} while discussing single-qubit gates.

For example, suppose having $n=3$ qubits, and consider the state $\ket{101}$ to be phase shifted to $e^{i\theta_3}\ket{101}$. Such transformation can be achieved with the following quantum circuit
\[
\begin{quantikz}[column sep = 0.2cm, row sep = 0.2cm]
& \qw & \ctrl{1}  & \qw &  &  & \qw & \ctrl{1} & \qw & \qw \\
& \qw & \octrl{1}  & \qw &
\push{\rule{.3em}{0em}=\rule{.3em}{0em}} & &
 \gate{X} & \ctrl{1} & \gate{X} & \qw \\
& \qw & \gate{\begin{bmatrix}1&0\\0&e^{i\theta_3}\end{bmatrix}} & \qw &  &  & \qw & \gate{\begin{bmatrix} 1 & 0 \\ 0 & e^{i\theta_3} \end{bmatrix} } & \qw & \qw 
\end{quantikz}
\]
whose action is indeed $\tilde{U}_3(\theta_3)\ket{101} = e^{i\theta_3}\ket{101}$, while leaving all other states of the computational basis untouched. Iterating a similar gate sequence for each state of the computational basis $\{\ket{i}\}$, one  eventually obtains the desired overall unitary operation $U(\bmt)$~\eqref{eq:Ui_matrix}. 

Summarising, we have shown how to build the data encoding quantum circuit $U_i$ that creates the quantum state $\ket{\psi_i}$~\eqref{eq:input} by means of the operation $U_i\ket{\bm{0}} \coloneqq U(\bm{\theta})H^{\otimes n}\ket{\bm{0}} = \ket{\psi_i}$, where the parameters $\bmt$ are the input classical data to be analysed~\eqref{eq:input}.\\

\paragraph{Inner product} The unitary $U_w$ can then be constructed in a similar fashion. First, one has to notice that the inner product $\braket{\psi_w}{\psi_i}$ resides in the overlap between the quantum state $\ket{\varphi_{i,w}} \coloneqq \qty( U(\bm{\varphi})H^{\otimes n} )^\dagger \ket{\psi_i}$ and the ground state $\ket{\bm{0}}$. In fact, by definition of $U(\bm{\phi})$ in Eq.~\eqref{eq:Ui_matrix}, it holds that $U({\bm{\varphi}})H^{\otimes n}\ket{\bm{0}} = \ket{\psi_w}$ and thus the scalar product is given as
\begin{equation}
\label{eq:mid_step}
\bra{\bm{0}}\overbrace{\qty(U(\bm{\varphi})H^{\otimes n})^\dagger\ket{\psi_i}}^{\ket{\varphi_{i,w}}} = \underbrace{\bra{\bm{0}}H^{\otimes n}U(\bm{\varphi})^\dagger}_{\bra{\psi_w}}\ket{\psi_i} = \braket{\psi_w}{\psi_i} \, . 
\end{equation}
Then, in order to extract the result, a final layer of Pauli-$X$ gates are applied to all encoding qubits, such that the desired coefficient now multiplies the state $\ket{\bm{1}}$ instead of $\ket{\bm{0}}$, namely
\begin{equation}
\label{eq:this_useless_bit}
    X^{\otimes n}\ket{\varphi_{i,w}} = \ket{\widetilde{\varphi}_{i,w}} = \sum_{k=0}^{2^n-2} c_k \ket{k} + c_{2^n-1}\ket{11\hdots 1}\quad \text{with}\quad c_{2^n-1}=\braket{\psi_w}{\psi_i}\,.
\end{equation}
Thus, by combining~\eqref{eq:mid_step} and~\eqref{eq:this_useless_bit}, one finds that the transformation $U_w$ of Fig.~\ref{fig:CQN_CQN} actually consists in the quantum operations $U_w \coloneqq X^{\otimes n}H^{\otimes n}U(\bm{\varphi})^\dagger$. \\

\paragraph{Measurement-induced activation function} 
By means of a multi-controlled \text{C}$^{n}$\text{NOT}, one can load the coefficient of interest $c_{2^n-1}$ on an ancillary qubit as follows
\begin{equation}
\label{eq:final_measurement_CQN}
\text{C}^{n}\text{NOT}~\ket{\widetilde{\varphi}_{i,w}} \otimes \ket{0}_\text{ancilla} = \sum_{k=0}^{2^n-1}c_k\, \ket{k}\otimes \ket{0}_\text{ancilla} + c_{2^n-1} \ket{11\ldots 1}\otimes \ket{1}_\text{ancilla}\ .
\end{equation}
In fact, a final measurement of the ancilla qubit will yield result $\ket{1}$, interpreted as a \textit{firing} neuron, with probability $ p_\text{out} = \abs{c_{2^n-1}}^2 = \abs{\braket{\psi_w}{\psi_i}}^2 = \abs{\bm{i} \cdot \bmw^*}^2/\qty(2^{2n})$, which is the quantum neuron activation function we introduced in Eq.~\eqref{eq:mod}. \\

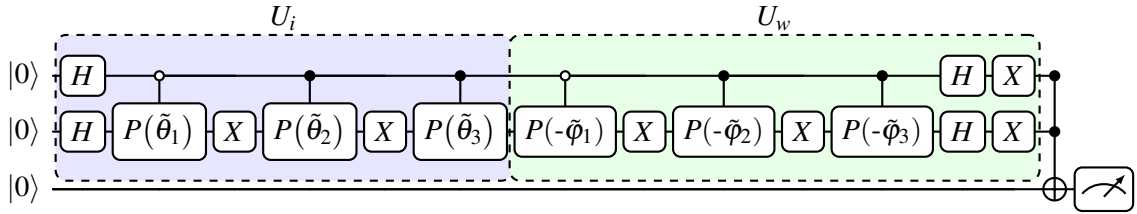
\begin{figure}[ht]
\centering
\begin{quantikz}[column sep = 0.1cm, row sep = 0.1cm]
\lstick{$\ket{0}$} & \gate{H}\gategroup[2, steps=6,style={dashed,
                   rounded corners,fill=blue!10, inner xsep=-2pt, xshift=0pt},
                   background]{$U_i$} & \octrl{1} & \qw & \ctrl{1} & \qw & \ctrl{1} & \octrl{1}\gategroup[2, steps=7, style={dashed,
                   rounded corners,fill=green!10, inner xsep=-2pt, xshift=0pt},
                   background]{$U_w$} & \qw & \ctrl{1} & \qw & \ctrl{1} & \gate{H} & \gate{X} & \ctrl{1} \\
\lstick{$\ket{0}$} & \gate{H} & \gate{P\qty(\tilde{\theta}_1)} & \gate{X} & \gate{P\qty(\tilde{\theta}_2)} & \gate{X} & \gate{P\qty(\tilde{\theta}_3)} & \gate{P\qty(\text{-}\tilde{\varphi}_1)} & \gate{X} & \gate{P\qty(\text{-}\tilde{\varphi}_2)} & \gate{X} & \gate{P\qty(\text{-}\tilde{\varphi}_3)} & \gate{H} & \gate{X} & \ctrl{1} \\
\lstick{$\ket{0}$} & \qw & \qw & \qw & \qw & \qw& \qw & \qw & \qw & \qw & \qw & \qw & \qw & \qw & \targ{0} & \meter{}
\end{quantikz}
\caption[Circuital implementation of the continuous quantum neuron.]{Circuital implementation of the continuous quantum neuron with $n=2$ qubits. The parameters are redefined as $\tilde{\theta}_i=\theta_i-\theta_0,\ \tilde{\varphi_i}=\varphi_i-\varphi_0$, as detailed in Eq.~\eqref{eq:few_gates}.}
\label{fig:CQN_QCirc}
\end{figure}

We now make a few remarks before proceeding with the application of the proposed neuron model to practical learning tasks. First of all, we note that the input vector $\bm{i} \in \mathbb{C}^d$~\eqref{eq:input} contains $d=2^n$ elements, while only $n+1$ qubits are required to implement the quantum neuron circuit in Fig.~\ref{fig:CQN_QCirc}. Moreover, the additional ancillary qubit can be easily removed by performing a joint measurement on all $n$ qubits in the state $\ket{\varphi_{i,w}}$ of Eq.~\eqref{eq:mid_step}, and now considering the probability of measuring $\ket{\bm{0}}$ instead. However, since machine learning techniques yield their full potential when used with connected structures of multiple single neurons, and having in mind the idea of implementing a quantum version of a feedforward neural network, it is essential to have a model for which information can be easily transferred from a neuron to another. This can be accomplished by using an ancilla qubit per artificial neuron, as discussed in~\cite{TacchinoQNN2020}. 

Regarding the time complexity, the number of operations required by this quantum circuit scales linearly with the dimension of the input vectors, but exponentially with the number of qubits. Indeed, the quantum circuit introduced above requires $\order{d} \equiv \order{2^n}$ operations to implement all the phase shifts necessary to build the LME states of Eq.~\eqref{eq:input}. Depending of the relation between the input data, $\theta_i \in \bmt \in \mathbb{R}^d$, other preparation schemes involving less operations could be devised~\cite{KrausLMEStates}, and we refer to Appendix~\ref{app:CQN_App_C} for a more in-depth discussion of this topic.

Finally, it is worth noticing that due to the global phase invariance of quantum states and probabilities, the activation function in Eq.~\eqref{eq:mod} can be recast as
\begin{equation} 
\label{eq:few_gates}
\abs{\braket{\psi_w}{\psi_i}}^2 = \frac{1}{2^{2n}}\abs{\displaystyle{\sum_{i=0}^{2^n-1}} e^{i(\theta_i-\varphi_i)}}^2 = \frac{1}{2^{2n}} \abs{ 1 + \displaystyle{\sum_{i=1}^{2^n-1}} e^{i\qty(\tilde{\theta}_i-\tilde{\varphi}_i)}}^2\ , 
\end{equation}
with $\tilde{\theta}_i=\theta_i-\theta_0,\ \tilde{\varphi_i}=\varphi_i-\varphi_0,~\forall\,i\geq 1$. By exploiting this redefinition of the parameters, it is possible to implement the same activation function but employing fewer gates, and it is equivalent to leaving the state $\ket{\bm{0}}$ unchanged without associating any phase to it. Depending on the actual quantum hardware and data, further simplifications to the circuit could be obtained at compiling time. In Fig.~\ref{fig:CQN_QCirc}, the scheme of a quantum circuit implementing the artificial neuron model is shown for the specific case involving $n=2$ qubits.

\section{Results}
The quantum neuron model introduced above is particularly suited to perform classification tasks involving gray-scale images. A gray-scale image consists of a grid of pixels whose intensities are usually represented by integer numbers in the range $[0, 255]$\footnote{This encoding of gray-scale images employs a single byte (i.e., 8 bits) per pixel on a classical computing register.}, as shown below in Fig.~\ref{fig:grayscale_img}.

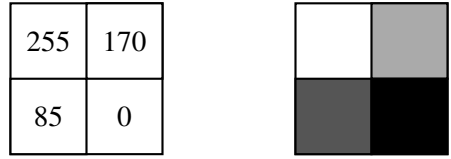
\begin{figure}[ht]
\centering
\begin{tikzpicture}[thick]
\begin{scope}[shift={(-7.25,1)},rotate=0,scale=1]
\draw [fill=white] (0,0) rectangle (2,2);
\draw [fill=white] (0,0) rectangle (1,1);
\draw [fill=white] (1,1) rectangle (2,2);
\node at (0.5,1.5)  {$255$};
\node at (1.5,1.5)  {$170$};
\node at (0.5,0.5)  {$85$};
\node at (1.5,0.5)  {$0$};
\end{scope}
\begin{scope}[shift={(-3.5,1)},rotate=0,scale=1]
\draw [fill=white] (0,0) rectangle (2,2);
\draw [fill=g170] (1,1) rectangle (2,2);
\draw [fill=g85] (0,0) rectangle (1,1);
\draw [fill=black] (1,0) rectangle (2,1);
\end{scope}
\end{tikzpicture}
\caption[A grey-scale image with pixels intensities.]{A grey-scale image with corresponding pixel intensities. This image can be encoded in the array $(255, 170, 85, 0)$, ordered from top-left to bottom-right. Highest value of intensity (255) corresponds to a white pixel, while black pixels are associated to minimum (0) intensity.}
\label{fig:grayscale_img}
\end{figure}

Since we make use of a phase encoding, all inputs (and weights) of the artificial neuron should be normalised in an appropriate angular range, such as $[0, 2\pi]$. However, in this work we further restrict this domain for two important reasons. First, the values in $[0, \pi]$ and $[\pi, 2\pi]$ are equivalent due to the periodicity in phase and the squared modulus in Eq.~\eqref{eq:mod}. Secondly, for the same reason, states with phase $\varphi = 0$ or $\varphi = \pi$ yield the same activation function, which in turn means that images with inverted colours (obtained by exchanging white with black) would be recognised as equivalent by this perceptron model. Hence, to distinguish a given image from its negative, we further restrict the input and weight elements to lie in the range $[0, \pi/2]$. Thus, an image such as the one reported in Fig.~\ref{fig:grayscale_img} is subject to the normalisation $(255, 170, 85, 0) \rightarrow \bmt = \frac{\pi/2}{255} (255, 170, 85, 0)$, before using it as an input vector to be encoded into the quantum neuron model.

We implemented and tested the quantum perceptron model both on simulators using \texttt{Qiskit}~\cite{Qiskit} and also on superconducting real quantum hardware provided by IBM~\cite{IBMQuantum}, and we now presents the results in what follows.

\subsection{Testing the quantum neuron for image recognition tasks}
\begin{figure}[ht]
\resizebox{\textwidth}{!}{
\begin{tikzpicture}[thick]
\begin{scope}[shift={(-8.8,-0.6)},rotate=0,scale=1]
\draw [fill=white] (0,0) rectangle (2,2);
\draw [fill=black] (1,1) rectangle (2,2);
\draw [fill=black] (0,0) rectangle (1,1);
\draw [fill=white] (1,0) rectangle (2,1);
\node at (1.,2.3)  {$\bm{\varphi}_{\text{target}}$};
\end{scope}

\begin{scope}[shift={(-6.6,1)},rotate=0,scale=1]
\draw [fill=white!91.76!black] (0,0) rectangle (2,2);
\draw [fill=white!14.11!black] (1,1) rectangle (2,2);
\draw [fill=white!12.15!black] (0,0) rectangle (1,1);
\draw [fill=white!33.72!black] (1,0) rectangle (2,1);
\node at (1,2.3)  {$9$};
\end{scope}

\begin{scope}[shift={(-4.4,1)},rotate=0,scale=1]
\draw [fill=g242] (0,0) rectangle (2,2);
\draw [fill=g60] (1,1) rectangle (2,2);
\draw [fill=g105] (0,0) rectangle (1,1);
\draw [fill=g242] (1,0) rectangle (2,1);
\node at (1.,2.3)  {$19$};
\end{scope}

\begin{scope}[shift={(-6.6,-1.8)},rotate=0,scale=1]
\draw [fill=white!6.66!black] (0,0) rectangle (2,2);
\draw [fill=white!90.19!black] (1,1) rectangle (2,2);
\draw [fill=white!76.86!black] (0,0) rectangle (1,1);
\draw [fill=white!43.13!black] (1,0) rectangle (2,1);
\node at (1,2.3)  {$7$};
\end{scope}

\begin{scope}[shift={(-4.4,-1.8)},rotate=0,scale=1]
\draw [fill=white!4.13!black] (0,0) rectangle (2,2);
\draw [fill=white!90.98!black] (1,1) rectangle (2,2);
\draw [fill=white!53.33!black] (0,0) rectangle (1,1);
\draw [fill=white!19.21!black] (1,0) rectangle (2,1);
\node at (1,2.3)  {$12$};
\end{scope}

\node at (3.1,0.8) {\scalebox{0.3}{\input{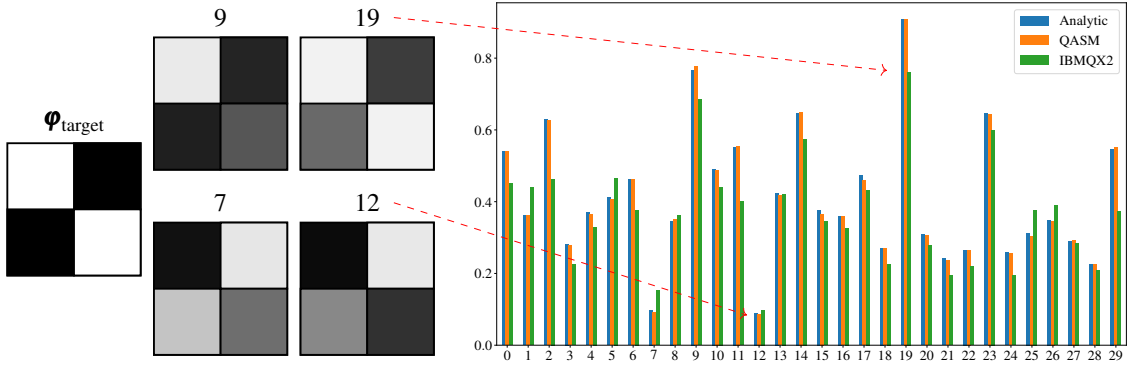}}};
\draw[dashed, thin,->, red] (-3.,3.3) -- (4.4,2.5);
\draw[dashed, thin,->, red] (-3.,0.5) -- (2.3,-1.2);
\end{tikzpicture}}
\caption[Image recognition task performed by the artificial quantum neuron]{Results of the image recognition task performed by a quantum artificial neuron, obtained by running the corresponding quantum circuit with either the \texttt{Qiskit QASM Simulator} backend, and with the \texttt{ibmqx2-yorktown} real quantum processor, and we also report the corresponding analytical values. The target weight vector $\bm{\varphi}_{\text{target}}$ is fixed, and $m=30$ random images are given as input vectors to the quantum artificial neuron. For each input, the corresponding quantum circuit is executed $n_\text{shots} = 8192$ times. The checkerboard-like image corresponds to the target weight vector $\bm{\varphi}_{\text{target}} = (\pi/2,\, 0,\, 0,\, \pi,2)$, while the images displaying respectively largest and lowest activation are the ones labelled as `19' and `12'. Input vectors labelled as `9' and `7' are examples of images with high and low activation, respectively.}
\label{fig:Comparison}
\end{figure}

To better appreciate the potentialities of the continuously valued quantum neuron, we analyse its performance in recognising similar images. Specifically, we fix the weight vector to $\bm{\varphi}_\text{target} = (\pi/2,\, 0,\, 0,\, \pi/2)$ corresponding to the checkerboard pattern represented in the left panel of Fig.~\ref{fig:Comparison}, and then generate a few random images to be used as inputs to the quantum neuron.

For each input, the circuit is executed with $n_\text{shots} = 8192$ measurement shots to gather enough statistics to recover an accurate estimation of the activation function~\eqref{eq:final_measurement_CQN}. With $m=30$ random generated images, the results of the classification are depicted in Fig.~\ref{fig:Comparison}, which includes the analytic results, the results of numerical simulations obtained with the \texttt{Qiskit QASM Simulator} that simulates stochastic measurement outcomes, and finally the results obtained by executing the quantum circuits on the \texttt{ibmqx2-yorktown} real device (accessed in March 2020). 

The images producing the largest activation are the ones corresponding to input vectors similar to the checkerboard-like weight vector, thus confirms the desired behaviour of the quantum neuron in recognising similar images. On the contrary, the images with lowest activation are similar to the negative of the target weight vector, as desired. Due to noise in the actual quantum processing device (see Sec.~\ref{sec:nisq}), the statistics of the outcomes from real experiments differ from either the simulated one and the analytic result. Nevertheless, the same overall behaviour can be easily recognised, thus showing that the quantum neuron model can be successfully implemented also on current quantum processors with reliable results, at least for such image recognition task. 

\section{Training the quantum neuron}
As we extensively discussed in the previous chapters, in the machine learning jargon the process of finding the appropriate value for the weights to implement a given task is called \textit{training} (or \textit{learning}), and it is generally based upon an optimization procedure in which a cost function is minimised by some gradient descent technique~\eqref{eq:gradient_descent_ML}. Ideally, the minimum of the cost function corresponds to the targeted solution.

A simple learning task for the proposed quantum neuron is to recognise a single given input. That is, starting from an input vector $\bm{\theta}_\text{target}$, we aim at finding a weight vector $\bm{\varphi}$ yielding an high activation. A naive yet efficient choice for the loss function driving the learning process is $L(\bm{\varphi}) = \qty(1-f(\bm{\theta}_\text{target}, \bm{\varphi}))^2$, where $f(\bm{\theta}_\text{target}, \bm{\varphi})$ is the activation function of the artificial neuron as in Eq.~\eqref{eq:mod2}, $\bm{\theta}_0$ is the input vector, and $\bm{\varphi}$ is the trainable vector of the weights. Clearly, the minimum of the loss $L(\bm{\varphi}_\text{opt}) = 0$ is achieved when the quantum neuron has full activation, that is $f(\bm{\theta}_\text{target}, \bm{\varphi}_\text{opt}) = 1$. Since the activation function implemented by the quantum perceptron is given in Eq.~\eqref{eq:mod}, we know that a perfect activation can be obtained when the input and weight vectors are exactly coincident, $\bm{\varphi}_\text{opt} = \bm{\theta}_\text{target}$, although other solutions may do equivalently well.

In our experiments, the minimization process is driven by the Simultaneous Perturbation Stochastic Approximation (SPSA)~\cite{Spall1998overview} algorithm, which is built for optimization processes characterized by the presence of noise and is thus particularly effective in the presence of probabilistic measurement outputs. Indeed, we implement all training procedures in the presence of shot noise by simulating the quantum neuron circuits with Qiskit \texttt{qasm{\_}simulator}. 
\begin{figure}[ht]
\resizebox{\textwidth}{!}{
\subfloat[\label{fig:learningSPSA}]{{\includegraphics{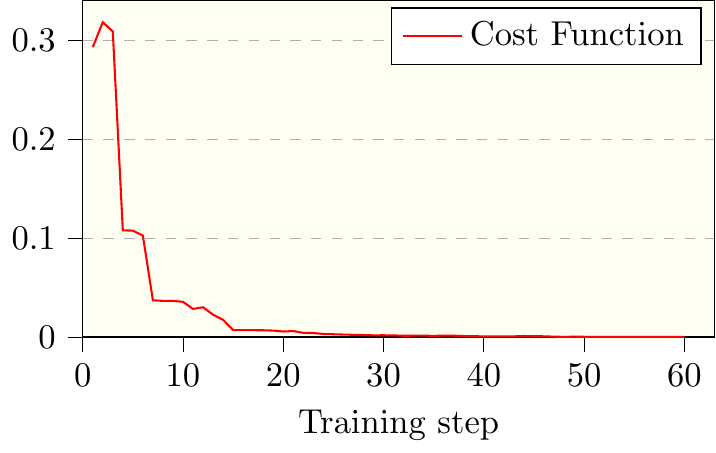}}}
\quad
\subfloat[\label{fig:finalW}]{
    \begin{tikzpicture}[thick]
    \node (pino) at (-6, -0.5) {};
    \begin{scope}[shift={(-8.6,1)},rotate=0,scale=1]
    \draw [fill=g102] (0,0) rectangle (2,2);
    \draw [fill=black] (1,1) rectangle (2,2);
    \draw [fill=g170] (0,0) rectangle (1,1);
    \draw [fill=g16] (1,0) rectangle (2,1);
    \node at (1,2.3) {$\bm{\theta}_{\text{target}}$};
    \end{scope}
    \begin{scope}[shift={(-6,1)},rotate=0,scale=1]
    \draw [fill=g32] (0,0) rectangle (2,2);
    \draw [fill=g162] (1,1) rectangle (2,2);
    \draw [fill=g65] (0,0) rectangle (1,1);
    \draw [fill=g211] (1,0) rectangle (2,1);
    \node at (1,2.3) {$\bm{\varphi}_{\text{start}}$};
    \node at (2.4,1) {$\rightarrow$};
    \end{scope}
    \begin{scope}[shift={(-3.2,1)},rotate=0,scale=1]
    \draw [fill=g204] (0,0) rectangle (2,2);
    \draw [fill=g28] (1,1) rectangle (2,2);
    \draw [fill=white] (0,0) rectangle (1,1);
    \draw [fill=g40] (1,0) rectangle (2,1);
    \node at (1,2.3) {$\bm{\varphi}_{\text{final}}$};
    \end{scope}
    \end{tikzpicture}}}
\caption[Training the quantum neuron]{Training the quantum neuron to recognise an input pattern. \textbf{(a)} Minimisation of the cost function $\mathcal{L}(\bm{\varphi}) = \qty(1-f(\bm{\theta}_\text{target}, \bm{\varphi}))^2$, as a function of the training steps using the SPSA optimiser and simulating the quantum circuits with Qiskit \texttt{qasm{\_}simulator}. \textbf{(b)} Images corresponding to the target input vector $\bm{\theta}_\text{target}$, the randomly initialised starting weight vector $\bm{\varphi}_\text{start}$, and the one obtained at the end of the optimisation procedure $\bm{\varphi}_\text{final}$.}
\label{fig:EasyLearning}
\end{figure}

In Fig.~\ref{fig:EasyLearning} we report the training results for the the task of recognising the input vector $\bmt_\text{target}=(\pi/5,\, 0,\, \pi/3,\, 0.1)$, whose equivalent grey-scale representation is shown in panel~\ref{fig:finalW}. As expected, the cost function is readily minimised by varying the weight vector, and it rapidly converges to values close to zero after a few iteration steps. In this case, the minimum of the loss is achieved at $\bm{\varphi}_\text{final} = (1.03,\, 0.19,\, 1.47,\, 0.61)$, whose grey-scale representation is plotted in Figure~\ref{fig:finalW}. Even though the input and weight vector are not numerically equivalent, we see that the final weight image actually looks very much like the target one. In fact, the two images retain almost the same shades of grey, with the optimised one being a bit shifted towards the brightest end of the spectrum. This is in agreement with the discussion in Sec.~\ref{ssec:color_invariance} regarding colour invariance, where we noticed that the neuron is blind to overall colour shifts.

\subsection{Classification tasks}
\label{ssec:CQN_classification}
We now move our attention to classification tasks, where we want the neuron to assign a given desired label to an input. Specifically, we consider supervised binary classification problems\footnote{This can be generalised in the case of a multi-class classification by adopting a \textit{one versus all} approach.}, where training inputs $\{\bm{\theta}_i\}_{i=1}^m$ are associated to binary labels, $\{y_i\}_{i=1}^m,~y_i \in \{0, 1\}$, and the artificial neuron is asked to reproduce such mapping between inputs and labels. 

In order to implement a binary dichotomy with a perceptron model, it is common practice to introduce a \textit{threshold} value $t$, such that if the activation of the neuron is higher than $t$, then the output of the neuron is $1$, and it is $0$ otherwise. Formally, given an input $\bmt_i$ and weight vector $\bm{\varphi}$, the label $\hat{y}_i$ predicted by the quantum neuron is 
\begin{equation}
\tilde{y}_i(\bm{\varphi}) = 
\begin{cases}
1\quad \text{if } f(\bm{\theta}_i, \bm{\varphi})> t \\
0\quad \text{otherwise}
\end{cases}\ .
\label{eq:assigned_label}
\end{equation}
where $f(\bmt_i,\bm{\varphi})$ is the activation of the neuron~\eqref{eq:mod}. Note that the the threshold $t$ is actually a \textit{hyperparameter} for our model, and in the following simulations it was heuristically optimized in order to achieve the best classification accuracy.

As with any supervised learning task, the training process is driven by empirical risk minimisation~\eqref{eq:empirical_risk}, where in this case we picked as a loss function an Mean Squared Error like distance between the correct label from the one predicted by the artificial neuron, namely
\begin{equation}
\label{eq:CQN_cost_function}
L(\bm{\varphi}) = \frac{1}{m}\sum_{i=1}^{m}(y_i - \tilde{y}_i(\bm{\varphi}))^2\ ,
\end{equation}
where $m$ is the number of samples in the training set, $y_i$ is the correct label associated to input data $\bm{\theta}_i$, and $\hat{y_i}(\bm{\varphi})$ is the label predicted by the neuron according to the decision rule~\eqref{eq:assigned_label}.  We now proceed discussing two specific classification tasks on which the quantum neuron was tested.

\subsubsection{Classification of two-dimensional data}
\begin{figure}[ht]
\centering
\resizebox{\textwidth}{!}{
\subfloat[\label{fig:2Da}]{\scalebox{0.52}{\input{Images/CQN/2Ddatas_original_NEW.pgf}}}\hspace{0em}%
\subfloat[\label{fig:2Db}]{\scalebox{0.52}{\input{Images/CQN/2D_decisionb_NEW.pgf}}}\hspace{0em}%
\subfloat[\label{fig:2D_learning}]{{\includegraphics{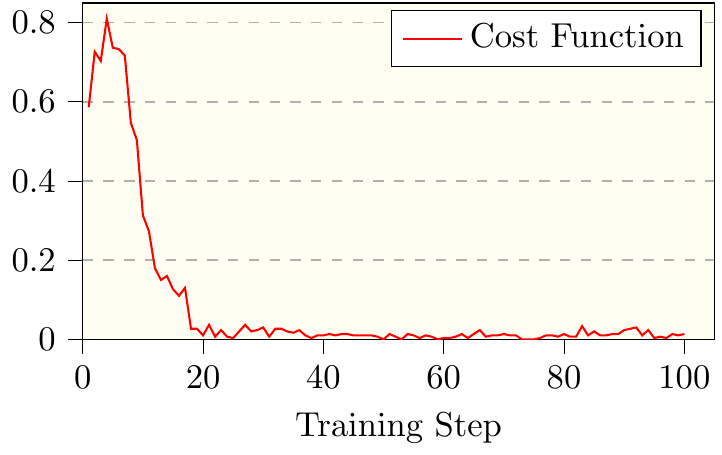}}}}
\caption[Classification of two dimensional data]{Classification of two dimensional data. \textbf{(a)} Input data used as training set. \textbf{(b)} Test set and decision boundary implemented by the optimised quantum neuron at the end of the learning procedure. The threshold used in the decision rule~\eqref{eq:assigned_label} is set to $t=0.95$. \textbf{(c)} Optimisation with the SPSA optimiser run on Qiskit \texttt{qasm{\_}simulator}.}
\label{fig:2Ddata_original}
\end{figure}

As a first example, we considered a classification problem of the form $\{(\bmt_i, y_i)\}_{i=1}^m$, in which inputs $\bmt_i = \qty(\theta^{(i)}_1, \theta^{(i)}_2) \in [0, \pi/2]^2$ are two dimensional input data, and $y_i \in \{0,1\}$ their labels, indicated by red ($y=0$) and blue ($y=1$) dots in Fig.~\ref{fig:2Ddata_original}. Since the data are two dimensional we only need a single qubit to encode the information in the quantum state, and also in this case all simulations were performed using Qiskit \texttt{qasm{\_}simulator}, thus taking into account the statistical noise due to finite number of shots.

The panel~\ref{fig:2Da} shows the data samples in the training set, while those shown in panel~\ref{fig:2Db} are those in the test set, along with the decision boundary that the quantum neuron learnt at the end of the optimisation procedure. The cost function in Eq.~\eqref{eq:CQN_cost_function} is minimised using the SPSA optimiser and its behaviour is reported in Fig.~\ref{fig:2D_learning}. Training proceeds towards a minimum of the empirical loss, and, as shown in Fig.~\ref{fig:2Db}, at the end of training the optimised neuron implements a perfect classification also on the test set.

\subsubsection{Classification of non separable data using a bias}
\begin{figure}[ht]
\centering
\resizebox{\textwidth}{!}{
\subfloat[\label{fig:2DcircleA}]{\scalebox{0.52}{\input{Images/CQN/2D_circle_NEW.pgf}}}\hspace{0em}
\subfloat[\label{fig:2DcircleB}]{\scalebox{0.52}{\input{Images/CQN/2DCircle_decisionb_NEW.pgf}}}\hspace{0em}
\subfloat[\label{fig:Circ_learning}]{\includegraphics{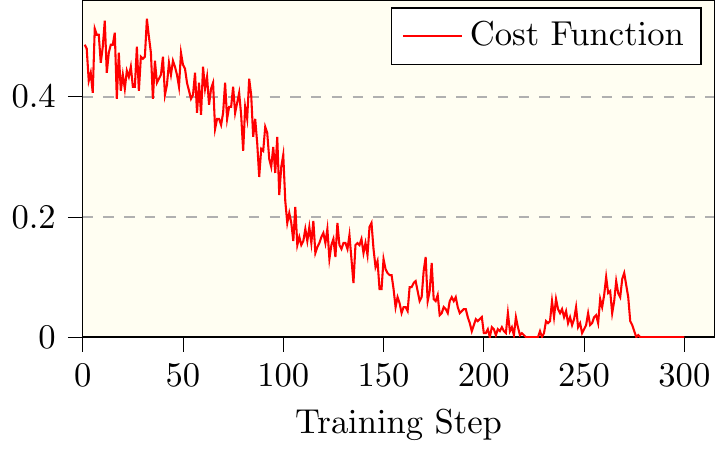}}}
\caption[Classification of two dimensional circles]{Classification of two dimensional circles. \textbf{(a)} Input data used as the training set. \textbf{(b)} Test set and decision boundary implemented by the optimised neuron at the end of the learning procedure. The threshold used in this example is $t=0.95$, and the bias $b=0.25$. \textbf{(c)} Optimisation by the SPSA optimiser run on the \texttt{QASM} Simulator. Each iteration in the optimisation procedure was performed calculating the loss over a batch of just 20 samples instead of the full training dataset, which explains why the error is not smooth but presents several spikes.}
\label{fig:Circ_data}
\end{figure}

We have just seen that a single neuron is capable of classifying some type of two dimensional data, but this procedure is not guaranteed to succeed on more structured dataset. Indeed, for a classification task with the data shown Fig.~\ref{fig:2DcircleA}, the single-qubit encoding of the previous model is not enough. 

However, such a problem can be tackled successfully by a quantum neuron using two qubits, which allows for additional degrees of freedom and thus increased expressivity. In fact, with $n=2$ qubits it is possible to encode $2^2=4$ parameters on the quantum states of interest: one can be kept fixed to zero\footnote{The activation function in Eq.~\eqref{eq:mod} only depends on the \textit{differences} between the parameters, thus fixing one of the parameters to a constant value can be thought just as choosing a reference point. In other words, the additional parameter is a global phase that can be ignored.}, two of these are used to encode the actual input data to be analysed, and the last free parameter can be interpreted as a \textit{bias} term, to be tuned accordingly to ensure good performances. 

Thus, a convenient encoding scheme is to consider a two-qubit quantum neuron whose input vectors are four-dimensional vectors of the form $\bm{\theta} = (0, \,\theta_1,\, \theta_2,\, 0)$, and the weight vector is given by $\bm{\varphi} = (0,\, \varphi_1,\, \varphi_2,\, b)$, where $b$ denotes the bias term, which in our simulations is an hyperparameter that we choose heuristically. In Fig.~\ref{fig:Circ_data} we show the results of training such model by minimising the empirical risk~\eqref{eq:CQN_cost_function} on the training data shown in panel~(a). After the learning procedure, reported in Fig.~\ref{fig:Circ_learning}, the neuron learns a decision boundary that achieves perfect score also on the test set, as shown in Fig.~\ref{fig:2DcircleB}.

\subsection{MNIST dataset}
\begin{figure}[ht]
\centering
\subfloat[]{
\includegraphics[width=0.225\textwidth]{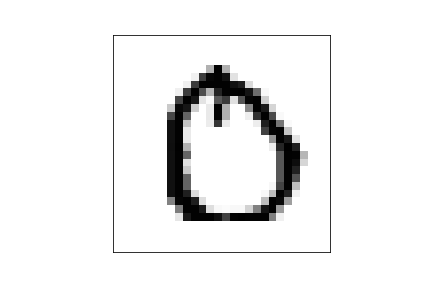}\hspace{-0.8cm}
\includegraphics[width=0.225\textwidth]{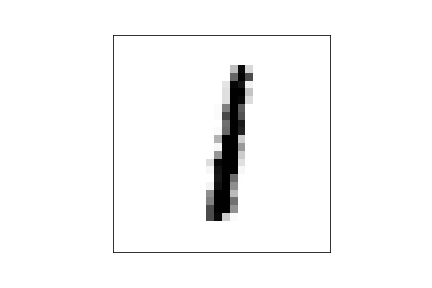}\vspace{5em}
}
\subfloat[]{\scalebox{0.6}{\input{Images/CQN/MNIST_confusion.pgf}}}
\caption[Application of the quantum neuron on the MNIST dataset]{Application of the quantum neuron on the MNIST dataset. \textbf{(a)} Examples of a `0' and a `1' drawn from the MNIST dataset. \textbf{(b)} Matrix containing the fidelities of some samples of ``one'' and ``zero'' images taken from the MNIST dataset, evaluated with the activation function in Eq.~\eqref{eq:mod} and implemented by our quantum neuron model.}
\label{fig:MNIST_total}
\end{figure}

As a concluding example, we briefly discuss how the proposed quantum neuron model could be used to analyse the widely known MNIST dataset~\cite{MNISTLecun2010}, which is a set containing $70.000$ gray-scale images of digits, from zero to nine. For simplicity, we limit our analysis to only images of zeros and ones, of which we show two examples in the left panel of Fig.~\ref{fig:MNIST_total}. Note that each image in the MNIST dataset is composed of $28 \times 28$ pixels, which is not of the form $2^{n/2} \times 2^{n/2}$ required to be encoded on the quantum state of an artificial neuron of $n$ qubits, admitting $d=2^n$ input data. Thus, we preprocess the images by adding a number of white pixels so that modified images have dimension $32 \times 32$ pixels, which can be given as inputs to a quantum neuron of $n = 10$ qubits.

The focus of the analysis is to check whether the activation function implemented by the proposed model~\eqref{eq:mod} is effective at discriminating between the encoded images of zeros and ones. With this goal in mind, one way to proceed is to fix the weight vector of the artificial neuron to a sample of a `1' selected from the MNIST dataset, and then calculate the activation function of the neuron when other images of zeros and ones are fed as inputs. By using such a quantum neuron with the decision rule in Equation~\eqref{eq:assigned_label} with threshold set to $t=0.85$, the cost function~\eqref{eq:CQN_cost_function} evaluated on a set of $m = 2060$ images amounts to $L \sim 0.02$, implying a remarkably good classification accuracy of $98\%$\footnote{True and predicted labels are $y_i,\, \hat{y_i} \in \qty{0, 1}$, and so the loss function~\eqref{eq:CQN_cost_function} effectively measures the number of misclassified data, that is the number samples for which the true and predicted labels are different.}. Moreover, in the right panel of Fig.~\ref{fig:MNIST_total}, we also report the activation function obtained for several pairs on zeros and ones. According to the artificial neuron, ones are more similar to each other with respect to the zeros, as the fidelities are higher generally much higher (top left corner of the matrix). Even if classical machine learning techniques can easily yield a classification accuracy above $99\%$, the present results show a remarkable degree of precision, also considering that in this particular example just a single quantum neuron has been used for the classification. 

In addition, we also tested a training-based procedure in which each image in the MNIST dataset is first compressed to a $4 \times 4$ image by means of a \textit{mean pooling} filter that aggregates values of neighbouring pixels to a single value, and then passed to the neuron. After training, the neuron reaches a best accuracy of about $80\%$, which, although far from perfect, shows the potential of the activation function implemented by the quantum neuron to be used also for recognition of complex patterns, such as numerical digits. 

Our quantum neuron model performs well when compared with other proposed quantum algorithms for the classification of the MNIST dataset. In fact, alternative algorithms have been proposed for this task, some of them using a hybrid classical-quantum approach, such as leveraging well established classical pre/post processing of data through classical machine learning techniques~\cite{Qiskit, Kerendinis2020}. These hybrid approaches may yield higher (although comparable) classification accuracy when compared to our quantum neuron model. However, we emphasise that in our case the artificial neuron model is fully quantum in nature. When compared to other works~\cite{Broughton2020TFQ, Farhi2018Classification} using only quantum resources and reporting accuracies of the order of 85\% to 98\%, our model seems to yield comparable or better results.

\section{Conclusions}
We have reported on a novel quantum algorithm allowing to implement a generalised perceptron model on a qubit-based quantum device that accepts and analyses continuously valued input data. The proposed algorithm is translated into a quantum circuit model to be readily run on existing quantum hardware. It takes full advantage of the exponentially large Hilbert space available to encode input data on the phases of large superposition states, known as locally maximally entanglable (LME) states. These LME states can be constructed with a bottom-up approach, by imprinting each single phase separately. However, it should be stressed that alternative and possibly more efficient strategies could directly yield such states as ground states of suitable Hamiltonians, or as stationary states from dissipative processes~\cite{KrausLMEStates}. 

The proposed continuously valued quantum neuron proves to be a good candidate for classification tasks and pattern recognition involving grey-scale images. In particular, the activation function implemented by the quantum neuron yields very high accuracy in the order of $~98\%$ when used to discriminate between images of zeros and ones from the MNIST dataset, thus indicating the ability to distinguish also complex patterns. Moreover, thanks to the phase encoding, the neuron can leverage a built-in ``colour translational" invariance, as well as significant noise resilience.

A further step would be to consider multiple layers of connected quantum neurons to build a continuous quantum feed-forward neural network, as proposed in~\cite{TacchinoQNN2020} for binary-valued quantum neurons. Another important investigation is the design of approximate methods to perform the weight unitary transformation in a way which scales more favourably with the number of encoding qubits, a topic which is at the core of the following Chapter~\ref{ch:VariationalQN}. Moreover, in Chapter~\ref{ch:Autoencoder} we will discuss how the application of phase encoding to other quantum machine learning techniques, namely quantum autoencoders.


\chapterimage{bg5.png} 
\chapterspaceabove{6.75cm} 
\chapterspacebelow{7.25cm} 

\chapter{Variational learning for quantum neural networks}\index{Variational Learning}
\label{ch:VariationalQN}
\startcontents[chapters]
\printcontents[chapters]{}{1}{}
\vspace*{0.5cm}

In this chapter\footnote{The content of this chapter is based on the author's work~\cite{TacchinoVariationalQNN2021}, and all the figures in this chapter are taken from, or are adaptations of, the figures present in such work.}, we first review a series of recent works describing the implementation of artificial neurons and feed-forward neural networks on quantum processors, and subsequently present an original realisation of efficient individual quantum nodes based on variational unsampling protocols. While keeping full compatibility with the overall memory-efficient feed-forward architecture, our constructions effectively reduce the quantum circuit depth required to determine the activation probability of single neurons upon input of the relevant data-encoding quantum states. This suggests a viable approach towards the use of quantum neural networks for pattern classification on near-term quantum hardware.

\section{Introduction}\index{}

This chapter is dedicated to the study and improvement of a recently proposed quantum algorithm implementing the activity of binary-valued artificial neurons for classification purposes, whose generalisation to continuous variables was the topic of the previous chapter~\ref{ch:CQN}. 

Although formally exact, this algorithm for a quantum neuron in general requires quite large circuit depth for the analysis of the input classical data. To mitigate for this effect we introduce a variational learning procedure, based on quantum unsampling techniques, aimed at critically reducing the quantum resources required for its realisation. Indeed, we investigate different learning strategies involving global and local layer-wise cost functions, and we assess their performances also in the presence of statistical measurement noise. By combining memory-efficient encoding schemes and low-depth quantum circuits for the manipulation and analysis of quantum states, the proposed methods suggest a practical route towards problem-specific instances of quantum computational advantage in machine learning applications.

\section{A model of quantum artificial neurons}
\label{sec:VAR_quantum_neuron}
The simplest formalisation of an artificial neuron can be given following the classical model proposed by McCulloch and Pitts~\cite{McCullochPitts1943}. In this scheme, a single node receives a set of binary inputs $\{i_0,\ldots,x_{d-1}\} \in \{-1,1\}^d$, which can either be signals from other neurons in the network or external data. The computational operation carried out by the artificial neuron consists in first weighting each input by a synapse coefficient $w_{j} \in \{-1,1\}$ and then providing a binary output $y \in \{-1,1\}$ denoting either an active or rest state of the node determined by an integrate-and-fire response
\begin{equation}
y = \begin{cases} 1 \quad & \text{if } \sum_j w_{j}x_j \geq t \\
-1 \quad & \text{otherwise }
 \end{cases}
 \label{eq:McP-operation}
\end{equation}
where $t$ represents some predefined threshold.

A quantum procedure closely mimicking the functionality of a binary valued McCulloch-Pitts artificial neuron can be designed by exploiting, on one hand, the superposition of computational basis states in quantum registers, and on the other hand the natural non-linear activation behaviour provided by quantum measurements. In this section, we will briefly outline a device-independent algorithmic procedure~\cite{TacchinoQN2019} designed to implement such a computational model on a gate-based quantum processor. More explicitly, we show how classical input and weight vectors of size $m$ can be encoded on a quantum hardware by using only $n = \log_2 d$ qubits~\cite{Schuld_Implementing_2017, RebentrostHopfield2018, TacchinoQNN2020}. For loading and manipulation of data, we describe a protocol based on the generation of quantum hypergraph states~\cite{rossi_quantum_2013}. This exact approach to artificial neuron operations will be used in the main body as a benchmark to assess the performances of approximate variational techniques designed to achieve more favourable scaling properties in the number of logical operations with respect to classical counterparts.

Let $\bm{i}$ and $\bm{w}$ be binary input and weight vectors of the form
\begin{equation}
\bm{i} = \begin{pmatrix}
    i_{0} \\
    i_{1} \\
    \vdots \\
    i_{d-1}
\end{pmatrix}\quad
\bm{w} = \begin{pmatrix}
    w_{0} \\
    w_{1} \\
    \vdots \\
    w_{d-1}
\end{pmatrix}
\label{eq:bin_vectors}
\end{equation}
with $i_j,w_j \in \{-1,1\}$ and $d = 2^{n}$. A simple and qubit-effective way of encoding such collections of classical data can be given by making use of the relative quantum phases (i.e.\ factors $\pm 1$ in our binary case) in equally weighted superpositions of computational basis states. We then  define the states
\begin{equation}
\label{eq:inputstate}
|\psi_i\rangle = \frac{1}{\sqrt{d}}\sum_{j = 0}^{d - 1} i_j \ket{j},\quad |\psi_w\rangle = \frac{1}{\sqrt{d}}\sum_{j = 0}^{d - 1} w_j \ket{j}
\end{equation}
where, as usual, we label computational basis states with integers $j\in\{0,\ldots,d-1\}$ corresponding to the decimal representation of the respective binary string. The set of all possible states which can be expressed in the form above is known as the class of hypergraph states~\cite{rossi_quantum_2013} and are a special case of the more general states treated in the previous Chapter~\ref{ch:CQN}, where we allowed the coefficients of the computational basis to be arbitrary phases $e^{i\theta}$ rather than just signs $\pm 1$.

According to Eq.~\eqref{eq:McP-operation}, the quantum algorithm must first perform the inner product $\bm{i} \cdot \bm{w}$. As shown earlier for the case of the continuous neuron, the inner product between inputs and weights is contained in the overlap $\braket{\psi_w}{\psi_i} = \bm{w}\cdot\bm{i}/{d}$~\cite{TacchinoQN2019}, and one can explicitly compute such overlap on a quantum register through a sequence of $\bm{i}$- and $\bm{w}$-controlled unitary operations. For clarity of exposition, we hereby summarise again the steps necessary to calculate the overlap of interest, and we refer to the previous chapter for a more detailed discussion. 

First, assuming that we operate on a $n$-qubit quantum register starting in the blank state $\ket{0}^{\otimes n}$, we can load the input-encoding quantum state $\ket{\psi_i}$ by performing a unitary transformation $U_i$ such that $U_i\ket{0}^{\otimes n}=\ket{\psi_i}$. It is important to mention that this preparation step would most effectively be replaced by, e.g., a direct call to a quantum memory~\cite{QRAM_Maccone}, or with the supply of data encoding states readily generated in quantum form by quantum sensing devices to be analyzed or classified. It is indeed well known that the interface between classical data and their representation on quantum registers currently constitutes one of the major bottlenecks for Quantum Machine Learning applications~\cite{Biamonte2017QML}. 

Let now $U_w$ be a unitary operator such that
\begin{equation}
\label{eq:UwConstraint}
U_w \ket{\psi_w} = \ket{1}^{\otimes n} = \ket{d-1}
\end{equation}
In principle, any $d\times d$ unitary matrix having the elements of $\bm{w}$ appearing in the last row satisfies this condition. If we apply $U_w$ after $U_i$, the overall $n$-qubits quantum state becomes
\begin{equation}
U_w \ket{\psi_i} = \sum_{j = 0}^{d - 1} c_j \ket{j} =: \ket{\varphi_{i,w}}
\label{eq:afterUs}
\end{equation}
Using Eq.~\eqref{eq:UwConstraint}, we then have
\begin{equation}
\label{eq:idotw}
\braket{\psi_w}{\psi_i} = \langle \psi_w | U_w^\dagger U_w | \psi_i\rangle = \langle m-1 |\phi_{i,w}\rangle = c_{m-1}
\end{equation}
We thus see that, as a consequence of the constraints imposed to $U_i$ and $U_w$, the desired result $\bm{i} \cdot \bm{w} \propto \braket{\psi_w}{\psi_i}$ is contained up to a normalisation factor in the coefficient $c_{d-1}$ of the final state $\ket{\varphi_{i,w}}$. 

The final step of the algorithm must access the computed input-weight scalar product and determine the activation state of the artificial neuron. In view of constructing a general architecture for feed-forward neural networks~\cite{TacchinoQNN2020}, it is useful to introduce an ancilla qubit $a$, initially set in the state $\ket{0}$, on which the $c_{d-1} \propto \braket{\psi_w}{\psi_i}$ coefficient can be written through a multi-controlled $\mathrm{CNOT}$ gate, where the role of controls is assigned to the $n$ encoding qubits~\cite{TacchinoQN2019}:
\begin{equation}
\label{eq:neuron-activation}
\ket{\phi_{i,w}}\ket{0}_\text{ancilla} \rightarrow \sum_{j = 0}^{d - 2} c_j \ket{j}\ket{0}_\text{ancilla} + c_{d-1}\ket{d-1}\ket{1}_\text{ancilla}
\end{equation}
At this stage, a measurement of ancillary qubit $a$ in the computational basis provides a probabilistic non-linear threshold activation behaviour, producing the output $\ket{1}$ state, interpreted as an active state of the neuron, with probability $|c_{d-1}|^2$. Although this form of the activation function is already sufficient to carry out elementary classification tasks and to realise a logical $\mathrm{XOR}$ operation~\cite{TacchinoQN2019}, more complex threshold behaviours can in principle be engineered once the information about the inner product is stored on the ancilla~\cite{CaoGuerreschiGuzik_Peceptron_2017, Torrontegui2019Perceptron}. Equivalently, the ancilla can be used, via quantum controlled operations, to pass on the information to other quantum registers encoding successive layers in a feed-forward network architecture~\cite{TacchinoQNN2020}. It is worth noticing that directing all the relevant information into the state of a single qubit, besides enabling effective quantum synapses, can be advantageous when implementing the procedure on real hardware on which readout errors constitute a major source of inaccuracy. Nevertheless, multi-controlled $\mathrm{NOT}$ operations, which are inherently non-local, can lead to complex decompositions into hardware-native gates especially in the presence of constraints in qubit-qubit connectivity. When operating a single node to carry out simple classification tasks or, as we will do in the following sections, to assess the performances of individual portions of the proposed algorithm, the activation probability of the artificial neuron can then equivalently be extracted directly from the register of $N$ encoding qubits by performing a direct measurement of $\ket{\varphi_{i,w}}$ targeting the $\ket{d-1} \equiv \ket{1}^{\otimes n}$ computational basis state.

\subsection{Exact implementation with quantum hypergraph states}
A general and exact realisation of the unitary transformations $U_i$ and $U_w$ can be designed by using the generation algorithm for quantum hypergraph states~\cite{TacchinoQN2019}. The latter have been extensively studied as useful quantum resources~\cite{rossi_quantum_2013, ghio_multipartite_2018}, and are formally defined as follows. Given a collection of $n$ vertices $V$, we call a $k$-hyper-edge any subset of exactly $k$ vertices. A hypergraph $g_{\leq n} = \{V,E\}$ is then composed of a set $V$ of vertices together with a set $E$ of hyper-edges of any order $k$, not necessarily uniform. Notice that this definition includes the usual notion of a mathematical graph if $k = 2$ for all (hyper)-edges. To any hypergraph $g_{\leq n}$ we associate a $n$-qubit quantum hypergraph state via the definition
\begin{equation}
|g_{\leq n}\rangle = \prod_{k=1}^n \prod_{\{q_{v_1},\dots,q_{v_k}\}\in E} \mathrm{C}^k \mathrm{Z}_{q_{v_1},\dots,q_{v_k}}|+\rangle^{\otimes n}
\label{eq:hyprgraphGeneral}
\end{equation}
where $q_{v_1},\dots,q_{v_k}$ are the qubits connected by a $k$-hyper-edge in $E$ and, with a little abuse of notation, we assume $\mathrm{C}^2 \mathrm{Z}\equiv\mathrm{CZ}$ and $\mathrm{C}^1 \mathrm{Z}\equiv \mathrm{Z} = R_{Z}(\pi)$. For $n$ qubits there are exactly $2^{2^n-1}$ different hypergraph states. We can make use of well known preparation strategies for hypergraph states to realise the unitaries $U_i$ and $U_w$ with at most a single $n$-controlled $\mathrm{C}^n\mathrm{Z}$ and a collection of $p$-controlled $\mathrm{C}^p\mathrm{Z}$ gates with $p<n$. It is worth pointing out already here that such an approach, while optimising the number of multi-qubit logic gates to be employed, implies a circuit depth which scales linearly in the size of the classical input, i.e.\ $\mathcal{O}(d) \equiv \mathcal{O}(2^n)$, in the worst case corresponding to a fully connected hypergraph~\cite{TacchinoQN2019}.

To describe a possible implementation of $U_i$, assume once again that the quantum register of $n$ encoding qubits is initially in the blank state $|0\rangle^{\otimes n}$. By applying parallel Hadamard gates ($\mathrm{H}^{\otimes n}$) we obtain the state $|+\rangle^{\otimes n}$, corresponding to a hypergraph with no edges. We can then use the target collection of classical inputs $\bm{i}$ as a control for the following iterative procedure:

\begin{algorithm*}
\label{alg:hypergraph}
\caption{Quantum hypergraph states generation routine~\cite{TacchinoQN2019}}
\For{$P=1, \,\hdots,\, n$}
{
    \For{$j=0,\, \hdots,\, d-1$}
    {
        \If{$\ket{j}$ has exactly $P$ qubits in $\ket{1}$~\textbf{and}~$i_j = -1$}
        { 
        Apply $\mathrm{C}^P \mathrm{Z}$ to those qubits\;
        Flip the sign of $i_k$ in $\bm{i}$\ \ $\forall k$ such that $\ket{k}$ has the same $P$ qubits in $\ket{1}$\;
        }
    }
}
\end{algorithm*}

Similarly, $U_w$ can be obtained by first performing the routine outlined above (without the initial parallel Hadamard gates) tailored according to the classical control $\bm{w}$: since all the gates involved in the construction are the inverse of themselves and commute with each other, this step produces a unitary transformation bringing $\ket{\psi_w}$ back to $\ket{+}^{\otimes n}$. The desired transformation $U_w$ is completed by adding parallel $H^{\otimes n}$ and $\mathrm{NOT}^{\otimes n}$ gates~\cite{TacchinoQN2019}. 

\section{Variational realisation of a quantum artificial neuron}
Although the implementation of the unitary transformations $U_i$ and $U_w$ outlined above is formally exact and optimises the number of multi-qubit operations to be performed by leveraging on the correlations between the $\pm 1$ phase factors, the overall requirements in terms of circuit depth pose in general severe limitations to their applicability in non error-corrected quantum devices. Moreover, although with such an approach the encoding and manipulation of classical data is performed in an efficient way with respect to memory resources, the computational cost needed to control the execution of the unitary transformations and to actually perform the sequences of quantum logic gates remains bounded by the corresponding classical limits. Therefore, the aim of this section is to explore conditions under which some of the operations introduced in our quantum model of artificial neurons can be obtained in more efficient ways by exploiting the natural capabilities of quantum processors.

In the following, we will mostly concentrate on the task of realising approximate versions of the weight unitary $U_w$ with significantly lower implementation requirements in terms of circuit depth. Although most of the techniques that we will introduce below could in principle work equally well for the preparation of encoding states $\ket{\psi_i}$, it is important to stress already at this stage that such approaches cannot be interpreted as a way of solving the long standing issue represented by the loading of classical data into a quantum register. Instead, they are pursued here as an efficient way of \textit{analysing} classical or quantum data presented in the form of a quantum state. Indeed, the variational approach proposed here requires ad-hoc training for every choice of the target vector $\bm{w}$ whose $U_w$ needs to be realised. To this purpose, we require access to many copies of the desired $\ket{\psi_w}$ state, essentially representing a quantum training set for our artificial neuron. As in our formulation a single node characterized by weight connections $\bm{w}$ can be used as an elementary classifier recognising input data sufficiently close to $\bm{w}$ itself~\cite{TacchinoQN2019}, the variational procedure presented here essentially serves the double purpose of training the classifier upon input of positive examples $\ket{\psi_w}$ and of finding an efficient quantum realisation of such state analyser.

\subsection{Global variational training}
\begin{figure*}[ht]
\centering
\begin{tikzpicture}
\node[inner sep=0pt] (russell) at (0,0) {\includegraphics[width=\textwidth]{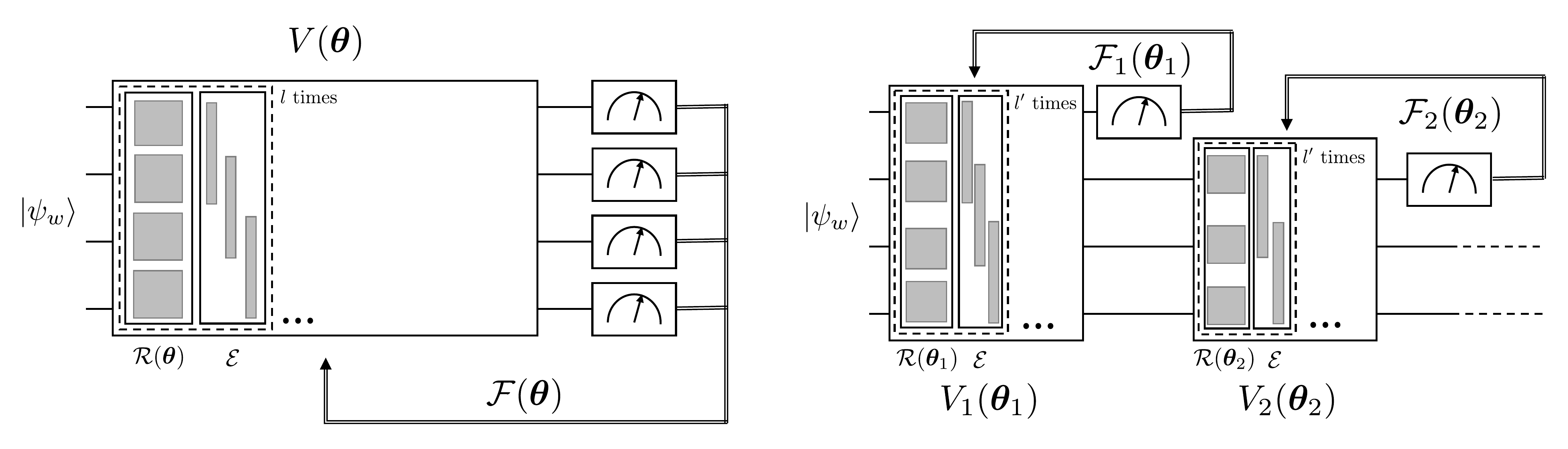}};
\node[inner sep=0pt] (russell) at (-7.3, 1.5) {\normalsize{\textbf{(a)}}};
\node[inner sep=0pt] (russell) at (0.3, 1.5) {\normalsize{\textbf{(b)}}};
\end{tikzpicture}
\caption[Variational learning via unsampling]{Variational learning via unsampling. \textbf{(a)} Global strategy, with optimization targeting  all qubits simultaneously. \textbf{(b)} Local qubit-by-qubit approach, in which each layer is used to optimize the operation for one qubit at a time.}
\label{fig:unsampling}
\end{figure*}

According to Eq.~\eqref{eq:UwConstraint}, the purpose of the transformation $U_w$ within the quantum artificial neuron implementation is essentially to reverse the preparation of a non-trivial quantum state $\ket{\psi_w}$ back to the relatively simple product state $\ket{1}^{\otimes n}$. Notice that in general the qubits in the state $\ket{\psi_w}$ share multipartite entanglement~\cite{ghio_multipartite_2018}. Here we discuss a promising strategy for the efficient approximation of the desired transformation satisfying the necessary constraints based on variational techniques. Inspired by the well known variational quantum eigensolver (VQE) algorithm~\cite{PeruzzoVQE2014}, and in line with a recently introduced unsampling protocol~\cite{carolan_variational_2020}, we define the following optimization problem: given access to independent copies of $\ket{\psi_w}$ and to a variational quantum circuit, characterized by a unitary operation $V(\bmt)$ and parametrized by a set of angles $\bmt$, we wish to find a set of values $\bmt_\text{opt}$ that guarantees a good approximation of $U_w$. The heuristic circuit implementation typically consists of sequential blocks of single-qubit rotations followed by entangling gates, repeated up to a certain number that guarantees enough freedom for the convergence to the desired unitary~\cite{KandalaHWE_2017}.

Once the solution $V(\bmt_\text{opt})$ is found, which in our setup corresponds to a fully trained artificial neuron, it would then provide a form of quantum advantage in the analysis of arbitrary input states $\ket{\psi_i}$ as long as the circuit depth for the implementation of the variational ansatz is sub-linear in the dimension of the classical data, i.e.\ sub-exponential in the size of the qubit register.
As it is customarily done in near-term VQE applications, the optimization landscape is explored by combining executions of quantum circuits with classical feedback mechanisms for the update of the $\bmt$ angles. 
In the most general scenario, and according to Eq.~\eqref{eq:UwConstraint}, a cost function can be defined as
\begin{equation}
    \mathcal{F}(\bmt) = 1 - |\langle 11\ldots 1| V(\bmt)|\psi_w\rangle|^2\,,
    \label{eq:global_cost_f}
\end{equation}
and the solution $\bmt_\text{opt}$ represented by
\begin{equation}
    \bmt_\text{opt} = \arg \min_{\bmt} \mathcal{F}(\bmt)
\end{equation}
which leads to $V(\bmt_\text{opt}) \simeq U_w$. We call this approach a \textit{global} variational unsampling as the cost function in Eq.~\eqref{eq:global_cost_f} requires all qubits to be simultaneously found as close as possible to their respective target state $\ket{1}$, without making explicit use of the product structure of the desired output state $\ket{1}^{\otimes n}$. It is indeed well known that VQE can lead in general to exponentially difficult optimization problems~\cite{McCleanBarren2018}, however the characteristic feature of the problem under evaluation may actually allow for a less complex implementation of the VQE for unsampling purposes~\cite{carolan_variational_2020}, as outlined in the following section.
A schematic representation of the global variational training is provided in Fig.~\ref{fig:unsampling}a. 

\subsection{Local variational training}
An alternative approach to the global unsampling task, particularly suited for the case we are considering in which the desired final state of the quantum register is fully unentangled, makes use of a local, qubit-by-qubit procedure. This technique, which was recently proposed and tested on a photonic platform as a route towards efficient certification of quantum processors~\cite{carolan_variational_2020}, is highlighted here as an additional useful tool within a general quantum machine learning setting.

In the local variational unsampling scheme, the global transformation $V(\bmt)$ is divided into successive layers $V_j(\bmt_j)$ of decreasing complexity and size. Each layer is trained separately, in a serial fashion, according to a cost function which only involves the fidelity of a single qubit to its desired final state. More explicitly, every $V_j(\bmt_j)$ operates on qubits ${j,\ldots,n}$ and has an associated cost function
\begin{equation}
    \mathcal{F}_j(\bmt_j) = 1 - \mel{1}{\Tr_{j+1,\ldots,n}[\rho_j]}{1}\,,
    \label{eq:local_cost_f}
\end{equation}
where the partial trace leaves only the degrees of freedom associated to the $j$-th qubit and, recursively, we define
\begin{equation}
    \rho_j = \begin{cases}
    V_j(\bmt_j)\rho_{j-1} V^\dagger_j(\bmt_j) \quad & j > 1 \\
    \dyad{\psi_w} \quad & j = 0
    \end{cases}\,.
\end{equation}
At step $j$, it is implicitly assumed that all the parameters $\bmt_{k}$ for $k = 1,\ldots,j-1$ are fixed to the optimal values obtained by the minimisation of the cost functions in the previous steps. Notice that, operationally, the evaluation of the cost function $\mathcal{F}_j$ can be automatically carried out by measuring the $j$-th qubit in the computational basis while ignoring the rest of the quantum register, as shown in Fig.~\ref{fig:unsampling}b.

The benefits of local variational unsampling with respect to the global strategy are mainly associated to the reduced complexity of the optimisation landscape per step. Indeed, the local version always operates on the overlap between single-qubit states, at the relatively modest cost of adding $n-1$ smaller and smaller variational ans\"{a}tze. In the specific problem at study, we thus envision the local approach to become particularly effective, and more advantageous than the global one, in the limit of large enough number of qubits, i.e.\ for the most interesting regime where the size of the quantum register, and therefore of the quantum computation, exceeds the current classical simulation capabilities.

\subsection{Case study: pattern recognition}
\label{sec:all2all}
\begin{figure*}[ht]
\centering
\resizebox{\textwidth}{!}
{
\begin{tikzpicture}
\node[inner sep=0pt] (russell) at (0,0){\includegraphics[width=\textwidth]{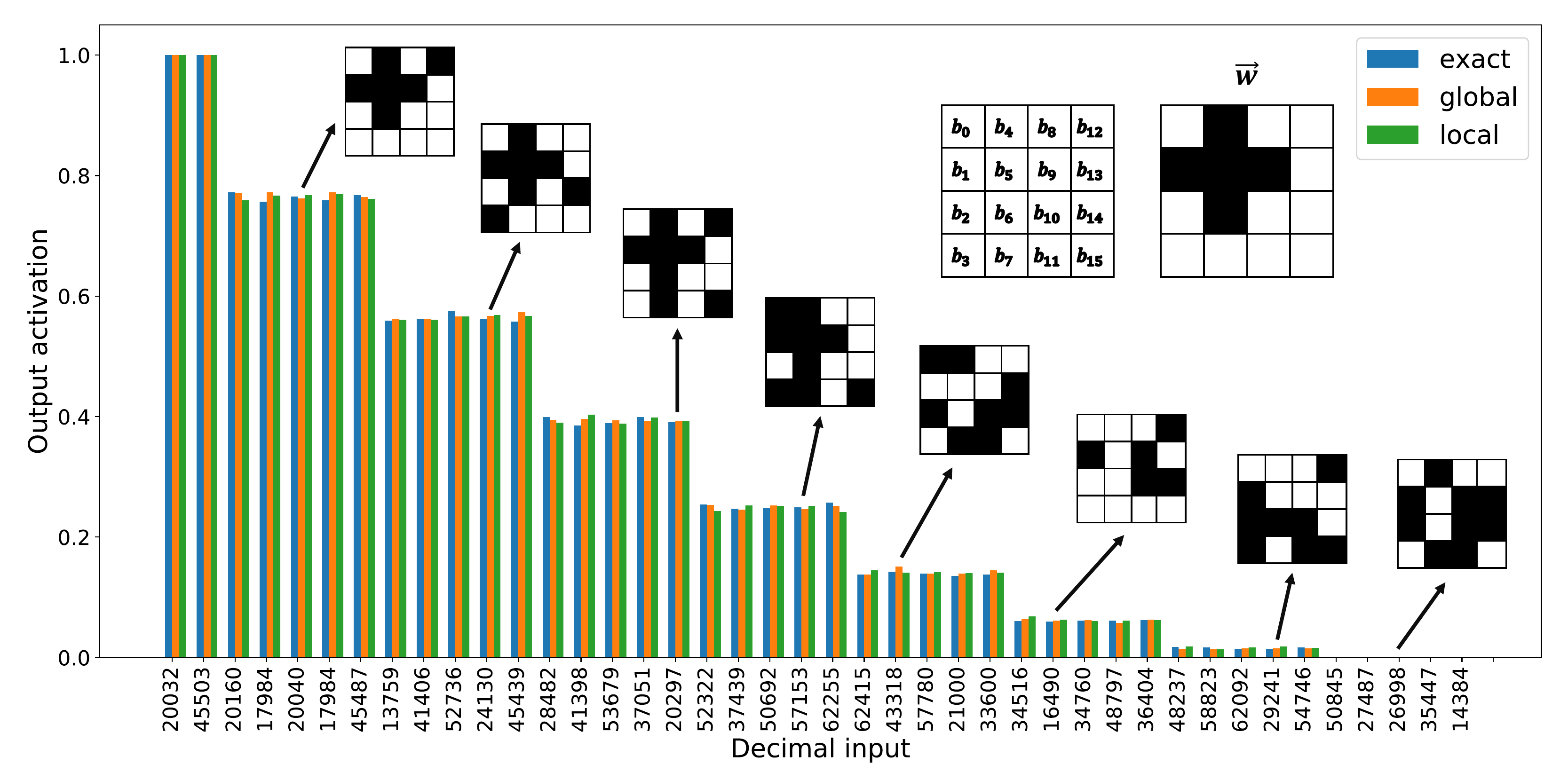}};
\node[inner sep=0pt] (russell) at (-6.25, 3.) {\normalsize{\textbf{(a)}}};
\node[inner sep=0pt] (russell) at (1.25, 3.) {\normalsize{\textbf{(b)}}};
\node[inner sep=0pt] (russell) at (3.5, 3.) {\normalsize{\textbf{(c)}}};
\fill [white] (4, 2.8) rectangle (5, 3.2);
\node[inner sep=0pt] (russell) at (4.15, 3.) {$\bmw$};
\end{tikzpicture}
}
\caption[Comparison of exact to approximate implementations]{Comparison of output activation $p_\text{out} = \abs{\braket{\psi_w}{\psi_i}}^2$ among the exact (hypergraph states routine), global ($n=3$) and local ($n'=2$) approximate implementations of $U_w$. The inset shows the general mapping of any $16$-dimensional binary vector $\vec{b}$ onto the $4\times 4$ binary image (b) and the cross-shaped $\vec{w}$ used in this example (c). The selected inputs on which the approximations are tested were chosen to cover all the possible cases for $p_\text{out}$, and are labelled with their corresponding integer $k_i$ (see main text).}
\label{fig:test_inputs}
\end{figure*}
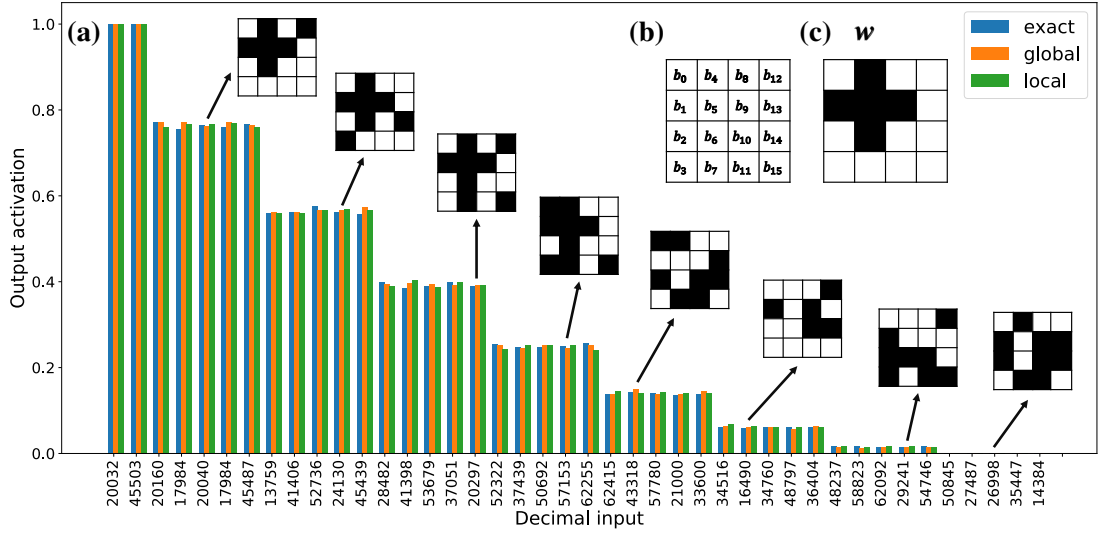

To show an explicit example of the proposed construction, let us fix $m = 16$, $n = 4$. Following Ref.~\cite{TacchinoQN2019}, we can visualise a 16-bit binary vector $\vec{b}$, see Eq.~\eqref{eq:bin_vectors}, as a $4\times 4$ binary pattern of black ($b_j = -1$) and white ($b_j = 1$) pixels. Moreover, we can assign to every possible pattern an integer label $k_b$ corresponding to the conversion of the binary string $\mathtt{k}_b = \mathtt{b}_{0}\ldots \mathtt{b}_{15}$, where $b_j = (-1)^{\mathtt{b}_j}$. We choose as our target $\bmw$ the vector corresponding to $k_w = 20032$, which represents a black cross on white background at the north-west corner of the 16-bit image, see Fig~\ref{fig:test_inputs}.

Starting from the global variational strategy, we construct a parametrized ansatz for $V(\bmt)$ as a series of entangling $\mathcal{E}$ and rotation $\mathcal{R}(\bmt)$ cycles:
\begin{equation}
    V(\bmt) = \qty(\prod_{c=1}^l \mathcal{R}(\theta_{c,1},\ldots,\theta_{c,4})\,\mathcal{E})\mathcal{R}(\theta_{0,1},\ldots,\theta_{0,4})\,,
\end{equation}
where $l$ is the total number of cycles which in principle can be varied to increase the expressibility of the ansatz by increasing its total depth. Rotations are assumed to be acting independently on the $n = 4$ qubits according to
\begin{equation}
    \mathcal{R}(\theta_{c,1}, \ldots, \theta_{c,4}) = \bigotimes_{q=1}^4 R_Y(\theta_{c,q}) = \bigotimes_{q=1}^4 \exp\qty(-i\frac{\theta_{c,q}}{2}Y^{(q)})\,,
\end{equation}
where $Y^{(q)}$ is the Pauli-$Y$ matrix acting on qubit $q$. At the same time, the entangling parts promote all-to-all interactions between the qubits according to\footnote{As discussed later in Chapter~\ref{ch:entanglement}, this all-to-all entangling scheme is actually equivalent to nearest-neighbour linear chain of CNOTs in reversed order. Thus, differences in performances between this entangling strategy $\mathcal{E}$ in Eq.~\eqref{eq:a2a_entangler} and the nearest-neighbour one $\mathcal{E}_{nn}$ in Eq.~\eqref{eq:nn_entangler} discussed in this chapter are likely to be attributed to the former one being more suited to the specific tasks analysed in this work, rather then to the creation of ``more" entanglement, as one would reasonably expect. Although the creation of nontrivial entanglement is necessary to avoid classical simulability (see Ch.~\ref{ch:entanglement}), these results confirm the importance of choosing problem-inspired variational ansätze for ensuring good performances.}
\begin{equation}
    \mathcal{E} = \prod_{q}\prod_{q' = q+1}^4 \mathrm{CNOT}_{qq'}\,,
    \label{eq:a2a_entangler}
\end{equation}
where $\mathrm{CNOT}_{qq'}$ is the usual controlled $\mathrm{NOT}$ operation between control qubit $q$ and target $q'$ acting on the space of all 4-qubits. For $l$ cycles, the total number of $\theta$-parameters, including the initial rotation layer $\mathcal{R}(\theta_{0,1}, \ldots, \theta_{0,4})$, is therefore $4 + 4l$. 

A qubit-by-qubit version of the ansatz can be constructed in a similar way by using the same structure of entangling and rotation cycles, decreasing the total number of qubits by one after each layer of the optimization. Here we choose a uniform number $l'$ of cycles per qubit (this condition will be relaxed afterwards, see Sec.~\ref{sec:new_part}), thus setting $\forall j\neq 4$
\begin{equation}
    V_j(\bmt_j) = \qty(\prod_{c=1}^{l'} \mathcal{R}(\theta_{c,j}, \ldots, \theta_{c,4})\mathcal{E})\mathcal{R}(\theta_{0,j}, \ldots. \theta_{0,4})
    \label{eq:local_unitary}
\end{equation}
For $j = 4$, we add a single general single-qubit rotation with three parameters
\begin{equation}
    \label{eq:u3}
    V_4(\alpha,\beta,\gamma) =  \exp \left(-i\frac{(\alpha,\beta,\gamma)\cdot\bm{\sigma}^{(4)}}{2}\right)
\end{equation}
where $\bm{\sigma} = (X, Y, Z)$ are again the usual Pauli matrices.

We implemented both versions of the variational training in Qiskit~\cite{Qiskit}, combining exact simulation of the quantum circuits required to evaluate the cost function with classical Nelder-Mead~\cite{NelderMead} and Cobyla~\cite{Cobyla} optimizers from the \texttt{scipy} Python library. We find that the values $l = 3$ and $l'=2$ allow the routine to reach total fidelities to the target state $|1\rangle^{\otimes n}$ well above $99.99\%$. As shown in Fig~\ref{fig:test_inputs}, this in turn guarantees a correct reproduction of the exact activation probabilities of the quantum artificial neuron with a quantum circuit depth of $19$ $(29)$ for the global (qubit-by-qubit) strategy, as compared to the total depth equal to $49$ for the exact implementation of $U_w$ using hypergraph states. This counting does not include the gate operations required to prepare the input state, i.e.\ it only evidences the different realisations of the $U_w$ implementation assuming that each $\ket{\psi_i}$ is provided already in the form of a wavefunction. Moreover, the multi-controlled $\mathrm{C}^P\mathrm{Z}$ operations appearing in the exact version were decomposed into single-qubit rotations and $\mathrm{CNOT}$s without the use of additional working qubits. Notice that these conditions are the ones usually met in real near-term superconducting hardware endowed with a fixed set of universal operations.

\subsection{Structure of the ansatz and scaling properties}
\label{sec:new_part}
In many practical applications, the implementation of the entangling block $\mathcal{E}$ could prove technically challenging, in particular for near term quantum devices based, e.g., on superconducting wiring technology, for which the available connectivity between qubits is limited. For this reason, it is useful to consider a more hardware-friendly entangling scheme, which we refer to as \textit{nearest neighbours}. In this case, each qubit is entangled only with at most two other qubits, essentially assuming the topology of a linear chain
\begin{equation}
    \mathcal{E}_{\text{nn}} = \prod_{q=1}^3 \text{CNOT}_{q,q+1}
    \label{eq:nn_entangler}
\end{equation}
This scheme may require even fewer two-qubit gates to be implemented with respect to the all-to-all scheme presented above. Moreover, this entangling unitary fits perfectly well on those quantum processors consisting of linear chains of qubits or heavy hexagonal layouts.

We implemented both global and local variational learning procedures with nearest neighbours entanglers in Qiskit~\cite{Qiskit}, using exact simulation of the quantum circuits with classical optimizers to drive the learning procedure. In the following, we report an extensive analysis of the performances and a comparison with the all-to-all strategy introduced in Sec.~\ref{sec:all2all} above. All the simulations are performed by assuming the same cross-shaped target weight vector $\bmw$ depicted in Fig.~\ref{fig:test_inputs}.

\begin{figure}[ht]
    \centering
    \includegraphics[width=0.75\textwidth]{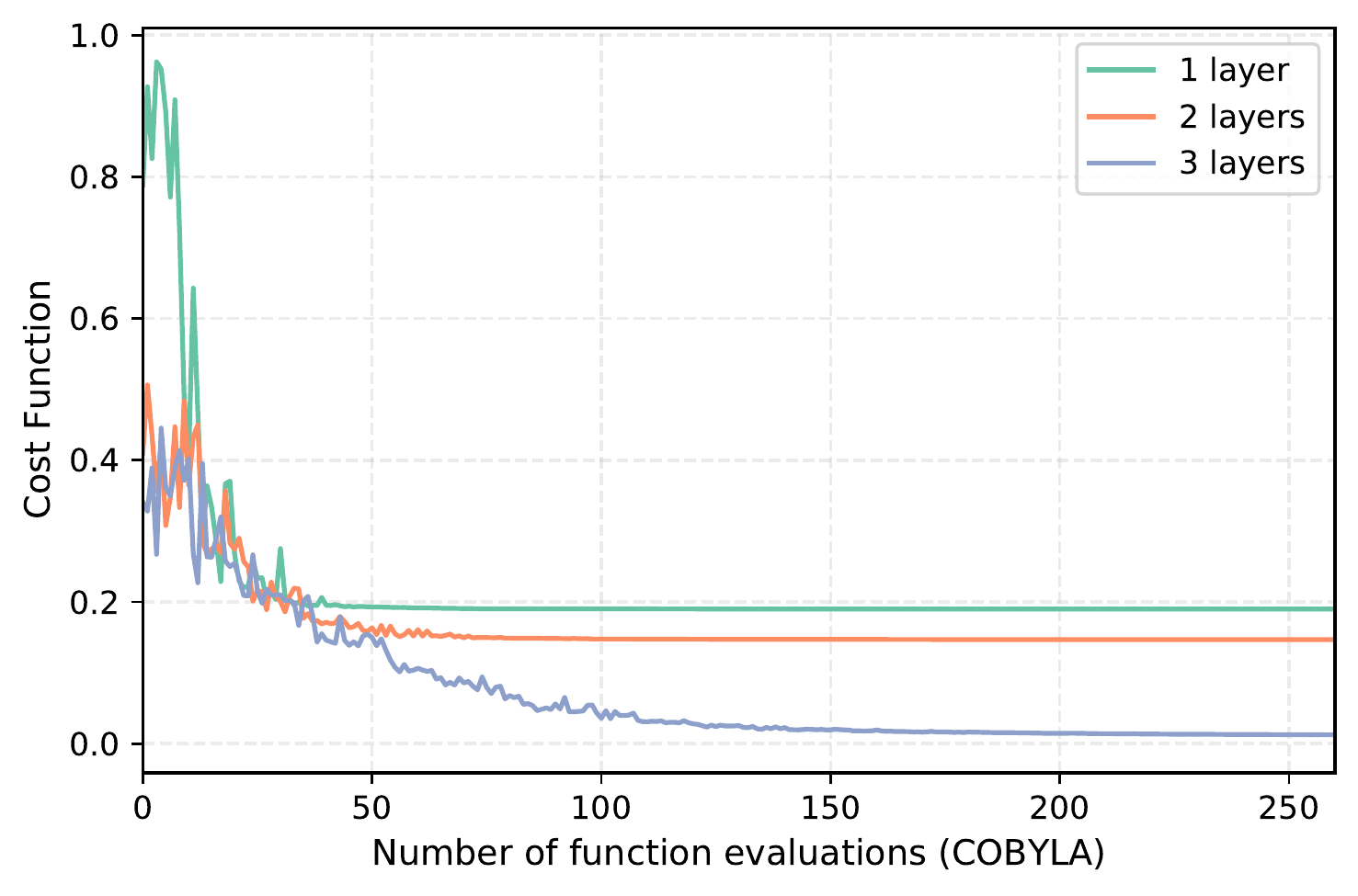}\hfill   
    \caption[Optimisation of the global unitary with nearest neighbours entanglement]{Optimisation of the global unitary with nearest neighbours entanglement for three different structures differing in the numbers of entangling blocks $l$. The cost function is $\abs{\mel{11\ldots 1}{V(\bmt)}{\psi_w}}^2 = 1 - \mathcal{F}(\bmt)$, see Eq.~\eqref{eq:global_cost_f}. Only for $l=3$ the learning model has enough expressibility to reach a good final fidelity. The classical optimiser used in this case was COBYLA~\cite{Cobyla}.}
    \label{fig:global_free}
\end{figure}

In Figure~\ref{fig:global_free} we show an example of the typical optimization procedure for three different choices of the ansatz depth (i.e.\ number of entangling cycles) $l=1,2,3$, assuming a global cost function. Here we find that $l=3$ allows the routine to reach a fidelity $\mathcal{F}(\bmt)$ to the target state $\ket{1}^{\otimes n}$ above $99\%$. 

In the local qubit-by-qubit variational scheme, we can actually introduce an  additional degree of freedom by allowing the number of cycles per qubit, $l'$, to vary between successive layers corresponding to the different stages of the optimization procedure. For example, we may want to use a deeper ansatz for the first unitary acting on all the qubits, and shallower ones for smaller subsystems. We thus introduce a different $l'_j$ for each $V_j(\bmt_j)$ in Eq.~\eqref{eq:local_unitary} and we name \textit{structure} the string `$l_1 l_2 l_3$'. The latter denotes a learning model consisting of three optimization layers: $V_1(\bmt_1)$ with  $l_1$ entangling cycles, $V_2(\bmt_2)$ with $l_2$ and $V_3(\bmt_3)$ with $l_3$, respectively. In the last step of the local optimization procedure, i.e.\ when a single qubit is involved, we always assume a single 3-parameter rotation, see Eq.~\eqref{eq:u3}. A similar notation will be also applied in the following when scaling up to $n>4$ qubits.

\begin{figure}[ht]
    \centering
    \subfloat[\label{fig:nn_a2a_entangler}]{\includegraphics[width=0.5\textwidth]{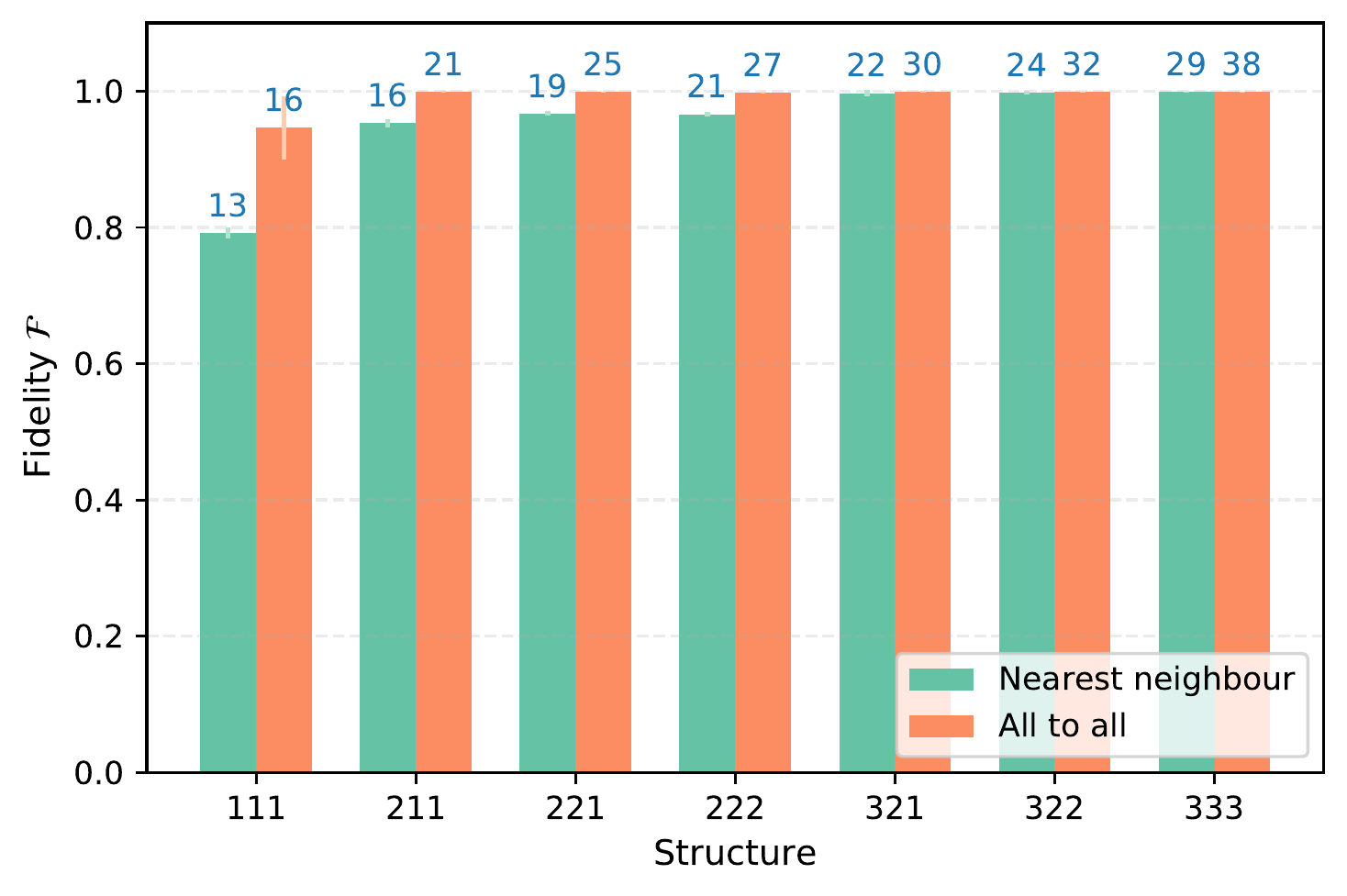}}
    \subfloat[\label{fig:structure_5}]{\includegraphics[width=0.5\textwidth]{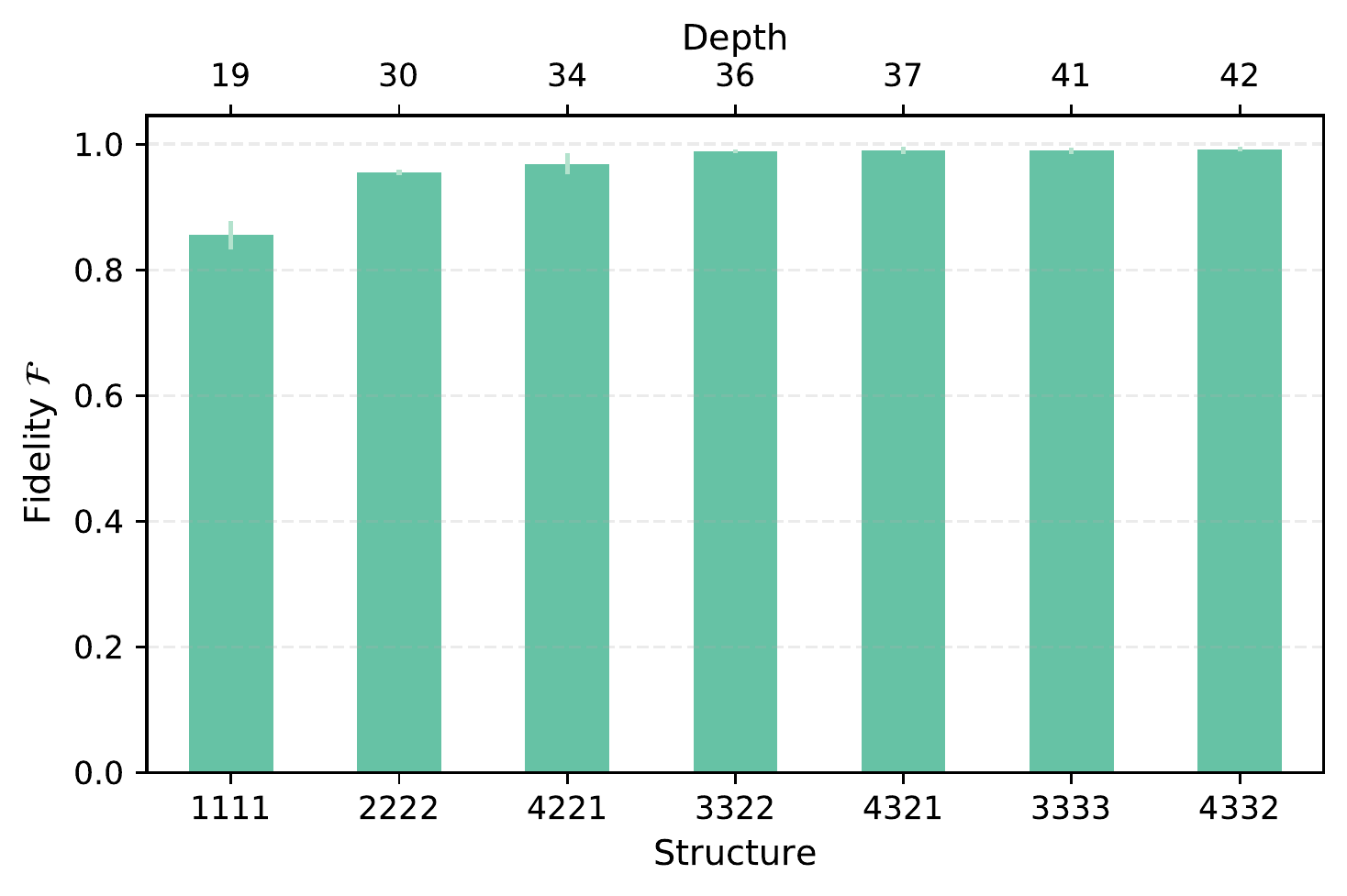}}
    \caption[Final fidelity for different structures and number of qubits]{Final fidelity for different structures and number of qubits. (a) Final fidelity obtained for the local variational training and using both the all-to-all entangler $\mathcal{E}$~\eqref{eq:a2a_entangler} and nearest neighbour $\mathcal{E}_{\text{nn}}$~\eqref{eq:nn_entangler}. On top of each rectangle, in light blue, it is reported the depth of the corresponding quantum circuit to implement that given structure with that particular entangling scheme. For clarity, a structure `$211$' corresponds  to a variational model having two repetitions ($l'_1 = 2$) for the first layer acting on all 4 qubits, and 1 cycle ($l'_1 = l'_2 = 1$) for the remaining two layers acting on $3$ and $2$ qubits respectively. Each bar was obtained executing the optimization process 10 times, and then evaluating the means and standard deviations (shown as error bars). The optimization procedure was performed using COBYLA~\cite{Cobyla}. (b) Final fidelities for different structures of the local variational learning model with a nearest neighbour entangler, for the case of $n=5$ qubits. Similarly to the case with $n=4$ qubits portrayed in Figure~\ref{fig:nn_a2a_entangler}, the most depth-efficient structure is the one consisting of constantly decreasing number of cycles.}
\end{figure}

The effectiveness of different structures is explored in Figure~\ref{fig:nn_a2a_entangler}. We see that, while the all-to-all entangling scheme typically performs better in comparison to the nearest neighbour one, this increase in performance comes at the cost of deeper circuits. Moreover, the stepwise decreasing structure `$321$' for the nearest neighbour entangler proves to be an effective solution to problem, achieving a good final accuracy (above $99\%$) with a low circuit depth. This trend is also confirmed for the higher dimensional case of $n=5$ qubits, which we report in Fig.~\ref{fig:structure_5}. Here, the dimension of the underlying pattern recognition task is increased by extending the original 16-bit weight vector $\bmw$ with extra 0s in front of the binary representation $\mathtt{k}_w$. In fact, it can easily be seen that, assuming directly nearest neighbours entangling blocks, the decreasing structure `$4321$'  gives the best performance-depth trade-off.

Such empirical fact, namely that the most efficient structure is typically the one consisting of decreasing depths, can be heuristically interpreted by recalling again that, in general, the optimization of a function depending on the state of a large number of qubits is a hard training problem~\cite{McCleanBarren2018}. Although we employ local cost functions, to complete our particular task each variational layer needs to successfully disentangle a single qubit from all the others still present in the register. It is therefore not totally surprising that the optimization carried out in larger subsystems requires more repetitions and parameters (i.e.\ larger $n'_j$) in order to make the ansatz more expressive.

\begin{figure}[ht]
    \centering
    \includegraphics[width=0.75\textwidth]{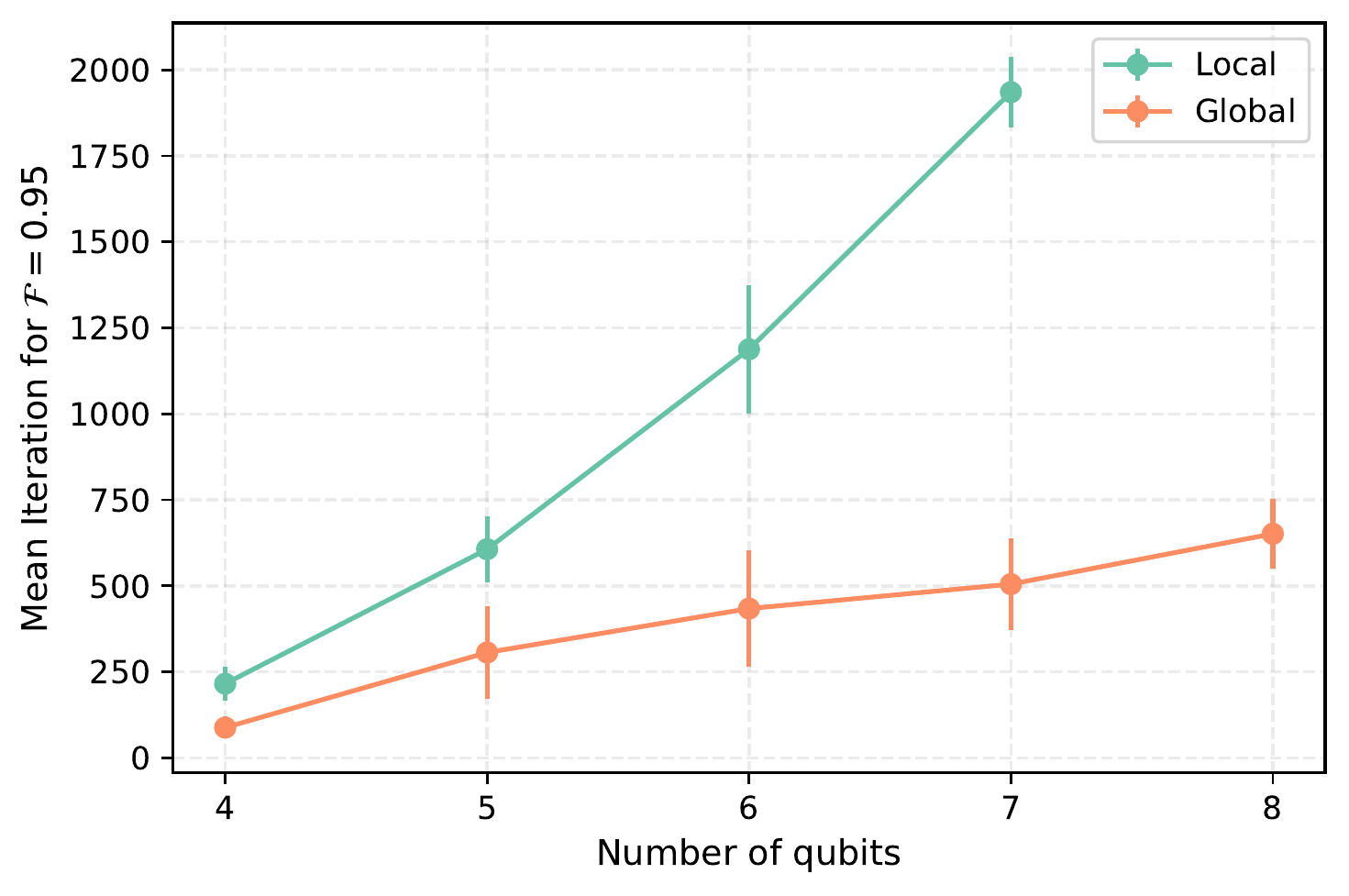}
    \caption[Number of iterations to reach target fidelity]{Number of iterations of the classical optimiser to reach a fidelity of $\mathcal{F}=95\%$. Each point in the plot is obtained by running the optimization procedure 10 times and then evaluating the mean and standard deviation (shown as error bars in the plot). All results refer to exact simulations of the quantum circuits in the absence of statistical measurement sampling or device noise, performed with Qiskit \texttt{statevector{\_}simulator}.}
    \label{fig:scalings}
\end{figure}

By assuming that the stepwise decreasing structure remains sufficiently good also for larger numbers of qubits, we studied the optimization landscape of global~\eqref{eq:global_cost_f} and local~\eqref{eq:local_cost_f} cost functions by investigating how the hardness of the training procedure scales with increasing $n$. As commented above for $n = 5$ qubits, we keep the same underlying target $\bmw$, which we expand by appending extra $0$s in the binary representation. To account for the stochastic nature of the optimization procedure, we run many simulations of the same learning task and report the mean number of iterations needed for the classical optimiser to reach a given target fidelity $\mathcal{F} = 95\%$, and we report simulation results in Figure~\ref{fig:scalings}. 

The most significant message is that the use of the aforementioned local cost function seems to require higher classical resources to reach a given target fidelity when the number of qubits increases. This actually should not come as a surprise, since the number of parameters to be optimised in the two cases is different. In fact, in the global scenarios there are $n+n\cdot l$ (the first $n$ is due to the initial layer of rotations) parameters to be optimised, while in the local case there are $n+n\cdot l'_1$ for the first layer, $(n-1) + (n-1)\cdot l'_2$ for the second and so on, for a total of 
\begin{equation}
    \#_{\text{local}} = \sum_{q=2}^n q + q\, l'_q + 3\,,
\end{equation}
where the final $3$ is due to the fact that the last layer always consist of a rotation on the Bloch sphere with three parameters, see Eq.~\eqref{eq:u3}. Using the stepwise decreasing structure, that is $n'_q=q-1$, we eventually obtain $\sum_{q=2}^n q + q(q-1) = \sum_{q=2}^n q^2  \sim O(n^3)$, compared to $\#_{\text{global}} \sim O(n^2)$. Here we are assuming a number of layers $l = n - 1$, consistently with the $n=4$ qubits case (see Figure~\ref{fig:global_free}). While in the global case the optimization makes full use of the available parameters to globally optimize the state towards $|1\rangle^{\otimes n}$, the local unitary has to go through multiple disentangling stages, requiring (at least for the cases presented here) more classical iteration steps. At the same time, it would probably be interesting to investigate other examples in which the number of parameters between the two alternative schemes remains fixed, as this would most likely narrow the differences and provide a more direct comparison. 

In agreement with similar investigations~\cite{SkolikLayerwise2021}, we can actually conclude that only modest differences between global and local layer-wise optimization approaches are present when dealing with exact simulations (i.e.\ free from statistical and hardware noise) of the quantum circuit. Indeed, both strategies achieve good results and a final fidelity $\mathcal{F}({\bmt})>99\%$. At the same time, it becomes interesting to investigate how the different approaches behave in the presence of noise, and specifically statistical noise coming from measurements operations. For this reason, we implemented the measurement sampling using Qiskit \texttt{qasm{\_}simulator} and employed a stochastic gradient descent (SPSA) classical optimization method. Each benchmark circuit is executed $n_\text{shots}= 1024$ times in order to reconstruct the statistics of the outcomes. Moreover, we repeat the stochastic optimization routine multiple times to analyse the average behaviour of the cost function.
\begin{figure}[ht]
    \centering
    \begin{tikzpicture}
    \node[inner sep=0pt] (russell) at (0,0){\includegraphics[width=0.75\textwidth]{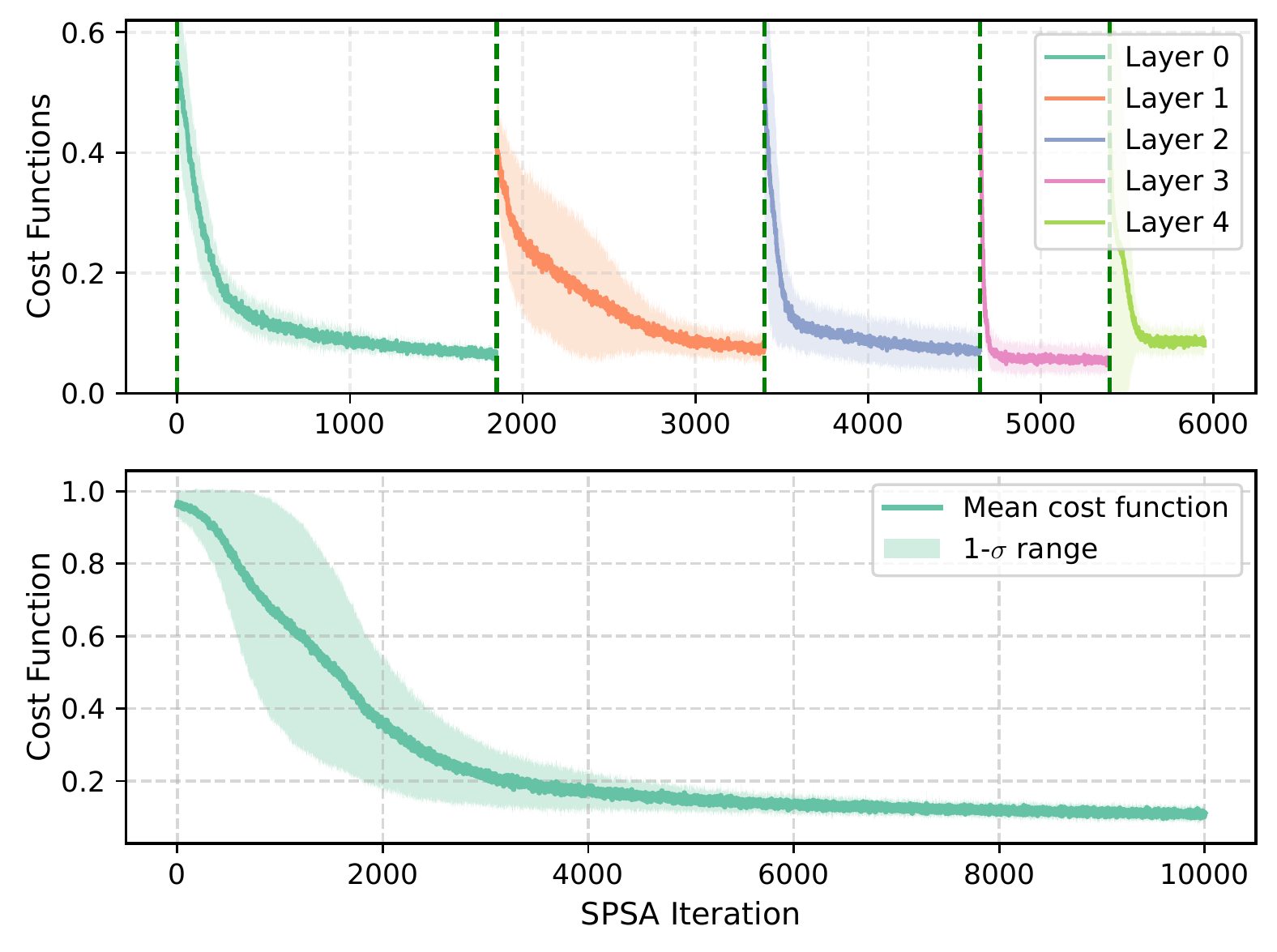}};
    \node[inner sep=0pt] (russell) at (0.1, 3.7) {\normalsize{(a)}};
    \node[inner sep=0pt] (russell) at (0.1,-0.25) {\normalsize{(b)}};
    \end{tikzpicture}
    \caption[Optimisation in presence of stochastic measurement outcomes]{Optimisation of cost functions for the local \textbf{(a)} and global \textbf{(b)} case in the presence of measurement noise for $n=5$ qubits. In each figure we plot the mean values averaged on 5 runs of the simulation. The shaded coloured areas denote one standard deviation. The number of measurement repetitions in each simulation was $n_\text{shots}=1024$. The final fidelity at the end of the training procedure in this case were $\mathcal{F_\text{local}} = 0.87 \pm 0.02$ and $\mathcal{F}_\text{global} = 0.89 \pm 0.02 $. Notice the difference in the horizontal axes bounds. \textbf{(a)} Optimisation of the local cost functions $V_j(\bmt_j)$ (see Eq.~\eqref{eq:local_cost_f}), plotted with different colours for clarity. The vertical dashed lines denotes the end of the optimization of one layer, and the start of the optimization for the following one. \textbf{(b)} Optimisation of the global cost function $V(\bmt)$ in Eq.~\eqref{eq:global_cost_f}}. 
    \label{fig:qasm_noise}
\end{figure}

In Figure~\ref{fig:qasm_noise} we show the optimization procedure for the local and global cost functions in the presence of measurement noise, with both of them reaching acceptable and identical final fidelities $\mathcal{F_\text{local}} = 0.87 \pm 0.02$ and $\mathcal{F}_\text{global} = 0.89 \pm 0.02 $. Notice that for the local case (Figure~\ref{fig:qasm_noise}a) each coloured line indicates the optimization of a $V_j(\bmt_j)$ from Eq.~\eqref{eq:local_cost_f}. We observe that the training for the local model generally requires fewer iterations, with an effective optimization of each single layer. On the contrary, in the presence of measurement noise the global variational training struggles to find a good direction for the optimization and eventually follows a slowly decreasing path to the minimum. These findings look to be in agreement, e.g., with results from Refs.~\cite{SkolikLayerwise2021, CerezoBarrenLocalCost2021}: with the introduction of statistical shot noise, the performances of the global model are heavily affected, while the local approach proves to be more resilient and capable of finding a good gradient direction in the parameters space~\cite{CerezoBarrenLocalCost2021}. In all these simulations, the parameters in the global unitary and in the first layer of the local unitary were initialised with a random distribution in $[0, 2\pi)$. All subsequent layers in the local model were initialised with all parameters set to zero in order to allow for smooth transitions from one optimization layer to the following. This strategy was actually suggested as a possible way to mitigate the occurrence Barren plateaus~\cite{SkolikLayerwise2021, Grant2019Initialization}. 

\begin{figure}[ht]
\centering
\includegraphics[width=0.75\textwidth]{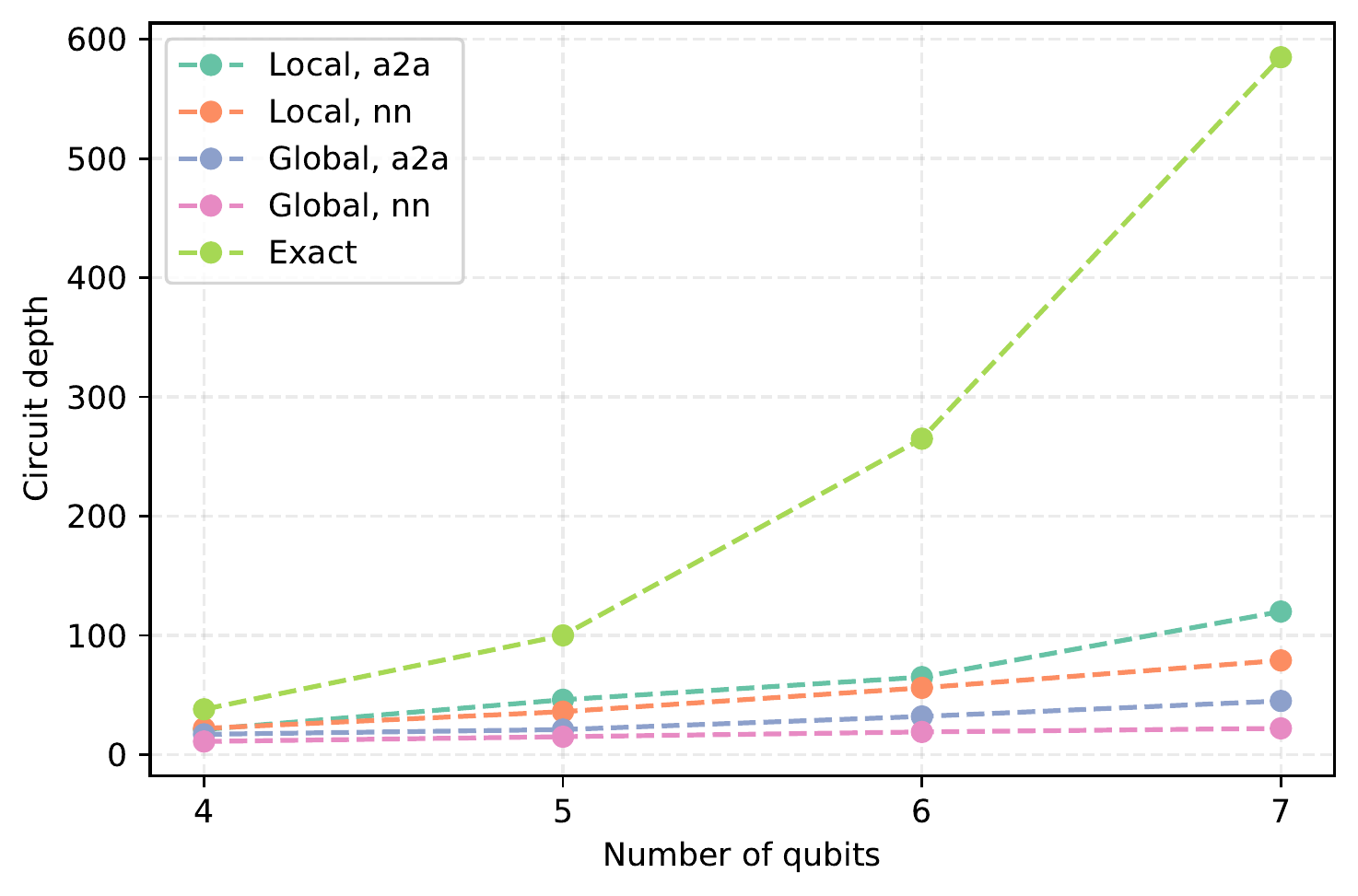}
\caption[Scaling of circuit depth]{Scaling of circuit depth for the implementation of $U_w$ computed with Qiskit. The labels \textit{locals} and \textit{global} refer to the local and global variational approaches, while \textit{a2a} and \textit{nn} refer to the all-to-all and nearest-neighbour entangling schemes respectively. The number of ansatz cycles used for both the global ($l$) and local/qubit-by-qubit ($l'$) variational constructions and for each entangling structure are increased with the number of qubits up to the minimum value guaranteeing a fidelity of the approximations above $98\%$.}
\label{fig:scaling_depth}
\end{figure}

We conclude the scaling analysis by reporting in Fig.~\ref{fig:scaling_depth} a summary of the quantum circuit depths required to implement the target unitary transformation with different strategies and for increasing sizes of the qubit register up to $n = 7$. As it can be seen, all the variational approaches scale much better when compared to the exact implementation of the target $U_w$, with the global ones requiring shallower depths in the specific case. In addition, we recall that the use of an all-to-all entangling scheme requires longer circuits due to the implementation of all the CNOTs, but generally needs less ansatz cycles (see Figure~\ref{fig:nn_a2a_entangler}). At last, while the global procedures seem to provide a better alternative compared to local ones in terms of circuit depth, they might be more prone to suffering from classical optimization issues~\cite{SkolikLayerwise2021, McCleanBarren2018} when trained and executed on real hardware, as suggested by the data reported in Fig.~\ref{fig:qasm_noise}. The overall promising results confirm the significant advantage brought by variational strategies compared to the exponential increase of complexity required by the exact formulation of the algorithm.

\section{Conclusions}
In this chapter we reviewed an exact model for the implementation of artificial neurons on a quantum processor and we introduced variational training methods for efficiently handling the manipulation of classical and quantum input data. Through extensive numerical analysis, we compared the effectiveness of different circuit structures and learning strategies, highlighting potential benefits brought by hardware-compatible entangling operations and by layerwise training routines. Our analysis suggests that quantum unsampling techniques represent a useful resource, upon input of quantum training sets, to be integrated in quantum machine learning applications. 

From a theoretical perspective, our proposed procedure allows for an explicit and direct quantification of possible quantum computational advantages for classification tasks. It is also worth pointing out that such a scheme remains fully compatible with recently introduced architectures for quantum feed-forward neural networks~\cite{TacchinoQN2019}, which are needed in general to deploy e.g.\ complex convolutional filters. Moreover, although the interpretation of quantum hypergraph states as memory-efficient carriers of classical information guarantees an optimal use of the available dimension of a $n$-qubit Hilbert space, the variational techniques introduced here can in principle be used to learn different encoding schemes designed, e.g., to include continuous-valued features or to improve the separability of the data to be classified~\cite{buhrman_quantum_2001, Havlicek2019QSVM, Schuld2019FeatureSpace}. 

In all envisioned applications, our proposed protocols are intended as an effective method for the analysis of quantum states as provided, e.g., by external devices or sensors, while it is worth stressing that the general problem of efficiently loading classical data into quantum registers still stands open. Finally, on a more practical level, a successful implementation on near-term quantum hardware of the variational learning algorithm introduced in this work will necessarily rely on a deeper analysis of the impact of realistic noise effects both on the training procedure and on the final optimised circuit. In particular, we anticipate that the reduced circuit depth produced via the proposed method could critically lessen the quality requirements for quantum hardware, eventually leading to meaningful implementation of quantum neural networks within the near-term regime.


\chapterimage{bg6.png} 
\chapterspaceabove{6.75cm} 
\chapterspacebelow{7.25cm} 

\chapter{Quantum autoencoder and classifier for an industrial use case}\index{AutoencoderENI}
\label{ch:Autoencoder}
\startcontents[chapters]
\printcontents[chapters]{}{1}{}
\vspace*{1cm}

Quantum computing technologies are in the process of moving from academic research to real industrial applications, with the first hints of quantum advantage demonstrated in recent months. In these early practical uses of quantum computers it is relevant to develop algorithms that are useful for actual industrial processes. In this chapter\footnote{The content of this chapter is based on the author's work~\cite{ManginiAutoencoder2022}, and all the figures in this chapter are taken from, or are adaptations of, the figures present in such work.}, we propose a quantum pipeline, comprising a quantum autoencoder followed by a quantum classifier, which are used to first compress and then label classical data coming from a separator, i.e., a machine used in one of Eni's Oil Treatment Plants. This study represents one of the first attempts to integrate quantum computing procedures in a real-case scenario of an industrial pipeline, in particular using actual data coming from physical machines, rather than pedagogical data from benchmark datasets.

\section{Introduction}
\label{sec:ch_Autoencoder_Intro} 

In this chapter we test the use of quantum machine learning algorithms on a specific industrial use case. In particular, we propose the application of a newly formulated quantum pipeline comprising a quantum autoencoder algorithm~\cite{RomeroAutoencoder2017, BravoPrietoAutoencoderReuploading2021, LamataAutoencoderAdder2018, KhoshamanVariationalAutoencoder2018} followed by a quantum classifier, applied to real data coming from a first stage water/oil separator of one of Eni’s oil treatment plant. This algorithm is compared to the performance of a classical autoencoder to compress the original data, which are then used to implement a classification task. It is particularly relevant to notice that these quantum autoencoding algorithms can be run on presently existing quantum hardware, thus making such quantum machine learning algorithm readily usable with actual input data coming from a realistic source of industrial interest. While various models of variational autoencoders in the quantum domain have been proposed in the literature, for example for generative modelling tasks~\cite{KhoshamanVariationalAutoencoder2018} and for the study of entanglement in quantum states~\cite{ChenDetectingEntanglement2021}, our implementation of the quantum autoencoder directly follows the architecture proposed by authors in~\cite{RomeroAutoencoder2017}, which is often studied as a prototypical model in the quantum machine learning literature~\cite{CerezoBarrenLocalCost2021}, and it was also even extended to feature input redundancy~\cite{Perez2020Reuploading}, as discussed in~\cite{BravoPrietoAutoencoderReuploading2021}.

The chapter is organised as follows. In Sec.~\ref{sec:case_study} we explain and give the specifics of the industrial case study considered in this work. In Sec.~\ref{sec:autoencoder} we introduce the classical neural network model of the autoencoder, and also discuss the clustering algorithm used to create the two classes for the classification problem. In Sec.~\ref{sec:quantumautoencoder} we review the quantum algorithm developed for a continuously valued input neuron   already discussed in Chapter~\ref{ch:CQN}~\cite{ManginiCQN2020}, from which the quantum algorithm for the quantum autoencoder is derived. In Sec.~\ref{sec:ch_Autoencoder_results} we show the results obtained for the data compression task, comparing them with those obtained with the purely classical autoencoder. At last in Sec.~\ref{sec:Autoencoder_classification}, we use the compressed data to implement a quantum classifier used to label the original data in a binary classification problem.

\section{\label{sec:case_study} Case study}
\begin{figure}[ht]
    \centering
    \includegraphics[width=0.75\textwidth]{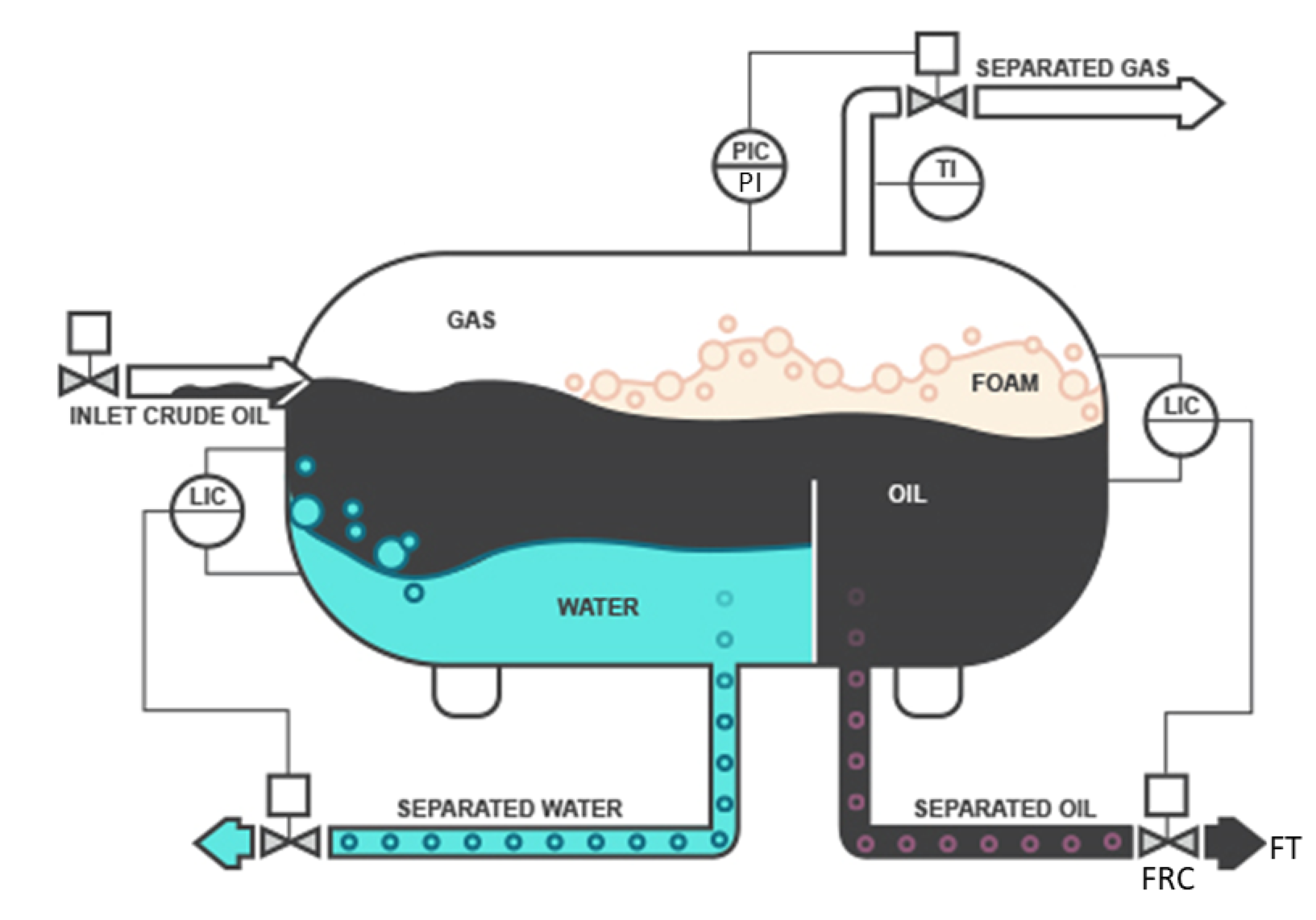}
    \caption[Snapshot of a separator]{Snapshot of the separator. The separator is regulated with three controllers: a pressure controller for the output gas stream, and two-level controllers for the water and the oil stream. The controllers use PID controller equations to regulate the opening of valves on the output streams.}
    \label{fig:separator}
\end{figure}

The industrial case study discussed in this Chapter aims at testing classical and quantum machine learning approaches to analyse data coming from an industrial equipment within one of Eni's Oil Treatment plants, showed in Fig.~\ref{fig:separator}. 
The equipment is a separator, i.e. a vessel receiving a stream of high pressure, high temperature crude oil (left part of the figure, indicated with a black stream), and exploits gravity to separate three output streams: Water (the heaviest component), indicated in the figure with a light blue stream; Oil (intermediate component), in the lower part of the figure indicated with a black stream; and Gas (lightest component), indicated with a light grey stream. The separator is regulated with three controllers: a pressure controller for the output gas stream, and two-level controllers for the water and the oil stream. Notice that the controllers use PID (proportional – integral – derivative) controller equations to regulate the opening of valves on the output streams. 

In a realistic machine learning problem, we might wish to use all the measurements coming from the sensors installed on this component, as well as on some of the components installed upstream, in order to predict if the behavior of the equipment is normal or faulty (i.e. working in a degraded mode). However, due to the limitation in the complexity of the problems that can currently be faced with quantum computing, we will focus on a simplified problem, involving only 4 variables, that are: the oil level (LIC), the oil output flow (FT), the pressure (PI), and the opening of the oil output valve (FRC). Sensor measurements are sampled every $10$ seconds and stored into data tables to be used for the training of the neural networks. 

The first step of the case study is the implementation of a dimensionality reduction procedure to compress the 4-dimensional input vector $\bmx=(x_\text{FRC},\, x_\text{FT},\, x_\text{LIC},\, x_\text{PI}) \in \mathbb{R}^4$ into a 2-dimensional vector. This is done both via a standard classical neural network autoencoder and a quantum autoencoder, introduced in Sec.~\ref{sec:autoencoder} and Sec.~\ref{sec:quantumautoencoder} respectively. 

\begin{figure*}[ht]
   \includegraphics[width=\textwidth]{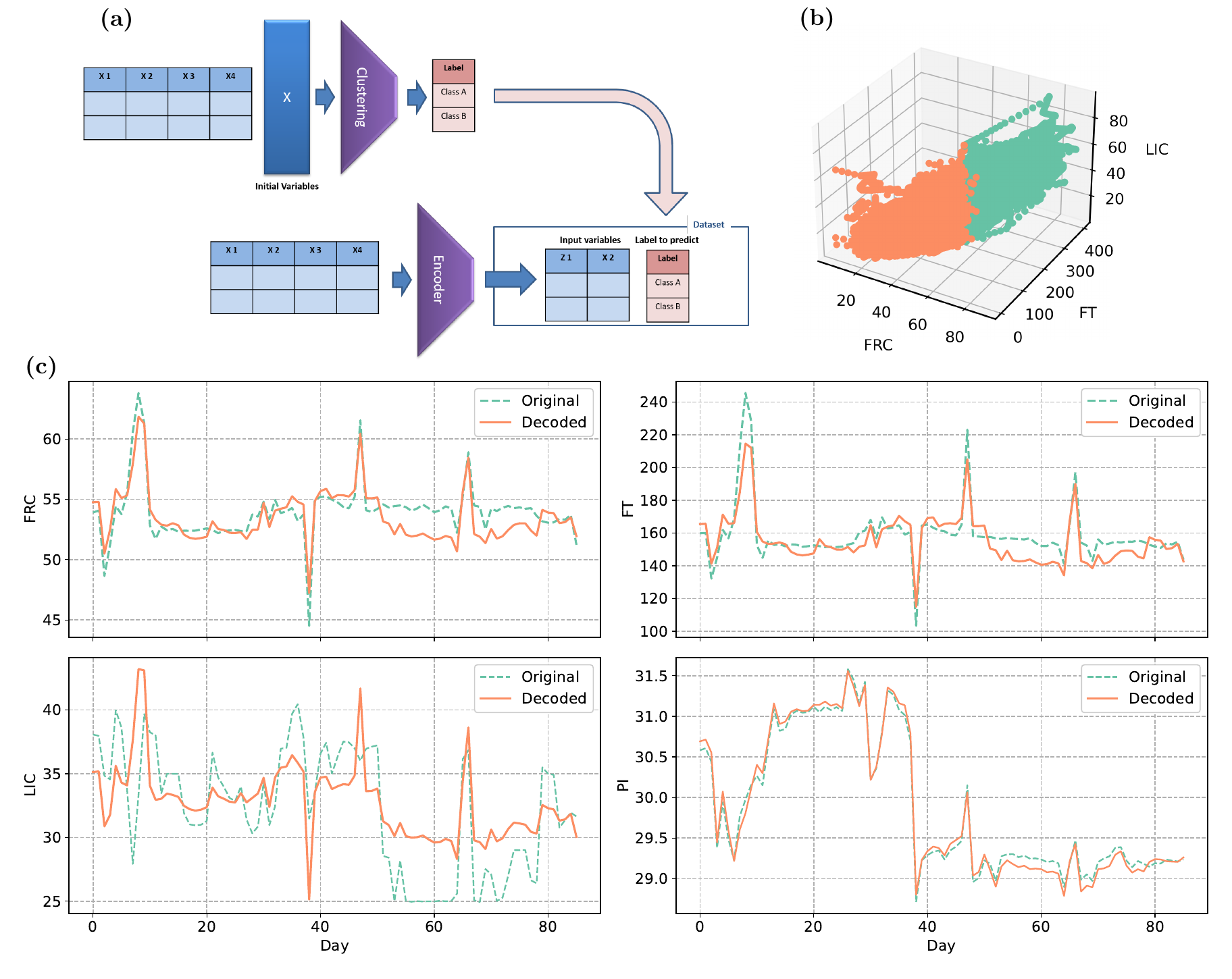}
    \caption[Summary of the approach followed in the work]{\textbf{(a)} The approach followed in this project: a clustering algorithm was used to define two categorical classes (Class A and Class B). Then, an autoencoder was used to reduce the dimensionality of the problem. Finally, a classifier was used to predict Class A and Class B identified with the clustering algorithm. \textbf{(b)} Results of the clustering algorithm KMeans on the input data. In particular, only the features FRC, FT and LIC are shown. The different colour indicates the different label (or class) assigned to the data. \textbf{(c)} Plot of the decoded data on top of the original validation data averaged by day. Here the features were rescaled to their original range.}
    \label{fig:Autoencoder_FIG3}
\end{figure*}

The second step is the implementation a classifier using the $2$-dimensional latent vector from the compression step to classify the status of the component. In order to do so, we need a labelled training dataset associating an input $\bmx_i$ to a label $y_i=\{0,1\}$ corresponding to the ``ok'' or ``faulty'' state respectively. However, since $4$ variables are too few to label the working status of the separator as ``ok'' or ``faulty'', we followed a different approach, as explained in the upper left panel of Fig.~\ref{fig:Autoencoder_FIG3}. We run a binary clustering algorithm on the initial variables, in order to identify two categorical states, named as ``Class A'' and ``Class B'', and then used these categorical states as the labels for the classification task. So, the latent vector from the encoder is used as input for the classifier, that is trained to correctly predict the ``Class A'' and ``Class B'' states. The clustering algorithm used is the KMeans algorithm as implemented in the \texttt{scikit-learn} library~\cite{scikit-learn}. This algorithm takes as input the desired number of clusters, in our case two, and tries to split the data in groups of equal variance. The centroids of the clusters were initialised uniformly at random. In Fig.~\ref{fig:Autoencoder_FIG3} we show the result of the clustering procedure, where for ease of plotting we show only three of the four variables. This categorical dataset is then used to train a classical and quantum classifier, whose implementation details and results are discussed in Sec.~\ref{sec:Autoencoder_classification}.

In Table~\ref{tab:Autoencoder_table_results} we summarise the findings of our work, showing the key figures (compression error and classification accuracy) for the classical and quantum pipelines considered in the case study. 

\begin{table}[ht]
\centering
\caption[Key figures of compression and classification with a quantum autoencoder]{\label{tab:Autoencoder_table_results} Key figures for the compression and classification tasks for the classical and quantum procedures considered. The compression task is implemented with classical and quantum autoencoders; the classification task is implemented with a KNeighborsClassifier and with a single qubit variational classifier. Compression error refers to the average reconstruction error defined in Eq.~\eqref{eq:recontruction_error}. Classification accuracy is defined as the percentage of correctly classified data.}
\begin{tabular}{ccc}
         \toprule
          & Compression error & Classification accuracy\\
          \hline\hline\\
Classical & 5\%         & 89.7\% \\
Quantum   & 5.4\%       & 87.4\% \\
Quantum hardware (\texttt{ibmq\_x2}) & --- & 82.3\% \\
        \bottomrule
\end{tabular}
\end{table}

\section{\label{sec:autoencoder} Neural network autoencoder}
\begin{figure}[ht]
    \centering
   \includegraphics[width=\textwidth]{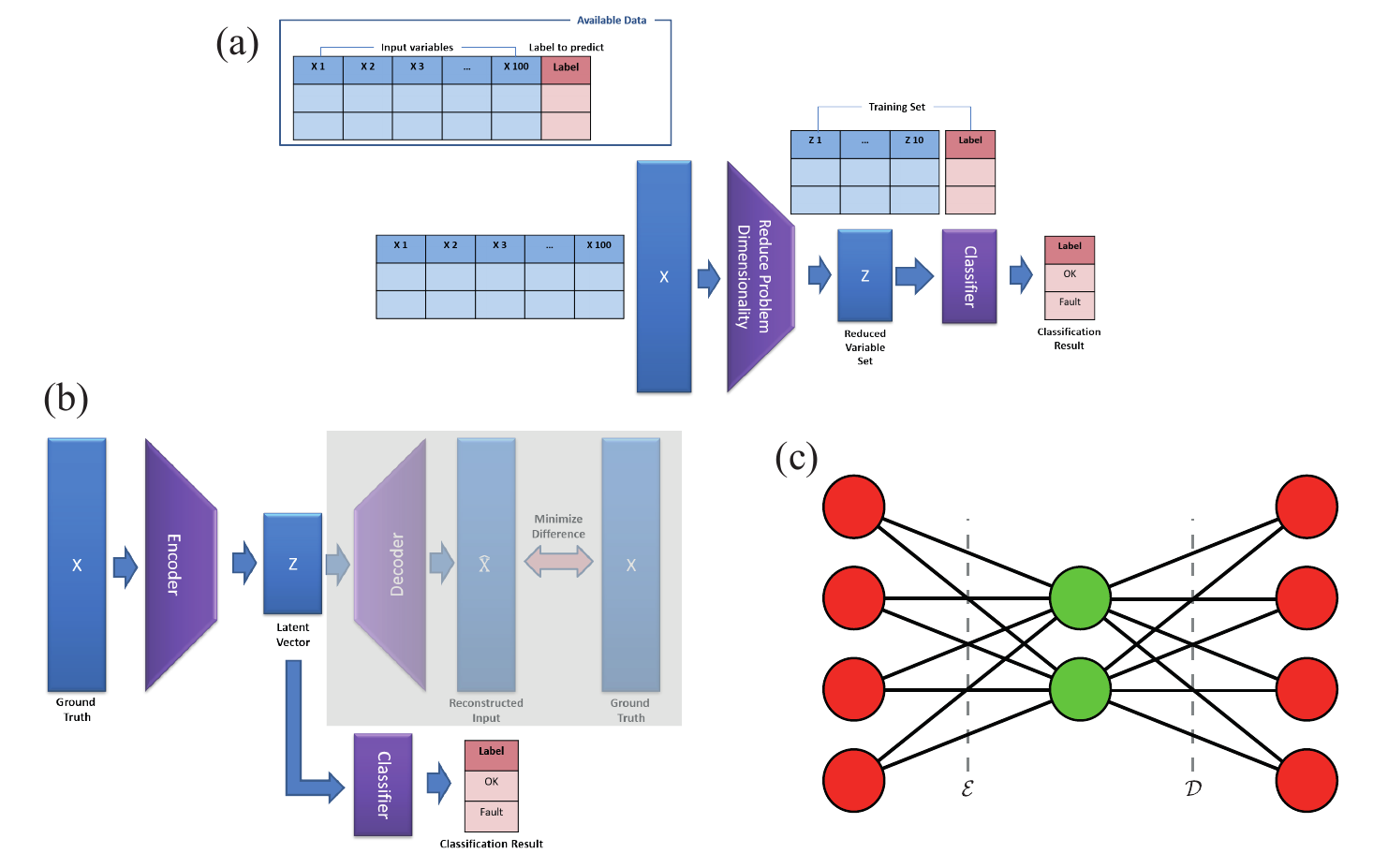}
    \caption[Details on the procedure]{\textbf{(a)} Reducing the dimensionality of a classification problem. \textbf{(b)} Using Autoencoders to reduce the dimensionality of a problem and solving the classification problem on the reduced variable set. \textbf{(c)} Schematic representation of the neural network autoencoder architecture. The input neurons in red are mapped to an hidden layer (in green) of lower dimension, storing the compressed information. Then, an output layer with the same number of neurons as the input one, tries to restore the original data with low error.}
    \label{fig:Autoencoder_FIG1}
\end{figure}

As extensively discussed in Sec~\ref{sec:ch_QML_Classical}, the most common use case of artificial neural networks is supervised learning, where the network is asked to learn a mapping from an input to an output space, by having access to an example set of input-output pairs. Specifically, for classification tasks the network is presented a labelled dataset $S = \qty{(\bmx_i, y_i) \subset \mathbb{R}^d \times \{0, 1 \ldots, c-1\}}_{i=1}^m$ consisting of a set of inputs $\bmx_i$ and the corresponding correct labels $y_j$, with $c$ being the total number of classes the inputs can be divided into (see upper left side of Fig.~\ref{fig:Autoencoder_FIG1}). Using this dataset, called \textit{training set}, a neural network can be trained in a supervised fashion to learn the relationship between the input variables and the expected classification results. When the training is complete, the neural network model can be used for \textit{inference}, that is for labelling previously unseen data. This property of neural networks, called \textit{generalisation}, is ultimately the key figure that distinguishes them from standard fitting techniques, making them incredibly powerful tools~\cite{Goodfellow2016DeepL, LeCun2015DeepL, Hastie2009Statistical, mnih2015human}.  

When dealing with real world problems, such as classifying the operational status of a plant as ``ok'' or ``faulty'' based on the measurements from the sensors installed on the plant, it is often the case  that a large number of input variables are available. In fact, measurements coming from tens of sensors need to be analyzed not only on their instantaneous values, but also on additional features computed on time intervals, such as moving averages, and minimal/maximal values trends. This leads to a situation where too many input variables are available in the dataset, and it is often ineffective to directly feed them into the neural network classifier. With such a large number of variables, correlation analysis and feature engineering are often performed to focus only on the most influencing variables, and only after these preprocessing steps the neural network can be used effectively. Another strategy is to use a \textit{dimensionality reduction} approach, consisting in computing a new set of variables, smaller than the initial one, incorporating most ---ideally all--- of the information contained in the original data. These new compressed data are then used as inputs to the classifier, as shown in Fig.~\ref{fig:Autoencoder_FIG1}a. 

In order to reduce the problem dimensionality, methods such as PCA (Principal Component Analysis) or SVD (Singular Value Decomposition)~\cite{Hastie2009Statistical} are typically used. However, these methods are based on linear decomposition of the initial variable space, and they could not be suitable when nonlinear relationships between the variables need to be kept into account.

\subsection{Classical Autoencoders}
An alternative method to reduce the dimensionality of the problem is to use so-called Autoencoders~\cite{Goodfellow2016DeepL}, as shown in Fig~\ref{fig:Autoencoder_FIG1}c. An autoencoder is a neural network composed of two modules, called \textit{encoder} and \textit{decoder}, designed in such a way that the subsequent application of the encoder and the decoder to the input data results into an output that is as close as possible to the input, i.e. the discrepancy between output and input is minimised. With such an approach, the encoder builds a compressed representation of the input data to be eventually used by the decoder to fully, and as faithfully as possible, reconstruct the input. This means that the compressed representation built by the encoder (often referred to as \textit{latent} vector) contains the same information of the initial input space, or at least that minimum information is lost. Once the autoencoder has been trained to reconstruct the input, the latent vector can be used as the input space for the classifier. Therefore, the classification problem can be described as shown in  Fig.~\ref{fig:Autoencoder_FIG1}b. 

In our case study, we consider a neural autoencoder as shown in Fig.~\ref{fig:Autoencoder_FIG1}c. The original input variables are fed to the input neurons, which are then passed to an intermediate hidden level (shown in green) consisting of a number of neurons much smaller than the input. Finally, there is an output layer (shown in red) with the same number of neurons as the input. The neural network is trained in an unsupervised fashion in order to generate an output that is as close as possible to the input. Thus, if it is possible to reconstruct the input (with a minimum loss of fidelity) starting from the inner layer, this means that the inner layer contains the same information as the input, and therefore we can use the compressed layer as an input for the classifier. The presence of non-linear activation functions within the neural network, such as the Rectified Linear Unit $\text{ReLU}(x) = \max(0, x)$, or sigmoid $s(x) = 1/(1+e^{-x})$ (see Eqs.~\eqref{eq:activation_functions}), ensures that the network can better capture non-linear relationships in the input variables compared to PCA or SVD.

\section{Quantum Data Compression}
\label{sec:quantumautoencoder} 
In order to use a quantum pipeline to analyse the classical data coming from the sensors, we need to encode such data on a quantum state to be used as the input of the quantum autoencoder. As discussed in Sec.~\ref{sec:ch_QML_Reuploading} regarding the Fourier expansion of quantum circuits, recent literature points out the importance of choosing a good encoding scheme, even though no standard procedure is yet available~\cite{AbbasPowerQNN2021, Lloyd2020Embeddings, Schuld2020Encoding, Theis2020Expressivity, LaRose2020Robust, Mitarai2018Learning}. 

Given the relatively simple and low dimensional nature of the data sets to be analyzed, we choose to use the phase encoding strategy introduced earlier when discussing models of quantum perceptrons~\cite{TacchinoQN2019, ManginiCQN2020}. This strategy provides an effective way to load classical data into a quantum state, and also already proved useful in other machine learning tasks such as pattern-recognition~\cite{TacchinoQN2019, ManginiCQN2020, TacchinoQNN2020, TacchinoVariationalQNN2021}. In particular, given a data sample $\bmx = (x_1, x_2, \ldots, x_d) \in \mathbb{R}^N$, this is encoded on the quantum state of $n = \log_2 d$ qubits as follows (see Eq.~\eqref{eq:inputstate})
\begin{equation}
\ket{\psi_{\bmx}} = \frac{1}{2^{n/2}}\sum_{i=0}^{2^n-1}e^{i\, x_i}\ket{i}
\label{eq:phase_encoding}
\end{equation}
where the data $\bmx$ are first re-scaled to fit into an appropriate range, such as $ x_i \in [0, \pi]$. We refer to Chapters~\ref{ch:CQN} and~\ref{ch:VariationalQN} for an extended discussion on this class of states for variational quantum procedures.

\subsection{\label{ssec:qae} Quantum Autoencoder}
Having fixed a data encoding strategy, we now build a variational quantum algorithm for data compression. In particular, borrowing from the classical machine learning literature, our goal is to implement a quantum autoencoder~\cite{RomeroAutoencoder2017, LamataAutoencoderAdder2018, BravoPrietoAutoencoderReuploading2021}. In classical autoencoders, the compression is built in the geometric structure of the neural network, since the input layer is followed by a much smaller hidden layer consisting of a number of neurons equal to the desired reduced dimension. This bottleneck forces the NN to learn a low dimensional representation of the inputs, which is stored in the intermediate hidden layer(s) of the network. However, this procedure cannot be straightforwardly applied to the quantum domain, because quantum computations follow a unitary, thus reversible, evolution. In fact, while classically it is possible to perform {\texttt{fan-in}}({\texttt{fan-out}}) operations, that is arbitrarily reducing (increasing) the number of classical bits in the computation, such operations are irreversible, which prevents their direct implementation on a quantum computer. Alternatively said, it is not possible to eliminate or create new qubits during the execution of a quantum computation. 

Nonetheless, it is possible to circumvent this issue as follows. Consider two quantum systems, denoted as system $A$ and system $B$, and be $\ket{\psi}_{AB}$ the quantum state of the composite quantum system $AB$. Our goal is to compress the information stored in the composite state in a lower dimensional representation, for example given by the state of subsystem $A$ only, with system $B$ being safely discarded. We can formalise this intuition in the following way: denote with $\mathcal{E}(\bm{\theta})$ a quantum encoding (in the sense of \textit{compressing}) operation depending on variational parameters $\bm{\theta}$, then the desired compression task consists in the operation
\begin{equation}
\mathcal{E}(\bm{\theta})\ket{\psi}_{AB}=\ket{\phi}_A \otimes \ket{\rm{trash}}_B \, ,
\label{eq:q_encoder}
\end{equation}
where the state $\ket{\psi}_{AB}$ of the composite system $AB$ is compressed on the state $\ket{\phi}_A$ of subsystem A only, and the system $B$ is mapped to a fixed reference state of choice, called \textit{trash} state, for example being the ground state $\ket{\rm{trash}}_B = \ket{0}^{\otimes |B|}$, with $\abs{B} = \text{dim}(\mathcal{H}_B)$ being the dimension of the Hilbert $\mathcal{H}_B$ space associated to system $B$. 

It is clear that the goal of the encoder is to \textit{disentangle} the two systems in such a way that one of them, the trash system, goes to the fixed reference state, while the other contains all the original information of the full quantum state. In order to recover the original quantum state $\ket{\psi}_{AB}$, it is then possible to act with a \textit{quantum decoder} operation $\mathcal{D}(\bm{\theta})$, defined as $\mathcal{D}(\bm{\theta}) \coloneqq \mathcal{E}(\bm{\theta})^{\dagger}$. Indeed, acting with the decoder on the compressed state yields the original state, namely
\begin{equation}
       \mathcal{D}(\bm{\theta})\qty(\ket{\phi}_A \otimes \ket{\rm{trash}}_B) = \mathcal{D}(\bm{\theta})\, \qty(\mathcal{E}(\bm{\theta}) \ket{\psi}_{AB}) = \qty(\mathcal{E}(\bm{\theta})^\dagger \mathcal{E}(\bm{\theta}))\, \ket{\psi}_{AB} = \ket{\psi}_{AB}\, .
\end{equation}

\begin{figure}[ht]
    \centering
   \includegraphics[width=0.75\textwidth]{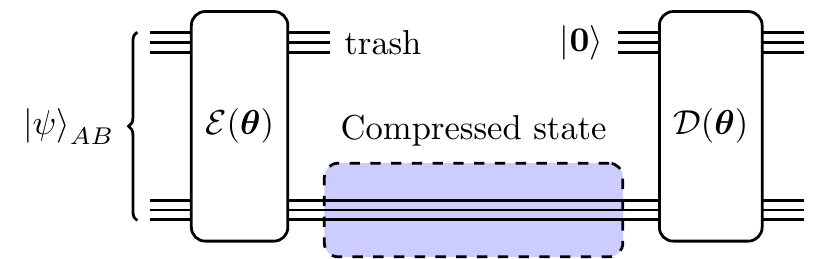}
    \caption[Schematic representation of a quantum autoencoder]{Schematic representation of the generic quantum autoencoder algorithm. The composite input quantum state $\ket{\psi}_{AB}$ is disentangled, so that system $A$ carries the compressed information, and system $B$, called ``trash'' system, is mapped to a reference quantum state of choice like $\ket{\text{trash}}_B = \ket{\bm{0}}_B$, and it can be discarded. Such procedure may then be reversed by coupling the information-carrying system $A$ with a new set of clean qubits, and then applying a joint quantum decoder operation $\mathcal{D}(\bm{\theta}) \coloneqq \mathcal{E}^\dagger(\bm{\theta})$ to retrieve the original state.}
    \label{fig:quantum_autoencoder}
\end{figure}

Thus, suppose having compressed the information stored in the quantum state of a composite system into one of its subsystems. Then, it is always possible to retrieve the original information by coupling such information-carrying system with some new qubits initialised in the $\ket{\rm{trash}}$ state, and then act on them with the quantum decoder operator, as schematically represented in Fig.~\ref{fig:quantum_autoencoder}. 

Of course, this only holds in the ideal case where the encoder perfectly manages to disentangle the subsystems $A$ and $B$, by obtaining the product state in Eq.\eqref{eq:q_encoder}. In practice, this is never the case since the input state $\ket{\psi}_{AB}$ depends on the classical input data via the phase encoding, and these states cannot be exactly disentangled, in general. In fact, after discarding the trash system $B$, the compressed state $A$ is no more a pure state, rather a mixed state given by the density matrix $\rho_A = \Tr_B [(\mathcal{E}(\bm{\theta})\ket{\psi_{AB}}) (\bra{\psi_{AB}})\mathcal{E}(\bm{\theta})^\dagger)]$. However, upon optimization of the variational parameters $\bm{\theta}$, the trained encoder creates a final state as close as possible to the target product state of Eq.~\eqref{eq:q_encoder}.\\

\paragraph*{Training the quantum autoencoder}
The initial quantum state $\ket{\psi}_{AB}$ is obtained by using phase encoding to load the classical information on the phase of the quantum state, with the following scheme. Be $ \qty{\bmx_i\,|\, \bmx_i \in \mathbb{R}^d, i = 1,\ldots, m}$ the set containing the classical data to be analyzed, then the quantum autoencoder is trained using the quantum states obtained as $\qty{\ket{\psi_{\bmx}} = \sum_i e^{i\,x_i}\ket{i}\,|\, \forall \bmx \in \mathcal{X}}$. In our specific case, the classical data are four dimensional $d=4$ and thus we only need $n=\log_2 d = 2$ qubits to encode the data. This in turn implies that the compressed system $A$ and the trash subsystem $B$ consist of a single qubit each.

\begin{figure}[ht]
    \centering
     \includegraphics[width=\textwidth]{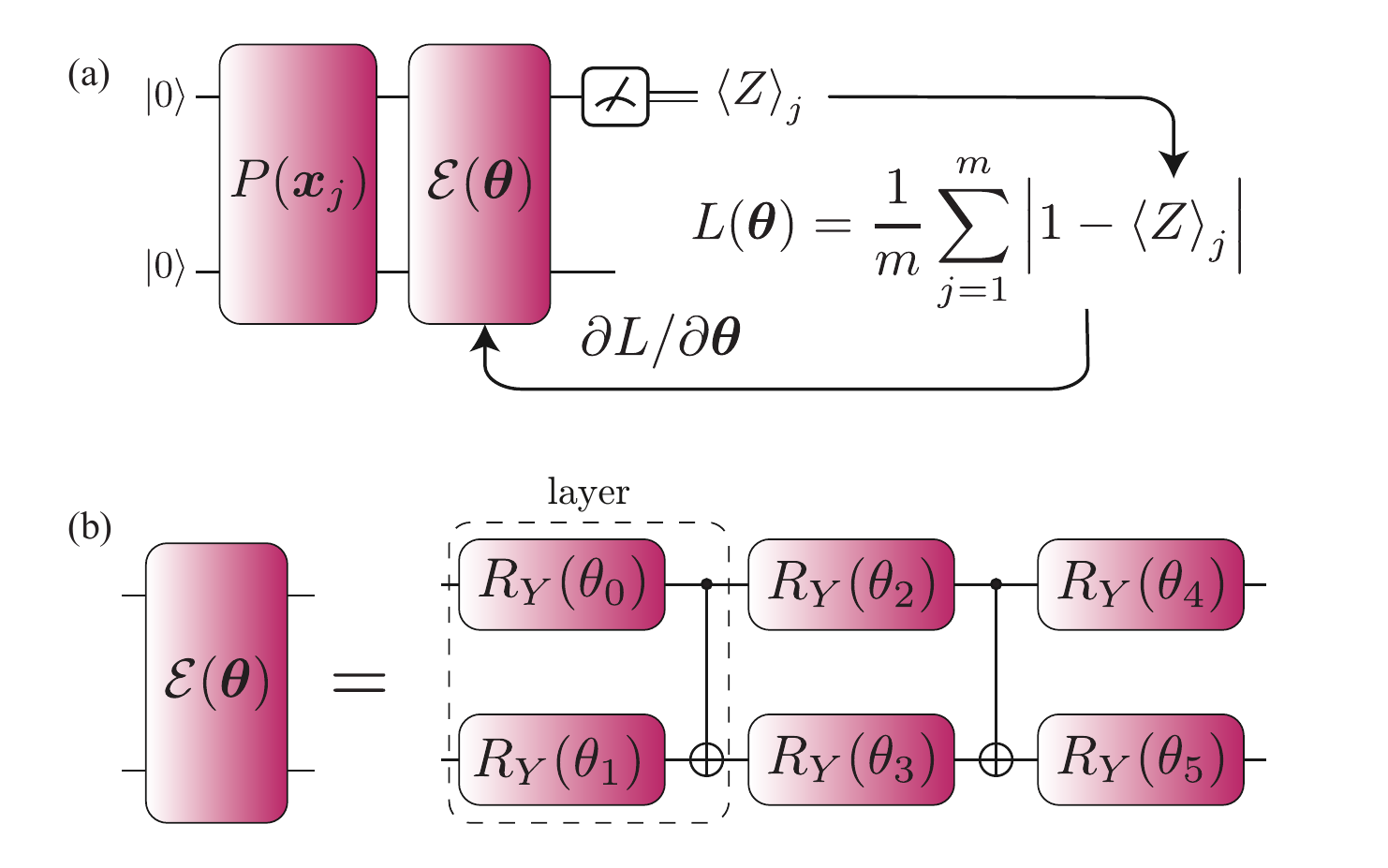}
    \caption[Quantum circuit for the quantum autoencoder]{\textbf{(a)} Quantum circuit used to train the quantum autoencoder. A register initialised in the ground state $\ket{00}$ is first subject to the phase encoding operation denoted by $P(\bmx)$, and then goes through the quantum encoder $\mathcal{E}(\bm{\theta})$. Then the trash qubit is measured, and the mean value of the Pauli operator $\expval{Z}$ is evaluated. Such value is then plugged into the loss function $\mathcal{L}(\bm{\theta})$ to drive the learning process. \textbf{(b)} Circuit representation of the quantum encoder $\mathcal{E}(\bm{\theta})$. Two layers of Pauli-$Y$ rotations and a CNOT, are followed by a final layer of Pauli-$Y$ rotations. In total, the circuit has 6 trainable parameters. The decoder $\mathcal{D}(\bm{\theta})=\mathcal{E}^\dagger(\bm{\theta})$ is obtained by reversing the order of the operations, and changing the sign of the rotations angles.}
    \label{fig:qae_training_circuit}
\end{figure}

Given the input data, the variational parameters $\bm{\theta}$ of the encoder $\mathcal{E}(\bmt)$ are optimised in order to rotate the trash qubit as close as possible to the target trash state, which we choose to be $\ket{\rm{trash}} \coloneqq \ket{0}$. This is achieved by means of a training procedure whose aim is to find optimal parameters $\bmt^*$ such that the loss function $L(\bmt)$ characterising the task is minimised. That is, the goal of training is to find 
\begin{equation}
\label{eq:loss_fn}
\bm{\theta}^* = \argmin_{\bm{\theta}} L(\bm{\theta}) \quad \text{with} \quad L(\bm{\theta}) = \frac{1}{m}\sum_{j=1}^{m}\abs{1-\expval{Z_B}_j}\, ,
\end{equation}
where we have defined
\begin{equation}
\expval{Z_B}_j \coloneqq \mel{\psi_{\bmx_j}}{\mathcal{E}^\dagger(\bm{\theta})(\mathbb{I}_A \otimes Z_B)\mathcal{E}(\bm{\theta})}{\psi_{\bmx_j}}
\end{equation}
as the mean value of the Pauli-$Z$ operator evaluated on the trash system $B$, after the encoder $\mathcal{E}(\bmt)$ acted on the input quantum state $\ket{\psi_{\bmx_j}}$ depending on the $j$-th sample $\bmx_j$. The loss function used in Eq.~\eqref{eq:loss_fn} is referred to as \textit{Mean Absolute Error} (MAE) in the classical machine learning literature, and together with the \textit{Mean Squared Error} (MSE) is the one of the most commonly employed loss functions in supervised regression tasks, which is also our case. Note that the loss function is faithful, in the sense that it reaches its global minimum $L(\bm{\theta}^\ast) = 0$, only when $\expval{Z_B}_j=1,\,\forall j=1,\ldots,m$, that is when the trash qubit is always and perfectly disentangled from the other qubit, and mapped to the target trash state $\ket{0}$. A schematic representation of the quantum circuit used for the training procedure is explicitly shown in Fig.~\ref{fig:qae_training_circuit}a.\\

\paragraph*{Variational ansatz}
The actual quantum circuit implementation of the encoder $\mathcal{E}(\bm{\theta})$, hence the decoder $\mathcal{D}(\bmt)$, is arbitrary, and in fact different variational ansätze have been proposed in the quantum machine learning literature~\cite{CerezoPQC2021Review, ManginiQNN, Bharti2021NIQSReview, McClean2016VQAs}, of which we gave an extended overview previously in Sec.~\ref{ssec:ch_VQA_ansatze}. In our case, we are dealing with only two qubits, and the most general ansatz consists of repeated applications of single qubit rotations and two-qubits entangling gates. In fact, having in mind to keep the parameters count and the overall circuit complexity low, we hereby propose a minimal yet efficient variational autoencoder consisting of two layers of Pauli-$Y$ rotations $R_Y(\theta) = e^{-i Y \theta/2}$~\eqref{eq:pauli_rotations} and a CNOT, followed by a final layer of rotations, as schematically depicted in Fig.~\ref{fig:qae_training_circuit}b.

\section{Experiments and Results}
\label{sec:ch_Autoencoder_results} 
In this section we discuss the experiments implementing the classical and quantum data analysis approaches described above for the data compression and classification tasks. 

\subsection{Data compression}
\paragraph*{Classical autoencoder}
The classical neural network autoencoder was implemented with the Keras library of TensorFlow~\cite{TensorFlow}, and it consists of two dense layers in a 4-2-4 structure as in Fig.~\ref{fig:Autoencoder_FIG1}c, with sigmoid activation function. 

The input data consists of a time series with 2873893 samples, 25\% of which are used as validation data, and the rest for training. Before training, features were transformed with a MinMax scaler, which scaled each feature to fit in the range $[0,1]$. After the learning phase, the average reconstruction error $\bar{e}$, defined as
\begin{equation}
\label{eq:recontruction_error}
\bar{e} \coloneqq \frac{1}{m}\sum_{i=1}^m \, \qty( \frac{1}{4}\sum_{j=1}^4\frac{\abs{x^{(i,j)}_\text{decoder}-x^{(i,j)}_\text{original}}}{\abs{x^{(i,j)}_\text{original}}} )
\end{equation}
amounts to 5\%, and in Fig.~\ref{fig:Autoencoder_FIG3} we show a comparison of the original against reconstructed data averaged by day, for the validation dataset. As we can see, the decoder shows quite good performance in the reconstruction of the input data for 3 of the 4 variables. For the ‘LIC’ variable, the median of the distribution of the reconstructed data coincides with the one of the original data, though the fluctuations are not very well described. There is no obvious a priori reason for the imperfect reconstruction of this particular variable, and this may well be a shortcoming of the autoencoding approach, which focuses more on the other variables to achieve a good-enough reconstruction scheme. 

In the following step we used the two latent variables from the compressed layer as input for a supervised classification algorithm, to predict the class assigned at the beginning through the clustering algorithm. We expect that, if the compressed vector is a suitable representation of the input data, a classification algorithm would be able to achieve very good performances.\\

\paragraph*{Quantum autoencoder}
The quantum autoencoder was simulated using a combination of PennyLane~\cite{Pennylane}, TensorFlow~\cite{TensorFlow} and Qiskit~\cite{Qiskit}, and the optimisation was performed using the automatic differentiation techniques implemented by these libraries. While automatic differentiation is only possible when performing a classical simulation of the quantum algorithm, in realistic scenarios of optimising a quantum circuit on real quantum hardware one can resort to parameter-shift rules~\eqref{eq:parameter_shift_rule} to estimate gradients and optimise the variational parameters~\cite{Schuld2019Gradients, Mitarai2018Learning}.

\begin{figure}[ht]
    \centering
    \includegraphics[width=0.75\columnwidth]{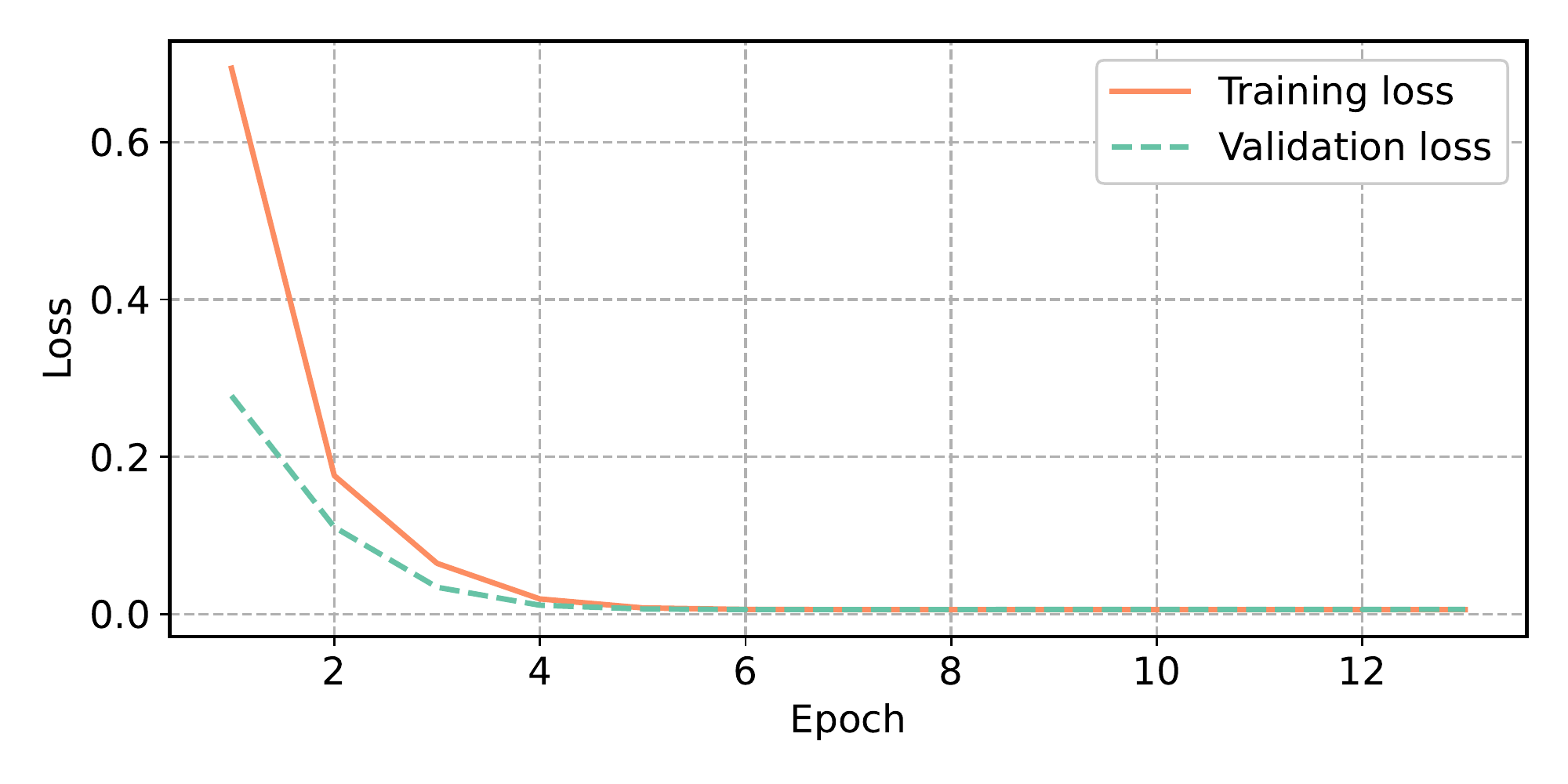}
    \caption[Optimisation of the quantum encoder]{Optimisation of the quantum encoder, $\mathcal{E}(\bm{\theta})$, showing the training and validation loss evaluated with data sets containing $10040$ and $520$ samples, respectively. The minimum of the loss at the end of training amounts to $L(\bm{\theta}^\ast)=0.0058$.}
    \label{fig:Autoencoder_training_loss}
\end{figure}

The variational circuit was trained using the Adam optimiser~\cite{KingmaAdam_2014} with learning rate set to $\eta = 0.001$, to update the six variational parameters $\bm{\theta} = (\theta_0, \theta_1, \theta_2, \theta_3, \theta_4, \theta_5)$. The training was performed using mini-batches of size $20$ for a total training set consisting of $m=10040$ samples. In Fig.~\ref{fig:Autoencoder_training_loss} it is shown the optimisation process across epochs of learning, both for the training loss, and for a validation set of $520$ samples. Before the phase encoding process, the classical data $\{\bmx_i\}_i$ were normalised as $\bmx_i\leftarrow \pi \cdot \bmx_i/||\bmx_i||$. It is clear that the quantum encoder is effectively trained, with the loss reaching the minimum value of $L(\bm{\theta}^*) = 0.0058$.

With a trained encoder, we can now proceed to investigate the quality of the data compression provided by the algorithm. The state of the qubits $A$ and $B$ after the quantum encoder operator consists of a general two-qubit state
\begin{equation}
\ket{\Psi}_{AB} =\, a\ket{0}_B\otimes\ket{0}_A + b\ket{0}_B\otimes\ket{1}_A + c\ket{1}_B\otimes\ket{0}_A + d\ket{1}_B\otimes\ket{1}_A \, ,
\end{equation}
where, if the encoder has been successfully trained, the probability of measuring qubit $B$ in state $\ket{1}$, $p_1 = |c|^2 + |d|^2$, is much smaller, ideally zero, than the probability of finding it in $\ket{0}$, namely $p_1\ll p_0 = |a|^2 + |b|^2$. Thus, in order to obtain a compressed pure state for qubit $A$ rather than a mixed one, we could post-select state $\ket{\Psi}_{AB}$ on measuring the trash qubit in state $\ket{0}$. In this case, let $\hat{\Pi}^B_0 = \dyad{0}_B \otimes \mathbb{I}_{A}$ be the projector on state $\ket{0}$ for system $B$, then the composite state is projected to
\begin{equation}
    \ket{\Psi}_{AB} \longrightarrow\, \frac{\hat{\Pi}^B_0 \dyad{\Psi}\hat{\Pi}^B_0}{\Tr[\hat{\Pi}^B_0 \dyad{\Psi}\hat{\Pi}^B_0]} = \dyad{0}_B \otimes \dyad{\psi_c}_A\,, \quad \ket{\psi_c}_{A} = \frac{a\ket{0}_A + b\ket{1}_A}{\sqrt{|a|^2+|b|^2}}\, .
\end{equation}

\begin{figure*}
    \centering
    \includegraphics[width=\textwidth]{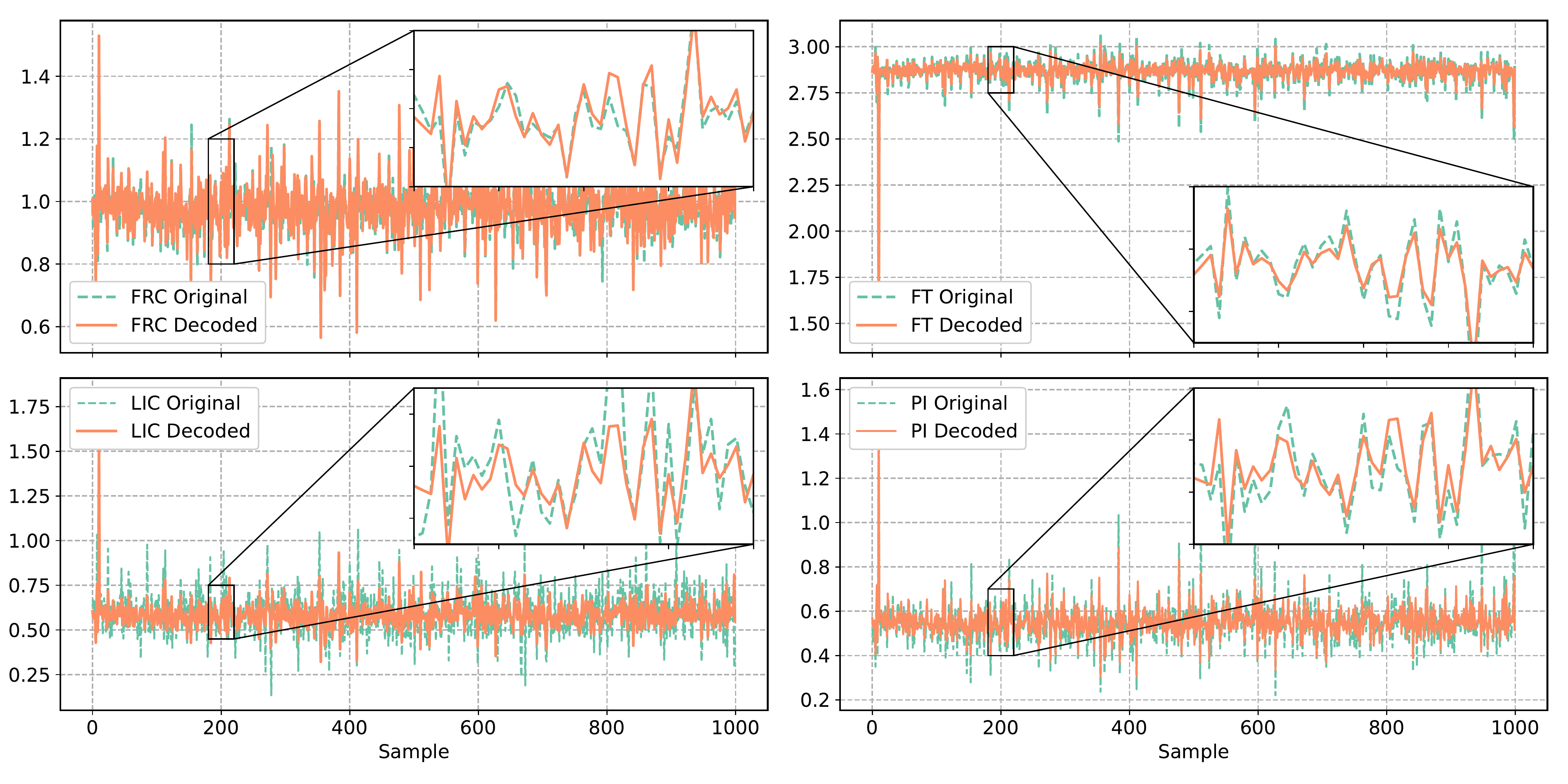}
    \caption[Performances of the quantum autoencoder]{Performances of the quantum autoencoder in a compression and decoding task. Each plot shows one of the input features labelled `FRC', `FT', `LIC', `PI', as reconstructed by the quantum autoencoder (`decoded') confronted with the original sample (`original'). This plots are evaluated on a test set consisting of $m=1000$ samples. The average reconstruction error $\bar{e}$ as defined in the main text in Eq.~\eqref{eq:recontruction_error}, amounts to $\bar{e}=5.4\%$. This results were obtained using the IBM Qiskit \texttt{statevector\_simulator}.}
    \label{fig:quantum_reconstruction}
\end{figure*}

If we wish to retrieve the original information, now stored in compressed form in the state $\ket{\psi_c}_{A}$ of system $A$ only, we can couple this system to a new qubit initialised in $\ket{0}$, and then apply the quantum decoder, as shown in Fig.~\ref{fig:quantum_autoencoder}. An example of this procedure is shown in Fig.~\ref{fig:quantum_reconstruction}, where the reconstruction performances of the quantum autoencoder are evaluated on a test set consisting of $1000$ samples coming from the original dataset. In the case of Fig.~\ref{fig:quantum_reconstruction}, the average reconstruction error defined in Eq.~\eqref{eq:recontruction_error} amounts to $\bar{e} = 5.4\%$, confirming that the quantum autoencoder can successfully compress and then retrieve information with low error.  

It is important to note that the results discussed in this section were obtained with an exact simulation of the wavefunction of quantum systems, using Qiskit's \texttt{statevector\_simulator}. This allows for a direct access to the amplitudes of the quantum states, and thus recover the final phases of the decoded state $\ket{\varphi_\text{decoder}} = \mathcal{D}(\bm{\theta})(\ket{0} \otimes \ket{\psi_c}_{A})$. However, in a real case scenario with a quantum hardware, it is not possible to perfectly retrieve the phases of the decoded state $\ket{\varphi_\text{decoder}}$, since one would need to perform quantum tomography of such state, and even in that case results could only be obtained up to an arbitrary constant, due to quantum measurement outcomes following Born's rule. Thus, while such reconstruction test would prove much harder to be performed on a real device, the results in Fig.~\ref{fig:quantum_reconstruction} obtained with the simulator are still relevant in checking the inner working of the quantum autoencoder, and that it is actually able to perform the task it was designed for, even if it is not currently accessible by a real experimenter. 

\begin{figure}[ht]
\includegraphics[width=\columnwidth]{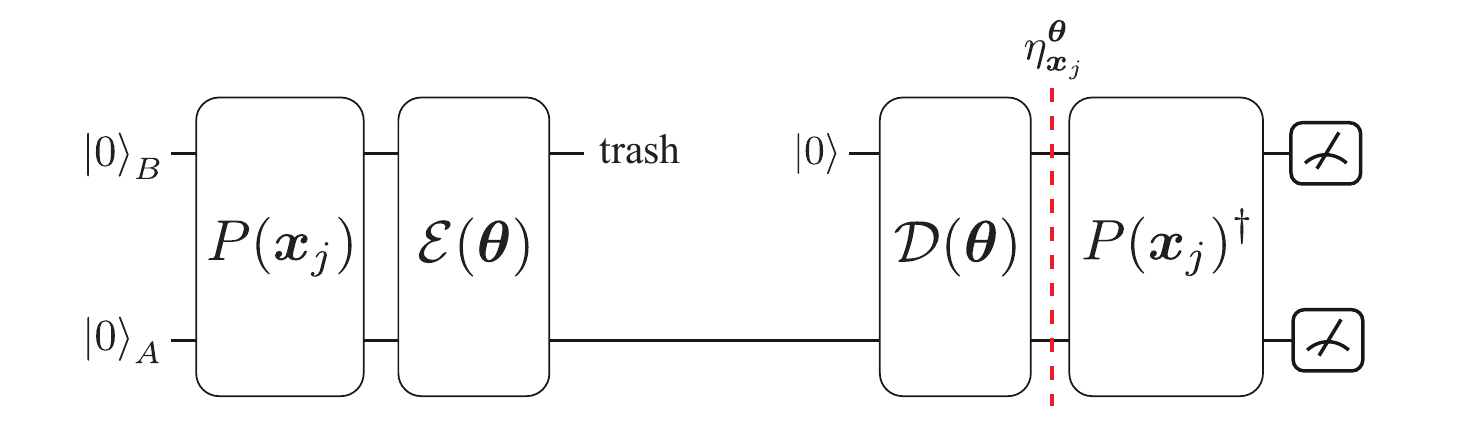}
\caption[Circuit to evaluate the fidelity]{Circuit to evaluate the fidelity $F(\rho_{\bmx},\eta^{\bm{\theta}}_{\bmx}) = \Tr[\rho_{\bmx}\, \eta^{\bm{\theta}}_{\bmx}]$ between the initial pure state $\rho_{\bmx} = \dyad{\psi_{\bmx}}$ and the generally mixed state $\eta^{\bm{\theta}}_{\bmx}$, obtained through the autoencoding procedure. The fidelity is obtained by counting the number of $\ket{00}$ outcomes at the end of the circuit. In fact, dropping the subscripts for simplicity, one has $p_0 = \Tr[P(\bmx)^\dagger \eta^{\bm{\theta}}_{\bmx} P(\bmx) \dyad{\bm{0}}] = \Tr[\eta^{\bm{\theta}}_{\bmx}\, P(\bmx) \dyad{\bm{0}}P(\bmx)^\dagger] = \Tr[\eta^{\bm{\theta}}_{\bmx} \dyad{\psi_{\bmx}}]=F(\rho_{\bmx},\eta^{\bm{\theta}}_{\bmx})$.}
\label{fig:FIG8_fidelity}
\end{figure}

We hereby discuss a second possible approach for measuring the faithfulness of the reconstruction, which albeit being indirect does not require state tomography and is thus more readily compatible with actual runs on quantum processors. The performances of the quantum autoencoder can be tested measuring the fidelity~\cite{WildeQuantumInfoBook2017} $F(\rho_{\bmx},\eta^{\bm{\theta}}_{\bmx})=\Tr[\rho_{\bmx}\,\eta^{\bm{\theta}}_{\bmx}]$ between the initial pure state $\rho_{\bmx} = \dyad{\psi_{\bmx}}$ obtained through phase encoding~\eqref{eq:phase_encoding}, and the generally mixed state obtained through the quantum circuit autoencoder of Fig.~\ref{fig:quantum_autoencoder}, defined as
\begin{equation}
    \eta^{\bm{\theta}}_{\bmx} \coloneqq {\sf D}(\bm{\theta})\qty[ \dyad{0} \otimes \Tr_B \qty[{\sf E}(\bm{\theta})\qty[\rho_{\bmx}] ] ]\, ,
\end{equation} 
where {\sf E}$(\bm{\theta})$ and {\sf D}$(\bm{\theta})$ represent the superoperators corresponding to the encoder $\mathcal{E}(\bm{\theta})$ and decoder $\mathcal{D}(\bm{\theta})$ operators, respectively. Clearly, the larger the fidelity the better, since it corresponds to the quantum autoencoder being able to recreate states that are very close to the initial ones.
Using this figure of merit, post-selecting on the trash subsystem $B$ is not necessary since qubit $A$ can be directly coupled to a new qubit initialised in $\ket{0}$, to then act with the decoder and with the evaluation of $\Tr[\rho_{\bmx}\,\eta^{\bm{\theta}}_{\bmx}]$. There are various techniques to evaluate state overlaps on quantum hardware~\cite{CincioStateOverlap2018, ManginiCQN2020}, the most common one being the SWAP test, and here we use the so-called \textit{compute-uncompute} method, whose circuit is shown in Fig.~\ref{fig:FIG8_fidelity}. 

Using a test set of $m=1000$ samples, a simulation of the trained quantum autoencoder, even including stochastic measurement outcomes with $n_\text{shots} = 10^4$ shots, yields an average fidelity
\begin{equation*}
\mathbb{E}\qty[\Tr[\rho_{\bmx} \, \eta^{\bm{\theta}}_{\bmx}]] = \frac{1}{m}\sum_{j=1}^m \Tr[\rho_{\bmx_j}\, \eta^{\bm{\theta}}_{\bmx_j}] = 0.975\pm 0.001\,,
\end{equation*}
which confirms again that the proposed variational quantum autoencoder is able to compress and later decode information. 

\subsection{Classification}
\label{sec:Autoencoder_classification} 
\paragraph*{Classical classifier}
The supervised classification algorithm used is the KNeighborsClassifier as implemented in \texttt{scikit-learn}. KNeighborsClassifier assigns the class to a point from a simple majority vote based on the $k$ nearest neighbours of that point. The number of nearest neighbours is a parameter of the algorithm, and after some trials we fixed it at $k=100$, which correspond to an optimal trade-off between performances and computational efficiency. 

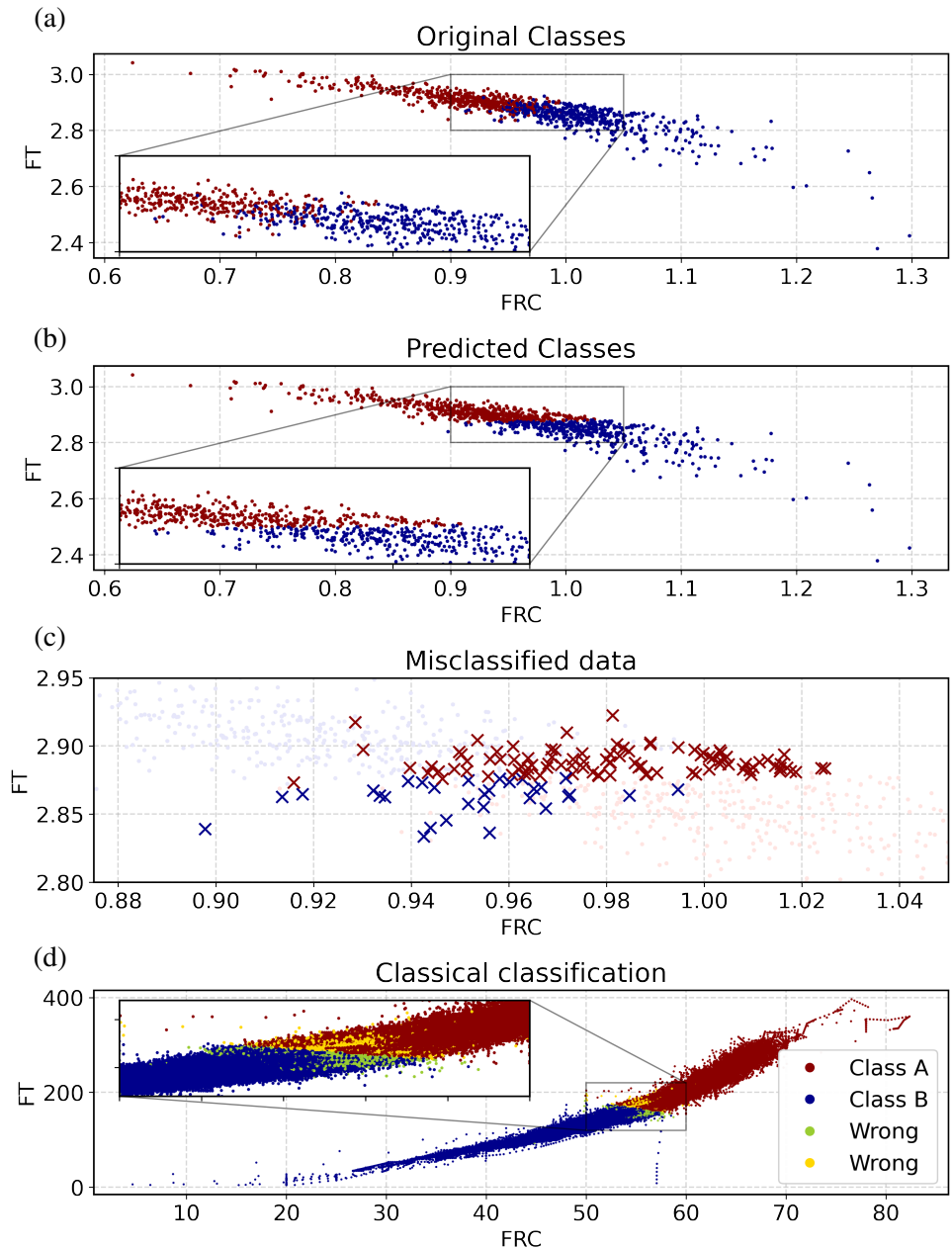
\begin{figure}[!ht]
    \centering
    \begin{tikzpicture}
    \node[inner sep=0pt] (russell) at (0,0){ \includegraphics[width=0.85\textwidth]{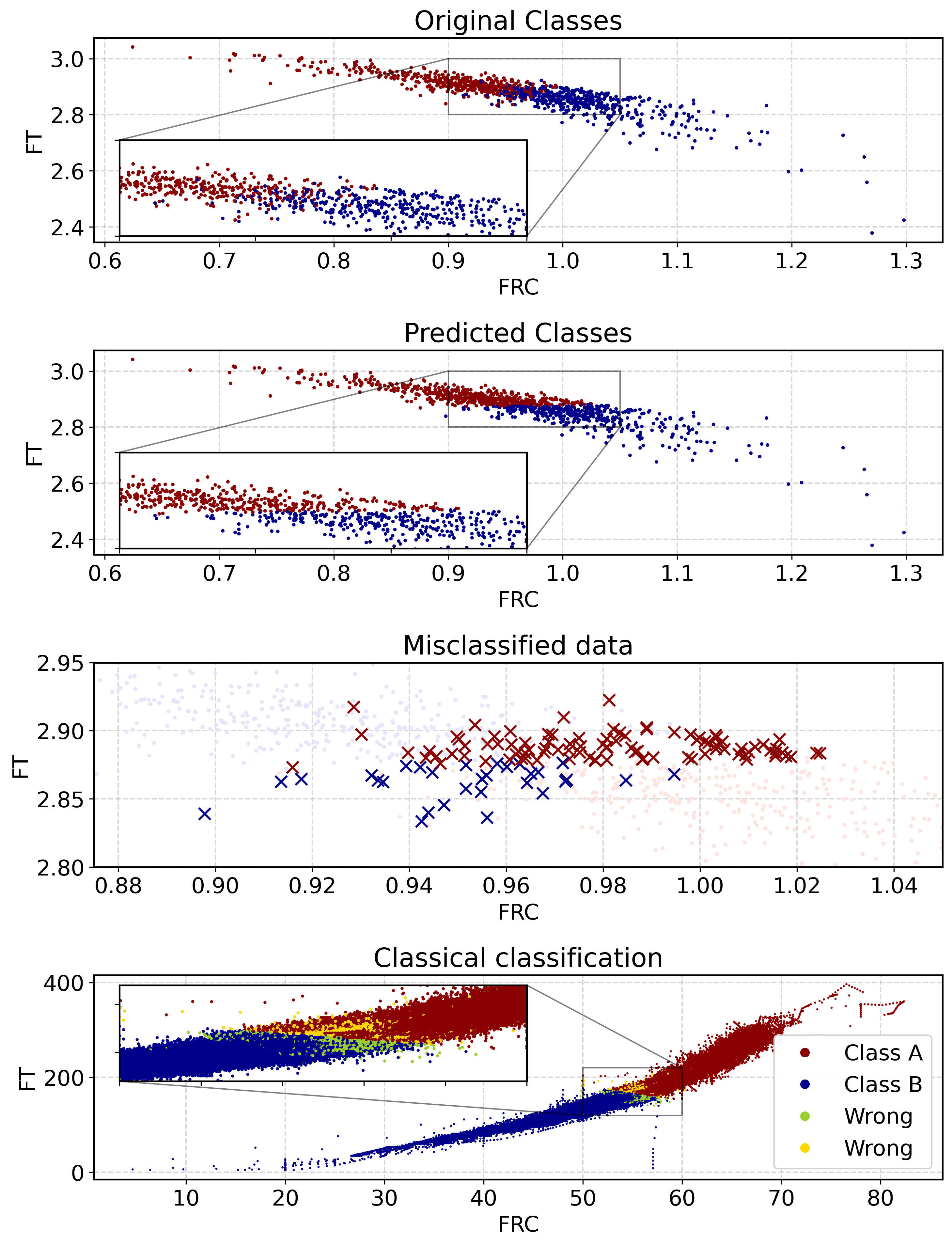}};
    \node[inner sep=0pt] (russell) at (-5.7, 8.3) {\normalsize{(a)}};
    \node[inner sep=0pt] (russell) at (-5.7, 4) {\normalsize{(b)}};
    \node[inner sep=0pt] (russell) at (-5.7, -0.) {\normalsize{(c)}};
    \node[inner sep=0pt] (russell) at (-5.7, -4.3) {\normalsize{(d)}};
    \end{tikzpicture}
    \caption[Results of the classification task]{Results of the classification task performed by the quantum classifier, for a test set of size $m=10^3$ samples. \textbf{(a)} Plot of the original data with the colour indicating the two different classes. Note that for simplicity, only the FRC and FT features are shown. \textbf{(b)} Label assigned by the trained quantum classifier. \textbf{(c)} Focus on the data that are mislabelled by the classifier. The colour indicates the label assigned by the quantum classifier, and the ``cross'' marker means that the data were misclassified. Note that these samples lay on the border of separating the two classes. The accuracy, evaluated as the percentage of correctly classified data, amounts to $87.4\%$. \textbf{(d)} Result of the classification using the classical autoencoder followed by a KNN clustering procedure. Note that the axis are different from the quantum case due to normalisation of the features. In this case the classification accuracy amounts to $89.7\%$}
    \label{fig:quantum_classification}
\end{figure}

The lowest panel of Fig.~\ref{fig:quantum_classification} shows the results of the classification, which is now anticipated but discussed later in comparison with the quantum algorithm results. In red and blue are the points that have been correctly classified, while in yellow and green are those which were misclassified. The classification accuracy, evaluated as the percentage of correctly classified data, reach a remarkably high value of $89.7\%$, indicating that the compressed vector is able to summarise the information carried by the input data. \\

\paragraph*{Single qubit quantum classifier}

Once the quantum autoencoder has been trained to learn a compressed representation of the original information, the compressed quantum state can be used as input for a classification task. We expect that, if the compressed information is a suitable representation of the input data, the classification algorithm would be able to learn the classes assigned to the full-size input data through the clustering algorithm described in Sec.~\ref{sec:case_study}. 

To do so, we can use the information-carrying qubit obtained with the encoder $\mathcal{E}(\bm{\theta})$, as input to a quantum classifier which is trained to learn the desired clustering of the original data. A quantum classifier is made of two parts: (\textit{i}) a trainable parametrized operation, $U(\bm{\varphi})$, which tries to map inputs belonging to different classes in two distant regions of the Hilbert space, and (\textit{ii}) a final measurement, which is used to extract and assign the label. 
Since we are dealing with a single qubit classifier, the most general transformation on a qubit is represented by the unitary matrix~\eqref{eq:u3}
\begin{equation}
\label{eq:Autoencoder_u3}
    U\left(\alpha,\beta,\gamma\right)=
    \begin{bmatrix}
    \cos(\alpha/2) & e^{-i\gamma}\sin(\alpha/2)\\
    e^{i\beta}\sin(\alpha/2) & e^{i(\beta+\gamma)}\cos(\alpha/2) \, 
    \end{bmatrix}\,.
\end{equation}
Thus, it is reasonable to use such operation as the trainable block of the classifier, since it ensures the greatest flexibility. Actually, as discussed later, the angle $\beta$ in Eq.~\eqref{eq:Autoencoder_u3} does not influence the measurement statistics of the qubit, hence it has no influence on the training of the classifier. For this reason, it is kept fixed at $\beta=0$, and the actual trainable gate used is $U(\alpha,0,\gamma)=U(\alpha, \gamma)$. 

As for the label assignment, since the measurement process of a qubit has only two possible outcomes, these are interpreted to be the two possible values for the labels, namely ``Class A'' and ``Class B'' which were described earlier in Sec.~\ref{sec:case_study}. Specifically, a label is assigned based on a majority vote on multiple shots of the same quantum circuit, that is an input is assigned to ``Class A'' if the majority of measurement gave $\ket{0}$ as outcome, and ``Class B'' otherwise. Formally, let $\rho^A_{\bmx} = \Tr_B [\mathcal{E}(\bm{\theta})(\dyad{\psi_{\bmx}})\mathcal{E}(\bm{\theta})^\dagger]$ be the compressed quantum qubit, then the label is assigned based on the decision rule
\begin{equation}
\hat{y}_i=
\begin{cases}
0\quad \text{if}\quad p_0=\Tr[\dyad{0}U(\bm{\varphi})\rho^A_{\bmx_i} U(\bm{\varphi})^\dagger]\geq0.5\\
1 \quad \rm{otherwise}
\end{cases}\,,
\end{equation}
where $p_0$ denotes the probability that the measurement yields $\ket{0}$ outcome. As mentioned earlier, one can check easily that $p_0$ does not depend on the angle $\beta$ of the unitary $U(\alpha, \beta, \gamma)$, and for this reason it is set to zero, yielding the variational unitary $U(\alpha,0,\gamma)=U(\alpha, \gamma)$.  

The loss function used to drive the training of the unitary $U(\alpha, \gamma)$ is the categorical cross entropy already introduced in Eq.~\eqref{eq:crossentropy}, defined as
\begin{equation}
\mathcal{L}(y_i,\hat{y}_i)=-(1-y_i)\log(1-\hat{y}_i)\ -\ y_i\log(\hat{y}_i) \, ,
\end{equation}
where $y_i$ is the correct label, and $\hat{y}_i$ is the label assigned by the quantum classifier, and the optimiser used is COBYLA~\cite{Cobyla} as implemented in SciPy's Python package~\cite{SciPy2020}. 

Figure~\ref{fig:quantum_classification} shows the results of the classification obtained after the optimization of the variational parameters $\bm{\varphi} = (\alpha, \gamma)$, for a test set of $m=10^3$ samples. The accuracy, measured as the ratio of correctly classified to total samples, is measured to be $87.4\%$ when evaluated with exact simulation of the quantum circuit. As clear from the figure, the misclassified data are only those located near the edge connecting the two classes. In fact, in this region, the samples are not neatly divided but rather a blurred border exists. On the contrary, given its relatively simple structure, the quantum classifier learns essentially a straight cut of the data in this region, thus committing some labelling errors. 

This should not come as a surprise since, as seen when discussing the Fourier representation of variational circuits in Sec.~\ref{ssec:ch_QML_1q_qnn}, a single-qubit classifier can only learn simple functions (i.e. sine functions) of the input data if there is not enough input redundancy~\cite{Schuld2020Encoding, Theis2020Expressivity, Perez2020Reuploading, LaRose2020Robust, GratseaPerceptronTeacher2022, Lloyd2020Embeddings}. However, note that the dependence of the classification on the original data is strictly non-linear, since the classical data first go through a classical preprocessing step, then are loaded onto the quantum states by means of rotations, and finally undergoes the encoding procedure which scrambles information even more.

It is interesting to notice that the classification performances remain stable even when including sources of noise, such as stochastic measurement outcomes. Indeed, a simulation of the circuit using $n_\text{shots}=1024$ on $m=10^3$ samples yields an accuracy of about $82.5\%$, which is only slightly lower than the exact case corresponding to an infinite number shots. In addition, the classifier proves robust even when tested on real quantum hardware. In fact, the circuit for the trained classifier was tested against IBM's \texttt{ibmq\_x2} quantum chip (accessed May 2021) but with a smaller test set of $m=75$ samples, due to limitations in the device usage. In this case, using $n_\text{shots} = 1024$ shots per circuit, and averaging over 5 executions with different test samples, the classification accuracy (evaluated again as the percentage of correctly classified data) was found to be $(82.3 \pm 1.3)\%$, indeed very close to the simulation including only measurement noise, and not much different from the noiseless result. 

\section{Conclusions}
\label{sec:conclusion}
We have presented a direct comparison between quantum and classical implementations of a neural network autoencoder, followed by a classifier algorithm, applied to sample real data coming from one of Eni's plants, in particular from a first stage separator. While the achievement of a clear quantum machine learning advantage with variational algorithms is still disputed~\cite{HuangInfoBounds2021, Huang2020Power, AbbasPowerQNN2021}, this work sets a milestone in the field of quantum machine learning, since it is one of the first examples of direct application of quantum computing software and hardware to analyse real data sets from industrial sources. 

As a first step, we have implemented and analyzed the performance of a variational quantum autoencoder to compress and subsequently recover the input data. We verified its performances using full simulation of the wavefunction, which allowed us to evaluate the average reconstruction error to about $\bar{e}=5\%$ ---essentially identical to the classical autoencoder--- thus confirming the capability of the quantum autoencoder to effectively store a compressed version of the original data set, and then being able to recover it. In addition, we also checked the correctness of the quantum autoencoding procedure by evaluating the quantum fidelity between original and decoded quantum states, which were again found to be very similar to each other, even in the presence of simulated stochastic measurement noise. 

Once the optimal parameters for the quantum autoencoder were determined during the training phase, we used the compressed quantum state as input to a quantum classifier, with the goal of  performing a binary classification task. The algorithm achieved an accuracy above $87\%$, absolutely comparable to that achieved in the classical setting using the neural network autoencoder followed by a nearest-neighbours classifier, thus indicating again that the quantum algorithm is able to correctly compress the relevant information of the input data. We also tested the performance of the full quantum pipeline (given by the quantum autoencoder plus the classifier) on actual and currently available IBM superconducting quantum hardware, obtaining a classification accuracy of $82\%$, which is only slightly smaller than the ideal result. 

The small size of current quantum devices and their relatively high noise levels make it hard to run actually relevant and large scale computations, thus making an effective quantum advantage out of reach. On the other hand, in this Chapter we provided a successful proof-of-concept demonstration that an original quantum autoencoder and a quantum classifier can actually reach the same level of accuracy as standard classical algorithms, on a data set that is sufficiently low dimensional to be handled on actual near-term quantum devices. In addition, it is worth emphasising that the quantum autoencoder allows to obtain results that are quantitatively comparable to the classical algorithm by using only 6 parameters instead of 16, thus displaying an increased efficiency in terms of number of trainable parameters already reached on NISQ devices. With continuing progress in quantum technologies and quantum information platforms, we envision the execution of the very same quantum algorithms on larger scales, possibly reaching the threshold for a classically intractable problem. We believe that these results take the first foundational steps towards the application of usable quantum algorithms on NISQ devices for industrial data. 

Although the use of quantum resources may offer computational advantages over purely classical methods, the latter are incredibly versatile tools, not only capable of giving rise to incredibly successful machine learning models, but also to provide an effective description of quantum system themselves. Thus, in order for a (variational) quantum algorithm to deliver a meaningful advantage, it must be hard to simulate via classical methods. The topic of the next Chapter is more fundamental and unrelated to practical use-cases, but instead addresses the classical simulability of quantum circuits under the lens of the entanglement, by studying the entanglement produced inside common variational quantum circuits.

\chapterimage{bg11.png} 
\chapterspaceabove{6.75cm} 
\chapterspacebelow{7.25cm} 

\chapter{Entanglement entropy production in quantum neural networks}\index{Entanglement in QNN}
\label{ch:entanglement}
\startcontents[chapters]
\printcontents[chapters]{}{1}{}
\vspace*{1cm}

In this chapter\footnote{The content of this chapter is based on the author's work~\cite{BallarinEntQNN_2022}, and all the figures in this chapter are taken from, or are adaptations of, the figures present in such work.
}, we use tensor networks techniques to characterise the entanglement features of several recent proposals for Quantum Neural Networks. Specifically, we study the production of entanglement in random parameterised quantum circuits of up to fifty qubits, showing that their entanglement, measured in terms of entanglement entropy between qubits, tends to that of Haar distributed random states as the depth of the QNN is increased. We certify the randomness of the quantum states also by measuring the expressibility of the circuits, as well as using tools from random matrix theory. We show a universal behaviour for the rate at which entanglement is created in any given QNN architecture, and consequently introduce a new measure to characterise the entanglement production in QNNs: the entangling speed. These results characterise the entanglement properties of quantum neural networks, and provides new evidence of the rate at which these approximate random unitaries.

\section{Introduction}

\begin{figure*}[ht!]
    \centering
    \includegraphics[width = \textwidth]{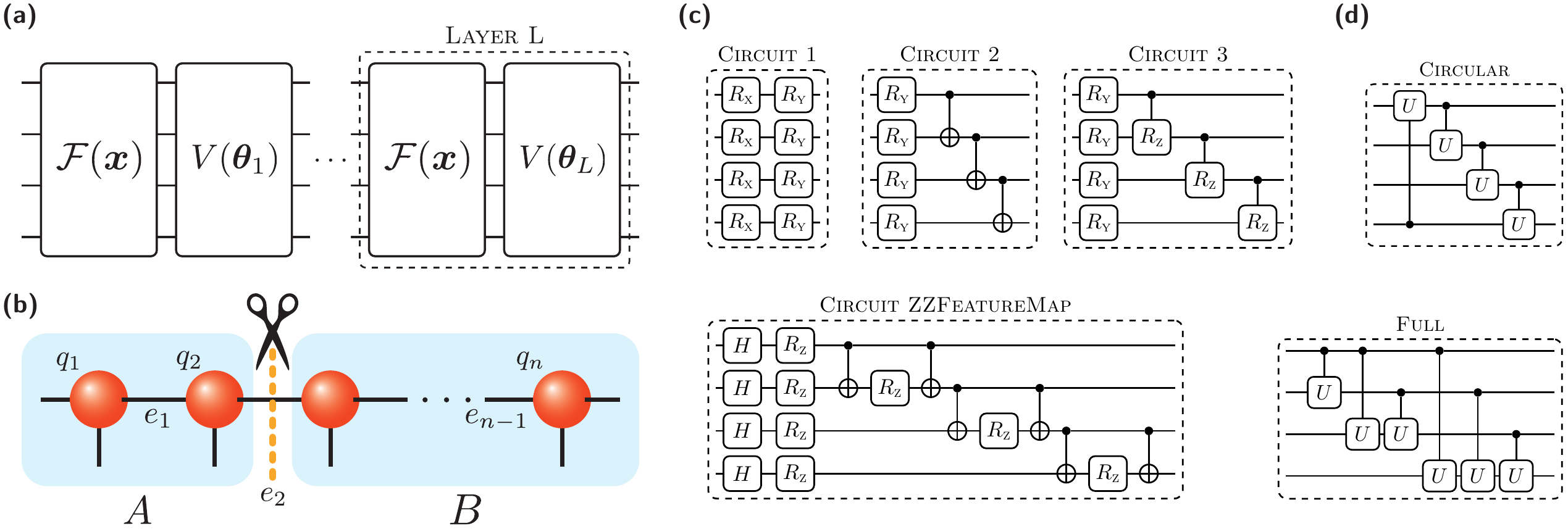}
    \caption[Graphical representation of QNN and MPS]{Graphical representation of QNN and MPS. \textbf{(a)} QNN structure with alternating feature map $\fmap$ and variational ansatz $\varans$. Note that the ansatz parameters are different in each layer, while the feature map parameters are the same throughout the whole circuit. \textbf{(b)} MPS diagram. Each sphere is a tensor, representing a qubit $q_j$. The entanglement entropy between bi-partitions $A$ and $B$ is computed by "cutting" the connecting edge $e_j$. \textbf{(c)} Circuits analysed in the present study, depicted with a linear entanglement topology, i.e. entangling gates are only applied between nearest neighbours on a line. \textbf{(d)} Different entanglement topologies: circular, with the first and last qubit of the line connected, and full, where the entangling gates are applied between each pair of qubits (see Appendix~\ref{app:entangling_maps} for a clear definition and discussion). When using parameterized two qubits gates, like the controlled rotations in circuit 3, the entanglement maps are generalised to their parameterized version by using the corresponding parameterized operation. Note that the circuit templates 2 and \textsc{ZZFeatureMap} are those used in the QNN of~\cite{AbbasPowerQNN2021}, and also that circuits 1, 2 and 3 share similarities with circuits 1, 15, and 13 of~\cite{SimPQCs2019}, respectively.}
    \label{fig:main_figure}
\end{figure*}

The topic of this chapter is the study of the entanglement properties of quantum neural networks when these initialised with random parameters. We employ methods from the tensor network literature, namely Matrix Product States (MPS), to study the entanglement generated in various QNNs architectures composed of up to 50 qubits. Since MPS are a very powerful tool for simulating quantum systems with bounded entanglement, if a quantum neural network can only access low entangled states, it can be easily simulated, which spoils any hope of achieving a concrete quantum advantage. Thus, using  entanglement entropy among qubits as a figure of merit, we evaluate the entanglement capabilities of some of the most common and promising QNN architectures~\cite{AbbasPowerQNN2021, SimPQCs2019}. We consider several QNNs with different combinations of feature maps $\fmap$ and variational forms $\varans$ and perform an extended numerical analysis varying: (\textit{i}) the number of qubits $n$, (\textit{ii}) the number of layers $L$ in the network, (\textit{iii}) the entangling topology of the circuit, (\textit{iv}) the data re-uploading~\cite{Perez2020Reuploading, Schuld2020Encoding} structure being either alternated or sequential. A summary of the circuit templates analysed in this study is shown in Fig.~\ref{fig:main_figure}.

For all the considered QNNs with nearest neighbour connectivity, as the number of layers $L$ is increased, the entanglement generated inside the circuit grows, eventually reaching a plateau when $L \approx n$, where $n$ is the number of qubits. This behaviour is associated with the typical entanglement of a random Haar-distributed quantum state. The choice of the entangling topology (nearest neighbours, circular, or all to all) clearly affects the rate of creation of entanglement in the circuit. We also point out that a careless definition of a full, i.e. all to all, connectivity map can effectively result in a linear nearest-neighbours interaction if unparameterized two qubits gates (CNOTs) are used, something apparently overlooked in the recent literature using this type of ansatz~\cite{AbbasPowerQNN2021, TacchinoVariationalQNN2021, Jaderberg2022SelfSupervised}.
By bounding the entanglement generated by the circuit, we are able to simulate QNNs with MPS up to $n=50$ qubits. It should be stressed that such simulations are exact up to a given number of layers, after which a truncation of the entanglement via MPS is applied. By appropriately normalising the entanglement produced we show that all the points for a given QNN architecture follow the same curve, independently from the number of qubits. Thus, we exploit this behaviour to define a universal figure of merit given the QNN architecture, the entangling speed. This figure of merit characterises how fast the entanglement is produced by the QNN, with respect to the number of layers $L$.

In addition, we evaluate the expressibility measure of the considered QNNs as defined in~\cite{SimPQCs2019} and argue that the optimality of the QNN introduced in~\cite{AbbasPowerQNN2021} may be related to its good trade-off between mild entanglement production and high expressibility. Finally, we employ tools from random matrix theory, specifically convergence to the Mar\v{c}enko-Pastur distribution, to further characterise the resemblance of the deep enough quantum neural networks to random unitary matrices. At last, we note that differently from~\cite{SimPQCs2019} which bases their analysis on the Meyer-Wallach entanglement measure~\cite{meyerGlobalEntanglementMultiparticle2002}, in this study we make use of the entanglement entropy among subsystems, which allows for a more careful analysis of the entanglement distribution in the system, and it is also readily accessed in an MPS simulation with no computational overhead.

The chapter is organised as follows. In Sec.~\ref{sec:ch_Entanglement_methods} we review the basis of tensor networks and MPS, and introduce the Von Neumann entropy as an entanglement measure. We then discuss the entanglement entropy properties of random quantum states. We proceed by discussing the most recent results on parameterized quantum circuits and QNNs, especially, on the relation between randomness, trainability, and entanglement found in these circuits. In Sec.~\ref{sec:ch_Entanglement_Results} we show the results of our analysis for various QNN architectures, and discuss the results in Sec.~\ref{sec:ch_Entanglement_Discussion}. Finally, we discuss the implications of our analysis and possible routes for future investigations in Sec.~\ref{sec:ch_Entanglement_Conclusion}. 

\section{Methods}
\label{sec:ch_Entanglement_methods}

\subsection{Tensor Networks and Matrix Product States}
\label{sec:MPS}
An $n$-qubit quantum state is defined in a Hilbert space $\mathcal{H}$ of dimension $\text{dim}(\mathcal{H})=2^n$. The exponential scaling of $\mathcal{H}$ with $n$ makes the classical description of quantum states an exponentially expensive task. This problem is widely known in many-body quantum physics, and many different techniques have been developed to alleviate the issue, like the Density Matrix Renormalization Group (DMRG) or Tensor Network (TN) techniques~\cite{tensor_anthology, Montangero_book}. 

In this study, we use Tensor Network methods to efficiently describe the $n$-qubit state. In particular, we employ Matrix Product States (MPS), which are a specific tensor network ansatz particularly suited to represent 1-dimensional (i.e. like atoms on a chain, as in Fig.~\ref{fig:main_figure}) quantum states~\cite{EisertEntanglementMPS}. The power of tensor networks lies in the assumption that we are only interested in a \textit{tiny subspace} of the entire Hilbert space, namely the states that display a limited amount of entanglement.

An $n$-qubit pure state $\ket{\psi} \in \mathcal{H}$ can be written as a MPS as follows~\cite{EisertEntanglementMPS}
\begin{equation}
    \label{eq:mps_representation}
    \ket{\psi} = \sum_{{s_1,\dots,s_n=0}}^1 \,\, \sum_{{\alpha_1, \dots, \alpha_n = 1}}^\chi  \textbf{M}_{1 \alpha_1}^{[1],s_1}\, \textbf{M}_{\alpha_1  \alpha_2}^{[2],s_2}\, \ldots\, \textbf{M}_{\alpha_{n-2} \alpha_{n-1}}^{[n-1],s_{n-1}}\, \textbf{M}_{\alpha_{n-1} 1}^{[n], s_n}\, \ket{s_1 s_2 \hdots s_n}\,.
\end{equation}
Each tensor $\textbf{M}_{\alpha_{i} \alpha_{i+1}}^{[i], s_i}$ is a local description for the $i$-th site, which allows one to apply a \textit{local} operator to a certain site without the need to change all the other coefficients. For a fixed $s_i$, $\textbf{M}_{\alpha_i \alpha_{i+1}}^{[i],s_i}$ is a $\chi \times \chi$ complex matrix, meaning that  Eq.~\eqref{eq:mps_representation} is the sum of basis elements weighted by matrix products. The integer $\chi$ is called the MPS \textit{bond dimension}, and a sufficiently high $\chi$ is needed to express a general $\ket{\psi}$ in such form. However, MPS with a \textit{lower} $\chi$ can still encode all the meaningful states, albeit clearly not \textit{all} possible states. In particular, to correctly describe any quantum state the bond dimension needed is $\chi=d^{\lfloor\frac{n}{2}\rfloor}$, where $d$ is the local dimension of the degrees of freedom ($d=2$ for qubits).
One can also efficiently evolve the state under the application of $2$-qubit gates, using an approach known in the literature as time-evolving block decimation~\cite{time_evolution_mps}, and perform measurements. Simulations using MPS are not bounded by the number of qubits in the system, but by the amount of entanglement generated inside it, as we explain in detail in Section~\ref{sec:Entanglement}.

Nonetheless, while the use of an MPS simulation imposes some constraints on the maximum entanglement that it is possible to represent, this issue is relevant only for very deep circuits involving many qubits. Indeed, we reliably simulate circuit instances involving up to $n=50$ qubits and moderate depth, which is already sufficient to provide clear insights on the entanglement entropy generated in such circuits. Moreover, as explained below in Sec.~\ref{sec:Entanglement}, during an MPS simulation one has constant access to the singular values of the quantum state, so the entanglement of the state can be calculated on the fly without any computational overhead. Thus, MPS are an effective tool to study the entanglement properties of quantum circuits, especially in regimes that cannot be easily accessed with a full-scale simulation of the statevector of the system.

\subsection{Entanglement measure in Matrix Product States}\label{sec:Entanglement}
Entanglement in quantum states can be evaluated using the so-called Von Neumann \textit{entanglement entropy}. Let $\rho = \dyad{\psi}$ be the quantum state of a system of $n$ qubits, and consider a bipartition $A\,,B$ of such system of qubits $n_A$ and $n_B=n-n_A$ respectively, like the one shown in Fig.~\ref{fig:main_figure}b. The entanglement entropy of the subsystem $A$ having reduced density matrix $\rho_A = \Tr_B[\rho]$, is defined as
\begin{equation}
\label{eq:ent_entropy}
    S(\rho_A) = -\Tr[\rho_A\log\rho_A]\,,
\end{equation}
and quantifies the amount of entanglement shared between the parties $A$ and its complement $B$\footnote{Note that throughout the whole Chapter we consider logarithms in natural base $e$.}. If $A$ and $B$ are in a product state then $S(\rho_A)=0$, while if the two subsystems share maximal entanglement one has $S(\rho_A) = n_A\,\log(2)$~\cite{EisertEntanglementMPS}. An important property of Eq.~(\ref{eq:ent_entropy}) is that the entanglement entropy of the two subsystems is equal, namely $S(\rho_A) = S(\rho_B)$, as it can be easily checked using the Schmidt decomposition of the pure global state $\rho = \dyad{\psi}$ (see below). 

It turns out that matrix product states are a natural tool to characterise the entanglement entropy of a quantum system. This can be illustrated by considering the simple case of a state of $n=2$ qubits. Indeed, the statevector 
\begin{equation}
    \ket{\psi} = \sum_{i,\,j=0}^1 c_{ij}\ket{ij} \quad \text{with}\quad \sum_{i,\,j=0}^1 |c_{ij}|^2=1,
\end{equation}
can be expressed in the Schmidt decomposition~\cite{NielsenChuang} as
\begin{equation}
\label{eq:schmidt}
    \ket{\psi} = \sum_{\alpha=1}^{\chi_s} \lambda_{\alpha}\ket{\xi_\alpha}_1 \otimes \ket{\eta_\alpha}_2,
\end{equation}
where $\chi_s$ is the \textit{Schmidt rank}, $\lambda_\alpha$ are the Schmidt coefficients, and $\{\ket{\xi_\alpha}_1\}_\alpha, \{\ket{\eta_\alpha}_2\}_\alpha$ are orthonormal bases in the space of the first and second qubit respectively. Using the decomposition~\eqref{eq:schmidt} in Eq.~\eqref{eq:ent_entropy}, the entanglement entropy between the two qubits then amounts to
\begin{equation}
    S\qty( \rho_A ) = -\sum_{\alpha=1}^{\chi_s} \lambda_a^2\log\lambda_\alpha^2\, .
\end{equation}

In an MPS simulation one always has access to a subset of the Schmidt coefficients, since such representation is built by iteratively applying the Singular Value Decomposition (SVD), a procedure equivalent to Schmidt-decomposing a quantum state. The reason why one has access only to subsets of the Schmidt coefficients is that the following conditions are imposed on them. Listing the coefficients in ascending order, i.e. $\lambda_0\geq \lambda_1 \geq\dots\geq \lambda_{\chi_s}$, then
\begin{itemize}
    \item Schmidt coefficients whose ratio with $\lambda_0$ is smaller than $\epsilon$ are discarded. The value of $\epsilon$ in this analysis is fixed at $\epsilon=10^{-9}$;
    \item only the first largest $\chi_\text{max}$ coefficient are retained. The value $\chi_\text{max}$ is called maximum \textit{bond dimension}.
\end{itemize}
The approximation we are performing is the optimal one in terms of the represented entanglement. Then, the measure of entanglement for the MPS now becomes
\begin{equation}
    S(\rho_A) = -\sum_{\alpha=1}^{\chi_{max}} \lambda_a^2\log\lambda_\alpha^2.
\end{equation}

As explained in detail in Appendix~\ref{app:convergence}, despite the approximations, the faithfulness of the simulation can be easily monitored. Finally, we remark that since we have constant access to the considered subset of Schmidt coefficients during the state evolution, we are able to compute the entanglement entropy of a quantum state without any computational overhead.

\subsection{Entanglement entropy in random quantum states}
\label{sec:HaarRandom}
In this section we briefly describe the entanglement features of uniformly distributed random pure quantum states, that is quantum states sampled according to the unique unitarily invariant probability distribution induced by the Haar measure. The Haar measure was already introduced and discussed previously in Chapter~\ref{ch:QC_and_VQAs} for deriving the Barren Plateau phenomenon~\ref{sec:ch_VQAs_BP}, and we hereby just recall some basic concepts necessary for the analysis presented in this Chapter. 

Denoting by $\mathbb{U}(n)$ the group of $2^n \times 2^n$ unitary matrices, there is a unique unitarily invariant probability measure $\mu(U)$ defined on the group, and such measure is called \textit{Haar measure}~\cite{HaydenAspectsGenericEntanglement2006, MeckesRandomMatrixTheory2019, EdelmanRandomMatrixTheory2005}. Unitary invariance corresponds to the requirement that the measure is invariant under translations in the space of unitary matrices, that is
\begin{equation}
    \mu(M U) = \mu(U M) =  \mu(U) \quad U,\, M \in \mathbb{U}(n)\, .
\end{equation}
The Haar measure induces a uniform probability distribution in the space of unitary matrices so that sampling a quantum state according to the Haar measure means randomly picking a state uniformly from the space of quantum states. We denote with $\mathcal{P}(n)$ such probability distribution.

We are interested in the entanglement features of random quantum states, particularly in the entanglement entropy. Let $\ket{\psi} \in (\mathbb{C}^2)^{\otimes{n}}$ be a quantum state of $n$ qubits sampled from the uniform distribution $\ket{\psi} \sim \mathcal{P}(n)$, and a bipartition of the $n$ qubits system in two subsystems $A$ and $B$, of size $n_A$ and $n_B = n - n_A$ respectively. Then, for $n_A \leq n_B$, the expectation value of the entanglement entropy~\eqref{eq:ent_entropy} corresponding to this cut amounts to the so-called Page value~\cite{HaydenAspectsGenericEntanglement2006, Page1993Entanglement}
\begin{equation}
\label{eq:haar_entanglement}
    \mathbb{E}[S(\rho_A)] = \sum_{j = d_B + 1}^{d_A d_B} \frac{1}{j} - \frac{d_A-1}{2d_{B}}\,,
\end{equation}
where $d_B = 2^{n_B}$, $d_A = 2^{n_A}$ are the local dimensions of the two subsystems, and the expectation value is over the uniform probability distribution $\mathbb{E}(\cdot) = \mathbb{E}_{\ket{\psi} \sim \mathcal{P}(n)}(\cdot)$. One can check that the entanglement is highest whenever the two partitions have equal size $n_A=n_B=n/2$ (for $n$ even, and similarly for $n$ odd, $n_A = {\lfloor\frac{n}{2}\rfloor}$ and $n_B = {\lceil\frac{n}{2}\rceil}$). 

From Eq.~\eqref{eq:haar_entanglement} it follows that $\mathbb{E}[S(\rho_A)] \geq \log d_A - d_A /2 d_B$~\cite{HaydenAspectsGenericEntanglement2006}, and since the maximum value of the entanglement entropy for such bi-partition is $\log d_A$, obtained if the subsystems $A$ and $B$ share maximal entanglement, one concludes that random states are generally highly entangled. Indeed, in ref.~\cite{HaydenAspectsGenericEntanglement2006} it was shown that the probability that a random pure state has entanglement entropy lower than $\log d_A - d_A /2 d_B$ is exponentially small. Thus, with very high probability, random quantum pure states are almost maximally entangled. 

\subsection{Quantum Neural Networks as Parameterised Quantum Circuits}\label{sec:QNN}
In this section we briefly recall the main ideas and nomenclature of Variational Quantum Algorithms (VQAs) and Quantum Neural Networks (QNNs) to present the analysis in this Chapter in a self-consistent manner, but we refer to Section~\ref{sec:VQAs} and Section~\ref{ssec:ch_QML_NeuralNetworks} for extended discussion on these topics. 

Variational quantum algorithms are based on PQCs, which are quantum circuits in which some of the unitary operations are characterized by \textit{variational} parameters to be adjusted in order to solve an optimization problem. The optimal parameters are found by minimizing a properly chosen cost (or loss) function encoding the task to be solved. Let $U_{\bm{\theta}}$ be the unitary evolution implemented by a quantum circuit with tunable parameters $\bm{\theta}$, and $O$ a Hermitian operator (an observable). The goal of variational quantum algorithms is to optimize the quantum circuit parameters $\bm{\theta}$ in order to minimize the expectation value (or variations thereof, see Sec.~\ref{sec:VQAs})
\begin{equation}
\label{eq:cost_function}
    f(\bm{\theta}) = \expval{O}_{\bm{\theta}} = \Tr[O\, U_{\bm{\theta}} \rho U^\dagger_{\bm{\theta}}]
\end{equation}
where $\rho$ is an initial quantum state, generally set to the ground state $\rho = \dyad{\bm{0}}$. This is achieved by means of an iterative hybrid quantum-classical approach where the quantum computer is used to estimate the cost function~\eqref{eq:cost_function}, and given such value, the classical computer proposes new variational parameters according to an optimization method, the most common one being gradient descent. 

There is freedom in the choice of the gate sequence defining the parameterized unitary $U_{\bm{\theta}}$, and a choice of its structure is referred to as variational \textit{ansatz}, of which we gave an extended overview in the dedicated Section~\ref{ssec:ch_VQA_ansatze}. For example, the unitary could be composed of a layer of Pauli rotations around the $X$-axis on each qubit $R_X(\theta) = \exp(-i\theta X/2)$, followed by a layer of CNOTs acting on pairs of neighbouring qubits. This is in fact the general blueprint of variational quantum circuits, as they are generally created by repeating single-qubits parameterized rotations followed by multi-qubits operations which introduce entanglement into the computation. Examples of parameterized quantum circuits are shown in Fig.~\ref{fig:main_figure}.\\

\paragraph*{Quantum Neural Networks} As it is often the case with learning tasks, either classical or quantum, the goal is to solve a problem given access to a dataset of inputs $\qty{\bmx_i}_i,\, \bmx_i \in \mathcal{X}$, representative of the task to be solved. As outlined in Definition~\ref{def:qml_model}, whenever data is involved, variational quantum circuits are often referred to as quantum neural networks. In this case, the quantum circuit of the ``neural network'' depends on two sets of parameters $\bm{x}$ and $\bm{\theta}$, the former being the input data to be analyzed, and the latter the variational parameters to be adjusted (i.e. the \textit{weights} of the neural network). In the quantum machine learning jargon, the encoding scheme used to load the input data onto the quantum computer is known as \textit{feature map}, and consists of a unitary operation parameterized by $\bm{x}$. We will denote such feature encoding gate with $\mathcal{F}(\bm{x})$, where $\bm{x}\in \mathcal{X}$. As with the variational unitary, there is no standard choice for a feature map, and one has to pick a specific ansatz, ideally biasing the choice towards architectures built using knowledge of the problem to be solved~\cite{skolik2022equivariant, MeyerSimmetriesQML_2022}. Summing up, a general QNN can be then expressed as
\begin{equation}
\label{eq:qnn_equation}
    U_{\text{QNN}}(\bm{x};\bm{\theta}) = \prod_{i=L}^1 V(\bm{\theta}_i)\mathcal{F}(\bm{x}) =  V(\bm{\theta}_L)\mathcal{F}(\bm{x})\cdots V(\bm{\theta}_1)\mathcal{F}(\bm{x})\,, 
\end{equation}
where $\mathcal{F}(\bm{x})$ is the feature map ansatz depending on the input data $\bm{x}$; $V(\bm{\theta}_i)$ is a variational ansatz depending on trainable parameters $\bm{\theta}_i\in \bm{\theta} = (\bm{\theta}_1, \cdots, \bm{\theta}_L)$ with $\bm{\theta} \in \mathbb{R}^p$; and $L$ is the number of repetitions (or \textit{layers}) of the such layered structure. 

As already discussed in Sec.~\ref{sec:ch_QML_Reuploading}, it was recently shown that uploading the input data multiple times throughout the circuit is essential for quantum neural networks to model higher-order functions of the inputs~\cite{Schuld2020Encoding, Theis2020Expressivity}, via a procedure now dubbed \textit{data-reuploading}~\cite{Torrontegui2019Perceptron, Perez2020Reuploading}. Indeed, note that the input data in the feature map in Eq.~\eqref{eq:qnn_equation} is the same in every layer, while the variational blocks $V$ use a different parameter vector in every layer. In Fig.~\ref{fig:main_figure}a we give a graphical representation of the general structure of QNNs. As for the explicit implementation of $\fmap$ and $\varans$, there is no fixed choice and these are usually composed of single qubit rotations followed by entangling operations, either fixed (e.g. CNOTs) or themselves parameterized (e.g. controlled rotations). See Fig.~\ref{fig:main_figure}c for some prototypical examples of parameterized blocks proposed in the literature~\cite{AbbasPowerQNN2021, SimPQCs2019}, which we will consider throughout the presented analysis.

\subsection{Randomness, Entanglement and Trainability}\label{par:2des_ent_bp}
As we now well know from the discussion in Sec.~\ref{sec:ch_VQAs_BP}, one of the hardest theoretical challenges affecting quantum machine learning models is the emergence of \textit{barren plateaus} (BP). Different sources can lead to the unfolding of barren plateaus, and these can be summarised as follows\footnote{In Chapter~\ref{ch:QC_and_VQAs} we anticipated that the four sources of BP shown in Fig.\ref{fig:barren_plateau_pictorial} actually can be reduced to three, since, as clearly explained in this section, expressibility-Induced BP and entanglement-induced BP both derive from an abundance of randomness inside the quantum circuit. For this reason, we hereby refer to these types of BPs as randomicity-induced BP.}: randomicity-induced BP~\cite{McCleanBarren2018, Holmes2021connecting, MarreroBarrenEntanglement2021, SackBPShadows2022}, BP induced by \textit{global} cost functions defined with observables having support on a large number of qubits~\cite{CerezoBarrenLocalCost2021}, and eventually noise-induced BP~\cite{WangNoiseinducedBarrenPlateaus2021}.

The analysis presented in this chapter concerns the former type of barren plateaus, that roughly occur when parameterized quantum circuits, when initialised with random parameters, resemble general random unitaries. Indeed, despite being quite limited in terms of qubits connectivity and gate operations, common instances of parameterized quantum circuits are often found to behave as unitary 2-designs~\cite{Dankert2des2009}, in which case, as proved in Th.~\ref{th:2des_grad}, the variance of the gradients of any cost function $f(\bm{\theta})$ defined on the circuit will vanish exponentially with the number of qubits $n$, namely~\cite{Holmes2021connecting}
\begin{equation}
    \label{eq:bp_vanishing}
    \text{Var}_{\bm{\theta}}[\partial_k f(\bm{\theta})] \in \order{c^{-n}}\, \quad c>1\,,
\end{equation}
where $f(\bm{\theta})$ is as in Eq.~\eqref{eq:cost_function}. Specifically, the cost function concentrates around its mean value and stays constant almost everywhere in parameter space~\cite{Arrasmith2021equivalence}, which makes training unfeasible. 

Vanishing gradients are used as a \textit{witness} to assess whether a parameterized quantum circuit resembles a unitary 2-designs. Of course, this is only \textit{necessary} but not \textit{sufficient} condition, as one can easily devise a circuit that is not a 2-design but has vanishing gradients, for example using a global cost with a shallow circuit~\cite{CerezoBarrenLocalCost2021}, as the one presented in Appendix~\ref{sec:app_global_cost}.

In addition to vanishing gradients, another witness of randomness is the entanglement generated inside the circuit~\cite{MarreroBarrenEntanglement2021}. Indeed, as discussed previously in Sec.~\ref{sec:HaarRandom}, random quantum states are almost maximally entangled, so one can use the maximality of entanglement generated by a parameterized circuit as an indicator of the resemblance to a random unitary evolution. As for vanishing gradients, the presence of large entanglement is however only a necessary but not sufficient condition for randomness, as a simple shallow circuit composed of Hadamards and CNOTs can create maximally entangled states (GHZ states), which are clearly not random. As discussed in~\cite{SackBPShadows2022}, the so-called entanglement-induced BPs~\cite{MarreroBarrenEntanglement2021, Patti2021EntDevisedBP} provide an alternative yet equivalent description of local cost barren plateaus (circuits with global costs always suffer of vanishing gradients~\cite{CerezoBarrenLocalCost2021}, regardless of randomicity), as they both stem from the proximity of parameterized quantum circuits to unitary 2-designs. 

Indeed, if a circuit is a unitary 2-design, then the average entanglement entropy of any subsystem $A$ of dimension $d_A$ ($d_A \leq d_B$) will be already very close to its maximal value~\cite{SackBPShadows2022, LiuEntanglementQuantumRandomness2018}
\begin{equation}
\label{eq:2des_vonn}
    \log d_A - 1 \leq \mathbb{E}_{\bm{\theta}}[S(\rho_A)] \leq \log d_A\,,
\end{equation}
and approaches the Page value~\eqref{eq:haar_entanglement} for truly Haar-random states. For completeness, we provide a proof of Eq.~\eqref{eq:2des_vonn} based on the Rényi 2-entropy in  Appendix~\ref{app:reny}.

To summarise, while the presence of entanglement is a necessary ingredient to avoid classical simulability, its uncontrolled growth is likely to signal the emergence of barren plateaus. The evaluation of the entangling capabilities of parameterized quantum circuits is then a valuable diagnostic tool to provide information both on the classical simulability and trainability issues of quantum machine learning models. At last, we remark that although various methods have been put forward to mitigate the occurrence of BPs~\cite{Grant2019Initialization, Volkoff2020Correlating, SkolikLayerwise2021}, including proposals based on entanglement control~\cite{SackBPShadows2022, Patti2021EntDevisedBP, Joonho2021EntBP}, these remain a bottleneck for scaling up quantum machine learning computations based on variational circuits. 

\section{Results}\label{sec:ch_Entanglement_Results}
We now proceed to analyse the entanglement production in various quantum neural network architectures with different feature maps and variational ansatz, obtained composing the circuit blocks shown in Fig.~\ref{fig:main_figure}. In particular, we take as a prototypical example the QNN introduced in \cite{AbbasPowerQNN2021}, argued as a good candidate for quantum machine learning applications in terms of capacity and expressibility, possibly achieving an advantage over classical counterparts. 

Such QNN model uses as feature-map $\mathcal{F}(\bm{x})$ the so-called $\textsc{ZZFeatureMap}$ firstly introduced in~\cite{Havlicek2019QSVM} as a classically-hard map to load classical data on a quantum state in a non-linear fashion. The variational block $V(\bm{\theta})$ is instead composed of single qubit rotations followed by entangling operations. In order to better understand the effect of every single operation in the quantum circuit, we also consider variations of the QNN introduced above, varying both the feature map, the variational form, and the entangling topology. All considered circuit blocks are graphically represented in Fig.~\ref{fig:main_figure}. 

Let $U_L(\bm{x}, \bm{\theta})$ be the unitary representing a specific quantum neural network with $L$ layers with input data $\bm{x}=(x_1, \hdots, x_d) \in \mathbb{R}^d$, and variational parameters $\bm{\theta}=(\theta_1, \hdots, \theta_p) \in \mathbb{R}^p$, see Eq.~\eqref{eq:qnn_equation}. We consider random instances of such QNN by sampling both the inputs and the variational parameters according to the uniform distribution $x_i, \theta_i \sim \text{Unif}[0, \pi]$, hence obtaining a collection of QNNs $U_L = \{U_L(\bm{x}_i, \bm{\theta}_i)\,, i=1, \hdots, M\}$. Then, we study the entanglement entropy properties of each of these instances and average the result over the $M$ trials (unless stated otherwise, we take $M=100$). Thus, when in the following we refer to the entanglement entropy of a quantum circuit, we are always denoting the average over $M$ realisations of that circuit. In order to evaluate the influence of the depth on the entanglement, we repeat this analysis by increasing the number of layers in the quantum neural network $L=1, \ldots, L_\text{max}$. 

Note that although the total number of parameters (inputs and parameters) depends on the specific feature map and variational form used, for the considered circuits such difference generally amounts to a constant and does not have a relevant impact on the results. In Tab.~\ref{tab:summary_circuit} we report the number of parameters in each circuit template analyzed in this work. We anticipate that while the number of parameters in the considered quantum circuits only scales polynomially with the system size $n$, these are found to be sufficient to reproduce some entanglement features of random unitaries, which are instead characterized by an exponential number parameters. This is in agreement with results on random quantum circuits that states that polynomial resources are sufficient to approximate unitary designs~\cite{HarrowRandomQuantumCircuits2009, HaferkampRandom}. We refer to Sec.~\ref{sec:ch_Entanglement_Discussion} for an extended discussion.
\begin{table}[ht]
    \centering
    \begin{tabular}{ccccc} \toprule
                                      & Abbr. &\multicolumn{3}{c}{Number of parameters}\\
                                      &       & \textsc{Linear} & \textsc{Circular} & \textsc{Full} \\ \midrule
        \textsc{Circuit 1}            & $\circuit{1}$   & $2n$  &   $2n$  & $2n$      \\     
        \textsc{Circuit 2}            & $\circuit{2}$   & $n$   & $n$  & $n$      \\
        \textsc{Circuit 3}            & $\circuit{3}$   & $2n-1$& $2n$ & $\frac{n^2+n}{2}$ \\
        \textsc{Circuit ZZFeatureMap} & $\circuit{zz}$  & $n$   & $n$ & $n$       \\
    \bottomrule
    \end{tabular}
    \caption[Number of parameters in considered ansätze]{Number of parameters for each considered circuit template and their relative entanglement topology. Notice that, while the number of parameters remains constant for  $\circuit{zz}$ as shown in the table, the number of parametric gates varies analogously to $\circuit{3}$.}
    \label{tab:summary_circuit}
\end{table}

\subsection{Alternating vs.~Sequential data reuploading}
As a first analysis, we study the difference in entanglement growth between a standard QNN using an \textit{alternated} repetitions of feature maps and variational forms (as in Fig.~\ref{fig:main_figure}), and one in which we have first $L$ repetitions of the feature map followed by $L$ repetitions of the variational form, which we refer to as a \textit{sequential} structure. The former leverages an alternated evolution of the quantum state which is typical of quantum neural networks using a data reuploading scheme~\cite{Perez2020Reuploading, Schuld2020Encoding, Theis2020Expressivity}. The latter instead uses an initial data-dependent evolution followed by a trainable unitary, thereby creating an architecture similar to quantum kernel machines~\cite{SchuldKernel2021}.

While the two structures (alternated and sequential) may be mapped to each other using ancillary qubits~\cite{Jerbi2022BeyondKernel}, they can have rather different performances, and we hereby show how they also create entanglement in a different way. Specifically, given the two unitary evolutions, namely the fixed input-dependent feature map $\mathcal{F}(\bm{x})$ and the varying parameterised variational form $V(\bm{\theta}_i)$, one expects the alternated dynamics
\[
U_\text{alt} = V(\bm{\theta}_L)\mathcal{F}(\bm{x})\, \ldots \,V(\bm{\theta}_1)\mathcal{F}(\bm{x})\,,
\]
to introduce randomness at a faster rate than the sequential process
\[
U_\text{seq} = \prod_{i=L}^1 V(\bm{\theta}_i)\,\prod_{i=L}^1\mathcal{F}(\bm{x})\,,
\]
and hence introduce more entanglement in the system. Such intuition is confirmed by the numerical results, and may be understood as a consequence of the universality of the alternating dynamics proved for example for QAOA circuits~\cite{LloydUniversalityQAOA, MoralesUniversalityQuantumApproximate2020}.

\begin{figure}[ht]
    \centering
    \includegraphics[width=0.7\textwidth]{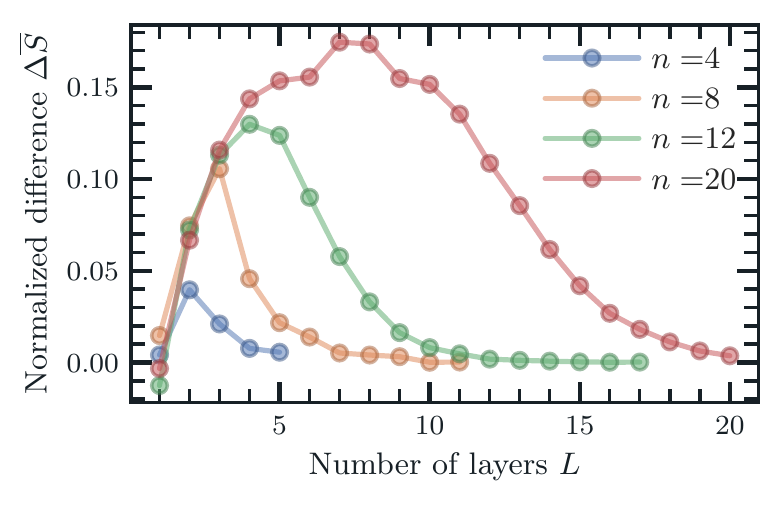}
    \caption[Normalised entanglement for various numbers of qubits]{Normalised entanglement difference $\Delta \overline{S}$, as defined in Eq.~\eqref{eq:ent_diff} for different numbers of qubits. The used QNN is defined by $\fmap = \circuit{zz}$ and $\varans = \circuit{2}$. All the data points are obtained by averaging over $10^3$ realisations.}
    \label{fig:ent_diff}
\end{figure}

Here we use $\mathcal{F}=\circuit{zz}$ and $V=\circuit{2}$, as defined in Fig.~\ref{fig:main_figure}, both with linear topology. Be $S^\text{alt}$ and $S^\text{seq}$ the entanglement entropy of the bipartition with an equal number of qubits, which is generally the highest, for the alternating and sequential structure, respectively. We define the normalized difference as 
\begin{align}
    \label{eq:ent_diff}
    \Delta \overline{S} = \frac{S^\text{alt}-S^\text{seq} }{(S^\text{alt}+S^\text{seq})/2}\,,
\end{align}
and study its behaviour as the depth of the quantum circuit is increased, as shown in Fig.~\ref{fig:ent_diff}. 

The metric is always positive and features a maximum, implying that the alternated structure is creating entanglement \textit{faster} (i.e. with fewer layers) than its non-alternated counterpart. Note that for $L=1$ layers the two structures are identical, so the generated entanglement is the same up to the statistical error, which explains why all the curves start around zero. At a high number of repetitions, the two structures tend to the same value, showing a $\Delta\overline{S}\simeq 0$, which can be understood in light of the results presented in the following sections: as the number of layers of a QNN is increased, the entanglement rapidly converges to that of a Haar-distributed random state, thus the alternated and non-alternated structure eventually converge to the same value. Given the higher entanglement production rate of the alternated structure, in the following analysis, we shall focus on this structure only.

\subsection{Entanglement distribution across bonds}\label{sec:EntanglementBond}
\begin{figure}[!ht]
    \centering
    \includegraphics[width=0.6\textwidth]{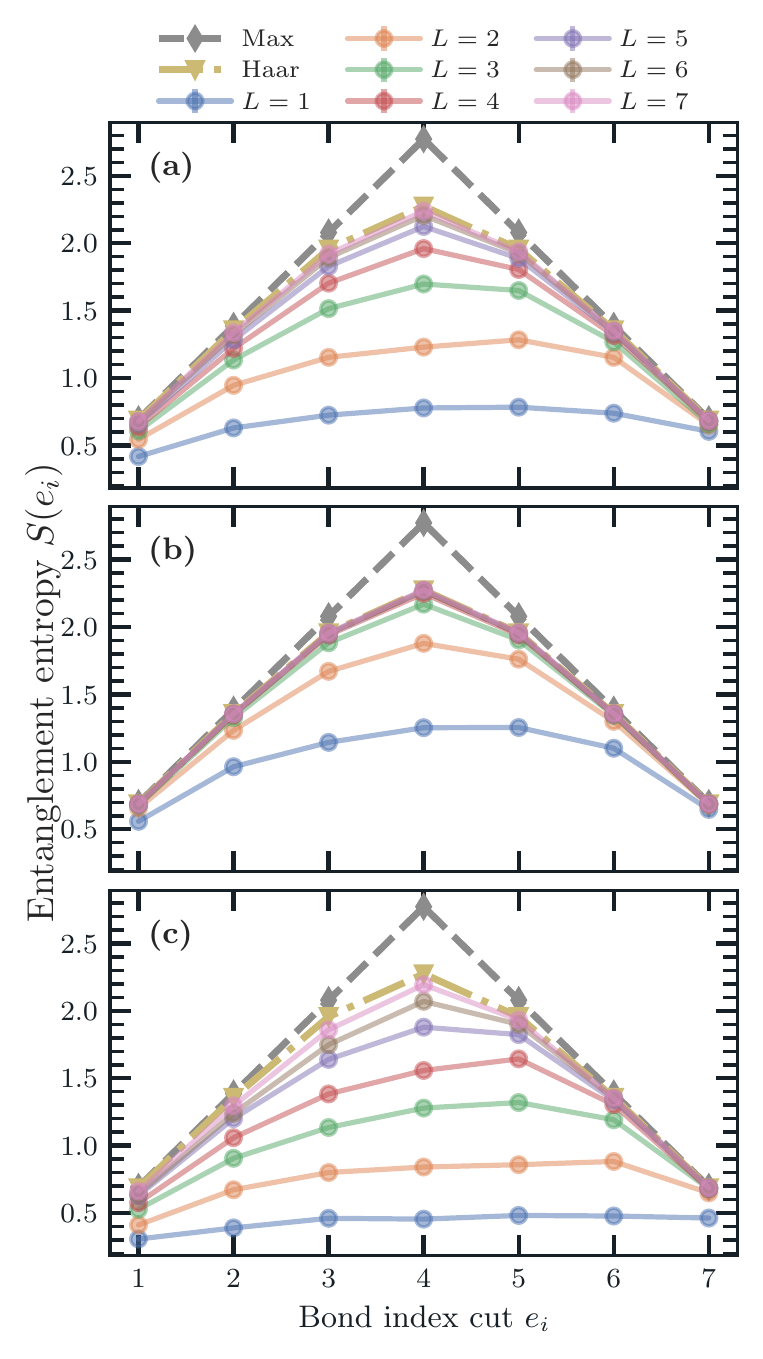}
    \caption[Average entanglement entropy across bonds]{Average entanglement entropy across bonds for a system composed of $n=8$ qubits, where $e_i$ is the bond connecting qubit $i$ and $i+1$, as in Fig.~\ref{fig:main_figure}. The curves represent different numbers of layers $L$ in the quantum neural network. \textbf{(a)} QNN with structure: $\mathcal{F} = \circuit{zz}$, $V = \circuit{2}$, both with linear entanglement. \textbf{(b)} QNN with structure: same as in (a) but with circular entanglement. \textbf{(c)} QNN with structure: $\mathcal{F} = \circuit{1}$ which has no entangling gates, and $V = \circuit{2}$ with linear entanglement.}
    \label{fig:bond_entanglement}
\end{figure}
It is natural to ask how the choice of the feature map, the variational form, and the entangling topology impact the growth of entanglement of the quantum state. In this section, we start to explore this question by studying how entanglement is distributed across all possible ordered bi-partitions of the $n$ qubits in the network. That is, given an MPS representation as in Fig.~\ref{fig:main_figure}(b), we study the entanglement entropy corresponding to each bond in the linear chain. Denoting with $e_i$ the bond connecting qubit $q_i$ and $q_{i+1}$, the entanglement entropy of that bond is (see Eq.~\eqref{eq:ent_entropy})
\begin{equation}
\label{eq:ent_entropy_bonds}
\begin{aligned}
    S(e_{i}) & = -\Tr[\rho_{[1:i]}\log \rho_{[1:i]}] \\
    \rho_{[1:i]} & = \Tr_{i+1, \hdots, n}[\rho]
\end{aligned}\,\,,
\end{equation}
where $\rho_{[1:i]}$ is the reduced density matrix of all the qubits up to the $i$-th one, and $\rho$ is the state obtained from the quantum neural network $\rho = U_L(\bm{x}, \bm{\theta}) \dyad{\bm{0}} U_L(\bm{x}, \bm{\theta})^\dagger$.

In Fig.~\ref{fig:bond_entanglement} we show the entanglement entropy distribution for the case of $n=8$ qubits using three different quantum neural networks architectures: in panel (a) the one proposed in~\cite{AbbasPowerQNN2021} with feature map $\mathcal{F}=\circuit{zz}$, variational ansatz $V=\circuit{2}$, both with linear entanglement; in (b) same as before but using a circular entanglement topology; and eventually in panel (c) a simpler circuit using a tensor product feature map $\mathcal{F}=\circuit{1}$ which encodes data independently on each qubit, followed by the same variational ansatz $V=\circuit{2}$, again with linear entanglement both. For reference, it is also shown the expectation value of the entanglement entropy for Haar-random quantum states evaluated with Eq.~\eqref{eq:haar_entanglement}, as well as an upper bound given by the highest possible entanglement $\log(\text{min}(d_A, d_B))$, obtained if the two partitions $A$ and $B$ were maximally entangled. Note that while we report only the simulation data for $n=8$, the discussion has general validity as identical results hold for all tested numbers of qubits, $n=2,\ldots, 20$.

First of all, the findings agree with the intuition that deeper circuits are able to create higher entangled states with respect to shallower ones, in accordance with results from~\cite{SimPQCs2019}. In particular, the entanglement entropy is higher at the centre of the chain. 
Clearly, depending on the specifics of the QNN, the entanglement grows faster in certain architectures with respect to others. Regarding the effect of the entangling topology, comparing panels (a) and (b) we see that circular connections produce greater entanglement compared to the nearest-neighbours interaction and that such entanglement grows at a faster rate as the number of layers is increased. As for the choice of the feature map, since the QNN in panel (c) produces entanglement only through the entangling gate in the variational blocks, its entanglement is lower and also grows slower with respect to the QNN in panel (a), even though it has twice the number of parameters in the feature map.

Interestingly, however, as the number of layers approaches the number of qubits $L\approx n$, all investigated QNNs converge to the same values, that is those obtained for random states sampled from the uniform Haar distribution. Deep enough QNNs are then flexible enough to reproduce the same entanglement spectrum of a random state, which, as discussed in section~\ref{sec:HaarRandom}, are very highly entangled. Again, even though the measure of entanglement is different, this is in agreement with the results presented in~\cite{SimPQCs2019}, where the convergence to the Haar distribution is encountered for various parameterized quantum circuits, and also with other results in the literature regarding the properties of random quantum circuits to approximate the Haar distribution~\cite{BrandaoLocalRandomQuantum2016, Harrow2018approximate}. We will discuss this more in detail in Sec.~\ref{sec:ch_Entanglement_Discussion}, while a more in-depth analysis of the convergence is the subject of the next section.

\subsection{Entanglement scaling with increasing depth}\label{sec:EntanglementScaling}
In order to better understand the entanglement scaling properties of QNNs, we introduce a new quantity, defined as the total entanglement entropy $S_{tot}$ created in the MPS chain
\begin{equation}
\label{eq:total_entanglement}
    S_\text{tot} = \sum_{i=1}^{n-1} S(e_i)\, ,
\end{equation}
which is the sum of the entanglement entropy of all the ordered bipartitions of the quantum state. We use this global measure to quantify how fast QNNs approach the Haar distribution in terms of overall entanglement production. In particular, we define a new figure of merit, the entangling layers $\widetilde{L}$, defined as the number of layers needed by an architecture to reach $90\%$ of the total entanglement of a Haar distributed state $S_{tot}^{\text{Haar}}$, namely
\begin{equation}
\label{eq:tildeL}
    \widetilde{L} = \text{min number of of layers } \quad \text{such that} \quad S_\text{tot} \geq 0.9\,S_\text{tot}^{\text{Haar}}.
\end{equation}
The choice of the $90\%$ threshold allows to select states that are already very close to the Haar-random value, and avoids undesired oscillating behaviours obtained when higher thresholds are used, e.g. $99\%$, which are caused by statistical fluctuations (recall that every QNN is sampled multiple times with different parameters to calculate averages).

\begin{figure}[ht]
    \centering
    \includegraphics[width=\textwidth]{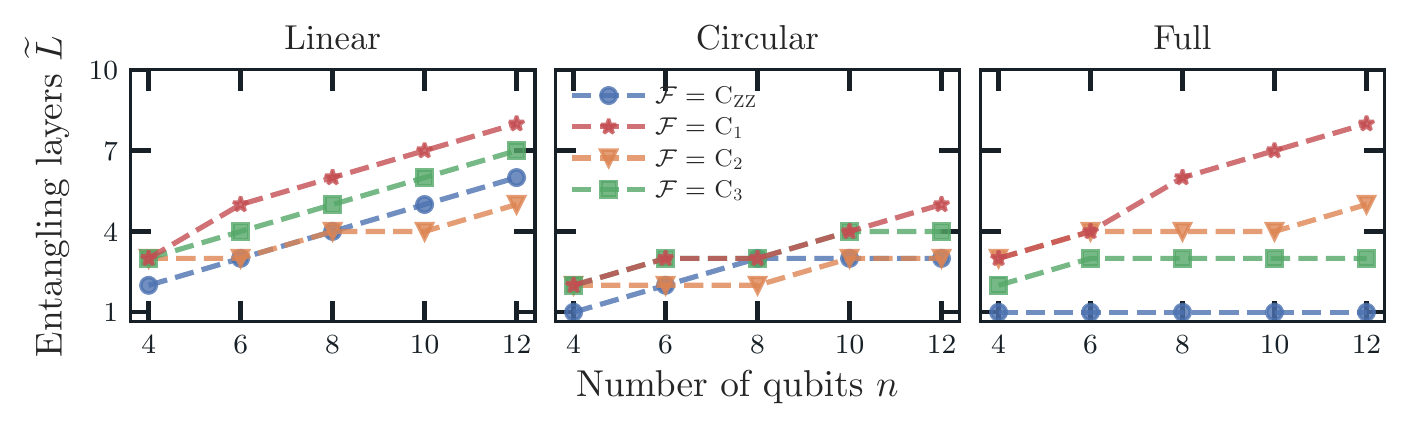}
    \caption[Number of layers to reach a target Haar-randomness]{Entangling layers $\widetilde{L}$, i.e number of layers to reach $90\%$ of the Haar entanglement, versus the number of qubits. The analysis is carried out over four different QNN architectures, each evaluated with different entangling topologies (\textit{linear}, \textit{circular}, and \textit{full}). The architectures leverage the same variational form $\varans$, while the feature map $\fmap$ is changed, as reported in the legend. }
    \label{fig:ent_scaling}
\end{figure}

In Fig.~\ref{fig:ent_scaling} we show the behaviour of $\widetilde{L}$ for four different QNNs as the number of qubits is increased. Note that each QNN is considered with all the three possible entangling topologies (\textit{linear}, \textit{circular} and \textit{full}, as defined in Fig.~\ref{fig:main_figure}). At last, note that all QNNs leverage the same variational form $\varans = \circuit{2}$, while the feature map is changed, as reported in the legend.

First, we observe that the entangling layers display a linear behaviour when a linear entanglement topology is used. This means that the number of layers needed to entangle the system scales linearly with the size of the system. The behaviour changes abruptly when we move to a circular or full entangling topology. All architectures display a faster entanglement production when passing from a linear to a circular topology, as can be seen from the lower slope of the curves. The all-to-all connectivity speeds up entanglement production only for $\fmap = \circuit{zz},\,\circuit{3}$, while the circuits $\fmap = \circuit{2},\,\circuit{1}$ show essentially the same behaviour of the linear case. We now proceed to discuss more in detail such results.

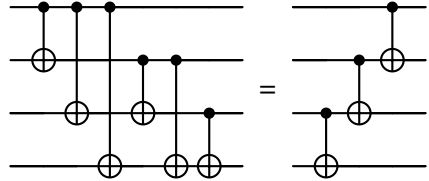
\begin{figure}[!ht]
    \centering
    \begin{quantikz}[column sep={12.4pt,between origins}, row sep={20pt,between origins}]
    & \ctrl{1} & \ctrl{2} & \ctrl{3} 	& \qw  		& \qw 	 & \qw		& \qw \\
    & \targ{}  & \qw		& \qw 			& \ctrl{1} & \ctrl{2} & \qw		& \qw \\ 
    & \qw 		& \targ{}  & \qw			& \targ{} 	& \qw		 & \ctrl{1}& \qw \\
    & \qw 		& \qw 		& \targ{}		 & \qw 		& \targ{} & \targ{}	& \qw 
    \end{quantikz}
    =
    \begin{quantikz}[column sep={12.4pt,between origins}, row sep={20pt,between origins}]
    & \qw 	 	& \qw		& \ctrl{1}&\qw \\
    & \qw 		& 	\ctrl{1}	& \targ{} & \qw \\ 
    & \ctrl{1}	& \targ{} & \qw		& \qw \\
    & \targ{} & \qw		& \qw 		& \qw 
    \end{quantikz}
    \caption[Equivalence of full and linear entangling topologies]{Circuital identity between a full (all-to-all) entangling map made of only CNOTs and the inverse of a linear (nearest-neighbours) entangling map.}
    \label{fig:cnot_equivalence}
\end{figure}

We start comparing the entangling capabilities of $\circuit{zz}$ vs. $\circuit{2}$. Both with linear and circular entangling topology, $\circuit{2}$ is able to produce entanglement essentially at the same rate as $\circuit{zz}$, despite $\circuit{2}$ being of a much simpler structure, with half the number of two-qubit gates. However, things change dramatically using a full entangling map, as the QNN reaches the $90\%$ threshold already at $\widetilde{L}=1$, while $\circuit{2}$ needs more layers, showing the same dependence of a linear connectivity. While counter-intuitive at first, is it easy to see that the entanglement generated by $\circuit{2}$ with a \textit{full} architecture is indeed equivalent to the \textit{linear} one. This is due to a simple circuit identity regarding networks of CNOTs reported in Fig.~\ref{fig:cnot_equivalence}. Such circuital identity holds for any number of qubits, which makes the full entangling map as shown in Fig.~\ref{fig:main_figure} just as a linear entangling map in disguise (in particular, it is the inverse of the linear entangling map). See Appendix~\ref{app:entangling_maps} for a more precise statement, discussion and proof. Such circuital identity thus explains the equivalence of the yellow ($\fmap = \circuit{2}$) and red ($\fmap = \circuit{1})$ curves between the first and last plot of Fig.~\ref{fig:ent_scaling}.

Such equivalence clearly does not hold if controlled rotations are used instead of CNOTs. Indeed, the feature map $\fmap = \circuit{3}$ uses controlled rotations with independent random parameters, and given that these gates do not cancel out, the entanglement is always increasing going from low to high connectivity. Note that such increase is mainly due to the feature map, as the variational ansatz $\varans = \circuit{2}$ is the same as other structures, suffering from the CNOTs cancellation issue described above. 

For comparison, we also show the performances of a QNN with the tensor product feature encoding $\fmap =  \circuit{1}$, using no entangling operations. Interestingly, even if this QNN uses two-qubit interactions only inside the variational blocks, these are sufficient to create entanglement similar to other considered QNNs, even at a slower yet comparable rate.

We report in Appendix \ref{app:ext_ent_scaling} the complete simulation results detailing the evolution of the entanglement with the depth of the circuit, for different numbers of qubits.

\subsection{Entanglement Speed}\label{sec:MPSSimulation}
So far we have presented numerical evidence for the entanglement production in QNNs up to a maximum of 20 qubits. In the following we extend the analysis leveraging MPS to simulate quantum systems of bigger size up to 50 qubits, with a maximum bond dimension of $\chi_{\max}=4096$. More importantly, we show how the entanglement growth follows a behaviour that is specific to each particular QNN architecture and the number of layers considered, but independent of the number of qubits in the circuit. We can thus uniquely assign an \textit{entanglement speed} value to each QNN, which, we stress again, only depends on the choice of the ansatz, and holds identically for any instantiation of that QNN with arbitrary number of qubits.

Taking into account the entanglement growth discussed in Sec.~\ref{sec:EntanglementScaling}, we restrict the analysis to a linear architecture, to increase as much as possible the number of layers we can correctly simulate with tensor networks techniques. Indeed, the entanglement production with a circular or full topology is too fast to allow for a convergent simulation with MPS for deep circuits.

Furthermore, we introduce the maximum Haar entanglement entropy, defined as the maximum across all bond entropies for a given number of qubits, as
\begin{equation}
\label{eq:haar_max}
    S^{\text{Haar}}_{n,\, \max} = \max_A\big( \mathbb{E}[S(\rho_A)] \big) \approx \frac{n}{2}\log 2 - \frac{1}{2}\quad \text{for}~n_A=\frac{n}{2} \gg 1,
\end{equation}
where the approximation in the second line has an error that scales as $\order{2^{-n/2}}$, see Appendix \ref{app:haar} for its derivation. Thus, for $n\geq30$ qubits, when the exact computation of the Haar entanglement entropy is unfeasible, we employ the approximated Eq.~\eqref{eq:haar_max}. Finally, we define the normalised entanglement entropy $\widetilde{S}_n$ as
\begin{align}
\label{eq:normalised_entanglement}
    \widetilde{S}_{n}=\frac{\max_{e_i}\left[S(e_i)\right]}{ S^{\text{Haar}}_{n, \max}}\,.
\end{align}
We stress that $\widetilde{S}_n$ is normalised to the maximum Haar entanglement for a fixed $n$, not to the real maximum of the entanglement, which would be $S=\frac{n}{2}\log 2$ for the equal size bipartition. 

\begin{figure}[ht]
    \centering
    \includegraphics[width=0.6\textwidth]{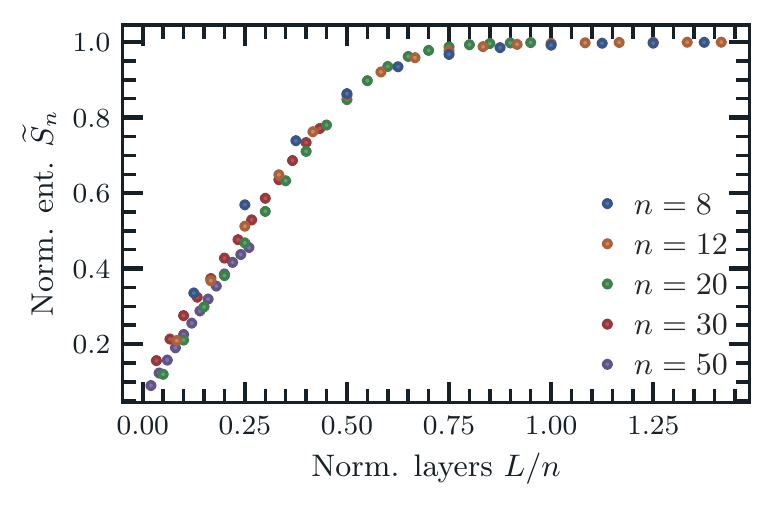}
    \caption[Normalised entanglement versus normalised number of layers]{Normalised entanglement $\widetilde{S}_n$~\eqref{eq:normalised_entanglement} versus the normalised number of layers $L/n$, for different number of qubits $n$, and for the QNN defined by $\fmap=\circuit{zz},~\varans=\circuit{2}$ with linear connectivity. All the normalised entanglement points follow the same curve, independently of the system size $n$. The points for $n=8,\, 12,\, 16,\, 20$ are obtained by averaging over $10^3$ samples, while for $n=30,\, 50$ the averages are over 10 samples.}
    \label{fig:normalized_entanglement}
\end{figure}

In Fig.~\ref{fig:normalized_entanglement} we show the evolution of $\widetilde{S}_{n}$ versus the normalised number of layers $L/n$ for $n \in \{8,\, 12,\, 16,\, 20,\, 30,\, 50\}$ qubits, for the QNN defined with $\fmap=\circuit{zz}, \; \varans=\circuit{2}$ with linear connectivity. We note that all the points, independently of the system size $n$, follow the same curve: an initial linear growth of the entanglement is followed by a saturation to the Haar-random value for the entanglement entropy~\eqref{eq:haar_entanglement}. In particular, we check this behaviour also at large system sizes with $n=30,\, 50$ qubits and circuits with up to $L=11$ layers, and confirm that such scaling is indeed size independent. See Appendix~\ref{app:convergence} for a discussion on the errors introduced by truncation in the MPS representation for simulations with $n=30,\, 50$ qubits.

\begin{table}[ht]
    \centering
    \begin{tabular}{ccc} \toprule
         Feature map $\fmap$      & Variational ansatz $\varans$ & Entangling speed $v_s$ \\ \midrule
         $\circuit{zz}$ & $\circuit{2}$   & $(1.8\pm0.1)$  \\ 
         $\circuit{zz}$  & $\circuit{3}$ & $(0.59\pm0.02)$\\ 
         $\circuit{1}$  & $\circuit{3}$   & $(0.316\pm0.006)$  \\ \bottomrule
    \end{tabular}
    \caption[Entangling speed]{Entangling speed, i.e. a measure of how fast the entanglement is created, for different QNN architectures. These results are obtained using up to $16$ qubits.}
    \label{tab:ent_speed}
\end{table}

Inspired by the behaviour of $\widetilde{S}_n$, we introduce a measure for the entanglement production which is specific to a given QNN architecture (feature map plus variational ansatz) and independent of the number of qubits. Borrowing from the literature on random quantum circuits, it is known that the entanglement of a system undergoing random evolution initially grows linearly in time (depth of the circuit) before reaching the plateau of Haar random states~\cite{Calabrese2005EntEntropyGrowth, TianciEntGrowth2019, NahumEntGrowth2017, Joonho2021EntBP}. Indeed, as clear from Fig.~\ref{fig:normalized_entanglement}, we observe the same initial linear growth, and thus we define the \textit{entangling speed} $v_s$ as
\begin{equation}\label{eq:ent_speed}
    \widetilde{S}_n \propto v_s \cdot \bigg(\frac{L}{n} \bigg) \quad \text{for} \quad \widetilde{S}_n\leq 0.5,
\end{equation}
where $0.5$ is a threshold such that the linear behaviour holds. 

The entangling speed can thus be obtained by fitting the curve in Fig.~\ref{fig:normalized_entanglement} with the linear model of Eq.~\eqref{eq:ent_speed} in the appropriate range. We report in Tab.~\ref{tab:ent_speed} the entangling speed for a subset of the inspected architectures, and notice that entanglement is produced at sensibly different rates. In agreement with the findings of Sec.~\ref{sec:EntanglementScaling}, we see that for a linear topology the circuit $\circuit{2}$ builds the entanglement at the fastest rate. Indeed, fixing the feature map to $\fmap=\circuit{zz}$, $\circuit{2}$ produces entanglement three times faster than $\circuit{3}$.

\begin{figure}[ht]
    \centering
    \includegraphics[width=0.6\textwidth]{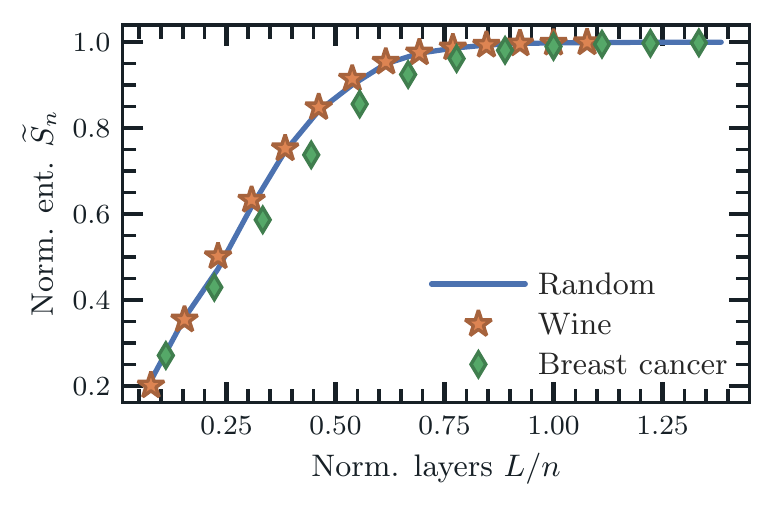}
    \caption[Normalised entanglement for real-world datasets]{Normalised entanglement using inputs from real-world datasets in the feature map, compared to the one obtained using random input data sampled from the uniform distribution $\text{Unif}[0, \pi]$, as shown also in Fig.~\ref{fig:normalized_entanglement}.}
    \label{fig:datasets}
\end{figure}

To further characterise the applicability of the entangling speed, we shown that the behaviour of Fig.~\ref{fig:normalized_entanglement} evaluated for random circuits also holds when the input data $\bm{x} \in \mathbb{R}^n$ in the feature map $\mathcal{F}(\bm{x})$ are not drawn from the uniform distribution, but rather from real world datasets. In particular, we select two common dataset in the machine learning literature, the wine~\cite{wine_dataset} and breast cancer~\cite{breast_cancer_dataset} datasets, and calculate the entanglement generated in the circuit when these data are fed into the feature maps (variational blocks are still populated with random parameters as before). The results presented in Fig.~\ref{fig:datasets} are obtained by rescaling all the features of the datasets in the interval $[0, \pi]$, and then averaging over all data points in the datasets. The wine dataset ($d=13$ features, hence $d=n=13$ qubits) follows perfectly the theoretical curve, and the breast cancer ($d=9$ features, $n=9$ qubits) only slightly deviates form it, producing entanglement at a smaller rate. We then conclude that the entangling speed depends primarily on the architecture of the circuit rather than the actual values of the parameters. Clearly this holds for reasonably distributed data features, that is excluding pathological cases of values being either zero or concentrating around it.

Thus, the entangling speed can be used as a good estimate of the entanglement generated in a QNN also in real use cases, especially at the start of optimisation, when trainable parameters are usually initialised at random. For example, one could measure the entangling speed of the architecture of interest on a random quantum circuit of just a few qubits, and then estimate the entanglement generated with the same architecture on an arbitrary number of qubits and circuit layers, especially in regimes where simulations are no longer computationally feasible.

\subsection{Expressibility}\label{sec:Expressibility}
\begin{figure}[ht]
    \centering
    \includegraphics[width=0.6\textwidth]{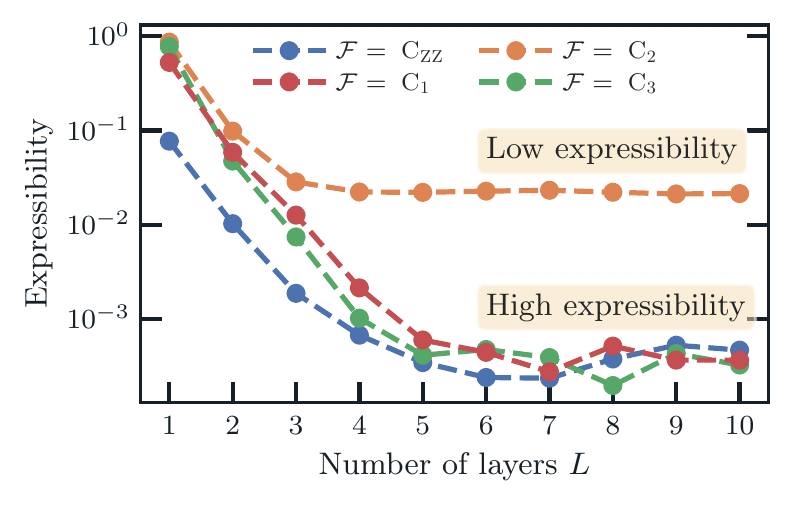}
    \caption[Expressibility of the quantum neural networks]{Expressibility of the QNNs analysed in Fig.~\ref{fig:ent_scaling}, for $n=8$ qubits with linear entanglement. The expressibility measures how well a variational circuit is able to address the unitary space (the lower, the better). All QNNs use the same variational form $\varans = \circuit{2}$, but with different feature maps. As the number of layers is increased, QNNs become more expressible, eventually reaching a plateau.}
    \label{fig:expressibility}
\end{figure}

In addition to entanglement, another useful quantity to characterise parametrized quantum circuits is the expressibility, as defined by authors in~\cite{SimPQCs2019}. Such measure quantifies how well the QNN is able to explore the Hilbert space by comparing the distribution of fidelities of states generated by the QNNs with that of randomly Haar-distributed ones, and we refer to Appendix~\ref{app:expressibility} for a formal definition and explanation. Thus, in order to have a comprehensive understanding of the factors at play in the behaviour of QNNs, in Fig.~\ref{fig:expressibility} we show the expressibility measure for the QNNs analysed in Fig.~\ref{fig:ent_scaling} with a linear connectivity. As one would expect, the expressibility increases as the number of layers is increased, up until a plateau is reached.

Interestingly, the structure with $\fmap = \varans = \circuit{2}$ turns out to be the least expressible of all the structures considered, even if it is the one producing  entanglement at the fastest rate, in agreement with the results reported in~\cite{SimPQCs2019}, as such QNN is indeed very similar to the parameterized circuit label ed `15' in~\cite{SimPQCs2019}. On the contrary, the QNN with $\fmap = \circuit{zz}$, and $\varans = \circuit{2}$ proposed in~\cite{AbbasPowerQNN2021} is able to reach high expressibility while producing entanglement at a controlled pace. As the presence of high entanglement is correlated with trainability issues~\cite{MarreroBarrenEntanglement2021}, this QNN attains an optimal balance of mild entanglement with high expressibility even at low depth, which could be related to its good performances in quantum machine learning task~\cite{AbbasPowerQNN2021, Havlicek2019QSVM}. However, a similar, yet less favourable balance, is achieved by the other two architectures, so further investigation is needed to discriminate where the optimality comes from. 

In this respect, the authors in~\cite{Hubregtsen2021EvaluationParameterizedQuantum} found the expressibility to be correlated with the classification accuracy of QNNs in supervised learning tasks, while weak correlation was found with the entanglement generated inside the circuit, in line with the observations regarding entanglement-induced barren plateaus~\cite{MarreroBarrenEntanglement2021}. As discussed earlier in Sec.~\ref{par:2des_ent_bp}, both expressibility and high entanglement are related to the resemblance of the circuit to a random unitary, but while the former provides a more direct evidence, the latter gives an indirect indication. Indeed, there are cases of circuits having low expressibility but high entanglement, indicating that such circuits selectively explore only some highly-entangled regions of the Hilbert space~\cite{SimPQCs2019}.

\subsection{Distribution of the singular values}\label{sec:mp}
The randomness of a quantum state can also be probed using tools from random matrix theory. Specifically, this can be done by studying the distribution of the eigenvalues of the reduced density matrices, which are known to follow the Mar\v{c}enko-Pastur (MP) law when pure random quantum states are considered~\cite{Znidaric_MP_2007, JaschkeQuantumGreen2023}. More in detail, let $\ket{\psi} \in \mathcal{H}_A \otimes \mathcal{H}_B$ be an Haar-random bipartite quantum state with Schmidt decomposition $\ket{\psi} = \sum_{i=1}^{d} \lambda_i~\ket{\xi_i}_A\otimes\ket{\eta_i}_B$, where $d=\text{min}(d_A, d_B)$ and $d_{A,B}$ is the dimension of the Hilbert space $\mathcal{H}_{A,B}$. The reduced density matrix $\rho_A = \Tr_B\qty[\dyad{\psi}]$ has eigenvalues $\lambda_i^2$ given by the square of the singular values, and for large system size their distribution is described by the MP distribution~\cite{MarcenkoPastur, PuchalaDistinguishability2016}. 

\begin{figure}[ht]
    \centering
    \includegraphics[width=0.6\textwidth]{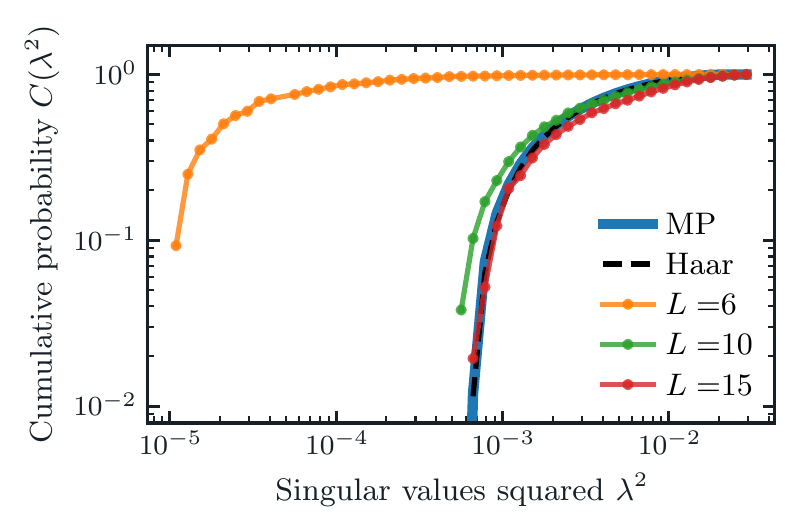}
    \caption[Convergence to the Mar\v{c}enko-Pastur (MP) distribution]{Convergence to the Mar\v{c}enko-Pastur (MP) distribution of the eigenvalues. The cumulative distribution $C(\lambda)$ of the eigenvalues  $\lambda^2$ corresponding to the central bond of a quantum state of $n=15$ qubits, generated with $\mathcal{F}=\circuit{zz}$, $V=\circuit{2}$, and linear connectivity. We show the behaviour for different numbers of layers $L$, and for truly Haar-random states which, as expected, exactly follows the MP curve.}
    \label{fig:mp_convergence}
\end{figure}

In Figure~\ref{fig:mp_convergence} we show the cumulative probability distribution of the eigenvalues $C(\lambda^2)$ of the reduced state of the first half of the qubits, obtained with a QNN with $n=15$ qubits, feature map $\mathcal{F} = \circuit{zz}$, variational ansatz $V=\circuit{2}$, and linear connectivity. The distribution of the singular values for the QNN is obtained by running the circuit $10^2$ times with different sets of parameters, and storing the singular values corresponding to the central cut. Then, we construct the cumulative distribution from the histogram of all the singular values obtained from the simulations. As the number of layers $L$ is increased, the distribution of the eigenvalues approaches the theoretical MP distribution, eventually matching it when the number of layers is equal to the number of qubits. This behaviour is displayed also by other QNN architectures. For completeness, we also show the distribution of the eigenvalues of a truly Haar-random quantum state, generated by sampling its entries independently from a normal complex distribution and then normalising it~\cite{Znidaric_MP_2007}, which, as expected, follows perfectly the MP curve.

\section{Discussion}\label{sec:ch_Entanglement_Discussion}
Moments of the Haar distribution can be approximated efficiently 
using local random quantum circuits of sufficient depth. Depending on the connectivity dimension $D$ of the qubits, defined as the number of other qubits that are connected to each qubit, order $\mathcal{O}(\text{poly}(t)\cdot n^{1/D})$-depth random circuits are sufficient to create approximate unitary $t$-designs~\cite{BrandaoLocalRandomQuantum2016, Harrow2018approximate, HaferkampRandom, HarrowRandomQuantumCircuits2009}, that is circuits that generate a distribution of unitaries which approximately matches moments of the Haar distribution up to order $t$~\cite{Dankert2des2009}. Numerical studies suggest that these results also hold for random parameterized quantum circuits of various forms~\cite{McCleanBarren2018, CerezoBarrenLocalCost2021, Holmes2021connecting, SimPQCs2019}.

We extend these results by showing similar results also for quantum neural networks featuring data re-uploading, both for random instances using random inputs and parameters, and also for real-world dataset when these are used as inputs in the feature map. In particular, for a linear connectivity, as the number of layers approaches to the number of qubits $L \approx n$, QNNs display the same entanglement entropy properties of Haar-distributed random states, a fact which can be taken as a proxy for QNNs approximating unitary designs. Such behaviour was also confirmed by studying the randomness of the circuits with other metrics, namely the expressibility and the convergence to the Mar\v{c}enko-Pastur distribution of the eigenvalues of the reduced states. In both cases, we find strong evidence of the QNN reproducing the same features of random quantum states as the number of layers approaches the system size using a linear connectivity.

Our analysis also underlines the importance of the entangling operations, as careless use of an all-to-all connection can result in unwanted simplifications, making the effective connectivity identical to a nearest-neighbours one. Parameterised two-qubit interactions can solve the problem, even though they may be challenging to implement on real hardware. A good trade-off is achieved with a circular entangling topology, which is immune to simplifications and shows remarkable entangling capabilities. 
Indeed, from the results of Fig.~\ref{fig:ent_scaling}, we see that such connectivity is able to create high multiparite entanglement between qubits already at shallow depth, and with only minor additional hardware resources compared to the linear connectivity. An all-to-all topology instead reaches typical values for entanglement of random states essentially at constant depth $L \in \order{1}$, independently of the system size, and the architecture used (when non-trivial feature maps and variational ans\"{a}tze are used).

While limiting the entanglement inside a quantum neural network may be necessary to ensure its trainability~\cite{MarreroBarrenEntanglement2021}, low entanglement makes the circuit prone to be simulated exactly with an MPS, as discussed in Sec.~\ref{sec:MPSSimulation}. Thus, we envision that a sweet spot should be found in order for QNNs to show signs of quantum advantage: not too high to preclude trainability, not to low to escape triviality.

At last, the introduction of the entangling speed $v_s$~\eqref{eq:ent_speed} can be used as a figure of merit for the entanglement production of a given QNN, independent of the size of the system. Indeed, the entangling speed can be studied and assigned to an architecture in the simulable regime (low number of qubits $n$), and then used to estimate the number of layers to achieve a well-determined quantity of the entanglement, for any system size. We also stress that $v_S$ characterises the most interesting interval of layers in a circuit. As discussed earlier, a value of the entanglement too high might be connected to barren plateaus, underlying the importance of exploring the regime where the entanglement has not saturated yet, and the linear regime still holds. 

We now briefly comment on future interesting investigation directions regarding entanglement and QNNs. The focus of this study was to carefully study the entanglement features of common quantum ansätze, specifically when they are initialised with random parameters and no optimisation has yet started. A natural followup is to ask whether entanglement plays any role also during the optimisation process, which is at core of variational quantum algorithms. While for some specific variational procedures like QAOA~\cite{DupontQAOAEntanglement} or VQE-based ground state solvers~\cite{Woitzik2020EntProdVQE} one has some knowledge of the structure of the target solution, and hence can infer the behaviour of the entanglement created in the circuit, this is not the case for quantum machine learning tasks, as they are usually very task-dependent. Indeed, current proposals for QML advocate for the use of constrained quantum ansätze specifically tailored to the problem under investigation~\cite{skolik2022equivariant, MeyerSimmetriesQML_2022, RagoneBraccia2022GQML}, and then one expects the depth of the circuit and the entanglement generated inside it to highly depend on the specific task to be solved, and dataset to fit, either classical or quantum~\cite{SharmaNFLEntanglement}. Moreover, while the use of deep QNN ansätze (with arguably more entanglement) could offer some optimisation advantages due to overparametrization~\cite{Larocca2021OverparametrizationQNN, KianiUnitariesOverparameterisation2020, AnschuetzLossTrapsQM2022}, the emergence of barren plateaus suggests using shallow circuits instead~\cite{CerezoBarrenLocalCost2021, WangNoiseinducedBarrenPlateaus2021}. The characterisation of the role played by entanglement in QNNs, and how it may be leveraged to achieve a quantum advantage over classical methods definitely deserved further explorations. For sake of exploration and clarity, in Appendix~\ref{app:Entanglement_Training_IRIS} we show some preliminary results on the study of the evolution of entanglement during training. 

\section{Conclusion}\label{sec:ch_Entanglement_Conclusion}
In this Chapter we discussed in detail the entanglement generated by different promising Quantum Neural Networks (QNNs) when these are initialised with random parameters, and showed that they reproduce the same properties of random quantum states under various measures. 

We employed a Matrix Product States (MPS) simulation of the quantum circuits, which guarantees an easy computation of the entanglement in the circuits, and let us study systems of large system size composed of up to $n=50$ qubits.

We showed that while all the architectures tend to a Haar entanglement distribution for a sufficiently high number of layers, the speed of convergence strongly depends on the specific circuit ansatz. This result highlights the universal behaviour of the normalised entanglement production~\eqref{eq:normalised_entanglement} for a given architecture, so we introduced a new measure to characterise a QNN in terms of its entanglement production, namely the entangling speed~\eqref{eq:ent_speed}. 

Finally, we argued that a trade-off between expressibility and entanglement is the key to a better understanding of QNN performances and an auspicious target for the search of quantum advantage. While high entanglement is a necessary condition to avoid classical simulability, a too-large entanglement is detrimental to the training procedure due to its tight connection with barren plateaus, as discussed in Sec.~\ref{sec:Entanglement}. A promising future direction is to extend the entanglement analysis of QNNs not only at initialisation but also during the training procedure~\cite{SackBPShadows2022, Joonho2021EntBP, Woitzik2020EntProdVQE}. These tests would help to understand if QNNs really are a suitable platform for proving quantum advantage.

Whilst entanglement is a necessary feature to bestow the computational power offered by quantum mechanics, on the other end of the spectrum is quantum noise, undoubtedly one of ---if not the most--- compelling challenges to be tackled to ensure the adoption and applicability of quantum computers. More generally, noise arise in many quantum information processing tasks, and it is at the core the next Chapter, where we discuss a technique to remove a wide class of noises when performing arbitrary measurements on qubit systems.

\chapterimage{bg9.png} 
\chapterspaceabove{6.75cm} 
\chapterspacebelow{7.25cm} 

\chapter{Noise deconvolution}\index{NoiseDeconvolution}
\label{ch:NoiseDeconvolutionChapter}
\startcontents[chapters]
\printcontents[chapters]{}{1}{}
\vspace*{1cm}

In this chapter\footnote{The content of this chapter is based on the author's work~\cite{ManginiDeconvolution2022}, and all the figures in this chapter are taken from, or are adaptations of, the figures present in such work.} we present a noise deconvolution technique to remove a wide class of noises when performing arbitrary measurements on qubit systems. In particular, we derive the inverse map of the most common single qubit noisy channels, and exploit it at the data processing step to obtain noise-free estimates of observables evaluated on a qubit system subject to known noise. We illustrate a self-consistency check to ensure that the noise characterisation is accurate providing simulation results for the deconvolution of a generic Pauli channel, as well as experimental evidence of the deconvolution of decoherence noise occurring on Rigetti quantum hardware.

\section{Introduction}\index{NoiseDeconvolution}
Quantum noise is currently the largest limiting factor in the adoption of quantum computation and quantum technology. Their theoretical performances are in fact hindered by the intrinsic fragility of quantum systems, and over the last years many proposal have been put forward to mitigate, ideally correct, the effect of noise and recover reliable results. On the computing side, as fault-tolerant quantum computers remain out of reach at the moment~\cite{FT_EC_Steane, FT_EC_Shor, knillQuantumComputingRealistically2005, Preskill2018NISQ}, various error mitigation techniques have been proposed to extend the capabilities of current small scale noisy quantum devices~\cite{SuzukiQEM_FT_2022, Cao_NISQMitigation, DynamicalDecoupling}. These ranges from correcting the readout noise via inversion of probability assignment matrix~\cite{ReadoutErrorMitigation}, extrapolating the noise in the device to the zero error case~\cite{ErrorMitigationMari, Temme_PEC, ZNE_QEM}, using a probabilistic sampling on specific circuits to approximate the noise free computation~\cite{Temme_PEC, Endo_Mitigation, ErrorMitigationMari}, to also using machine learning approaches to learn how to recover ideal results~\cite{ML_QEM}.

While these methods are concerned with mitigating noise occurring in a computation, here we instead focus on the more generic task of correcting the expectation value of arbitrary observables evaluated on a system which is subject to a known noise happening before the measurement stage. Such a scenario is relevant in quantum communication and quantum tomography tasks~\cite{Vikesh_MitigationTomography}. 

Noise in quantum systems is described by means of quantum channels~\cite{NielsenChuang}
\begin{equation}
\label{eq:Kraus}
    \rho \longrightarrow \mathcal{E}(\rho) = \sum_k A_k \rho A_k^\dagger\,,
\end{equation}
where $\rho$ is the density matrix representing the state of a quantum system, and $A_k$ are operators acting on the system, which are called Kraus operators. While the effect of unitary dynamics can be reversed using realisable operations, quantum channels cannot be undone, and one can only hope to find operations which only approximately invert the noise process at hand. Examples of this approach leverages for example Petz recovery maps~\cite{LautenbacherRecoveryMaps2022, GilyenPetz2022, WildeQuantumInfoBook2017}, or unitaries which, on average, are able to best reverse the noise based on given distance measures~\cite{Benatti_Inversion, Zyczkowski_Inversion, Aurell_2015}.

\begin{figure}[t]
    \centering
    \begin{tikzpicture}
    \node at (-3,0.5) {\textbf{(a)}};
    \node at (0,0) (circ) {
        \begin{quantikz}
        \lstick{$\rho$} & \gate{M}\gategroup[1,steps=2,style={dashed,
                       rounded corners,fill=blue!5, inner xsep=0pt},
                       background]{$O$} &[-0.25cm] \meter{} & \cw \rstick{$\Tr[O\rho]$}
        \end{quantikz}};
    \node at (4, -0.25) {Ideal};
    \node at (-3,-1.5) {\textbf{(b)}};
    \node at (0,-2) {
        \begin{quantikz}
         \lstick{$\rho$} & \gate[style={fill=yellow!20}]{\mathcal{N}}  &\gate{M}\gategroup[1,steps=2,style={dashed,
                       rounded corners,fill=blue!5, inner xsep=0pt},
                       background]{$O$} &[-0.25cm] \meter{} & \cw \rstick{$\Tr[O\mathcal{N}(\rho)]$}
        \end{quantikz}};
    \node at (4, -2.25) {Noisy};
    \node at (-3,-3.5) {\textbf{(c)}};
    \node at (0,-4) {
        \begin{quantikz}
        \lstick{$\rho$} &  \gate[style={fill=yellow!20}]{\mathcal{N}} & \gate{M}\gategroup[1,steps=2,style={dashed,
                       rounded corners,fill=blue!5, inner xsep=0pt},
                       background]{$\mathcal{N}^{-1}(O)$} &[-0.25cm] \meter{} & \cw \rstick{$\expval{O}$}
        \end{quantikz}};
    \node at (4, -4.25) {Deconvolution};
    \node at (-3,-5.5) {\textbf{(d)}};
    \node at (0, -6) {
        \begin{quantikz}
         \lstick{$\rho$} & \gate[style={fill=yellow!20}]{\mathcal{N}_0} & \gate{U} & \gate[style={fill=yellow!20}]{\mathcal{N}_1} & \meter{$O$} & \cw
        \end{quantikz}};
    \node at (4, -6.) {Only $\mathcal{N}^{-1}_1$};
    \end{tikzpicture}
    \caption[Summary of the noise deconvolution process]{General scheme for the noise deconvolution process applied to a qubit. \textbf{(a)} Ideal estimation of an observable $O$ on a single qubit in state $\rho$. The operator $M \in \{\mathbb{I}, H, H S^\dagger\}$, with $H$ and $S$ being the Hadamard and phase gate, used to select a measurement basis in $\{\sigma_z, \sigma_x, \sigma_y\}$ respectively, and thus reconstruct a generic observable $O$, using Eq.~\eqref{eq:qubit_tomography}. \textbf{(b)} Noise (indicated with a yellow box) happening before measurement leads to noisy estimates of the expectation values. \textbf{(c)} Noise deconvolution approach: measurements of the noise-inverted observables $\mathcal{N}^{-1}(O)$ on the noisy state leads to the mitigated ideal result $\expval{O}$. \textbf{(d)} The noise deconvolution approach can be used to mitigate the effects of $\mathcal{N}_1$ only. However, the full noise ($\mathcal{N}_0$ and $\mathcal{N}_1$) can be mitigated either if the unitary can be easily inverted as well, or if the noise processes commutes with the interleaving unitary, as is the case for depolarizing noise. }
    \label{fig:noise_deconvolution_summary}
\end{figure}
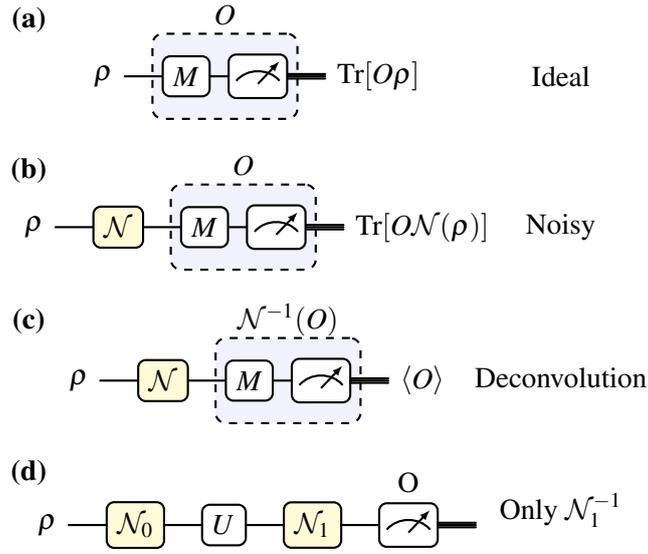

Here instead we show that noise can be eliminated by means of a \textit{deconvolution} process, provided that the noise map describing the process is known and invertible. In fact, we drop the requirements of the inverse transformation being itself a quantum channel, since the transformation is not applied to the quantum system itself,  but to the outcome statistics as a classical post-processing step. We derive the inversion maps of the most common single-qubit noisy channels (both unital and non-unital), and show how to use these to remove the effect of noise from the expectation values of general observables.  In Figure~\ref{fig:noise_deconvolution_summary} we schematically summarise the noise deconvolution idea. The mitigation is effectively obtained by multiplying the noisy estimates by a factor depending on the noise $\expval{O}_\text{mitig} \sim c \expval{O}_\text{noisy}$, which comes at the cost of increasing the variance of the estimation, as $\text{Var}[\expval{O}_\text{mitig}] \sim c^2 \text{Var}[\expval{O}_\text{noisy}]$, so one needs to gather more statistic to reach a target precision. A related post processing technique specialised for quantum many-body systems and quantum field theory is put forward in ref.~\cite{benyQuantumDeconvolution2017}. 

In addition, we provide both numerical simulations of the noise deconvolution process, as well as evidence of deconvolution of decoherence noise occurring on the superconducting quantum computer ``Aspen-9'' provided by Rigetti, accessed using the Quantum Cloud Services (QCS)~\cite{Karalekas_2020}. We show how simple self-consistency checks can test whether the known noise map is accurate and how a feedback scheme can be used to adjust the noise parameters.

Our contributions then include: (\textit{i}) formalisation and discussion of CPTP (namely, \textit{completely positive trace preserving}) noise deconvolution of expectation values through (mathematical) inversion of the noise map; (\textit{ii}) explicit derivation of the inverse map of the most common single qubit noise channels; (\textit{iii}) numerical and experimental application of the ideas introduced before. 

Before continuing, we briefly describe the relation of the proposed noise deconvolution idea to probabilistic error mitigation (PEC)~\cite{Temme_PEC, Endo_Mitigation}, a quantum error mitigation technique aimed at correcting noisy operations during a quantum computation. Given a characterisation of the noise, PEC works by using the inverse noise map of the operations to build an ensemble of suitably generated quantum circuits. These are sampled according to specific weights, and the results combined to build an approximation of the action of the noise-free quantum circuit. In particular, the mitigation procedure is \textit{active}, in the sense that the experimenter need to generate new quantum circuits and run them against the quantum device. On the contrary, we are instead concerned with the correction of expectation values evaluated on a noisy state, with no computation or dynamics involved. In addition, within our framework, the mitigation is \textit{passive}, in the sense that the mitigation happens classically as a post-processing step, and no action on the quantum system is necessary. Appropriately limiting PEC to the specific case of measurement error mitigation, and realising that sampling on quantum circuits is no longer a necessary step, then one can recover the noise deconvolution procedure presented here, whose regime of application is not restricted to quantum computation, but applies to a general quantum mechanical measurement scenario. As such, some of the results presented here can be recovered also with the techniques proposed in~\cite{Temme_PEC, Endo_Mitigation}. That said, the explicit calculations presented here for the general noise maps we analyse have not been presented elsewhere in full generality, e.g. see Table~\ref{tab:tablesummary} below.

The chapter is organised as follows. In Sec.~\ref{sec:ch_Noise_Preliminaries} we recall some basic concepts about quantum channels and the Pauli transfer matrix formalism, and the idea of noise deconvolution in Sec.~\ref{sec:NoiseDeconvolution}. In Sec.~\ref{sec:MapInversion} we leverage the Pauli transfer matrix formalism to explicitly derive the inverse map of the most common single qubit noise channels, and use inside the noise deconvolution procedure to obtain noise-free estimates. In Table~\ref{tab:tablesummary} we summarise all the maps taken in consideration as well as their inverse. In Sec.~\ref{sec:Experiments} we show by means of simulations that the noise deconvolution process can be used to cancel out the effect of a general Pauli channel, and also provide experimental evidence of the deconvolution of decoherence noise as performed on a real quantum device by Rigetti. 

\section{Methods}
\label{sec:ch_Noise_Preliminaries}
In this section we introduce the notation and the theoretical tools used to derive the main results of the work. As in the previous chapters, we denote with $\mathcal{H}$ the Hilbert space of the quantum system under investigation, and with $\mathcal{L}(\mathcal{H})$ the space of squared linear operators acting on $\mathcal{H}$. Also, in this chapter we use the standard quantum information ---rather than computation--- notation for Pauli matrices, that is we use $\qty{\sigma_0, \sigma_x, \sigma_y, \sigma_z}$ instead of the previous $\qty{\mathbb{I}, X, Y, Z}$ defined in Eq.~\eqref{eq:Paulimats}, which is more convenient and appropriate for the topics treated in this chapter. In addition, we refer to Appendix~\ref{app:KarusDecomposition} for a brief overview of quantum channels and Kraus decomposition~\cite{NielsenChuang}.

\subsection{Quantum channels}
\label{subsec:Channels} 
In general quantum channels cannot be physically inverted, as there is no quantum evolution capable of reversing their actions. Formally stated, let $\mathcal{E}$ be a CPTP map, it is not possible to find another CPTP map $\mathcal{D}=\mathcal{E}^{-1}$, such that $(\mathcal{D \circ \mathcal{E}})(\rho) = \rho~\forall\, \rho$. The only trivial case when this is possible, is for maps having only a single Kraus operator, in which case they reduce to standard unitary evolution $\mathcal{E}(\rho) = U\rho U^\dagger$, with the inverse given by $\mathcal{D}(\cdot) = U^\dagger\,(\cdot)\,U$.

The CPTP conditions impose hard constraints to the operatorial form that physically realisable evolutions must match, namely the Kraus representation. However, the requirement for admitting a more general operator-sum representation are looser. In fact, any Hermiticity preserving map, i.e. a map such that $\Phi(\rho)^\dagger = \Phi(\rho)$ for $\rho=\rho^\dagger$, admits an operator-sum representation as~\cite{JohndepillisLinearTransformationsWhich1967, BourdonUnitalQuantumOperations2004} 
\begin{equation}
\label{eq:general_map}
    \Phi(\rho) = \sum_{k} \lambda_k\, A_k \rho A_k^\dagger \quad\text{with}\quad \lambda_k \in \qty{\pm 1}\,.
\end{equation}
Clearly, if all the coefficients are $\lambda_k = 1~\forall k$, then the map $\Phi(\cdot)$ is also completely positive, since it is in standard Kraus form~\eqref{eq:Kraus}. Another useful characterisation is the following.
\begin{corollary}[Corollary II.2 of ~\cite{BourdonUnitalQuantumOperations2004}]
Let $\mathcal{C}_N$ be the space of complex $N\times N$ matrices. Suppose $\Phi:\mathcal{C}_N\rightarrow \mathcal{C}_N$ is a completely positive map having the form
\begin{equation}
\label{eq:corollary}
    \Phi(\rho) = \sum_{k}\beta_k A_k \rho A_k ^\dagger\,,
\end{equation}
where $\qty{A_k}_k$ are linearly independent in $\mathcal{C}_N$, and $\beta_k \in \mathbb{R}~\forall k$. Then $\beta_k \geq 0~ \forall k$.
\end{corollary}
Conversely, if a map has the form~\eqref{eq:corollary} with linearly independent operators $\qty{A_k}_k$ but has \textit{some} of the coefficients $\beta_k<0$, then the map is not completely positive. This result is readily applied to maps acting on qubit systems where $\mathcal{C}_N = \mathbb{C}^{2\times2}$. In fact, Pauli matrices $\sigma_x, \sigma_y$ and $\sigma_z$ together with the identity $\sigma_0 = \mathbb{I}_2$, form a linearly independent set in the space of $2\times2$ complex matrices, and then any map of the form 
\begin{equation}
\label{eq:general_NON_CPTP_map}
    \mathcal{E}(\rho) = \beta_0 \sigma_0 \rho \sigma_0 + \beta_1 \xox{\rho} + \beta_2 \yoy{\rho} + \beta_3 \zoz{\rho}\,,
\end{equation}
having some negative coefficients is not a CP map, thus it is not a physically realisable channel. In the following we will derive many inverse maps having this form, for which this result holds. Of course, we already know that a quantum channel cannot be inverted (apart from the trivial unitary case), so that if an inversion map is found, then it is certainly not CP. Nonetheless, this result is still of interest because it gives a nice and clear condition that can be used to quickly assess the nature of the maps under investigation. In addition, as shown in ref.~\cite{Jiang2021_QuantumMaps}, if a CPTP map is invertible, then its inverse is Hermitian preserving (HP), and so it admits an operator-sum form of Eq.~\eqref{eq:general_map}.

\subsection{Qubit systems and Pauli Transfer Matrix formalism} 
Our analysis is focused on quantum systems made of qubits ($\mathcal{H} = \mathbb{C}^2$), and we now briefly review some useful results on qubit channels. The identity and the Pauli matrices $\{\mathbb{I}, \sigma_x, \sigma_y, \sigma_z\}$ form a basis on $\mathcal{L}(\mathcal{H})=\mathbb{C}^{2\times 2}$, and so any operator $O \in \mathcal{L}(\mathcal{H})$ can be expressed in this basis as 
\begin{equation}
O = \sum_{i=0}^{3} o_i\, \sigma_i = o_0\, \mathbb{I} + \bm{o}\cdot{\bm{\sigma}}\,, \quad o_i = \Tr[\sigma_i\, O]
\end{equation}
where we have introduced the vector of Pauli matrices $\bm{\sigma} \coloneqq (\sigma_x, \sigma_y, \sigma_z)$, and the vector of coefficients $\bm{o} \coloneqq (o_1, o_2, o_3) \in \mathbb{C}^3$. Similarly, as discussed in Sec.~\ref{par:ch_QC_BlochSphere}, density operators are expressed in this basis in terms of their Bloch vector as $\rho = (\mathbb{I}+\bm{r}\cdot\bm{\sigma})/2$, with $\bm{r}=(r_x, r_y, r_z) \in \mathbb{R}^3$, and $|\bm{r}|\leq 1$, where equality holds only for pure states $\rho = \dyad{\psi}$~\cite{NielsenChuang}.

Since any operator $O$ is completely specified by its components in the Pauli basis, we define its vector of coefficients as the column vector $|O\rrangle \coloneqq (o_0, o_1, o_2, o_3)^\mathsf{T}$, but we refer to refs.~\cite{RoncalloMultiDeconvolution2023, DarianoBook2017} for a detailed discussion on the correspondence between operators and vectors.

In addition, every linear map $\Phi: \mathcal{L}(\mathbb{C}^2) \rightarrow \mathcal{L}(\mathbb{C}^2)$ can be represented in this basis as a $4\times4$ matrix $\Gamma$, whose action is given by~\cite{BourdonUnitalQuantumOperations2004,RuskaiAnalysisCompletelypositiveTracepreserving2002,KingMinimalEntropyStates2001, GreenbaumQuantumGate2015}
\begin{equation}
\begin{aligned}
\label{eq:matrix_channel}
    & \Phi(O) \longrightarrow \Gamma |O\rrangle = 
    \begin{bmatrix}
    \gamma_{0} & \bm{\gamma} \\
    \bm{t} & T \\
    \end{bmatrix}
    \begin{bmatrix}
    o_0 \\
    \bm{o}
    \end{bmatrix}
    = \begin{bmatrix}
    \gamma_0\, o_0 + \bm{\gamma}\cdot\bm{o}\\
    o_0\, \bm{t} + T\bm{o}
    \end{bmatrix}\,,\\
    \qq{or} & \Phi(O) = (\gamma_0\, o_0 + \bm{\gamma}\cdot\bm{o})\, \mathbb{I} + (o_0\, \bm{t}+T\bm{o})\cdot \bm{\sigma}\, ,
\end{aligned}
\end{equation}
where $\bm{\gamma}$ and $\bm{t}$ are row and column vectors respectively, and $T$ is a $3\times3$ matrix. The matrix $\Gamma$ associated to the map $\Phi$ is called \textit{Pauli Transfer Matrix} (PTM), and its elements are given by
\begin{equation}
\label{eq:PTM}
    \Gamma_{ij} = \frac{1}{2}\Tr[\sigma_i~\Phi(\sigma_j)] \quad i,j \in \qty{0,1,2,3}\,.
\end{equation}
If we restrict to trace-preserving maps, then $\bm{\gamma} = \bm{0}$ and $\gamma_0 = 1$, so the $\Gamma$ matrix reduces to the simpler form 
\begin{equation}
\label{eq:TracePreserving_PTM}
    \Gamma = \mqty[1 & \bm{0} \\ \bm{t} & T]\, .
\end{equation}
Furthermore, if the map is also unital, that is it preserves the identity matrix $\Phi(\mathbb{I}) = \mathbb{I})$), then also $\bm{t} = \bm{0}$. As an example, the quantum bit-flip channel described by the map
\begin{equation}
    \mathcal{N}_x(\rho) = (1-p)\rho + p\sigma_x \rho \sigma_x\, , 
\end{equation}
has a corresponding PTM representation as
\begin{equation}
\mathcal{N}_x \longrightarrow \Gamma_x = 
    \begin{bmatrix}
    1 & 0 & 0 & 0 \\
    0 & 1 & 0 & 0 \\
    0 & 0 & 1-2p & 0 \\
    0 & 0 & 0 & 1-2p \\
    \end{bmatrix}\, .
\end{equation}

\subsection{Quantum tomographic reconstruction}
\label{subsec:Deconvolution} 
Quantum tomography~\cite{DarianoTomography, DARIANO_Tomography, Dariano_Tomography_1, DMacconePaini2003Spin} is a method to estimate the ensemble average of any arbitrary operator by using measurement outcomes of a quorum of observables. The goal of a tomographic reconstruction of an observable is to identify a set of observables $\{Q_\lambda\}$, called \textit{quorum}~\cite{Universal_DAriano}, such that the mean value $\expval{O} = \Tr[O \rho]$ of any observable $O \in \mathcal{L}(\mathcal{H})$, for all states $\rho$, can be reconstructed by using measurements outcomes of the quorum observables. A tomographic reconstruction formula for an operator $O$ is obtained by using a spectral decomposition of the identity in the operator Hilbert space~\cite{Universal_DAriano,ParisQuantumStateEstimation2004,DMacconeParis2001Quorum, Bisio2009OptimalTomography}, namely
\begin{equation}
\label{eq:TomographicRecontructionFormula}
    O = \int_{\Lambda} d\lambda \Tr [C_\lambda^\dagger O] C_\lambda\, ,
\end{equation}
where $\lambda$ is a parameter living in either a continuous or discrete manifold $\Lambda$, and operators $C_\lambda$ depend on the quorum observables. The term $\mathbb{E}[O](Q_\lambda) := \Tr [C^\dagger_\lambda O]\, C_\lambda$ is called \textit{quantum estimator} of the operator $O$, and given a quantum state $\rho$, the expectation value $\expval{O}$ on such state amounts to
\begin{equation}
\label{eq:tomography_mean}
\begin{aligned}
\langle O \rangle = \Tr [O\rho]  & =  \int_{\Lambda} d\lambda\, \Tr[OC^\dagger_\lambda]\, \Tr[C_\lambda \rho] = \int_{\Lambda} d\lambda\, \Tr[\mathbb{E}[O](Q_\lambda)\, \rho] \\
 & = \int_{\Lambda} d\lambda\, \expval{\mathbb{E}[O](Q_\lambda)}.
\end{aligned}
\end{equation}

For qubit systems, the most common choice (but non unique, e.g.~\cite{DMacconePaini2003Spin}) for the quorum are the Pauli matrices $\{Q_\lambda\}_{\lambda} = \{\sigma_x, \sigma_y, \sigma_z\}$, and the tomographic reconstruction formula results in the standard expansion in the Pauli basis, albeit with a slightly different notation (see Appendix~\ref{app:QubitTomographicFormula} for the explicit derivation), that is
\begin{equation}
\label{eq:qubit_tomography}
\begin{aligned}
    & \expval{O} = \sum_{\alpha=x,y,z}\frac{1}{3}\, \expval{\mathbb{E}[O](\sigma_\alpha)}\,,\\
    & \mathbb{E}[O](\sigma_\alpha) = \left( \frac{3\Tr[O\sigma_\alpha]}{2}\sigma_{\alpha} + \frac{\Tr[O]}{2}\mathbb{I} \right)\, .
\end{aligned}
\end{equation}
Note that the quantum tomographic reconstruction can be straightforwardly applied to multipartite quantum systems by simply using as a quorum the tensor product of single-system quorums~\cite{Universal_DAriano}.

\section{\label{sec:NoiseDeconvolution} Noise Deconvolution}
The tomographic reconstruction formula can be used whenever one has access to the quantum state $\rho$ and measurements of the quorum observables. In practical scenarios however, estimations are performed in the presence of noise and one generally deals with noisy quantum states $\rho \rightarrow \tilde{\rho} = \mathcal{N}({\rho})$ which then leads to noisy estimates $\expval{O}_{\tilde{\rho}} = \Tr[O\, \mathcal{N}({\rho})]$. The idea of noise deconvolution is to correct the errors by considering a new quorum of observables taking into account the noise, and then use a noise inverted quantum estimator to recover the ideal estimates, namely the ones that we would obtain in the absence of noise.

Suppose the noise map $\mathcal{N}$ acting on the quantum system can be formally inverted, that is there exist a linear (not CP) map $\mathcal{N}^{-1}$ such that $(\mathcal{N}^{-1}\circ\mathcal{N})(\rho)=\rho~\forall\rho$. Then, we say that the noise can be \textit{deconvolved} in the following sense: instead of measuring the original observable $O$, we can evaluate the expectation value of the noise-inverted operator $\hat{\mathcal{N}}^{-1}(O)$, thus obtaining as a result the desired noise-free ideal result $\expval{O}$, that is
\begin{equation}
\label{eq:noise_inv_informal}
    \expval{\hat{\mathcal{N}}^{-1}(O)}_{\tilde{\rho}} = \Tr[\hat{\mathcal{N}}^{-1}(O)\,\mathcal{N}(\rho)] \Tr[O\, \mathcal{N}^{-1}(\mathcal{N}(\rho))] = \Tr[O\, \rho] = \expval{O}\, ,
\end{equation}
where $\hat{\mathcal{N}}^{-1}(\cdot)$ denotes the adjoint of the inverse map $\mathcal{N}^{-1}(\cdot)$, and in the second line we made explicit use of the definition of the adjoint map (see Appendix~\ref{app:AmplitudeDamping} for a formal definition of adjoint map). The conditions for \textit{deconvolving} the effect of a noise channel $\mathcal{N}$ are~\cite{ParisQuantumStateEstimation2004, Universal_DAriano}:
\begin{itemize}
    \item the inverted noise map exists, that is there is a $\mathcal{N}^{-1}$ such that $(\mathcal{N}^{-1}\circ\mathcal{N})(O) = O~\forall O \in \mathcal{L}(\mathcal{H})$. 
    \item the quantum estimator $\mathbb{E}[O](Q_\lambda)$ is in the domain of $\mathcal{N}^{-1}$.
    \item the map $\mathcal{N}^{-1}(\mathbb{E}[O](Q_\lambda))$ is a function of $Q_\lambda$.
\end{itemize}
If these hold, then one can substitute the quantum estimator in Eq.~\eqref{eq:tomography_mean}, with the deconvolved quantum estimator $\hat{\mathcal{N}}^{-1}(\mathbb{E}[O](Q_\lambda))$, yielding
\begin{equation}
\label{eq:inv_noise}
\begin{aligned}
 \int_{\Lambda} d\lambda\, \Tr [\hat{\mathcal{N}}^{-1}(\mathbb{E}[O](Q_\lambda))\, \mathcal{N}(\rho)] & = \int_{\Lambda} d\lambda\, \Tr [\mathbb{E}[O](Q_\lambda)\, \mathcal{N}^{-1}(\mathcal{N}(\rho))]\\
 & = \int_{\Lambda} d\lambda\, \Tr [\mathbb{E}[O](Q_\lambda)\, \rho] = \Tr[O\rho] = \expval{O}\,.
\end{aligned}
\end{equation}

This procedure yields the ideal expectation value of any observable $O$ on the state $\rho$, even if having access only to a noisy version of it and provided that the noise map is known (and invertible). Note that this definition is similar to that recently reported in ref.~\cite{Cao_NISQMitigation}, regarding invertible noise channels with non-CPTP inverse. Specialising it for qubits, using Eq.~\eqref{eq:inv_noise} in~\eqref{eq:qubit_tomography}, leads to (see Appendix \ref{app:QubitNoiseDeconvolution} for further details)
\begin{equation}
\label{eq:qubit_deconvolution}
    \expval{O} = \frac{1}{2}\Tr[O] + \frac{1}{2}\sum_{\alpha = x,y,z}\Tr[O\, \sigma_\alpha]\expval{\hat{\mathcal{N}}^{-1}(\sigma_\alpha)}_{\tilde{\rho}}\, .
\end{equation}

Similarly to standard tomographic reconstruction, noise deconvolution can be applied also to multi qubits systems~\cite{RoncalloMultiDeconvolution2023}, in which case the mitigated tomographic estimates can be obtained considering the tensor product of the deconvolved quantum estimator of each subsystem. In addition, generally non-invertible maps could still be deconvoluted if one restricts the attention only to a subset of states of interest upon which the given map is invertible~\cite{FaithfulStates_Dariano_Presti, Calibration_Dariano_Maccone_Presti}. 

As shown later, the correction of the expectation value of a Pauli matrix is obtained by multiplying the noisy estimate ---the one the experimenter has access to--- by a constant depending on the noise, i.e. $\expval{\sigma_\alpha}_\text{mitig} = c\expval{\sigma_\alpha}_\text{noisy}$. This clearly increases the variance of the estimation, since $\text{Var}[\expval{\sigma_\alpha}_\text{mitig}] = c^2 \text{Var}[\expval{\sigma_\alpha}_\text{noisy}] \sim c^2 / M$, where $M$ is the number of measurement shots performed on the system, and thus the experimenter need to increase the outcome statistics proportionally to $c^2$ to reach a desired target precision.

We now proceed discussing how the deconvolution behaves in the presence of multiple noise channels. Consider two noise processes $\mathcal{N}_0$ and $\mathcal{N}_1$ separated by a unitary gate $\mathcal{U}(\cdot) = U \cdot U^\dagger$, as shown in Fig.~\ref{fig:noise_deconvolution_summary}(d). The action of the circuit is
\begin{equation}
    \big(\mathcal{N}_1 \circ \mathcal{U} \circ \mathcal{N}_0 \big)(\rho) = \mathcal{N}_1 \bigg(U \mathcal{N}_0 \big(\rho\big)U^\dagger\bigg) = \mathcal{N}_1 (\tilde{\rho}_{U})\,,
\end{equation}
with $\tilde{\rho}_{U} = U\mathcal{N}_0 \big(\rho\big)U^\dagger$. Using~\eqref{eq:inv_noise}, it is possible to deconvolve the outermost noise $\mathcal{N}_1$ with 
\begin{equation}
    \Tr[\hat{\mathcal{N}}^{-1}_1(O)\mathcal{N}_1(\tilde{\rho}_U)]\,,
\end{equation}
but not $\mathcal{N}_0$, since the unitary $U$ is in the way. Actually, one could decide to deconvolve the unitary as well, using the trivial inverse $\mathcal{U}_1^{-1}(\cdot) = U_1^\dagger \cdot U_1$, and thus making it possible to deconvolve also the first noise channel $\mathcal{N}_0$, as 
\begin{align}
    & \Tr[\hat{\mathcal{N}}^{-1}_1\big((U\hat{\mathcal{N}}^{-1}_0(O)U^\dagger\big)\mathcal{N}_1(\tilde{\rho}_U)] = \Tr[\big(U\hat{\mathcal{N}}^{-1}_0(O)U^\dagger\big) \,U\mathcal{N}_0\big(\rho\big)U^\dagger)]\\
    &= \Tr[\hat{\mathcal{N}}^{-1}_0(O)\mathcal{N}_0(\rho)] = \Tr[O\rho]\,.
\end{align}
However, this procedure cannot be employed to invert the noise that happens before a generic unitary $U$, since it essentially offloads the computation from the quantum computer to the classical one, by simulating the inverse evolution of the quantum system.

A more interesting case is obtained when the error map happens to commute with all the operations in the computation, as is the case for the depolarizing noise, described by the map
\begin{equation}
    \mathcal{N}_{\text{dep}}(\rho) = \frac{p\mathbb{I}}{2} + (1-p)\rho\, ,
\end{equation}
for which it is easy to see that $\big(\mathcal{N}_{\text{dep}}\circ \mathcal{U}\big)(\rho) = \big(\mathcal{U}\circ \mathcal{N}_{\text{dep}}\big)(\rho)\, \forall\,\, \mathcal{U}(\cdot)=U\cdot U^\dagger$. Suppose one is performing a quantum computation given by a sequence of operations $U_i$, each one followed by depolarizing noise
\begin{equation}
    \rho = \bigg(\prod_{i=1}^{d} \mathcal{N}^{(i)}_{\text{dep}}\circ \mathcal{U}_i\,\bigg) (\rho_0) = \bigg(\prod_{i=1}^{d}\mathcal{N}^{(i)}_{\text{dep}} \circ \prod_{i=1}^{d}\mathcal{U}_i\bigg)(\rho_0) = \mathcal{N}_{\text{dep}}^{\text{tot}}(\rho_U)\, ,
\end{equation}
with $\mathcal{N}_{\text{dep}}^{\text{tot}} = \prod_i \mathcal{N}_{\text{dep}}^{(i)}$ the composition of all the depolarizing channels, and $\rho_U = \prod \mathcal{U}_i(\rho_0)$ the state obtained by the ideal noise-free computation. Most importantly, one can check that the composition of multiple depolarizing channels is still a depolarizing channel with probability parameter $1-p_{\text{tot}} = \prod_i (1-p_i)$, where $p_i$ is the probability associated with each depolarizing noise. In such a case, it is possible to deconvolve all noise at once, using the deconvolution formula for the depolarizing noise with the total noise parameter $p_\text{tot}$ (see Eq.~\eqref{eq:depol_deconv}). 

Similarly, this also holds for computations involving multi qubits subject to \textit{global} depolarizing errors. The authors in ref.~\cite{GlobalDepolarizingInversion} leverage this property to perform a simple yet effective error mitigation technique for quantum computers, based on the assumption that noise in quantum circuits is well described by global depolarizing error channels. While exact depolarizing errors (either local or global) are hardly found in realistic quantum circuits where errors are both due to \textit{coherent} (i.e. unitary) and \textit{incoherent} noise (i.e. interaction), Pauli twirling and randomised compiling techniques~\cite{RandomizedCompiling, HashimRandomizedCompilincExperiments2021, RB, PauliRandomizationRC_Ware} can be used to approximately tailor noise to stochastic Pauli channels, preferably depolarizing noise, and then use the procedure above to mitigate it~\cite{Ville2021TailoringDepolarizing}.

\section{\label{sec:MapInversion} Inversion of common noise maps}
We now proceed by explicitly evaluating the inverse maps of some of the most common noisy channels, leveraging the Pauli Transfer Matrix formalism introduced in Sec~\ref{sec:ch_Noise_Preliminaries}. The general method for finding the inverse map goes as follows: we first evaluate the matrix representation~\eqref{eq:matrix_channel} of the channel, we then invert this matrix, and from this recover the operator sum representation of the inverse channel whenever this exists. We start from simpler cases to build some intuition on the construction of the inverse maps, and then proceed towards more complicated cases. In Table~\ref{tab:tablesummary} we summarise the results obtained in this section, comprising all noise channels considered in this analysis together with their inverse maps. 
\begin{table}[ht]
\centering
\begin{tabular}{l@{\hskip -22pt}c@{\hskip -12pt}c}
    \toprule
    & Noise Channel $\mathcal{N}(\rho)$ & Inverse Map $\mathcal{N}^{-1}(O)$ \\ \hline\hline \\[0pt]
    
    Bit-Flip   &  $(1-p)\rho + p\sx\rho \sx$   &   $\displaystyle{\frac{1-p}{1-2p}O - \frac{p}{1-2p}\sx O \sx}$ \\[10pt]
    
    Phase-Flip (\textit{dephasing})   &  $(1-p)\rho + p\sz\rho \sz$   &  $\displaystyle{\frac{1-p}{1-2p}O - \frac{p}{1-2p}\sz O \sz}$ \\[10pt]
    
    Bit-Phase-Flip  &  $(1-p)\rho + p\sy\rho \sy$   &   $\displaystyle{\frac{1-p}{1-2p}O - \frac{p}{1-2p}\sy O \sy}$ \\[10pt]
    
    Depolarizing    &  $(1-p)\rho + p\displaystyle{\frac{\mathbb{I}}{2}}$  &  $\displaystyle{\frac{1}{1-p}\left(O-\frac{p}{2}\Tr[O]\mathbb{I} \right)}$ \\[10pt]
    
    \begin{tabular}{@{}l@{}}General Pauli Channel \\ (see Eq.~\eqref{eq:inv_paulichannel}) \end{tabular} &  $p_0 \rho + \displaystyle{\sum_{k=x,y,z} p_k \sigma_k\rho\sigma_k}$ & $\beta_0 O +  \displaystyle{\sum_{k=x,y,z}\beta_ k \sigma_k O \sigma_k}$ \\[10pt] 
    
    Amplitude Damping & $V_0\rho V_0 + V_1 \rho V_1^\dagger$  & $K_0 O K_0 - K_1 O K_1^\dagger$ \\[0pt]
     & $\begin{array} {lcl} & V_0 = \dyad{0} + \sqrt{1-\gamma}\dyad{1} \\[5pt] & V_1 = \sqrt{\gamma}\ketbra{0}{1}\end{array}$ & $\begin{array} {lcl} & K_0 = \dyad{0} + \sqrt{\frac{1}{1-\gamma}}\dyad{1} \\[5pt] & K_1 = \sqrt{\frac{\gamma}{1-\gamma}}\ketbra{0}{1}\end{array}$\\[20pt]
    
    \begin{tabular}{@{}l@{}}2-Kraus Channel\\(see Eq.~\eqref{eq:two_kraus_inverse})\end{tabular}   & $A_0\rho A_0 + A_1 \rho A_1^\dagger$  &  $B_0 O B_0^\dagger - B_1 O B_1^\dagger$ \\[0pt]
     & $\begin{array} {lcl} & A_0 = \cos\alpha\dyad{0} + \cos\beta\dyad{1}\\[5pt] & A_1 = \sin\beta\ketbra{0}{1}+\sin\alpha\ketbra{1}{0}\end{array}$ & $\begin{array} {lcl} & B_0 = \sqrt{h_{\alpha\beta}}(\cos\beta\dyad{0} + \cos\alpha\dyad{1}) \\[5pt] & B_1 = \sqrt{h_{\alpha\beta}}(\sin\beta\ketbra{0}{1} + \sin\alpha\ketbra{1}{0})\end{array}$ \\[20pt]
     \bottomrule
\end{tabular}
\caption[Noise channels and their inverse maps]{\label{tab:tablesummary} Table summarising the results of the present analysis, consisting of some of the most common single-qubit noisy channels $\mathcal{N}$, along with their inverse noise maps $\mathcal{N}^{-1}$, defined as the map such that $(\mathcal{N}^{-1}\circ\mathcal{N})(\rho) = \rho~\forall\, \rho$. Clearly, all noise channels are CPTP maps, while the inverse channels are not, yet they admit an operator-sum representation. All the noise maps except for amplitude damping and 2-Kraus channel have trivial adjoint channels, so one must pay attention in using the adjoint channel inside the deconvolution formula~\eqref{eq:qubit_deconvolution}. As discussed in Eq.~\eqref{eq:two_kraus_inverse}, the coefficient in the inverse 2-Kraus channel is $h_{\alpha\beta} = 2/(\cos2\alpha+\cos2\beta)$.}
\end{table}

\subsubsection{Bit-flip, phase-flip and bit-phase-flip}
The bit-flip, phase-flip and bit-phase-flip channels are described by the Kraus operators, $A_0 = \sqrt{p}\mathbb{I}$ and $A_{1,\alpha}$ = $\sqrt{1-p}\,\sigma_\alpha$, with $\sigma_\alpha \in \{\sigma_x,\sigma_z,\sigma_y\}$ respectively. For simplicity, in the following we focus only on the bit-flip channel (generated by $\sigma_x$), but the results hold equivalently also for the other two channels. The bit-flip channel acts as
\begin{equation}
\label{eq:bit-flip}
\mathcal{N}_x(\rho) = (1-p)\rho + p\sigma_x\rho \sigma_x\,,
\end{equation}
and its PTM is given by 
\begin{equation}
\Gamma_x = 
    \begin{bmatrix}
    1 & 0 & 0 & 0 \\
    0 & 1 & 0 & 0 \\
    0 & 0 & 1-2p & 0 \\
    0 & 0 & 0 & 1-2p \\
    \end{bmatrix}\, .
\end{equation}
In order to find an operator-sum expression for the inverse map $\mathcal{N}_x^{-1}$, consider the inverse matrix
\begin{equation}
\Gamma_x^{-1} = 
    \begin{bmatrix}
    1 & 0 & 0 & 0 \\
    0 & 1 & 0 & 0 \\
    0 & 0 & \frac{1}{(1-2p)} & 0 \\
    0 & 0 & 0 & \frac{1}{(1-2p)} \\
    \end{bmatrix}\, .
\end{equation}
It's clear that $\Gamma_x$ can be inverted provided that $p \neq 1/2$, since in that case $\text{det}\,\Gamma_x=0$. This is not a problem for real case scenarios, where the probability of errors are usually small, and can safely assume $0<p<1/2$. We now proceed using a derivation similar to that proposed in~\cite{BourdonUnitalQuantumOperations2004}.

Note that $\Gamma_x^{-1}$ is diagonal in the Pauli basis, thus has eigenvectors $\{|\mathbb{I}\rrangle, |\sigma_x\rrangle\,|\sigma_y\rrangle\,|\sigma_z\rrangle\}$ with eigenvalues $\bm{\lambda} = (1, 1, (1-2p)^{-1}, (1-2p)^{-1})$ respectively. Now, consider the generic map 
\begin{equation}
\label{eq:op-sum}
    \mathcal{E}(O) = \sum_{k=0}^{3}\beta_k\, \sigma_k O \sigma_k\, .
\end{equation}
Also this map has eigenvectors $\{\mathbb{I}, \bm{\sigma}\}$, but with eigenvalues $\bm{\beta}=(\beta_0, \beta_1, \beta_2, \beta_3)$. Since two maps are equal if they have the same action on a basis, if we can find a way to match the two sets of eigenvalues $\bm{\lambda}$ and $\bm{\beta}$, we would then recover the operator-sum representation for $\Gamma_x^{-1}$. 

By evaluating the PTM $\Gamma_{\mathcal{E}}$ of $\mathcal{E}(\cdot)$~\eqref{eq:op-sum}, we can relate the coefficients in the operator-sum representation~\eqref{eq:op-sum}, with those appearing in the expression for $\Gamma_x^{-1}$ (see Appendix~\ref{app:inverse_maps} for a derivation). In particular, we want these to hold
\begin{align}
(\Gamma_x^{-1})_{11} = \beta_0 + \beta_1 - \beta_2 - \beta_3\,,\\
(\Gamma_x^{-1})_{22} = \beta_0 - \beta_1 + \beta_2 - \beta_3\,,\\
(\Gamma_x^{-1})_{33} = \beta_0 - \beta_1 - \beta_2 + \beta_3\,,
\end{align}
plus the trace-preserving condition $1 = \beta_0 + \beta_1 + \beta_2 + \beta_3$, that the inverse map must satisfy because the direct map is trace-preserving. This condition is inherently satisfied by $\Gamma_x^{-1}$, since its first row has the form $(1, 0, 0, 0)$. This system has solutions $\beta_0 = (1-p)/(1-2p)$, $\beta_1=-p/(1-2p)$, and $\beta_2=\beta_3=0$, and substituting them back into Eq.~\eqref{eq:op-sum}, we obtain the operator-sum representation of the inverse bit-flip map
\begin{equation}
\label{eq:bit-flip-inv}
    \mathcal{N}_x^{-1}(O) = \frac{1-p}{1-2p}O - \frac{p}{1-2p}\sigma_x O \sigma_x\, .
\end{equation}
By virtue of Corollary II, and noticing that the coefficients appearing in the expression above have always opposite signs, we are sure that this map is not CP, as expected, yet it possesses an operator-sum representation. Note how similar the direct and inverse map are, a feature which we will encounter in all the cases discussed here. 

The same procedure can be applied to phase-flip (also referred to as \textit{dephasing}, generated by $\sigma_z$), and bit-phase-flip (generated by $\sigma_y$) channels, yielding inverse maps
\begin{align}
    \mathcal{N}_z^{-1}(O) & = \frac{1-p}{1-2p}O - \frac{p}{1-2p}\sigma_z O \sigma_z\,,\\
    \mathcal{N}_y^{-1}(O) & = \frac{1-p}{1-2p}O - \frac{p}{1-2p}\sigma_y O \sigma_y\, .
\end{align}

We can plug these inversion maps in the deconvolution formula~\eqref{eq:qubit_deconvolution} to obtain noise-free expectation values of observables. In particular, assume we are measuring a Pauli matrix $O = \sigma_\alpha$, and that the system is subject to one of the noise processes $\rho \rightarrow \rho_\beta = \mathcal{N}_\beta(\rho)$ with $\beta = \{x,y,z\}$. Then the ideal expectation values $\expval{\sigma_\alpha}_{\rho} = \Tr[\sigma_\alpha\rho]$ can be expressed in compact form as (see Appendix~\ref{app:inverse_maps} for the explicit derivation)
\begin{equation}
\label{eq:pauli_deconv}
    \expval{\sigma_\alpha}  = \delta_{\alpha \beta}\, \expval{\sigma_\alpha}_{\rho_\beta} + (1-\delta_{\alpha \beta})\,  \frac{1}{1-2p}\, \expval{\sigma_\alpha}_{\rho_\beta}\, ,
\end{equation}
where $\delta_{\alpha\beta}$ is a Kronecker delta. It is then clear that if the noise happens along the measurement direction ($\alpha=\beta$), then the noise does not affect the measurement statistics, as the ideal and noisy value coincide. While for orthogonal directions ($\alpha\neq\beta$), these are equally contracted by a factor $1-2p$, thus recovering the usual pictorial representation of the contracting Bloch sphere on the plane orthogonal to the noise~\cite{NielsenChuang}. 

\subsubsection{Depolarizing noise}
The depolarizing noise channel is defined as
\begin{equation}
    \mathcal{N}_{\text{dep}}(\rho) = (1-p)\rho + \frac{p\mathbb{I}}{2}\, ,
\end{equation}
whose action is to leave the state untouched with probability $1-p$, and sends it to the completely mixed state $\mathbb{I}/2$ with probability $p$. The channel can be expressed in Kraus form in multiple ways, one of them being~\cite{NielsenChuang}
\begin{equation}
    \mathcal{N}_{\text{depol}}(\rho) = \bigg(1-\frac{3p}{4}\bigg)\rho + \frac{p}{4}\bigg(\sigma_x \rho \sigma_x + \sigma_y \rho \sigma_y + \sigma_z \rho \sigma_z\bigg)\, ,
\end{equation}
with corresponding Kraus operators $A_0 = \sqrt{1-3p/4}\,\mathbb{I},\, A_1 = \sqrt{p}\,\sigma_x/2\,, A_2 = \sqrt{p}\,\sigma_y/2\,, A_3 = \sqrt{p}\,\sigma_z/2$. Following the same procedure outlined before for the bit-flip channel, one can easily recover the inverse linear map (see Appendix~\ref{app:inverse_maps} for explicit derivation)
\begin{equation}
\label{eq:depol_inv}
\mathcal{N}_{\text{depol}}^{-1}(O) = \frac{1}{1-p}\left( O-\frac{p}{2}\Tr[O]\mathbb{I} \right)\, .
\end{equation}
While this is already a known result in the literature~\cite{DarianoDepolarizingInversion, Bisio2009OptimalTomography, Universal_DAriano, HuangPredicting2020, Temme_PEC}, it is presented without an explicit constructive derivation, as given here. 

Using this formula in the deconvolution tomographic reconstruction~\eqref{eq:qubit_deconvolution}, we find
\begin{equation}
\label{eq:depol_deconv}
    \expval{O} = \frac{1}{2}\Tr[O] + \sum_{\alpha}\frac{\Tr[O\sigma_\alpha]}{1-p}\expval{\sigma_\alpha}_{\mathcal{N}_{\text{dep}}(\rho)}\, ,
\end{equation}
where it is clear that to counterbalance the effect of the depolarizing channel, whose effect on the Bloch sphere is to contract it uniformly, one needs perform an expansion of the same amount, obtained dividing by $1-p$. 

While our analysis is focused only on single qubit systems, it is worth noticing that a similar approach can be used to correct correlated and asymmetric depolarizing channels acting on multi-qubits systems~\cite{CafaroAsymmetricDepol2010}, as recently shown in~\cite{RoncalloMultiDeconvolution2023}.

\subsubsection{General Pauli channel}
A more general and interesting case is that of general Pauli channels, where noise acts with different strength along the three Pauli axes, defined as
\begin{equation}
\label{eq:general_pauli_channel}
\mathcal{N}_{\bm{p}}(\rho) = p_0 O + p_x \xox{\rho} + p_y \yoy{\rho} + p_z \zoz{\rho}\, .
\end{equation}
The channel is parametrized by the probabilities $\boldsymbol{p}=(p_0, p_x, p_y, p_z)$, with the trace-preserving condition implying $p_0 = 1-p_x-p_y-p_z$. Importantly, upon choosing appropriate values for $\boldsymbol{p}$, this channel reduces to all noise maps treated before. Though of considerable more general structure, the inverse map of this channel is derived using the same machinery developed above, and eventually one obtains
\begin{equation}
\label{eq:inv_paulichannel}
\begin{alignedat}{1}
& \mathcal{N}^{-1}_{\bm{p}}(O) = \beta_0 O + \beta_1 \xox{O} + \beta_2 \yoy{O} + \beta_3 \zoz{O}\,,\quad \text{with}\\
& \beta_0 =\frac14\bigg(1+\frac{1}{1-2(p_x+p_y)}+\frac{1}{1-2(p_x+p_z)}+\frac{1}{1-2(p_y+p_z)}\bigg)\,,\\
& \beta_1 = \frac14\bigg(1-\frac{1}{1-2(p_x+p_y)}-\frac{1}{1-2(p_x+p_z)}+\frac{1}{1-2(p_y+p_z)}\bigg)\,,\\ 
& \beta_2 = \frac14\bigg(1-\frac{1}{1-2(p_x+p_y)}+\frac{1}{1-2(p_x+p_z)}-\frac{1}{1-2(p_y+p_z)}\bigg)\,,\\
& \beta_3 = \frac14\bigg(1+\frac{1}{1+2(p_x+p_y)}-\frac{1}{1-2(p_x+p_z)}-\frac{1}{1-2(p_y+p_z)}\bigg)\,.
\end{alignedat}
\end{equation}
One can check that varying $\bm{p}$ it is possible to recover the inverse maps of all the cases treated before. For example, for $\bm{p} = (1-p, p, 0, 0)$ corresponding to the bit-flip channel in Eq.~\eqref{eq:bit-flip}, one gets $\beta_0 = (1-p)/(1-2p)$ and $\beta_1=-p/(1-2p)$, as in Eq.~\eqref{eq:bit-flip-inv}.

The noise deconvolution applied to measurements of Pauli matrices $O \in \{\sigma_x, \sigma_y, \sigma_z\}$, leads to the following relations
\begin{equation}
\label{eq:deconvolution_gpc}
\begin{aligned}
    \expval{\sigma_x} & = \frac{1}{1-2(p_y+p_z)}\expval{\sigma_x}_{\mathcal{N}_{\bm{p}}(\rho)}\,,\\
    \expval{\sigma_y} & = \frac{1}{1-2(p_x+p_z)}\expval{\sigma_y}_{\mathcal{N}_{\bm{p}}(\rho)}\,,\\
    \expval{\sigma_z} & = \frac{1}{1-2(p_x+p_y)}\expval{\sigma_z}_{\mathcal{N}_{\bm{p}}(\rho)}\,,
\end{aligned}
\end{equation}
which can be used together with Eq.~\eqref{eq:qubit_deconvolution} to reconstruct the expectation value of a general observable $O$. As before, we see that the noise disturbs the estimation along orthogonal directions. Note that the explicit inversion of the general Pauli channel was also recently reported in ref.~\cite{SuzukiQEM_FT_2022}.

\subsubsection{Amplitude Damping}
The amplitude damping (AD) channel describes the energy loss of a quantum system, for example obtained through relaxation from the excited to the ground state. Its Kraus representation is
\begin{equation}
\label{eq:AmplitudeDamping}
     \mathcal{N}_{\text{AD}}(\rho) = V_0\rho V_0^\dagger + V_1\rho V_1^\dagger\,,\quad
     V_0 = \begin{bmatrix}
    1 & 0  \\ 
    0 & \sqrt{1-\gamma}
    \end{bmatrix}, \quad
    V_1 = \begin{bmatrix}
    0 & \sqrt{\gamma}\\
    0 & 0
    \end{bmatrix}\, ,
\end{equation}
where $\gamma \in [0,1]$ is a parameter that encodes the strength of the energy loss process, which for real systems is often expressed in terms of characteristic decay times, as discussed in Sec.~\ref{sec:Experiments}.

While still being trace preserving (TP), amplitude damping channel is not unital, since $\mathcal{N}_{\text{AD}}(\mathbb{I}) = \mathbb{I}+\gamma Z$. This in turn implies that the Pauli Transfer Matrix $\Gamma_{\text{AD}}$ is not diagonal, but has an addition nonzero element in the last row of first column. This changes the derivation of the inverse map with respect to the previous cases, but it can still be carried out without major changes, as shown in Appendix~\ref{app:AmplitudeDamping}. The inverse linear map in operator-sum representation is then found to be
\begin{equation}
\label{eq:AD_inv}
 \mathcal{N}_{\text{AD}}^{-1}(\rho) = K_0 O K_0^\dagger - K_1 O K_1^\dagger\, \quad
 K_0 = 
\begin{bmatrix}
1 & 0 \\
0 & \frac{1}{\sqrt{1-\gamma}}
\end{bmatrix}, \quad 
K_1 = 
\begin{bmatrix}
0 & \sqrt{\frac{\gamma}{1-\gamma}} \\ 
0 & 0
\end{bmatrix}\, .
\end{equation}
Up until now, all noisy channels (and their inverse maps) had trivial \textit{adjoint} map, since all Kraus operators were Hermitian. However this is not the case for amplitude damping, since both $V_1\neq V_1^\dagger$ and $K_1\neq K_1^\dagger$. Thus, one must be careful in applying the adjoint inverse $\hat{\mathcal{N}}^{-1}$ in Eq.~\eqref{eq:qubit_deconvolution}, and not just $\mathcal{N}^{-1}$ of \eqref{eq:AD_inv} (see Appendix~\ref{app:AmplitudeDamping} for an extended discussion).
Deconvolution of amplitude damping for measurements of the Pauli matrices leads to 
\begin{equation}
\label{eq:ad_sigmas}
\begin{aligned}
\expval{\sigma_x} & = \frac{1}{\sqrt{1-\gamma}}\langle \sigma_x \rangle_{\mathcal{N}_{\text{AD}}(\rho)}\,, \\
\expval{\sigma_y} & = \frac{1}{\sqrt{1-\gamma}}\langle \sigma_y \rangle_{\mathcal{N}_{\text{AD}}(\rho)}\,, \\
\expval{\sigma_z} & = \frac{1}{1-\gamma}\qty(\langle \sigma_z \rangle_{\mathcal{N}_{\text{AD}}(\rho)} -\gamma)\, .
\end{aligned}
\end{equation}

Similarly, one can also obtain the inverse map of the \textit{generalised amplitude damping} (GAD) channel, used to model the interaction of a qubit with an environment at a finite temperature~\cite{CafaroAmplitude2014, NielsenChuang}. Such channel is parameterised by two parameters $\gamma$ and $p$, and it is defined as
\begin{equation}
\begin{aligned}
&\mathcal{N}_{\text{GAD}}(\rho) = A_0 \rho A_0^\dagger + A_1 \rho A_1^\dagger + A_2 \rho A_2^\dagger + A_3 \rho A_3^\dagger \\
&A_0 = \sqrt{p}
\begin{bmatrix}
1 & 0\\ 
0 & \sqrt{1-\gamma}
\end{bmatrix},\quad
A_1 = \sqrt{p}
\begin{bmatrix}
0 & \sqrt{\gamma} \\ 
0 & 0
\end{bmatrix},\\
&A_2 = \sqrt{1-p}
\begin{bmatrix}
\sqrt{1-\gamma} & 0\\ 
0 & 1
\end{bmatrix},\quad
A_3 = \sqrt{1-p}
\begin{bmatrix}
0 &  \\ 
\sqrt{\gamma} & 0
\end{bmatrix}.
\end{aligned}
\end{equation}
One can check that the following map is the inverse of the GAD channel
\begin{equation}
\begin{aligned}
&\mathcal{N}^{-1}_{\text{GAD}}(\rho) = B_0 \rho B_0^\dagger - B_1 \rho B_1^\dagger + B_2 \rho B_2^\dagger - B_3 \rho B_3^\dagger \\
&B_0 = \sqrt{p}
\begin{bmatrix}
1 & 0\\ 
0 & \sqrt{\frac{1}{1-\gamma}}
\end{bmatrix},\quad
B_1 = \sqrt{p}
\begin{bmatrix}
0 & \sqrt{\frac{\gamma}{1-\gamma}} \\ 
0 & 0
\end{bmatrix},\\
&B_2 = \sqrt{1-p}
\begin{bmatrix}
\sqrt{\frac{1}{1-\gamma}} & 0\\ 
0 & 1
\end{bmatrix},\quad
B_3 = \sqrt{1-p}
\begin{bmatrix}
0 &  \\ 
\sqrt{\frac{\gamma}{1-\gamma}} & 0
\end{bmatrix},
\end{aligned}
\end{equation}
with corresponding noise deconolved Pauli expectation values given by
\begin{equation}
\label{eq:gad_sigmas}
\begin{aligned}
\expval{\sigma_x} & = \frac{1}{\sqrt{1-\gamma}}\langle \sigma_x \rangle_{\mathcal{N}_{\text{GAD}}(\rho)}\,, \\
\expval{\sigma_y} & = \frac{1}{\sqrt{1-\gamma}}\langle \sigma_y \rangle_{\mathcal{N}_{\text{GAD}}(\rho)}\,,\\
\expval{\sigma_z} & = \frac{1}{1-\gamma}\qty(\langle \sigma_z \rangle_{\mathcal{N}_{\text{GAD}}(\rho)} -\gamma(2p-1))\,.
\end{aligned}
\end{equation}

\subsubsection{Two-Kraus channels}
We conclude the analysis of single-qubit noise channels by considering the set of channels generated by two parametrized Kraus operators, namely
\begin{equation}
\mathcal{N}_{\text{two}}(\rho) = \sum_{i=1, 2}A_i\rho A_i^\dagger\, ,  
\end{equation}
with $A_1 = \cos\alpha\dyad{0}+\cos\beta\dyad{1}$, and $A_2 = \sin\beta\ketbra{0}{1}+\sin\alpha\ketbra{1}{0}$. This channel reduces to bit-flip for $\alpha=\beta$, and to amplitude damping for $\alpha=0$. Following a procedure similar to the amplitude damping case, the inverse map of the two-Kraus channels is found to be
\begin{equation}
\label{eq:two_kraus_inverse}
\begin{aligned}
& \mathcal{N}_{\text{two}}(O)^{-1} = B_1 O B_1^\dagger - B_2 O B_2^\dagger\,,\\
&
B_1 = \begin{bmatrix}
\frac{\sqrt{2}\cos\beta}{\sqrt{\cos2\alpha+\cos2\beta}} & 0 \\ 
0 & \frac{\sqrt{2}\cos\alpha}{\sqrt{\cos2\alpha+\cos2\beta}}
\end{bmatrix}\,, \\
& 
B_2 = \begin{bmatrix}
0 & \frac{\sqrt{2}\sin\beta}{\sqrt{\cos2\alpha+\cos2\beta}} \\
\frac{\sqrt{2}\sin\alpha}{\sqrt{\cos2\alpha+\cos2\beta}} & 0
\end{bmatrix}\,.
\end{aligned}
\end{equation}
Similarly to amplitude damping, one of the generators ($B_2$) is not Hermitian, thus we must employ the adjoint inverse map when evaluating the deconvolved expectation values. By straightforward calculations the following holds
\begin{equation}
\begin{aligned}
\expval{\sigma_x} & = \frac{1}{\cos(\alpha-\beta)}\expval{\sigma_x}_{\mathcal{N}_{\text{two}}(\rho)}\,,\\
\expval{\sigma_y} & = \frac{1}{\cos(\alpha+\beta)}\expval{\sigma_y}_{\mathcal{N}_{\text{two}}(\rho)}\,,\\
\expval{\sigma_z} & = h_{\alpha\beta}\big(\cos^2\beta+\sin^2\alpha-1+\expval{\sigma_z}_{\mathcal{N}_{\text{two}}(\rho)}\big)\, ,
\end{aligned}
\end{equation}
with $h_{\alpha\beta} = 2 / (\cos(2\alpha)+\cos(2\beta))$. Note that upon varying the parameters $\alpha$ and $\beta$, the formulas above correctly reduce to the other limiting channels. For example, setting $\alpha=0$ leads to amplitude damping channel in Eq.~\eqref{eq:ad_sigmas} with $\cos(\beta):=\sqrt{1-\gamma}$.

\section{\label{sec:Experiments} Experimental deconvolution}
In this section we provide some concrete applications of the noise deconvolution procedures for qubit tomography outlined above. In particular, we show both numerically and by experimentation on superconducting quantum hardware by Rigetti how to address a \textit{decoherence} noise model, and we also provide numerical evidence for the deconvolution of the general Pauli channel~\eqref{eq:general_pauli_channel}. All simulations are performed using PyQuil and the real quantum device used is ``Aspen-9", accessed via Rigetti's Quantum Cloud Services (QCS)~\cite{smith2016practical, Karalekas_2020}. 

\subsection{Decoherence noise model}
The concurrent action of a dephasing channel followed by amplitude damping is referred to as \textit{decoherence} noise, which is an effective way to describe the noisy evolution a qubit undergoes due to uncontrolled interaction with its external environment. Using the definitions~\eqref{eq:bit-flip} and~\eqref{eq:AmplitudeDamping}, one obtains
\begin{equation}
\label{eq:decoherence_noise}
\begin{aligned}
\mathcal{N}_{\text{dec}}(\rho) & = \qty( \mathcal{N}_{\text{AD}}(\gamma) \circ \mathcal{N}_z (p) ) \qty(\mqty[a & b \\ c & 1-a]) \\
& = \mqty[ 1-(1-a)(1-\gamma) & (1-2p)\sqrt{1-\gamma}\,b \\ (1-2p)\sqrt{1-\gamma}\,c & (1-\gamma)\,(1-a)] \\
& = \mqty[1-(1-a)e^{-t/T_1} & e^{-t/T_2}\,b \\ e^{-t/T_2}\,c & e^{-t/T_1}\,(1-a)]\,,
\end{aligned}
\end{equation}
where we have introduced the relaxation times $T_1$ and $T_2$ characterising the ``quality" of the physical qubits. These are related to the noise parameters $\gamma$ and $p$ through the following relations
\begin{equation}
\label{eq:noise_and-Ts}
    \gamma = 1-e^{-t/T_1}\,,\quad p = \frac{1}{2}\big(1-e^{-(t/T_2-t/2T_1)}\big)\,,
\end{equation}
where $t$ is a time parameter indicating the duration of the noise process. 

Since the correction terms in the deconvolution formulas for dephasing~\eqref{eq:ad_sigmas} and amplitude damping~\eqref{eq:pauli_deconv} are multiplicative, for a decoherence channel these combine as
\begin{equation}
\label{eq:decoherence_sigmas}
\begin{aligned}
\expval{\sigma_x} & = \frac{1}{(1-2p)}\frac{1}{\sqrt{1-\gamma}}\langle \sigma_x \rangle_{\mathcal{N}_{\text{dec}}(\rho)}\,,\\
\expval{\sigma_y} & = \frac{1}{(1-2p)}\frac{1}{\sqrt{1-\gamma}}\langle \sigma_y \rangle_{\mathcal{N}_{\text{dec}}(\rho)}\,,\\
\expval{\sigma_z} & = \frac{1}{1-\gamma}\qty(\langle \sigma_z \rangle_{\mathcal{N}_{\text{dec}}(\rho)} -\gamma)\,.
\end{aligned}
\end{equation}
Additionally, if the quantum system under investigation is subject to repeated applications of a decoherence noise channel, i.e. $\mathcal{N}_{\text{dec}}^{\circ m}(\rho) =\mathcal{N}^{(1})_{\text{dec}}\circ \mathcal{N}^{(2)}_{\text{dec}} \cdots \circ \mathcal{N}^{(m)}_{\text{dec}}(\rho)$, then the ideal expectation values are obtained through the following equations
\begin{equation}
\label{eq:decoherence_sigmas_multiple}
\begin{aligned}
\expval{\sigma_x} & = \frac{1}{\qty((1-2p)\sqrt{1-\gamma})^m}\langle \sigma_x \rangle_{\mathcal{N}_{\text{dec}}(\rho)}\,,\\
\expval{\sigma_y} & = \frac{1}{\qty((1-2p)\sqrt{1-\gamma})^m}\langle \sigma_y \rangle_{\mathcal{N}_{\text{dec}}(\rho)}\,,\\
\expval{\sigma_z} & = \frac{1}{(1-\gamma)^m}\qty(\langle \sigma_z \rangle_{\mathcal{N}_{\text{dec}}(\rho)} -1+(1-\gamma)^m)\, .
\end{aligned}
\end{equation}

\begin{figure}[!ht]
    \centering
    \includegraphics[width=\textwidth]{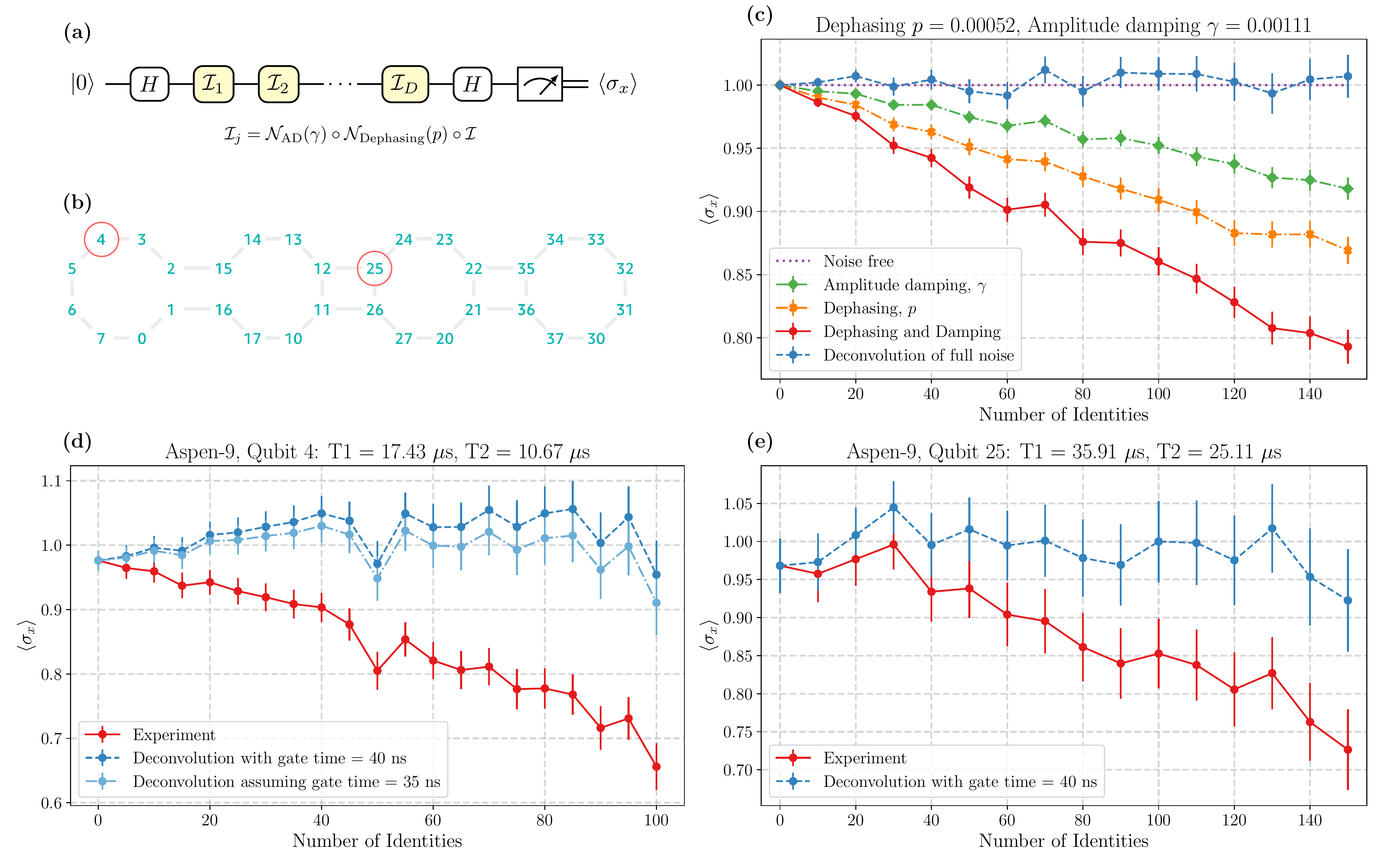}
    \caption[Deconvolution of decoherence noise on real hardware]{Deconvolution of decoherence noise both on a simulator and the real quantum device Aspen-9 by Rigetti. \textbf{(a)} Scheme of the quantum circuit used in the simulations and runs on the actual quantum device. A qubit is prepared in the superposition state and then it is left to decohere for a certain amount of time, dependent on the number $D$ of identities in the circuit. Eventually the qubit is measured in the $x$-basis to estimate the expectation value $\sigma_x$. \textbf{(b)} Scheme of Aspen-9, the real quantum device by Rigetti used to run the quantum circuit. \textbf{(c)} Simulation of the decoherence noise for dephasing ($p$) and damping ($\gamma$) intensities equal to those characterising qubit $25$ of Aspen-9, with gate duration of $40$~ns. For comparison, the effect of the action of these channels alone is also showed. Using the deconvolution formulas for decoherence noise~\eqref{eq:decoherence_sigmas_multiple}, it is possible to mitigate the decay caused by the noise, and recover the ideal result. Each expectation value is estimated evaluating the mean over $n_{\text{shots}} = 2048$ measurement outcomes, and the error bars showed are the statistical error of the mean. \textbf{(d)} Results obtained from running the circuit on qubit $4$ of Aspen-9, characterised by relaxation times $T_1 = 17.43\cdot 10^{-6}$~s and $T_2 = 10.67\cdot 10^{-6}$~s, with $n_{\text{shots}} = 2048$, and the error bars are twice the error of the mean. See main text for comments on the results. \textbf{(e)} Results obtained from running the circuit on qubit $25$ of Aspen-9, characterised by relaxation times $T_1 = 35.91\cdot 10^{-6}$~s and $T_2 = 25.11\cdot 10^{-6}$~s, with $n_{\text{shots}} = 1024$. Also in this case the error bars are equals to twice the error of the mean. See main text for comments on the results.}
    \label{fig:decoherence_noise}
\end{figure}

In Figure~\ref{fig:decoherence_noise} we show the application of these formulas to deconvolve the decoherence noise occurring on a qubit. The specific quantum circuit used for the experiments is showed in Figure~\ref{fig:decoherence_noise}a: first the system is prepared in the superposition state $\ket{+}=H\ket{0}=(\ket{0}+\ket{1})/\sqrt{2}$, then we let qubit decohere for a certain amount of time dictated by the number $D$ of (noisy) identities each of which supposedly takes a time $t$, and at last we measure the expectation value of the operator $\sigma_x$. Clearly, in a noise-free scenario, the result would always be $\expval{\sigma_x}=1$, independent of the depth $D$. Figure~\ref{fig:decoherence_noise}c shows a simulation of these circuits with stochastic measurement outcomes for different values of $D$, and for a given choice of noise parameters $p$ and $\gamma$. For comparison, the individual effect of dephasing and amplitude damping channels alone are also showed. Thanks to Eq.~\eqref{eq:decoherence_sigmas_multiple} we can invert the effect of the decoherence noise, and so retrieve the ideal noise-free results.

We also tested this procedure on real superconducting quantum hardware provided by Rigetti, in particular on the device ``Aspen-9'', whose topology is reported in Fig.~\ref{fig:decoherence_noise}b. The device comes with the calibration data reporting the $T_1$ and $T_2$ parameters for any qubit, as well as the time duration of a single gate. Identities in the circuits are used to introduce time delays, and thus let the qubit decohere for longer intervals of time, depending on the depth $D$. Differently from the previous simulations where only the identities are supposed to introduce (decoherence) noise, in the real case scenario noise happens along the whole computation, including state preparation, application of all gates in the circuit (both Hadamards and Identities), and finally measurement errors. Of these, the most detrimental are undoubtedly readout errors, and we addressed them by using the standard mitigation technique of calibrating the device and inverting the assignment probability matrix to recover readout mitigated results. Calibration data reports that the time it takes to execute a single qubit identity gate is $t=40$~ns, and together with $T_1$ and $T_2$, these are used to calculate the parameters $p$ and $\gamma$ of the decoherence noise, using relations~\eqref{eq:noise_and-Ts}. These are in turn used inside the deconvolution formulas to recover the noise-free results. Figures~\ref{fig:decoherence_noise}d and \ref{fig:decoherence_noise}e show the results of the execution of circuit Fig.~\ref{fig:decoherence_noise}a on qubits $4$ and $25$, respectively. 

The noise mitigation procedure on qubit $4$ shown in panel~\ref{fig:decoherence_noise}d yields slightly unphysical results, in that the mitigated expectation values exceeds one at times. A naive solution to this problem could be to impose that the mitigation results are in the physical range $\expval{\sigma_\alpha} \in [-1,+1]$, so that if the result exceed the limits, it should be substituted with the appropriate physical bound. Though, assuming a gate time duration of $t=35$~ns instead of standard $40$~ns, yields results which are more in agreement with the expected theoretical behaviour for decoherence noise, as the deconvoluted results are compatible with one, as expected. This hints that either the quality of the qubit is better then reported in the available calibration data (either due to shorter gate times $t$,
or larger $T_1$ and $T_2$), or that the decoherence model alone poorly describes the noise happening on idle qubit $4$ left interacting with the external environment. However, the good accordance between the deconvoluted results with $t=35$~ns and the experiments suggests the first hypothesis to hold. 

Such conclusion is also corroborated by the experimental results obtained with qubit $25$. In fact, using the deconvolution formulas with reported $T_1$, $T_2$ and standard gate time ($t=40$~ns), we are able to mitigate the effect of noise with good accuracy, as showed in Fig.~\ref{fig:decoherence_noise}e, hinting that indeed the decay law of the qubit is well described through a decoherence noise model of Eq.~\eqref{eq:decoherence_noise}. Also, note that the simulation in Fig.~\ref{fig:decoherence_noise}c is tuned with the same noise parameters $p$ and $\gamma$ characterising qubit $25$. Apart from fluctuations due to, e.g imperfect readout, stochastic measurement outcomes, and noisy Hadamards, there is good agreement between the simulated (red curve in panel (c)) and experimental result (red curve in panel (e)). We do not report analogues experiments using other qubits in the device that produced obviously biased data.

\subsection{Arbitrary Pauli channel}
\begin{figure}[!ht]
    \centering
    \includegraphics[width=0.7\textwidth]{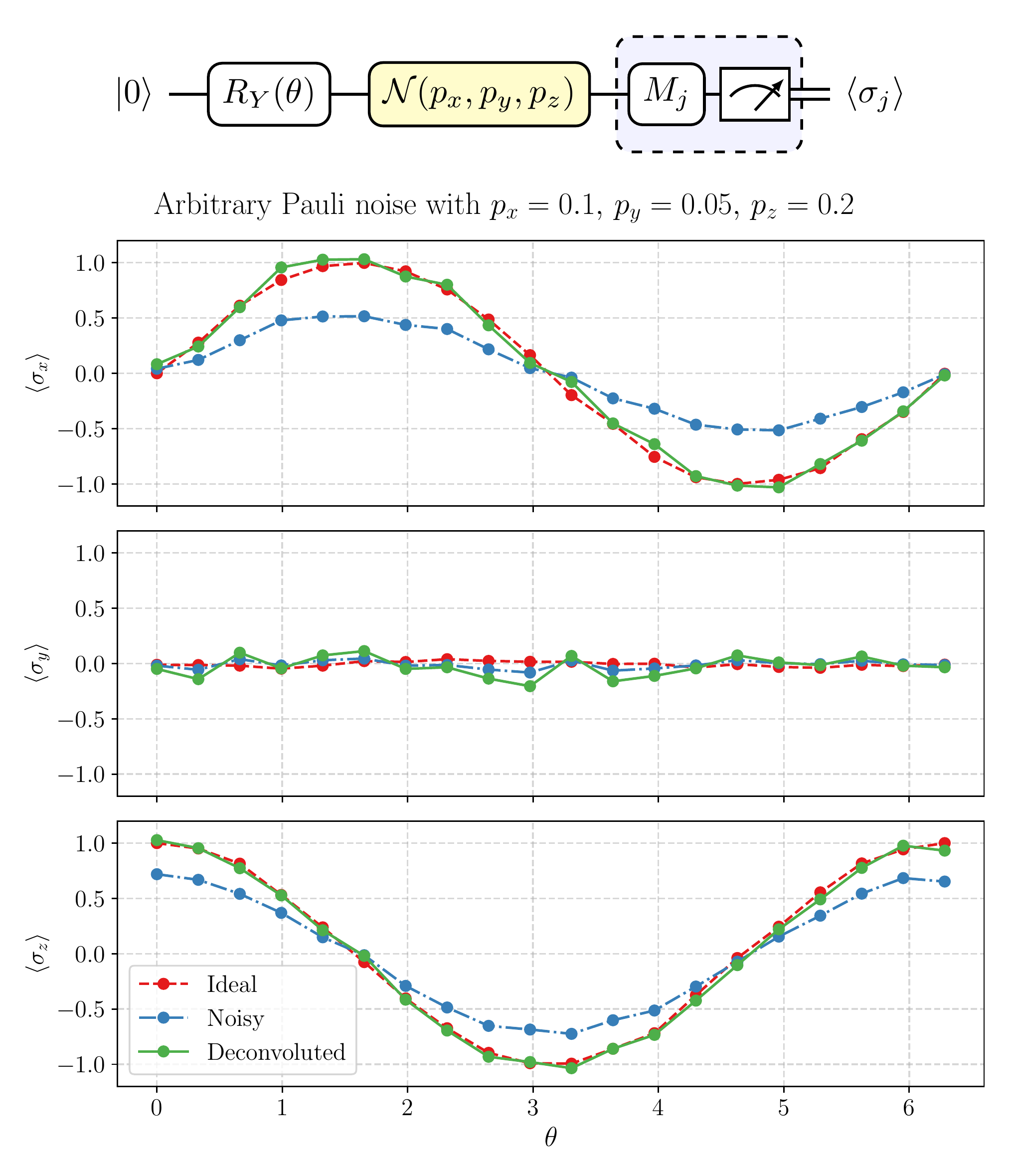}
    \caption[Deconvolution of Pauli channels]{Simulation of the deconvolution process for the general Pauli channel $\mathcal{N}_{\bm{p}}$~\eqref{eq:general_pauli_channel}. The noise parameters along the three Pauli axes are set to $p_x=0.1, p_y=0.05, p_z=0.2$. The results are obtained simulating the circuit portrayed on top of the image for $n_{\text{shots}} = 1024$ shots and for multiple values of the angle $\theta$. Then, the deconvolution formulas~\eqref{eq:deconvolution_gpc} are used to retrieve the ideal noise-free result. It is clear that the deconvolution effectively mitigates the Pauli noise yielding a final result which is much closer to the ideal noise-free one, up to differences due to stochastic measurement outcomes. In particular, the estimation of $\sigma_y$ is dominated by the statistical error, which is amplified by the correction factor $1/(1-2(p_x+p_z)) = 2.5$.}
    \label{fig:GPC_deconvolution}
\end{figure}

We implemented a simulation of the noise deconvolution of the general Pauli channel~\eqref{eq:general_pauli_channel}, using the quantum virtual machine (QVM) simulator provided with PyQuil~\cite{smith2016practical}. The simulated circuit is showed on top of Figure~\ref{fig:GPC_deconvolution}. A qubit starting in the ground state is rotated in the Bloch sphere around the $y$ axis via $R_Y(\theta) = e^{-i\theta\sigma_y/2}$, and then it is subject to the general Pauli noise (yellow box), simulated applying a Pauli transformation chosen randomly with probabilities  $p_x,\,p_y$ and $p_z$. At last, we estimate the expectation value of the three Pauli matrices by appending the appropriate change of basis gate, i.e $M_j \in \{\mathbb{I}, H, H S^\dagger\}$ for $\{\sigma_z, \sigma_x, \sigma_y\}$ respectively. 

The noise parameters $(p_x, p_y, p_z)$ are used within the deconvolution formulas~\eqref{eq:deconvolution_gpc} to recover the mitigated results (green curve), which are, as expected, in perfect agreement with the ideal noise-free ones, obtained from executing the quantum circuit without the noisy channel (red curve).

\section{\label{sec:Conclusions} Conclusions}
In conclusion we have shown how mathematically invertible noise maps can always be removed from the final measurement stage, so that one can obtain unbiased expectation values of general observables provided that the noise process is known. We illustrated the method on most known qubit noise maps, and systematically derived their inverse maps (see Table~\ref{tab:tablesummary}). We simulated the noise deconvolution procedure for the case of a general Pauli channel (Fig.~\ref{fig:GPC_deconvolution}) and illustrated our method on noise on actual quantum hardware (Fig.~\ref{fig:decoherence_noise}).

\chapterimage{bg7bis.png} 
\chapterspaceabove{6.75cm} 
\chapterspacebelow{7.25cm} 

\chapter{Conclusions}\index{Conclusions and Outlooks}
\label{ch:Conclusion}
\epigraph{\textit{The greatest success of our field [quantum computation and informatics] will not be to speed up calculations or communicate in secrecy, but to help people understand that this world is quantum-mechanical.}}{Charles Henry Bennett\\ \footnotesize{as reported by Simone Severini in his book ``Nella terra dei qubit''~\cite{SeveriniNellaTerraQubit}}.}.
\vspace*{0.5cm}
\startcontents[chapters]

In this Thesis we have covered several topics regarding variational quantum algorithms and quantum machine learning, providing several compelling examples of how parameterized quantum circuits can be understood as machine learning models. 

Indeed, we have started in Chapter~\ref{ch:QC_and_VQAs} with a bird's eye view on the new and exciting field of variational quantum algorithms, that is that ensemble of procedures that leverage parameterized quantum circuits and classical computing power in tandem to take full advantage of near-term quantum computing devices. Before future experimental and theoretical advancements pave the way toward the construction of large-scale noise-resilient quantum computers, near-term devices (NISQ) allow experimenting with quantum information processing, with an eye also on the possibility of achieving some sort of useful quantum advantage already with this new paradigm of hybrid quantum-classical computation.

The analysis of variational quantum algorithms then culminated in Chapter~\ref{ch:QML} where the field of Quantum Machine Learning was thoroughly characterised. Importantly, we have found how parameterized quantum models can be effectively described using tools from classical machine learning, for example when discussing kernel methods, expressing the output of data-dependent quantum circuits as truncated Fourier series, and last but not least how the classical statistical learning framework can be applied to derive statements about the generalisation performances of quantum neural networks.

The discussion then moved on to some concrete examples of quantum learning models, presenting some novel contributions to the field. In Chapter~\ref{ch:CQN} we have reported on a novel quantum algorithm implementing a generalised perceptron model on a qubit-based quantum device that accepts and analyses continuously valued input data. The proposed algorithm can be readily run on existing quantum hardware, and it takes full advantage of the exponentially large Hilbert space available to encode input data on the phases of large superposition states, known as locally maximally entanglable (LME) states. In addition, we saw how the proposed model can be used to implement classification tasks and pattern recognition involving grey-scale images.

In Chapter~\ref{ch:VariationalQN} we reviewed a discrete version of the continuous quantum neuron discussed in Ch.~\ref{ch:CQN} and we introduced variational training methods for efficiently handling the manipulation of classical and quantum input data. Through extensive numerical analysis, we compared the effectiveness of different circuit structures and learning strategies, highlighting potential benefits brought by hardware-compatible entangling operations and by layerwise training routines that use local, instead of global, cost functions. In all envisioned applications, our proposed protocols are intended as an effective method for the analysis of quantum states as provided, e.g., by external devices or sensors, while it is worth stressing that the general problem of efficiently loading classical data into quantum registers still stands open. 

Building on the phase encoding strategy introduced for the continuous quantum neuron, in Chapter~\ref{ch:Autoencoder} we have presented a toy model for a quantum pipeline comprising a quantum autoencoder and a classifier for analysing data coming from an industrial power plant. Specifically, we have implemented a variational quantum autoencoder that can compress information stored on a multipartite quantum state onto just some of its constituents. The compressed quantum states coming from the trained quantum autoencoder were then used as inputs to a quantum classifier to perform a binary classification task. For both tasks, the quantum procedures performed equally well to comparable classical counterparts, and they were also tested on real superconducting quantum hardware provided by IBM. While the achievement of a clear quantum advantage is still out of reach, this approach sets a milestone in the field of quantum machine learning, since it is one of the first examples of a direct application of quantum computing software and hardware to analyse real data sets from industrial sources.

Finally, in the last Chapters, we broadened our analysis to include more quantum information-related topics, discussing entanglement and noise. Specifically, in Chapter~\ref{ch:entanglement} we discussed in detail the Entanglement generated by different promising Quantum Neural Networks (QNNs) when these are initialised with random parameters, and showed that they reproduce the same properties of Haar random quantum states under various measures. Employing tensor network methods (MPS) we could simulate wide quantum circuits of up to 50 qubits, and introduced a new measure, the entangling speed, to characterise the rate of production of entanglement of a given circuit ansatz as its depth is increased.

At last, Chapter~\ref{ch:NoiseDeconvolutionChapter} moved our attention to discussing the impact of quantum noise on the estimation of expectation values of observables. Indeed, we have shown how mathematically invertible noise maps can always be removed from the final measurement stage so that one can obtain unbiased expectation values of general observables provided that the noise process is known. We illustrated the method on most known qubit noise maps, systematically derived their inverse maps, and also provided simulation and experimental application of the method on actual superconducting quantum hardware provided by Rigetti.

Quantum computing research is currently experiencing a surge of interest, with a significant amount of effort devoted to not only studying the long-term goal of universal fault-tolerant quantum devices, but also maximising the potential of current-generation quantum computers, both scientifically and technologically. In this thesis, we provided an extensive description of the state-of-the-art of variational quantum algorithms and machine learning, as well as several compelling original contributions to the field, some of which are more applied and others more fundamental.

Although the scientific value of near-term quantum computers is undeniable, the field is still too immature to confidently assess when and how a useful quantum advantage will be achieved. This is especially true when it comes to Quantum Machine Learning, which is the convergence of quantum physics and computing with artificial intelligence and deep learning, two notoriously complex and theoretically difficult fields.

Ultimately, only time will tell whether we will discover a useful and indisputable computational advantage from quantum computing and/or quantum machine learning. In the meantime, we have the privilege of witnessing the evolution of this exciting field, and we should enjoy the journey as much as the destination.

\part*{References}



\chapterimage{} 
\chapterspaceabove{2.5cm} 
\chapterspacebelow{2cm} 


\chapter*{Bibliography}
\markboth{\sffamily\normalsize\bfseries Bibliography}{\sffamily\normalsize\bfseries Bibliography} 
\addcontentsline{toc}{chapter}{\textcolor{ocre}{Bibliography}} 





\printbibliography[heading=bibempty] 



\part*{Appendices}

\chapterimage{orange2.jpg} 
\chapterspaceabove{6.75cm} 
\chapterspacebelow{7.25cm} 

\begin{appendix}

\renewcommand{\chaptername}{Appendix} 

\chapterimage{} 
\chapterspaceabove{2.5cm} 
\chapterspacebelow{2cm} 
\chapter{Variational Quantum Algorithms}
\label{app:proofs_VQA}

\section{Global and local cost functions}
\label{sec:app_global_cost}
In this Appendix we report the toy model introduced in~\cite{CerezoBarrenLocalCost2021} to show that the global cost functions can lead to the emergence of vanishing gradients, hence the use of local costs is advisable to ensure trainability of the parameterised quantum circuit. Consider a simple tensor-product parameterised ansatz
\begin{eqnarray}
    U(\bm{\theta}) = \bigotimes_{i=1}^n e^{-i \theta_i X / 2}\,,
\end{eqnarray}
and the global and local observables
\begin{equation}
    O_G = \mathbb{I}^{\otimes n} - \dyad{\bm{0}}\,, \quad 
    O_L = 1 - \frac{1}{n}\sum_{i=1}^n \mathbb{I}^{(1)}\otimes \cdots \otimes \dyad{0}^{(i)}\otimes \cdots \otimes \mathbb{I}^{(n)}\,,
\end{equation}
where $\dyad{\bm{0}} = \dyad{0}^{\otimes n}$ is the ground state of quantum system of $n$ qubits, and the local cost is given by the sum of terms composed of single-qubit operators $\dyad{0}^{(i)}$ acting on the $i$-th site, and trivially on the others. Then, consider the corresponding cost functions
\begin{equation}
    C_G(\bm{\theta}) = \Tr[O_G\, V(\bm{\theta})\dyad{\bm{0}}V(\bm{\theta})^\dagger]\,,\quad C_L(\bm{\theta}) = \Tr[O_L\, V(\bm{\theta})\dyad{\bm{0}}V(\bm{\theta})^\dagger]\, ,
\end{equation}
which has the same vanish on the same solution, that is $C_G(\vecparam) = 0 \iff C_L(\vecparam)=0$. By direct calculation the cost functions can be easily shown to be
\begin{equation}
      C_G(\bm{\theta}) = 1- \prod_{i=1}^n\cos^2\frac{\theta_i}{2}\,,\quad C_L(\bm{\theta}) = 1-\frac{1}{n}\sum_{i=1}^n \cos^2\frac{\theta_i}{2}\, .  
\end{equation}
whose partial derivatives amount to
\begin{equation}
      \frac{\partial C_G(\bm{\theta})}{\partial \theta_k} = \frac{\sin\theta_k}{2}\,\prod_{i\neq k}^n\cos^2\frac{\theta_i}{2}\,,\quad 
      \frac{\partial C_L(\bm{\theta})}{\partial \theta_k} = \frac{\sin\theta_k}{2n}\, .  
\end{equation}

Assuming that each variational angle is sampled independently from the uniform distribution $\theta_i \sim \text{Unif}[0, 2\pi]$, then the expectation value of the partial derivative vanishes
\begin{equation}
    \ee{\frac{\partial C_G(\bm{\theta})}{\partial \theta_k}} = \ee{\frac{\sin\theta_k}{2}}\prod_{i\neq k}^n \ee{\cos^2\frac{\theta_i}{2}} = \int_{0}^{2\pi} \frac{d\theta_k }{2\pi} \frac{\sin \theta_k}{2} \cdot \qty(\int_{0}^{2\pi} \frac{d\theta}{2\pi}\cos^2\frac{\theta}{2})^{n-1} = 0\,,
\end{equation}
and similarly for the local cost $\ee{\partial_k C(\vecparam)} = 0$. The variance of the gradients can be calculated in a similar way, thus obtaining
\begin{align}
    \vv{\frac{\partial C_G(\bm{\theta})}{\partial \theta_k}} &= \ee{\qty(\frac{\partial C_G(\bm{\theta})}{\partial \theta_k})^2} = \int_{0}^{2\pi} \frac{d\theta_k }{2\pi} \frac{\sin^2 \theta_k}{4} \cdot \qty(\int_{0}^{2\pi} \frac{d\theta}{2\pi}\cos^4\frac{\theta}{2})^{n-1}\\
    & = \frac{1}{8}\cdot \qty(\frac{3}{8})^{n-1} \xrightarrow[n \rightarrow \infty]{} 0\,.
\end{align}
which is \textit{exponentially} vanishing with the number of qubits, the hallmark of barren plateaus. On the contrary, the local cost has a variance which vanish only \textit{polynomially} with the number of qubits, since
\begin{equation}
    \vv{\frac{\partial C_L(\bm{\theta})}{\partial \theta_k}} = \ee{\qty(\frac{\sin\theta_k}{2n})^2} = \int_{0}^{2\pi} \frac{d\theta_k }{2\pi} \frac{\sin^2 \theta_k}{4n^2} = \frac{1}{8 n^2}\,.
\end{equation}

This example makes it clear that a careful choice of the cost function is necessary to ensure trainability of the circuit. Indeed, in this case we showed that even a depth one circuit suffer of exponentially vanishing gradients if a global cost function is used. Thus, the use of a local cost function, that is one that only uses local measurements on a subset of the qubits, can ameliorate the barren plateau phenomenon, at least for shallow circuits whose depth $L$ scales logarithmically with the system size $n$, that is $L \sim \order{\log n}$~\cite{CerezoBarrenLocalCost2021}.

\section{Variance of gradients}
\label{sec:app_variance_gradients}
We are interested in evaluating the following expectation values
\begin{align}
\label{eq:app_2des_deriv}
    \text{Var}[\partial_k C(\vecparam)] & =  \mathbb{E}\qty[(\partial_k C(\vecparam))^2] = -\frac{1}{4}\mathbb{E}\qty[\Tr[U_A^\dagger O U_A\,\qty[P_k, U_B\rho U_B^\dagger]]^2]\, ,
\end{align}
under the assumption that the unitary ensembles $\mathbb{U}_{A,B}$ generated by $U_{A,B}$ are 2-designs, so that we can apply the explicit formula for integration over random unitary matrices reported in Eqs.~\eqref{eq:1haar},~\eqref{eq:2haar_alternative}, and~\eqref{eq:2haar}, which we recall also here for $d=2^n$
\begin{align}
\mathbb{E}_{U}[U A U^\dagger] &= \int d\mu(U)\,U A U^\dagger = \frac{\Tr[A]\, \mathbb{I}}{d}\, \label{eq:app_1haar}\\[1em]
\mathbb{E}_{U}[A U B U^\dagger C U D U^\dagger] &= \int d\mu(U)\,A U B U^\dagger C U D U^\dagger \nonumber\\
&= \frac{\Tr[BD]\Tr[C]A + \Tr[B]\Tr[D]AC}{d^2-1} \label{eq:app_2haar} \\ & \quad -\frac{\Tr[BD]AC + \Tr[B]\Tr[C]\Tr[D]A}{d(d^2-1)}\nonumber \\[1em]
\mathbb{E}_U\qty[\Tr[UAU^\dagger B]\Tr[UCU^\dagger D]] &= \int d\mu(U)\,\Tr[UAU^\dagger B]\Tr[UCU^\dagger D] \nonumber \\
&= \frac{\Tr[A]\Tr[B]\Tr[C]\Tr[D] + \Tr[AC]\Tr[BD]}{d^2-1} \label{eq:app_2haar_alternative} \\ & \quad -\frac{\Tr[AC]\Tr[B]\Tr[D] + \Tr[A]\Tr[C]\Tr[BD]}{d(d^2-1)}\nonumber
\end{align}

\paragraph{$\mathbb{U}_A$ is a 2-design}
First, we start with the case when $\mathbb{U}_A$ is a 2-design. Then, by direct application of~\eqref{eq:app_2haar_alternative} to~\eqref{eq:app_2des_deriv}, and setting $Q = \qty[P_k, U_B\rho U_B^\dagger]$, one has
\begin{align}
\label{eq:app_A2des}
    -4\text{Var}_{\mathbb{U}_A}\qty[\partial_k C(\vecparam)] &= \mathbb{E}_{\mathbb{U}_A}\qty[\Tr[U_A^\dagger O U_A Q]\Tr[U_A^\dagger O U_A Q]] \nonumber \\
    &= \frac{\Tr[O]^2\Tr[Q]^2 + \Tr[O^2]\Tr[Q^2]}{2^{2n}-1} -\frac{\Tr[O^2]\Tr[Q]^2 + \Tr[O]^2\Tr[Q^2]}{2^n(2^{2n}-1)}\nonumber\\
    & = \frac{\Tr[O^2]\Tr[Q^2]}{2^{2n}-1} -\frac{\Tr[O]^2\Tr[Q^2]}{2^n(2^{2n}-1)}\nonumber\\
    & = \frac{1}{2^{2n}-1}\qty(\Tr[O^2]-\frac{\Tr[O]^2}{2^n})\Tr[\qty[P_k, U_B\rho U_B^\dagger]^2]\, ,
\end{align}
where in the second line we made use of Eq.~\eqref{eq:app_2haar_alternative}, and in the third line the terms having $\Tr[Q]^2$ vanish because $\Tr[Q]=0$ since $Q$ is a commutator.\\

\paragraph{$\mathbb{U}_B$ is a 2-design}
The procedure is very similar if $\mathbb{U}_B$ is a 2-design. First, we rearrange the partial derivative as 
\begin{equation}
    \partial_k C(\vecparam) = -\frac{i}{2}\Tr[U_A^\dagger O U_A\,[P_k,\,U_B\rho U_B^\dagger]] = -\frac{i}{2}\Tr[U_B\rho U_B^\dagger\,[U_A^\dagger O U_A,\, P_k]]\,
\end{equation}
and then apply again Eq.~\eqref{eq:2haar_alternative}. Setting $Q = \qty[U_A^\dagger O U_A,\, P_k]$, one then obtains
\begin{align}
\label{eq:app_B2des}
    -4\text{Var}_{\mathbb{U}_B}\qty[\partial_k C(\vecparam)] &= \mathbb{E}_{\mathbb{U}_B}\qty[\Tr[U_B\rho U_B^\dagger\,Q]\Tr[U_B\rho U_B^\dagger\,Q]] \nonumber \\
    &= \frac{\Tr[\rho]^2\Tr[Q]^2 + \Tr[\rho^2]\Tr[Q^2]}{2^{2n}-1} -\frac{\Tr[\rho^2]\Tr[Q]^2 + \Tr[\rho]^2\Tr[Q^2]}{2^n(2^{2n}-1)}\nonumber\\
    & = \frac{1}{2^{2n}-1}\qty(\Tr[\rho^2]-\frac{1}{2^n})\Tr[\qty[U_A^\dagger O U_A,\, P_k]^2]\, ,
\end{align}
where as before terms with $\Tr[Q]$ vanish because $Q$ is a commutator, and in the last line we used $\Tr[\rho]=1$ by definition of density matrix.\\

\paragraph{$\mathbb{U}_{A,B}$ are both 2-designs}
Finally, we analyse what happens when both $\mathbb{U}_A$ and $\mathbb{U}_B$ are 2-designs. First, by setting again $O_A = U_A^\dagger O U_A$, we rewrite explicitly the squared commutator as
\begin{align}
\label{eq:app_commutator2}
    \Tr[\qty[O_A, P_k]^2] &= \Tr[O_A P_k O_A P_k - O_A P_k^2 O_A - P_k O_A^2 P_k + P_k O_A P_k O_A] \\
    & = 2\Tr[O_A P_k O_A P_k] - 2\Tr[O_A^2 P_k^2]\, ,
\end{align}
and then use Eq.~\eqref{eq:app_2haar} to calculate the expectation values of these quantities. The first term amounts to
\begin{align}
\label{eq:app_piece1_haar}
    &\quad\mathbb{E}_{\mathbb{U}_A}\qty[\Tr[U_A^\dagger O U_A P_k U_A^\dagger O U_A P_k]] = \Tr[\mathbb{E}_{\mathbb{U}_A}\qty[U_A^\dagger O U_A P_k U_A^\dagger O U_A] P_k] \nonumber\\ &= \Tr[\frac{\Tr[O^2]\Tr[P_k]\mathbb{I}+\Tr[O]^2 P_k}{2^{2n}-1}P_k] - \Tr[\frac{\Tr[O^2]P_k + \Tr[O]^2\Tr[P_k]\mathbb{I}}{2^n(2^{2n}-1)}P_k] \nonumber\\
    &=\frac{\Tr[O^2]\Tr[P_k]^2+\Tr[O]^2\Tr[P_k^2]}{2^{2n}-1} - \frac{\Tr[O^2]\Tr[P_k^2]+\Tr[O]^2\Tr[P_k]^2}{2^n(2^{2n}-1)},
\end{align}
and the second term can be calculated using the formula for first-order moments of Haar-random matrices~\eqref{eq:app_1haar}
\begin{align}
\label{eq:app_piece2_haar}
    \mathbb{E}_{\mathbb{U}_A}\qty[\Tr[O_A^2P_K^2]] &= \Tr[\mathbb{E}_{\mathbb{U}_A}\qty[U_A^\dagger O^2 U_A]P_k^2] = \Tr[\frac{\Tr[O^2]\mathbb{I}}{2^n}P_k^2] = \frac{\Tr[O^2]\Tr[P_k^2]}{2^n}\, .
\end{align}
Thus, one can eventually compute the expectation value over $\mathbb{U}_{A,B}$ by combining Eqs.~\eqref{eq:app_piece1_haar} and~\eqref{eq:app_piece2_haar} with~\eqref{eq:app_commutator2}, and plugging these into Eq.~\eqref{eq:app_B2des}, obtaining
\begin{align}
-4\text{Var}_{\mathbb{U}_{A,B}}\qty[\partial_k C(\vecparam)] &= -4\mathbb{E}_{\mathbb{U}_A}\qty[\mathbb{E}_{\mathbb{U}_B}\qty[\qty(\partial_k C(\vecparam))^2]] \nonumber\\ &= \frac{1}{2^{2n}-1}\qty(\Tr[\rho^2]-\frac{1}{2^n})\mathbb{E}_{\mathbb{U}_A}\qty[\Tr[\qty[U_A^\dagger O U_A,\, P_k]^2]]\nonumber\\
&= \frac{2}{2^{2n}-1}\qty(\Tr[\rho^2]-\frac{1}{2^n})\bigg(\frac{\Tr[O^2]\Tr[P_k]^2+\Tr[O]^2\Tr[P_k^2]}{2^{2n}-1}\nonumber \\
&\quad\quad\quad - \frac{\Tr[O^2]\Tr[P_k^2]+\Tr[O]^2\Tr[P_k]^2}{2^n(2^{2n}-1)} - \frac{\Tr[O^2]\Tr[P_k^2]}{2^n} \bigg)\label{eq:app_AB2des}
\end{align}

Summing up, we summarise Equations~\eqref{eq:app_A2des},~\eqref{eq:app_B2des}, and~\eqref{eq:app_AB2des}, as follows. 
\begin{gather}
    \text{If $\mathbb{U}_A$ is at least a 2-design, then $\forall~k$:}\nonumber\\
    \quad\text{Var}_{\mathbb{U}_A}[\partial_k C(\vecparam)] = -\frac{1}{4}\frac{1}{2^{2n}-1}\qty(\Tr[O^2]-\frac{\Tr[O]^2}{2^n})\Tr[\qty[P_k, U_B\rho U_B^\dagger]^2]\label{eq:app_A2des_total}
\\[0.5em]
    \text{If $\mathbb{U}_B$ is at least a 2-design, then $\forall~k$:}\nonumber\\
    \quad\text{Var}_{\mathbb{U}_B}[\partial_k C(\vecparam)] = -\frac{1}{4}\frac{1}{2^{2n}-1}\qty(\Tr[\rho^2]-\frac{1}{2^n})\Tr[\qty[U_A^\dagger O U_A, P_k]^2]\label{eq:app_B2des_total}\\[0.5em]
    \text{If $\mathbb{U}_{A,B}$ are both at least 2-designs, then $\forall~k$:}\nonumber\\
    \quad\begin{aligned}
    \text{Var}_{\mathbb{U}_{A, B}}[\partial_k C(\vecparam)] &= -\frac{1}{4}\frac{2}{2^{2n}-1}\qty(\Tr[\rho^2]-\frac{1}{2^n})\bigg(\frac{\Tr[O^2]\Tr[P_k]^2+\Tr[O]^2\Tr[P_k^2]}{2^{2n}-1}\\
    &\quad\quad\quad - \frac{\Tr[O^2]\Tr[P_k^2]+\Tr[O]^2\Tr[P_k]^2}{2^n(2^{2n}-1)} - \frac{\Tr[O^2]\Tr[P_k^2]}{2^n} \bigg)
    \end{aligned}
\end{gather}

Note that despite the minus signs in front of the expressions, one can verify that all these variances are correctly positive. Indeed, for Eq.~\eqref{eq:app_A2des_total}, by Cauchy–Schwarz it holds
\[\Tr[O\,\mathbb{I}]^2 \leq \Tr[O^\dagger O] \Tr[\mathbb{I}] = 2^n\Tr[O^2] \implies \Tr[O^2] \geq \frac{\Tr[O]^2}{2^n}\,\]
and similarly for the term involving the purity $\Tr[\rho^2]$ in~\eqref{eq:app_B2des_total}, where the latter can also be seen as a consequence of the minimum of the purity of a quantum state being achieved on the completely mixed state. On the contrary, the trace of squared commutators are always negative values, as for any matrices $A,B$ it holds $\abs{\Tr[ABAB]} \leq \Tr[A^\dagger A B B^\dagger]$~\cite{Petz1994TraceInequalities, Bhatia1996matrix}. Applying such inequality for the specific case of $A,B$ being Hermitian matrices, yields
\begin{align}
    \Tr[[A,B]^2] = 2\qty(\Tr[ABAB]-\Tr[A^2B^2]) \leq 2\qty(\Tr[A^2 B^2]-\Tr[A^2B^2]) = 0\,,
\end{align}
where in our case $A,B$ are the Hermitian operators $P_k$, $U_B\rho U_B^\dagger$, and $U_A^\dagger O U_A$, as both $P_k$ and $O$ are Hermitian.

The expressions for the variance take a simpler form when evaluated for the most common case of parameterised gates generated by Pauli rotations $P_k$, when the observable $O$ is a Pauli string, and the initial state is a pure state, for example $\rho = \dyad{0}$. In this case, one has $\Tr[P_k] = \Tr[O] = 0$, and $\Tr[P_k^2] = \Tr[O^2] = \Tr[\mathbb{I}] = 2^n$, because Pauli matrices are traceless and involutory, and also $\Tr[\rho^2] = \Tr[\rho] = 1$, because $\rho$ is a pure state. Then, the equations above greatly simplifies to
\begin{gather}
    \text{If $\mathbb{U}_A$ is at least a 2-design, then $\forall~k$:}\nonumber\\
    \quad\text{Var}_{\mathbb{U}_A}[\partial_k C(\vecparam)] = \frac{1}{2}\frac{2^n}{2^{2n}-1}\qty(1-\Tr[P_k U_B\rho U_B^\dagger P_k U_B\rho U_B^\dagger])\label{eq:app_2desA_simplified}\\[0.5em]
    \text{If $\mathbb{U}_B$ is at least a 2-design, then $\forall~k$:}\nonumber\\
    \quad\text{Var}_{\mathbb{U}_B}[\partial_k C(\vecparam)] = \frac{1}{2}\frac{1}{2^n(2^{n}+1)}\qty(2^n - \Tr[P_k U_A^\dagger O U_A P_k U_A^\dagger O U_A])\label{eq:app_2desB_simplified}\\[0.5em]
    \text{If $\mathbb{U}_{A,B}$ are both at least 2-designs, then $\forall~k$:}\nonumber\\
    \quad\text{Var}_{\mathbb{U}_{A, B}}[\partial_k C(\vecparam)] = \frac{2^{2n}}{2(2^n+1)(2^{2n}-1)}\label{eq:app_2desAB_simplified}\,.
\end{gather}
For all the cases treated above it is clear that the variance of the gradients vanish exponentially with the number of qubits, namely $\text{Var}[\partial_k C(\vecparam)] \in \order{2^{-n}}$.

\chapterimage{} 
\chapterspaceabove{2.5cm} 
\chapterspacebelow{2cm} 
\chapter{Quantum Machine Learning}
\label{app:proofs_QML}

\section{\label{app:qnn_generalization} Generalisation bound for data-reuploading quantum neural networks}
In this Appendix we show how to derive the generalisation bound for reuploading quantum neural networks as the one shown in Eq.~\eqref{eq:qnn_generalization}, using Rademacher complexity as a measure of uniform convergence. In particular, we first start introducing definitions and tools on Rademacher complexity and show how to apply them to linear models with features in Sec.~\ref{ssec:rademacher_linear_models}. Then, in Sec.\ref{ssec:rademacher_qnn_models} we show how to apply these results for the case of data reuploading quantum neural networks.  

\subsection{Rademacher complexity and generalisation error}
We now proceed introducing the Rademacher Complexity measure and the so-called Concentration Lemma, needed for the derivation of the generalisation bound. Then, we state the main theorem connecting the Rademacher complexity of an hypothesis class with its generalisation error. 

\begin{definition}[Empirical Rademacher Complexity]
\label{def:empirical_rademacher}
Let $\mathcal{F}$ a family of functions mapping a data space $\mathcal{Z}$ to $[a, b]$, and $S = (z_1, \hdots, z_m) \subset \mathcal{Z}^m$ a fixed sample of size $m$ with elements in $\mathcal{Z}$. The empirical Rademacher complexity of class $\mathcal{F}$ with respect to the sample $S$ is defined as
\begin{equation}
    \mathcal{R}_S(\mathcal{F}) \coloneqq \frac{1}{m}\, \mathbb{E}_{\bm{\sigma}}\bigg[\sup_{f \in \mathcal{F}} \,\sum_{i=1}^m \sigma_i f(z_i)\bigg]\,,
\end{equation}
where $\bm{\sigma} = (\sigma_1,\hdots,\sigma_m) \in \{\pm 1\}^m$, with $\sigma_i \sim \mathrm{Unif}\{+1,-1\}$ are independent uniformly distributed binary random variables. The random variables $\sigma_i$ are called \textit{Rademacher} variables. 
\end{definition}
More generally, given a set of vectors $A = \{\bm{a}_1, \bm{a}_2, \hdots\, |\, \bm{a}_i \in \mathbb{R}^m\} \subset \mathbb{R}^{m}$, one defines the Rademacher complexity of the set $A$ as
\begin{equation}
    \mathcal{R}(A) \coloneqq \frac{1}{m}\, \mathbb{E}_{\bm{\sigma}}\bigg[\sup_{\bm{a} \in A} \,\sum_{i=1}^m \sigma_i\, a_i\bigg]\,.
\end{equation}
We now state the Contraction lemma regarding the Rademacher complexity of composed functions.

\begin{lemma}[Contraction Lemma (lemma 26.9 in ref.~\cite{ShalevBenDavid2014})]
\label{lem:contraction_lemma}
For each $i \in 1, \hdots, m$, let $\ell _i: \mathbb{R} \rightarrow \mathbb{R}$ be a $L$-Lipschitz function, namely for all $\alpha,\,\beta \in \mathbb{R}$ we have $|\ell_i(\alpha) - \ell_i(\beta)| \leq L |\alpha - \beta|$. For $\bm{a} \in \mathbb{R}^m$ let $\bm{\ell}(\bm{a})$ denote the vector $\bm{\ell}(\bm{a}) = (\ell_1(a_1), \hdots, \ell_m(a_m))$. Let $\bm{\ell} \circ A = \{\bm{\ell}(\bm{a})\,|\, \bm{a} \in A\}$. Then,
\begin{equation}
    \mathcal{R}(\bm{\ell} \circ A) \leq L\, \mathcal{R}(A)
\end{equation}
\end{lemma}

Let $S = \{z_1, \hdots, z_m\} \subset \mathcal{Z}^{m}$ be a collection of \textit{iid} variables sampled from a distribution $\mathcal{D}$ on $\mathcal{Z}$. In supervised learning scenarios one has $\mathcal{Z} =\mathcal{X} \times \mathcal{Y}$, where elements $z_i = (x_i, y_i)$ are pairs of input data $x_i \in \mathcal{X}$ and corresponding output $y_i \in \mathcal{Y}$. Let $\mathcal{M}$ denote an hypothesis class containing mappings form input to output space $\mathcal{M} \subset \{h\, |\, h: \mathcal{X} \rightarrow \mathcal{Y}\}$, and let $\ell$ denote a loss function $\ell: \mathcal{Y} \times \mathcal{Y} \rightarrow \mathbb{R}$. In addition, let 
\begin{equation}
    \mathcal{G} = \ell \circ \mathcal{M} : =\qty{z = (x,y) \mapsto \ell(h(x), y) \, |\, h \in \mathcal{M} }
\end{equation}
be the class of mappings from the data space $\mathcal{Z}$ to $\mathbb{R}$, obtained by combining the action of the models $h \in \mathcal{M}$ in the hypothesis class with the loss function $\ell$. With $g \in \mathcal{G}$, we define the true risk $L_\mathcal{D}(g)$ and the empirical risk $L_{S}(g)$, as
\begin{equation}
    L_\mathcal{D}(g) \coloneqq \mathbb{E}_{z\sim \mathcal{D}} [g (z)] \quad , \quad L_{S}(g) \coloneqq \frac{1}{m}\sum_{i=1}^m g(z_i)\, ,
\end{equation}
which are defined in terms of the probability distribution $\mathcal{D}$ on data space $\mathcal{Z}$, the sample set $S$ consisting of $m$ data points, the hypothesis class $\mathcal{M}$ representing the parameterised learning model, and finally the loss function $\ell$ used to train the model. With these definitions, we are now ready to state the well-known theorem in classical statistical learning theory which bounds the generalisation error of a parameterised model with its Rademacher complexity. 

\begin{theorem}[Generalisation Bound via Rademacher Complexity (th. 26.5 in Ref.~\cite{ShalevBenDavid2014}, th. 1.15 in Ref.~\cite{WolfML}] 
Be $\mathcal{Z} = \mathcal{X} \times \mathcal{Y}$ a data space with arbitrary inputs and outputs spaces, $\mathcal{X}$ and $\mathcal{Y}$. Consider an hypothesis class $\mathcal{M} \subset \{h: \mathcal{X} \rightarrow \mathcal{Y}\}$, a loss function $\ell: \mathcal{Y} \times \mathcal{Y} \rightarrow [0,c]$, and define $\mathcal{G} := \{z \mapsto \ell(h(x), y)\, |\, h \in \mathcal{M}\}$. For any $\delta > 0$, and probability measure $\mathcal{D}$ over $\mathcal{Z}$, with probability at least $1-\delta$ over the draw of a training set $S \in \mathcal{Z}^m$ of size m, for all $g \in \mathcal{G}$:
\begin{equation}
\label{eq:th_rad_generalization}
    L_\mathcal{D}(g) - L_{S}(g) < 2\mathcal{R}_S (\mathcal{G}) + 3c\,\sqrt{\frac{\log 2/\delta}{2m}}
\end{equation}
\end{theorem}

\subsection{\label{ssec:rademacher_linear_models} Rademacher complexity of Linear Classes}
In the following we derive the Rademacher complexity of linear --- with respect to the trainable parameters --- models where we allow the inputs to be transformed with a feature map. The derivation follows that found in ref.~\cite{ShalevBenDavid2014}, with the difference that we here allow inputs to go through a feature map, and we use a different final bound on the norm. Consider the hypothesis class of parametric linear models
\begin{equation}
    \mathcal{M} = \{\bmx \mapsto \bm{w} \cdot \bm{\phi}(\bmx)~|~\norm{\bm{w}}_2 \leq B\}\,,
\end{equation}
where $\bm{\phi}: \mathbb{R}^d \rightarrow \mathbb{R}^M$ is a general feature map mapping inputs to feature vectors $\bm{x} \in \mathbb{R}^d \mapsto \bm{\phi}(\bmx) \in \mathbb{R}^M$, and $\norm{\cdot}_2$ denotes the standard Euclidean norm. 

Let $S = ((\bmx_1, y_1), \hdots, (\bmx_m, y_m)) \in (\mathbb{R}^d \times \mathbb{R})^{m}$ be a training set, and $\ell: \mathbb{R} \times \mathbb{R} \rightarrow \mathbb{R}$ a $L$-lipschitz loss function. Then, consider the class
\begin{equation}
\label{eq:linear_model_class}
    \mathcal{G} = \{(\bmx_i, y_i) \mapsto \ell(h(\bmx_i), y_i)\, |\, h \in \mathcal{M}\}\,.
\end{equation}
The empirical Rademacher complexity of class $\mathcal{G}$ on set $S$ is defined as (see Definition~\ref{def:empirical_rademacher})
\begin{eqnarray}
    \mathcal{R}_S(\mathcal{G}) &=& \frac{1}{m}\, \mathbb{E}_{\bm{\sigma}}\bigg[\sup_{g \in \mathcal{G}} \,\sum_{i=1}^m \sigma_i\, g(z_i)\bigg]\\
    &=& \frac{1}{m}\, \mathbb{E}_{\bm{\sigma}}\bigg[\sup_{h \in \mathcal{M}} \,\sum_{i=1}^m \sigma_i\, \ell(h(\bmx_i), y_i)\bigg]\\
    &\leq& \frac{L}{m}\, \mathbb{E}_{\bm{\sigma}}\bigg[\sup_{h \in \mathcal{M}} \,\sum_{i=1}^m \sigma_i\, h(\bmx_i)\bigg]\,,
\end{eqnarray}
where in the last line we made use of the Contraction lemma~\ref{lem:contraction_lemma} to remove the dependence on the loss function, by defining the map $\bm{\ell}(\bm{a}) \mapsto (\ell_1(a_1), \hdots, \ell_m(a_m)) = (\ell(a_1, y_1), \hdots, \ell(a_m, y_m))$ with $\bm{a} = (h(\bmx_1), \hdots, h(\bmx_m))$, and assuming that $\ell(a_i, y_i)$ is $L$-Lipschitz for all $a_i\,, y_i\,,  i=1,\hdots, m$.

Substituting the definition of the linear model~\eqref{eq:linear_model_class} $h(\bmx_i) = \bm{w} \cdot \bm{\phi}(\bmx_i)$, we then have
\begin{eqnarray}
    \mathcal{R}_S(G) &\leq& \frac{L}{m}\, \mathbb{E}_{\bm{\sigma}}\bigg[\sup_{\norm{\bm{w}}_2 \leq B} \,\sum_{i=1}^m \sigma_i\, \bm{w} \cdot \bm{\phi}(\bmx_i)\bigg]\\
    &=& \frac{L}{m}\, \mathbb{E}_{\bm{\sigma}}\bigg[\sup_{\norm{\bm{w}}_2 \leq B} \, \bm{w} \cdot \sum_{i=1}^m \sigma_i\, \bm{\phi}(\bmx_i)\bigg] \\  
    \textrm{(Cauchy-Schwartz)}&\leq& \frac{L}{m}\, \mathbb{E}_{\bm{\sigma}}\qty[\sup_{\norm{\bm{w}}_2 \leq B} \, \|\bm{w}\|_2\, \norm{\sum_{i=1}^m \sigma_i\, \bm{\phi}(\bmx_i)}_2] \label{eq:proof_cauchy}\\
    &\leq& \frac{B\, L}{m}\, \mathbb{E}_{\bm{\sigma}}\qty[\norm{\sum_{i=1}^m \sigma_i\, \bm{\phi}(\bmx_i)}_2]\\
    &=& \frac{B\, L}{m}\, \mathbb{E}_{\bm{\sigma}}\qty[ \qty({\norm{\sum_{i=1}^m \sigma_i\, \bm{\phi}(\bmx_i)}_2^2})^{1/2} ]\\
    \textrm{(Jensen's inequality)}&\leq& \frac{B\, L}{m}\, \qty( \mathbb{E}_{\bm{\sigma}}\qty[\norm{\sum_{i=1}^m \sigma_i\, \bm{\phi}(\bmx_i)}_2^2] )^{1/2}\, \,, \label{eq:proof_jensen}
\end{eqnarray}
where in~\eqref{eq:proof_cauchy} we used Cauchy-Schwartz inequality $\bm{w} \cdot \bm{q} \leq \|\bm{w}\|_2 \|\bm{q}\|_2$ followed by $\|\bm{w}\|_2\leq B$, and in~\eqref{eq:proof_jensen} we used Jensen's inequality to move the square root out of the expectation value.

Now, since the Rademacher variables $\sigma_i$ are independent, the expectation can be calculated as follows
\begin{eqnarray}
    \mathbb{E}_{\bm{\sigma}}\qty[\norm{\sum_{i=1}^m \sigma_i\, \bm{\phi}(\bmx_i)}_2^2] &=& \mathbb{E}_{\bm{\sigma}}\qty[\sum_{i,j=1}^m \sigma_i\sigma_j\, \bm{\phi}(\bmx_i)\cdot \bm{\phi}(\bmx_j)] \\
    &=& \sum_{i\neq j}^m \bm{\phi}(\bmx_i)\cdot \bm{\phi}(\bmx_j)\, \underbrace{\mathbb{E}_{\bm{\sigma}}\qty[\sigma_i\sigma_j]}_{=0} + \sum_{i=1}^m \bm{\phi}(\bmx_i)\cdot \bm{\phi}(\bmx_i)\, \underbrace{\mathbb{E}_{\bm{\sigma}}\qty[\sigma_i^2]}_{=1}\\
    &=& \sum_{i=1}^m \|\bm{\phi}(\bmx_i)\|_2^2 \leq m \max_i \|\bm{\phi}(\bmx_i)\|_2^2 \\
    &\leq& m\, M \, \max_i \|\bm{\phi}(\bmx_i)\|_{\infty}^2 \label{eq:fin1}
\end{eqnarray}
where in the last line we used that for any vector $\bm{\phi} \in \mathbb{R}^M$ one can bound the 2-norm with the infinity norm as $\|\bm{\phi}\|_2 \leq \sqrt{M} \|\bm{\phi}\|_{\infty}$, where the infinity norm of a vector is defined as the maximum absolute value of its components, namely $\|\bm{\phi}\|_{\infty} = \max(|\phi_1|, \hdots, |\phi_M|)$.

Putting it all together, substituting Eq.~\eqref{eq:fin1} into Eq.~\eqref{eq:proof_jensen}, we have shown that the linear model class~\eqref{eq:linear_model_class} has 
\begin{equation}
\label{eq:final_res_linear_model}
    \mathcal{R}_S(\mathcal{G}) \leq B\,L\,\max_i \|\bm{\phi}(\bmx_i)\|_{\infty}^2\,\sqrt{\frac{M}{m}}\, .
\end{equation}

\subsection{\label{ssec:rademacher_qnn_models} Generalisation bound of Quantum Neural Networks}
Using the result on linear models with features, we are now ready to prove the generalisation bound shown in Eq.~\eqref{eq:qnn_generalization} for quantum neural networks.

We first show that a data reuploading quantum model can be written in terms of a ``linear" model with trigonometric features. A quantum model can be expressed as 
\begin{equation}
    \Tr[\rho(\bmx; \bm{\theta})\, O] = f(\bm{x};\, \bm{\theta}) = \sum_{\bm{\omega} \in \Omega} c_{\bm{\omega}}(\bm{\theta})\, e^{-i \bm{\omega}\cdot\bm{x}}
\end{equation}
where the coefficients are such that $c_{-\bmo} = c_{\bmo}^*$, and the spectrum the spectrum $\Omega$ consists of positive and negative frequencies 
\begin{equation}
    \Omega = \{\Omega^-\} \cup \{\bm{0}\} \cup \{\Omega^+\}\quad \textrm{with}\quad  \Omega^- = \{-\bmo\,|\, \bmo \in \Omega^+\}
\end{equation}
so that $|\Omega| = 2|\Omega^+| + 1$. Dropping the dependence of the coefficients on $\bm{\theta}$ for ease of notation, the model can be expressed in terms of real coefficients and trigonometric features as
\begin{eqnarray}
    f(\bm{x};\, \bm{\theta}) &=& \sum_{\bm{\omega} \in \Omega} c_{\bm{\omega}}\, e^{-i \bm{\omega}\cdot\bm{x}} \label{eq:qnn_fourier}\\
    &=& \sum_{\bm{\omega} \in \Omega^-} c_{\bm{\omega}}\, e^{-i \bm{\omega}\cdot\bm{x}} + c_0 + \sum_{\bm{\omega} \in \Omega^+} c_{\bm{\omega}}\, e^{-i \bm{\omega}\cdot\bm{x}} \\
    &=& \sum_{\bm{\omega} \in \Omega^+} c_{\bm{\omega}}^*\, e^{i \bm{\omega}\cdot\bm{x}} + c_0 + \sum_{\bm{\omega} \in \Omega^+} c_{\bm{\omega}}\, e^{-i \bm{\omega}\cdot\bm{x}} \\
    &=& c_0 + \sum_{\bm{\omega} \in \Omega^+} \qty(c_{\bm{\omega}}^*\, e^{i \bm{\omega}\cdot\bm{x}} + c_{\bm{\omega}}\, e^{-i \bm{\omega}\cdot\bm{x}}) \\
    &=& c_0 + \sum_{\bm{\omega} \in \Omega^+} \qty((a_{\bm{\omega}}-i b_{\bmo})\, e^{i \bm{\omega}\cdot\bm{x}} +(a_{\bm{\omega}}-i b_{\bmo})\, e^{-i \bm{\omega}\cdot\bm{x}}) \\
    &=& c_0 + \sum_{\bmo \in \Omega^{+}} 2 a_{\bmo} \cos{(\bmo\cdot\bmx)} + \sum_{\bmo \in \Omega^{+}} 2 b_{\bmo} \sin{(\bmo\cdot\bmx)} \\
    &=& \bm{w} \cdot \bm{\phi}({\bmx})\, , \label{eq:qnn_linear_version}
\end{eqnarray}
where we have introduced  a Fourier-like feature map $\bm{\phi}: \mathbb{R}^d \rightarrow \mathbb{R}^{|\Omega|}$ and coefficients vector defined as
\begin{equation}
\label{eq:qnn_trig_features}
    \bmx \longmapsto \bm{\phi}(\bmx) = \qty[ 1,\, \cos(\bmo_1\cdot\bmx),\,\hdots, \cos(\bmo_{|\Omega^+|}\cdot\bmx),\, \sin(\bmo_1\cdot\bmx), \hdots,\, \sin(\bmo_{|\Omega^+|}\cdot\bmx) ]\,,
\end{equation}
and the vector of coefficients $\bm{w} \in \mathbb{R}^{|\Omega|}$
\begin{equation}
\label{eq:qnn_trig_coeffs}
    \bm{w} = \qty[c_0, 2a_{\bmo_1},\hdots, 2a_{\bmo_{|\Omega^+|}}, 2b_{\bmo_1},\hdots, 2b_{\bmo_{|\Omega^+|}} ]\,.
\end{equation}
Denote with $\bm{c} = [c_{\bmo}\,, \bmo \in \Omega]$ the vector of complex coefficients in Eq.~\eqref{eq:qnn_fourier}, then the norm of the coefficient vectors $\bm{c}$ and $\bm{w}$ amounts to
\begin{eqnarray}
    \|\bm{c}\|_2^2 &=& c_0 ^ 2 + 2 \sum_{\bmo \in \Omega^+} a_{\bmo}^2 + b_{\bmo}^2 \\
    \|\bm{w}\|_2^2 &=& c_0 ^ 2 + 4 \sum_{\bmo \in \Omega^+} a_{\bmo}^2 + b_{\bmo}^2 = 2 \|\bm{c}\|_2^2 - c_0^2 \leq 2 \|\bm{c}\|_2^2\, .
\end{eqnarray}

Now, let $\bm{x} \in [0, 2\pi]^{d}$, and assume that the frequency spectrum is made of integer frequencies only, that is $\bmo \in \mathbb{Z}^d,~\forall\,\bmo \in \Omega$. In this case, one can prove the so-called \textit{Parseval's equality}, which connects the integral of a function to the the norm of the coefficients in the Fourier expansion of the function. Indeed, by dropping again the dependence on $\bm{\theta}$ for ease of notation, one can show that
\begin{eqnarray}
\label{eq:parseval_proof}
\int_{[0,2\pi]^d} d\bmx \, |f(\bmx)|^2 &=& \int_{[0,2\pi]^d} d\bmx \sum_{\bm{\omega},\bm{\nu} \in \Omega} c^*_{\bm{\nu}} c_{\bm{\omega}}\, e^{-i (\bm{\omega}-\bm{\nu})\cdot\bm{x}} \\
& = & \sum_{\bm{\omega},\bm{\nu} \in \Omega} c^*_{\bm{\nu}} c_{\bm{\omega}}\, \int_{[0,2\pi]^d} d\bmx\,  e^{-i (\bm{\omega}-\bm{\nu})\cdot\bm{x}} \\
&=& \sum_{\bm{\omega},\bm{\nu} \in \Omega} c^*_{\bm{\nu}} c_{\bm{\omega}}\, \prod_{j=1}^d\, \int_0^{2\pi} d x^{(j)}\, e^{-i (\omega^{(j)}-\nu^{(j)})\,x^{(j)}} \\
&=& \sum_{\bm{\omega},\bm{\nu} \in \Omega} c^*_{\bm{\nu}} c_{\bm{\omega}}\, (2\pi)^d\delta_{\bmo, \bm{\nu}} = (2\pi)^d \sum_{\bmo \in \Omega} |c_{\bmo}|^2 = (2\pi)^d \|\bm{c}\|_2^2\, ,
\end{eqnarray}
where $\delta_{\bmo, \bm{\nu}}$ is a Kronecker delta which is one if $\bmo = \bm{\nu}$, and zero otherwise.

Since the output of the quantum model is the expectation value of the observable $O$, the maximum value it can attain is bounded by the operator norm (i.e. the largest eigenvalue) of the observable, namely
\begin{equation}
|f(\bm{x};\, \bm{\theta})| = |\Tr[\rho(\bmx; \bm{\theta})\, O]| \leq \|\rho(\bmx; \bm{\theta})\|_1 \|O\|_{\infty} =  \|O\|_{\infty}.
\end{equation}  
where $\|\rho(\bmx; \bm{\theta})\|_1 = \Tr[\rho(\bmx; \bm{\theta})]= 1$ because $\rho(\bmx; \bm{\theta})$ is a density matrix. Eventually, one can bound the 2-norm of the Fourier coefficient vector $\bm{c}$, hence $\bm{w}$, as follows
\begin{eqnarray}
    \int_{[0,2\pi]^d} d\bmx \, |f(\bmx)|^2 \leq \int_{[0,2\pi]^d} d\bmx \, \|O\|^2_{\infty} = (2\pi)^d\, \|O\|^2_{\infty}\, \\
    \implies \|\bm{c}\|_2 \leq \|O\|_{\infty}\quad \textrm{and} \quad \|\bm{w}\|_2 \leq 2\|O\|_{\infty}\, .
\end{eqnarray}

Summing up, we have shown how to write a quantum reuploading model as a linear model with trigonometric features $f(\bmx) = \bm{w} \cdot \bm{\phi}(\bmx)~\eqref{eq:qnn_linear_version}$, whose parameter have bounded 2-norm $\|\bm{w}\|_2 \leq 2\|O\|_{\infty}$, so that the results from the previous section on linear models readily apply also to the case of quantum neural networks. 

Let $\mathcal{M}_{\textrm{QNN}}$ denote the model class implemented by a quantum neural network with an encoding scheme generating an integer-valued spectrum $\Omega$, obtained by measuring an observable $O$. Such class and its composition with a $L$-lipschitz loss function $\bm{\ell}$ can then be expressed respectively as (see~\ref{ssec:rademacher_linear_models})
\begin{equation}
\begin{aligned}
    & \mathcal{M}_{\textrm{QNN}} =  \left\{ \bmx \mapsto \bmw \cdot \bm{\phi}({\bmx})~|~ \norm{\bm{w}}_2 \leq 2 \norm{O}_{\infty},\,\, \bmw \in \mathbb{R}^{|\Omega|}\textrm{ as in~\eqref{eq:qnn_trig_coeffs}}, \right.\\
    & \left. \quad \quad \quad \quad \quad \quad \quad \quad \quad
    \quad \quad \quad \quad \quad \quad \quad \quad
    \bm{\phi}: [0, 2\pi]^d \rightarrow \mathbb{R}^{|\Omega|} \textrm{ as in~\eqref{eq:qnn_trig_features}} \right\} \,,
\end{aligned}
\end{equation}
and 
\begin{equation}
    \mathcal{G}_{\textrm{QNN}} = \bm{\ell} \circ \mathcal{H_{\textrm{QNN}}} = \{(\bmx_i, y_i) \mapsto \ell(h(\bmx_i), y_i)\, |\, h \in \mathcal{M}_{\textrm{QNN}}\}\, .
\end{equation}
Since the feature map $\bm{\phi}$ consists of trigonometric functions of the input data, it holds that $\bm{\phi}(\bmx_i) \in [-1, 1]^{|\Omega|}\, \forall i=1, \hdots, m$. Hence $\max_{i} \|\bm{\phi}(\bmx_i)\|_{\infty} = 1$, achieved on the first element of the vector, it being the constant feature 1. Thus, finally, using~\eqref{eq:final_res_linear_model} on class $\mathcal{G}_{\textrm{QNN}}$, one has
\begin{equation}
    \label{eq:final_res_qnn_model}
    \mathcal{R}_S(\mathcal{G}_{\textrm{QNN}}) \leq 2\|O\|_{\infty}\,L\, \sqrt{\frac{|\Omega|}{m}}\,,
\end{equation}
and plugging this in the generalisation theorem~\eqref{eq:th_rad_generalization}, one finally has the desired generalisation bound. 

\begin{theorem}[Generalisation Bound for Quantum Neural Networks (see also Theorem~6 in ref.~\cite{Caro2021encodingdependent})] 
Be $\mathcal{Z} = [0,2\pi]^d \times \mathbb{R}$ a data space of inputs and outputs. Consider a data reuploading quantum circuit whose encoding scheme generates an integer-valued spectrum $\Omega$, whose model class is $\mathcal{M}_{\textrm{QNN}} := \{\bm{x} \mapsto \Tr[\rho(\bm{x};\bm{\theta})\, O] = \sum_{\bmo \in \Omega} c_{\bmo} e^{-i \bmo \cdot \bmx}\}$. Be $\ell: \mathbb{R} \times \mathbb{R} \rightarrow [0,c]$ an $L$-Lipschitz loss function and define $\mathcal{G}_{\textrm{QNN}} := \{z = (x, y) \mapsto \ell(h(x), y)\, |\, h \in \mathcal{M}_{\textrm{QNN}}\}$. For any $\delta > 0$ and probability measure $\mathcal{D}$ over $\mathcal{Z}$, with probability at least $1-\delta$ over the drawn of a training set $S \in \mathcal{Z}^m$ of size m, for all $g \in \mathcal{G}_{\textrm{QNN}}$:
\begin{equation}
    L_\mathcal{D}(g) - L_{S}(g) < 4\|O\|_{\infty}\,L\, \sqrt{\frac{|\Omega|}{m}}\, + 3c\,\sqrt{\frac{\log 2/\delta}{2m}}
\end{equation}
\end{theorem}

\chapterimage{} 
\chapterspaceabove{2.5cm} 
\chapterspacebelow{2cm} 
\chapter{Continuous Quantum Neuron}
\label{app:proofs_CQN}

\section{Proof of the activation function of the quantum neuron}
\label{app:CQN_App_A}
Consider a collection of complex numbers $z_i = r_i e^{i\gamma_i} \in \mathbb{C},\, i = 1, \hdots, N$. The squared modulus of their sum can be explicitly calculated as follows
\begin{align}
\abs{\sum_{i=1}^N z_i}^2 & = \qty(\sum_{i=1}^N z_i) \qty(\sum_{j=1}^N z^*_j) = \sum_{i,j=1}^N z_i  z^*_j = \sum_{i=j}^N |z_i|^2 + \sum_{i \neq j}^N r_i r_j e^{i(\gamma_i-\gamma_j)} \\
& = \sum_{i=j}^N r_i^2 + 2 \sum_{i<j}^N r_i r_j \cos(\gamma_j - \gamma_i)\ ,
\end{align}
where in the last line the following relation has been applied
\begin{equation}
e^{i x}+e^{-ix} = 2 \cos(x)\ .
\end{equation}
Setting $r_i=1/N$ and $\gamma_i = \theta_i-\phi_i$, one obtains
\begin{equation}
 \left| \sum_{i=1}^N\frac{e^{i(\theta_i-\varphi_i)}}{N}\right|^2  = \frac{1}{N} + \frac{2}{N^2}\sum_{i<j}^{N}\cos((\theta_j - \varphi_j) - (\theta_i - \varphi_i))\ ,
\end{equation}
which correctly reduces to Eq.~\eqref{eq:mod} in the main text upon substituting $N=2^n$ and shifting the summation indices to start from zero.

\section{Noise resilience}
\label{app:CQN_App_B}
Consider an input vector $\bm{\theta}$ equal to the weight vector $\bm{\varphi} = \bm{\theta}$, and suppose that the input is corrupted by some noise and transformed into $\bm{\theta} \rightarrow \bm{\theta} + \bm{\Delta} = \bm{\varphi} + \bm{\Delta}$, with $\Delta = (\Delta_0,\, \Delta_1,\, \hdots,\, \Delta_{2^n-1})$. The activation function in Eq.~\eqref{eq:mod} thus reduces to
\begin{equation}
f(\bm{\Delta}) = \frac{1}{2^n} + \frac{1}{2^{2n-1}}\sum_{i<j}^{2^n-1}\cos(\Delta_j-\Delta_i)\ .
\end{equation}
Assuming that the noise values $\Delta_i$ are sampled from the uniform distribution
\begin{equation}
p(\Delta_i) = \frac{1}{a}\quad \text{for } \Delta_i \in \left[-\frac{a}{2}, \frac{a}{2}\right]\ ,
\quad a \in \mathbb{R}\ ,
\label{eq:CQN_Uniform}
\end{equation}
it is possible to evaluate the \textit{average} activation function as
\begin{align}
\label{eq:App_noise0}
\mathbb{E}_{\bm{\Delta}}[f(\bm{\Delta})] & = \frac{1}{2^n}+\frac{1}{2^{2n-1}}\sum_{i<j}^{2^n-1} \mathbb{E}_{\bm{\Delta}}[\cos(\Delta_j-\Delta_i)] \nonumber \\
& = \frac{1}{2^n}+\frac{1}{2^{2n-1}}\frac{2^n(2^n-1)}{2}\langle \cos(\Delta_j-\Delta_i) \rangle
\end{align}
where in the last line used that $\mathbb{E}_{\bm{\Delta}}[\cos(\Delta_j-\Delta_i)]$ is the same for all $i, j$. Averaging then yields
\begin{align}
\mathbb{E}_{\bm{\Delta}}[\cos(\Delta_j-\Delta_i)] & \coloneqq \int_{-\frac{a}{2}}^{\frac{a}{2}} \int_{-\frac{a}{2}}^{\frac{a}{2}} \cos(\Delta_j-\Delta_i) \frac{d\Delta_i d \Delta_j}{a^2}\nonumber \\
& = 2\left( \frac{1-\cos(a)}{a^2}\right)\ .
\end{align}
Substituting back into \eqref{eq:App_noise0} one obtains
\begin{equation}
\mathbb{E}_{\bm{\Delta}}[f(\bm{\Delta})] = \frac{1}{2^n}+\frac{2^n-1}{2^{n-1}}\qty(\frac{1-\cos(a)}{a^2})\ .
\end{equation}

Consider now the case where the input and weight vectors are different $\bm{\theta} \neq \bm{\varphi}$, and we ask again how much does the activation function change if the input is corrupted by the presence of noise. As before, considering an input $\bm{\theta} \rightarrow \bm{\theta}+\bm{\Delta}$, the activation function reads
\begin{equation}
f(\bm{\theta},\bm{\phi},\bm{\Delta}) = \frac{1}{2^n} + \frac{1}{2^{2n-1}}\sum_{i<j}^{2^n-1}\cos(A_{ij} + D_{ij})\ .
\end{equation}
with $A_{ij}=(\theta_j-\phi_j)-(\theta_i-\phi_i)$, and $D_{ij}=\Delta_j-\Delta_i$. Since $\cos(A_{ij}+D_{ij}) = \cos(A_{ij})\cos(D_{ij})+\sin(A_{ij})\sin(D_{ij})$, and $\mathbb{E}_{\bm{\Delta}}[\sin(D_{ij})] = 0$ for uniformly distributed noise~\eqref{eq:CQN_Uniform}, the activation function finally results in
\begin{equation}
\langle f(\bm{\theta},\bm{\phi},\bm{\Delta})\rangle = \frac{1}{2^n} + \frac{D}{2^{2n-1}}\sum_{i<j}^{2^n-1}\cos(A_{ij})\ .
\end{equation}
with $D=2(1-\cos(a))/a^2$. A more realistic noise model would consist of a Gaussian distribution centred in zero with width $\sigma = a/2$, namely
\begin{equation}
p(\Delta_i) = \frac{1}{\sqrt{2 \pi (a/2)^2}}e^{-\frac{\Delta_i^2}{2(a/2)^2}}\,.
\end{equation}
By repeating the same procedure above one obtains
\begin{equation}
    \mathbb{E}_{\bm{\Delta}}[\cos(\Delta_i-\Delta_j)] =  \frac{2}{\pi a^2}\int_{-\infty}^\infty \int_{-\infty}^\infty e^{-\sqrt{2}(\Delta_i+\Delta_i)^2/a^2} \cos(\Delta_i-\Delta_j)d\Delta_i d\Delta_j = e^{-a^2/4}\,,
\end{equation}
which is comparable to the uniform distribution noise model, but less effective in terms of noise resilience due to the Gaussian tails having the net effect of lowering the mean activation. Nonetheless, this qualitative behaviour proves the quantum neuron to have some internal degree of noise resilience.

\section{Alternative schemes for the data encoding operations}
\label{app:CQN_App_C}
Several strategies can be envisioned to reduce the time complexity of the proposed quantum neuron algorithm. First, the encoding of input data could be effectively replaced by a direct call to a quantum memory, such as a qRAM~\cite{QRAM_Maccone}. In this case, the information to be analyzed would be directly stored in the form of quantum states coming, e.g. from quantum internet applications or quantum simulators. 

Alternatively, one could make use of some specific properties of the LME states arising from the phase encoding procedure used in the main text. Indeed, let $U_{\psi}$ be the unitary operation whose action is to create an LME state from a blank register, i.e $U_{\psi}\ket{\bm{0}} = \ket{\psi}$. It is then easy to check that, for $W_k = U_{\psi}Z_k U_{\psi}^\dagger $, where $Z_k$ is the Pauli-$Z$ operator acting on the $k$-th qubit, it holds that $W_k \ket{\phi} = \ket{\phi}\ \forall k$if and only if $\ket{\phi} = \ket{\psi}$. This means that the operators $\{W_1, W_2, \hdots, W_n\}$ stabilise the state $\ket{\psi}$. Depending on the values of the phases $\{\alpha_i\}_{i=0}^{2^{n-1}}$, these operators $\{W_k\}_{k=0}^{n-1}$ may be quasi-local, meaning that they only act on a smaller subsystem of the whole $n$-qubit register. In this case, it can be shown \cite{KrausLMEStates, KrausPreparationEntangledStates_2008} that there exists a quantum dissipative process for which $\ket{\psi}$ is the only stationary state. Of course, this property strongly depends on the nature of the phases $\alpha_i$ of the target LME state, i.e., correlations in the phases are directly related to a specific preparation scheme. However, it might be the case that for some special classes of incoming inputs, a clear a priori correlation exists between the phases, which would allow to replace the ``brute force" approach used in the main text with a more efficient preparation scheme. 

Finally, it is worth mentioning that more general strategies to load probability distributions and classical datasets on quantum states are known in the literature~\cite{GroverRudolph_LoadingProbability_2002, BookSchuldQML}, whose application could be investigated also in the present case. 

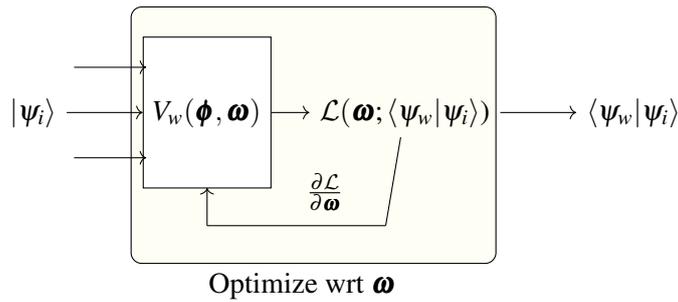
\begin{figure}[ht]
\centering
\begin{tikzpicture}
     \draw  [rounded corners, fill = yellow!5!] (-1,-2) rectangle (3.8, 1.4);
     \node (1) [draw, minimum height=2cm, fill = white!100!] {$V_w(\bm{\phi}, \bm{\omega})$};
     \node (2) [left = of 1] {$\ket{\psi_i}$};
     \draw [->] (2) -- (1);
     \draw [->] (-1.75, 0.6) -- (-0.8, 0.6);
     \draw [->] (-1.75, -0.6) -- (-0.8, -0.6);
     \node (6) [right = 0.5cm of 1] {$\mathcal{L}(\bm{\omega};\braket{\psi_w}{\psi_i}$)};
     \draw  [->] (1) -- (6);
     \draw (6) -- (2.35, -1.5);
     \draw (2.35, -1.5) -- (0, -1.5);
     \draw  [->] (0, -1.5) -- (0, -1);
     \node (11) at (1.25, -2.3) {Optimize wrt $\bm{\omega}$};
     \node (15) at (1.55, -1.05) {$\frac{\partial \mathcal{L}}{\partial \bm{\omega}}$};
     \node (12) [right = of 6] {$\braket{\psi_w}{\psi_i}$};
     \draw [->] (6) -- (12);
\end{tikzpicture}
\caption[Variational approach for the quantum neuron]{Scheme of a variational approach for implementing the $U_w$ operation in the quantum neuron model shown in Fig.~\ref{fig:CQN_CQN}.}
\label{fig:var_circ}
\end{figure}

Additional room for improvement could be represented by a more efficient implementation of the inner product operation $U_w$~\ref{fig:CQN_CQN}. In this case, instead of simply inverting the preparation circuit, one could devise a variational circuit optimised to output the desired result~\eqref{eq:innerprod} using the approach shown in Figure~\ref{fig:var_circ}. There $V_w(\bm{\phi}, \bm{\omega})$ is an operator approximating the unitary $U_w$ depending on variational parameters $\bm{\omega}$ and the weights of neuron $\bm{\phi}$, and $L(\bm{\omega}; \braket{\psi_w}{\psi_i})$ is a cost function evaluating the distance between the output of a the circuit using $V_w(\bm{\phi}, \bm{\omega})$, and the desired inner product $\braket{\psi_w}{\psi_i}$. Optimisation of the trainable parameters $\bm{\omega}$ could yield an approximate yet efficient implementation for evaluating the inner product gate $U_w$. 

It is also interesting to notice that such optimization procedure could in principle be carried out in combination with a supervised learning approach in order to simultaneously train the value of the weight vector and the actual quantum circuit realisation of the required operation. Finally, if an efficient preparation scheme exists for both the quantum input state $\ket{\psi_i}$ and the quantum weight state $\ket{\psi_w}$, their inner product could also be evaluated by means of specialised algorithms, such as the SWAP test or the Bell basis algorithm~\cite{CincioStateOverlap2018}.

\chapterimage{} 
\chapterspaceabove{2.5cm} 
\chapterspacebelow{2cm} 
\chapter{Entanglement of Quantum Neural Networks}
\label{app:app_Entanglement}

\section{Lower bound on entanglement entropy for unitary 2-designs\label{app:reny}}
The presented derivation is a straightforward application of known results on the entanglement of random states and properties of Rényi-entropies~\cite{LiuEntanglementQuantumRandomness2018, SackBPShadows2022}. The Rényi $\alpha$-entropies of a density operator $\rho$ are defined as
\begin{equation}
    S_\alpha(\rho) \coloneqq \frac{1}{1-\alpha}\log \Tr[\rho^\alpha]\,,
\end{equation}
where $\lim_{\alpha \rightarrow 1}S_{\alpha} (\rho) = S(\rho)$ is the Von Neumann entropy of Eq.~\eqref{eq:ent_entropy}, and it holds that $S_\beta(\rho) \leq S_\alpha(\rho)$ for $\beta \geq \alpha$. Of particular interest is the Rényi 2-entropy $S_2(\rho) = -\log \Tr[\rho^2]$ depending on the purity $\Tr[\rho^2]$ of the system, which is much easier to computer and it can be used to lower bound the Von Neumann entropy via $S(\rho) > S_2(\rho)$. 

Indeed, let $\ket{\psi} \in (\mathbb{C}^2)^{\otimes n}$ be the state of a composite system made of subsystems $A$ and $B$ with dimensions $d_A = 2^{n_A}$ and $d_B = 2^{n - n_A}$ respectively, and suppose that $\ket{\psi}$ is a random state $\ket{\psi} = U\ket{\psi_0}$, where $U$ is sampled from an ensemble of unitaries that constitutes at least a unitary 2-design. Then, the average value of the purity of the reduced density matrix $\rho_A = \Tr_B[\dyad{\psi}]$ amounts to~\cite{LiuEntanglementQuantumRandomness2018, SackBPShadows2022}
\begin{equation}
    \mathbb{E}_{\text{2-design}} \Tr[\rho_A^2] = \frac{d_A + d_B}{d_Ad_B+1}\, .
\end{equation}
By the convexity of Rényi-entropies with respect to $\Tr[\rho^\alpha]$, and using Jensen's inequality (that is, $\mathbb{E}~f \geq f~\mathbb{E}$), one can lower bound the average Rényi 2-entropy as 
\begin{equation}
    \mathbb{E}_{\text{2-design}}[S_2(\rho_A)] \geq -\log\mathbb{E}_{\text{2-design}}[\rho_A^2]
\end{equation}
and consequently
\begin{equation}
\label{eq:app_s123}
    \mathbb{E}_{\text{2-design}}[S_2(\rho_A)] \geq - \log \frac{d_A + d_B}{d_A d_B + 1} >  \log d_A - \log\frac{d_A + d_B}{d_B} > \log d_A - 1\,.
\end{equation}
Then, since $S(\rho) = S_1(\rho) \geq S_2(\rho)~\forall~\rho$, taking the expectation value on both sides yields a lower bound on the average Von Neumann entropy of $\rho_A$, namely
\begin{equation}
\label{eq:app_s21}
     \log d_A - 1 < \mathbb{E}_{\text{2-design}}[S_2(\rho_A)] \leq \mathbb{E}_{\text{2-design}}[S(\rho_A)] \leq \log d_A.
\end{equation}
which is the bound shown in Eq.~\eqref{eq:2des_vonn} in the main text.

If the state $\ket{\psi}$ is instead a truly Haar-random state, that is $U$ is sampled from the uniform Haar distribution and not just from a 2-design, the entanglement entropy is given by the Page value of Eq.~\eqref{eq:haar_entanglement} in the main text, which is itself lower bounded by~\cite{HaydenAspectsGenericEntanglement2006}
\begin{equation}
\label{eq:app_s22}
    \mathbb{E}_{\text{Haar}}[S(\rho_A)] > \log d_A - \frac{1}{2}\frac{d_A}{d_B} > \log d_A - \frac{1}{2} \quad\quad\quad (d_A < d_B)\, .
\end{equation}

Summarising, for $d_A < d_B$, putting together the bounds~\eqref{eq:app_s21} and~\eqref{eq:app_s22} one has
\begin{gather}
    \log d_A - 1 < \mathbb{E}_{\text{2-design}}[S(\rho_A)] < \log d_A\\
    \log d_A - \frac{1}{2} < \mathbb{E}_{\text{Haar}}[S(\rho_A)] < \log d_A\,,
\end{gather}
Alternatively, in the limit when the subsystem $B$ is much larger than $A$, $d_B \gg d_A$, then by approximating the logarithm $\log(1+x)\approx x$ in~\eqref{eq:app_s123} one also has
\begin{gather}
    \log d_A - \frac{d_A}{d_B} < \mathbb{E}_{\text{2-design}}[S(\rho_A)] < \log d_A\\
    \log d_A - \frac{1}{2}\frac{d_A}{d_B} < \mathbb{E}_{\text{Haar}}[S(\rho_A)] < \log d_A\,,
\end{gather}

Thus, the entanglement entropy of a state sampled from a 2-design is close to that of a truly Haar-random state, with both achieving near-maximal entanglement. Of course, one also expects the Von Neumann entropy of a general $t$-design to be upper bounded by the Page value, $\mathbb{E}_{t\text{-design}}[S(\rho_A)] < \mathbb{E}_{\text{Haar}}[S(\rho_A)]$, with equality obtained in the limit $t \gg 1$.

\section{Details on Haar entanglement}
\label{app:haar}
While Eq.~\eqref{eq:haar_entanglement} is the theoretical definition of the Haar entanglement entropy, it is not possible to exactly compute it due to the exponential number of terms in the sum. However, it is possible to exploit the similarity of the sum with the harmonic series to obtain a good approximation.

Indeed, first denote with $H_n$ the truncated harmonic series
\begin{align}
    H_n \coloneqq \sum_{k=1}^n\frac{1}{k}\,.
\end{align}
Then, rewrite the sum in Eq.~\eqref{eq:haar_entanglement} in a more convenient form
\begin{align}
    \sum_{j = d_B + 1}^{d_A d_B} \frac{1}{j} = \sum_{j=1}^{d_Ad_B}\frac{1}{j} - \sum_{j=1}^{d_B}\frac{1}{j}=H_{d_Ad_B}-H_{d_B}\,,
\end{align}
where each term can be approximated effectively using a well-known result for truncated Harmonic series~\cite{harmonic_number}, namely
\begin{align}
    H_n = \log{n} + \gamma + \frac{1}{2n} - \epsilon_n\,,
\end{align}
where $\gamma\simeq 0.5772$ is the Euler-Mascheroni constant, and $0\leq \epsilon_n\leq 1/8n^2$. Thus, the correction $\epsilon_n$ goes to zero as the number of terms in the sum $n$ increases, allowing for a meaningful approximation of the value. Using this technique, we are able to estimate the Haar entanglement entropy of a $50$-qubit state with an error of the order $~10^{-16}$.

We now proceed to compute the maximum and average of the distribution with a fixed number of qubits $n$. Using Eq.~\ref{eq:haar_entanglement} and recalling $d_{A(B)}=2^{n_{A(B)}}$, $n_B= n-n_A$, $n_A\in [1, n/2]$ we can write:
\begin{align}
    \mathbb{E}[S(\psi_A)] &= H_{d_Ad_B}-H_{d_B} - \frac{d_A-1}{2d_{B}} \\
    &= H_{2^n} - H_{2^{n-n_A}}-\frac{2^{n_A}-1}{2^{n-n_A+1}} \\
    &= \log 2^{n_A}-\frac{2^{n_A}-1}{2^{n-{n_A}+1}} + O\left(\frac{1}{2^{n-n_A}}\right).
\end{align}
We are now interested in the maximum and average of the distribution. It is easy to see that the maximum is achieved for $n_A=n/2$. In this scenario, when $2^{n_A} \gg 1$, one then has
\begin{align}
    \max_A\big( \mathbb{E}[S(\psi_A)] \big) = \frac{n}{2}\log 2 - \frac{1}{2} + O\left(\frac{1}{2^{n/2}}\right).
\end{align}
Taking into account that for an $n$-qubit system the maximum of the entanglement entropy is $S=\frac{n}{2}\log 2$ we can state that, in the large $n$ limit, a Haar state presents a maximally entangled bond. 

\section{Triviality of the full entangling map} \label{app:entangling_maps}
\begin{algorithm}[!ht]
\caption{Full (or all-to-all) entangling map}
\label{alg:full}
\KwData{$q_1, \dots q_n$, qubits} 
\KwResult{Quantum circuit}
\For{$i=1, \ldots, n$}
{
    \For{$j=i, \ldots, n$}
    {
        CNOT$_{q_i, q_j}$;\\
    }
}
\end{algorithm}

\begin{figure}[ht]
    \centering
    \includegraphics[width=\textwidth]{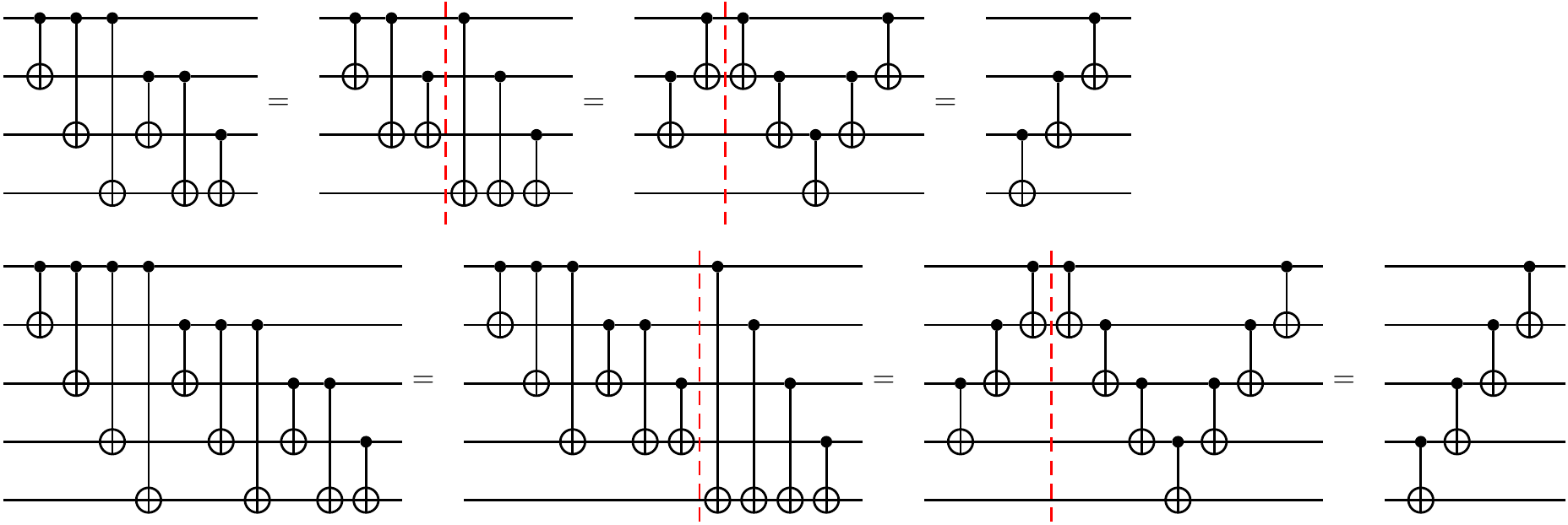}
    \caption[Equivalence of full and linear entangling maps]{Equivalence of the full entangling map with a nearest-neighbours scheme. Using the circuit identities discussed in the main text, it is straightforward to check that the all-to-all entangling scheme as defined in Alg.~\ref{alg:full} is equivalent to a nearest-neighbours interaction.}
    \label{fig:triviality_full}
\end{figure}

The full (or all-to-all) entangling map defined in Alg.~\ref{alg:full} can be shown to be equivalent to a nearest neighbours entangling map with the gates in reversed order, see Fig.~\ref{fig:triviality_full}. The proof is straightforward and obtained by direct evaluation, making use of some circuit identities for networks of CNOTs~\cite{Garciaescartin2011equivalent}. In particular, (\textit{i}) a CNOT can be distributed into four CNOTs acting on an additional intermediate qubit
\[
\begin{quantikz}[column sep={12.4pt,between origins}, row sep={20pt,between origins}]
& \ctrl{3} & \qw \\
& \qw  		& \qw \\ 
& \qw 		& \qw \\
& \targ{} 	& \qw 
\end{quantikz}
=
\begin{quantikz}[column sep={12.4pt,between origins}, row sep={20pt,between origins}]
& \ctrl{1} 	& \qw 			& \ctrl{1}	& \qw 		& \qw  \\
& \targ{}   	& \ctrl{2}		& \targ{} 	& \ctrl{2} & \qw \\ 
& \qw 			& \qw  			& \qw		& \qw 		& \qw 	\\
& \qw 			& \targ{} 		& \qw		& \targ{} & \qw
\end{quantikz}
\]
(\textit{ii}) CNOTs having different controls and targets commute with each other
\[
\begin{quantikz}[column sep={12.4pt,between origins}, row sep={20pt,between origins}]
& \ctrl{3} 	& \qw 		& \qw\\
& \qw   		& \ctrl{1}	& \qw\\ 
& \qw 			& \targ{}  & \qw\\
& \targ{} 		& \qw 		& \qw
\end{quantikz}
=
\begin{quantikz}[column sep={12.4pt,between origins}, row sep={20pt,between origins}]
& \qw 		& \ctrl{3}	& \qw\\
& \ctrl{1} & \qw		& \qw\\ 
& \targ{} 	& \qw  		& \qw\\
& \qw 		& \targ{} 	& \qw
\end{quantikz}
\]
(\textit{iii}) a cascade of CNOTs can be decomposed as
\[
\begin{quantikz}[column sep={12.4pt,between origins}, row sep={20pt,between origins}]
& \ctrl{3}  & \qw 	 & \qw		& \qw \\
& \qw 		 & \ctrl{2} & \qw		& \qw \\ 
& \qw 		 & \qw		 & \ctrl{1}& \qw \\
& \targ{0}& \targ{} & \targ{}	& \qw 
\end{quantikz}
=
\begin{quantikz}[column sep={12.4pt,between origins}, row sep={20pt,between origins}]
& \ctrl{1} 	& \qw 		& \qw 		& \qw  		& \ctrl{1}& \qw \\
& \targ{}   	& \ctrl{1}	 & \qw 		& \ctrl{1} & \targ{} & \qw \\ 
& \qw 			& \targ{}  & \ctrl{1}	& \targ{} 	& \qw		& \qw \\
& \qw 			& \qw 		& \targ{}	& \qw 		& \qw		& \qw 
\end{quantikz}
\]

The full entangling map can be highly simplified using these three rules, reducing it to a simple sequence of nearest-neighbors interactions. For example, for $n=3$ qubits, using (\textit{i}) to distribute the long-range CNOT, one obtains
\[
\begin{quantikz}[column sep={12.4pt,between origins}, row sep={20pt,between origins}]
& \ctrl{1} & \ctrl{2} & \qw & \qw \\
& \targ{}  & \qw 		& \ctrl{1} & \qw\\ 
& \qw 		& \targ{} &  \targ{} & \qw
\end{quantikz}
=
\begin{quantikz}[column sep={12.4pt,between origins}, row sep={20pt,between origins}]
&  \ctrl{1}\border{2}{2}  	& \ctrl{1} & \qw		& \ctrl{1} & \qw 		 & \qw &  \qw  \\
& \targ{}   	& \targ{}	& \ctrl{1} & \targ{} & \ctrl{1}  & \ctrl{1} &\qw \\ 
& \qw 			& \qw 		& \targ{}	& \qw 	& \targ{} &	\targ{} & \qw
\end{quantikz}
=
\begin{quantikz}[column sep={12.4pt,between origins}, row sep={20pt,between origins}]
& \qw 			& \ctrl{1}	& \qw  \\
& \ctrl{1} 	& \targ{}	& \qw 	\\
& \targ{}   	& \qw 		& \qw 
\end{quantikz}
\]
The simplification process can be iterated for a higher number of qubits by first commuting long range CNOTs at the end of the circuit to create a final cascade, and then making use of the result from the lower dimension case. In Fig.~\ref{fig:triviality_full} the simplification process for $n=4,\,5$ qubits is explicitly shown, and it is directly generalised for all numbers of qubits.

Clearly, these results only hold for networks composed of plain CNOTs, and do not apply for general two-qubit interactions made of controlled unitaries. 

\section{Expressibility of Parameterised Quantum Circuits}\label{app:expressibility}
The expressibility introduced in~\cite{SimPQCs2019} quantifies how well the QNN is able to explore the unitary space by comparing the distribution of fidelities of states generated by the QNN with that of randomly Haar-distributed ones.

Let $U(\bm{\varphi})$ be the unitary operation implemented by a parameterized quantum circuit with parameters $\bm{\varphi}$, and let $\ket{\psi_{\bm{\varphi}}} = U(\bm{\varphi})\ket{\bm{0}}$. Given two realizations of the quantum circuit with parameters $\bm{\varphi}_1$ and $\bm{\varphi}_2$, consider the fidelity $F=\abs{\braket{\psi_{\bm{\varphi}_1}}{\psi_{\bm{\varphi}_2}}}^2$. By repeatedly sampling two sets of parameters and evaluating the corresponding fidelity $F$, one can construct a histogram approximating the probability distribution $\hat{\mathcal{P}}(F)$ of the fidelity for states generated by the considered parameterised quantum circuit. 

For truly Haar random quantum states, the probability density function of fidelity is known and amounts to $P_{\text{Haar}}(F) = (N-1)(1-F)^{N-1}$, where $N=2^n$ is the dimension of the Hilbert space~\cite{ZyczkowskiRandomState}. The expressibility is then defined as the Kullback–Leibler divergence $D_{KL}$ between the estimated fidelity distribution and that of a Haar-distributed ensemble, namely
\begin{equation}
    \text{Expressibility} := D_{KL}\qty(\hat{P}_{\text{PQC}}(F) || P_{\text{Haar}}(F))\, .
\end{equation}

\section{Entanglement scaling with increasing depth}
\label{app:ext_ent_scaling}
\begin{figure*}[t]
    \centering
    \includegraphics[width=\textwidth]{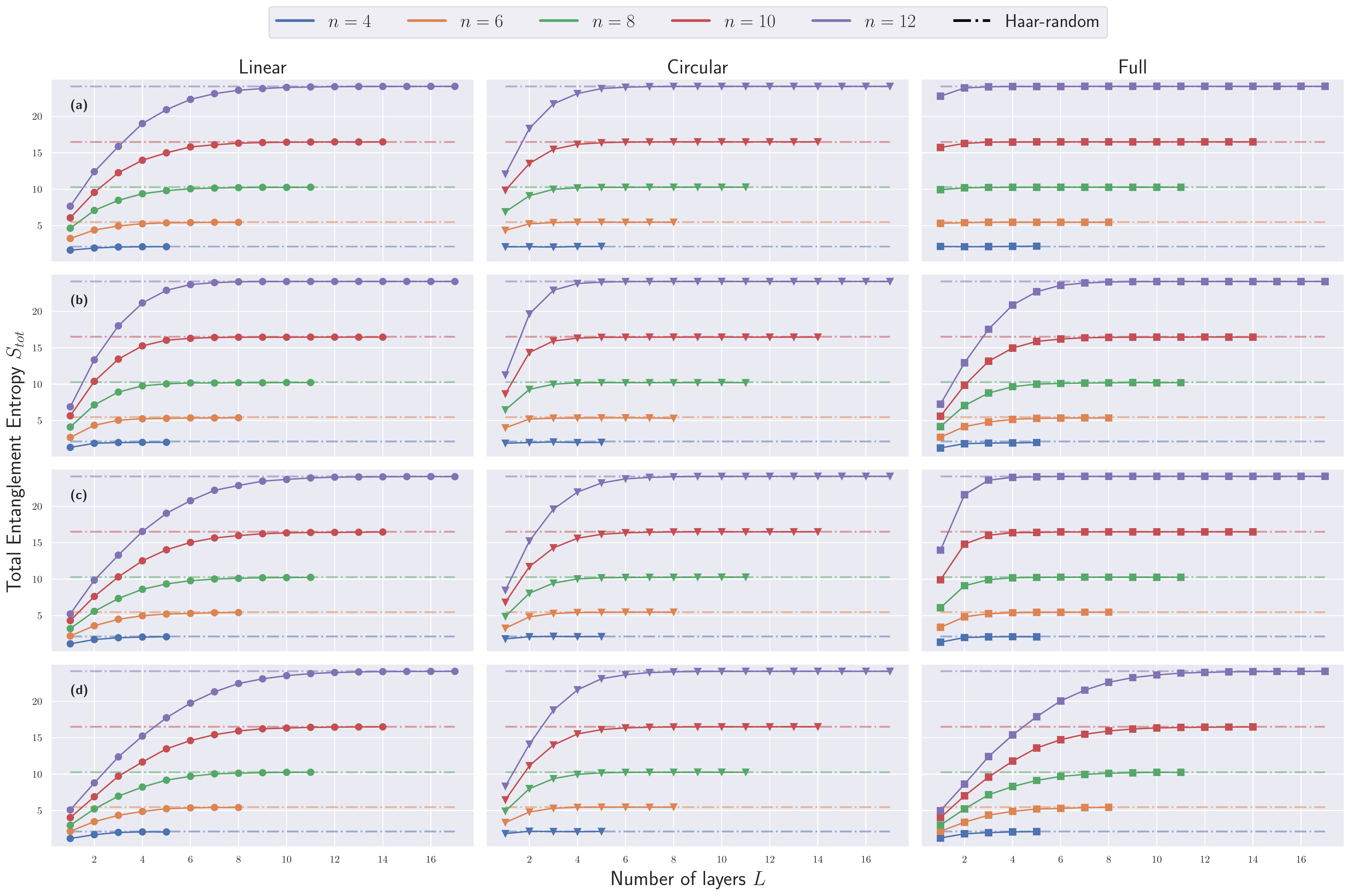}
    \caption[Total entanglement for various quantum neural networks]{Total entanglement $S_{tot}$~\eqref{eq:total_entanglement} for four different QNN architectures, each evaluated with different entangling topologies (\textit{linear}, \textit{circular} and \textit{full}), shown for increasing number of layers $L$ and for several numbers of qubits $n$. QNNs architecture given by: \textbf{(a)} $\fmap = \circuit{zz}$, $\varans = \circuit{2}$; \textbf{(b)} $\fmap = \circuit{2}$, $\varans = \circuit{2}$; \textbf{(c)} $\fmap = \circuit{3}$, $\varans = \circuit{2}$; \textbf{(d)} $\fmap = \circuit{1}$, $\varans = \circuit{2}$. Note that QNNs leverage the same variational form $\varans$, while the feature map $\fmap$ is changed. See the main text for a discussion of the results.}
    \label{fig:extensive_ent_scaling}
\end{figure*}

In Figure~\ref{fig:extensive_ent_scaling} we show the behaviour of the total entanglement $S_\text{tot}$ defined in Eq.~\eqref{eq:total_entanglement} for four different QNNs, and as the depth of the quantum circuit is increased. Note that each QNN is considered with all the three possible entangling topologies (\textit{linear}, \textit{circular} and \textit{full} as defined in Fig.~\ref{fig:main_figure}), and the results are shown for several numbers of qubits $n=4,6,8,10,12$. At last, note that all QNNs leverage the same variational form $\varans = \circuit{2}$, while the feature map is changed, $\fmap = \circuit{zz}, \circuit{2}, \circuit{3}, \circuit{1}$ for panels (a), (b), (c) and (d), respectively. See main text for comments on results.

\section{Convergence of MPS simulations}\label{app:convergence}
Using tensor network methods, MPS in this case, it is possible to approximately simulate large qubit systems, and in our analysis we go up to simulating circuits of $n=50$ qubits. 

The error introduced by the approximations can be monitored and so one has always an estimate of the faithfulness of the tensor network simulation~\cite{JaschkeQuantumGreen2023}. Let $\ket{\psi_\text{exact}}$ be the true state of the quantum system after the $i$-th two qubit gates in the circuit is applied (one qubit gates do not imply approximation errors), and let $\ket{\psi_\text{trunc}}$ denote the truncated quantum state represented by the MPS. The fidelity between these two states evaluated on the $i$-th step of the computation is
\begin{align}
F_i &=\abs{\bra{\psi_\text{exact}}\ket{\psi_\text{trunc}}}^2= \abs{\sum_{\alpha=1}^{\chi_\text{exact}}\lambda_\alpha\bra{\xi_\alpha}_1\otimes \bra{\eta_\alpha}_2 ~ \sum_{\beta=1}^{\chi_{s}}\lambda_\beta\ket{\xi_\beta}_1\otimes \ket{\eta_\beta}_2}^2 \\
&=\left|\sum_{\alpha=1}^{\chi_{s}} \lambda_\alpha^2\right|^2 = \left|1-\sum_{\alpha=\chi_s+1}^{\chi_\text{exact}}\lambda_\alpha^2\right|^2,
\end{align}
where we represented the states in the Schmidt decomposition with respect to the bond where the $i$-th two-qubit gate was applied, and $\chi_s$ is the bond dimension of the MPS state. The fidelity $F_t$ of the simulation after application of the $t$-th two-qubit gate is lower bounded by the product of the previous fidelities $F_i$, as~\cite{JaschkeQuantumGreen2023}
\begin{align}
\label{eq:fid_mps}
    F_t\geq\prod_{i=1}^{t-1} F_i.
\end{align}
where we note that the single step fidelities $F_i$ are readily accessed during the MPS simulation, since one calculates the fidelity before the truncation of the singular values takes place. Equation~\eqref{eq:fid_mps} gives a lower bound to the error introduced by truncation in terms of the fidelity between the true state and the one evolved using an MPS simulation, and one can then control the faithfulness of the simulation at any given time step of the circuit.

\begin{figure}[ht]
    \centering
    \includegraphics{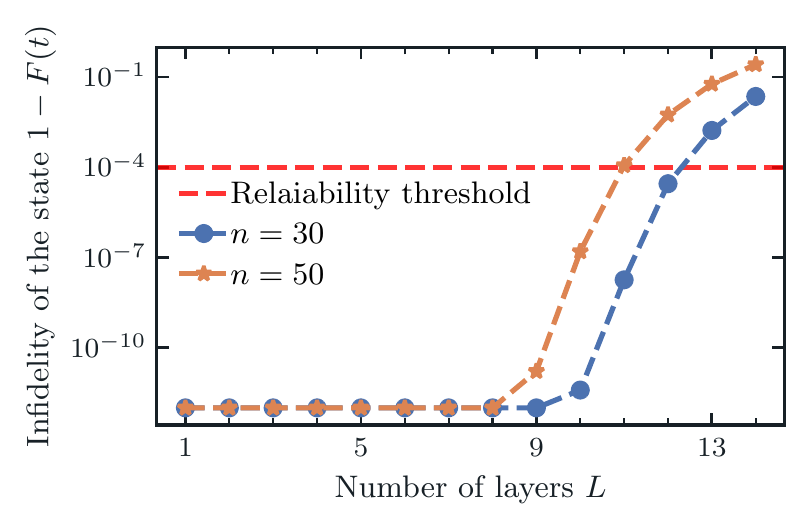}
    \caption[Approximation errors of MPS]{We show the infidelity of the state as a function of the number of layers for $n=30,\, 50$ qubits. The results are reliable up to $L=11$ layers for $n=50$ and up to $L=12$ for $n=30$.}
    \label{fig:mps_convergence}
\end{figure}

In Figure~\ref{fig:mps_convergence} we show the infidelity $1-F$ of the final state from the circuit for $n=30, 50$ with a maximum bond dimension $\chi_s=4096$. The plotted result is the average over $M=10$ realisation of the quantum circuit with different sets of parameters. Defining reliable results with the infidelity of at most $1-F=10^{-4}$ we observe that, for $n=50$, we describe reliably circuits up to $L=11$ layers, while for $n=30$ we can reach $L=12$ layers.

\section{Entanglement evolution during training}
\label{app:Entanglement_Training_IRIS}
In this section we discuss some preliminary investigation regarding the evolution of entanglement entropy while training a quantum neural network. Indeed, the focus of Chapter~\ref{ch:entanglement} was the study of the relationship between entanglement growth and depth in common parameterized quantum ans\"{a}tze, specifically when they are initialised with uniform random parameters and no optimization has yet started. Though the key ingredient in variational quantum algorithms is the learning procedure, and it is natural to ask what role plays the entanglement --- if any --- during the optimization process. 

In combinatorial problems based on QAOA~\cite{FarhiQAOA_2014} the final state of the quantum circuit should correspond to the specific bitstring solving the binary optimization problem at hand. One then expects that as the number of layers in QAOA is increased, the entanglement may first grow but then decrease when the circuit is deep enough to find the correct solution~\cite{DupontQAOAEntanglement}. While theoretically sound, this situation is not always met in real instances~\cite{DupontQAOAEntanglement, ChenEntanglementQAOA}, and the role of entanglement for the performance and classical simulability of QAOA is still an active area of research~\cite{DupontQAOAEntanglementSimulability, SreedharQAOAEntanglement}. 

However, differently from binary optimization problems or ground state solver~\cite{Woitzik2020EntProdVQE}, in general, there is no \textit{a priori} structure that can be used to assess the entanglement properties of states coming from optimised quantum neural networks, independently of their depth. While the use of deep QNNs could offer some optimization advantages due to overparametrization~\cite{Larocca2021OverparametrizationQNN, KianiUnitariesOverparameterisation2020, AnschuetzLossTrapsQM2022}, randomicity-induced barren plateaus can hinder the training of these circuits altogether~\cite{McCleanBarren2018, CerezoBarrenLocalCost2021}. Current proposals then advocate for the use of constrained quantum ans\"{a}tze specifically tailored to the problem under investigation~\cite{skolik2022equivariant, MeyerSimmetriesQML_2022, LaroccaGQML2022}, and then one expects the impact of depth to be highly dependant on the specific task to be solved, and dataset to fit, either classical or quantum~\cite{SharmaNFLEntanglement}. The general impact of circuit depth on the accuracy quantum machine learning model is still not fully established, let alone the entanglement features of the quantum states generated by the model. 

As we are specifically focused on the entanglement properties of quantum neural networks, in the following we start to shed light on the relation between entanglement and optimization by considering a fixed depth circuit and studying how entanglement entropy evolves during training. Specifically, we reproduce the classification task of the well-known IRIS dataset~\cite{IrisDataset} proposed by Abbass et al.~in~\cite{AbbasPowerQNN2021} to study the expressivity of quantum machine learning models, but instead focus our attention on its entanglement features.

\begin{figure}[ht]
    \centering
    \includegraphics[width=\textwidth]{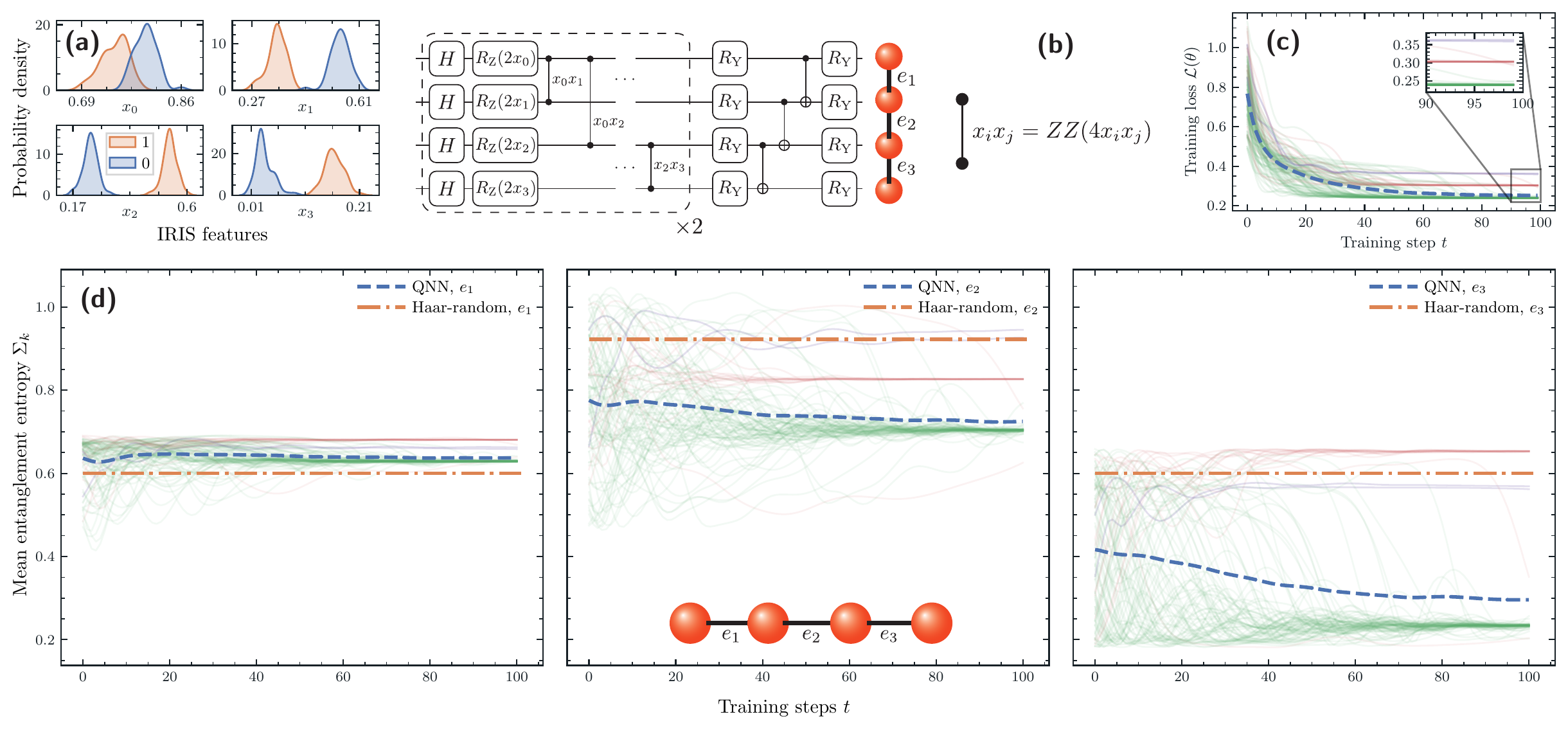}
    \caption[Evolution of entanglement entropy during training]{Evolution of entanglement entropy while training a parameterized quantum circuit to classify the IRIS dataset, using the same setup proposed in~\cite{amira_abbas_2021_4732856, AbbasPowerQNN2021}. Extended details on the preprocessing of the data, the optimiser, the quantum circuit, and the loss function for training the classifier can be found in Appendix~\ref{app:Entanglement_training_details}. \textbf{(a)} Plot of the distribution of the IRIS features after normalisation of the dataset. \textbf{(b)} Parameterised quantum circuit used in~\cite{AbbasPowerQNN2021}, which uses one qubit per input data feature. Two repetitions of a \textsc{ZZfeatureMap} with an all-to-all connectivity are used to encode the data, followed by a variational ansatz with 8 trainable parameters. The two-qubit operations in the feature map are shown in Eq.~\ref{eq:zzfmap}. \textbf{(c)} Training loss for $10^2$ training runs starting with different initialisation of the variational parameters. The dashed line is the average over such training trajectories. Curves are grouped into three classes denoted by the colours (green, red, purple), depending on the final loss they achieve at the end of training. Such colour-coding is used in panel (d) to distinguish the training trajectories and study how the final loss relates to the entanglement created in the circuit. \textbf{(d)} Mean entanglement entropy~\eqref{eq:mean_ent_training} averaged over the full training dataset, for the bipartitions occurring at bonds $e_1$, $e_2$, and $e_3$. The curves are coloured according to the final loss obtained at the end of the training, as shown in panel (c). Dashed lines are obtained by averaging over all the trajectories.}
    \label{fig:training_qnn_entanglement}
\end{figure}

In Fig.~\ref{fig:training_qnn_entanglement} we show the results of training the QNNs shown in panel (c), to classify the first two classes of the IRIS dataset, consisting of $m=100$ samples of normalised four-dimensional inputs vectors, whose features distribution is shown in panel (a). We refer to Sec.~\ref{app:Entanglement_training_details} for extended details on the preprocessing of the dataset, the choice of the ansatz, and the optimization process. The training procedure is run $100$ times starting from different initialisations of the parameters, thus obtaining multiple training trajectories which are plotted in Fig.~\ref{fig:training_qnn_entanglement}b. 

At the end of the training, trajectories cluster around three possible values of the final loss, and each curve is coloured depending on the cluster they belong to, that is the final loss they achieve at the end of training. Such colour coding scheme is then used to distinguish trajectories in panel (d), which shows the evolution of the mean (over the dataset) entanglement entropy Eq.~\eqref{eq:mean_ent_training} during training, across the three bipartitions corresponding to bonds $e_1$, $e_2$, and $e_3$, indicated in the Figure. Finally, dashed lines in panels (b) and (d) are obtained by taking the mean over all curves in the corresponding plots. 

The mean entanglement entropy $\Sigma_k(t)$ is defined as the average over the full training dataset of the entanglement entropy corresponding to bond $e_k$, when the parameter vector is $\bm{\theta}_t$ at training step $t$, namely
\begin{equation}
    \label{eq:mean_ent_training}
    \Sigma_k(t) = \frac{1}{m}\sum_{i=1}^m S(e_k; \bm{x}_i, \bm{\theta}_t)\,,
\end{equation}
where $m=100$ is the size of the dataset, $\bm{x}_i$ is an input vector from the dataset, and $S(e_k; \bm{x}_m, \bm{\theta}_t)$ is the entanglement entropy of the bipartition corresponding to cut $e_k$ of the quantum state $\rho = U(\bm{x}_i, \bm{\theta}_t) \dyad{\bm{0}} U(\bm{x}_i, \bm{\theta}_t)^\dagger$, see Eq.~\eqref{eq:ent_entropy_bonds}.\\

\paragraph{Entanglement at $t=0$}
Various observations can be drawn from the simulation results reported in Fig.~\ref{fig:training_qnn_entanglement}. First of all, the mean entanglement at the start of training $\Sigma_k(t=0)$, when the circuit is initialized with random parameters, is very different on single qubit cuts corresponding to bonds $e_1$ and $e_3$. While the first qubit starts with a highly entangled state with the rest of the system, even more than a typical Haar-random state, qubit four has instead a much lower initial entanglement. This can be understood as a consequence of the structure of the quantum circuit, as well as the actual numerical values of the features. Indeed, the two-qubits gates used in the feature map are parameterized $ZZ$-interactions of the form 
\begin{equation}
\label{eq:zzfmap}
    ZZ_{ij}(\phi_{ij}) = \exp\qty(-i\,\phi_{ij}\,Z_i\otimes Z_j / 2)\,,
\end{equation}
where the rotation amount is a function on the numerical value of the features, in our case $\phi_{ij} = \phi(x_i, x_j) = 2x_i\cdot 2x_j$. 

If the rotation angle is small $\phi \ll 1$, then $ZZ(\phi) \sim \mathbb{I}$, and so little entanglement is created between the interested interacting qubits. Since the feature $x_3$ is particularly smaller than the other features (especially for class zero, see Fig.~\ref{fig:training_qnn_entanglement}a), then the $ZZ$ interactions involving the fourth qubit will be closer to the identity, and hence the entanglement generated in the feature map between the fourth qubit and the rest of the system will be consequently smaller. 

In addition, the variational part of the circuit consists of a reversed linear network of CNOT gates, which has a different impact on different qubits. For example, only one CNOT is inside the past light cone of the fourth qubit (corresponding to the feature $x_3$), while the state of the first qubit (corresponding to feature $x_0$) is affected by all the CNOTs. 

Thus, overall due to the specific structure of the feature map and the variational ansatz, one can expect the fourth qubit to have little entanglement with the rest of the system, which is indeed confirmed by the numerical results. At last, it is worth noticing that while the circuit uses two repetitions of an all-to-all feature map, these are arranged in a sequential manner, and thus entanglement at the central bond $e_2$ has not yet reached the Haar-random value, as one would expect from a two-layered all-to-all circuit, see App.~\ref{app:ext_ent_scaling}.\\

\paragraph{Entanglement at $t>0$}
We now move our attention to the dynamics of entanglement during training. First, looking at the average dashed lines in panel (d) of Fig.~\ref{fig:training_qnn_entanglement}, entanglement among qubits generally decreases as training progresses, especially for cut $e_3$, but also for $e_2$ to a lesser extent. The training trajectories that reach better solutions (lower loss, indicated in green) are those that correspond to lowest entanglement, thus indicating that for this specific setup a decrease in entanglement is beneficial to reach good performances. This is particularly true for the fourth qubit, as the training procedure drives the system towards states where the qubit is further disentangled from the rest of the system. 

Specifically, in Sec.~\ref{app:Entanglement_training_details} we show how the entanglement evolves for the two classes separately, that is evaluating $\Sigma_k$ in Eq.~\eqref{eq:mean_ent_training} not on the full dataset but on each class separately. We see that elements in class $0$ are almost disentangled from the rest of the system as training proceeds towards good minima (green curves). The specific choice of the variational ansatz impacts the shape of the loss landscape, and it is such that continuing to disentangle the fourth qubit is the most effective strategy to ensure good performance.\\

\paragraph{A few concluding remarks}
The dependence between the entanglement generated in the data encoding part of the quantum circuit and the numerical values of the features underlines the importance of preprocessing the classical data, which is known to have a profound impact on the class of functions the parameterized quantum model have access to, so its performances~\cite{Schuld2020Encoding, Theis2020Expressivity}. Moreover, the trainable part of the circuit has a profound impact on how training can increase or diminish entanglement among qubits. 

Finally, we stress that as the quantum model is trained to minimize the loss function, the entanglement entropy in the circuit must not be a monotone function of the training step, and in fact, we check numerically that varies a lot both within a single training run, as well as across several training trajectories. It would be interesting to study whether the exploration of different regimes of entanglement is related to the performances of a quantum machine learning model. 

\subsection{Details on the classification procedure}
\label{app:Entanglement_training_details} 
For the optimization task of the IRIS dataset we use a custom Pennylane~\cite{Pennylane} implementation based on the code~\cite{amira_abbas_2021_4732856}, accompanying the results presented in Abbass et al.~\cite{AbbasPowerQNN2021}. \\

\paragraph{IRIS dataset} The IRIS dataset~\cite{IrisDataset} consists of 150 samples of four dimensional real input vectors $\bm{x}_i = \qty(x^{(i)}_0, x^{(i)}_1, x^{(i)}_2, x^{(i)}_3) \in \mathbb{R}^4$, each one paired with a corresponding class label $\lambda_i \in \{0,\,1,\,2\}$. We consider a classification task involving only the first two classes $\lambda_i \in \{0,\,1\}$, so the actual dataset considered for our analysis consists of $m=100$ samples\footnote{The full IRIS dataset comprises 50 samples per class, for a total of 150 samples.}, namely
\begin{equation}
    S = \qty{(\bm{x}_i, y_i) \in \mathbb{R}^4 \times \{0,\,1\}~\big|~\norm{\bm{x}_i} = 1,~i=1,\hdots, 100},
\end{equation}
where each input vector was normalised $\bm{x}_i \rightarrow \bm{x}_i / \norm{\bm{x}_i}$, and we defined the desired output label as $y_i = 1 - \lambda_i$, that is $y_i = 1$ if the sample is in class $\lambda_i = 0$, and zero otherwise. 

\begin{figure}[ht]
    \centering
    \includegraphics[width=\textwidth]{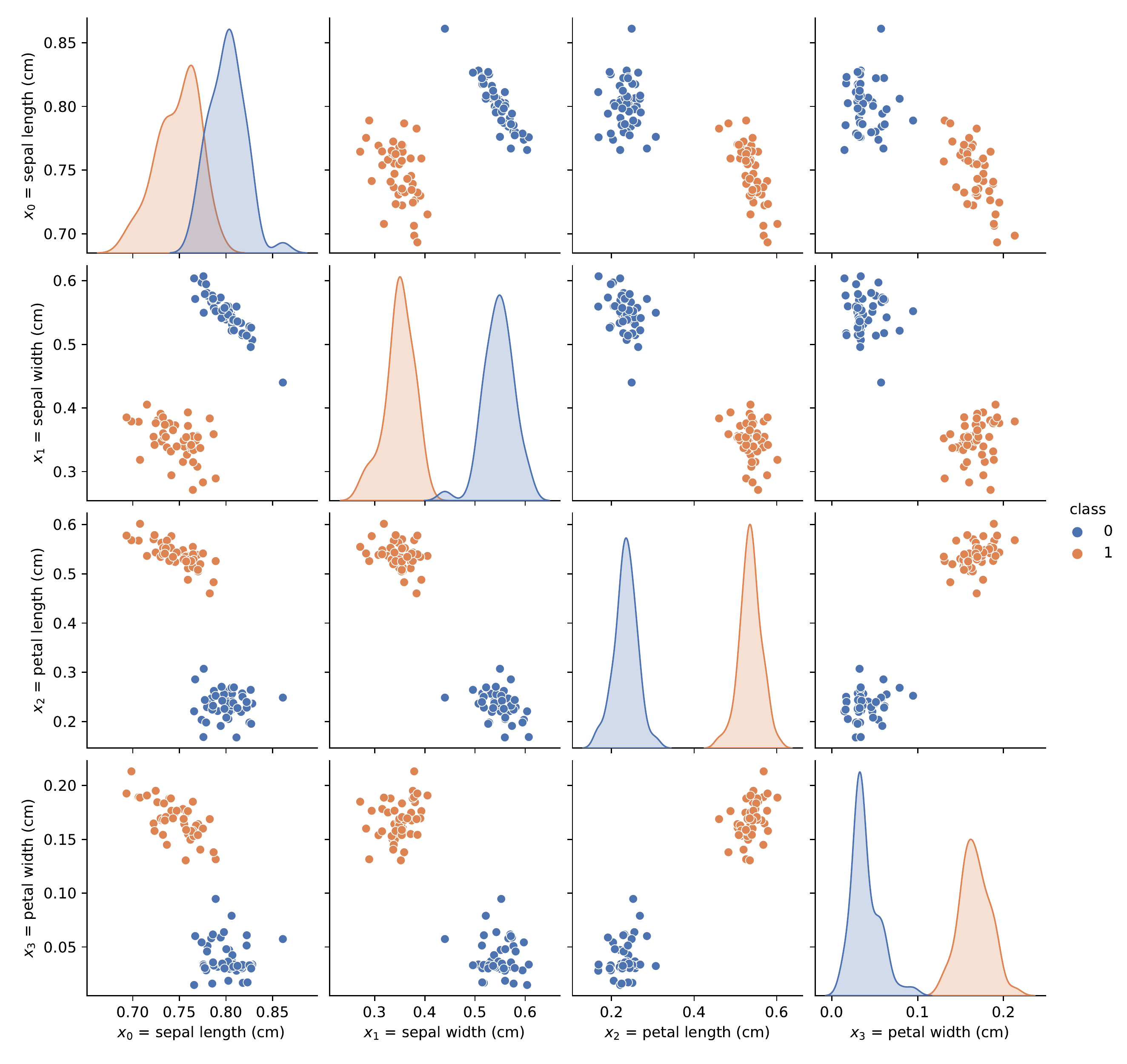}
    \caption[Preprocessing of the IRIS dataset]{Summary of the preprocessed IRIS dataset used in our simulations to reproduce the classification task proposed in~\cite{AbbasPowerQNN2021}. The plots on the diagonal of the grid correspond to a smoothed version of the histogram of the frequencies of each feature built with \texttt{seaborn}~\cite{SeabornWaskom2021}, and are the same plots shown in Fig.~\ref{fig:training_qnn_entanglement} in the main text.}
    \label{fig:iris}
\end{figure}

In Fig.~\ref{fig:iris} we summarise the properties of the IRIS dataset used in our simulations. Note that the plots on the diagonal are those shown in Fig.~\ref{fig:training_qnn_entanglement}a, and represent the continuous probability density of the corresponding features, obtained by smoothing the histogram of the frequencies~\cite{SeabornWaskom2021}.\\

\paragraph{Quantum neural network circuit and optimisation} 
The quantum neural network ansatz, denoted with $U(\bm{x}_i;\,\bm{\theta})$, acts on $n=4$ qubits and it is showed in Fig.~\ref{fig:training_qnn_entanglement}c. It consists of two layers of the \textsc{ZZFeatureMap} with an all-to-all (full) entangling connectivity which encodes the inputs $\bm{x}_i$ into the circuit, where one qubit per feature is used. 

The encoding part of the circuit is followed by a variational ansatz with eight trainable parameters $\bm{\theta} \in \mathbb{R}^8$, and a network of CNOT gates. Note that we used a reversed linear network of CNOTs, as this is equivalent to the full entangling connectivity originally used in~\cite{AbbasPowerQNN2021, amira_abbas_2021_4732856}, as shown previously in Appendix~\ref{app:entangling_maps}. 

Given an input $\bm{x}_i$ and trainable parameters $\bm{\theta}$, the output of the circuit is obtained by measuring a Pauli-$Z$ operator on all qubits $\expval{O}_{\bm{x}_i;\,\bm{\theta}} = \mel{\bm{0}}{U(\bm{x}_i; \bm{\theta})^\dagger\,Z^{\otimes 4}\,U(\bm{x}_i; \bm{\theta})}{\bm{0}}$, and by constructing the predicted class probability
\begin{equation}
    \tilde{y}_i(\bm{\theta}) = \frac{1+\expval{O}_{\bm{x}_i;\,\bm{\theta}}}{2} \in [0,\,1]\,,
\end{equation}
which is equivalent to measuring the parity of the bitstrings at the output of the circuit. 

The circuit is optimized through gradient descent with Adam optimiser~\cite{KingmaAdam_2014} with learning rate set to $\eta = 0.1$, which was used to to minimize the cross-entropy loss function averaged over the whole training dataset
\begin{equation}
    L(\bm{\theta}) = \frac{1}{m}\sum_{i=1}^{m} \ell_i(y_i, \hat{y}_i(\bm{\theta})) = -\frac{1}{100}\sum_{i=1}^{100} \qty(y_i\log \hat{y}_i + (1-y_i)\log(1-\hat{y}_i))\,.
\end{equation}
We run the training process $100$ times, each time initialising the circuit with variational parameters sampled randomly from the uniform distribution $\theta_i \sim \text{Unif}[0, 2\pi)$.\\

\begin{figure}[ht]
    \centering
    \includegraphics[width=\textwidth]{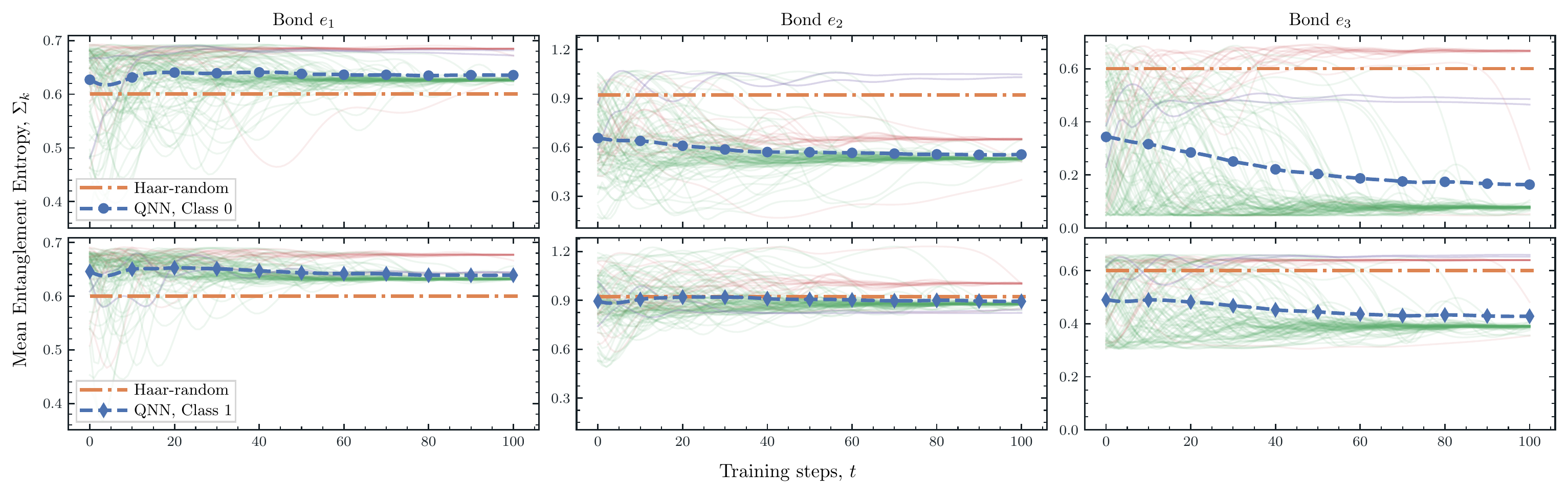}
    \caption[Mean entanglement entropy per IRIS class]{Mean entanglement entropy $\Sigma_k$~\eqref{eq:mean_ent_training} averaged over elements belonging to the different classes, for the three bonds $e_1$, $e_2$, $e_3$. Dashed lines represent averages over the various training trajectories.}
    \label{fig:ent_per_classes}
\end{figure}

\paragraph{Mean entanglement entropy per class}
In Fig.~\ref{fig:ent_per_classes} we report the mean entanglement entropy $\Sigma_k(t)$ defined in Eq.~\eqref{eq:mean_ent_training}, where the average is taken separately over elements belonging to the two different classes, namely
\begin{align}
    \Sigma^{(0)}_k(t) = \frac{1}{50}\sum_{m=1}^{50} S(e_k; \bm{x}_m, \bm{\theta}_t) \quad \text{for } \lambda_m = 0\,, \\
    \Sigma^{(1)}_k(t) = \frac{1}{50}\sum_{m=1}^{50} S(e_k; \bm{x}_m, \bm{\theta}_t) \quad \text{for } \lambda_m = 1\, .
\end{align}
As discussed in the main text, the entanglement entropy for cut $e_3$ for elements belonging to class 0 is smaller in general, due to the feature $x_3$ for Class 0 being roughly one order of magnitude smaller than the remaining features, see Fig.~\ref{fig:iris}. 

\chapterimage{} 
\chapterspaceabove{2.5cm} 
\chapterspacebelow{2cm} 
\chapter{Noise Deconvolution}
\label{app:app_Noise}

\section{Kraus Decomposition}
\label{app:KarusDecomposition} 
A quantum physical evolution is represented by (\textit{i}) \textit{linear}, (\textit{ii})\textit{completely-positive} and (\textit{iii}) \textit{trace-preserving} (CPTP) maps taking quantum density operators to quantum density operators. A map satisfying these three properties is called a \textit{quantum channel}, and can be interpreted as a quantum evolution obtained through the interaction of the system with an external environment. A map is a quantum channel if and only if it admits a Kraus (or \textit{operator-sum}) representation as 
\begin{equation}
\label{eq:Kraus_App}
    \rho \longrightarrow \mathcal{E}(\rho) = \sum_{k} A_k \rho A_k^\dagger\,,
\end{equation}
with the trace preserving condition requiring
\begin{equation}
    \Tr[\mathcal{E}(\rho)] = \Tr[\rho] \implies \sum_{k} A_k^\dagger A_k = \mathbb{I}\, .
\end{equation}
The operators $\{A_k\}_k$ are called the Kraus operators of the channel, which are however non-unique~\cite{NielsenChuang}. Such channels are often referred to as \textit{stochastic} channels~\cite{KingMinimalEntropyStates2001, RuskaiAnalysisCompletelypositiveTracepreserving2002}, and if they also preserve identity ($\mathcal{E}(\mathbb{I}) = \mathbb{I}$), they are called \textit{unital} (or \textit{bistochastic}). Unitality corresponds to the requirement that also $\sum_{k}A_kA_k^\dagger = \mathbb{I}$, from which it is clear that a sufficient condition for a CPTP map to be unital is that its Kraus operators be self-adjoint $A_k = A_k^\dagger \, \forall k$.

\section{Tomographic reconstruction formula for qubits}
\label{app:QubitTomographicFormula}
In this Appendix, we show how the tomographic reconstruction formula for systems made of qubits $\mathcal{H} = \mathbb{C}^2$ can be recovered starting from the standard basis expansion in terms of the Pauli matrices~\cite{DMacconePaini2003Spin}. The set of matrices $\{\mathbb{I}, \sigma_x, \sigma_y, \sigma_z\}$ forms an orthonormal set and constitutes a basis for the space of $2\times2$ complex matrices $\mathcal{L}(\mathcal{H})=\mathbb{C}^{2\times 2}$. So, given an operator $O \in \mathcal{L}(\mathcal{H})$, it can be expressed as
\begin{align}
O  & =  \frac{\mathbb{I}\Tr[O] + \sigma_x \Tr[O \sigma_x] + \sigma_y \Tr [O \sigma_y] + \sigma_z \Tr[O \sigma_z]}{2}\nonumber\\
& =  \frac{\Tr[O]}{2}\, \mathbb{I}+\sum_{\alpha=x,y,z}\frac{\Tr[O\sigma_\alpha]}{2}\sigma_{\alpha}\nonumber\\ 
& =  \sum_{\alpha = x,y,z}\frac{1}{3}\left( \frac{3\Tr[O\sigma_\alpha]}{2}\sigma_{\alpha} + \frac{\Tr[O]}{2}\, \mathbb{I} \right)\nonumber\\
& =  \sum_{\alpha=x,y,z}\frac{1}{3}\, \mathcal{E}[O](\sigma_\alpha)\label{eq:TomographyQubit}\ ,
\end{align}
where we defined
\begin{equation}
\mathcal{E}[O](\sigma_\alpha) = \left( \frac{3\Tr[O\sigma_\alpha]}{2}\sigma_{\alpha} + \frac{\Tr[O]}{2}\mathbb{I} \right)\, 
\end{equation}
which is the desired \textit{quantum estimator}, with $\{\sigma_x,\sigma_y,\sigma_z\}$ constituting the \textit{quorum} of observables of the tomographic reconstruction. The equation~\eqref{eq:TomographyQubit} has the same form of the tomographic reconstruction formula in Eq.~\eqref{eq:TomographicRecontructionFormula}, with substitutions
\begin{equation*}
\int_{\Lambda}  \longrightarrow  \sum_\lambda\,,\quad
d\lambda  \longrightarrow  1/3\,,\quad
\lambda  \longrightarrow  \{x,y,z\}\,,\quad
\{Q_\lambda\}  \longrightarrow \{\sigma_x,\sigma_y,\sigma_z\}\, ,
\end{equation*}
which account for the fact that we are dealing with a discrete, and not continuous, basis expansion. 

Also, note that~\eqref{eq:TomographyQubit} is not the unique choice for the tomographic formula. In fact, one could use a continuous parameterisation of the group $SU(2)$ given by the operator $\mathcal{R}(\vec{n},\psi )=e^{i\vec{s}\cdot\vec{n}\psi}$,
where $\vec{s}$ is the spin of the particle ($\vec{s}=\vec{\sigma}/2$ for qubits), $\vec{n}=(\cos\phi\sin\theta,\sin\phi\sin\theta,\cos\phi)$, $\theta \in [0,\pi]$ and $\phi,\, \psi \in [0,2\pi]$~\cite{DMacconePaini2003Spin}.

\section{\label{app:QubitNoiseDeconvolution} Noise deconvolution for qubits}
In this appendix, we derive the noise deconvolution formula for qubits. Let $\rho$ be a quantum state, and $\mathcal{N}$ be a noise channel that admits an inverse map $\mathcal{N}^{-1}$, and $\hat{\mathcal{N}}^{-1}$ its adjoint map. Then, using Eq.~\eqref{eq:inv_noise} in~\eqref{eq:qubit_tomography}, one obtains
\begin{equation}
\begin{aligned}
\label{eq:DeconvolvedQubits}
\expval{O} & = \sum_\alpha \frac{1}{3}\Tr [\hat{\mathcal{N}}^{-1}(\mathbb{E}(O)[\sigma_\alpha])\, \mathcal{N}(\rho)]\\
& = \sum_\alpha \frac{1}{3}\Tr \qty[ \bigg(\frac{3}{2}\Tr[O\sigma_\alpha] \hat{\mathcal{N}}^{-1}(\sigma_\alpha) + \frac{1}{2}\Tr [O]\, \hat{\mathcal{N}}^{-1}(\mathbb{I}) \bigg)\mathcal{N}(\rho) ]\\
& = \sum_\alpha \frac{1}{3} \qty(\frac{3}{2}\Tr [O\sigma_\alpha]\langle\hat{\mathcal{N}}^{-1}(\sigma_\alpha)\rangle_{\mathcal{N}(\rho)} + \frac{1}{2}\Tr [O] 
\underbrace{\Tr[ \hat{\mathcal{N}}^{-1}(\mathbb{I})\mathcal{N}(\rho)]}_{=\Tr[\mathbb{I}\rho]=1} )\\ 
& = \frac{1}{2}\Tr [O]+\frac{1}{2}\sum_{\alpha=x,y,z}\Tr[O\sigma_\alpha]\langle \hat{\mathcal{N}}^{-1}(\sigma_\alpha)\rangle_{\mathcal{N}(\rho)}\, .
\end{aligned}
\end{equation}
This equation~\eqref{eq:DeconvolvedQubits} lets us deconvolve the effect of noise by evaluating the expectation value of the noise-inverted Pauli matrices $\sigma_\alpha$ on the noisy state $\mathcal{N}(\rho)$. In particular, note that the formula remains valid whether the noise is unital or not. In fact, in the second line we can always move the adjoint inverse noise $\hat{\mathcal{N}}^{-1}$ from the identity to the noisy state $\mathcal{N}(\rho)$, thus obtaining $\Tr[\hat{\mathcal{N}}^{-1}(\mathbb{I})\mathcal{N}(\rho)]=\Tr[\mathbb{I}\, \mathcal{N}^{-1}(\mathcal{N}(\rho))] = \Tr[\rho]=1$.

\section{Inverse maps of Noise channels}
\label{app:inverse_maps} 
We hereby report the explicit calculations to derive the inverse maps of the noise channels considered in the main text. 

\subsection{Bit-flip, phase-flip, and bit-phase-flip channels}
We focus on the bit-flip channel, but the calculations are identical for the phase-flip and bit-phase-flip channels as well. The bit-flip channel is described by Kraus operators $A_0 = \sqrt{1-p}\mathbb{I}$ and $A_{1}$ = $\sqrt{p}\,\sigma_x$, so that its action is given by
\begin{equation}
\label{eq:bit-flipApp}
\mathcal{N}_x(\rho) = (1-p)\rho + p\sigma_x\rho \sigma_x\, .
\end{equation}
The Pauli Transfer Matrix $\Gamma_x$ is defined as
\begin{equation}
\label{eq:ptm_x_app}
    (\Gamma_x)_{ij} = \frac{1}{2}\Tr[\sigma_i~\mathcal{N}_x(\sigma_j)]\, ,
\end{equation}
and by straightforward calculation one obtains
\begin{align*}
    (\Gamma_x)_{11} &= (1-p) + p = 1\,,\\
    (\Gamma_x)_{22} &= (1-p) - p = 1-2p\,,\\
    (\Gamma_x)_{33} &= (1-p) - p = 1-2p\,,\\
    (\Gamma_x)_{ij} &= 0\,,\quad \text{for}~i\neq j\,,
\end{align*}
thus yielding
\begin{equation}
\Gamma_x = 
    \begin{bmatrix}
    1 & 0 & 0 & 0 \\
    0 & 1 & 0 & 0 \\
    0 & 0 & (1-2p) & 0 \\
    0 & 0 & 0 & (1-2p) \\
    \end{bmatrix}\, ,
\end{equation}
whose inverse is trivially 
\begin{equation}
\Gamma_x^{-1} = 
    \begin{bmatrix}
    1 & 0 & 0 & 0 \\
    0 & 1 & 0 & 0 \\
    0 & 0 & \frac{1}{(1-2p)} & 0 \\
    0 & 0 & 0 & \frac{1}{(1-2p)} \\
    \end{bmatrix}\, .
\end{equation}
The eigenvectors of such Pauli Transfer Matrix are clearly the Pauli matrices $\mathbb{I}\rrangle$, $|\sigma_x\rrangle$, $|\sigma_y\rrangle$, $|\sigma_z\rrangle$, with eigenvalues $\bm{\lambda} = (1,\,1,\,1/(1-2p),\,1/(1-2p))$, respectively. 

The operator sum representation of $\mathcal{N}_x^{-1}$ can be reconstructed by noticing that the map 
\begin{equation}
\label{eq:general_map2}
    \mathcal{E}(O) = \sum_{j=0}^3 \beta_j \sigma_j O \sigma_j\, .
\end{equation}
has also the Pauli matrices as eigenvectors, but with eigenvalues $\bm{\beta} = (\beta_0, \beta_1, \beta_2, \beta_3)$. Since two maps are equal if they have the same action on a basis, we can find the operator-sum representation of $\mathcal{N}_x^{-1}$ by finding those $\beta_j$ such that $\bm{\lambda} = \bm{\beta}$. If we can find such a mapping, then inserting those values into~\eqref{eq:general_map2}, we recover the operator sum of the inverse map. The PTM matrix $\Gamma_{\mathcal{E}}$ of $\mathcal{E}$ amounts to
\begin{align}
    \Gamma_{\mathcal{E}} = \text{diag}(& \beta_0 + \beta_1 + \beta_2 + \beta_3,~\beta_0 + \beta_1 - \beta_2 - \beta_3, \\
    & \beta_0 - \beta_1 + \beta_2 - \beta_3,~\beta_0 - \beta_1 - \beta_2 + \beta_3)\, ,
\end{align}
The equality $\Gamma_x^{-1} = \Gamma_\mathcal{E}$ correspond to the system of equations
\begin{equation}
\label{eq:system}
\begin{cases}
1 = \beta_0 + \beta_1 + \beta_2 + \beta_3\\
1 = \beta_0 + \beta_1 - \beta_2 - \beta_3\\
\frac{1}{1-2p} = \beta_0 - \beta_1 + \beta_2 - \beta_3\\
\frac{1}{1-2p} = \beta_0 - \beta_1 - \beta_2 + \beta_3
\end{cases}
\end{equation}
where the first equation is the trace-preserving condition, dictated by the fact that the direct map is TP, and so the inverse map has to be. This condition is also evident from the expression of $\Gamma_x^{-1}$ and $\Gamma_\mathcal{E}$, since the first row is of the form $(1,0,0,0)$. The system of equations~\eqref{eq:system} has solutions
\begin{equation*}
    \beta_0 = \frac{1-p}{1-2p}\,,\quad \beta_1 = -\frac{p}{1-2p}\,,\quad \beta_2 = \beta_3 = 0\,,
\end{equation*}
and substituting these values in Eq.~\eqref{eq:general_map2} leads to the desired operator-sum representation
\begin{equation}
    \mathcal{N}_x^{-1}(O) = \frac{1-p}{1-2p}O - \frac{p}{1-2p}\sigma_x O \sigma_x\, .
\end{equation}

Similarly, the same procedure can be carried out for the dephasing (generated by $\sigma_z$) and bit-phase-flip channel (generated by $\sigma_y$), leading to
\begin{align}
    \mathcal{N}_z^{-1}(O) & = \frac{1-p}{1-2p}O - \frac{p}{1-2p}\sigma_z O \sigma_z\,, \\
    \mathcal{N}_y^{-1}(O) & = \frac{1-p}{1-2p}O - \frac{p}{1-2p}\sigma_y O \sigma_y\,.
\end{align}
Note that for all these three cases, the adjoint channels are equal to the direct ones, i.e. $\hat{\mathcal{N}}^{-1} = \mathcal{N}^{-1}$, since the generating operators are all Hermitian (see Appendix~\ref{app:AmplitudeDamping} for a case where this is not true).

We now proceed to evaluate the explicit form of the deconvolution formula. Let $\beta \in \{x,y,z\}$ index one of the noise channels $\mathcal{N}_\beta \in \{\mathcal{N}_x, \mathcal{N}_y, \mathcal{N}_z\}$, the action of the inverse map on a Pauli matrix $\sigma_\alpha$ amounts to
\begin{equation}
\mathcal{N}^{-1}_\beta(\sigma_\alpha) = \frac{1}{1-2p}\bigg((1-p)\, \sigma_\alpha -p\, \sigma_\beta\sigma_\alpha\sigma_\beta \bigg) = \frac{1-2\delta_{\alpha \beta}\,p}{1-2p}\sigma_\alpha\,,
\end{equation}
where in the second line we made use of the fact that $\sigma_\beta\sigma_\alpha\sigma_\beta = (2\delta_{\alpha\beta}-1)\sigma_\alpha$. Substituting this in Eq~\eqref{eq:DeconvolvedQubits}, one obtains
\begin{equation}
\label{eq:deconv_flips}
\langle O \rangle_\beta = \frac{1}{2}\Tr [O]+\frac{1}{2(1-2p)}~\sum_{\alpha=x,y,z}\Tr[O\sigma_\alpha]\big(1-2\delta_{\alpha \beta}\,p\big)\expval{\sigma_\alpha}_{\mathcal{N}_\beta(\rho)}\, ,
\end{equation}
where the subscript $\beta$ in $\expval{O}_{\beta}$ is just used to denote that we are deconvolving with respect to noise $\mathcal{N}_\beta$, but remember that it correspond to the mitigated noise-free result. Clearly, when the observable to be measured is itself a Pauli matrix $O=\sigma_\gamma$, this further simplifies to
\begin{align}
\expval{\sigma_\gamma}_\beta  & = \frac{1}{2(1-2p)}~\sum_{\alpha=x,y,z}\underbrace{\Tr[\sigma_\gamma\sigma_\alpha]}_{=2\delta_{\gamma\alpha}}\big(1-2\delta_{\alpha \beta}\,p\big)\expval{\sigma_\alpha}_{\mathcal{N}_\beta(\rho)}\\ 
& = \frac{1-2\delta_{\gamma\beta}\,p}{1-2p}\expval{\sigma_\gamma}_{\mathcal{N}_\beta(\rho)}\, .
\end{align}

\subsection{Depolarizing channel}
The depolarizing noise channel is represented by the map
\begin{equation*}
    \mathcal{N}_{\text{dep}}(\rho) = \bigg(1-\frac{3p}{4}\bigg)\rho + \frac{p}{4}\bigg(\sigma_x \rho \sigma_x + \sigma_y \rho \sigma_y + \sigma_z \rho \sigma_z\bigg)\, ,
\end{equation*}
having Kraus operators $A_0 = \sqrt{1-3p/4}\,\mathbb{I}$, $A_1 = \sqrt{p}\,\sigma_x/2$, $A_2 = \sqrt{p}\,\sigma_y/2$, and $A_3 = \sqrt{p}\,\sigma_z/2$. By straightforward calculation, the Pauli Transfer Matrix of the depolarising channel as well as its inverse, amounts to
\begin{align}
    \Gamma_{\text{dep}} &= 
    \begin{bmatrix}
    1 & 0 & 0 & 0 \\
    0 & 1-p & 0 & 0 \\
    0 & 0 & 1-p & 0 \\
    0 & 0 & 0 & 1-p \\
    \end{bmatrix}\, ,\\
    \Gamma^{-1}_{\text{dep}} &= 
    \begin{bmatrix}
    1 & 0 & 0 & 0 \\
    0 & \frac{1}{1-p} & 0 & 0 \\
    0 & 0 & \frac{1}{1-p} & 0 \\
    0 & 0 & 0 & \frac{1}{1-p} \\
    \end{bmatrix}\, .
\end{align}
Following the same procedure used for the bit-flip channel, one arrives at the system of equations
\begin{equation}
    \begin{cases}
    1 = \beta_0 + \beta_1 + \beta_2 + \beta_3\\
    \frac{1}{1-p} = \beta_0 + \beta_1 - \beta_2 - \beta_3\\
    \frac{1}{1-p} = \beta_0 - \beta_1 + \beta_2 - \beta_3\\
    \frac{1}{1-p} = \beta_0 - \beta_1 - \beta_2 + \beta_3
    \end{cases}
\end{equation}
which has solutions $\beta_0 = (4-p)/4(1-p)$ and $\beta_1=\beta_2=\beta_3 = -p/4(1-p)$. Substituting these values in~\eqref{eq:general_map2}, and using the relation $2\Tr[O]\,\mathbb{I} = O + \sigma_x O \sigma_x + \sigma_y O \sigma_y + \sigma_z O \sigma_z$, one obtains 
\begin{equation}
    \mathcal{N}_{\text{depol}}^{-1}(O) = \frac{1}{1-p}\left( O-\frac{p}{2}\Tr[O]\mathbb{I} \right)\, .
\end{equation}
Plugging this in the tomographic deconvolution formula~\eqref{eq:DeconvolvedQubits}, leads to:
\begin{equation}
    \expval{O} = \frac{1}{2}\Tr[O] + \frac{1}{2}\sum_{\alpha}\frac{\Tr[O\sigma_\alpha]}{1-p}\expval{\sigma_\alpha}_{\mathcal{N}_{\text{dep}}(\rho)}\, ,
\end{equation}
from which it is clear that whenever a Pauli matrix is to be measured, $O=\sigma_k$, then the expectation values are contracted by a factor $1-p$, i.e. $\expval{\sigma_k} = \expval{\sigma_k}_\text{dep}/(1-p)$.

\subsection{General Pauli channel}
The most general channel involving only Pauli operators is the channel given by 
\begin{equation}
\mathcal{N}_{\bm{p}}(\rho) = p_0 \rho + p_x \xox{\rho} + p_y \yoy{\rho} + p_z \zoz{\rho}\, 
\end{equation}
characterized by probabilities $\boldsymbol{p}=(p_0, p_x, p_y, p_z)$, with the trace-preserving condition implying $p_0 = 1-p_x-p_y-p_z$. 
The PTM of this map is diagonal
\begin{align}
    \Gamma_{\bm{p}} = \text{diag}(& 1,\, p_0 + p_x - p_y -p_z, \nonumber\\
    &\,\, p_0 - p_x + p_y - p_z, \\
    &\,\, p_0 - p_x - p_y + p_z)\, ,\nonumber
\end{align}
and has trivial inverse 
\begin{align}
    \Gamma_{\bm{p}}^{-1} = 
   \text{diag}(& 1,\, (p_0 + p_x - p_y -p_z)^{-1}\nonumber \\
    &\,\, (p_0 - p_x + p_y - p_z)^{-1}, \\
    &\,\, (p_0 - p_x - p_y + p_z)^{-1})\nonumber\, .
\end{align}

Again, using the same procedure as before, one arrives at the system of equations: 
\begin{equation}
    \begin{cases}
    1 = \beta_0 + \beta_1 + \beta_2 + \beta_3\\
    \frac{1}{p_0 + p_x -p_y -p_z} = \beta_0 + \beta_1 - \beta_2 - \beta_3\\
    \frac{1}{p_0 - p_x +p_y -p_z} = \beta_0 - \beta_1 + \beta_2 - \beta_3\\
    \frac{1}{p_0 - p_x -p_y +p_z} = \beta_0 - \beta_1 - \beta_2 + \beta_3
    \end{cases}\, ,
\end{equation}
whose solution is reported in Eq.~\eqref{eq:inv_paulichannel} in the main text. The action of the inverse map on the Pauli matrix $\sigma_x$ is
\begin{align}
   \mathcal{N}_{\bm{p}}^{-1}(\sigma_x) & = \beta_0\sigma_x+\beta_1\xox{\sigma_x}+\beta_2\yoy{\sigma_x}+\beta_3\zoz{\sigma_x} \\
   & = (\beta_0 + \beta_1 - \beta_2 - \beta_3)\sigma_x \\
   & = \frac{1}{1-2(p_y+p_z)}\sigma_x\,,
\end{align}
and a similar expression also hold for $\sigma_y$ and $\sigma_z$, from which one obtains the deconvolution formulas in Eq.~\eqref{eq:deconvolution_gpc}.

\subsection{Amplitude Damping}
\label{app:AmplitudeDamping}

The amplitude damping channel is given by the map
\begin{equation}
\begin{aligned}
    & \mathcal{N}_{\text{AD}}(\rho) = K_0\rho K_0^\dagger + K_1\rho K_1^\dagger\, ,\\
    & K_0 = \begin{bmatrix}
    1 & 0 \\ 
    0 & \sqrt{1-\gamma}
    \end{bmatrix}\quad
    K_1 = \begin{bmatrix}
    0 & \sqrt{\gamma}\\
    0 & 0
    \end{bmatrix}\, .
\end{aligned}
\end{equation}
Differently from all the other cases treated above, this channel is not generated by coupled sigma matrices, and in addition one of its generators is not Hermitian. This has two consequences: first, we cannot straightforwardly apply the same eigenvalue matching procedure used above, second one must consider the adjoint channel when deconvolving. 

The PTM of the amplitude damping channel is
\begin{equation}
    \Gamma_{\text{AD}} = 
    \begin{bmatrix}
    1 & 0 & 0 & 0 \\
    0 & \sqrt{1-p} & 0 & 0 \\
    0 & 0 & \sqrt{1-p} & 0 \\
    p & 0 & 0 & 1-p
    \end{bmatrix}
\end{equation}
whose inverse is
\begin{equation}
    \Gamma_{\text{AD}}^{-1} = 
    \begin{bmatrix}
    1 & 0 & 0 & 0 \\
    0 & \frac{1}{\sqrt{1-p}} & 0 & 0 \\
    0 & 0 & \frac{1}{\sqrt{1-p}} & 0 \\
    \frac{-p}{1-p} & 0 & 0 & \frac{1}{1-p}
    \end{bmatrix}
\end{equation}
In this case the eigenvalues of $\Gamma_{\text{AD}}$ and $\Gamma_{\text{AD}}^{-1}$ are not the Pauli matrices, and so we cannot use the eigenvalue matching with the general map in~\eqref{eq:general_map}. However, the two PTMs have the same structure, so one may easily guess that the operator-sum representation of the two maps share the same operators, something that also always happened in all previous cases. 
Let us then suppose that the inverse map $\mathcal{N}_{\text{AD}}^{-1}$ has the form 
\begin{equation}
\label{eq:ad_mid_app}
    \mathcal{N}_{\text{AD}}^{-1}(\cdot) = \tilde{K}_0 \cdot \tilde{K}_0^{\dagger} - \tilde{K}_1 \cdot \tilde{K}_1^{\dagger}  
\end{equation}
with $\tilde{K}_0 = \dyad{0}+\kappa\dyad{1}$, and $\tilde{K_1} = \tau\ketbra{0}{1}$, with $\kappa\,,\tau$ free parameters to be determined. This map has corresponding PTM
\begin{equation}
    \Gamma(\kappa,\tau) = 
    \begin{bmatrix}
    \frac{1+\kappa^2-\tau^2}{2} & 0 & 0 & 0 \\
    0 & \kappa & 0 & 0 \\
    0 & 0 & \kappa & 0 \\
    \frac{1-\tau^2-\kappa^2}{2} & 0 & 0 & \frac{1+\tau^2+\kappa^2}{2}
    \end{bmatrix}\, ,
\end{equation}
and by requiring that $\Gamma(\kappa,\tau)=\Gamma_{\text{AD}}^{-1}$, we obtain
\begin{equation}
\kappa = \frac{1}{\sqrt{1-\gamma}}\,,\quad \tau = \sqrt{\frac{\gamma}{1-\gamma}}\,, 
\end{equation}
thus recovering the inverse map
\begin{equation}
\begin{aligned}
& \mathcal{N}_{\text{AD}}^{-1}(O) = \tilde{K}_0 O \tilde{K}_0^\dagger - \tilde{K}_1 O \tilde{K}_1^\dagger\,\quad
& \tilde{K}_0 = 
\begin{bmatrix}
1 & 0 \\
0 & \frac{1}{\sqrt{1-\gamma}}
\end{bmatrix}
\,,\,
\tilde{K}_1 = 
\begin{bmatrix}
0 & \sqrt{\frac{\gamma}{1-\gamma}} \\ 
0 & 0
\end{bmatrix}\, .
\end{aligned}
\end{equation}

In order to evaluate the deconvolution formula, we first need to calculate the adjoint of the inverse channel. Let $\Phi(\cdot)$ be a linear map acting on the space of operators $\mathcal{L}(\mathcal{H})$, its adjoint $\hat{\Phi}$ is defined as the unique map satisfying the following relation
\begin{equation}
\langle A, \Phi(B)\rangle = \langle \hat{\Phi}(A), B \rangle_{HS}\, . 
\end{equation}
where $\langle \cdot, \cdot \rangle_{HS}$ denotes the Hilbert-Schmidt inner product $\langle A, B\rangle \equiv \Tr[A^\dagger B]$. Let's consider a generic linear map of the form 
\begin{equation}
\label{eq:general_map_for_adjoint}
\Phi(A) = \sum_{k}\alpha_k\, V_k A V^\dagger_k\, , \quad  \alpha_k \in \mathbb{R}\, .
\end{equation}
which is, in general, neither CP nor TP, since we make no further hypothesis on $\alpha_k$ and $V_k$. By direct application of the definition of adjoint map, we obtain
\begin{align}
\langle A, \Phi(B)\rangle & \equiv \Tr[A^\dagger \Phi(B)] = \Tr\left[ A^\dagger \sum_k \alpha_k\, V_k B V_k^\dagger \right] = \Tr[\sum_k \alpha_k\,  V_k^\dagger A^\dagger V_k\, B]\\
& = \Tr \left[ \left(\sum_k \alpha_k V_k^\dagger A V_k\right)^\dagger B\right] = \left\langle \sum_k \alpha_k V_k^\dagger A V_k, B \right\rangle \\
& \quad \Rightarrow \hat{\Phi}(A) = \sum_k \alpha_k V_k^\dagger A V_k\, ,
\end{align}
where we used the linearity and cyclic property of the trace, as well as the fact that the coefficients are real, $\alpha_k^*=\alpha_k \in \mathbb{R}$. We see that for any map of the form~\eqref{eq:general_map_for_adjoint}, its adjoint is obtained by simply substituting the operators with their adjoint, i.e. $V_k \rightarrow V_k^\dagger$. If the map $\Phi(\cdot)$ leverages only Hermitian operators $V_k=V_k^\dagger$, as it happens with every Pauli noise channel, than the adjoint and the direct map of course coincides, $\hat{\Phi}(\cdot) = \Phi(\cdot)$. However, the Amplitude Channel uses non Hermitian generators $V_k$, thus has a non-trivial, yet simple, adjoint map.

Straightforward application of the deconvolution formula then leads to the deconvolved expectation values
\begin{equation}
\begin{aligned}
\expval{\sigma_x} & = \frac{1}{\sqrt{1-\gamma}}\langle \sigma_x \rangle_{\mathcal{N}_{\text{AD}}(\rho)}\,,\\
\expval{\sigma_y} & = \frac{1}{\sqrt{1-\gamma}}\langle \sigma_y \rangle_{\mathcal{N}_{\text{AD}}(\rho)}\,,\\
\expval{\sigma_z} & = \frac{1}{1-\gamma}\qty(\langle \sigma_z \rangle_{\mathcal{N}_{\text{AD}}(\rho)} -\gamma)\, .
\end{aligned}
\end{equation}

\subsection{2-Kraus channel}
\label{app:TwoKraus}
The set of channels considered here is generated by two parametrized Kraus operators
\begin{equation}
\mathcal{N}_{\text{two}}(\rho) = \sum_{i=1, 2}A_i\rho A_i^\dagger\, ,  
\end{equation}
with $A_1 = \cos\alpha\dyad{0}+\cos\beta\dyad{1}$, and $A_2 = \sin\beta\ketbra{0}{1}+\sin\alpha\ketbra{1}{0}$. The PTM of this channel and its inverse are respectively

\begin{align}
\Gamma_{\text{two}} & =
\begin{bmatrix}
1 & 0 & 0 & 0 \\
0 & \cos(\alpha-\beta) & 0 & 0 \\
0 & 0 & \cos(\alpha+\beta) & 0 \\
\frac{\cos(2\alpha)-\cos(2\beta)}{2} & 0 & 0 & \frac{\cos(2\alpha)+\cos(2\beta)}{2}
\end{bmatrix}, \label{eq:2k_ptm}\\
\Gamma_{\text{two}}^{-1} & =
\begin{bmatrix}
1 & 0 & 0 & 0 \\
0 & \frac{1}{\cos(\alpha-\beta)} & 0 & 0 \\
0 & 0 & \frac{1}{\cos(\alpha+\beta)} & 0 \\
\frac{\cos(2\beta)-\cos(2\alpha)}{\cos(2\alpha)+\cos(2\beta)} & 0 & 0 & \frac{2}{\cos(2\alpha)+\cos(2\beta)}
\end{bmatrix}\, .
\end{align}
Using the trigonometric relation 
\begin{align}
    \cos(2\alpha)+\cos(2\beta) &= 2\cos(\frac{2\alpha-2\beta}{2})\cos(\frac{2\alpha+2\beta}{2})\\
    &= 2\cos(\alpha-\beta)\cos(\alpha+\beta)
\end{align}
we can rewrite the elements of $\Gamma_{\text{two}}^{-1}$ as
\begin{align}
    (\Gamma_{\text{two}}^{-1})_{11} &= h_{\alpha\beta} \cos(\alpha+\beta)\,,\\
    (\Gamma_{\text{two}}^{-1})_{22} &= h_{\alpha\beta} \cos(\alpha-\beta)\,,\\
    (\Gamma_{\text{two}}^{-1})_{33} &= h^2_{\alpha\beta} \frac{\cos(2\alpha)+\cos(2\beta)}{2}\,,\\
    (\Gamma_{\text{two}}^{-1})_{30} &= h_{\alpha\beta}\frac{\cos(2\beta)-\cos(2\alpha)}{2}\,,
\end{align}
with $h_{\alpha\beta} = 2/(\cos(2\alpha)+\cos(2\beta))$. 

Expressed in this manner, these matrix elements are very similar to those in the Pauli transfer matrix of the direct channel $\Gamma_{\text{two}}$ in Eq.~\eqref{eq:2k_ptm}. The differences are in the presence of the pre-factor $h_{\alpha\beta}$, as well as in the signs of the angles in elements $(\Gamma_{\text{two}}^{-1})_{11}$ and $(\Gamma_{\text{two}}^{-1})_{22}$, and in the sign in the difference in element $(\Gamma_{\text{two}}^{-1})_{30}$.
This suggests that the operator-sum representation of the inverse map can be obtained starting from the direct one with some small changes, as it happened with the amplitude damping channel. First of all, we can multiply the Kraus operators by $\sqrt{h_{\alpha\beta}}$ to introduce the pre-factor, then, to account for the difference in elements $(\Gamma_{\text{two}}^{-1})_{11}$ and $(\Gamma_{\text{two}}^{-1})_{22}$, we can subtract the two operators instead of summing them. At last, element $(\Gamma_{\text{two}}^{-1})_{30}$ can be fixed by changing $\alpha \leftrightarrow \beta$ in the first Kraus operator $A_1$.  Incidentally, these changes also fix the $(\Gamma_{\text{two}}^{-1})_{33}$ element to the correct value. Thus, eventually, implementing these changes leads to the definition of the operators
\begin{align}
    & B_1  = \sqrt{h_{\alpha\beta}}\cos(\beta)\dyad{0} + \sqrt{h_{\alpha,\beta}}\cos(\alpha)\dyad{1}\,,\\
    & B_2  = \sqrt{h_{\alpha\beta}}\sin(\beta)\ketbra{0}{1}+\sqrt{h_{\alpha,\beta}}\sin(\alpha)\ketbra{1}{0}\,,\\
    & h_{\alpha\beta} = \frac{2}{\cos(2\alpha)+\cos(2\beta)}\, ,
\end{align}
to be used within the inverse map 
\begin{equation}
\mathcal{N}_{\text{two}}^{-1}(\cdot) = B_1 \cdot B_1^\dagger - B_2 \cdot B_2^\dagger\, .
\end{equation}
One can check that this map has the desired Pauli Transfer Matrix $\Gamma_{\text{two}}^{-1}$. As with the amplitude damping case, one the generators ($B_2$) is not Hermitian, thus one must be careful in considering the adjoint inverse map when evaluating the deconvolved mean values. By explicit calculations the following holds
\begin{align}
\expval{\sigma_x} & = \frac{1}{\cos(\alpha-\beta)}\expval{\sigma_x}_{\mathcal{N}_{\text{two}}(\rho)}\,,\nonumber\\
\expval{\sigma_y} & = \frac{1}{\cos(\alpha+\beta)}\expval{\sigma_y}_{\mathcal{N}_{\text{two}}(\rho)}\,,\\
\expval{\sigma_z} & = h_{\alpha\beta}\big(\cos^2(\beta)+\sin^2(\alpha)-1+\expval{\sigma_z}_{\mathcal{N}_{\text{two}}(\rho)}\big)\,.\nonumber
\end{align}

\end{appendix}


\end{document}